%% file: main.tex
\documentclass[10pt,letterpaper]{article}
\input{preamble.tex}

\title{Where to weight? 
Estimating population causal effects with weighted double score matching in complex surveys}
\author{Yukang Zeng\textsuperscript{1,2,3},
Guangyu Tong\textsuperscript{1,2,3},
Jiaqi Tong\textsuperscript{1,3},\\
Haidong Lu\textsuperscript{4,5},
Bhramar Mukherjee\textsuperscript{1,5},
Fan Li\textsuperscript{1,2,3}\thanks{Corresponding author: Fan Li, PhD, Department of Internal Medicine and Biostatistics, Yale School of Medicine and Yale School of Public Health, New Haven, Connecticut, USA. Email: \href{mailto:fan.f.li@yale.edu}{fan.f.li@yale.edu}.}\\
\small\textsuperscript{1}Department of Biostatistics, Yale School of Public Health, Connecticut, United States\\
\small\textsuperscript{2}Section of Cardiovascular Medicine, Department of Internal Medicine{,\ }\\
\small Yale School of Medicine, Connecticut, United States\\
\small\textsuperscript{3}Center for Methods in Implementation and Prevention Science{,\ }\\
\small Yale School of Public Health, Connecticut, United States\\
\small\textsuperscript{4}Section of General Internal Medicine, Department of Internal Medicine{,\ }\\
\small Yale School of Medicine, Connecticut, United States\\
\small\textsuperscript{5}Department of Chronic Disease Epidemiology, Yale School of Public Health, Connecticut, United States}
\date{\small Preprint. This manuscript has not been peer reviewed.}
\begin{document}
\maketitle
\begin{abstract}
Double score matching is an emerging method for causal inference that combines propensity and prognostic scores to achieve dimension reduction and model robustness. However, when the observational data are drawn from a complex survey, the existing literature does not provide sufficient guidance on the proper use of survey weights in matching strategies for population-level causal inference. We develop weighted double score matching (WDSM) that provides a unified framework for incorporating survey weights throughout the score estimation, matching, and outcome analysis under both retrospective (informative) and prospective (non-informative) sampling designs. We establish asymptotic normality of the resultant effect estimates under both known and estimated scores and develop a linearization-based bootstrap procedure for variance estimation that preserves the survey-weighted matching structure.  
Simulation studies show that WDSM substantially reduces bias relative to unweighted matching when either the propensity score model or the prognostic score models are correctly specified. Under correct specification, WDSM also improves efficiency and confidence interval coverage relative to survey-weighted propensity score matching, with particularly pronounced gains under poor overlap.
Applications to early marijuana use and mental health in the National Survey on Drug Use and Health and to sleep duration and glycemic control in the National Health and Nutrition Examination Survey illustrate the practical value of WDSM for population-level inference.
We conclude with a discussion of our findings and provide practical guidance on incorporating survey weights into matching-based analyses of observational data.

\end{abstract}
\medskip
\section{Introduction}
\label{sec:intro}

Observational data are frequently used to estimate the causal effects of treatments across medical and social science applications, particularly when randomized trials are infeasible. One key limitation of observational data is the lack of randomization, which can introduce systematic differences in baseline characteristics between treatment conditions and may bias the estimated treatment effect.\cite{Rosenbaum2002} The propensity score, defined as the probability of receiving treatment given observed covariates,\cite{Li2018, rosenbaum1983central} is an important device to address baseline confounding in observational studies. The propensity score can be incorporated to improve baseline balance through three main approaches, including matching,\cite{Stuart2010} weighting,\cite{Li2018} and stratification,\cite{Lunceford2004} all of which help form comparable exposed and unexposed groups to reduce bias in the estimated treatment effect.

Large-scale observational studies often use complex survey designs involving stratification, cluster sampling, and deliberate oversampling of selected population subgroups.\cite{LaVange2001Applying,lumley2011complex} National surveys such as the National Health and Nutrition Examination Survey (NHANES) and the National Survey on Drug Use and Health (NSDUH) are frequently used to study the effects of exposures and health behaviors. Survey weights account for unequal selection probabilities and nonresponse, allowing analyses to target the population represented by the survey. 
The proper incorporation of survey weights is especially critical in retrospective sampling designs, where survey weights may depend on both exposure status and covariates.\cite{Ridgeway2015,salerno2025whatsweight} The need to combine complex survey designs with propensity score methods for valid population-level causal inference has motivated substantial methodological development. \cite{ZanuttoLuHornik2005,Zanutto2006,DuGoff2014,Ridgeway2015,Yang2023,salerno2025whatsweight,zeng2025movingbestpracticeusing} For example, previous work by Zanutto et al.\cite{ZanuttoLuHornik2005,Zanutto2006} adapted propensity score subclassification to complex survey data, recommending unweighted propensity score estimation followed by survey-weighted within-subclass outcome regression. DuGoff et al.\cite{DuGoff2014} broadened the investigation to cover matching, subclassification, and weighting; they recommended that survey weights should be included as predictors in propensity score models and incorporated into outcome analysis. For propensity score weighting, Ridgeway et al.\cite{Ridgeway2015} recommended incorporating survey weights at both the propensity score estimation and outcome analysis stages, and Yang et al.\cite{Yang2023} compared alternative strategies for binary outcomes. Salerno et al.\cite{salerno2025whatsweight} developed identification and estimation methods for average controlled differences when survey weights vary with the group variable under comparison. Zeng et al.\cite{zeng2025movingbestpracticeusing} recently embedded survey weights for weighting and augmented weighting estimators, under both retrospective and prospective sampling designs, reinforcing the need to include survey weights in multiple nuisance models for valid estimation and inference for a class of weighted average treatment effect estimands. 

Despite the aforementioned developments, guidance for propensity score matching in complex survey studies has been far less consistent because two distinct weighting decisions are involved. Through simulation studies, Austin et al.\cite{austin2018propensity} compared an unweighted model, a survey-weighted model, and an unweighted model that included the survey weight as an additional covariate, for propensity score estimation. The choice among these specifications had little effect on covariate balance, while their relative performance in treatment effect estimation varied across simulation settings. Lenis et al.\cite{Lenis2019} extended this comparison to different nonresponse mechanisms and similarly found little difference in bias across these approaches, as long as adequate population-level balance in confounders had been achieved between the treated and comparison groups. Beyond propensity score estimation, matching raises the additional question of which survey weights should be assigned to matched comparison units after matches are formed. Austin et al. found that retaining the comparison units' original survey weights, referred to as the \emph{natural weights}, generally resulted in better covariate balance and lower bias. Lenis et al., however, provided theoretical justification for transferring each treated unit's survey weight to its matched comparison unit when estimating the population average treatment effect on the treated, so that the weighted covariate distribution among comparison units represents that of the treated population. Their simulations further showed that such weight transfer can be beneficial when nonresponse depends on the treatment status. These prior findings, however, leave two gaps. First, there is no general guidance on how survey weights should be incorporated when constructing matches and estimating treatment effects from the matched sample. Second, existing work has largely examined propensity score matching under prospective sampling, leaving other matching methods and informative retrospective sampling, in which selection depends on both covariates and treatment status, less well understood.

Among different matching methods, double score matching (DSM) has emerged as a powerful extension of single-score matching by combining propensity and prognostic scores, which summarize covariate information related to treatment assignment and potential outcomes, respectively, to achieve both improved efficiency and double robustness against model misspecification. Leacy and Stuart\cite{LeacyStuart2014} first demonstrated the empirical benefits of matching jointly on propensity and prognostic scores, showing that the use of both scores improves covariate balance and reduces bias compared to propensity score matching alone. Antonelli et al.\cite{Antonelli2018} provided theoretical foundations by establishing consistency and convergence rates under partial model misspecification, proving that DSM leads to a consistent average causal effect estimator when either the propensity score model or the prognostic score model is correctly specified. Yang and Zhang\cite{yang_multiply_2023} formalized DSM within a complete asymptotic framework that includes multiply robust matching and bias-corrected estimation, and demonstrated that matching on treatment-specific double scores achieves bias of order $O(n^{-1/2})$ that is comparable to parametric rates. These developments provide a foundation for robust matching-based estimation, but do not address how survey weights should enter double score construction, matched-set imputation, and variance estimation for population-level inference.

We address these gaps by proposing weighted double score matching (WDSM) for observational data from complex surveys. First, we study where and how survey weights should be incorporated into score estimation, matching, and outcome analysis under retrospective (informative) and prospective (non-informative) sampling designs. We focus on the population average treatment effect (PATE) and extend the approach to the population average treatment effect on the treated (PATT), thus covering two essential estimands of interest in survey observational data. 
Second, under the corresponding regularity conditions, we establish consistency and asymptotic normality of the WDSM estimators and show that they retain double robustness when either the propensity score or prognostic score model is correctly specified. Third, we develop a linearization-based bootstrap procedure for variance estimation that preserves the survey-weighted matching structure.
Finally, we evaluate the methods across overlap conditions, model specifications, and sampling designs, and illustrate their use in two applications with national survey data.

The remainder of this paper is organized as follows. Section~\ref{sec: notation and assumption} introduces the notation and assumptions under the potential outcomes framework in the context of survey observational data, distinguishing retrospective from prospective sampling designs. Section~\ref{sec:method} develops the WDSM estimator for PATE through survey-weighted imputation within match sets, derives the bias decomposition and asymptotic theory under both known and estimated scores, and presents the bootstrap variance estimation procedure. Section~\ref{sec:PATT extension} extends the framework to PATT. Section~\ref{sec:simulation} evaluates the performance of different matching estimators incorporating survey weights across varying overlap conditions and sampling designs through simulations. Section~\ref{sec:case studies} applies the proposed methods to two real-world data examples, investigating whether marijuana initiation before age 15 affects the population prevalence of any mental illness and serious mental illness among U.S. young adults using the 2023 NSDUH, and whether short sleep duration affects glycated hemoglobin levels among U.S. adults aged 40--70 without a reported diabetes diagnosis using the 2015--2018 NHANES. Section~\ref{sec:discussion} concludes.

\section{Notation and Assumptions}\label{sec: notation and assumption}

\subsection{Setup and estimands}

Consider an independent and identically distributed sample of size $N$ from a target superpopulation $\Omega$ with law $P_\Omega$, where the generic full-data vector is $\{Y(0),Y(1),\mathbf{X},Z,S\}$ and $S=1$ indicates selection into the survey sample. Let $n$ denote the number of selected units, indexed by $i=1,\ldots,n$. 
For each selected individual, we observe a $d$-dimensional vector of pre-treatment covariates $\mathbf{X}_i \in \mathcal{X} \subseteq \mathbb{R}^d$ and a binary treatment indicator $Z_i \in \{0,1\}$, with $Z_i = 1$ for the treated group and $Z_i = 0$ for the control group. We use \emph{treatment} more generically; depending on the application, $Z_i$ may represent an intervention, exposure, behavior, or group membership (e.g., a demographic subgroup). Under the Stable Unit Treatment Value Assumption (SUTVA), the potential outcomes $Y_i(z)$ for $z \in \{0,1\}$ are well-defined, and the observed outcome for each individual is $Y_i = Z_i Y_i(1) + (1-Z_i) Y_i(0)$. Conditional on $n$, the observed sample $\{(\mathbf{X}_i,Z_i,Y_i)\}_{i=1}^n$ consists of independent and identically distributed draws from the selected-sample law $P_s(\cdot)=P_\Omega(\cdot\mid S=1)$, with treatment-specific sample sizes $n_1=\sum_{i=1}^n Z_i$ and $n_0=\sum_{i=1}^n(1-Z_i)$.
We denote the population-level propensity score by $e(\mathbf{X}_i) = P(Z_i = 1 \mid \mathbf{X}_i)$, the outcome mean functions $\mu_z(\mathbf{X}_i) = E[Y_i(z) \mid \mathbf{X}_i]$ for $z \in \{0,1\}$, the conditional average treatment effect $\tau(\mathbf X_i)=\mu_1(\mathbf X_i)-\mu_0(\mathbf X_i)$, and the conditional variance function $\sigma_z^2(\mathbf{X}_i) = \operatorname{Var}(Y_i(z) \mid \mathbf{X}_i)$ for $z \in \{0,1\}$, where all expectations are taken under $P_\Omega$.

Our primary causal estimands are population-level causal parameters: the population average treatment effect (PATE) $\tau_{\text{PATE}} = E[Y_i(1) - Y_i(0)]$ and the population average treatment effect among the treated (PATT) $\tau_{\text{PATT}} = E[Y_i(1) - Y_i(0) \mid Z_i = 1]$. These differ from their survey sample analogs, and recovering them from the survey sample requires both causal identification conditions and proper incorporation of survey weights.\cite{DuGoff2014,Ridgeway2015} For causal identification, we assume \textit{treatment unconfoundedness}: $(Y_i(0), Y_i(1)) \perp Z_i \mid \mathbf{X}_i$, 
and \textit{strong positivity}: there exists a constant $c_e\in(0,1/2]$ such that $c_e\le e(\mathbf{X}_i)\le1-c_e$ almost surely.  
For PATT estimation, unconfoundedness can be relaxed to the one-sided condition $Y_i(0) \perp Z_i \mid \mathbf{X}_i$, 
and positivity to the one-sided condition that there exists a constant $c_{e,0}>0$ such that $1-e(\mathbf{X}_i)\ge c_{e,0}$ almost surely in the treated population.\cite{Heckman1997,yang_multiply_2023}
The conditions for the survey weight are developed in Section~\ref{subsec:survey_design}. Below, we present our methodology for estimating PATE under retrospective (informative) sampling as the general case and discuss extensions to PATT in Section~\ref{sec:PATT extension}. The developments under prospective (non-informative) sampling are provided in Web Appendix A.7 and A.8.

\subsection{Survey designs and survey weights} \label{subsec:survey_design}

\begin{figure}[htbp]
    \centering
    \includegraphics[width=0.65\textwidth]{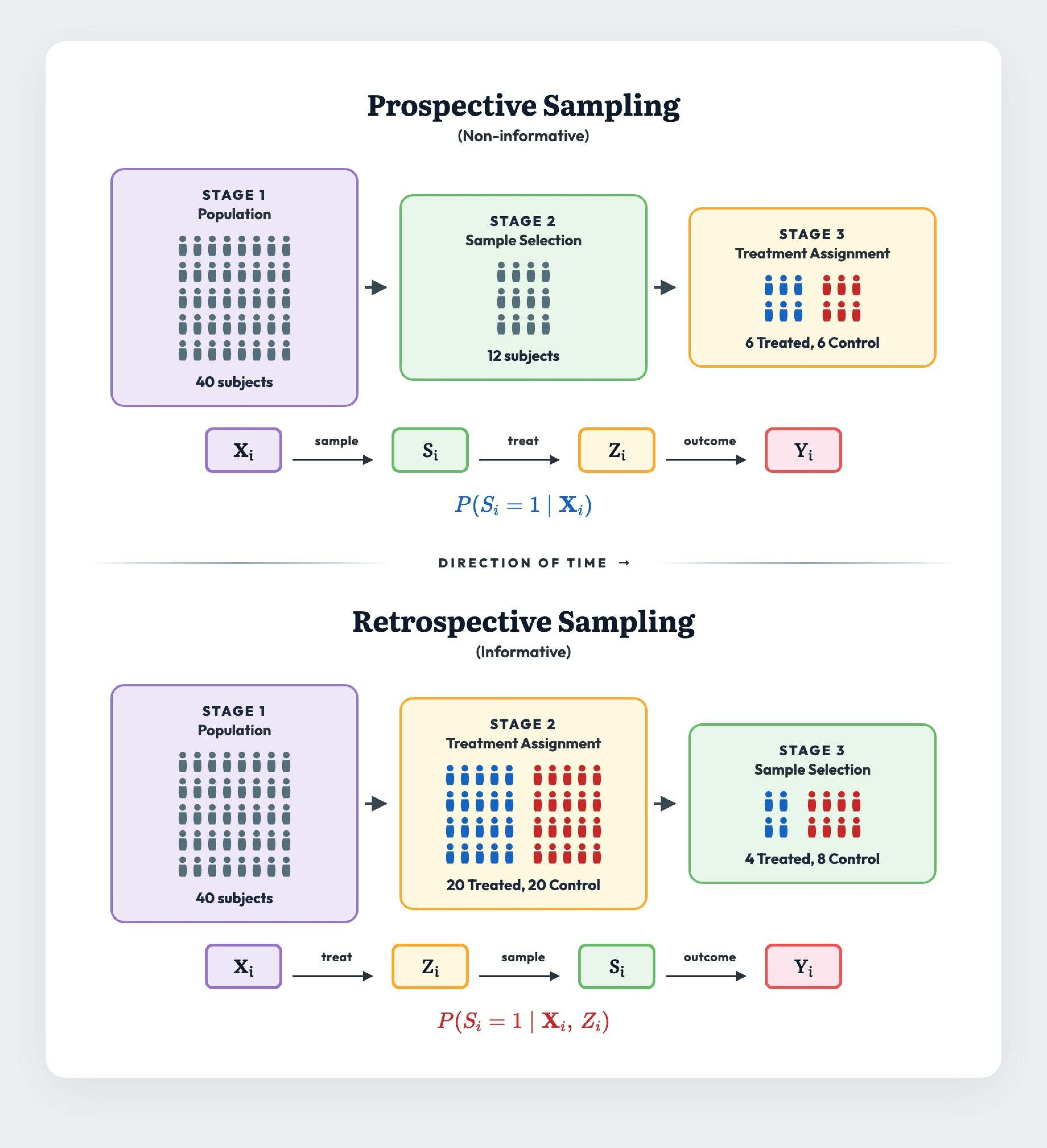}
    \caption{Prospective and retrospective sampling designs. The temporal ordering of sampling relative to when treatment (or exposure) status is realized determines the sampling probability structure and the corresponding survey weight requirements.}
    \label{fig:sampling_diagram}
\end{figure}

Population-level causal inference requires accounting for the mechanism that determines survey inclusion. 
Following Zeng et al.\cite{zeng2025movingbestpracticeusing} and Salerno et al.,\cite{salerno2025whatsweight} we distinguish retrospective and prospective frameworks according to whether survey selection may depend on treatment after conditioning on the covariates. We present the retrospective framework as the general case and the prospective framework as a special case, as illustrated in Figure~\ref{fig:sampling_diagram}. Retrospective designs arise when treatment (or exposure) status is realized before sampling and the sampling mechanism depends on both treatment status and covariates. Examples include surveys where the survey weights are tied to the comparison variable,\cite{salerno2025whatsweight} such as NHANES oversampling by race when the comparison of interest is across racial subgroups, and surveys whose design weights carry features associated with the treatment variable.\cite{DuGoff2014} To identify population-level treatment effects under this design, we require the following conditions on the sampling mechanism.

\begin{assumption}[Retrospective Survey Ignorability]
\label{ass:retro_ignorability}
Under retrospective sampling: (a) sampling probabilities depend on both covariates and treatment status, $\pi_z(\mathbf{x}_i) = P(S_i = 1 \mid \mathbf{X}_i = \mathbf{x}_i, Z_i = z)$ for $z \in \{0,1\}$, and we denote the treatment-specific survey weights as $w_z(\mathbf{x}_i) = 1/\pi_z(\mathbf{x}_i)$, which are known from the survey design; (b) strong survey positivity holds: there exists a constant $c_\pi\in(0,1/2]$ such that $c_\pi\leq\pi_z(\mathbf{X}_i)\leq1-c_\pi$ almost surely for $z\in\{0,1\}$; 
and (c) selection is conditionally independent of potential outcomes given covariates and treatment, $\{Y_i(0), Y_i(1)\} \perp S_i \mid \mathbf{X}_i, Z_i$.
\end{assumption}

Assumption~\ref{ass:retro_ignorability}(c) excludes outcome-dependent sampling, such as case-control designs in which selection depends directly on the realized outcome $Y_i$ (for example, oversampling respondents with a mental-health diagnosis when that diagnosis is the outcome). 
Because the sampling probabilities may differ across treatment groups for individuals with the same covariates, treatment-specific weights are necessary to recover population quantities from the survey sample.
In contrast, prospective designs arise when survey selection takes place before treatment (or exposure) status is realized. 
In this case, the sampling mechanism depends only on pre-treatment covariates. 
This setting can be viewed as a generalizability problem with known survey weights.\cite{dahabreh2020extending,stuart2011use}
For completeness, we also list the following condition required for identifying population-level treatment effects under the prospective design. 

\begin{assumption}[Prospective Survey Ignorability]
\label{ass:prosp_ignorability}
Under prospective sampling: (a) sampling probabilities depend only on pre-treatment covariates, $\pi(\mathbf{x}_i) = P(S_i = 1 \mid \mathbf{X}_i = \mathbf{x}_i)$, and we denote the common survey weights as $w(\mathbf{x}_i) = 1/\pi(\mathbf{x}_i)$, which are known from the survey design; (b) strong survey positivity holds: there exists a constant $c_\pi\in(0,1/2]$ such that $c_\pi\leq\pi(\mathbf{X}_i)\leq1-c_\pi$ almost surely; 
and (c) selection is conditionally independent of potential outcomes given covariates, $\{Y_i(0), Y_i(1)\} \perp S_i \mid \mathbf{X}_i$.
\end{assumption}

For prospective PATE, we further assume $Z_i \perp \{Y_i(0), Y_i(1), S_i\} \mid \mathbf{X}_i$.\cite{stuart2011use}
This treatment assignment condition implies $\pi_1(\mathbf{x}_i) = \pi_0(\mathbf{x}_i) = \pi(\mathbf{x}_i)$ and that the sample-level propensity score coincides with the population-level propensity score defined in Section~\ref{sec: notation and assumption}, $P(Z_i = 1 \mid \mathbf{X}_i, S_i = 1) = P(Z_i = 1 \mid \mathbf{X}_i) = e(\mathbf{X}_i)$.

\subsection{Constructing the double score and balancing properties}

Following Yang and Zhang,\cite{yang_multiply_2023} we construct treatment-specific matching variables using population-level scores. We define the prognostic score $\Psi_z(\mathbf{X}_i)$ as a sufficient statistic for $Y_i(z)$ in the sense that $Y_i(z) \perp \mathbf{X}_i \mid \Psi_z(\mathbf{X}_i)$. We use the mean score specification $\Psi_z(\mathbf{X}_i)=\mu_z(\mathbf{X}_i)$, which is a valid prognostic score whenever the conditional distribution of $Y_i(z)$ given $\mathbf{X}_i$ depends on $\mathbf{X}_i$ only through $\mu_z(\mathbf{X}_i)$, as occurs under a location-shift family with conditional density $f_z\{y-\mu_z(\mathbf{X}_i)\}$.\cite{Hansen2008,yang_multiply_2023} 
Combining the population-level propensity score and arm-specific prognostic scores, we define the treatment-specific double score for PATE as
\[
\mathbf{D}_z(\mathbf{X}_i) = \left(e(\mathbf{X}_i),\, \Psi_z(\mathbf{X}_i)\right) \in \mathbb{R}^2, \qquad z \in \{0,1\},
\]
and for PATT only the one-sided control score $\mathbf{D}_0(\mathbf{X}_i) = (e(\mathbf{X}_i),\, \Psi_0(\mathbf{X}_i))$ is required. Rather than matching on potentially high-dimensional full covariates, the treatment-specific double score reduces the matching dimension to two.

Matching on the double score is justified by the following balancing property, established by Antonelli et al.\cite{Antonelli2018} and refined by Yang and Zhang\cite{yang_multiply_2023} for treatment-specific scores. 

\begin{lemma}[Treatment-Specific Double Score Balancing Properties; Antonelli et al.\cite{Antonelli2018}, Yang and Zhang\cite{yang_multiply_2023}]\label{lem:ds_balancing}
Under treatment unconfoundedness $\{Y_i(0), Y_i(1)\} \perp Z_i \mid \mathbf{X}_i$, we have (a) $Y_i(z) \perp Z_i \mid \{h(\mathbf{X}_i),\, \Psi_z(\mathbf{X}_i)\}$, and (b) $Y_i(z) \perp Z_i \mid \{e(\mathbf{X}_i),\, h(\mathbf{X}_i)\}$ for any $h(\mathbf{X}_i)$ and $z \in \{0,1\}$. Together (a) and (b) yield (c) $Y_i(z) \perp Z_i \mid \mathbf{D}_z(\mathbf{X}_i)$ for $z \in \{0,1\}$. A double score containing either the population propensity score or the conditional mean $\mu_z(\mathbf{X}_i)$ satisfies\[E\{Y_i(z)\mid Z_i,\mathbf D_z(\mathbf X_i)\}=E\{Y_i(z)\mid\mathbf D_z(\mathbf X_i)\}.\]For PATT, under the one-sided condition $Y_i(0) \perp Z_i \mid \mathbf{X}_i$, similarly we have $Y_i(0) \perp Z_i \mid \mathbf{D}_0(\mathbf{X}_i)$.
\end{lemma}

Lemma~\ref{lem:ds_balancing}(c) implies that confounding due to $\mathbf{X}_i$ is removed within strata defined by the two-dimensional score $\mathbf{D}_z(\mathbf{X}_i)$, and hence matching on the double score achieves the treatment-specific outcome balance required for identification, with $E[Y_i(z)] = E\{E[Y_i \mid Z_i = z, \mathbf{D}_z(\mathbf{X}_i)]\}$ for $z \in \{0,1\}$, so that each individual's counterfactual outcome can be imputed from matched individuals with the same score value in the opposite arm. In survey observational data, the population-level scores must be properly estimated from the survey sample, and the matched-set imputation should ideally incorporate survey weights for valid causal inference. This motivates our development of the survey-weighted matching estimator in Section~\ref{sec:method}. For brevity of the main manuscript, additional regularity conditions on the double score are detailed in Web Appendix A.2.4.

\section{Weighted Double Score Matching for Survey Observational Studies}\label{sec:method} 

This section develops the WDSM methodology for retrospective PATE as the general case. Extensions to PATT are provided in Section~\ref{sec:PATT extension}, and prospective sampling counterparts are collected in Web Appendix A.7 and A.8.

\subsection{Double score model specification}\label{subsec:score_models}

In practice, the population-level double scores $\mathbf{D}_z(\mathbf{X}_i) = (e(\mathbf{X}_i), \Psi_z(\mathbf{X}_i))$ defined in Section~\ref{sec: notation and assumption} are unknown and must be estimated from the observed data. 
For notational simplicity, the model design vectors in the linear predictors and estimating equations are also denoted by $\mathbf X$ and include an intercept and the specified covariate terms.
We consider a logistic regression model for the propensity score, $e(\mathbf{X}; \boldsymbol{\alpha}) = \operatorname{expit}(\mathbf{X}^T \boldsymbol{\alpha})$, and linear regression models for the prognostic scores, $\Psi_z(\mathbf{X}; \boldsymbol{\beta}_z) = \mathbf{X}^T \boldsymbol{\beta}_z$ for $z \in \{0,1\}$, though flexible semiparametric specifications can also be accommodated. We collect all score parameters into $\boldsymbol{\theta} = (\boldsymbol{\alpha}^T, \boldsymbol{\beta}_0^T, \boldsymbol{\beta}_1^T)^T$. A key question in the survey setting is whether the estimation of each score component requires survey weights, which depends on the sampling design as follows.

Under retrospective sampling, the sample-level propensity score $P(Z_i = 1 \mid \mathbf{X}_i, S_i = 1)$ differs from the population propensity score $e(\mathbf{X}_i)$ due to treatment-dependent sampling (Assumption~\ref{ass:retro_ignorability}), and survey-weighted estimation is required to recover the population propensity score; also see Zeng et al. for a justification in the propensity score weighting context.\cite{zeng2025movingbestpracticeusing} The estimator $\widehat{\boldsymbol{\alpha}}$ solves the survey-weighted estimating equation
\begin{align}\label{eq:ps_ee}
\sum_{i=1}^n w_{Z_i}(\mathbf{X}_i)\, \mathbf{X}_i \left\{Z_i - e(\mathbf{X}_i; \boldsymbol{\alpha})\right\} = \mathbf{0}.
\end{align}
In contrast, for prognostic score estimation, Assumption~\ref{ass:retro_ignorability}(c) and treatment unconfoundedness together imply $E[Y_i \mid \mathbf{X}_i, Z_i = z, S_i = 1] = E[Y_i \mid \mathbf{X}_i, Z_i = z] = \Psi_z(\mathbf{X}_i)$, so the population prognostic score can be consistently estimated directly from the observed sample data within each treatment arm without survey weights. The estimator $\widehat{\boldsymbol{\beta}}_z$ solves
\begin{align}\label{eq:pgs_ee}
\sum_{i=1}^n \mathbf{1}(Z_i=z)\mathbf{X}_i \left\{Y_i - \Psi_z(\mathbf{X}_i; \boldsymbol{\beta}_z)\right\} = \mathbf{0}.
\end{align}
Under the prospective sampling and treatment assignment conditions in Section~\ref{subsec:survey_design}, the sample-level and population-level propensity scores coincide, and both score components can be estimated without survey weights. 

\subsection{Constructing the WDSM estimator}\label{subsec:simple estimator}

Given the estimated double scores $\widehat{\mathbf{D}}_z(\mathbf{X}_i) = \left(\widehat{e}(\mathbf{X}_i), \widehat{\Psi}_z(\mathbf{X}_i)\right)$ obtained in Section~\ref{subsec:score_models}, we construct the WDSM estimator through matching, survey-weighted imputation, and Hájek-normalized aggregation. Following Yang and Zhang,\cite{yang_multiply_2023} we estimate $\tau_{\mathrm{PATE}} = E[Y(1)] - E[Y(0)]$ by estimating $E[Y(z)]$ separately for $z = 0, 1$, aggregating potential outcomes across individuals in both treatment groups. 
For each individual $i$, if $Z_i = z$, the potential outcome $Y_i(z) = Y_i$ is directly observed; if $Z_i = 1-z$, the potential outcome $Y_i(z)$ is not observed and needs to be imputed from individuals in arm $z$ with similar covariates, which is achieved by matching on the arm-specific double score $\widehat{\mathbf{D}}_z(\mathbf{X})$, since Lemma~\ref{lem:ds_balancing}(c) and SUTVA jointly imply $E[Y(z) \mid Z = 1-z, \mathbf{D}_z(\mathbf{X}_i)] = E[Y \mid Z = z, \mathbf{D}_z(\mathbf{X}_i)]$.

For each individual $i$ with $Z_i\neq z$, we select the $M$ nearest neighbors from arm $z$ using Euclidean distance in the standardized double score space $\widehat{\mathbf D}_z(\mathbf X)$. Each score component is standardized to mean zero and unit variance for numerical stability.\cite{abadie2006} We denote the matched index set by $\mathcal J_{\widehat{\mathbf D}_z}(i)$. Matching is performed with replacement, with $M$ fixed.
In practice, we recommend assessing sensitivity to a small, prespecified set of matching ratios, such as $M=1,3,5$, alongside covariate balance and matching distances. 
The observed or imputed value of $Y_i(z)$ for each individual $i$ is therefore given by
\begin{align}\label{eq:imputation}
\widehat{Y}_i^{R}(z) =
\begin{cases}
Y_i, & Z_i = z, \\[2mm]
\displaystyle \frac{\sum_{j \in \mathcal{J}_{\widehat{\mathbf{D}}_z}(i)} w_z(\mathbf{X}_j)\, Y_j}{\sum_{j \in \mathcal{J}_{\widehat{\mathbf{D}}_z}(i)} w_z(\mathbf{X}_j)}, & Z_i \neq z,
\end{cases}
\end{align}
where the superscript $R$ denotes retrospective sampling and the survey weights $w_z(\mathbf{X}_j)$ are those of the matched individuals from arm $z$. 
In standard unweighted matching, each matched individual contributes equally with weight $1/M$.\cite{abadie2006,yang_multiply_2023} 
In survey observational data, \eqref{eq:imputation} weights matched individuals in proportion to their survey weights, reflecting the number of individuals they represent in the target population.

Once both potential outcomes $\widehat{Y}_i^{R}(1)$ and $\widehat{Y}_i^{R}(0)$ are obtained for all individuals, the retrospective PATE estimator aggregates the arm-specific imputed means through Hájek normalization,
\begin{align}\label{eq:pate_estimator}
\widehat{\tau}_{\mathrm{PATE}}^{R} = \widehat{\mu}_1^{R} - \widehat{\mu}_0^{R} = \frac{\sum_{i=1}^n w_{Z_i}(\mathbf{X}_i) \{\widehat{Y}_i^{R}(1) - \widehat{Y}_i^{R}(0)\}}{\sum_{i=1}^n w_{Z_i}(\mathbf{X}_i)} ,
\end{align}
where $\widehat{\mu}_z^{R} = \sum_{i=1}^n w_{Z_i}(\mathbf{X}_i) \widehat{Y}_i^{R}(z) / \sum_{i=1}^n w_{Z_i}(\mathbf{X}_i)$ estimates each arm-specific potential outcome mean separately.\cite{yang_multiply_2023}

To study the asymptotic properties of $\widehat{\tau}_{\mathrm{PATE}}^{R}$, we rewrite the estimator in terms of matching weights that express each observed outcome's total contribution. In standard unweighted matching, the reuse structure is summarized by the count 
$K_M(i)=\sum_{l:Z_l\neq Z_i}\mathbf{1}\{i\in\mathcal{J}_{\widehat{\mathbf D}_{Z_i}}(l)\}$ 
recording the number of times individual $i$ is used as a match, and the total matching weight on $Y_i$ is $(2Z_i - 1)(1 + M^{-1}K_M(i))$.\cite{abadie2006,yang_multiply_2023} Under survey-weighted matching, equal within-match contributions are replaced by normalized survey weights, and the reuse count generalizes to the survey-weighted reuse coefficient; that is, each time individual $i$ in arm $z$ is used as a match for a target individual $l$ in arm $1-z$, individual $i$ contributes its normalized survey weight $w_z(\mathbf{X}_i)/\sum_{k \in \mathcal{J}_{\widehat{\mathbf{D}}_z}(l)} w_z(\mathbf{X}_k)$ to the imputation of $Y_l(z)$, weighted by the target individual's survey weight $w_{1-z}(\mathbf{X}_l)$. For estimating PATE under the retrospective design, the survey-weighted reuse coefficient for individual $i$ in arm $z$ is
\begin{align}\label{eq:reuse_freq}
K_z^{R,\mathrm{PATE}}(i)
=
\sum_{l:Z_l=1-z}
\frac{w_{1-z}(\mathbf{X}_l)w_z(\mathbf{X}_i)\,\mathbf{1}\left\{i\in\mathcal{J}_{\widehat{\mathbf{D}}_z}(l)\right\}}
{\sum_{k\in\mathcal{J}_{\widehat{\mathbf{D}}_z}(l)} w_z(\mathbf{X}_k)}, \qquad z \in \{0,1\}.
\end{align}
The arm-specific imputed mean hence admits the matching weight representation
\begin{align}\label{eq:arm_weight_rep}
\widehat{\mu}_z^{R}
=
\frac{\sum_{i:Z_i=z}\left\{w_z(\mathbf{X}_i)+K_z^{R,\mathrm{PATE}}(i)\right\}Y_i}{\sum_{i=1}^n w_{Z_i}(\mathbf{X}_i)}, \qquad z \in \{0,1\},
\end{align}
and the final WDSM estimator can be written compactly as
\begin{align}\label{eq:matching_weight_rep}
\widehat{\tau}_{\mathrm{PATE}}^{R} = \widehat{\mu}_1^{R} - \widehat{\mu}_0^{R}
=
\frac{\sum_{i=1}^n (2Z_i - 1)\left\{w_{Z_i}(\mathbf{X}_i)+K_{Z_i}^{R,\mathrm{PATE}}(i)\right\}Y_i}{\sum_{i=1}^n w_{Z_i}(\mathbf{X}_i)}.
\end{align}
This matching weight representation parallels the expressions in Abadie and Imbens\cite{abadie2006} and Yang and Zhang,\cite{yang_multiply_2023} and provides the foundation for the bias decomposition in Section~\ref{subsec:decomposition and bias correction}. The PATT counterpart is given in Section~\ref{sec:PATT extension}, and the counterparts under the prospective design are collected in Web Appendix A.7 and A.8.

\subsection{Bias decomposition and correction}\label{subsec:decomposition and bias correction}

Following Abadie and Imbens\cite{abadie2006} and Yang and Zhang,\cite{yang_multiply_2023} we decompose the estimation error of $\widehat{\tau}_{\mathrm{PATE}}^{R}$ into three components that isolate the sources of variability and bias, in the presence of survey weights. Let $\boldsymbol{\theta}^*$ denote the probability limit of the first-stage score estimator and let $\mathbf D_z^*(\mathbf X)$ denote the probability limit of the standardized estimated double score $\widehat{\mathbf D}_z(\mathbf X)$. The population conditional means below are indexed by $\mathbf D_z^*$, whereas the feasible match sets are constructed using the standardized estimated scores $\widehat{\mathbf D}_z$.
For $z\in\{0,1\}$, let $m_z(\mathbf d)=E\{Y(z)\mid\mathbf D_z^*(\mathbf X)=\mathbf d\}$ denote the population conditional outcome mean given the double score. Whenever $\mathbf D_z^*$ satisfies the balancing property in Lemma~\ref{lem:ds_balancing}, including when either the propensity score or the corresponding prognostic score is correctly specified, we have $m_z(\mathbf d)=E\{Y\mid Z=z,\mathbf D_z^*(\mathbf X)=\mathbf d\}$, and Assumption~\ref{ass:retro_ignorability} further identifies this quantity from the selected sample as $m_z(\mathbf d)=E\{w_z(\mathbf X)Y\mid Z=z,\mathbf D_z^*(\mathbf X)=\mathbf d,S=1\}/E\{w_z(\mathbf X)\mid Z=z,\mathbf D_z^*(\mathbf X)=\mathbf d,S=1\}$. 
Let $\tau_{\mathbf D}(\mathbf X_i)=m_1\{\mathbf D_1^*(\mathbf X_i)\}-m_0\{\mathbf D_0^*(\mathbf X_i)\}$ denote the treatment-effect contrast based on the double scores.
Define the corresponding residual $\varepsilon_{z,i}=Y_i-m_z\{\mathbf D_z^*(\mathbf X_i)\}$ for each individual in arm $z$. The estimation error decomposes as
\begin{align}\label{eq:bias_decomp}
\widehat{\tau}_{\mathrm{PATE}}^{R} - \tau_{\mathrm{PATE}}
=
H_{R,\mathrm{PATE}}
+
B_{R,\mathrm{PATE}}
+
E_{R,\mathrm{PATE}},
\end{align}
where
\begin{align}
H_{R,\mathrm{PATE}}
&=
\frac{1}{\sum_{i=1}^n w_{Z_i}(\mathbf{X}_i)}
\sum_{i=1}^n
w_{Z_i}(\mathbf{X}_i)
\left[
\tau_{\mathbf D}(\mathbf{X}_i)
-
\tau_{\mathrm{PATE}}
\right],
\label{eq:H_term}
\\
B_{R,\mathrm{PATE}}
&=
\frac{1}{\sum_{i=1}^n w_{Z_i}(\mathbf{X}_i)}
\sum_{i=1}^n
w_{Z_i}(\mathbf{X}_i)(2Z_i - 1)
\left[
m_{1-Z_i}\{\mathbf D_{1-Z_i}^*(\mathbf X_i)\}
-
\frac{
\sum_{j \in \mathcal{J}_{\widehat{\mathbf{D}}_{1-Z_i}}(i)}
w_{1-Z_i}(\mathbf{X}_j)\,
m_{1-Z_i}\{\mathbf D_{1-Z_i}^*(\mathbf X_j)\}
}{
\sum_{j \in \mathcal{J}_{\widehat{\mathbf{D}}_{1-Z_i}}(i)}
w_{1-Z_i}(\mathbf{X}_j)
}
\right],
\label{eq:B_term}
\\
E_{R,\mathrm{PATE}}
&=
\frac{1}{\sum_{i=1}^n w_{Z_i}(\mathbf{X}_i)}
\sum_{i=1}^n
(2Z_i - 1)
\{w_{Z_i}(\mathbf{X}_i) + K_{Z_i}^{R,\mathrm{PATE}}(i)\}
\,\varepsilon_{Z_i,i}.
\label{eq:E_term}
\end{align}
The term $H_{R,\mathrm{PATE}}$ is a Hájek-type heterogeneity term that captures the survey-weighted deviation of the individual conditional mean treatment-effect contrast $\tau_{\mathbf D}(\mathbf X_i)$ from the population average, and its numerator has mean zero under the selected-sample distribution by the Hájek recovery identity from Assumption~\ref{ass:retro_ignorability}. The term $B_{R,\mathrm{PATE}}$ is the matching discrepancy between the target individual's conditional mean for the missing treatment arm and the survey-weighted mean of the same function among its matched individuals. For known scores, the two-dimensional matching conditions in Web Appendix A.2.4 yield $B_{R,\mathrm{PATE}} = O_p(n^{-1/2})$. The term $E_{R,\mathrm{PATE}}$ aggregates matching-weight-scaled residuals and contributes to the asymptotic variance under the residual-centering and stabilization conditions in Web Appendix A.2.2 and A.2.5.

Correcting the matching discrepancy can improve finite-sample performance.
Following Abadie and Imbens,\cite{AbadieImbens2011} we construct a bias-corrected (BC) imputation by subtracting a fitted conditional mean from each matched outcome and adding back its fitted value at the target individual's double-score value. Let 
$\widehat m_z(\mathbf d)$ be an estimator of the population conditional mean $m_z(\mathbf d)$ obtained by survey-weighted regression of $Y$ on the estimated double score $\widehat{\mathbf D}_z(\mathbf X)$ within arm $z$.\cite{chen2007sieve}
For each individual $i$ with $Z_i \neq z$, the BC imputed potential outcome is
\begin{align}\label{eq:bc_imputation}
\widehat{Y}_{i}^{R,\mathrm{BC}}(z)
=
\widehat m_z\{\widehat{\mathbf D}_z(\mathbf X_i)\}
+
\frac{\sum_{j \in \mathcal{J}_{\widehat{\mathbf{D}}_z}(i)} w_z(\mathbf{X}_j)\{Y_j - \widehat m_z\{\widehat{\mathbf D}_z(\mathbf X_j)\}\}}{\sum_{j \in \mathcal{J}_{\widehat{\mathbf{D}}_z}(i)} w_z(\mathbf{X}_j)},
\end{align}
Set $\widehat{Y}_{i}^{R,\mathrm{BC}}(z)=Y_i$ when $Z_i=z$.
The bias-corrected WDSM estimator becomes
\begin{align}\label{eq:bc_estimator}
\widehat{\tau}_{\mathrm{PATE}}^{R,\mathrm{BC}}
= \widehat{\tau}_{\mathrm{PATE}}^{R} - \widehat{B}_{R,\mathrm{PATE}}
=
\frac{\sum_{i=1}^n w_{Z_i}(\mathbf{X}_i) \{\widehat{Y}_i^{R,\mathrm{BC}}(1) - \widehat{Y}_i^{R,\mathrm{BC}}(0)\}}{\sum_{i=1}^n w_{Z_i}(\mathbf{X}_i)},
\end{align}
where $\widehat{B}_{R,\mathrm{PATE}}$ is obtained from \eqref{eq:B_term} by replacing each regression value $m_z\{\mathbf D_z^*(\mathbf X)\}$ with $\widehat m_z\{\widehat{\mathbf D}_z(\mathbf X)\}$, 
while keeping the match sets and survey weights unchanged. 
Consequently, the bias-corrected estimator satisfies $\widehat{\tau}_{\mathrm{PATE}}^{R,\mathrm{BC}} - \tau_{\mathrm{PATE}} = H_{R,\mathrm{PATE}} + E_{R,\mathrm{PATE}} + (B_{R,\mathrm{PATE}} - \widehat{B}_{R,\mathrm{PATE}})$, where $B_{R,\mathrm{PATE}} - \widehat{B}_{R,\mathrm{PATE}} = o_p(n^{-1/2})$ under the corresponding bias-correction remainder conditions in Web Appendix A.5.3 for known scores and A.5.4 for estimated scores, so that the matching discrepancy $B_{R,\mathrm{PATE}}$ is eliminated up to an asymptotically negligible remainder.
An additional conditionally stabilized estimator is presented in Web Appendix A.9.
The retrospective PATT formulas are given in Section~\ref{sec:PATT extension}, and the prospective analogues are recorded in Web Appendix A.7 and A.8. Throughout the proposed WDSM procedure, survey weights are used in different steps. For ease of reference, Table~\ref{tab:weight_requirements} summarizes where survey weights need to be incorporated under retrospective and prospective sampling designs.

\begin{table}[htbp]
\centering
\caption{Summary of components that incorporate survey weights in WDSM point estimation and variance estimation under retrospective and prospective sampling designs.}
\label{tab:weight_requirements}
\setlength{\tabcolsep}{4pt}
\begin{tabular}{p{0.12\textwidth}p{0.42\textwidth}p{0.17\textwidth}p{0.17\textwidth}}
\arrayrulecolor{white}
\noalign{\global\arrayrulewidth=0.5pt}
\rowcolor[rgb]{0.92,0.92,0.92}
\textbf{Stage} & \textbf{Component} & \textbf{Retrospective} & \textbf{Prospective} \\[2pt]
\arrayrulecolor{black}
\midrule
\multirow{6}{*}{\textit{\shortstack[l]{Point\\Estimation}}}
& \cellcolor[rgb]{0.92,0.92,0.92} Propensity score estimation, $\widehat e(\mathbf X)$ & \cellcolor[rgb]{0.92,0.92,0.92} $\cmark$\;\,($w_z(\mathbf X)$) & \cellcolor[rgb]{0.92,0.92,0.92} $\times$ \\
& Prognostic score regression, $\widehat\Psi_z(\mathbf X)$ & $\times$ & $\times$ \\
& \cellcolor[rgb]{0.92,0.92,0.92} Match set construction, $\mathcal J_{\widehat{\mathbf D}_z}(i)$ & \cellcolor[rgb]{0.92,0.92,0.92} $\times$ & \cellcolor[rgb]{0.92,0.92,0.92} $\times$ \\
& Within-match imputation of $Y_i(z)$ & $\cmark$\;\,($w_z(\mathbf X)$) & $\cmark$\;\,($w(\mathbf X)$) \\
& \cellcolor[rgb]{0.92,0.92,0.92} H\'{a}jek normalization & \cellcolor[rgb]{0.92,0.92,0.92} $\cmark$\;\,($w_z(\mathbf X)$) & \cellcolor[rgb]{0.92,0.92,0.92} $\cmark$\;\,($w(\mathbf X)$) \\
& Bias-correction regression, $\widehat m_z(\mathbf d)$ & $\cmark$\;\,($w_z(\mathbf X)$) & $\cmark$\;\,($w(\mathbf X)$) \\
\midrule
\multirow{5}{*}{\textit{\shortstack[l]{Variance\\Estimation}}}
& \cellcolor[rgb]{0.92,0.92,0.92} Bootstrap weight generation, $w_i^{*,b}$ & \cellcolor[rgb]{0.92,0.92,0.92} $\cmark$\;\,($w_z(\mathbf X)$) & \cellcolor[rgb]{0.92,0.92,0.92} $\cmark$\;\,($w(\mathbf X)$) \\
& \quad Propensity score replication, $\widehat e^{*,b}(\mathbf X)$ & $\cmark$\;\,($w_i^{*,b}$) & $\times$ \\
& \cellcolor[rgb]{0.92,0.92,0.92} \quad Prognostic score replication, $\widehat\Psi_z^{*,b}(\mathbf X)$ & \cellcolor[rgb]{0.92,0.92,0.92} $\times$ & \cellcolor[rgb]{0.92,0.92,0.92} $\times$ \\
& Bias-correction regression replication, $\widehat m_z^{*,b}(\mathbf d)$ & $\cmark$\;\,($w_i^{*,b}$) & $\cmark$\;\,($w_i^{*,b}$) \\
& \cellcolor[rgb]{0.92,0.92,0.92} Linearized bootstrap estimate, $\widehat\tau_{\mathrm{PATE}}^{*,b}$, $\widehat\tau_{\mathrm{PATT}}^{*,b}$ & \cellcolor[rgb]{0.92,0.92,0.92} $\cmark$\;\,($w_i^{*,b}$) & \cellcolor[rgb]{0.92,0.92,0.92} $\cmark$\;\,($w_i^{*,b}$) \\
\end{tabular}
\vspace{1ex}

\parbox{\textwidth}{\footnotesize Note: Survey weights are $w_z(\mathbf X)=1/\pi_z(\mathbf X)$ for individuals in arm $z$ under retrospective sampling, and $w(\mathbf X)=1/\pi(\mathbf X)$ under prospective sampling. When imputing $Y_i(z)$ for an individual with $Z_i\neq z$, survey weights are those of the matched individuals in arm $z$. Match set construction uses unweighted Euclidean distance in the standardized estimated double score space $\widehat{\mathbf D}_z$, and prognostic score models are fitted without survey weights. Replication weights $w_i^{*,b}$ combine multinomial perturbations with the original survey weights across $b=1,\ldots,B$ replicates.}
\end{table}

\subsection{Asymptotic results}\label{subsec:asymptotics}

We establish the asymptotic properties of the bias-corrected WDSM estimator $\widehat{\tau}_{\mathrm{PATE}}^{R,\mathrm{BC}}$, 
which depends on the standardized estimated double scores $\widehat{\mathbf D}_z(\mathbf X_i;\widehat{\boldsymbol\theta})$ constructed from $\left(\widehat e(\mathbf X_i;\widehat{\boldsymbol\alpha}),\widehat\Psi_z(\mathbf X_i;\widehat{\boldsymbol\beta}_z)\right)$ 
with parameter estimates $\widehat{\boldsymbol{\theta}} = (\widehat{\boldsymbol{\alpha}}^\top, \widehat{\boldsymbol{\beta}}_0^\top, \widehat{\boldsymbol{\beta}}_1^\top)^\top$ obtained in Section~\ref{subsec:score_models}. 
We first derive the large-sample distribution when $\boldsymbol{\theta}^*$ is treated as known, which we refer to as the oracle setting, and then derive the asymptotic results with variance adjustment from first-step estimation of $\boldsymbol{\theta}^*$.
Throughout the oracle analysis, match sets and reuse coefficients are evaluated using $\mathbf D_z^*$.

Under the bias-correction remainder condition stated in Section~\ref{subsec:decomposition and bias correction} and the joint residual-centering condition in Assumption~A1(i) in Web Appendix A.2.2, the bias-corrected WDSM estimator can be written as $\widehat{\tau}_{\mathrm{PATE}}^{R,\mathrm{BC}} - \tau_{\mathrm{PATE}} = H_{R,\mathrm{PATE}} + E_{R,\mathrm{PATE}}+o_p(n^{-1/2})$. 
Adapting the matching decomposition and two-stage martingale representation of Abadie and Imbens and Yang and Zhang,\cite{abadie2006,abadie2016matching,yang_multiply_2023} we establish in Web Appendix A.5.3 that $\sqrt n\,H_{R,\mathrm{PATE}}$ and $\sqrt n\,E_{R,\mathrm{PATE}}$ converge jointly to a mean-zero normal distribution with component variances $V_{H,R}$ and $V_{E,R}$ and zero covariance.
The oracle asymptotic variance of $\sqrt{n}\{\widehat{\tau}_{\mathrm{PATE}}^{R,\mathrm{BC}}(\boldsymbol{\theta}^*) - \tau_{\mathrm{PATE}}\}$ therefore decomposes additively as $V_R = V_{H,R} + V_{E,R}$. The expressions for $V_{H,R}$ and $V_{E,R}$ and their derivations are provided in Web Appendix A.5.3.

When the scores in $\mathbf{D}_z$ are estimated in the first stage as described in Section~\ref{subsec:score_models}, the oracle variance $V_R$ requires adjustment. Because matching estimators are non-smooth functionals of the matching variables,\cite{abadie2016matching} this adjustment cannot be derived by standard Taylor expansion and instead requires the local experiment framework of Andreou and Werker,\cite{AndreouWerker2012} as adapted by Yang and Zhang\cite{yang_multiply_2023} for double score matching, applied to the stacked individual estimating function contribution $U(\boldsymbol{\theta}) = (U_{\boldsymbol{\alpha}}(\boldsymbol{\alpha})^\top, U_{\boldsymbol{\beta}_0}(\boldsymbol{\beta}_0)^\top, U_{\boldsymbol{\beta}_1}(\boldsymbol{\beta}_1)^\top)^\top$ corresponding to the summands in Equations \eqref{eq:ps_ee}--\eqref{eq:pgs_ee}. 
The corresponding regularity conditions and proofs are provided in Web Appendix A.5.3--A.5.4.

\begin{theorem}[Asymptotic Normality of Retrospective PATE Estimators]
\label{thm:main_asymptotics}
Under Assumption~\ref{ass:retro_ignorability} and the matching and bias-correction regularity conditions 
in Web Appendix A.5.1--A.5.3, if either the propensity score model is correctly specified or both arm-specific prognostic score models are correctly specified:

\textbf{(i) Known scores.} The bias-corrected WDSM estimator satisfies
\begin{align}\label{eq:clt_known}
\sqrt{n}\left(\widehat{\tau}_{\mathrm{PATE}}^{R,\mathrm{BC}}(\boldsymbol{\theta}^*)-\tau_{\mathrm{PATE}}\right)
\Rightarrow
\mathcal{N}\!\left(0,\,V_R\right),
\end{align}
where $V_R = V_{H,R} + V_{E,R}$ is defined above.

\textbf{(ii) Estimated scores.} Under the additional first-stage and local-matching regularity conditions in Web Appendix A.5.4, 
\begin{align}\label{eq:clt_estimated}
\sqrt{n}\left\{
\widehat{\tau}_{\mathrm{PATE}}^{R,\mathrm{BC}}(\widehat{\boldsymbol{\theta}})
-
\tau_{\mathrm{PATE}}
\right\}
\Rightarrow
\mathcal{N}(0,V_{R,\mathrm{adj}}),
\end{align}
where
\begin{align}\label{eq:V_adj}
V_{R,\mathrm{adj}}
=
V_R
-\gamma_1^\top\Sigma_U^{-1}\gamma_1
+\widetilde\gamma_2^\top
\Sigma_{\boldsymbol{\theta}^*}\widetilde\gamma_2,
\end{align}
$\Sigma_U = E[U(\boldsymbol{\theta}^*) U(\boldsymbol{\theta}^*)^\top \mid S=1]$ is the second moment matrix of the stacked estimating function, $\Sigma_{\boldsymbol{\theta}^*} = \Gamma^{-1}\Sigma_U(\Gamma^{-1})^\top$ is the Godambe sandwich covariance\cite{Binder1983} of $\sqrt{n}(\widehat{\boldsymbol{\theta}} - \boldsymbol{\theta}^*)$ with Jacobian $\Gamma = E[\partial U(\boldsymbol{\theta}^*)/\partial\boldsymbol{\theta}^\top \mid S=1]$, 
$\gamma_1$ is the limiting cross-covariance vector of $n^{-1/2}\sum_{i=1}^n U_i(\boldsymbol\theta^*)$ with the root-$n$-scaled oracle matching error, and $\widetilde\gamma_2=d_R-\Gamma^\top\Sigma_U^{-1}\gamma_1$, where $d_R$ is the local matching drift coefficient defined in Web Appendix A.5.4.

\end{theorem}

With equal survey weights, the estimator reduces to unweighted double score matching.\cite{yang_multiply_2023} In Equation~\eqref{eq:V_adj}, $-\gamma_1^\top\Sigma_U^{-1}\gamma_1$ is the nonpositive projection adjustment associated with the correlation between the oracle matching statistic and first-stage score estimation,\cite{abadie2016matching} whereas $\widetilde\gamma_2^\top\Sigma_{\boldsymbol{\theta}^*}\widetilde\gamma_2$ accounts for the remaining first-stage variation. The overall adjustment can therefore increase, decrease, or leave unchanged the asymptotic variance.

\subsection{Variance estimation}\label{subsec:variance estimation}

Constructing valid confidence intervals for $\widehat{\tau}_{\mathrm{PATE}}^{R,\mathrm{BC}}$ requires a consistent variance estimator. 
Although Theorem \ref{thm:main_asymptotics} characterizes the asymptotic variance through a heterogeneity component, a residual quadratic-variation limit, and an estimated-score local-experiment adjustment, direct plug-in estimation is impractical because the residual limit and the covariance vector $\gamma_1$ do not have simple closed-form estimators, as Yang and Zhang\cite{yang_multiply_2023} noted for unweighted double score matching.
Resampling provides an alternative, but the standard bootstrap is inapplicable because it cannot reproduce the distribution of matching reuse counts for fixed $M$.\cite{abadie2008failure} To overcome this, Otsu and Rai\cite{otsu2017bootstrap} developed a wild bootstrap based on the asymptotic linear representation for unweighted covariate matching, and Yang and Zhang\cite{yang_multiply_2023} extended the linearization-based replication to double score matching by constructing bootstrap replicates from the linearized estimator while retaining the original matching structure. In the survey-weighted setting, the variance depends not only on matching reuse coefficients but also on how survey weights are distributed within each match set, as this determines the relative contribution of individuals from different subpopulations to each imputed potential outcome. Extending Yang and Zhang\cite{yang_multiply_2023} to survey observational studies, we develop a linearization-based replication procedure that retains the original matching structure and survey-weighted reuse coefficients.
 
Building on the limiting score representation in Web Appendix A.5.3 and the estimated score result in Theorem~\ref{thm:main_asymptotics}, we construct feasible bootstrap replicates for $\widehat{\tau}_{\mathrm{PATE}}^{R,\mathrm{BC}}(\widehat{\boldsymbol{\theta}})$ by evaluating the linearized terms at replicate-specific nuisance estimates while retaining the matching structure constructed from the original estimated scores.\cite{yang_multiply_2023}
The replication algorithm proceeds as follows.

\begin{description}
\item[Step 1.(Bootstrap weight generation)] Generate multinomial weights $\mathbf m^{*,b}=(m_1^{*,b},\ldots,m_n^{*,b})\sim\operatorname{Multinomial}(n,\mathbf 1_n/n)$ and construct multiplicative bootstrap weights $w_i^{*,b} = m_i^{*,b}w_{Z_i}(\mathbf{X}_i)$.

\item[Step 2.(Score parameter replication)] Obtain a bootstrap replicate $\widehat{\boldsymbol{\theta}}^{*,b}$ of the score parameters by solving the weighted estimating equations with bootstrap weights. Let $\widehat e^{*,b}(\mathbf X)$ and $\widehat\Psi_z^{*,b}(\mathbf X)$ denote the propensity and arm-specific prognostic scores obtained from $\widehat{\boldsymbol\theta}^{*,b}$. For retrospective sampling, the propensity score replicate solves $\sum_{i=1}^n w_i^{*,b} \mathbf{X}_i \{Z_i - e(\mathbf{X}_i; \boldsymbol{\alpha})\} = 0$, while for prospective sampling the unweighted multinomial weights $m_i^{*,b}$ replace $w_i^{*,b}$, consistent with the design-specific score estimation in Section \ref{subsec:score_models}. For both designs, prognostic score replicates solve $\sum_{i: Z_i=z} m_i^{*,b} \mathbf{X}_i \{Y_i - \Psi_z(\mathbf{X}_i; \boldsymbol{\beta}_z)\} = 0$. 
Construct the bootstrap double scores $\widehat{\mathbf D}_z^{*,b}(\mathbf X_i)$ from $\widehat{\boldsymbol\theta}^{*,b}$, recomputing their standardization in each replicate using the original rule. Within each treatment arm, refit the bias-correction regression using weights $w_i^{*,b}$ and the bootstrap double scores to obtain $\widehat m_z^{*,b}$, for example, using regression with polynomial terms\cite{yang_multiply_2023} or polynomial sieve methods\cite{chen2007sieve}.

\item[Step 3.(Linearized bootstrap estimate)] Compute a bootstrap replicate of $\widehat{\tau}_{\mathrm{PATE}}^{R,\mathrm{BC}}$ as
\begin{align}\label{eq:boot_tau}
\widehat{\tau}_{\mathrm{PATE}}^{*,b} = \frac{1}{\sum_{i=1}^n w_i^{*,b}} \sum_{i=1}^n \Big[&w_i^{*,b}
\left[
\widehat m_1^{*,b}\{\widehat{\mathbf D}_1^{*,b}(\mathbf X_i)\}
-
\widehat m_0^{*,b}\{\widehat{\mathbf D}_0^{*,b}(\mathbf X_i)\}
\right]
\notag\\
&+ m_i^{*,b}(2Z_i - 1)\{w_{Z_i}(\mathbf{X}_i) + K_{Z_i}^{R,\mathrm{PATE}}(i)\}
\left[
Y_i-
\widehat m_{Z_i}^{*,b}
\{\widehat{\mathbf D}_{Z_i}^{*,b}(\mathbf X_i)\}
\right]
\Big],
\end{align}
where the survey-weighted reuse coefficients $K_{Z_i}^{R,\mathrm{PATE}}(i)$ are computed from the original estimated score matching structure and held fixed across replicates.

\item[Step 4.(Variance calculation)] Repeat Steps 1--3 for $b = 1, \ldots, B$ replicates and estimate the variance by $\widehat{\operatorname{Var}}(\widehat{\tau}_{\mathrm{PATE}}^{R,\mathrm{BC}}) = B^{-1}\sum_{b=1}^B (\widehat{\tau}_{\mathrm{PATE}}^{*,b} - \bar{\tau}^*)^2$, with $\bar{\tau}^* = B^{-1}\sum_{b=1}^B \widehat{\tau}_{\mathrm{PATE}}^{*,b}$.
\end{description}

We construct 95\% Wald confidence intervals using the replication-based standard errors and normal approximation. Sufficient conditions for replication validity and Wald inference are given in Propositions~A16 and~A18 in Web Appendix A.4.3 and A.4.5.
The variance estimation rows of Table \ref{tab:weight_requirements} summarize the survey weight requirements for each component of this procedure.

\section{Extension to population average treatment effect on the treated}\label{sec:PATT extension}

Because the PATT averages treatment effects over the treated population, $Y_i(1)=Y_i$ is observed for every individual with $Z_i=1$ by consistency. 
Thus, only the missing control potential outcome $Y_i(0)$ must be imputed. We therefore match each treated individual to control individuals on the control-side double score $\mathbf{D}_0(\mathbf{X}) = (e(\mathbf{X}), \Psi_0(\mathbf{X}))$, which satisfies $Y(0) \perp Z \mid \mathbf{D}_0(\mathbf{X})$ by Lemma \ref{lem:ds_balancing} under the one-sided identification condition $Y(0) \perp Z \mid \mathbf{X}$. For each treated individual, the counterfactual control outcome is imputed from $M$ nearest control individuals using the estimated double score $\widehat{\mathbf D}_0$:
\begin{align}\label{eq:patt_imputation}
\widehat{Y}_i^{R}(0)
=
\frac{\sum_{j\in \mathcal{J}_{\widehat{\mathbf D}_0}(i)}
w_0(\mathbf{X}_j) Y_j}{\sum_{k\in \mathcal{J}_{\widehat{\mathbf D}_0}(i)} w_0(\mathbf{X}_k)},
\qquad
Z_i=1.
\end{align}
Control individuals may appear in multiple treated match sets, and the survey-weighted reuse coefficient for the control arm is defined as
\begin{align}\label{eq:patt_reuse_freq}
K_0^{R,\mathrm{PATT}}(j)
=
\sum_{i:Z_i=1}
\frac{w_1(\mathbf{X}_i)w_0(\mathbf{X}_j)\mathbf{1}\{j\in\mathcal{J}_{\widehat{\mathbf D}_0}(i)\}}{\sum_{k\in\mathcal{J}_{\widehat{\mathbf D}_0}(i)} w_0(\mathbf{X}_k)}.
\end{align}
With these components, the retrospective PATT estimator uses Hájek normalization with treated-side survey weights and has the matching weight representation
\begin{align}\label{eq:patt_estimator}
\widehat{\tau}_{\mathrm{PATT}}^{R}
&=
\frac{\sum_{i:Z_i=1} w_1(\mathbf{X}_i)\{Y_i - \widehat{Y}_i^{R}(0)\}}{\sum_{i:Z_i=1} w_1(\mathbf{X}_i)}
\nonumber\\
&=
\frac{\sum_{i:Z_i=1} w_1(\mathbf{X}_i) Y_i - \sum_{j:Z_j=0} K_0^{R,\mathrm{PATT}}(j) Y_j}{\sum_{i:Z_i=1} w_1(\mathbf{X}_i)}.
\end{align}
We hence use the bias-corrected estimator $\widehat{\tau}_{\mathrm{PATT}}^{R,\mathrm{BC}}=\widehat{\tau}_{\mathrm{PATT}}^R-\widehat B_{R,\mathrm{PATT}}$, paralleling Equation~\eqref{eq:bc_estimator}. Here, $\widehat B_{R,\mathrm{PATT}}$ is constructed by replacing $m_0\{\mathbf D_0^*(\mathbf X)\}$ with $\widehat m_0\{\widehat{\mathbf D}_0(\mathbf X)\}$ in $B_{R,\mathrm{PATT}}$, defined in Web Appendix A.6.2, using the same matched sets and survey weights as the original estimator. 
The asymptotic results for known and estimated scores are provided in Web Appendix A.6.3 and A.6.4, respectively.

For variance estimation of $\widehat{\tau}_{\mathrm{PATT}}^{R,\mathrm{BC}}$, the linearization-based replication procedure of Section \ref{subsec:variance estimation} is adapted by re-estimating the propensity score and control prognostic score parameters and the control-side bias-correction regression in each replicate, using only the control-side reuse coefficients $K_0^{R,\mathrm{PATT}}(j)$, and restricting Hájek normalization to treated individuals. 
Retaining the original estimated score matching structure yields the feasible bootstrap replicate
\begin{align}\label{eq:patt_boot_tau}
\widehat{\tau}_{\mathrm{PATT}}^{*,b} = \frac{1}{\sum_{i:Z_i=1} w_i^{*,b}}
\left[
\sum_{i:Z_i=1} w_i^{*,b} \{Y_i - \widehat m_0^{*,b}\{\widehat{\mathbf D}_0^{*,b}(\mathbf X_i)\}\}
-
\sum_{j:Z_j=0} m_j^{*,b} K_0^{R,\mathrm{PATT}}(j) \{Y_j - \widehat m_0^{*,b}\{\widehat{\mathbf D}_0^{*,b}(\mathbf X_j)\}\}
\right],
\end{align}
where $K_0^{R,\mathrm{PATT}}(j)$ is computed from the original estimated score matching structure and held fixed across replicates. 
The variance estimator is $\widehat{\operatorname{Var}}(\widehat{\tau}_{\mathrm{PATT}}^{R,\mathrm{BC}}) = B^{-1}\sum_{b=1}^B (\widehat{\tau}_{\mathrm{PATT}}^{*,b} - \bar{\tau}_{\mathrm{PATT}}^*)^2$. Here, $\bar{\tau}_{\mathrm{PATT}}^*=B^{-1}\sum_{b=1}^B\widehat{\tau}_{\mathrm{PATT}}^{*,b}$. 
The corresponding sufficient conditions for replication validity and Wald inference are given in Propositions~A17 and~A18 in Web Appendix A.4.3 and A.4.5. The prospective PATT analogue is presented in Web Appendix A.8.5.

\section{Simulation Studies}\label{sec:simulation}

\subsection{Simulation design}

We carried out a set of simulation studies to evaluate the finite-sample performance of our proposed WDSM approach and demonstrate the necessity for correctly incorporating survey weights for population-level causal effect estimation. Our data-generating process builds upon Zeng et al.,\cite{zeng2025movingbestpracticeusing} who adapted the multi-stage strategy from Austin et al.\cite{austin2018propensity} and Lenis et al.\cite{Lenis2019}. We simulated a population of $1{,}000{,}000$ individuals, organized into 10 strata $(j = 1, \ldots, 10)$. Each stratum was further divided into 20 clusters, resulting in a total of 200 clusters indexed globally as $(k = 1, \ldots, 200)$, with 5000 individuals in each cluster $(i = 1, \ldots, 5000)$. 
For $l=1,\ldots,6$, the baseline covariates were generated from the linear mixed model $
X_{l,ijk}
=\nu_{l,j}^{\mathrm{stratum}}
+\nu_{l,k}^{\mathrm{cluster}}
+\eta_{l,ijk},
$
where $\nu_{l,j}^{\mathrm{stratum}}\sim N(0,0.35^2)$, $\nu_{l,k}^{\mathrm{cluster}}\sim N(0,0.15^2)$, and $\eta_{l,ijk}\sim N(0,1)$. The random effects and individual errors were generated independently, with each stratum or cluster effect shared by individuals in that stratum or cluster. For the treatment and outcome models below, we index individuals consecutively by $i=1,\ldots,N$ and write $\mathbf X_i=(X_{1i},\ldots,X_{6i})^\top$.
We specified the true population-level propensity score for each individual using a logistic regression:
\begin{align}
\operatorname{logit}\{e(\mathbf{X}_i)\} = a_0 + \psi(\boldsymbol{a}_{1:6}^\top \mathbf{X}_i + a_7 X_{1i} X_{2i}), \nonumber
\end{align}
where $\boldsymbol{a}_{1:6} = (\log(1.1), \log(1.25), \log(1.5), \log(1.75), \log(2), \log(2.5))^\top$ and $a_7 = \log(1.1)$. Treatment assignment $Z_i$ was simulated from a Bernoulli distribution: $Z_i \sim \operatorname{Bernoulli}(e(\mathbf{X}_i))$. The scalar $\psi$ reflects the strength of confounding and dictates the level of overlap in the propensity score.\cite{Hu2021} We set $\psi = 0.6$ or $2$ to mimic scenarios with good or poor overlap in the full population. 
For both scenarios, the overall treatment prevalence was approximately 30\%, with $a_0$ set to $\log(35/80)$ for good overlap (31.79\%) and $\log(20/80)$ for poor overlap ((31.85\%)), allowing us to examine the impact of overlap when evaluating estimator performance.

We simulated continuous outcomes for each individual $i$ in the population using a linear combination of covariates, treatment effects, and interaction terms, with
\begin{align}
Y_i = b_0 + \delta_0(\boldsymbol{b}_{1:6}^\top \mathbf{X}_i + b_7 X_{1i} X_{2i}) + Z_i\{\delta_1 + \delta_2(\boldsymbol{b}_{1:6}^\top \mathbf{X}_i + b_8 X_{1i} X_{2i})\}+ \varepsilon_i, \nonumber
\end{align}
where $\varepsilon_i \sim \mathcal{N}(0, 1)$. We set $b_0 = 0$ and $\boldsymbol{b}_{1:6} = (2.5, -2, 1.75, -1.25, 1.5, 1.1)^\top$. The baseline treatment effect was represented by $\delta_1 = 1$, while $\delta_0 = 0.3$ and $\delta_2 = 0.2$ introduced heterogeneity in the control and treatment conditions, respectively. Two interaction terms were included, with $b_7 = 2.5$ and $b_8 = 1.5$, to capture the more complex influence of baseline covariates on outcomes across both conditions. The true PATE and PATT were calculated by averaging $Y_i(1)-Y_i(0)$ over all individuals and treated individuals in the generated population, respectively. The true PATE was 0.8664 in both overlap settings, while the true PATT was 1.0351 under good overlap and 1.1819 under poor overlap.

For each overlap setting, we held the generated population fixed and drew 1,000 random samples under each of the two sampling frameworks to evaluate performance conditional on that population. Under retrospective sampling (our primary focus), the sampling mechanism depended on both baseline covariates and treatment assignment. The sampling probability $\pi_z(\mathbf{x}_i) = P(S_i = 1 \mid Z_i, \mathbf{X}_i)$ was computed as
\begin{align}
\operatorname{logit}(\pi_z(\mathbf{x}_i)) = c_0 + \delta_z^{(s)} Z_i + \mathbf{c}^\top \mathbf{X}_i, \nonumber
\end{align}
where $c_0 = \log(0.005/0.995)$ defines the baseline sampling rate of 0.5\%, and $\delta_z^{(s)} = \log(0.9)$ slightly reduces the sampling probability for the treatment group. The coefficient vector was specified as $\mathbf{c} = (\log(1.05), \log(1.10), \log(1.15), \log(1.10), \log(1.05), \log(1.10))^\top$. Treatment-specific survey weights were computed as $w_z(\mathbf{x}_i) = 1/\pi_z(\mathbf{x}_i)$. Under prospective sampling, samples of 5000 individuals each (representing 0.5\% of the overall population) were drawn using a multi-stage design where the sampling mechanism was conditionally independent of the treatment assignment mechanism, consistent with the survey-weighted propensity score matching simulation settings in Austin et al.\cite{austin2018propensity} and Lenis et al.\cite{Lenis2019}. Sample sizes were allocated across the 10 strata as $\{850, 750, 700, 650, 600, 400, 350, 300, 250, 150\}$, ensuring that the observed sample would have a different distribution from the population. Within each stratum, five clusters were selected using simple random sampling, and from each selected cluster, an equal number of individuals were sampled. For each sampled individual, we calculated the true survey weights as $w_i = 1/\pi(\mathbf{x}_i)$ by taking the inverse of the multi-stage inclusion probability.

For each sample within each scenario, we compared five matching methods: (a) unweighted propensity score matching (PSM), (b) unweighted prognostic score matching (PGM), (c) unweighted double score matching (DSM) via the \texttt{dsmatch} package (version 1.7.1),\cite{yang_multiply_2023} (d) survey-weighted propensity score full matching (SW-PSM) via \texttt{MatchIt},\cite{Ho2011} and (e) the proposed weighted double score matching (WDSM). 
Our aim was to compare the methods as they would be used in practice.
Using the supported implementations and settings of each package, PSM, PGM, and DSM used $M = 1$ nearest neighbor matching with replacement and SW-PSM used optimal full matching (matching all individuals in variable-ratio sets to minimize total within-set distances), while our WDSM was evaluated at $M \in \{1, 3, 5\}$ nearest neighbors with replacement, with $M = 5$ shown in Figure \ref{fig:retro_overlap_comparison} and all three reported in Web Tables 1--4. 
We recommend $M=5$ as a practical choice for WDSM in settings similar to those studied here, given its efficiency gains over $M=1$ with generally similar bias when at least one working score model is correctly specified.
For point and variance estimation, WDSM incorporated survey-weighted regression with linear, squared, and interaction terms in the standardized double scores for bias correction following the regression approach of Yang and Zhang\cite{yang_multiply_2023} and the linearization-based bootstrap variance estimation of Section \ref{subsec:variance estimation} with $B = 200$ replications. We assessed performance under four model misspecification scenarios: (i) correct propensity score model and correct prognostic score model (Cor|Cor), (ii) misspecified propensity score model but correct prognostic score model (Mis|Cor), (iii) correct propensity score model but misspecified prognostic score model (Cor|Mis), and (iv) both models misspecified (Mis|Mis). Correctly specified models included all six main effects plus the $X_{1i}X_{2i}$ interaction, while misspecified models omitted this interaction. These labels refer to the population model terms, not to the use of survey weights. SW-PSM retained a correctly specified treatment-interacted postmatching outcome regression under all four score-model specifications. This regression is distinct from the prognostic score model and can also be misspecified in practice. We did not examine its misspecification here because the simulations focused on the score models. These score-model misspecification scenarios parallel those of Yang and Zhang,\cite{yang_multiply_2023} who evaluated standard unweighted double score matching under similar scenarios.

Given the complexity of the design, we note the total number of analysis configurations considered. We examined two levels of overlap (good versus poor), two sampling designs (retrospective versus prospective), five matching methods (PSM, PGM, DSM, SW-PSM, WDSM), two estimands (PATE, PATT), and four model specification scenarios (Cor|Cor, Mis|Cor, Cor|Mis, Mis|Mis). This combination yields $2 \times 2 \times 5 \times 2 \times 4 = 160$ distinct configurations in the main analysis. Additional configurations with varying $M$ increase this number further. Table \ref{tab:scenario_counts} outlines these factors and their levels for quick reference.

\begin{table}[!htbp]
\centering
\caption{Summary of factors and number of analysis configurations in the main simulation design.}
\label{tab:scenario_counts}
\setlength{\tabcolsep}{6pt}
\renewcommand{\arraystretch}{1.2}
\begin{tabular}{l c c}
\rowcolor[rgb]{0.92,0.92,0.92}
\textbf{Factor} & \textbf{Levels} & \textbf{Count} \\[6pt]
Overlap scenario & Good ($\psi = 0.6$) vs.\ Poor ($\psi = 2.0$) & 2 \\
\rowcolor[rgb]{0.92,0.92,0.92}
Sampling design & Retrospective vs.\ Prospective & 2 \\
Matching method & PSM, PGM, DSM, SW-PSM, WDSM & 5 \\
\rowcolor[rgb]{0.92,0.92,0.92}
Estimand & PATE, PATT & 2 \\
Model specification & Cor|Cor, Mis|Cor, Cor|Mis, Mis|Mis & 4 \\
\midrule
\rowcolor[rgb]{0.92,0.92,0.92}
Total analysis configurations & & 160 \\
\bottomrule
\end{tabular}
\end{table}

Performance metrics included relative bias, relative efficiency, and empirical coverage of the 95\% confidence intervals. 
Relative bias was calculated as $100(\overline{\widehat{\tau}}-\tau)/|\tau|$, where $\overline{\widehat{\tau}}$ is the mean estimate across 1,000 samples and $\tau$ is the corresponding true population effect.
Web Tables 1--4 report signed relative bias, whereas the figure displays its absolute value. Relative efficiency was determined as the ratio of the Monte Carlo variance of unweighted PSM in the ``Cor|Cor'' scenario
to that of each evaluated estimator. 
Empirical coverage was the proportion of 95\% Wald confidence intervals containing the true population treatment effect. Standard errors for unweighted PSM, PGM, and DSM were obtained using the linearized bootstrap implemented in \texttt{dsmatch}.\cite{yang_multiply_2023} For SW-PSM, we used delta method standard errors based on a sandwich covariance matrix clustered by matched subclass, implemented in \texttt{marginaleffects} following the \texttt{MatchIt} documentation.\cite{Greifer2026MatchItEffects,ArelBundock2024Marginaleffects} For WDSM, we used the survey-weighted replication procedure in Section~\ref{subsec:variance estimation}.
The simulation code and the proposed WDSM estimator are available online at \url{https://github.com/ykzeng-yale/SW_DSM}.

\subsection{Simulation results}

\begin{figure}[htbp]
\centering
\includegraphics[width=0.95\textwidth]{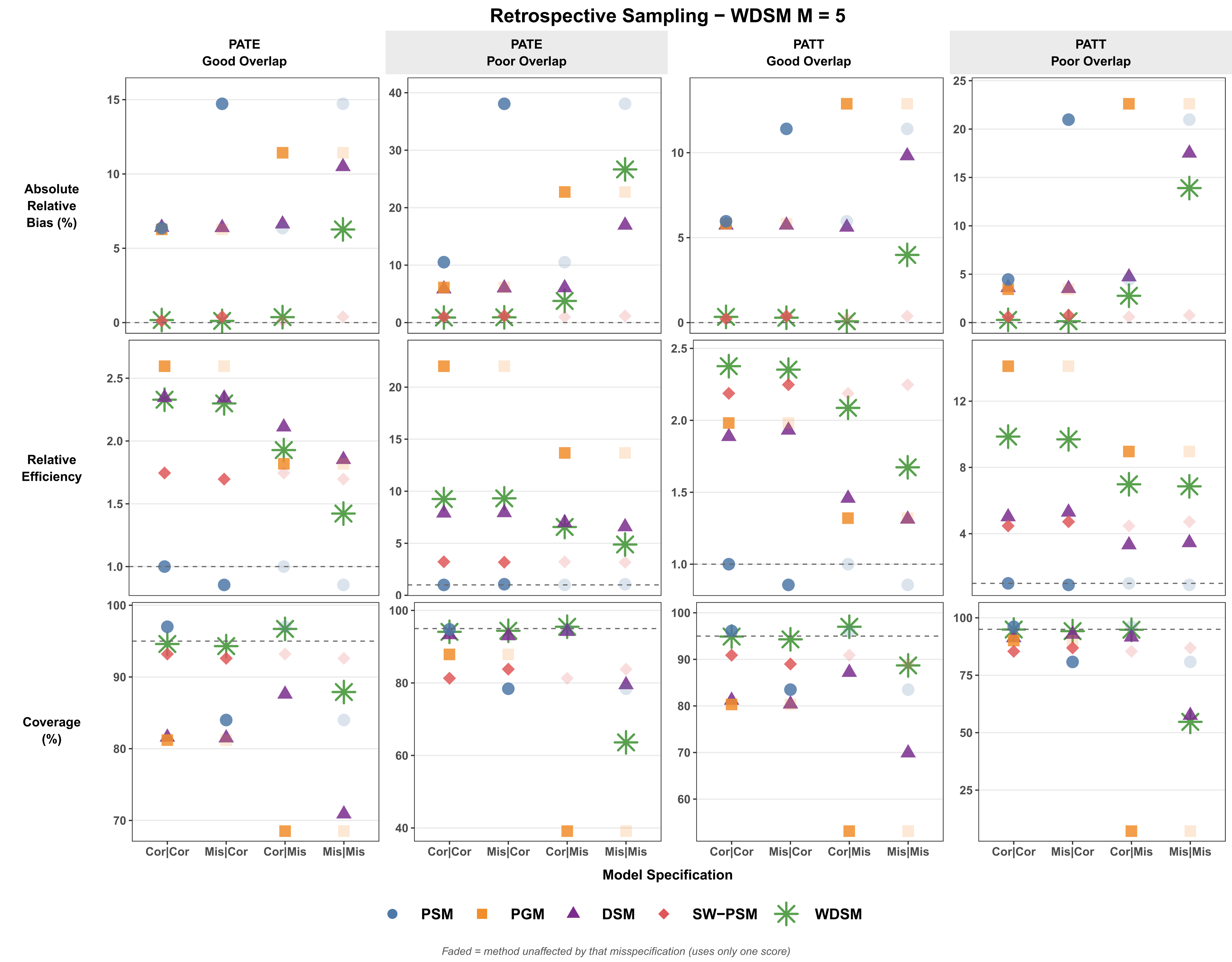}
\caption{Performance of matching estimators under retrospective sampling across overlap conditions and model specifications (1,000 replications). Columns correspond to estimand and overlap combinations: PATE and PATT under good overlap ($\psi = 0.6$) and poor overlap ($\psi = 2.0$). Rows display absolute relative bias (\%), relative efficiency (variance ratio relative to PSM under Cor|Cor), and coverage of 95\% Wald confidence intervals. The vertical-axis ranges differ across coverage panels. Dashed reference lines indicate zero bias, unit efficiency, and the nominal 95\% coverage level. Faded markers denote scenarios where misspecification does not affect the method, as these methods use only one score model. Following the default implementations of each package, PSM, PGM, and DSM use $M = 1$ nearest neighbor matching with replacement, SW-PSM uses optimal full matching, and WDSM uses $M = 5$ nearest neighbors with replacement. PSM = propensity score matching, PGM = prognostic score matching, DSM = double score matching, SW-PSM = survey-weighted propensity score full matching, WDSM = weighted double score matching with polynomial regression bias correction and linearization-based bootstrap variance estimation. Detailed numerical results are provided in Web Tables 1 and 2.}
\label{fig:retro_overlap_comparison}
\end{figure}

Figure \ref{fig:retro_overlap_comparison} presents the simulation results under retrospective sampling, comparing absolute relative bias, relative efficiency, and coverage for PATE and PATT with good and poor overlap. 
In the figure, WDSM uses $M = 5$ nearest neighbors with replacement, while unweighted PSM, PGM, and DSM use $M = 1$ nearest neighbor matching as implemented in \texttt{dsmatch}, and SW-PSM uses optimal full matching. 
Unless otherwise stated, the numerical comparisons below use WDSM with $M=5$ and report signed relative bias. 
Detailed numerical results including WDSM with $M \in \{1, 3, 5\}$ are provided in Web Tables 1 and 2 for retrospective sampling and Web Tables 3 and 4 for prospective sampling. We pursue the answers to four questions: (1) whether incorporating survey weights is essential for population-level causal inference via matching; (2) how WDSM compares to existing methods when both working models are correctly specified; (3) how the WDSM estimator performs when one working model is misspecified; and (4) how the degree of overlap affects the performance of each method.

\subsubsection{Is survey weight incorporation necessary for population-level causal inference via matching?}

Our simulation results support the theoretical conclusion that incorporating survey weights throughout the matching pipeline as identified in Table~\ref{tab:weight_requirements} is essential for consistent estimation of population-level treatment effects. Here, we focus on the scenario where both models are correctly specified to examine the role of survey weights without introducing score model misspecification.

For the relative bias, as shown in the top row of Figure \ref{fig:retro_overlap_comparison} (the ``Cor|Cor'' scenario), all three unweighted methods exhibit 6--7\% relative bias for PATE under retrospective design and good overlap, and unweighted PSM bias increases to 10.5\% under poor overlap. Even the unweighted DSM, which achieves double robustness under simple random sampling, shows bias of 6.4\% for estimating PATE under good overlap and 5.8\% under poor overlap. In contrast, both survey-weighted methods, WDSM and SW-PSM, achieve near-zero bias across both overlap conditions. 
For WDSM with $M=1$, the relative bias for estimating PATE is $-0.2$\% under good overlap and 0.3\% under poor overlap (Web Tables 1 and 2).
The bias observed under the ``Cor|Cor'' scenario indicates that correct score specification alone does not eliminate the discrepancy between sample and population targets for the unweighted methods in these settings.

The efficiency comparison reinforces this conclusion. As shown in the middle row of Figure \ref{fig:retro_overlap_comparison} (the ``Cor|Cor'' scenario), unweighted PSM serves as the reference (efficiency = 1.00), while both survey-weighted methods achieve higher efficiency: 2.33 for WDSM and 1.75 for SW-PSM under good overlap, increasing to 9.25 and 3.22 under poor overlap. For WDSM, prognostic information can improve outcome similarity within matched sets, while averaging over multiple matches can reduce imputation noise.

Finally, in the bottom row of Figure \ref{fig:retro_overlap_comparison} (the ``Cor|Cor'' scenario), WDSM (green asterisk) consistently achieves coverage close to the nominal 95\% level across both estimands and overlap conditions, whereas coverage for unweighted PGM and DSM is approximately 80--82\% under good overlap. 
Under prospective sampling (Web Tables 3 and 4), where sampling is conditionally independent of treatment assignment given the covariates, unweighted methods show comparable bias. Unweighted PSM exhibits 6.4\% and 10.6\% bias for PATE under good and poor overlap, nearly identical to the retrospective values of 6.4\% and 10.5\%, consistent with bias arising from the distributional discrepancy between sample and population under both sampling designs.
Both survey-weighted methods achieve near-zero bias under prospective sampling as well, with WDSM maintaining 93.2\% coverage for PATE under good overlap and 94.5\% under poor overlap (Web Tables 3 and 4). In summary, across both sampling designs, unweighted methods retain bias for the population targets, whereas the survey-weighted methods substantially reduce it.

\subsubsection{How does WDSM compare to existing matching methods under correct model specification?}

Having established the necessity of survey weight incorporation, we now compare WDSM with SW-PSM and with unweighted DSM under correct specification to evaluate the additional benefit of WDSM. In terms of bias, both WDSM ($M = 5$) and SW-PSM achieve near-zero bias under the ``Cor|Cor'' scenario across all four combinations of estimand and overlap in Figure \ref{fig:retro_overlap_comparison}, while unweighted DSM (purple triangle) shows persistent 3.6--6.4\% bias despite correct model specification. WDSM and SW-PSM are comparable on this metric, consistent with the benefit of incorporating survey weights. 

The efficiency comparison reveals the primary advantage of WDSM over SW-PSM. As shown in the middle row of Figure \ref{fig:retro_overlap_comparison} (the ``Cor|Cor'' scenario), WDSM achieves higher efficiency than SW-PSM across all conditions: 2.33 versus 1.75 for PATE under good overlap, 9.25 versus 3.22 under poor overlap, and 9.86 versus 4.47 for PATT under poor overlap. 
The correctly specified prognostic score directs matching toward individuals with similar expected outcomes, which can improve precision when propensity score matching alone leaves substantial outcome heterogeneity, especially when propensity score distributions are widely separated under poor overlap scenarios.\citep{yang_multiply_2023}
The efficiency advantage is also evident for PATT, where the control prognostic score guides imputation of the missing untreated outcomes. 
Compared to unweighted DSM, WDSM achieves higher efficiency for PATE under retrospective sampling and poor overlap (9.25 versus 7.90). Even with $M=1$, WDSM has relative efficiency of 6.67, slightly lower than unweighted DSM (7.90) but higher than unweighted PSM (1.00) and SW-PSM (3.22) in this setting (Web Table 2). Under prospective sampling, WDSM achieves relative efficiency of 2.37 versus 1.89 for SW-PSM under good overlap and 11.66 versus 4.57 under poor overlap for PATE (Web Tables 3 and 4).

The coverage comparison confirms the advantage of WDSM over SW-PSM in these settings. In the bottom row of Figure \ref{fig:retro_overlap_comparison} (the ``Cor|Cor'' scenario), WDSM achieves 94.6\% coverage for PATE under good overlap compared to 93.2\% for SW-PSM, and 94.9\% versus 90.9\% for PATT. Under poor overlap, the gap widens substantially: 94.1\% versus 81.3\% for PATE and 94.9\% versus 85.4\% for PATT. The consistent under-coverage of SW-PSM under poor overlap, observed across both retrospective and prospective sampling designs (82.8\% for PATE under prospective sampling and poor overlap in Web Table 4), further favors WDSM for inference under limited propensity score overlap. Unweighted DSM achieves 93.3\% coverage for PATE under retrospective sampling and poor overlap but only 81.8\% under prospective sampling and good overlap (Web Table 3), as persistent bias pushes its confidence intervals away from the true population effect.

\subsubsection{How does WDSM perform when one working model is misspecified?}

When the propensity score is misspecified but the prognostic score is correct (the ``Mis|Cor'' scenario in Figure \ref{fig:retro_overlap_comparison}), WDSM maintains near-zero bias by leveraging the correctly specified prognostic score, demonstrating clear advantages over both SW-PSM and unweighted single-score methods. In the top row of Figure \ref{fig:retro_overlap_comparison} (the ``Mis|Cor'' scenario), WDSM remains near the zero-bias dashed line across all four columns, while unweighted PSM has bias of 38.1\% for PATE under poor overlap and 21.0\% for PATT. 
SW-PSM, which uses only the propensity score for matching but retains a correctly specified outcome regression, shows modest PATE bias ($-0.4$\% to $-1.2$\%) but its coverage degrades under poor overlap.
Under retrospective sampling and good overlap, WDSM achieves $-0.1$\% bias with 94.3\% coverage for PATE compared to $-0.4$\% with 92.6\% for SW-PSM (Web Table 1). Under poor overlap, WDSM achieves 0.9\% bias with 94.4\% coverage versus $-1.2$\% with 83.8\% for SW-PSM (Web Table 2), and for PATT 0.1\% with 94.2\% versus -0.8\% with 86.9\%. 
Under prospective sampling (Web Tables 3 and 4), WDSM maintains similar robustness with $-0.6$\% bias and 93.3\% coverage for PATE under good overlap and $-0.1$\% with 94.7\% under poor overlap.

When the prognostic score is misspecified but the propensity score is correct (the ``Cor|Mis'' scenario in Figure \ref{fig:retro_overlap_comparison}), the impact on WDSM varies with the overlap condition and sampling design. Under retrospective sampling and poor overlap, WDSM shows 3.8\% bias with 95.5\% coverage for PATE (Web Table 2), compared with $-0.4$\% bias and 96.7\% coverage under good overlap (Web Table 1). 
For PATT, WDSM shows $-0.1$\% bias with 97.0\% coverage under retrospective sampling and good overlap, compared with 2.8\% bias and 94.8\% coverage under poor overlap.
Under prospective sampling, WDSM has $-0.3$\% bias with 96.4\% coverage for estimating PATE under good overlap and 2.7\% bias with 96.2\% coverage under poor overlap (Web Tables 3 and 4). 
Correct propensity score specification provides the balancing property underlying double robustness, but does not ensure close expected outcomes for individual matches, particularly when limited propensity score overlap makes comparable matches even more scarce given a misspecified prognostic model.

By contrast, unweighted PGM, which relies entirely on the prognostic score, is severely affected by prognostic misspecification regardless of overlap or sampling design. In the ``Cor|Mis'' scenario shown in Figure \ref{fig:retro_overlap_comparison}, unweighted PGM (orange square) shows 22.7\% bias with 39.1\% coverage for PATE under retrospective sampling and poor overlap and 22.6\% with 7.1\% for PATT (Web Table 2). 
Under prospective sampling and poor overlap, coverage for estimating PATT is 5.2\% for unweighted PGM, compared with 93.4\% for WDSM (Web Table 4). WDSM has substantially smaller bias and coverage closer to the nominal 95\% level than PGM in these settings, consistent with the doubly robust protection afforded by the correctly specified propensity score.

\subsubsection{How does overlap quality affect the robustness of each method?}

Comparing across the good and poor overlap columns in Figure \ref{fig:retro_overlap_comparison}, the relative advantage of WDSM over competing methods depends on which score models are correctly specified, but grows as overlap deteriorates when prognostic score models are correct. 
In terms of bias, WDSM maintains near-zero values across both overlap conditions under the ``Cor|Cor'' and ``Mis|Cor'' scenarios, while unweighted methods show increasing bias as overlap worsens (e.g., unweighted PSM from 6.4\% to 10.5\% for PATE from good to poor overlap in the top row of Figure \ref{fig:retro_overlap_comparison}). 
Under the ``Cor|Mis'' scenario, the absolute PATE bias increases from 0.4\% under good overlap to 3.8\% under poor overlap for retrospective sampling. This pattern is consistent with greater finite-sample sensitivity when limited overlap reduces the availability of comparable matches and the misspecified prognostic score omits relevant outcome structure.
Under the ``Mis|Cor'' scenario, WDSM maintains near-zero bias under both overlap conditions.

In terms of efficiency, under the ``Cor|Cor'' scenario, the advantage of WDSM over SW-PSM grows with deteriorating overlap. Under good overlap (middle row, first column of Figure \ref{fig:retro_overlap_comparison}), the efficiency ratio between WDSM and SW-PSM is $2.33/1.75 \approx 1.33$ for PATE, but under poor overlap this ratio increases to $9.25/3.22 \approx 2.87$, consistent with the stabilizing role of the prognostic score when propensity score distributions are separated. Under prospective sampling, the pattern is similar: the efficiency ratio increases from $2.37/1.89 \approx 1.25$ under good overlap to $11.66/4.57 \approx 2.55$ under poor overlap (Web Tables 3 and 4).

Comparing the bottom row of Figure \ref{fig:retro_overlap_comparison} across columns, the coverage gap between WDSM and SW-PSM is modest under good overlap (94.6\% versus 93.2\% for PATE) but widens substantially under poor overlap (94.1\% versus 81.3\%), with a similar pattern under prospective sampling (93.2\% versus 92.5\% under good overlap, 94.5\% versus 82.8\% under poor overlap in Web Tables 3 and 4). 
These findings are consistent with the potential precision gains from incorporating prognostic information when propensity score overlap is limited.

When both models are misspecified (the ``Mis|Mis'' scenario in Figure \ref{fig:retro_overlap_comparison}), WDSM achieves smaller bias and higher coverage than unweighted DSM under good overlap. Under retrospective sampling and good overlap, WDSM shows 6.3\% bias with 87.9\% coverage for PATE versus 10.5\% with 70.9\% for unweighted DSM (Web Table 1). Under poor overlap, WDSM has larger PATE bias and lower coverage than DSM, with respective values of 26.7\% and 63.6\% versus 16.9\% and 79.5\% under retrospective sampling (Web Table 2), and 25.8\% and 65.1\% versus 17.3\% and 82.4\% under prospective sampling (Web Table 4). For PATT under retrospective sampling and poor overlap, WDSM retains smaller bias (13.9\% versus 17.5\%), although coverage is slightly lower (54.7\% versus 57.5\%). Under prospective design and poor overlap, WDSM achieves both smaller bias (12.6\% versus 18.7\%) and higher coverage (70.4\% versus 54.2\%) than DSM (Web Table 4) for estimating PATT.

The bias-variance tradeoff across $M \in \{1, 3, 5\}$ provides additional practical guidance (Web Tables 1--4). In the ``Cor|Cor'' scenario under retrospective sampling and poor overlap, relative efficiency for estimating PATE increases from 6.67 ($M=1$) to 9.25 ($M=5$), with bias of 0.3\% and 0.9\%, respectively. 
In the ``Mis|Mis'' scenario under the same sampling and overlap conditions, increasing $M$ from 1 to 5 raises relative efficiency from 3.48 to 4.88 but also increases bias from 24.2\% to 26.7\% and reduces coverage from 77.5\% to 63.6\%. 

In summary, WDSM generally achieves small bias and near-nominal coverage when at least one score model is correctly specified, across both sampling designs, both estimands, and both overlap conditions. Survey weight incorporation is essential for population-level inference regardless of sampling design, and WDSM provides superior coverage and efficiency compared to SW-PSM under the ``Cor|Cor'' scenario, with the advantage especially pronounced under poor overlap. 
Compared to unweighted DSM, WDSM substantially reduces bias when at least one score model is correctly specified, consistent with its theoretical double robustness property.

\section{Case Studies}\label{sec:case studies}

\subsection{Application to early marijuana use and mental health outcomes (prospective sampling)}

\subsubsection{Data and research question}

We illustrate WDSM using data from the 2023 National Survey on Drug Use and Health (NSDUH),\cite{SAMHSA2024} an annual survey conducted by the Substance Abuse and Mental Health Services Administration (SAMHSA) that provides nationally representative estimates of substance use and mental health among the U.S. civilian, noninstitutionalized population. 
The NSDUH employs a complex multistage probability sampling design with deliberate oversampling of adolescents aged 12--17 and young adults aged 18--25.\cite{SAMHSA2024codebook} We use the prospective formulation under the assumption that survey selection is conditionally independent of early marijuana initiation given the adjustment covariates.

Our analytic sample consists of all 17,471 young adults aged 18--29 in the released data, weighted to represent approximately 51.8 million individuals. We used the edited and imputed variables provided by SAMHSA.\cite{SAMHSA2024codebook}
Treatment is defined as marijuana use initiation before age 15 ($n=1{,}877$, 9.3\% weighted) versus first use at age 15 or later or never used ($n=15{,}594$, 90.7\% weighted). This age threshold reflects developmental research identifying early adolescence as a critical period for substance use vulnerability and brain development.\cite{Squeglia2009,Lubman2015,Volkow2016effects,Albaugh2021} The endocannabinoid system reaches peak expression and activity throughout the brain during adolescence,\cite{Hurd2019} and cannabis exposure during this neurodevelopmental window may disrupt normal brain maturation processes.\cite{Filbey2014}

The primary outcome is any mental illness in the past year (AMIPY), classified using NSDUH prediction models aligned with DSM-IV criteria.\cite{SAMHSA2024} 
The secondary outcome is serious mental illness (SMIPY), indicating disorders with substantial functional impairment. 
Longitudinal studies have linked adolescent cannabis use to depression and suicidal ideation in young adulthood.\cite{Gobbi2019}
Covariates include demographics (gender, race/ethnicity, age groups), socioeconomic factors (income levels), geographic indicators (metropolitan status), and other substance use patterns (alcohol initiation before age 15 and lifetime cigarette use).

We implement the prospective WDSM estimator described in Web Appendix A.7, fitting the propensity score by unweighted logistic regression and the arm-specific prognostic scores by unweighted linear regression.
Matching uses $M=5$ nearest neighbors with replacement in the standardized double score space, with bootstrap variance estimation using $B=200$ replications. Distance ties are resolved at random for each individual being matched. We also assess sensitivity to tie allocation by repeating the analysis under 20 additional random allocations for each outcome. For comparison, we implement unweighted propensity score matching (PSM), prognostic score matching (PGM), and double score matching (DSM) with a 1:1 matching ratio and bootstrap inference, and survey-weighted propensity score full matching (SW-PSM) with 
standard errors clustered by full-matching subclass.\cite{Ho2011}

\subsubsection{Results}

\begin{figure}[htbp]
\centering
\includegraphics[width=\textwidth]{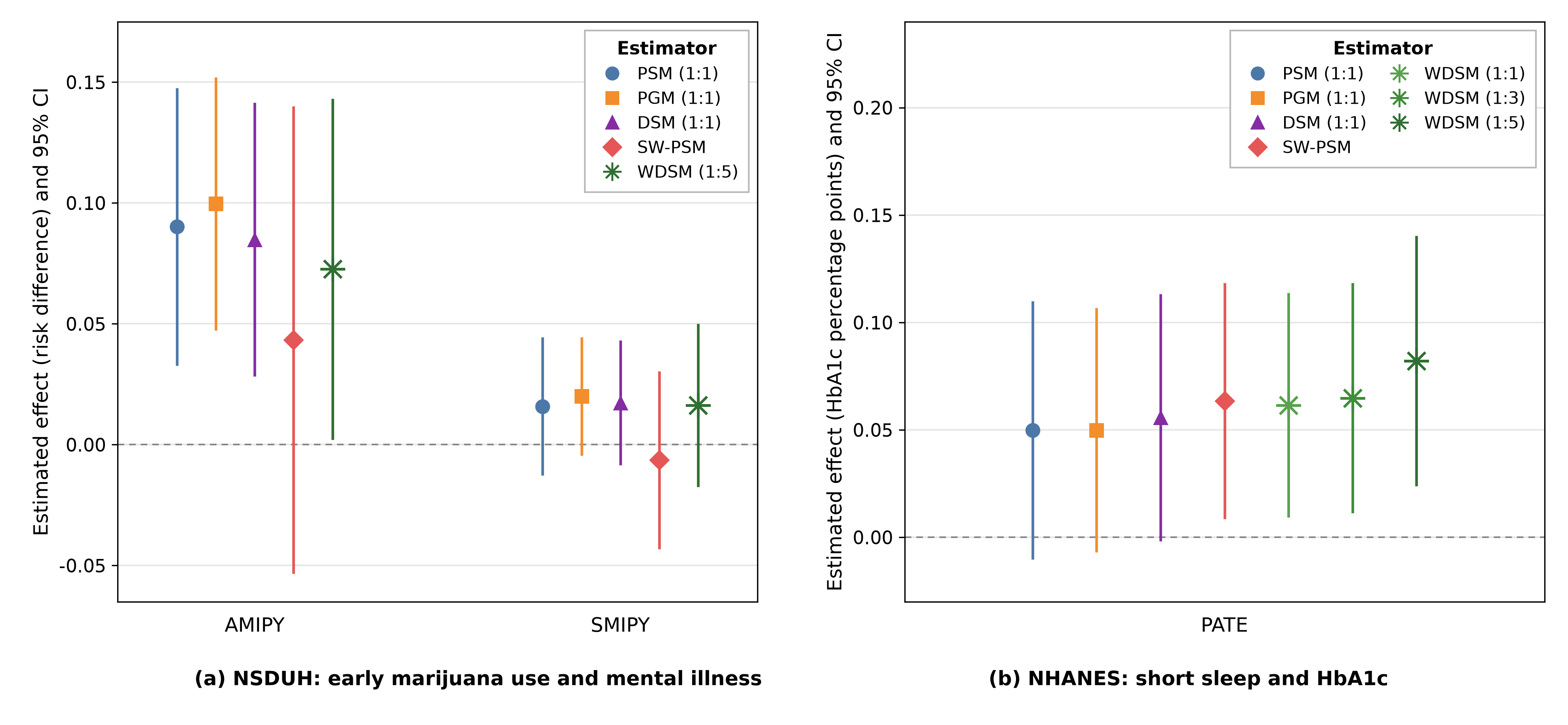}
\caption{Estimated population average treatment effects with 95\% confidence intervals across five matching estimators in two case studies. Panel (a) displays the effect of early marijuana use (before age 15) on prevalence of any mental illness (AMIPY) and serious mental illness (SMIPY) among 17,471 young adults aged 18--29 from NSDUH 2023. In panel (a), a risk difference of 0.01 corresponds to 1 percentage point. Panel (b) displays the effect of short sleep duration ($<7$ hours on weekdays) on HbA1c (\%) among 3,783 adults aged 40--70 without a reported diabetes diagnosis from NHANES 2015--2018, with WDSM evaluated at $M = 1, 3, 5$ to assess sensitivity to the matching ratio. 
WDSM, PSM, PGM, and DSM use bootstrap standard errors with $B=200$ replications for NSDUH and $B=2{,}000$ for NHANES, whereas SW-PSM uses standard errors clustered by full-matching subclass. 
The horizontal reference line indicates zero effect.}
\label{fig:casestudy_pate}
\end{figure}

Figure \ref{fig:casestudy_pate}(a) presents the estimated population average treatment effects of early marijuana use on mental health outcomes. 
For the primary outcome (AMIPY), WDSM estimates an increase of 7.26 percentage points in mental illness prevalence (95\% CI: 0.20, 14.32; $p=0.044$).
The unweighted methods (PSM, PGM, DSM) produce larger point estimates ranging from 8.49 to 9.97 percentage points with confidence intervals excluding zero. 
SW-PSM yields a smaller estimate of 4.34 percentage points with a wider confidence interval (95\% CI: $-5.33$, 14.00). 
The WDSM point estimate lies between those from the unweighted methods and survey-weighted PS full matching, with a narrower confidence interval than the latter. Across 20 additional random allocations of distance ties, the WDSM estimates for AMIPY ranged from 6.94 to 7.40 percentage points, and five of the corresponding 95\% confidence intervals included zero.

For serious mental illness (SMIPY), all methods show smaller effects near the null. 
WDSM estimates a 1.63 percentage point increase (95\% CI: $-1.75$, 5.01), while unweighted methods range from 1.59 to 2.00 percentage points. SW-PSM produces a small negative estimate ($-0.64$ percentage points). All confidence intervals include zero.

WDSM incorporates survey weights throughout the matching pipeline to target population-level effects. In this application, it yields narrower confidence intervals than SW-PSM for both mental health outcomes.
The estimated 7.26 percentage point increase in mental illness prevalence, if causal, would be of substantial public health relevance among U.S. young adults.

\subsection{Application to sleep duration and glycemic control in NHANES (retrospective sampling)}

\subsubsection{Data and research question}

We apply WDSM to investigate whether insufficient sleep duration affects glycemic regulation in adults without a reported diabetes diagnosis using NHANES 2015--2018 data. With approximately one-third of U.S. adults habitually sleeping less than the recommended 7 hours per night,\cite{liu2016prevalence} understanding sleep's metabolic consequences has substantial implications for diabetes prevention. Laboratory studies demonstrate that experimental sleep restriction impairs glucose tolerance and insulin sensitivity,\cite{spiegel1999impact,buxton2010sleep} but epidemiological evidence for population-level effects remains limited by residual confounding and sample selection issues. 
We use the retrospective WDSM estimator, fitting the propensity score by survey-weighted logistic regression and the arm-specific prognostic scores by unweighted linear regression.

We identified 4,304 adults aged 40--70 with observed sleep duration and HbA1c after excluding reported diabetes diagnoses and recorded pregnancies. We further excluded 521 participants (12.1\%) with missing covariates, most commonly the poverty-income ratio, leaving an analytic sample of 3,783.
We dichotomize usual weekday sleep at the clinical threshold of 7 hours,\cite{watson2015recommended} comparing short sleepers ($<7$ hours) to those sleeping $\geq 7$ hours. 
In the comparison group, median sleep duration is 8 hours, with an interquartile range of 7.5--9 hours and a range of 7--14.5 hours.
The outcome is HbA1c measured as a percentage via standardized methods,\cite{little2001national} with values between 5.7\% and 6.4\% indicating prediabetes.\cite{ada2024diagnosis} 
Covariates include age, sex, race/ethnicity, education, poverty-income ratio, BMI, and smoking status. We implement WDSM with $M=1, 3, 5$ matches to assess sensitivity to the matching ratio, while unweighted methods use 1:1 matching. WDSM and the unweighted methods employ bootstrap inference with $B=2{,}000$ replications, while SW-PSM uses standard errors clustered by full-matching subclass.

\subsubsection{Results}

Figure \ref{fig:casestudy_pate}(b) presents estimated effects of short sleep duration on HbA1c levels. 
The unweighted methods yield small positive estimates ranging from 0.050 to 0.056 percentage points, with confidence intervals including zero. Survey-weighted PS full matching estimates an increase of 0.064 percentage
points (95\% CI: 0.009, 0.119). The WDSM estimates range from 0.062 to 0.082 percentage points, with confidence intervals excluding zero for all three matching ratios.

For $M=1$ and $M=3$, WDSM yields similar estimates of 0.062 (95\% CI: 0.009, 0.114) and 0.065 (95\% CI: 0.011, 0.118) percentage points, respectively. 
SW-PSM yields a similar point estimate and confidence interval to WDSM with $M=1$ in this example.
Increasing the matching ratio to $M=5$ yields a larger estimate of 0.082 percentage points (95\% CI: 0.024, 0.140), with a wider confidence interval.

Although modest in absolute terms, the WDSM estimates suggest a small population-level increase in mean HbA1c associated with short sleep. If causal, such a shift could have population health relevance given the high prevalence of short sleep.\cite{liu2016prevalence}
These results illustrate the risk of ignoring survey weights. Unweighted matching yields estimates closer to the null and confidence intervals including zero, potentially obscuring a population association between short sleep and higher HbA1c. In contrast, both WDSM and SW-PSM yield larger estimates with confidence intervals excluding zero. These findings highlight the importance of incorporating survey weights when assessing population effects in this application.

\section{Discussion}\label{sec:discussion}

We develop weighted double score matching for population-level causal inference from complex survey data. The framework incorporates survey weights throughout the matching pipeline, including score estimation, within-match averaging, Hájek normalization, and bias correction, under both retrospective and prospective sampling designs. We establish asymptotic normality for both known and estimated scores (Theorem~\ref{thm:main_asymptotics} and Web Appendix A.6.3--A.6.4), demonstrate that WDSM preserves double robustness under survey-weighted estimation subject to the stated regularity condition, and provide a computationally efficient linearization-based bootstrap procedure for inference that avoids re-matching in each replicate. 
WDSM estimation and inference for PATE and PATT are implemented in the \texttt{wdsmatch} R package (version 0.2.1), available from CRAN at \url{https://CRAN.R-project.org/package=wdsmatch}.

Several lessons have emerged from our investigation. First, incorporating survey weights at the design-specific stages identified in Table~\ref{tab:weight_requirements} 
is essential for consistent estimation of population-level treatment effects in survey observational studies. Consistent with the findings of Ridgeway et al.\cite{Ridgeway2015} and Zeng et al.\cite{zeng2025movingbestpracticeusing} for propensity score weighting, our simulation results confirm that omitting survey weights restricts covariate balance to within the sample,\cite{Zanutto2006,DuGoff2014} introducing persistent bias when generalizing to the target population. Under retrospective sampling where $P(Z=1\mid\mathbf X,S=1)\neq e(\mathbf X)$, survey-weighted propensity score estimation is necessary for recovering the population propensity score as established in Section \ref{subsec:score_models}. Under prospective sampling where these quantities coincide, either approach is theoretically valid, but incorporating survey weights remains advisable as a safeguard against potential violations of the conditional independence between sampling and treatment.\cite{DuGoff2014}

Second, the use of both propensity and prognostic scores in the matching variable provides substantial advantages over single-score matching. When the propensity score model is misspecified, the prognostic score preserves consistency through double robustness, and vice versa.
Double robustness is an asymptotic property and does not imply symmetric finite-sample performance under propensity and prognostic score misspecification. The greater sensitivity to prognostic misspecification observed in some of our scenarios is consistent with asymmetric finite-sample behavior reported for other doubly robust procedures.\cite{li2013multilevel,li2022generalizing} 
The efficiency gains from double score matching are especially pronounced under poor overlap, where WDSM achieves relative efficiency of 9.25 for PATE and 9.86 for PATT compared to 3.22 and 4.47 for survey-weighted propensity score matching. 
The prognostic score can improve precision by guiding matches toward individuals with similar expected outcomes, complementing propensity-score matching when treatment probabilities are widely separated.\cite{yang_multiply_2023}
The primary limitation of double score matching relative to single-score propensity matching emerges under joint model misspecification with poor overlap, where misspecification of the prognostic score may amplify the matching errors.

Third, practical implementation of WDSM requires attention to several diagnostic and modeling considerations. Individual covariates may remain imbalanced despite overall double score balance, and we recommend post-matching balance assessment using standardized mean differences\cite{Austin2009balance} to verify adequate covariate balance within the matched sample. When residual imbalance is detected, regression-based bias correction\cite{AbadieImbens2011} incorporating survey weights or caliper restrictions on the matching distance can mitigate the impact. 
The choice of matching ratio $M$ presents a bias-variance tradeoff.\cite{Austin2009} We recommend assessing sensitivity to a small, prespecified set of matching ratios, such as $M=1,3,5$, alongside covariate balance and matching distances.

Our case studies illustrate the practical consequences of methodological choice. In the NSDUH analysis of early marijuana use and mental health, WDSM estimates an increase of 7.26 percentage points, compared with 8.49 for unweighted DSM and 4.34 for survey-weighted propensity score full matching, with a narrower confidence interval than the latter. In the NHANES analysis of sleep duration and glycemic control, both WDSM and survey-weighted propensity score full matching yield confidence intervals excluding zero, whereas the unweighted methods do not. 
At $M=1$ and $M=3$, WDSM yields slightly narrower confidence intervals than survey-weighted propensity score full matching.
These applications illustrate the practical value of combining propensity and prognostic scores with survey weights for population-level estimation.

The validity of our findings rests on several key assumptions. Like all matching methods, WDSM cannot address unmeasured confounding, with the additional complexity that survey design itself may relate to unmeasured factors affecting treatment and outcomes.\cite{rosenbaum1987sensitivity,vanderweele2017sensitivity} Sensitivity analysis methods for assessing robustness to unmeasured confounding, such as the E-value framework of VanderWeele and Ding\cite{vanderweele2017sensitivity} and the flexible sensitivity analysis of Lu and Ding,\cite{LuDing2023} have been developed for weighting and regression estimators but have not been extended to the survey-weighted matching setting. Developing sensitivity analysis tools for WDSM represents an important direction for future research. Our analysis also assumes that survey weights are known and derived from well-designed surveys. If this assumption is violated through inadequate non-response adjustment or flawed post-stratification, the resulting survey weights may exacerbate rather than correct selection bias. Our asymptotic theory treats the observed units as independent draws from the selected-sample distribution. Extending inference to dependent observations arising from multistage sampling requires further development.

Several extensions of WDSM merit further development. Current limitations to binary treatments motivate extensions to multiple treatments building on the generalized propensity score framework,\cite{yang2020propensity,Li2019b} survival outcomes,\cite{AustinSmall2014} and heterogeneous treatment effects.\cite{imai2013estimating} Each extension requires careful integration of survey weights with specific methodological requirements. WDSM could also be extended to matching in a general fixed dimension, drawing on Chen and Han\cite{chen2024limiting} to derive explicit expressions for the asymptotic variance. Our analysis relies on parametric models for propensity score and prognostic score estimation, reflecting their widespread use in applied practice. Data-adaptive machine learning approaches offer greater flexibility in capturing nonlinear relationships and complex interactions, and extending WDSM to incorporate double machine learning estimators\cite{chernozhukov2018double} for score estimation in the survey-weighted setting represents a promising direction.

\section*{Funding}

Haidong Lu's contribution was supported by the National Institute on Drug Abuse of the National Institutes of Health under award number R00DA057487. The content is solely the responsibility of the authors and does not necessarily represent the official views of the National Institutes of Health.

\bibliography{references}
\end{document}


\renewcommand{\thesection}{\Alph{section}}
\renewcommand{\thesubsection}{\thesection.\arabic{subsection}}
\renewcommand{\thesubsubsection}{\thesubsection.\arabic{subsubsection}}

\begin{center}
{\Large\bfseries Supporting Information: Web Appendix\par}
\medskip
{\large\bfseries Where to weight? Estimating population causal effects
with weighted double score matching in complex surveys\par}
\smallskip
Zeng et al.
\end{center}

\makeatletter
\begingroup
\let\addcontentsline\@gobblethree
\tableofcontents
\endgroup
\makeatother
\clearpage

\section{Technical details}\label{app:technical-details}

We first restate and prove the main-text results. Sections~A.1--A.4 develop the identification, matching, estimated-score, and replication arguments used in the four combinations of sampling design and estimand in Sections~\ref{app:retrospective-pate}--\ref{app:prospective_patt}.

\noindent We denote expectations under the superpopulation and selected-sample laws by $E_{\Omega}$ and $E_s$, respectively, where $P_s(\cdot)=P_{\Omega}(\cdot\mid S=1)$. For retrospective sampling, we use $w_z(\mathbf X)=\pi_z(\mathbf X)^{-1}$ and $w_Z(\mathbf X)=Z w_1(\mathbf X)+(1-Z)w_0(\mathbf X)$. For prospective sampling, the weight is $w(\mathbf X)=\pi(\mathbf X)^{-1}$. We treat the inclusion-probability functions as known and use $A=R$ and $A=P$ for the retrospective and prospective regimes, respectively.

We write $e(\mathbf X)=e_\Omega(\mathbf X)=P_\Omega(Z=1\mid\mathbf X)$ for the population propensity and $e_s(\mathbf X)=P_s(Z=1\mid\mathbf X)$ for the selected-sample propensity. The retrospective survey-weighted propensity fit consistently estimates $e_\Omega$ when its working model is correctly specified.

For the estimated-score results, $e^*(\mathbf X)=e(\mathbf X;\boldsymbol\alpha^*)$ denotes the limiting working propensity and need not equal $e_\Omega(\mathbf X)$. The limiting double scores $\mathbf D_z^*$ are the standardized versions of $(e^*,\Psi_z^*)$. Their decompositions use the population reduced means $m_z^A(\mathbf D_z^*)$. The specialization $m_z^A(\mathbf D_z^*)=\mu_z(\mathbf X)$ holds when the corresponding prognostic mean is correctly specified.

Throughout the asymptotic arguments, the observed individuals are independent and identically distributed under $P_s$. We use a fixed measurable, outcome-independent rule to resolve distance ties. At fixed oracle scores, the matching sets and their coefficients are measurable in the matching-information fields defined below.

Let \(n_z=\sum_{i=1}^n\mathbf1(Z_i=z)\). We state the finite-sample matching identities on \(\{n_0\ge M,n_1\ge M\}\) for PATE and \(\{n_0\ge M,n_1\ge1\}\) for PATT. For the oracle moment and limit calculations, we adopt the convention that the matching counts, coefficients, radii, and \(H,B,E\) components are zero outside the corresponding event. We use the same convention for inverse-denominator expressions in these calculations. Under the stated i.i.d. sampling conditions with positive arm probabilities and fixed \(M\), the excluded probability decreases exponentially in \(n\).

\subsection*{Main-text results and proofs}
\phantomsection \addcontentsline{toc}{subsection}{Main-text results and proofs}
\label{app:main-text-results}

The following statements reproduce the main-text sampling assumptions, balancing lemma, and PATE theorem. A corresponding PATT result follows. The restatements retain the main-text numbering. Results specific to this appendix are numbered A1, A2, and so on within each result type.

\begin{maintextassumption}[Retrospective Survey Ignorability]
\label{ass:retro_ignorability}
Under retrospective sampling: (a) sampling probabilities depend on both covariates and treatment status, $\pi_z(\mathbf{x}_i) = P(S_i = 1 \mid \mathbf{X}_i = \mathbf{x}_i, Z_i = z)$ for $z \in \{0,1\}$, and we denote the treatment-specific survey weights as $w_z(\mathbf{x}_i) = 1/\pi_z(\mathbf{x}_i)$, which are known from the survey design; (b) strong survey positivity holds: there exists a constant $c_\pi\in(0,1/2]$ such that $c_\pi\leq\pi_z(\mathbf{X}_i)\leq1-c_\pi$ almost surely for $z\in\{0,1\}$; 

and (c) selection is conditionally independent of potential outcomes given covariates and treatment, $\{Y_i(0), Y_i(1)\} \perp S_i \mid \mathbf{X}_i, Z_i$.
\end{maintextassumption}

\begin{maintextassumption}[Prospective Survey Ignorability]
\label{ass:prosp_ignorability}
Under prospective sampling: (a) sampling probabilities depend only on pre-treatment covariates, $\pi(\mathbf{x}_i) = P(S_i = 1 \mid \mathbf{X}_i = \mathbf{x}_i)$, and we denote the common survey weights as $w(\mathbf{x}_i) = 1/\pi(\mathbf{x}_i)$, which are known from the survey design; (b) strong survey positivity holds: there exists a constant $c_\pi\in(0,1/2]$ such that $c_\pi\leq\pi(\mathbf{X}_i)\leq1-c_\pi$ almost surely; and (c) selection is conditionally independent of potential outcomes given covariates, $\{Y_i(0), Y_i(1)\} \perp S_i \mid \mathbf{X}_i$.
\end{maintextassumption}

For prospective PATE, the main text additionally assumes \(Z_i\perp\{Y_i(0),Y_i(1),S_i\}\mid\mathbf X_i\). For prospective PATT, the corresponding one-sided joint conditions are stated in Section~\ref{app:prospective_patt}. As in the main text, a prognostic score \(\Psi_z(\mathbf X)\) is distributionally sufficient when \(Y(z)\perp\mathbf X\mid\Psi_z(\mathbf X)\). A correct conditional mean provides mean balance, without requiring distributional sufficiency. The population double score is \(\mathbf D_z(\mathbf X)=\{e(\mathbf X),\Psi_z(\mathbf X)\}\).

\begin{maintextlemma}[Treatment-Specific Double Score Balancing Properties; Antonelli et al.\cite{Antonelli2018}, Yang and Zhang\cite{yang_multiply_2023}]\label{lem:ds_balancing}
Under treatment unconfoundedness $\{Y_i(0), Y_i(1)\} \perp Z_i \mid \mathbf{X}_i$, we have (a) $Y_i(z) \perp Z_i \mid \{h(\mathbf{X}_i),\, \Psi_z(\mathbf{X}_i)\}$, and (b) $Y_i(z) \perp Z_i \mid \{e(\mathbf{X}_i),\, h(\mathbf{X}_i)\}$ for any $h(\mathbf{X}_i)$ and $z \in \{0,1\}$. Together (a) and (b) yield (c) $Y_i(z) \perp Z_i \mid \mathbf{D}_z(\mathbf{X}_i)$ for $z \in \{0,1\}$. A double score containing either the population propensity score or the conditional mean $\mu_z(\mathbf{X}_i)$ satisfies\[E\{Y_i(z)\mid Z_i,\mathbf D_z(\mathbf X_i)\}=E\{Y_i(z)\mid\mathbf D_z(\mathbf X_i)\}.\]For PATT, under the one-sided condition $Y_i(0) \perp Z_i \mid \mathbf{X}_i$, similarly we have $Y_i(0) \perp Z_i \mid \mathbf{D}_0(\mathbf{X}_i)$.
\end{maintextlemma}
\begin{proof}
Apply Lemma~\ref{lem:armwise-double-score-balancing} to \(\{h(\mathbf X),\Psi_z(\mathbf X)\}\) in part (a) and to \(\{e(\mathbf X),h(\mathbf X)\}\) in part (b). Taking \(h=e\) in (a), or \(h=\Psi_z\) in (b), gives (c). The same lemma proves the displayed conditional-mean identity and the one-sided conclusion for \(z=0\).
\end{proof}

For the PATE restatement, retain the main-text notation
\[
V_{H,R}=V_{H,R,\mathrm{PATE}},\qquad
V_{E,R}=V_{E,R,\mathrm{PATE}},\qquad V_R=V_{H,R}+V_{E,R}.
\]
The heterogeneity variance is defined in Section~\ref{exact-known-score-asymptotics}. The residual variance is
\begin{equation}\label{eq:V_E}
V_{E,R}
=\operatorname*{plim}_{n\to\infty}
\frac{n\displaystyle\sum_{z=0}^1\sum_{j:Z_j=z}
 \{w_z(\mathbf X_j)+K_z^{R,\mathrm{PATE}}(j)\}^{2}v_{z,j}}
 {\left\{\displaystyle\sum_{i=1}^n w_{Z_i}(\mathbf X_i)\right\}^{2}},
\end{equation}
where \(v_{z,j}=\operatorname{Var}_s\{Y_j\mid Z_j=z,\mathbf D_0^*(\mathbf X_j),\mathbf D_1^*(\mathbf X_j), w_z(\mathbf X_j)\}\). The main-text \(K_z^{R,\mathrm{PATE}}\) is \(K_{n,z}^{R,\mathrm{PATE}}\) in the technical proofs. Proposition~\ref{prop:primitive-residual-qv-limit} establishes the finite, deterministic limit. The theorem also requires the stated residual-centering and bias-correction conditions for known and estimated scores.

\begin{maintexttheorem}[Asymptotic Normality of Retrospective PATE Estimators]
\label{thm:main_asymptotics}
Under Main-text Assumption~\ref{ass:retro_ignorability} and the matching and bias-correction regularity conditions 

in Web Appendix A.5.1--A.5.3, if either the propensity score model is correctly specified or both arm-specific prognostic score models are correctly specified:

\textbf{(i) Known scores.} The bias-corrected WDSM estimator satisfies
\begin{align}\label{eq:clt_known}
\sqrt{n}\left(\widehat{\tau}_{\mathrm{PATE}}^{R,\mathrm{BC}}(\boldsymbol{\theta}^*)-\tau_{\mathrm{PATE}}\right)
\Rightarrow
\mathcal{N}\!\left(0,\,V_R\right),
\end{align}
where $V_R = V_{H,R} + V_{E,R}$ is defined above.

\textbf{(ii) Estimated scores.} Under the additional first-stage and local-matching regularity conditions in Web Appendix A.5.4, 
\begin{align}\label{eq:clt_estimated}
\sqrt{n}\left\{
\widehat{\tau}_{\mathrm{PATE}}^{R,\mathrm{BC}}(\widehat{\boldsymbol{\theta}})
-
\tau_{\mathrm{PATE}}
\right\}
\Rightarrow
\mathcal{N}(0,V_{R,\mathrm{adj}}),
\end{align}
where

\begin{align}\label{eq:V_adj}
V_{R,\mathrm{adj}}
=
V_R
-\gamma_1^\top\Sigma_U^{-1}\gamma_1
+\widetilde\gamma_2^\top
\Sigma_{\boldsymbol{\theta}^*}\widetilde\gamma_2,
\end{align}
$\Sigma_U = E[U(\boldsymbol{\theta}^*) U(\boldsymbol{\theta}^*)^\top \mid S=1]$ is the second moment matrix of the stacked estimating function, $\Sigma_{\boldsymbol{\theta}^*} = \Gamma^{-1}\Sigma_U(\Gamma^{-1})^\top$ is the Godambe sandwich covariance\cite{Binder1983} of $\sqrt{n}(\widehat{\boldsymbol{\theta}} - \boldsymbol{\theta}^*)$ with Jacobian $\Gamma = E[\partial U(\boldsymbol{\theta}^*)/\partial\boldsymbol{\theta}^\top \mid S=1]$, $\gamma_1$ is the limiting cross-covariance vector of $n^{-1/2}\sum_{i=1}^n U_i(\boldsymbol\theta^*)$ with the root-$n$-scaled oracle matching error, and $\widetilde\gamma_2=d_R-\Gamma^\top\Sigma_U^{-1}\gamma_1$, where $d_R$ is the local matching drift coefficient defined in Web Appendix A.5.4.

\end{maintexttheorem}
\begin{proof}
For part (i), the matching and moment conditions are those of Assumption~\ref{ass:retrospective-pate-oracle}, and the bias-correction condition is \(\sqrt n(B_{n,R,\mathrm{PATE}}-\widehat B_{n,R,\mathrm{PATE}}) \to_p0\). Theorem~\ref{thm:retrospective-pate-oracle} therefore gives \eqref{eq:clt_known}, with the variance notation identified above.

For part (ii), apply Theorem~\ref{thm:retrospective-pate-estimated-scores}(i) under all conditions of Theorem~\ref{thm:pate-two-arm-actual-joint-law}, including the bias-correction remainder conditions for deterministic local parameters and fitted scores. Identify
\[
\boldsymbol\Gamma=\boldsymbol\Gamma_R,\qquad \boldsymbol\Sigma_U=\boldsymbol\Sigma_{R,U},\qquad
\boldsymbol\gamma_1=\boldsymbol\gamma_{R,1},\qquad
\boldsymbol\Sigma_{\boldsymbol\theta^*}=\boldsymbol\Sigma_{R,\theta},
\]
and write the main-text coefficient \(d_R\) as \(\boldsymbol b_{R,\mathrm{loc}}\) in the technical proofs. The limiting random variable is \(G_{\tau,R}+\boldsymbol b_{R,\mathrm{loc}}^\top\mathbf G_{\theta,R}\), where \(\boldsymbol b_{R,\mathrm{loc}}\) is given in \eqref{eq:pate-two-arm-actual-drift}. The joint martingale limit gives
\[
\operatorname{Cov}(\mathbf G_{\theta,R},G_{\tau,R})
=-\boldsymbol\Gamma^{-1}\boldsymbol\gamma_1,\qquad
\operatorname{Var}(\mathbf G_{\theta,R})
=\boldsymbol\Gamma^{-1}\boldsymbol\Sigma_U\boldsymbol\Gamma^{-\top}.
\]
Consequently,
\[
\begin{aligned}
\operatorname{Var}(G_{\tau,R}+\boldsymbol b_{R,\mathrm{loc}}^\top\mathbf G_{\theta,R})
&=V_R-2\boldsymbol b_{R,\mathrm{loc}}^\top\boldsymbol\Gamma^{-1}\boldsymbol\gamma_1
       +\boldsymbol b_{R,\mathrm{loc}}^\top\boldsymbol\Sigma_{\boldsymbol\theta^*}\boldsymbol b_{R,\mathrm{loc}}\\
&=V_R-\boldsymbol\gamma_1^\top\boldsymbol\Sigma_U^{-1}\boldsymbol\gamma_1
 +(\boldsymbol b_{R,\mathrm{loc}}-\boldsymbol\Gamma^\top\boldsymbol\Sigma_U^{-1}\boldsymbol\gamma_1)^\top
   \boldsymbol\Sigma_{\boldsymbol\theta^*}
  (\boldsymbol b_{R,\mathrm{loc}}-\boldsymbol\Gamma^\top\boldsymbol\Sigma_U^{-1}\boldsymbol\gamma_1).
\end{aligned}
\]
This is \eqref{eq:V_adj} and proves \eqref{eq:clt_estimated}.
\end{proof}

For PATT, the common control score is \(\mathbf D_0^*\), and
\[
\begin{aligned}
m_{1\mid1}(\mathbf d)
&=E_\Omega\{Y(1)\mid Z=1,\mathbf D_0^*(\mathbf X)=\mathbf d\},\\
m_0^R(\mathbf d)
&=E_\Omega\{Y(0)\mid\mathbf D_0^*(\mathbf X)=\mathbf d\},\\
\tau_{\mathrm{PATT}}(\mathbf d)&=m_{1\mid1}(\mathbf d)-m_0^R(\mathbf d).
\end{aligned}
\]
For PATT, use \(\boldsymbol\theta=(\boldsymbol\alpha^\top,\boldsymbol\beta_0^\top)^\top\) and the stacked propensity and control prognostic score estimating functions in Section~\ref{retrospective-patt-estimated-score-results}. \phantomsection\label{thm:patt_asymptotics-entry}
\begin{pattcounterpart}[Asymptotic Normality of Retrospective PATT Estimators]
\label{thm:patt_asymptotics}
Under Main-text Assumption~\ref{ass:retro_ignorability}, the one-sided identification condition $Y(0)\perp Z\mid\mathbf X$, and the matching and bias-correction regularity conditions in Web Appendix A.6.1--A.6.3, if either the propensity score model or the control prognostic score model is correctly specified:

\textbf{(i) Known scores.} The bias-corrected PATT estimator satisfies
\begin{align}\label{eq:patt_clt_known}
\sqrt{n}\left(\widehat{\tau}_{\mathrm{PATT}}^{R,\mathrm{BC}}(\boldsymbol{\theta}^*) - \tau_{\mathrm{PATT}}\right)
\Rightarrow
\mathcal{N}\!\left(0,\, V_{R,\mathrm{PATT}}\right),
\end{align}
where
\begin{align}\label{eq:V_PATT_oracle}
V_{R,\mathrm{PATT}}
&= V_{H,R,\mathrm{PATT}} + V_{E,R,\mathrm{PATT}}
\nonumber\\
&=
\frac{1}{\left(E[Z w_1(\mathbf{X}) \mid S = 1]\right)^2}
\operatorname{Var}\!\left(Z w_1(\mathbf{X})\{\tau_{\mathrm{PATT}}\{\mathbf D_0^*(\mathbf X)\} - \tau_{\mathrm{PATT}}\} \mid S = 1\right)
\nonumber\\
&\quad
+
\operatorname*{plim}_{n\to\infty}
\frac{n}{\left(\sum_{i:Z_i=1}w_1(\mathbf{X}_i)\right)^{2}}
\Bigg[
\sum_{i:Z_i=1}w_1^2(\mathbf{X}_i)\,v_{1,i}^{\mathrm{PATT}}
+
\sum_{j:Z_j=0}\{K_0^{R,\mathrm{PATT}}(j)\}^{2}\,v_{0,j}^{\mathrm{PATT}}
\Bigg].
\end{align}
For individuals with $Z_i=z$, $v_{z,i}^{\mathrm{PATT}}=\operatorname{Var}\{Y_i\mid Z_i=z,\mathbf D_0^*(\mathbf X_i),w_z(\mathbf X_i),S_i=1\}$. We use the linearization-based bootstrap below for inference.

\textbf{(ii) Estimated scores.} Under the additional first-stage and local-matching regularity conditions in Web Appendix A.6.4,
\begin{align}
\sqrt n\left\{
\widehat{\tau}_{\mathrm{PATT}}^{R,\mathrm{BC}}(\widehat{\boldsymbol\theta})
-\tau_{\mathrm{PATT}}
\right\}
\Rightarrow
\mathcal N(0,V_{R,\mathrm{PATT},\mathrm{adj}}),
\end{align}
where
\begin{align}\label{eq:V_PATT_adj}
V_{R,\mathrm{PATT},\mathrm{adj}}
=
V_{R,\mathrm{PATT}}
-\gamma_1^\top\Sigma_U^{-1}\gamma_1
+\widetilde\gamma_2^\top
\Sigma_{\boldsymbol{\theta}^*}\widetilde\gamma_2.
\end{align}
Here, $\Sigma_U=E\{U(\boldsymbol\theta^*)U(\boldsymbol\theta^*)^\top\mid S=1\}$, $\Gamma=E\{\partial U(\boldsymbol\theta^*)/\partial\boldsymbol\theta^\top\mid S=1\}$, and $\Sigma_{\boldsymbol\theta^*}=\Gamma^{-1}\Sigma_U(\Gamma^{-1})^\top$ are defined for the PATT-specific nuisance vector $\boldsymbol\theta=(\boldsymbol\alpha^\top,\boldsymbol\beta_0^\top)^\top$. The vector $\gamma_1$ is the limiting cross-covariance of the normalized estimating-function sum with the root-$n$-scaled PATT oracle matching error, and $\widetilde\gamma_2=-j_R-\Gamma^\top\Sigma_U^{-1}\gamma_1$, where $j_R$ is the local matching coefficient defined in Web Appendix A.6.4.
\end{pattcounterpart}
\begin{proof}
For part (i), apply the bias-corrected conclusion of Theorem~\ref{thm:retro-patt-oracle}, under Assumption~\ref{ass:retro-patt-oracle-regularity} and its stated BC remainder. The theorem's heterogeneity variance and conditional residual variance give exactly \eqref{eq:V_PATT_oracle}: the normalizer is the total treated survey weight and the controls contribute through \(K_{n,0}^{R,\mathrm{PATT}}\). Thus \eqref{eq:patt_clt_known} follows.

For part (ii), retain all conditions of Theorem~\ref{thm:retrospective-patt-estimated-scores}, including the actual-score conditions of Theorem~\ref{thm:patt-sufficient-branch-actual-score-law}. Define the PATT coefficient by
\[
\mathbf j_R=\left\{\frac{\Delta_R(\mathbf e_r)}
                   {\bar w_{R,\mathrm{PATT}}}\right\}_{r=1}^{p_R},
\qquad
\bar w_{R,\mathrm{PATT}}=E_s\{Zw_1(\mathbf X)\},
\]
where \(\mathbf e_r\) is the \(r\)-th coordinate vector and \(\Delta_R(h)\) is the control-side linear functional in \eqref{eq:patt-drift-fixed-law-geometric-limit}. Equation~\eqref{eq:patt-verified-local-mean-coefficient} then gives \(\widetilde{\boldsymbol \gamma}_2=-\mathbf j_R-\boldsymbol\Gamma^\top\boldsymbol\Sigma_U^{-1}\boldsymbol\gamma_1\). The actual-score conclusion and \eqref{eq:patt-verified-geometric-grid-variance} give \eqref{eq:V_PATT_adj}, with \(V_{R,\mathrm{PATT}}=V_{H,R,\mathrm{PATT}}+V_{E,R,\mathrm{PATT}}\).
\end{proof}

\subsection{Population identities and score relations}\label{population-representation-identities-and-score-relations}

We assume finite potential-outcome means in the identities below and \(P_\Omega(Z=1)>0\) for PATT.

\subsubsection{Sampling identities}
\label{retrospective-population-representation-identities}
\label{prospective-population-representation-identities}
\label{selected-sample-propensity-relations}

\begin{lemma}[Population recovery and propensity identities]
\label{lem:retrospective-population-recovery}
\label{lem:prospective-population-recovery}
\label{lem:retrospective-propensity-bayes}
\label{lem:prospective-propensity-identity}
Let \(H=H(\mathbf X,Z,Y(0),Y(1))\) be integrable under the population law.
\begin{enumerate}
\item In retrospective sampling, suppose \(0<\pi_z(\mathbf X)\le1\) on the relevant population support, \(w_z(\mathbf X)=\pi_z(\mathbf X)^{-1}\), and \((Y(0),Y(1))\perp S\mid\mathbf X,Z\), where \(\pi_z(\mathbf X)=P_\Omega(S=1\mid\mathbf X,Z=z)\). Then
\[E_s\!\left[w_Z(\mathbf X)H\right]
=
\frac{E_{\Omega}[H]}{P_{\Omega}(S=1)}.\] Consequently,
\[\tau_{\mathrm{PATE}}
=
\frac{E_s\!\left[w_Z(\mathbf X)\{Y(1)-Y(0)\}\right]}{E_s\!\left[w_Z(\mathbf X)\right]},\]
and \[\tau_{\mathrm{PATT}}
=
\frac{E_s\!\left[Z\,w_1(\mathbf X)\{Y(1)-Y(0)\}\right]}{E_s\!\left[Z\,w_1(\mathbf X)\right]}.\]

If additionally \(0<e(\mathbf X)<1\), then
\[
\frac{e(\mathbf X)}{1-e(\mathbf X)}
=\frac{w_1(\mathbf X)}{w_0(\mathbf X)}
 \frac{e_s(\mathbf X)}{1-e_s(\mathbf X)},
\qquad
e(\mathbf X)=
\frac{w_1(\mathbf X)e_s(\mathbf X)}
 {w_1(\mathbf X)e_s(\mathbf X)+w_0(\mathbf X)\{1-e_s(\mathbf X)\}}.
\]
\item In prospective sampling, suppose \(0<\pi(\mathbf X)\le1\), \(w(\mathbf X)=\pi(\mathbf X)^{-1}\), and \((Z,Y(0),Y(1))\perp S\mid\mathbf X\), where \(\pi(\mathbf X)=P_\Omega(S=1\mid\mathbf X)\). Then
\[E_s\!\left[w(\mathbf X)H\right]
=
\frac{E_{\Omega}[H]}{P_{\Omega}(S=1)}.\] Consequently,
\[\tau_{\mathrm{PATE}}
=
\frac{E_s\!\left[w(\mathbf X)\{Y(1)-Y(0)\}\right]}{E_s\!\left[w(\mathbf X)\right]},\]
and \[\tau_{\mathrm{PATT}}
=
\frac{E_s\!\left[Z\,w(\mathbf X)\{Y(1)-Y(0)\}\right]}{E_s\!\left[Z\,w(\mathbf X)\right]}.\]
Moreover, \(e_s(\mathbf X)=e(\mathbf X)\).
\end{enumerate}

\end{lemma}
\begin{proof}
For retrospective sampling, conditioning on the population variables and using sampling ignorability and \(w_Z\pi_Z=1\) gives
\[
\begin{aligned}
E_s[w_Z(\mathbf X)H]
&=\frac{E_\Omega[S w_Z(\mathbf X)H]}{P_\Omega(S=1)}\\
&=\frac{E_\Omega[w_Z(\mathbf X)H
 E_\Omega\{S\mid\mathbf X,Z,Y(0),Y(1)\}]}{P_\Omega(S=1)}\\
&=\frac{E_\Omega[H]}{P_\Omega(S=1)}.
\end{aligned}
\]
Applying this identity to \(H=Y(1)-Y(0),1\) gives the PATE representation. Applying it to \(H=Z\{Y(1)-Y(0)\},Z\) gives the PATT representation. For PATE, the separate sampling and treatment conditions in Assumption~\ref{ass:prospective-pate-oracle}(i) imply the joint independence in part~(ii). This joint condition implies \((Y(0),Y(1))\perp S\mid\mathbf X,Z\) and \(\pi_z(\mathbf X)=\pi(\mathbf X)\) for both arms. The same calculation with \(w_0=w_1=w\) proves the prospective identities. Bayes' rule gives
\[
e_s(\mathbf X)
=\frac{e(\mathbf X)\pi_1(\mathbf X)}
 {e(\mathbf X)\pi_1(\mathbf X)+\{1-e(\mathbf X)\}\pi_0(\mathbf X)}.
\]
Taking odds and using \(\pi_z=w_z^{-1}\) gives
\[
\frac{e_s(\mathbf X)}{1-e_s(\mathbf X)}
=\frac{\pi_1(\mathbf X)}{\pi_0(\mathbf X)}
 \frac{e(\mathbf X)}{1-e(\mathbf X)}.
\]
Rearrangement gives the retrospective propensity identities. In prospective sampling, \(Z\perp S\mid\mathbf X\), so \(P_\Omega(Z=1\mid\mathbf X,S=1)=P_\Omega(Z=1\mid\mathbf X)\).
\end{proof}

\subsubsection{Treatment-specific double-score balancing}
Let \(\boldsymbol\theta^*\) denote the probability limit of the first-stage parameter estimator. The limiting double score \(\mathbf D_z^*(\mathbf X)\) is a fixed nonsingular standardization of \(\{e(\mathbf X;\boldsymbol\alpha^*), \Psi_z(\mathbf X;\boldsymbol\beta_z^*)\}\).

\begin{lemma}[Arm-specific double-score balancing]
\label{lem:armwise-double-score-balancing}
Under PATE treatment ignorability
\[
(Y(0),Y(1))\perp Z\mid\mathbf X,
\]
the distributional statement
\[
Y(z)\perp Z\mid\mathbf D_z^*(\mathbf X)
\]
holds if either the propensity model is correct or the limiting prognostic score is distributionally sufficient:
\[
Y(z)\perp\mathbf X\mid\Psi_z(\mathbf X;\boldsymbol\beta_z^*).
\]
If only the prognostic conditional mean is correct, so that \(\mu_z(\mathbf X)=E_\Omega\{Y(z)\mid\mathbf X\}\) is measurable in \(\mathbf D_z^*(\mathbf X)\), the required conclusion is instead
\[
E_\Omega\{Y(z)\mid Z,\mathbf D_z^*(\mathbf X)\}
=E_\Omega\{Y(z)\mid\mathbf D_z^*(\mathbf X)\}.
\]
This mean-balancing conclusion also follows from the distributional statement. Under the one-sided PATT condition \(Y(0)\perp Z\mid\mathbf X\), the same respective conclusions hold for \(z=0\), without imposing ignorability of \(Y(1)\).
\end{lemma}
\begin{proof}
If the propensity model is correct, \(e(\mathbf X)\) is measurable in \(\mathbf D_z^*\), and
\[
P_\Omega(Z=1\mid\mathbf X,\mathbf D_z^*)=e(\mathbf X)
=P_\Omega(Z=1\mid\mathbf D_z^*).
\]
Thus \(\mathbf X\perp Z\mid\mathbf D_z^*\). For every bounded measurable \(f\), treatment ignorability gives
\[
\begin{aligned}
E_\Omega\{f(Y(z))\mid Z,\mathbf D_z^*\}
&=E_\Omega\{E_\Omega[f(Y(z))\mid\mathbf X,Z]
                                       \mid Z,\mathbf D_z^*\}\\
&=E_\Omega\{E_\Omega[f(Y(z))\mid\mathbf X]
                                       \mid Z,\mathbf D_z^*\}\\
&=E_\Omega\{E_\Omega[f(Y(z))\mid\mathbf X]
                                       \mid\mathbf D_z^*\}\\
&=E_\Omega\{f(Y(z))\mid\mathbf D_z^*\}.
\end{aligned}
\]
If the prognostic score is distributionally sufficient, we instead obtain
\[
\begin{aligned}
E_\Omega\{f(Y(z))\mid Z,\mathbf D_z^*\}
&=E_\Omega\{E_\Omega[f(Y(z))\mid\mathbf X]
                                      \mid Z,\mathbf D_z^*\}\\
&=E_\Omega\{E_\Omega[f(Y(z))\mid\Psi_z]
                                      \mid Z,\mathbf D_z^*\}\\
&=E_\Omega\{f(Y(z))\mid\Psi_z\}\\
&=E_\Omega\{f(Y(z))\mid\mathbf D_z^*\},
\end{aligned}
\]
since \(\Psi_z\) is measurable in \(\mathbf D_z^*\).

For a correct prognostic mean,
\[
\begin{aligned}
E_\Omega\{Y(z)\mid Z,\mathbf D_z^*\}
&=E_\Omega\{E_\Omega[Y(z)\mid\mathbf X,Z]\mid Z,\mathbf D_z^*\}\\
&=E_\Omega\{\mu_z(\mathbf X)\mid Z,\mathbf D_z^*\}\\
&=\mu_z(\mathbf X)\\
&=E_\Omega\{Y(z)\mid\mathbf D_z^*\}.
\end{aligned}
\]
Only \(Y(0)\perp Z\mid\mathbf X\) is used when \(z=0\).
\end{proof}

For retrospective sampling, we define the within-match coefficients by
\[
\lambda_{i\to j}^{R,(z)}
=
\frac{w_z(\mathbf X_j)\mathbf 1\{j\in\mathcal J_{R,z}(i)\}}
{\sum_{k\in\mathcal J_{R,z}(i)} w_z(\mathbf X_k)},
\]
and for prospective sampling,
\[
\lambda_{i\to j}^{P,(z)}
=
\frac{w(\mathbf X_j)\mathbf 1\{j\in\mathcal J_{P,z}(i)\}}
{\sum_{k\in\mathcal J_{P,z}(i)} w(\mathbf X_k)}.
\]
When the sampling regime is fixed, we suppress the superscript \(A\in\{R,P\}\). For PATT, we use \(z=0\) and the treated-to-control matched set.

\subsubsection{Survey-weighted conditional means}\label{survey-weighted-matched-set-conditional-means}

Define the population conditional mean given the limiting double score by
\[
m_z^A(d):=E_\Omega\{Y(z)\mid\mathbf D_z^*(\mathbf X)=d\},
\qquad A\in\{R,P\}.
\]

For PATT, we write
\[
m_{1\mid1}(d):=E_\Omega\{Y(1)\mid Z=1,\mathbf D_0^*(\mathbf X)=d\}.
\]

\begin{theorem}[Weighted conditional means and identification]
\label{lem:weighted-score-population-mean}
\label{thm:four-scenario-score-identification}
\label{lem:score-space-rewrite-mean-score}
Fix an arm \(z\in\{0,1\}\), assume consistency and \(E_\Omega|Y(z)|<\infty\), and use the fixed limiting score \(\mathbf D_z^*(\mathbf X)\). The relevant observed-arm score support covers the target support, so the following conditional ratios are defined where they are used.

Assume the relevant treatment mean-balancing identity
\[
E_\Omega\{Y(z)\mid Z,\mathbf D_z^*(\mathbf X)\}
=E_\Omega\{Y(z)\mid\mathbf D_z^*(\mathbf X)\}.
\]

\emph{(i) Weighted conditional means.} For retrospective sampling, under the conditions of Lemma~\ref{lem:retrospective-population-recovery}(i),
\[
m_z^{R}(d)
=
\frac{E_s[w_z(\mathbf X)Y\mid Z=z,\mathbf D_z^*(\mathbf X)=d]}{E_s[w_z(\mathbf X)\mid Z=z,\mathbf D_z^*(\mathbf X)=d]}.
\]

For prospective sampling, under the conditions of Lemma~\ref{lem:prospective-population-recovery}(ii),
\[
m_z^{P}(d)
=
\frac{E_s[w(\mathbf X)Y\mid Z=z,\mathbf D_z^*(\mathbf X)=d]}{E_s[w(\mathbf X)\mid Z=z,\mathbf D_z^*(\mathbf X)=d]}.
\]

\emph{(ii) Identification.} Assume the conditions in part (i) for both arms in PATE and only the control arm in PATT. For PATT also assume \(E_\Omega\{|Y(1)|\mid Z=1\}<\infty\) and \(P_\Omega(Z=1)>0\).

For either sampling regime \(A\in\{R,P\}\), the PATE conditions give
\[
\begin{aligned}
E_\Omega[Y(z)]
&=E_\Omega\!\left[m_z^A\{\mathbf D_z^*(\mathbf X)\}\right],
\qquad z=0,1,\\
\tau_{\mathrm{PATE}}
&=E_\Omega\!\left[m_1^A\{\mathbf D_1^*(\mathbf X)\}
                  -m_0^A\{\mathbf D_0^*(\mathbf X)\}\right].
\end{aligned}
\]
Under the PATT conditions,
\[
\tau_{\mathrm{PATT}}
=E_\Omega\!\left[m_{1\mid1}\{\mathbf D_0^*(\mathbf X)\}
                  -m_0^A\{\mathbf D_0^*(\mathbf X)\}\mid Z=1\right].
\]

\end{theorem}
\begin{proof}
Write \(D=\mathbf D_z^*(\mathbf X)\). Under retrospective sampling, consistency and sampling ignorability give
\[
\begin{aligned}
\frac{E_s[w_z(\mathbf X)Y\mid Z=z,D]}
     {E_s[w_z(\mathbf X)\mid Z=z,D]}
&=\frac{E_\Omega[S w_z(\mathbf X)Y(z)\mid Z=z,D]}
        {E_\Omega[S w_z(\mathbf X)\mid Z=z,D]}\\
&=\frac{E_\Omega[w_z(\mathbf X)\pi_z(\mathbf X)
                  E_\Omega\{Y(z)\mid\mathbf X,Z=z\}\mid Z=z,D]}
        {E_\Omega[w_z(\mathbf X)\pi_z(\mathbf X)\mid Z=z,D]}\\
&=E_\Omega\{Y(z)\mid Z=z,D\}\\
&=E_\Omega\{Y(z)\mid D\}=m_z^R(D).
\end{aligned}
\]
Treatment mean balancing gives the last equality. Taking \(\pi_0=\pi_1=\pi\) and \(w_0=w_1=w\) gives the prospective representation of \(m_z^P\).

For \(A\in\{R,P\}\), the tower property gives
\[
E_\Omega\{m_z^A(\mathbf D_z^*)\}=E_\Omega\{Y(z)\}.
\]
Subtracting the two arm identities gives the PATE result. For PATT, control mean balancing on the common score gives
\[
\begin{aligned}
\tau_{\mathrm{PATT}}
&=E_\Omega\!\left[
 E_\Omega\{Y(1)-Y(0)\mid Z=1,\mathbf D_0^*\}\mid Z=1\right]\\
&=E_\Omega\!\left[
 m_{1\mid1}(\mathbf D_0^*)-E_\Omega\{Y(0)\mid\mathbf D_0^*\}
 \mid Z=1\right]\\
&=E_\Omega\!\left[
 m_{1\mid1}(\mathbf D_0^*)-m_0^A(\mathbf D_0^*)\mid Z=1\right].
\end{aligned}
\]

Under consistency, treatment ignorability, overlap, and the corresponding sampling conditions, a correctly specified prognostic mean \(\Psi_z(\mathbf X;\boldsymbol\beta_z^*)=\mu_z(\mathbf X)\) gives, for \(m_{s,z}(d):=E_s[Y\mid Z=z,\mathbf D_z^*(\mathbf X)=d]\),
\[
\begin{aligned}
m_{s,z}(\mathbf D_z^*)
&=E_s[Y\mid Z=z,\mathbf D_z^*]\\
&=E_s\{E_s[Y(z)\mid\mathbf X,Z=z]\mid Z=z,\mathbf D_z^*\}\\
&=E_s\{\mu_z(\mathbf X)\mid Z=z,\mathbf D_z^*\}\\
&=\mu_z(\mathbf X).
\end{aligned}
\]
Overlap extends this version to the target score support. Its population analogue is \(E_\Omega\{Y(z)\mid\mathbf D_z^*\}=\mu_z(\mathbf X)\). When both prognostic means are correct,
\[
\begin{aligned}
\tau_{\mathbf D}(\mathbf X)
&:=E_\Omega\{Y(1)\mid\mathbf D_1^*\}
  -E_\Omega\{Y(0)\mid\mathbf D_0^*\}\\
&=\mu_1(\mathbf X)-\mu_0(\mathbf X).
\end{aligned}
\]
For PATT, correctness of the control prognostic mean gives
\[
\begin{aligned}
\tau_{\mathrm{PATT}}(\mathbf D_0^*)
&:=m_{1\mid1}(\mathbf D_0^*)
  -E_\Omega\{Y(0)\mid\mathbf D_0^*\}\\
&=m_{1\mid1}(\mathbf D_0^*)-\mu_0(\mathbf X),\qquad Z=1.
\end{aligned}
\]
For either PATE arm, or for the PATT control arm, the matched-set discrepancy is
\[
\begin{aligned}
&\mu_z(\mathbf X_i)
 -\sum_{j=1}^n\lambda_{i\to j}^{(z)}\mu_z(\mathbf X_j)\\
&\quad=m_{s,z}\{\mathbf D_z^*(\mathbf X_i)\}
 -\sum_{j=1}^n\lambda_{i\to j}^{(z)}
                     m_{s,z}\{\mathbf D_z^*(\mathbf X_j)\}.
\end{aligned}
\]
\end{proof}

\subsection{Matching estimators and oracle limits}
\label{common-matching-algebraic-tools}

\subsubsection{Survey-weighted within-match coefficients}\label{exact-within-match-survey-weight-normalized-within-match-coefficients}

For \(A\in\{R,P\}\), write
\[
w_{A,z}(\mathbf X):=
\begin{cases}
w_z(\mathbf X),& A=R,\\
w(\mathbf X),& A=P.
\end{cases}
\]

\begin{definition}[Survey-weighted within-match coefficient]
Let \(\mathcal J_{\mathbf D_z}(i)\) contain the \(M\) nearest observations in arm \(z\) to unit \(i\), using the specified score distance. For \(A\in\{R,P\}\), define
\[
\lambda_{i\to j}^{A,(z)}
:=
\frac{w_{A,z}(\mathbf X_j)\mathbf1\{j\in\mathcal J_{\mathbf D_z}(i)\}}
{\sum_{k\in\mathcal J_{\mathbf D_z}(i)}w_{A,z}(\mathbf X_k)}.
\]
PATE uses \(z=0,1\), whereas PATT uses \(z=0\) for treated individuals. We include the estimand in the superscript when both are considered and suppress the sampling-design superscript when the design is fixed. For each individual with a defined matched set,
\[
\sum_{j=1}^n\lambda_{i\to j}^{(z)}=1.
\]
\end{definition}

\subsubsection{Oracle double-score decomposition}
\label{oracle-score-space-decomposition-behind-the-double-robust-theorems}

We decompose the matching error into heterogeneity, residual variation, and matching discrepancy, as in Abadie and Imbens\citep{Abadie2006} and Yang and Zhang.\citep{yang_multiply_2023} The oracle decomposition uses matches on the fixed scores \(\mathbf D_z^*\) and the reduced means identified in Theorem~\ref{lem:weighted-score-population-mean}(i). For an observation in arm \(z\), define the corresponding residual
\[
\varepsilon_{z}^{A}:=Y-m_z^A\{\mathbf D_z^*(\mathbf X)\}.
\]
For PATE, we write
\[
\tau_A^D(\mathbf X)
:=
m_1^A\{\mathbf D_1^*(\mathbf X)\}
-
m_0^A\{\mathbf D_0^*(\mathbf X)\}.
\]

By Theorem~\ref{lem:weighted-score-population-mean}(i),
\[
E_s[w_{A,z}(\mathbf X)\varepsilon_z^A
       \mid Z=z,\mathbf D_z^\ast(\mathbf X)]=0.
\]

For the oracle PATE decomposition, we use the initial sigma-field
\[
\mathcal G_{A,n}^{\mathrm{match}}
:=\sigma\!\left\{
Z_i,\mathbf D_0^*(\mathbf X_i),\mathbf D_1^*(\mathbf X_i),
w_{A,Z_i}(\mathbf X_i):1\le i\le n
\right\},\qquad A\in\{R,P\}.
\]
For PATT, we define the treated residual on the common control score by
\[
m_{1\mid1}(\mathbf d)
:=E_\Omega\{Y(1)\mid Z=1,\mathbf D_0^*(\mathbf X)=\mathbf d\},
\qquad
\varepsilon_{1\mid1,i}
:=Y_i-m_{1\mid1}\{\mathbf D_0^*(\mathbf X_i)\},\quad Z_i=1.
\]
In this scenario the oracle initial sigma-field is
\[
\mathcal G_{A,n}^{\mathrm{match}}
:=\sigma\!\left\{
Z_i,\mathbf D_0^*(\mathbf X_i),w_{A,Z_i}(\mathbf X_i):1\le i\le n
\right\}.
\]

\begin{assumption}[Conditional residual centering]
\label{ass:residual-centering-survey-normalized}
In addition to population recovery and treatment mean balancing, assume the following conditions for the estimand under consideration.
\begin{enumerate}
\item For PATE, for \(z=0,1\),
\[
E_s\!\left[
Y-m_z^A\{\mathbf D_z^*(\mathbf X)\}
\mid Z=z,\mathbf D_0^*(\mathbf X),\mathbf D_1^*(\mathbf X),
w_{A,z}(\mathbf X)
\right]=0.
\]
\item For PATT,
\[
\begin{aligned}
E_s\!\left[
Y-m_{1\mid1}\{\mathbf D_0^*(\mathbf X)\}
\mid Z=1,\mathbf D_0^*(\mathbf X),w_{A,1}(\mathbf X)
\right]&=0,\\
E_s\!\left[
Y-m_0^A\{\mathbf D_0^*(\mathbf X)\}
\mid Z=0,\mathbf D_0^*(\mathbf X),w_{A,0}(\mathbf X)
\right]&=0.
\end{aligned}
\]
\end{enumerate}
\end{assumption}

For PATE, fix \(A\in\{R,P\}\). Let \(\lambda_{i\to j}^{A,(1-Z_i)}\) be the survey-normalized within-match coefficient used to impute the missing potential outcome for unit \(i\). Define
\[
\bar w_{A,n}:=\frac1n\sum_{i=1}^n w_{A,Z_i}(\mathbf X_i),
\]
and let
\[
K_{n,z}^{A}(j)
:=
\sum_{i:Z_i=1-z}
w_{A,1-z}(\mathbf X_i)\lambda_{i\to j}^{A,(z)},
\qquad Z_j=z,
\]
be the total weighted contribution assigned to that donor.

We define the three decomposition terms by
\[
H_{n,A}^D
:=
\bar w_{A,n}^{-1}\frac1n\sum_{i=1}^n
w_{A,Z_i}(\mathbf X_i)
\{\tau_A^D(\mathbf X_i)-\tau_{\mathrm{PATE}}\},
\]
\[
E_{n,A}^D
:=
\bar w_{A,n}^{-1}\frac1n
\sum_{z=0}^1(2z-1)
\sum_{j:Z_j=z}
\{w_{A,z}(\mathbf X_j)+K_{n,z}^{A}(j)\}
\varepsilon_{z,j}^A,
\]
and
\[
B_{n,A}^D
:=
\bar w_{A,n}^{-1}\frac1n\sum_{i=1}^n
w_{A,Z_i}(\mathbf X_i)(2Z_i-1)
\sum_{j=1}^n\lambda_{i\to j}^{A,(1-Z_i)}
\left[
m_{1-Z_i}^A\{\mathbf D_{1-Z_i}^*(\mathbf X_i)\}
-
m_{1-Z_i}^A\{\mathbf D_{1-Z_i}^*(\mathbf X_j)\}
\right].
\]

For PATT, we collect the matched control outcomes on the common control-score scale. In the scenario under consideration, we write
\[
\tau_{\mathrm{PATT}}(\mathbf d)
:=m_{1\mid1}(\mathbf d)-m_0^A(\mathbf d).
\]
For the outer normalization, we use
\[
\begin{aligned}
\bar w_{A,1,n}
&:=\frac1n\sum_{i=1}^n Z_iw_{A,1}(\mathbf X_i)\\
&=\frac1n\sum_{i:Z_i=1}w_{A,1}(\mathbf X_i)\\
&=\frac{n_1}{n}
  \left\{\frac1{n_1}\sum_{i:Z_i=1}w_{A,1}(\mathbf X_i)\right\}.
\end{aligned}
\]
We define the control reuse coefficient by
\[
K_{n,0}^{A,\mathrm{PATT}}(j)
:=\sum_{i:Z_i=1}w_{A,1}(\mathbf X_i)
\lambda_{i\to j}^{A,(0)},\qquad Z_j=0.
\]

The corresponding decomposition terms are
\[
H_{n,A}^{D,\mathrm{PATT}}
:=\frac{1}{n\bar w_{A,1,n}}
\sum_{i:Z_i=1}w_{A,1}(\mathbf X_i)
\left[\tau_{\mathrm{PATT}}\{\mathbf D_0^*(\mathbf X_i)\}
-\tau_{\mathrm{PATT}}\right],
\]
\[
E_{n,A}^{D,\mathrm{PATT}}
:=\frac{1}{n\bar w_{A,1,n}}
\left[\sum_{i:Z_i=1}w_{A,1}(\mathbf X_i)\varepsilon_{1\mid1,i}
-\sum_{j:Z_j=0}K_{n,0}^{A,\mathrm{PATT}}(j)
\varepsilon_{0,j}^A\right],
\]
and
\[
B_{n,A}^{D,\mathrm{PATT}}
:=\frac{1}{n\bar w_{A,1,n}}
\sum_{i:Z_i=1}w_{A,1}(\mathbf X_i)
\sum_j\lambda_{i\to j}^{A,(0)}
\left[m_0^A\{\mathbf D_0^*(\mathbf X_i)\}
-m_0^A\{\mathbf D_0^*(\mathbf X_j)\}\right].
\]

\begin{proposition}[Matching decompositions and conditional centering]
\label{prop:exact-pate-score-space-decomposition}
\label{lem:residual-centering-survey-normalized}
\label{cor:oracle-h-e-orthogonality}
\emph{(i) PATE.} The oracle PATE WDSM estimator satisfies the finite-sample identity
\[
\widehat\tau_{\mathrm{PATE}}^{A}-\tau_{\mathrm{PATE}}
=H_{n,A}^D+E_{n,A}^D+B_{n,A}^D,
\]

\emph{(ii) PATT.} The estimator satisfies
\[
\widehat\tau_{\mathrm{PATT}}^{A}-\tau_{\mathrm{PATT}}
=H_{n,A}^{D,\mathrm{PATT}}+E_{n,A}^{D,\mathrm{PATT}}
+B_{n,A}^{D,\mathrm{PATT}},
\]

\emph{(iii) Conditional centering.} Under Assumption~\ref{ass:residual-centering-survey-normalized}, for \(A\in\{R,P\}\),
\[
E_s\!\left[K_{n,z}^{A}(j)\varepsilon_{z,j}^{A}
                  \mid\mathcal G_{A,n}^{\mathrm{match}}\right]=0
\]
for \(z=0,1\) in PATE and for the matched control arm in PATT. The residual term in either decomposition has conditional mean zero given its corresponding \(\mathcal G_{A,n}^{\mathrm{match}}\). Consequently the heterogeneity and residual terms are uncorrelated whenever they have finite second moments.
\end{proposition}
\begin{proof}
For PATE, since \(\sum_j\lambda_{i\to j}^{A,(1-Z_i)}=1\),
\[
\begin{aligned}
&\widehat Y_i^{A}(1)-\widehat Y_i^{A}(0)\\
&=(2Z_i-1)\left[Y_i-\sum_j
\lambda_{i\to j}^{A,(1-Z_i)}Y_j\right]\\
&=(2Z_i-1)
\left[
m_{Z_i}^A\{\mathbf D_{Z_i}^*(\mathbf X_i)\}
-\sum_j\lambda_{i\to j}^{A,(1-Z_i)}
m_{1-Z_i}^A\{\mathbf D_{1-Z_i}^*(\mathbf X_j)\}
\right]\\
&\quad+
(2Z_i-1)
\left[
\varepsilon_{Z_i,i}^A
-\sum_j\lambda_{i\to j}^{A,(1-Z_i)}
\varepsilon_{1-Z_i,j}^A
\right]\\
&=\tau_A^D(\mathbf X_i)\\
&\quad+
(2Z_i-1)\sum_j\lambda_{i\to j}^{A,(1-Z_i)}
\left[
m_{1-Z_i}^A\{\mathbf D_{1-Z_i}^*(\mathbf X_i)\}
-m_{1-Z_i}^A\{\mathbf D_{1-Z_i}^*(\mathbf X_j)\}
\right]\\
&\quad+
(2Z_i-1)
\left[
\varepsilon_{Z_i,i}^A
-\sum_j\lambda_{i\to j}^{A,(1-Z_i)}
\varepsilon_{1-Z_i,j}^A
\right].
\end{aligned}
\]
Reindexing the weighted residual sum by the matched arm gives
\[
\begin{aligned}
&\sum_i w_{A,Z_i}(\mathbf X_i)(2Z_i-1)
 \left\{\varepsilon_{Z_i,i}^A
 -\sum_j\lambda_{i\to j}^{A,(1-Z_i)}
                \varepsilon_{1-Z_i,j}^A\right\}\\
&=\sum_{z=0}^1(2z-1)\sum_{j:Z_j=z}
                 w_{A,z}(\mathbf X_j)\varepsilon_{z,j}^A
 -\sum_{z=0}^1(1-2z)\sum_{i:Z_i=1-z}\sum_{j:Z_j=z}
  w_{A,1-z}(\mathbf X_i)\lambda_{i\to j}^{A,(z)}
                       \varepsilon_{z,j}^A\\
&=\sum_{z=0}^1(2z-1)\sum_{j:Z_j=z}
 \left\{w_{A,z}(\mathbf X_j)
 +\sum_{i:Z_i=1-z}w_{A,1-z}(\mathbf X_i)
                   \lambda_{i\to j}^{A,(z)}\right\}
                  \varepsilon_{z,j}^A\\
&=\sum_{z=0}^1(2z-1)\sum_{j:Z_j=z}
 \{w_{A,z}(\mathbf X_j)+K_{n,z}^{A}(j)\}
                  \varepsilon_{z,j}^A.
\end{aligned}
\]
Division by \(n\bar w_{A,n}\) and subtraction of \(\tau_{\mathrm{PATE}}\) yield the stated decomposition.

For PATT, identification and iterated expectation express the target on the common control score:
\[
\begin{aligned}
\tau_{\mathrm{PATT}}
&=E_\Omega\!\left[Y(1)-m_0^A\{\mathbf D_0^*(\mathbf X)\}\mid Z=1\right]\\
&=E_\Omega\!\left[
E_\Omega\{Y(1)\mid Z=1,\mathbf D_0^*(\mathbf X)\}
-m_0^A\{\mathbf D_0^*(\mathbf X)\}\mid Z=1\right]\\
&=E_\Omega\!\left[
m_{1\mid1}\{\mathbf D_0^*(\mathbf X)\}
-m_0^A\{\mathbf D_0^*(\mathbf X)\}\mid Z=1\right]\\
&=E_\Omega\!\left[
\tau_{\mathrm{PATT}}\{\mathbf D_0^*(\mathbf X)\}\mid Z=1\right].
\end{aligned}
\]
For a treated unit \(i\), expanding the outcomes about their reduced means and using the normalized within-match coefficients gives
\[
\begin{aligned}
Y_i-\sum_j\lambda_{i\to j}^{A,(0)}Y_j
&=m_{1\mid1}\{\mathbf D_0^*(\mathbf X_i)\}
+\varepsilon_{1\mid1,i}
-\sum_j\lambda_{i\to j}^{A,(0)}Y_j\\
&=m_{1\mid1}\{\mathbf D_0^*(\mathbf X_i)\}
+\varepsilon_{1\mid1,i}
-\sum_j\lambda_{i\to j}^{A,(0)}
\left[m_0^A\{\mathbf D_0^*(\mathbf X_j)\}+\varepsilon_{0,j}^A\right]\\
&=\tau_{\mathrm{PATT}}\{\mathbf D_0^*(\mathbf X_i)\}
+m_0^A\{\mathbf D_0^*(\mathbf X_i)\}
+\varepsilon_{1\mid1,i}\\
&\quad-\sum_j\lambda_{i\to j}^{A,(0)}
\left[m_0^A\{\mathbf D_0^*(\mathbf X_j)\}+\varepsilon_{0,j}^A\right]\\
&=\tau_{\mathrm{PATT}}\{\mathbf D_0^*(\mathbf X_i)\}
+\varepsilon_{1\mid1,i}
-\sum_j\lambda_{i\to j}^{A,(0)}\varepsilon_{0,j}^A\\
&\quad+\sum_j\lambda_{i\to j}^{A,(0)}
\left[m_0^A\{\mathbf D_0^*(\mathbf X_i)\}
-m_0^A\{\mathbf D_0^*(\mathbf X_j)\}\right].
\end{aligned}
\]
Multiplying by \(Z_iw_{A,1}(\mathbf X_i)\), summing over \(i\), and collecting each control residual gives the PATT decomposition.

For conditional centering, independence of the rows gives a factorization after conditioning on their respective score and weight coordinates. Under the PATE field,
\[
E_s\!\left[\prod_{i=1}^n f_i(Y_i)\mid\mathcal G_{A,n}^{\mathrm{match}}\right]
=\prod_{i=1}^n E_s\!\left[
 f_i(Y_i)\mid Z_i,\mathbf D_0^*(\mathbf X_i),
 \mathbf D_1^*(\mathbf X_i),w_{A,Z_i}(\mathbf X_i)\right]
\]
for bounded measurable \(f_i\). For PATT omit \(\mathbf D_1^*\). By the rowwise centering assumption and coefficient measurability,
\[
E_s[K_{n,z}^{A}(j)\varepsilon_{z,j}^A\mid\mathcal G_{A,n}^{\mathrm{match}}]
=K_{n,z}^{A}(j)E_s(\varepsilon_{z,j}^A\mid\mathcal G_{A,n}^{\mathrm{match}})=0.
\]
For a treated PATT observation, \(E_s\{w_{A,1}(\mathbf X_i)\varepsilon_{1\mid1,i}\mid\mathcal G_{A,n}^{\mathrm{match}}\}=0\). For either decomposition, measurability of \(H\), the coefficients and the denominator consequently gives
\[
\begin{aligned}
E_s(E\mid\mathcal G_{A,n}^{\mathrm{match}})&=0,\\
E_s(E)&=E_s\{E_s(E\mid\mathcal G_{A,n}^{\mathrm{match}})\}=0,\\
\operatorname{Cov}_s(H,E)
&=E_s\{H E_s(E\mid\mathcal G_{A,n}^{\mathrm{match}})\}=0.
\end{aligned}
\]
In particular, for every valid matched set,
\[
\begin{aligned}
&E_s\!\left[
 \sum_{j\in\mathcal J_{\mathbf D_z^*}(i)}
 \lambda_{i\to j}^{A,(z)}\varepsilon_{z,j}^A
 \,\middle|\,\mathcal G_{A,n}^{\mathrm{match}}\right]\\
&=\sum_{j\in\mathcal J_{\mathbf D_z^*}(i)}
 \lambda_{i\to j}^{A,(z)}
 E_s(\varepsilon_{z,j}^A\mid\mathcal G_{A,n}^{\mathrm{match}})=0.
\end{aligned}
\]

\end{proof}

\subsubsection{Conditional martingale limits}\label{a-common-martingale-array-clt-lemma-for-survey-weighted-matched-set-coefficient-residual-sums}

Following Abadie and Imbens,\citep{AbadieImbens2012} we collect each individual's repeated residual contributions and reveal the observations in a fixed order. The conditioning field in Assumption~\ref{ass:residual-centering-survey-normalized} determines the survey-weighted coefficients for both estimands and sampling designs.

\begin{lemma}[Martingale and conditional central limit theorems]
\label{lem:conditional-residual-stable-clt}
\label{lem:conditional-residual-martingale-clt}
Let \((\mathcal F_{n,\ell})_{\ell=0}^{m_n}\) be a filtration with \(\mathcal F_{n,0}=\mathcal G_n\). For each \(n\), let \(\xi_{n,\ell}\) be square-integrable and \(\mathcal F_{n,\ell}\)-measurable. Assume that for every \(\ell\),
\[E(\xi_{n,\ell}\mid \mathcal F_{n,\ell-1})=0.\] Define the predictable
quadratic variation by
\[V_n:=\sum_{\ell=1}^{m_n}E(\xi_{n,\ell}^2\mid \mathcal F_{n,\ell-1}).\]
Assume that there exists \(\delta>0\) and a constant \(V\in[0,\infty)\)
such that \[V_n\to_p V,\] and
\[\sum_{\ell=1}^{m_n}E\bigl(|\xi_{n,\ell}|^{2+\delta}\mid \mathcal F_{n,\ell-1}\bigr)\to_p 0.\]
Then \[\sum_{\ell=1}^{m_n}\xi_{n,\ell}\Rightarrow N(0,V).\]

If, in addition, \(\xi_{n,1},\ldots,\xi_{n,m_n}\) are conditionally independent given \(\mathcal F_{n,0}=\mathcal G_n\), and the filtration is \(\mathcal F_{n,\ell} =\mathcal G_n\vee\sigma(\xi_{n,1},\ldots,\xi_{n,\ell})\), put
\[
M_n:=\sum_{\ell=1}^{m_n}\xi_{n,\ell}.
\]
Then, for every \(t\in\mathbb R\),
\[
E\!\left(e^{itM_n}\mid\mathcal F_{n,0}\right)
\xrightarrow{p}e^{-t^2V/2}.
\]
For every sequence of \(\mathcal F_{n,0}\)-measurable random variables \(H_n\) satisfying \(\sup_n\|H_n\|_\infty<\infty\),
\[
E\!\left[H_ne^{itM_n}\right]
-E(H_n)e^{-t^2V/2}\longrightarrow0.
\]
Consequently, if \(W_n\) is \(\mathcal F_{n,0}\)-measurable and \(W_n\Rightarrow W\), then
\[
(W_n,M_n)\Rightarrow (W,G),
\qquad
G\sim N(0,V),
\qquad
G\perp W.
\]
\end{lemma}
\begin{proof}
The conditional moment bound implies Lindeberg's condition: for every \(\varepsilon>0\),
\[
\sum_{\ell=1}^{m_n}
E\{\xi_{n,\ell}^2\mathbf1(|\xi_{n,\ell}|>\varepsilon)
       \mid\mathcal F_{n,\ell-1}\}
\le \varepsilon^{-\delta}
\sum_{\ell=1}^{m_n}
E(|\xi_{n,\ell}|^{2+\delta}\mid\mathcal F_{n,\ell-1})
\to_p0.
\]
Corollary~3.1 and its constant-variance remark in Hall and Heyde\citep{HallHeyde1980} (pp.~58--59) give the martingale CLT when \(V>0\). For \(V=0\), write \(V_{n,k}=\sum_{\ell\le k}E(\xi_{n,\ell}^2\mid \mathcal F_{n,\ell-1})\). For fixed \(\eta>0\), the increments \(\xi_{n,k}\mathbf1\{V_{n,k}\le\eta\}\) remain martingale differences and their sum has second moment at most \(\eta\). Consequently, for every \(\epsilon>0\),
\[
P\left(\left|\sum_k\xi_{n,k}\right|>\epsilon\right)
\le P(V_n>\eta)+\eta/\epsilon^2.
\]
First letting \(n\to\infty\) and then \(\eta\downarrow0\) proves convergence to the degenerate normal law.

For the conditional conclusion with \(V>0\), the independent-array Lyapunov CLT (Theorem~27.3 of Billingsley\citep{Billingsley1995}), applied along almost-surely convergent further subsequences, gives
\[
E(e^{itM_n}\mid\mathcal G_n)\to_p e^{-t^2V/2}.
\]
When \(V=0\), conditional Cauchy--Schwarz gives
\[
\left|E(e^{itM_n}\mid\mathcal G_n)-1\right|
\le |t|E(|M_n|\mid\mathcal G_n)
\le |t|\sqrt{V_n}\to_p0,
\]
which gives the same conclusion. The conditional characteristic functions are bounded by one, so
\[
\left|E[H_ne^{itM_n}]-E(H_n)e^{-t^2V/2}\right|
\le\|H_n\|_\infty
E\left|E(e^{itM_n}\mid\mathcal G_n)-e^{-t^2V/2}\right|\to0.
\]
Taking \(H_n=e^{isW_n}\) yields
\[
\begin{aligned}
E(e^{isW_n+itM_n})
&=E(e^{isW_n})e^{-t^2V/2}+o(1)\\
&\longrightarrow E(e^{isW})e^{-t^2V/2}.
\end{aligned}
\]
L\'evy's continuity theorem proves the joint limit.
\end{proof}
\subsubsection{Matching discrepancy and score radii}\label{a-common-bias-order-lemma-from-exact-within-match-survey-weight-normalized-coefficients-and-two-dimensional-score-radii}

\begin{lemma}[Matching distances and bias]
\label{lem:oracle-score-discrepancy}
\label{lem:oracle-score-radius}
Suppose the observed rows are independent and identically distributed under \(P_s\), with positive probabilities for both treatment arms. Fix a known-score matching problem for either estimand and sampling design, and let \(\mathbf D_z(\mathbf X)\) denote its two-dimensional double score. Under the selected-sample law, assume that the donor score support \(\mathcal D_z\subset\mathbb R^2\) is compact, convex, and full-dimensional, with density \(g_z\) satisfying
\[
0<\underline g_z\le g_z(d)\le \overline g_z<\infty,
\qquad d\in\mathcal D_z.
\]
Assume that the score support \(\mathcal R_z\) of target individuals matched to arm \(z\) is contained in \(\mathcal D_z\), and that constants \(c_{\mathcal D}>0\) and \(r_{\mathcal D}>0\) exist such that
\[
\lambda_2\{B(d,r)\cap\mathcal D_z\}\ge c_{\mathcal D}r^2,
\qquad d\in\mathcal R_z,\quad 0<r\le r_{\mathcal D}.
\]
Let
\[
r_{n,z}^{\mathrm{match}}(i):=
\max_{j\in\mathcal J_{\mathbf D_z}(i)}
\|\mathbf D_z(\mathbf X_i)-\mathbf D_z(\mathbf X_j)\|.
\]
The target individual has \(Z_i=1-z\), which is included in the conditioning below. If fewer than \(M\) donors are available, set \(r_{n,z}^{\mathrm{match}}(i)=0\). Set \(r_{n,z}^{\mathrm{match}}(i)=0\) also when \(Z_i=z\). Then there is a finite constant \(C\), depending only on \(M\), the matched-arm probability, and the score-space regularity constants, such that
\[
\sup_i E_s\!\left[nr_{n,z}^{\mathrm{match}}(i)^2\mid \mathbf D_z(\mathbf X_i)\right]\le C.
\]
Consequently, for every nonnegative index-unit weight sequence \(a_i\) satisfying
\[
\frac1n\sum_{i=1}^n a_i^2=O_p(1),
\]
it holds that
\[
\sqrt n\,\frac1n\sum_{i=1}^n a_ir_{n,z}^{\mathrm{match}}(i)=O_p(1).
\]

For matching bias, take \(\mathbf D_z=\mathbf D_z^*\) and suppose \(m_z\) is Lipschitz on the relevant score support. On \(n_z\ge M\),
\[
\left|m_z\{\mathbf D_z^*(\mathbf X_i)\}
-\sum_{j=1}^n\lambda_{i\to j}^{(z)}
m_z\{\mathbf D_z^*(\mathbf X_j)\}\right|
\le L_zr_{n,z}^{\mathrm{match}}(i).
\]
With bounded outer survey weights and a denominator converging to a positive constant, the corresponding matching-bias term is \(O_p(n^{-1/2})\).
\end{lemma}
\begin{proof}
Conditional on \(Z_i=1-z\), the score \(\mathbf D_z(\mathbf X_i)=d\in\mathcal R_z\), and the arm size \(n_z\ge M\), the matched-arm scores are \(n_z\) independent draws from \(g_z\). The uniform distance-moment bound in Lemma~2 of Abadie and Imbens,\citep{Abadie2006} applied in two dimensions, gives
\[
\sup_{n_z\ge M}\ \sup_{d\in\mathcal R_z}
E_s\{n_zr_{n,z}^{\mathrm{match}}(i)^2\mid n_z,d,Z_i=1-z\}\le C_0.
\]

Let \(q_z=P_s(Z=z)>0\). Given this opposite-arm observation, \(n_z\sim\operatorname{Binomial}(n-1,q_z)\), so
\[
\begin{aligned}
E_s\{nr_{n,z}^{\mathrm{match}}(i)^2\mid d\}
&\le C_0E_s\!\left[\frac n{n_z}\mathbf1(n_z\ge M)\,\middle|\,d\right]\\
&\le C_0\{2/q_z+(n/M)C_1e^{-C_2n}\}\\
&\le C.
\end{aligned}
\]
The bound implies \(n^{-1}\sum_i nr_{n,z}^{\mathrm{match}}(i)^2=O_p(1)\) by Markov's inequality. Hence Cauchy--Schwarz yields
\[
\begin{aligned}
\sqrt n\,\frac1n\sum_{i=1}^n a_ir_{n,z}^{\mathrm{match}}(i)
&\le
\left(\frac1n\sum_{i=1}^n a_i^2\right)^{1/2}
\left(\frac1n\sum_{i=1}^n nr_{n,z}^{\mathrm{match}}(i)^2\right)^{1/2}\\
&=O_p(1).
\end{aligned}
\]

For the matching-bias bound, work on the event that every required donor arm contains at least \(M\) observations. Its probability tends to one. The Lipschitz condition and the nonnegative coefficients summing to one give
\[
\begin{aligned}
\left|m_z\{\mathbf D_z^*(\mathbf X_i)\}
 -\sum_j\lambda_{i\to j}^{(z)}
            m_z\{\mathbf D_z^*(\mathbf X_j)\}\right|
&\le\sum_j\lambda_{i\to j}^{(z)}
 \left|m_z\{\mathbf D_z^*(\mathbf X_i)\}
       -m_z\{\mathbf D_z^*(\mathbf X_j)\}\right|\\
&\le L_zr_{n,z}^{\mathrm{match}}(i).
\end{aligned}
\]
For PATE, take \(a_i=w_{A,Z_i}(\mathbf X_i)\) and \(\bar w_{A,n}=n^{-1}\sum_i a_i\). The preceding bounds give
\[
\begin{aligned}
|B_{n,A,\mathrm{PATE}}|
&\le\frac1{\bar w_{A,n}}\sum_{z=0}^1 L_z
 \frac1n\sum_{i:Z_i=1-z}a_i r_{n,z}^{\mathrm{match}}(i)\\
&\le\frac1{\bar w_{A,n}}\sum_{z=0}^1 L_z
 \left(\frac1n\sum_{i:Z_i=1-z}a_i^2\right)^{1/2}
 \left(\frac1n\sum_{i:Z_i=1-z}
       \{r_{n,z}^{\mathrm{match}}(i)\}^2\right)^{1/2}\\
&=O_p(n^{-1/2}).
\end{aligned}
\]
For PATT, retain only \(z=0\), take \(a_i=Z_iw_{A,1}(\mathbf X_i)\), and use the treated normalizer \(\bar w_{A,n}=n^{-1}\sum_i a_i\). The same bound gives \(B_{n,A,\mathrm{PATT}}=O_p(n^{-1/2})\). In both cases, the normalizer converges to a positive constant.
\end{proof}

\subsubsection{Conditional residual variance}\label{residual-quadratic-variation-stabilization}

Fix the estimand and sampling design \(A\in\{R,P\}\). Using the conditioning field \(\mathcal G_{A,n}^{\mathrm{match}}\) in Assumption~\ref{ass:residual-centering-survey-normalized}, write the oracle residual term as
\[
E_{n,A}
=
\frac{1}{n\bar w_{A,n}}
\sum_{\ell=1}^nc_{n,A,\ell}\varepsilon_{n,A,\ell},
\]
where the coefficients \(c_{n,A,\ell}\) are measurable with respect to \(\mathcal G_{A,n}^{\mathrm{match}}\), and \(E_s(\varepsilon_{n,A,\ell}\mid\mathcal G_{A,n}^{\mathrm{match}})=0\). Conditional independence gives
\[
\operatorname{Var}_s(\sqrt n E_{n,A}\mid\mathcal G_{A,n}^{\mathrm{match}})
=
\frac{1}{n\bar w_{A,n}^{2}}
\sum_{\ell=1}^n
c_{n,A,\ell}^{2}
\operatorname{Var}_s(\varepsilon_{n,A,\ell}\mid\mathcal G_{A,n}^{\mathrm{match}}).
\]

The matching directions are \(z=0,1\) for PATE and \(z=0\) for PATT. We write \(q_z=P_s(Z=z)\) and let \(f_{a\mid z}\) denote the density of \(\mathbf D_z^*(\mathbf X)\) conditional on \(Z=a\). The residuals below are those in Assumption~\ref{ass:residual-centering-survey-normalized}, including \(\varepsilon_{1\mid1}\) for treated PATT observations.

\begin{assumption}[Score densities and moment conditions]
\label{ass:residual-qv-stabilization}
Assumption~\ref{ass:residual-centering-survey-normalized} holds. The following additional conditions concern the fixed limiting-score graph.
\begin{enumerate}
\item \(q_z\in(0,1)\). The score metric is fixed and two-dimensional Euclidean distance, with any deterministic nonsingular standardization included in \(\mathbf D_z^*\), and \(M\) is a fixed positive integer. Ties are resolved by a fixed, measurable, outcome-independent rule.
\item For each direction \(z\) used for matching, the donor score \(\mathbf D_z^*(\mathbf X)\mid Z=z\) has a density \(f_{z\mid z}\) on a compact convex set \(\mathcal S_z\subset\mathbb R^2\) with nonempty interior. This density is bounded above and away from zero on \(\mathcal S_z\). The query density \(f_{1-z\mid z}\), namely the density of \(\mathbf D_z^*(\mathbf X)\mid Z=1-z\), is bounded above and supported on \(\mathcal S_z\).
\item For fixed constants \(0<w_{\min}\le w_{\max}<\infty\), \(w_{\min}\le w_{A,Z}(\mathbf X)\le w_{\max}\) almost surely. For some \(\delta>0\), each observed-arm residual used in the decomposition has a finite selected-sample \((2+\delta)\)-moment.
\end{enumerate}
\end{assumption}

The weighted reuse coefficient is
\begin{equation}\label{eq:qv-exact-incoming-coefficient}
K_{n,z}^{A}(j)
=\sum_{i:Z_i=1-z}w_{A,1-z}(\mathbf X_i)
\frac{w_{A,z}(\mathbf X_j)
\mathbf1\{j\in\mathcal J_{\mathbf D_z^*}(i)\}}
{\sum_{k\in\mathcal J_{\mathbf D_z^*}(i)}w_{A,z}(\mathbf X_k)}.
\end{equation}
For PATT only \(K_{n,0}^{A}\) is used. The signed coefficients and normalizing factors in the general representation are
\begin{equation}\label{eq:qv-four-scenario-coefficients}
\begin{aligned}
c_{n,A,j}^{\mathrm{PATE}}
&=(2Z_j-1)\{w_{A,Z_j}(\mathbf X_j)
                    +K_{n,Z_j}^{A}(j)\},\\
\bar w_{A,n}^{\mathrm{PATE}}
&=\frac1n\sum_{i=1}^n w_{A,Z_i}(\mathbf X_i),\\
c_{n,A,j}^{\mathrm{PATT}}
&=Z_jw_{A,1}(\mathbf X_j)
  -(1-Z_j)K_{n,0}^{A}(j),\\
\bar w_{A,n}^{\mathrm{PATT}}
&=\frac1n\sum_{i=1}^n Z_iw_{A,1}(\mathbf X_i).
\end{aligned}
\end{equation}
A term multiplied by \(1-Z_j\) is evaluated only for a control.

For observation \(j\), let \(\mathcal G_{A,j}^{\mathrm{row}}\) contain its treatment, limiting double scores, and observed weight for the specified estimand. Thus
\[
\mathcal G_{A,n}^{\mathrm{match}}
=\bigvee_{j=1}^n\mathcal G_{A,j}^{\mathrm{row}}.
\]
Let \(\varepsilon_{A,j}\) denote the observed residual for the chosen estimand. Independence of the rows and the centering assumption give
\begin{equation}\label{eq:qv-own-row-variance-mark}
\begin{aligned}
v_{A,j}
&:=\operatorname{Var}_s(\varepsilon_{A,j}\mid\mathcal G_{A,j}^{\mathrm{row}})\\
&=E_s(\varepsilon_{A,j}^{2}\mid\mathcal G_{A,j}^{\mathrm{row}})\\
&=\operatorname{Var}_s(\varepsilon_{A,j}\mid\mathcal G_{A,n}^{\mathrm{match}}).
\end{aligned}
\end{equation}

\begin{proposition}[Conditional variance and residual limit]
\label{lem:residual-qv-clt}
\label{prop:primitive-residual-qv-limit}
Under Assumption~\ref{ass:residual-qv-stabilization}, the coefficients in \eqref{eq:qv-four-scenario-coefficients} satisfy \(\bar w_{A,n}\to_p\bar w_A>0\) and
\[
\operatorname{Var}_s(\sqrt n E_{n,A}\mid\mathcal G_{A,n}^{\mathrm{match}})
\to_{L^1}V_{E,A},\qquad 0\le V_{E,A}<\infty.
\]
They also satisfy the conditional Lyapunov condition
\begin{equation}\label{eq:qv-derived-lyapunov}
\frac{1}{n^{1+\delta/2}\bar w_{A,n}^{2+\delta}}
\sum_{\ell=1}^n
|c_{n,A,\ell}|^{2+\delta}
E_s\!\left(|\varepsilon_{n,A,\ell}|^{2+\delta}\mid\mathcal G_{A,n}^{\mathrm{match}}\right)
\to_p0 .
\end{equation}

Consequently,
\[
\sqrt n\,E_{n,A}\Rightarrow N(0,V_{E,A})
\]
for the fixed estimand and regime \(A\), with \(N(0,0)\) interpreted as the point mass at zero.
\end{proposition}
\begin{proof}
\par\smallskip\noindent\emph{Conditional reuse moments.}\quad
The weighted coefficients are bounded by ordinary reuse counts, whose moments follow from the catchment-set argument of Abadie and Imbens.\citep{Abadie2006} For matching on \(\mathbf D_z^*\), write
\[
K_{M,z}(j):=\sum_{i:Z_i=1-z}
\mathbf1\{j\in\mathcal J_{\mathbf D_z^*}(i)\}.
\]
The bounded survey weights give
\begin{equation}\label{eq:qv-coefficient-count-bound}
\begin{aligned}
0\le K_{n,z}^{A}(j)
&\le\frac{w_{\max}^{2}}{Mw_{\min}}K_{M,z}(j),\\
|c_{n,A,j}|&\le C\{1+K_{M,Z_j}(j)\}
\end{aligned}
\end{equation}
for matched arms. The PATT treated coefficient is bounded directly.

To apply the catchment argument in Lemma~3 of Abadie and Imbens,\citep{Abadie2006} we bound the catchment radius using the donor density lower bound and integrate using the query density upper bound. Conditional on \(Z_j=z\), \(\mathbf D_z^*(\mathbf X_j)=d\), and the arm sizes \((n_z,n_{1-z})\), let \(A_{M,n_z}(d)\) denote the order-\(M\) catchment set of observation \(j\), and define
\[
p_{M,n_z}(d):=\int_{A_{M,n_z}(d)}f_{1-z\mid z}(x)\,dx.
\]
Convexity and the donor density bounds imply, uniformly in \(x\in\mathcal S_z\),
\[
br^2\le\int_{B(x,r)\cap\mathcal S_z}f_{z\mid z}(y)\,dy\le Br^2,
\qquad 0<r\le\operatorname{diam}(\mathcal S_z).
\]
For query positions \(x_1,\ldots,x_r\), \(r\in\mathbb N\), set \(R=\max_a\|x_a-d\|\). If all queries select the root, the farthest has at most \(M-1\) closer donors. Thus, for \(n_z\ge2M\), the binomial tail bound gives
\[
P_s\{x_1,\ldots,x_r\in A_{M,n_z}(d)\mid d,n_z\}
\le C\{1+(n_zR^2)^{M-1}\}e^{-cn_zR^2}.
\]
Integrating this bound against the bounded query densities gives
\begin{equation}\label{eq:qv-catchment-moment-bound}
\begin{aligned}
E_s\{p_{M,n_z}(d)^r\mid d,n_z\}
&=\int_{\mathcal S_z^r}
 P_s\{x_1,\ldots,x_r\in A_{M,n_z}(d)\mid d,n_z\}
 \prod_{a=1}^r f_{1-z\mid z}(x_a)\,d(x_1,\ldots,x_r)\\
&\le C_r\int_0^\infty
       \{1+(n_zt^2)^{M-1}\}e^{-cn_zt^2}t^{2r-1}\,dt\\
&=C'_r n_z^{-r}.
\end{aligned}
\end{equation}
Increasing \(C'_r\) covers \(M\le n_z<2M\). Given valid arm sizes and the donor scores, the reuse count is binomial with parameters \(n_{1-z}\) and \(p_{M,n_z}(d)\). Hence
\[
\begin{aligned}
E_s\{K_{M,z}(j)^r\mid d,n_z,n_{1-z}\}
&\le\sum_{a=1}^r S(r,a)(n_{1-z})_a
             E_s\{p_{M,n_z}(d)^a\mid d,n_z\}\\
&\le C_r\sum_{a=1}^r(n_{1-z}/n_z)^a,
\end{aligned}
\]
where \(S(r,a)\) and \((n_{1-z})_a\) denote Stirling numbers and falling factorials. Conditional on \(Z_j=z\), \(n_z-1\sim\operatorname{Binomial}(n-1,q_z)\). Splitting at \(n_z=nq_z/2\) and applying a Chernoff bound yields, for every real \(r>0\),
\begin{equation}\label{eq:qv-uniform-conditional-reuse}
\sup_{n\ge1}\ \mathop{\rm ess\,sup}_{d\in\mathcal S_z}
E_s\!\left[\{1+K_{M,z}(j)\}^{r}
       \mid Z_j=z,\mathbf D_z^*(\mathbf X_j)=d\right]<\infty.
\end{equation}

\par
Write \(p=1+\delta/2>1\). Conditional on the root's treatment and matching score, \(K_{M,z}(j)\) depends only on the other observations and is independent of the root's remaining variables, including \(v_{A,j}\). Conditional Jensen's inequality gives
\[
v_{A,j}^{p}
\le E_s(|\varepsilon_{A,j}|^{2+\delta}\mid\mathcal G_{A,j}^{\mathrm{row}}).
\]
We denote the arm-specific conditional-variance contribution by \(V_{n,A,z,j}:=\mathbf1\{Z_j=z\}c_{n,A,j}^2v_{A,j}\). Equations~\eqref{eq:qv-coefficient-count-bound} and \eqref{eq:qv-uniform-conditional-reuse} now give
\begin{equation}\label{eq:qv-summand-uniform-integrability}
\begin{aligned}
E_s(V_{n,A,z,j}^{p})
&\le C E_s\!\left[\mathbf1\{Z_j=z\}
             \{1+K_{M,z}(j)\}^{2p}v_{A,j}^{p}\right]\\
&=C E_s\!\left[\mathbf1\{Z_j=z\}v_{A,j}^{p}
 E_s\!\left\{\{1+K_{M,z}(j)\}^{2p}
              \mid Z_j,\mathbf D_z^*(\mathbf X_j)\right\}\right]\\
&\le C' E_s\!\left[\mathbf1\{Z_j=z\}v_{A,j}^{p}\right]\\
&\le C' E_s\!\left[\mathbf1\{Z_j=z\}
                         |\varepsilon_{A,j}|^{2+\delta}\right]
<\infty,
\end{aligned}
\end{equation}
uniformly in \(n\). For treated PATT observations, boundedness of the direct coefficient gives the same uniform-integrability bound.

\par\smallskip\noindent\emph{Local limit.}\quad
For direction \(z\), mark each score location by its treatment, survey weight and conditional residual variance. The arm subdensities and joint conditional mark laws are
\begin{equation}\label{eq:qv-local-mark-law}
\begin{aligned}
h_{a\mid z}(d)&:=q_a f_{a\mid z}(d),\qquad a=0,1,\\
\mathsf K_{A,a\mid z}(d,\cdot)
&:=\mathcal L_s\!\left\{
 (w_{A,a}(\mathbf X),v_A)
 \mid Z=a,\mathbf D_z^*(\mathbf X)=d\right\}.
\end{aligned}
\end{equation}
For almost every donor location \(d\), we place a root at the origin with weight and variance \((w,v)\) distributed according to \(\mathsf K_{A,z\mid z}(d,\cdot)\). Independently of this pair, let the other donors and the queries be independent homogeneous marked Poisson processes on \(\mathbb R^2\) with respective intensity measures
\[
h_{z\mid z}(d)\,dy\,\mathsf K_{A,z\mid z}(d,dw',dv'),
\qquad
h_{1-z\mid z}(d)\,dy\,
             \mathsf K_{A,1-z\mid z}(d,dw',dv').
\]
Let \(\mathcal J_z^\infty(y)\) be the \(M\) nearest donors to query \(y\), including the possible root. With \(W_0=w\), define
\begin{equation}\label{eq:qv-poisson-exact-coefficient}
K_{A,z}^\infty(d;w,v)
:=\sum_{y\text{ a query}}W_y
\frac{w\,\mathbf1\{0\in\mathcal J_z^\infty(y)\}}
     {\sum_{k\in\mathcal J_z^\infty(y)}W_k}.
\end{equation}

Let \(g\) be a nonnegative continuous test function with bounded spatial support. For the other \(n-1\) rows, conditional on the root location and its associated variables, the Laplace functional of the locations scaled by \(\sqrt n\) about \(d\) is
\begin{equation}\label{eq:qv-marked-local-laplace}
\left[
1-\frac1n\sum_{a=0}^1\int_{\mathbb R^2}
h_{a\mid z}(d+y/\sqrt n)
\int\{1-e^{-g(y,a,w',v')}\}
\mathsf K_{A,a\mid z}(d+y/\sqrt n,dw',dv')\,dy
\right]^{n-1}.
\end{equation}
Lebesgue differentiation, applied to a countable determining class of the joint position--mark measures, gives, for almost every \(d\), the limit
\[
\exp\!\left[
-\sum_{a=0}^1 h_{a\mid z}(d)
 \int_{\mathbb R^2}\int\{1-e^{-g(y,a,w',v')}\}
 \mathsf K_{A,a\mid z}(d,dw',dv')\,dy
\right].
\]
This Laplace functional identifies the marked Poisson limit. Its two treatment restrictions are independent by Theorems~3.9 and~5.2 of Last and Penrose.\citep{LastPenrose2017}

\phantomsection\label{app:qv-localization} To pass from bounded windows to the incoming coefficient, cover the plane by finitely many cones of angular diameter less than \(\pi/3\) and choose \(R\) so that each cone contains \(M\) nonroot donors within distance \(R\). For a query at distance \(t>R\) and a donor in its cone at distance \(s\le R\),
\[
\begin{aligned}
\|y-k\|^2
&=t^2+s^2-2ts\cos\theta\\
&<t^2+s^2-ts\\
&\le t^2.
\end{aligned}
\]
Every query in \(B_R\) has \(M\) donors within distance \(2R\). Consequently,
\[
0\in\mathcal J_z^\infty(y)
\quad\Longrightarrow\quad
y\in B_R,\qquad \mathcal J_z^\infty(y)\subset B_{3R},
\]
so every query contributing to \eqref{eq:qv-poisson-exact-coefficient}, and every donor in its normalizing sum, lies in a bounded window. For cones \(C_1,\ldots,C_J\), the probability that one contains fewer than \(M\) donors in \(B_R\) is bounded by
\[
\sum_{\nu=1}^J
 \exp\{-h_{z\mid z}(d)\lambda_2(C_\nu\cap B_R)\}
 \sum_{k=0}^{M-1}
 \frac{\{h_{z\mid z}(d)\lambda_2(C_\nu\cap B_R)\}^k}{k!}
\longrightarrow0\qquad(R\to\infty).
\]
On the event that every cone contains at least \(M\) donors in \(B_R\), the coefficient is continuous outside distance ties and window boundaries, which have probability zero under the limiting law. The continuous mapping theorem gives convergence on this event. Letting \(R\to\infty\) removes the window restriction and yields the conditional root-summand limit. Integrating over the root law gives unconditional convergence, because the support boundary and exceptional differentiation set have zero root probability.

The corresponding mean limits are
\begin{equation}\label{eq:qv-poisson-mean-limits}
\begin{aligned}
\Lambda_{A,z}^{\mathrm{PATE}}
&:=\int_{\mathcal S_z}h_{z\mid z}(d)
 \int E\!\left[\{w+K_{A,z}^\infty(d;w,v)\}^{2}v\right]
                \mathsf K_{A,z\mid z}(d,dw,dv)\,dd,\\
\Lambda_{A,0}^{\mathrm{PATT}}
&:=\int_{\mathcal S_0}h_{0\mid0}(d)
 \int E\!\left[\{K_{A,0}^\infty(d;w,v)\}^{2}v\right]
                \mathsf K_{A,0\mid0}(d,dw,dv)\,dd,\\
\Lambda_{A,1}^{\mathrm{PATT}}
&:=E_s\!\left[Z w_{A,1}(\mathbf X)^2v_A\right].
\end{aligned}
\end{equation}
The inner expectations condition on the root mark. Uniform integrability in \eqref{eq:qv-summand-uniform-integrability} gives
\begin{equation}\label{eq:qv-expectation-limit}
E_s(V_{n,A,z,1})\longrightarrow\Lambda_{A,z},
\qquad 0\le\Lambda_{A,z}<\infty.
\end{equation}
The direct treated PATT term follows by the law of large numbers.

\par\smallskip\noindent\emph{Concentration.}\quad
Concentration follows by applying the two-location and truncation argument of Penrose and Yukich\citep{PenroseYukich2003} to the joint mark laws in \eqref{eq:qv-local-mark-law}. For two distinct donor roots, fixed rescaled neighborhoods are eventually disjoint. The joint conditional Laplace functional then satisfies
\[
\{1-(b_{n,1}+b_{n,2})/n\}^{n-2}\longrightarrow e^{-b_1}e^{-b_2}.
\]
Here \(b_{n,r}\) denotes the integral in \eqref{eq:qv-marked-local-laplace} for the test function around root \(r\). This factorization, followed by localization and integration over the independent roots, gives independent copies of the one-root limit. Conditioning on the two root treatments includes the treatment indicators.

For \(T>0\), let \(V_{n,A,z,j}^{[T]}=\min(V_{n,A,z,j},T)\). Boundedness and the two-root convergence imply
\[
\operatorname{Cov}_s
 (V_{n,A,z,1}^{[T]},V_{n,A,z,2}^{[T]})\longrightarrow0.
\]
Exchangeability then gives
\begin{equation}\label{eq:qv-truncated-concentration}
\begin{aligned}
\operatorname{Var}_s\!\left(\frac1n\sum_{j=1}^n
                                 V_{n,A,z,j}^{[T]}\right)
&=\frac1n\operatorname{Var}_s(V_{n,A,z,1}^{[T]})
 +\frac{n-1}{n}\operatorname{Cov}_s
   (V_{n,A,z,1}^{[T]},V_{n,A,z,2}^{[T]})\\
&\longrightarrow0.
\end{aligned}
\end{equation}
The tail error satisfies
\[
\begin{aligned}
E_s\!\left[\frac1n\sum_{j=1}^n
       \{V_{n,A,z,j}-V_{n,A,z,j}^{[T]}\}\right]
&=E_s\{V_{n,A,z,1}-V_{n,A,z,1}^{[T]}\}\\
&\le T^{1-p}E_s(V_{n,A,z,1}^{p})\\
&\le C T^{1-p}.
\end{aligned}
\]
Combining these bounds with \eqref{eq:qv-expectation-limit},
\[
\begin{aligned}
E_s\left|\frac1n\sum_j V_{n,A,z,j}-\Lambda_{A,z}\right|
&\le 2CT^{1-p}
 +\left\{\operatorname{Var}_s\left(
       \frac1n\sum_jV_{n,A,z,j}^{[T]}\right)\right\}^{1/2}\\
&\quad+\left|E_sV_{n,A,z,1}-\Lambda_{A,z}\right|.
\end{aligned}
\]
We first let \(n\to\infty\) and then \(T\to\infty\) to obtain
\begin{equation}\label{eq:qv-armwise-lln}
\frac1n\sum_{j=1}^n V_{n,A,z,j}
\longrightarrow_{L^1}\Lambda_{A,z}.
\end{equation}
For PATT the treated term is an ordinary sample average. For PATE, apply this argument separately to the two directions.

\par
The ordinary law of large numbers gives
\[
\bar w_A^{\mathrm{PATE}}=E_s\{w_{A,Z}(\mathbf X)\}>0,
\qquad
\bar w_A^{\mathrm{PATT}}
=E_s\{Zw_{A,1}(\mathbf X)\}
=P_s(Z=1)E_s\{w_{A,1}(\mathbf X)\mid Z=1\}>0.
\]
It follows from \eqref{eq:qv-armwise-lln} that
\begin{equation}\label{eq:qv-primitive-deterministic-limit}
\operatorname{Var}_s(\sqrt n E_{n,A}\mid\mathcal G_{A,n}^{\mathrm{match}})\to_p
V_{E,A}:=\bar w_A^{-2}\sum_{z=0}^1\Lambda_{A,z}.
\end{equation}
To strengthen convergence to \(L^1\), it remains to control the random denominator. For PATT, \(\bar w_{A,n}\ge w_{\min} n_1/n\) on \(n_1\ge1\). For every finite \(s>0\), splitting at \(n_1=nq_1/2\) gives
\[
E_s\!\left[(n/n_1)^s\mathbf1\{n_1\ge1\}\right]
\le(2/q_1)^s+n^sP_s(0<n_1<nq_1/2)=O(1).
\]
Jensen's inequality and \eqref{eq:qv-summand-uniform-integrability} give a uniform \(p\)-moment for \(n^{-1}\sum_jc_{n,A,j}^2v_{A,j}\). For any \(1<r<p\), H\"older's inequality and the preceding inverse moments therefore give
\[
\begin{aligned}
E_s\!\left[\left\{\operatorname{Var}_s
(\sqrt n E_{n,A}\mid\mathcal G_{A,n}^{\mathrm{match}})\right\}^{r}\right]
&\le
 \left[E_s\!\left\{\left(\frac1n\sum_j
                     c_{n,A,j}^2v_{A,j}\right)^p\right\}\right]^{r/p}
 \left[E_s\{\bar w_{A,n}^{-2rp/(p-r)}\}\right]^{(p-r)/p}\\
&=O(1).
\end{aligned}
\]
For PATE the inverse denominator is bounded directly. Hence uniform integrability of the conditional variance strengthens \eqref{eq:qv-primitive-deterministic-limit} to convergence of expectations and convergence in \(L^1\).

Every term in \eqref{eq:qv-poisson-mean-limits} is nonnegative. For PATE,
\[
V_{E,A}
\ge\bar w_A^{-2}w_{\min}^2E_s(v_A)>0
\quad\text{if }P_s(v_A>0)>0.
\]
For PATT,
\[
V_{E,A}
\ge\bar w_A^{-2}w_{\min}^2E_s(Zv_A)>0
\quad\text{if }P_s(Z=1,v_A>0)>0.
\]
Alternatively, its control contribution is strictly positive if
\[
\int_{\mathcal S_0}h_{0\mid0}(d)
\mathbf1\{h_{1\mid0}(d)>0\}
\mathsf K_{A,0\mid0}(d,\{(w,v):v>0\})\,dd>0.
\]
Conditional on such a root and its donor process, a neighborhood of the root lies in its catchment set. A positive query intensity then gives a positive incoming coefficient with positive probability.

For the Lyapunov condition, use the conditional reuse bound with exponent \(2+\delta\) and the conditional independence of the unweighted count from the remaining root variables:
\[
\begin{aligned}
E_s\!\left[\frac1n\sum_j|c_{n,A,j}|^{2+\delta}
E_s(|\varepsilon_{A,j}|^{2+\delta}\mid\mathcal G_{A,j}^{\mathrm{row}})\right]
&\le C E_s\!\left[E_s(|\varepsilon_{A,1}|^{2+\delta}
\mid\mathcal G_{A,1}^{\mathrm{row}})\right]\\
&=C E_s(|\varepsilon_{A,1}|^{2+\delta})
<\infty.
\end{aligned}
\]
Markov's inequality and \(\bar w_{A,n}\to_p\bar w_A>0\) give \eqref{eq:qv-derived-lyapunov}.

By Proposition~\ref{lem:residual-centering-survey-normalized}(iii), the residuals are conditionally independent and centered given \(\mathcal G_{A,n}^{\mathrm{match}}\). The variance and Lyapunov limits therefore allow us to apply Lemma~\ref{lem:conditional-residual-stable-clt} when \(V_{E,A}>0\). If \(V_{E,A}=0\), conditional Chebyshev instead gives, for \(\epsilon>0\),
\[
P_s(|\sqrt n E_{n,A}|>\epsilon)
\le \epsilon^{-2}E_s\!\left[
\operatorname{Var}_s(\sqrt n E_{n,A}\mid\mathcal G_{A,n}^{\mathrm{match}})\right]
\longrightarrow0.\qedhere
\]
\end{proof}

\begin{corollary}[Bias-correction remainder on fixed scores]
\label{cor:oracle-regression-bc}
Suppose the conditions of Lemma~\ref{lem:oracle-score-radius} hold for the fixed oracle scores in the relevant scenario, with bounded outer survey weights and an outer normalizer converging to a positive constant. For each matching direction, suppose that \(m_z\) and \(\widehat m_z\) are continuously differentiable on the convex score support and
\[
\max_z\|\nabla(\widehat m_z-m_z)\|_\infty=o_p(1),
\]
where PATT requires only \(z=0\). Construct \(\widehat B_{n,A}\) by replacing \(m_z\) with \(\widehat m_z\) in \(B_{n,A}\), retaining the same scores, match sets, outer weights and normalized donor weights. Then
\[
\sqrt n(B_{n,A}-\widehat B_{n,A})\to_p0.
\]
\end{corollary}

\begin{proof}
Work on the event that every required donor arm contains at least \(M\) observations, whose probability tends to one. Write \(\mathbf D_{z,i}^*=\mathbf D_z^*(\mathbf X_i)\). Convexity and the fundamental theorem of calculus give
\[
\begin{aligned}
&(\widehat m_z-m_z)(\mathbf D_{z,i}^*)
 -(\widehat m_z-m_z)(\mathbf D_{z,j}^*)\\
&\quad=\int_0^1
 \nabla(\widehat m_z-m_z)
 \{\mathbf D_{z,j}^*+t(\mathbf D_{z,i}^*-\mathbf D_{z,j}^*)\}^{\top}
 (\mathbf D_{z,i}^*-\mathbf D_{z,j}^*)\,dt.
\end{aligned}
\]
Consequently,
\[
\begin{aligned}
&|(\widehat m_z-m_z)(\mathbf D_{z,i}^*)
 -(\widehat m_z-m_z)(\mathbf D_{z,j}^*)|\\
&\quad\le
 \|\nabla(\widehat m_z-m_z)\|_\infty
 \|\mathbf D_{z,i}^*-\mathbf D_{z,j}^*\|.
\end{aligned}
\]
The normalized donor weights are nonnegative and sum to one. Hence each directional discrepancy satisfies
\[
\begin{aligned}
&\left|\frac1{\sqrt n}\sum_{i:Z_i=1-z}w_{A,1-z}(\mathbf X_i)
 \sum_j\lambda_{i\to j}^{A,(z)}
 \left\{(\widehat m_z-m_z)(\mathbf D_{z,i}^*)
       -(\widehat m_z-m_z)(\mathbf D_{z,j}^*)\right\}\right|\\
&\le
 \|\nabla(\widehat m_z-m_z)\|_\infty
 \frac1n\sum_{i:Z_i=1-z}
 w_{A,1-z}(\mathbf X_i)\sqrt n\,r_{n,z}^{\mathrm{match}}(i)\\
&=o_p(1)\,O_p(1)\\
&=o_p(1),
\end{aligned}
\]
where Lemma~\ref{lem:oracle-score-radius} bounds the weighted average matching radius. Combining the required directions and dividing by the outer denominator, whose probability limit is positive, proves the result.
\end{proof}

This derivative condition follows the bias-correction argument of Abadie and Imbens.\citep{AbadieImbens2011} For a correctly specified fixed basis with bounded continuous derivatives, coefficient consistency suffices. A growing series must also satisfy uniform derivative approximation.

\subsubsection{Reuse bounds for perturbed scores}
\label{app:local-score-reuse}

Suppose observations are independent and identically distributed under \(P_s\), and the fixed-\(M\) and score-density conditions in Assumption~\ref{ass:residual-qv-stabilization} hold. Let \(S_z(O)\) be a matrix with \(q\) fixed columns and \(\sup_O\|S_z(O)\|\le L_S<\infty\). We write \(\mathbf D_{z,i}^\ast=\mathbf D_z^\ast(\mathbf X_i)\), use \(r_{n,z}^{\mathrm{match}}(i)\) for the \(M\)-th nearest-neighbor distance on \(\mathbf D_z^\ast\), and define
\[
\mathbf D_{z,n,h}(O)
=\mathbf D_z^\ast(\mathbf X)+n^{-1/2}S_z(O)h,
\qquad h\in\mathbb R^q.
\]
Let \(\mathcal J_{z,n,h}(i)\) be the corresponding match set. For fixed \(C<\infty\), every \(j\in\mathcal J_{z,n,h}(i)\), with \(\|h\|\le C\), satisfies
\[
\begin{aligned}
\|\mathbf D_{z,j}^\ast-\mathbf D_{z,i}^\ast\|
&\le
\|\mathbf D_{z,n,h}(O_j)-\mathbf D_{z,n,h}(O_i)\|
+\frac{2L_SC}{\sqrt n}\\
&\le
\max_{k\in\mathcal J_{\mathbf D_z^\ast}(i)}
\|\mathbf D_{z,n,h}(O_k)-\mathbf D_{z,n,h}(O_i)\|
+\frac{2L_SC}{\sqrt n}\\
&\le r_{n,z}^{\mathrm{match}}(i)+\frac{4L_SC}{\sqrt n}.
\end{aligned}
\]
The second inequality uses the baseline \(M\) nearest neighbors as candidates for the perturbed match. If
\[
\widehat{\mathbf D}_{z,i}
=\mathbf D_{z,i}^\ast
 +n^{-1/2}\{S_z(O_i)H_n+\mathbf r_{n,z,i}\},
\]
the same calculation gives the bound \(r_{n,z}^{\mathrm{match}}(i)+(4L_SC+4)/\sqrt n\) on \(\{\|H_n\|\le C,\max_i\|\mathbf r_{n,z,i}\|\le1\}\).

For fixed \(a\ge0\), we define
\[
K_{M,z}^{+,a}(j)
:=\sum_{i:Z_i=1-z}
\mathbf1\!\left\{
\|\mathbf D_{z,j}^\ast-\mathbf D_{z,i}^\ast\|
\le r_{n,z}^{\mathrm{match}}(i)+a/\sqrt n
\right\}.
\]
For every \(q'>0\),
\begin{equation}\label{eq:stabilizer-inflated-reuse}
\sup_n\mathop{\rm ess\,sup}_{d\in\mathcal S_z}
E_s\{(1+K_{M,z}^{+,a}(j))^{q'}\mid Z_j=z,\mathbf D_{z,j}^\ast=d\}
<\infty.
\end{equation}
To prove this bound, we condition on \(Z_j=z\), \(\mathbf D_{z,j}^*=d\), and the arm sizes \((n_z,n_{1-z})\). We write \(q_z=P_s(Z=z)>0\). For \(r\) opposite-arm positions, let \(R\) be their largest distance from \(d\). If all belong to the inflated catchment and \(R>2a/\sqrt n\), the farthest position has \(M\)-th neighbor distance greater than \(R/2\). The binomial rank-tail bound gives
\[
P_s(\text{all }r\text{ positions are in the catchment}\mid d,n_z)
\le\mathbf1_{\{R\le2a/\sqrt n\}}
 +C\{1+(n_zR^2)^{M-1}\}e^{-cn_zR^2}.
\]
Let \(p_{M,n_z;n}^{+,a}(d)\) be the conditional probability, given the matched-arm observations, that an independent opposite-arm observation belongs to this catchment. On \(n_z\ge nq_z/2\), integration over the bounded opposite-arm densities yields
\[
\begin{aligned}
E_s\{p_{M,n_z;n}^{+,a}(d)^r\mid d,n_z\}
&\le C_r(a/\sqrt n)^{2r}
 +C_r\int_0^\infty\{1+(n_zt^2)^{M-1}\}e^{-cn_zt^2}t^{2r-1}\,dt\\
&\le C_{r,a}n^{-r}.
\end{aligned}
\]
Conditional on the matched-arm observations, the count is binomial with success probability \(p_{M,n_z;n}^{+,a}(d)\). Its factorial moments are therefore uniformly bounded. The event \(n_z<nq_z/2\) has exponentially small probability, and the count is at most \(n\). Combining the factorial-moment bound with this tail bound proves \eqref{eq:stabilizer-inflated-reuse} for integer moments. Monotonicity of moment norms gives every finite \(q'>0\), uniformly up to the support boundary.

\subsubsection{Local score limits and nearest-neighbor gaps}
\label{app:local-score-marked-limits}

\paragraph{Local convergence.}
Let \(\eta\) collect \(S_z\), the survey weight, and the fixed finite-dimensional variables contributed by each observation. Write \(\mathsf K_{A,a\mid z}(d,d\eta)\) for their joint conditional law given \(Z=a,\mathbf D_z^\ast=d\), and \(h_{a\mid z}=q_af_{a\mid z}\). The calculation in \eqref{eq:qv-marked-local-laplace} applies to these joint position--mark measures at almost every query root \((d,\eta_0)\). Bounded \(S_z\) confines each perturbed bounded window to a fixed enlarged baseline window. The frozen intensity is unchanged by the mark-specific translation:
\[
\begin{aligned}
&\int_{\mathbb R^2}\int
 [1-e^{-g(y+\{S_z(\eta)-S_z(\eta_0)\}h,a,\eta)}]\,
 \mathsf K_{A,a\mid z}(d,d\eta)\,dy\\
&\qquad=
\int_{\mathbb R^2}\int[1-e^{-g(u,a,\eta)}]\,
 \mathsf K_{A,a\mid z}(d,d\eta)\,du .
\end{aligned}
\]
By the mapping theorem\citep{LastPenrose2017} (Theorem~5.1), the translated process has the baseline homogeneous Poisson law, including the joint distribution of \(\eta\). For two distinct query roots, the enlarged windows are eventually disjoint. Their joint Laplace functional factors in the limit as above, giving independent local processes. Query-score absolute continuity excludes coincident roots, support boundaries and exceptional differentiation points. Positive donor intensity and absence of ties extend the finite-window limit to the nearest \(M\) donors.

\paragraph{Nearest-neighbor gaps.}
Let \(R_{n,z,k}(i;h)\) be the distance from observation \(i\) to its \(k\)-th nearest neighbor in arm \(z\), on \(\mathbf D_{z,n,h}\). For \(t>0\), let
\[
G_{n,z,i}(h,t)
:=
\mathbf1\{Z_i=1-z\}
\mathbf1\!\left\{
\sqrt n\{R_{n,z,M+1}(i;h)-R_{n,z,M}(i;h)\}\le t
\right\}.
\]
We take this indicator to be zero when arm \(z\) has fewer than \(M+1\) observations. The bounded indicators and the one- and two-query marked limits give
\[
\frac1n\sum_iG_{n,z,i}(h,t)\to_p g_z(t).
\]
The limit is independent of \(h\), because the homogeneous marked spatial law is unchanged by the displacement. Conditional on a query location, put \(\lambda=h_{z\mid z}(d)>0\). For a homogeneous planar Poisson process,
\[
\begin{aligned}
P(R_{M+1}-R_M\le t\mid R_M=r)
&=1-\exp\{-\pi\lambda[(r+t)^2-r^2]\}\\
&\le \pi\lambda(2rt+t^2).
\end{aligned}
\]
Since \(\pi\lambda R_M^2\) has the Gamma\((M,1)\) law,
\[
E R_M=
\frac{\Gamma(M+1/2)}{\Gamma(M)\sqrt{\pi\lambda}}.
\]
The donor intensity is bounded above. Integrating the last two displays over query locations therefore gives
\[
0\le g_z(t)\le C(t+t^2).
\]
The limiting gap has no atom at \(t\). The independent two-query limit and boundedness give the asserted convergence in probability by the second-moment identities
\[
\begin{aligned}
E_s\!\left\{\frac1n\sum_iG_{n,z,i}(h,t)\right\}&\to g_z(t),\\
E_s\!\left[\left\{\frac1n\sum_iG_{n,z,i}(h,t)\right\}^2\right]&\to g_z(t)^2.
\end{aligned}
\]

For a finite \(\delta\)-net \(\mathcal L_{C,\delta}\) of the ball \(\{\|h\|\le C\}\), hold \(\delta>0\) fixed. Applying the preceding convergence at each of its finitely many points gives
\[
\max_{h'\in\mathcal L_{C,\delta}}
\frac1n\sum_iG_{n,z,i}(h',t)
\to_p g_z(t)\le C(t+t^2).
\]

\subsection{Estimated-score asymptotics}

\label{estimated-score-extensions-fixed-target-and-changing-law}

Fix an estimand and sampling design \(A\in\{R,P\}\). Let \(\mathbf U_A\) denote its estimating function, \(\widehat{\boldsymbol\theta}_A\) the estimated parameter, \(\boldsymbol\Gamma_A\) the expected Jacobian, and \(\boldsymbol\Sigma_{A,U}\) the estimating-function covariance. The estimand is specified in each result and omitted from these subscripts. The sandwich covariance is \(\boldsymbol\Sigma_{A,\theta}=\boldsymbol\Gamma_A^{-1}\boldsymbol\Sigma_{A,U}\boldsymbol\Gamma_A^{-T}\). Following the local-likelihood and discretization arguments of Andreou and Werker,\citep{AndreouWerker2012} Abadie and Imbens,\citep{AbadieImbens2016} and Yang and Zhang,\citep{yang_multiply_2023} we first establish the score-estimator expansion and its joint limit with the matching statistic.

\subsubsection{Score-parameter estimation}\label{primitive-weighted-m-estimation-block-in-section-10}

Fix an estimand and a sampling regime \(A\in\{R,P\}\). Let \(\mathbf U_A(O;\boldsymbol\theta)\in\mathbb R^{p_A}\) denote the stacked estimating function for that scenario, and define the sample equation
\[
\mathbb U_{A,n}(\boldsymbol\theta):=\frac1n\sum_{i=1}^n \mathbf U_A(O_i;\boldsymbol\theta).
\]
We denote the population Jacobian by
\[
 \boldsymbol\Gamma_A
 :=
 E_s\!\left[
 \frac{\partial \mathbf U_A(O;\boldsymbol\theta_A^\ast)}{\partial\boldsymbol\theta^\top}
 \right].
\]

\begin{assumption}[Score estimating equations]
\label{ass:ordinary-weighted-m-estimation}
The rows \(O_i\) are independent and identically distributed under the selected-sample law \(P_s\). Assume the following.

\begin{enumerate}
\item \(\boldsymbol\theta_A^\ast\in \operatorname{int}(\Theta_A)\) and \(\widehat{\boldsymbol \theta}_A\to_p\boldsymbol\theta_A^\ast\).
\item \(E[\mathbf U_A(O;\boldsymbol\theta_A^\ast)]=0\), and there exists \(\delta>0\) and a neighborhood \(\mathcal N_A\) of \(\boldsymbol\theta_A^\ast\) such that
 \[
 \sup_{\boldsymbol\theta\in\mathcal N_A}
 E\!\left[\|\mathbf U_A(O;\boldsymbol\theta)\|^{2+\delta}\right]
 <\infty.
 \]
\item \(\mathbf U_A(O;\boldsymbol\theta)\) is continuously differentiable in \(\boldsymbol\theta\) on \(\mathcal N_A\), and \(\boldsymbol\Gamma_A\) is nonsingular. On a closed ball about \(\boldsymbol\theta_A^\ast\) contained in \(\mathcal N_A\), the derivative has an integrable envelope:
 \[
 E_s\!\left[
 \sup_{\boldsymbol\theta\in\mathcal N_A}
 \left\|\frac{\partial \mathbf U_A(O;\boldsymbol\theta)}{\partial\boldsymbol\theta^\top}\right\|
 \right]<\infty,
 \]
 where \(\mathcal N_A\) may be reduced to that ball.
\item The estimator solves to first order the sample equation:
 \[
 \mathbb U_{A,n}(\widehat{\boldsymbol \theta}_A)=o_p(n^{-1/2}).
 \]
\end{enumerate}
\end{assumption}

The normalized estimating equation and its covariance are
\[
\sqrt n\,\mathbb U_{A,n}(\boldsymbol\theta_A^\ast)
=
\frac1{\sqrt n}\sum_{i=1}^n \mathbf U_A(O_i;\boldsymbol\theta_A^\ast),
\]
\[
\boldsymbol\Sigma_{A,U}
:=
E\!\left[
\mathbf U_A(O;\boldsymbol\theta_A^\ast)\mathbf U_A(O;\boldsymbol\theta_A^\ast)^\top
\right],
\]
and
\[
\boldsymbol\Sigma_{A,\theta}
:=
\boldsymbol\Gamma_A^{-1}\boldsymbol\Sigma_{A,U}\boldsymbol\Gamma_A^{-T}.
\]
Here \(\mathbf U_A(O;\boldsymbol\theta)\) is the contribution of one observation, \(\mathbb U_{A,n}(\boldsymbol\theta)\) is its sample average, and \(\sqrt n\,\mathbb U_{A,n}(\boldsymbol\theta_A^*)\) is the normalized sum. The expected Jacobian \(\boldsymbol\Gamma_A\) and covariance \(\boldsymbol\Sigma_{A,U}\) determine the parameter estimator's asymptotic covariance below.

\begin{lemma}[Asymptotic linearity of the score estimator]
\label{lem:ordinary-weighted-first-step}
Under Assumption~\ref{ass:ordinary-weighted-m-estimation},
\[
\sqrt n\,\mathbb U_{A,n}(\boldsymbol\theta_A^\ast)\Rightarrow N(0,\boldsymbol\Sigma_{A,U}),
\]
\[
\sqrt n(\widehat{\boldsymbol \theta}_A-\boldsymbol\theta_A^\ast)
=
-\boldsymbol\Gamma_A^{-1}\sqrt n\,\mathbb U_{A,n}(\boldsymbol\theta_A^\ast)+o_p(1),
\]
and therefore
\[
\sqrt n(\widehat{\boldsymbol \theta}_A-\boldsymbol\theta_A^\ast)\Rightarrow N(0,\boldsymbol\Sigma_{A,\theta}).
\]

\end{lemma}
\begin{proof}
The finite-dimensional Z-estimator expansion follows from local uniform convergence of the Jacobian. See Sections~5.3 and~5.6 of van der Vaart.\citep{vanderVaart2000} The derivative envelope and continuity give, for every deterministic \(r_n\downarrow0\),
\[
\sup_{\|\boldsymbol\theta-\boldsymbol\theta_A^\ast\|\le r_n}
\left\|\frac{\partial\mathbb U_{A,n}(\boldsymbol\theta)}{\partial\boldsymbol\theta^\top}
             -\boldsymbol\Gamma_A\right\|\to_p0.
\]
Consistency and the integral Taylor formula therefore give
\[
\begin{aligned}
o_p(1)
&=\sqrt n\,\mathbb U_{A,n}(\widehat{\boldsymbol \theta}_A)\\
&=\sqrt n\,\mathbb U_{A,n}(\boldsymbol\theta_A^\ast)
 +\left\{\int_0^1
 \frac{\partial\mathbb U_{A,n}
 (\boldsymbol\theta_A^\ast+t(\widehat{\boldsymbol \theta}_A-\boldsymbol\theta_A^\ast))}
      {\partial\boldsymbol\theta^\top}\,dt\right\}
 \sqrt n(\widehat{\boldsymbol \theta}_A-\boldsymbol\theta_A^\ast)\\
&=\sqrt n\,\mathbb U_{A,n}(\boldsymbol\theta_A^\ast)
 +\{\boldsymbol\Gamma_A+o_p(1)\}\sqrt n(\widehat{\boldsymbol \theta}_A-\boldsymbol\theta_A^\ast).
\end{aligned}
\]
Since \(\boldsymbol\Gamma_A\) is nonsingular, the iid CLT and Slutsky's theorem yield
\[
\begin{aligned}
\sqrt n(\widehat{\boldsymbol \theta}_A-\boldsymbol\theta_A^\ast)
&=-\{\boldsymbol\Gamma_A+o_p(1)\}^{-1}
       \{\sqrt n\,\mathbb U_{A,n}(\boldsymbol\theta_A^\ast)+o_p(1)\}\\
&=-\boldsymbol\Gamma_A^{-1}\sqrt n\,\mathbb U_{A,n}(\boldsymbol\theta_A^\ast)+o_p(1)\\
&\Rightarrow N(0,\boldsymbol\Gamma_A^{-1}\boldsymbol\Sigma_{A,U}\boldsymbol\Gamma_A^{-T}).
\end{aligned}
\]
\end{proof}

For multinomial replication, we draw
\[
(m_1^{\ast,b},\ldots,m_n^{\ast,b})
\sim\operatorname{Multinomial}(n;1/n,\ldots,1/n)
\]
independently of the observations, and solve
\[
\mathbb U_{A,n}^{\ast,b}(\boldsymbol\theta)
:=\frac1n\sum_{i=1}^n m_i^{\ast,b}\mathbf U_A(O_i;\boldsymbol\theta)=0.
\]
Thus the retrospective propensity weights are \(m_i^{\ast,b}w_{Z_i}(\mathbf X_i)\), the prospective propensity weights are \(m_i^{\ast,b}\), and the prognostic equations use \(m_i^{\ast,b}\mathbf1(Z_i=z)\) in both designs. We use a fixed finite measurable completion when any required unique finite root is unavailable, as in the original fit.

Let \(\mathcal F_n=\sigma(O_1,\ldots,O_n)\), including the measurable original fits, and let \(P^\ast,E^\ast\) denote probability and expectation conditional on \(\mathcal F_n\). All conditional convergence below is in \(P_s\)-probability. In particular, \(R_n^{\ast,b}=o_{p^\ast}(1)\) means that, for every \(\varepsilon,\eta>0\),
\[
P_s\{P^\ast(\|R_n^{\ast,b}\|>\varepsilon)>\eta\}\longrightarrow0.
\]

{In the regression estimating equations below, \(\mathbf X\) also denotes the corresponding model design vector, including an intercept and the specified covariate terms.}

\begin{lemma}[Logistic and linear score estimators and their replicates]
\label{lem:primitive-logistic-linear-roots}
\label{lem:primitive-score-root-bootstrap}
For a fixed estimand and sampling regime \(A\in\{R,P\}\), let the covariate dimension be fixed and define the propensity-equation weight
\[
a_A(O):=
\begin{cases}
w_Z(\mathbf X),&A=R,\\
1,&A=P.
\end{cases}
\]
Suppose the selected observations are i.i.d., \(0<c\le a_A(O)\le C<\infty\), and, for some \(\delta>0\),
\begin{equation}\label{eq:primitive-score-root-moments}
E_s\|\mathbf X\|^{4+2\delta}
+E_s\{\|\mathbf X\|^{2+\delta}|Y|^{2+\delta}\}<\infty.
\end{equation}
Assume that the population logistic equation has a finite solution \(\boldsymbol\alpha^\ast\in\mathbb R^q\):
\[
E_s\!\left[
a_A(O)\mathbf X
\{Z-\operatorname{expit}(\mathbf X^\top\boldsymbol\alpha^\ast)\}
\right]=0,
\]
and that its negative score derivative
\[
\mathcal H_A(\boldsymbol\alpha^\ast)
:=
E_s\!\left[
a_A(O)e(\mathbf X;\boldsymbol\alpha^\ast)
\{1-e(\mathbf X;\boldsymbol\alpha^\ast)\}
\mathbf X\mathbf X^\top
\right]
\]
is positive definite. For each prognostic block present in the scenario, assume that
\[
J_z:=E_s\{\mathbf1(Z=z)\mathbf X\mathbf X^\top\}
\]
is positive definite, and define
\[
\boldsymbol\beta_z^\ast
:=J_z^{-1}E_s\{\mathbf1(Z=z)\mathbf X Y\}.
\]
PATE includes \(z=0,1\). PATT includes \(z=0\) only.

\emph{(i) Original sample.} With probability tending to one, the empirical weighted logistic equation has a unique finite root, and each indicated unweighted arm OLS equation has a unique root. These roots are consistent for the displayed population roots. With any fixed finite measurable completion when a required unique finite root is unavailable, their stack satisfies Assumption~\ref{ass:ordinary-weighted-m-estimation}, and hence Lemma~\ref{lem:ordinary-weighted-first-step}.

\emph{(ii) Multinomial replication.} The conditional probability that all required unique finite replicate roots exist tends to one, and
\[
\widehat{\boldsymbol \theta}_A^{\ast,b}-\widehat{\boldsymbol \theta}_A=o_{p^\ast}(1).
\]
Moreover,
\begin{equation}\label{eq:primitive-score-bootstrap-linearization}
\begin{aligned}
\sqrt n(\widehat{\boldsymbol \theta}_A^{\ast,b}-\widehat{\boldsymbol \theta}_A)
&=-\boldsymbol\Gamma_A^{-1}\frac1{\sqrt n}
 \sum_{i=1}^n(m_i^{\ast,b}-1)\mathbf U_A(O_i;\boldsymbol\theta_A^\ast)
 +o_{p^\ast}(1)\\
&=\frac1{\sqrt n}\sum_{i=1}^n(m_i^{\ast,b}-1)
 \{-\boldsymbol\Gamma_A^{-1}\mathbf U_A(O_i;\boldsymbol\theta_A^\ast)\}
 +o_{p^\ast}(1).
\end{aligned}
\end{equation}
For the class \(\operatorname{BL}_1(\mathbb R^{p_A})\) of real functions bounded by one and with Lipschitz constant at most one,
\begin{equation}\label{eq:primitive-score-bootstrap-law}
\sup_{f\in\operatorname{BL}_1(\mathbb R^{p_A})}
\left|
E^\ast f\{\sqrt n(\widehat{\boldsymbol \theta}_A^{\ast,b}-\widehat{\boldsymbol \theta}_A)\}
-\int f(u)\,N(0,\boldsymbol\Sigma_{A,\theta})(du)
\right|\longrightarrow_p0.
\end{equation}
The Gaussian law may be singular.
\end{lemma}

\begin{proof}
\emph{(i)} We first establish existence and consistency of the logistic root. The empirical logistic Hessian is
\[
\mathcal H_{A,n}(\alpha)
:=\frac1n\sum_i a_A(O_i)e(\mathbf X_i;\alpha)
       \{1-e(\mathbf X_i;\alpha)\}\mathbf X_i\mathbf X_i^\top.
\]
Its expectation is \(\mathcal H_A(\alpha)\). For sufficiently small \(\epsilon>0\), continuity and the uniform LLN give
\[
\begin{aligned}
\inf_{\|\alpha-\alpha^\ast\|\le\epsilon}
 \lambda_{\min}\{\mathcal H_A(\alpha)\}&\ge2c>0,\\
\sup_{\|\alpha-\alpha^\ast\|\le\epsilon}
 \|\mathcal H_{A,n}(\alpha)-\mathcal H_A(\alpha)\|&=o_p(1),\\
\left\|\frac1n\sum_i a_A(O_i)\mathbf X_i
                \{Z_i-e(\mathbf X_i;\alpha^\ast)\}\right\|&=o_p(1).
\end{aligned}
\]
On the event where the sample Hessian is bounded below by \(cI_q\), the outward derivative on the boundary satisfies
\[
\begin{aligned}
&\inf_{\|u\|=1}u^\top\frac1n\sum_i a_A(O_i)\mathbf X_i
          \{e(\mathbf X_i;\alpha^\ast+\epsilon u)-Z_i\}\\
&=\inf_{\|u\|=1}\left[u^\top\frac1n\sum_i a_A(O_i)\mathbf X_i
          \{e(\mathbf X_i;\alpha^\ast)-Z_i\}
 +\epsilon\int_0^1u^\top
             \mathcal H_{A,n}(\alpha^\ast+t\epsilon u)u\,dt\right]\\
&\ge-\left\|\frac1n\sum_i a_A(O_i)\mathbf X_i
                   \{Z_i-e(\mathbf X_i;\alpha^\ast)\}\right\|
  +\epsilon\inf_{\|u\|=1}\int_0^1
          u^\top\mathcal H_{A,n}(\alpha^\ast+t\epsilon u)u\,dt\\
&\ge-\left\|\frac1n\sum_i a_A(O_i)\mathbf X_i
                   \{Z_i-e(\mathbf X_i;\alpha^\ast)\}\right\|
  +\epsilon c.
\end{aligned}
\]
With probability tending to one, this outward derivative is positive. The continuous convex logistic criterion therefore attains its minimum inside the ball. Positive definiteness of \(\mathcal H_{A,n}(\alpha^\ast)\) implies full sample design rank and strict convexity at every finite \(\alpha\). The root is unique. Since \(\epsilon\) can be arbitrarily small, \(\widehat\alpha\to_p\alpha^\ast\).

For each included arm, the LLN and \(J_z>0\) give
\[
\frac1n\sum_i\mathbf1(Z_i=z)\mathbf X_i\mathbf X_i^\top\to_pJ_z,
\]
and, on the event of invertibility,
\[
\begin{aligned}
\widehat\beta_z-\beta_z^\ast
&=\left\{\frac1n\sum_i\mathbf1(Z_i=z)
                      \mathbf X_i\mathbf X_i^\top\right\}^{-1}
  \frac1n\sum_i\mathbf1(Z_i=z)\mathbf X_i
                      (Y_i-\mathbf X_i^\top\beta_z^\ast)\\
&\to_p J_z^{-1}
 E_s\{\mathbf1(Z=z)\mathbf X(Y-\mathbf X^\top\beta_z^\ast)\}\\
&=0.
\end{aligned}
\]
On a fixed parameter neighborhood, the required envelopes satisfy
\[
\begin{aligned}
\sup_{\boldsymbol \theta}\|\mathbf U_A(O;\boldsymbol\theta)\|^{2+\delta}
&\le C\{1+\|\mathbf X\|^{4+2\delta}
                 +\|\mathbf X\|^{2+\delta}|Y|^{2+\delta}\},\\
\sup_{\boldsymbol \theta}\left\|\frac{\partial \mathbf U_A(O;\boldsymbol\theta)}
                              {\partial\boldsymbol\theta^\top}\right\|
&\le C\|\mathbf X\|^2,\\
\|\mathcal H_{A,n}(\alpha)-\mathcal H_{A,n}(\alpha')\|
&\le C\|\alpha-\alpha'\|\frac1n\sum_i\|\mathbf X_i\|^3.
\end{aligned}
\]
Their integrability follows from \eqref{eq:primitive-score-root-moments}. The derivative blocks are \(-\mathcal H_A(\alpha^\ast)\) and \(-J_z\), so Assumption~\ref{ass:ordinary-weighted-m-estimation} holds and Lemma~\ref{lem:ordinary-weighted-first-step} applies.

\emph{(ii)} For every fixed scalar row array \(a_i\), multinomial counts satisfy
\[
E^\ast\!\left[\left\{\frac1n\sum_i(m_i^\ast-1)a_i\right\}^{2}\right]
=\frac1{n^2}\sum_i(a_i-\bar a_n)^2
\le\frac1{n^2}\sum_i a_i^2.
\]
This covariance identity controls the score and Hessian entries at each fixed parameter. A finite net on a small coefficient ball extends the Hessian bound uniformly, using the preceding modulus and the replicate envelope:
\[
E^\ast\!\left(\frac1n\sum_i m_i^\ast\|\mathbf X_i\|^3\right)
=\frac1n\sum_i\|\mathbf X_i\|^3=O_p(1).
\]
The net bound and conditional Markov inequality give
\[
\begin{aligned}
\sup_{\|\alpha-\alpha^\ast\|\le\epsilon}
\left\|\frac1n\sum_i(m_i^\ast-1)a_A(O_i)e(\mathbf X_i;\alpha)
 \{1-e(\mathbf X_i;\alpha)\}\mathbf X_i\mathbf X_i^\top\right\|
&=o_{p^\ast}(1),\\
\frac1n\sum_i m_i^\ast\mathbf1(Z_i=z)\mathbf X_i\mathbf X_i^\top
&=J_z+o_p(1)+o_{p^\ast}(1).
\end{aligned}
\]
The resulting Hessian and gradient bounds allow the argument in part (i) to be applied conditionally, proving existence and consistency of the replicate roots.

Taylor expansion of the replicate estimating equation about \(\widehat{\boldsymbol \theta}_A\), followed by convergence of its Jacobian, gives
\[
\sqrt n(\widehat{\boldsymbol \theta}_A^\ast-\widehat{\boldsymbol \theta}_A)
=-\boldsymbol\Gamma_A^{-1}\frac1{\sqrt n}\sum_i(m_i^\ast-1)
 \mathbf U_A(O_i;\widehat{\boldsymbol \theta}_A)+o_{P^\ast}(1).
\]
The derivative envelope and consistency give
\[
\frac1n\sum_i\|\mathbf U_A(O_i;\widehat{\boldsymbol \theta}_A)
                 -\mathbf U_A(O_i;\boldsymbol\theta_A^\ast)\|^2
\le C\|\widehat{\boldsymbol \theta}_A-\boldsymbol\theta_A^\ast\|^2
                    \frac1n\sum_i\|\mathbf X_i\|^4=o_p(1).
\]
Consequently, the multinomial covariance bound yields
\[
\begin{aligned}
&E^\ast\left\|\frac1{\sqrt n}\sum_i(m_i^\ast-1)
 \{\mathbf U_A(O_i;\widehat{\boldsymbol \theta}_A)-\mathbf U_A(O_i;\boldsymbol\theta_A^\ast)\}\right\|^2\\
&\le\frac1n\sum_i\|\mathbf U_A(O_i;\widehat{\boldsymbol \theta}_A)
                         -\mathbf U_A(O_i;\boldsymbol\theta_A^\ast)\|^2\\
&=o_p(1),
\end{aligned}
\]
which proves \eqref{eq:primitive-score-bootstrap-linearization}. The empirical-mean bootstrap CLT in Section~23.2 of van der Vaart\citep{vanderVaart2000} now gives \eqref{eq:primitive-score-bootstrap-law}, also for a singular covariance.
\end{proof}

Let \(b_z(O)\) be a known measurable standardization weight and \(q_z=E_sb_z(O)\). This choice specifies whether the moments are pooled or arm-specific and whether survey weights are used. Apply the same rule to the original fit and each replicate.

Write the two unstandardized coordinates as
\[
r_{z1}(\mathbf X;\boldsymbol\theta)=e(\mathbf X;\boldsymbol\alpha),
\qquad
r_{z2}(\mathbf X;\boldsymbol\theta)=\mathbf X^\top\boldsymbol\beta_z.
\]
Their gradients below are with respect to the entire scenario vector \(\boldsymbol\theta\), with zeros in absent coordinates. Define
\[
\ell_{zk}:=\frac{E_s\{b_z(O)r_{zk}(\mathbf X;\boldsymbol\theta_A^\ast)\}}{q_z},
\qquad
v_{zk}:=\frac{E_s[b_z(O)
 \{r_{zk}(\mathbf X;\boldsymbol\theta_A^\ast)-\ell_{zk}\}^2]}{q_z},
\qquad s_{zk}:=\sqrt{v_{zk}},
\]

We use the corresponding empirical ratios
\[
\widehat\ell_{zk}
:=\frac{\sum_i b_z(O_i)r_{zk}(\mathbf X_i;\widehat{\boldsymbol \theta}_A)}
        {\sum_i b_z(O_i)},
\qquad
\widehat v_{zk}
:=\frac{\sum_i b_z(O_i)
 \{r_{zk}(\mathbf X_i;\widehat{\boldsymbol \theta}_A)-\widehat\ell_{zk}\}^2}
        {\sum_i b_z(O_i)},
\qquad
\widehat s_{zk}:=\sqrt{\widehat v_{zk}}.
\]
In the replicate insert \(m_i^{\ast,b}\) into every sum and use \(\widehat{\boldsymbol \theta}_A^{\ast,b}\). The matching coordinates for this chosen normalization are
\[
D_{zk}^\ast(\mathbf x)
:=\frac{r_{zk}(\mathbf x;\boldsymbol\theta_A^\ast)-\ell_{zk}}{s_{zk}},
\qquad
\widehat D_{zk}(\mathbf x)
:=\frac{r_{zk}(\mathbf x;\widehat{\boldsymbol \theta}_A)-\widehat\ell_{zk}}
        {\widehat s_{zk}}.
\]
On samples with a zero moment denominator or variance, assign fixed finite locations and positive scales.

Let \(\psi_A(O):=-\boldsymbol\Gamma_A^{-1}\mathbf U_A(O;\boldsymbol\theta_A^\ast)\), and define the influence functions
\begin{equation}\label{eq:score-standardization-influence}
\begin{aligned}
\mathcal I_{\ell,zk}(O)
:={}&\frac{b_z(O)}{q_z}
 \{r_{zk}(\mathbf X;\boldsymbol\theta_A^\ast)-\ell_{zk}\}\\
&+\frac{E_s\{b_z(O)\nabla_{\boldsymbol \theta}
                  r_{zk}(\mathbf X;\boldsymbol\theta_A^\ast)\}^{\top}}{q_z}
 \psi_A(O),\\
\mathcal I_{v,zk}(O)
:={}&\frac{b_z(O)}{q_z}
 \big[\{r_{zk}(\mathbf X;\boldsymbol\theta_A^\ast)-\ell_{zk}\}^2-v_{zk}\big]\\
&+\frac{2E_s[b_z(O)
 \{r_{zk}(\mathbf X;\boldsymbol\theta_A^\ast)-\ell_{zk}\}
 \nabla_{\boldsymbol \theta} r_{zk}(\mathbf X;\boldsymbol\theta_A^\ast)]^{\top}}{q_z}
 \psi_A(O),\\
\mathcal I_{s,zk}(O)
:={}&\frac{\mathcal I_{v,zk}(O)}{2s_{zk}}.
\end{aligned}
\end{equation}

\begin{lemma}[Empirical score location and scale]
\label{lem:primitive-score-standardization}
Assume the conditions and use the score roots of Lemma~\ref{lem:primitive-logistic-linear-roots}. For each score used by the scenario, fix a known measurable standardization weight \(b_z(O)\) with \(0\le b_z(O)\le C_b<\infty\) and \(q_z:=E_s b_z(O)>0\). Assume \(v_{zk}>0\) for each coordinate actually matched on. The exceptional samples just described have probability tending to zero under both the sampling and conditional replicate laws.

The influence rows defined above have mean zero. Then, jointly over all required coordinates,
\begin{equation}\label{eq:score-standardization-sampling}
\sqrt n
\begin{pmatrix}
\widehat\ell_{zk}-\ell_{zk}\\
\widehat v_{zk}-v_{zk}\\
\widehat s_{zk}-s_{zk}
\end{pmatrix}
=
\frac1{\sqrt n}\sum_i
\begin{pmatrix}
\mathcal I_{\ell,zk}(O_i)\\
\mathcal I_{v,zk}(O_i)\\
\mathcal I_{s,zk}(O_i)
\end{pmatrix}
+o_p(1),
\end{equation}
and, conditionally in probability,
\begin{equation}\label{eq:score-standardization-bootstrap}
\sqrt n
\begin{pmatrix}
\widehat\ell_{zk}^{\ast,b}-\widehat\ell_{zk}\\
\widehat v_{zk}^{\ast,b}-\widehat v_{zk}\\
\widehat s_{zk}^{\ast,b}-\widehat s_{zk}
\end{pmatrix}
=
\frac1{\sqrt n}\sum_i(m_i^{\ast,b}-1)
\begin{pmatrix}
\mathcal I_{\ell,zk}(O_i)\\
\mathcal I_{v,zk}(O_i)\\
\mathcal I_{s,zk}(O_i)
\end{pmatrix}
+o_{p^\ast}(1).
\end{equation}
Both expansions hold jointly with the score-parameter expansion. Their Gaussian covariance is that of the stacked influence functions under \(P_s\), and may be singular.

The common location cancels from each pairwise score difference:
\begin{equation}\label{eq:standardized-pairwise-location-cancellation}
\begin{aligned}
&\widehat D_{zk}(\mathbf x)-\widehat D_{zk}(\mathbf y)\\
&=\frac{r_{zk}(\mathbf x;\widehat{\boldsymbol \theta}_A)-\widehat\ell_{zk}}
        {\widehat s_{zk}}
 -\frac{r_{zk}(\mathbf y;\widehat{\boldsymbol \theta}_A)-\widehat\ell_{zk}}
        {\widehat s_{zk}}\\
&=\frac{r_{zk}(\mathbf x;\widehat{\boldsymbol \theta}_A)
       -r_{zk}(\mathbf y;\widehat{\boldsymbol \theta}_A)}{\widehat s_{zk}}.
\end{aligned}
\end{equation}
If additionally \(\|\mathbf X\|\le C_X\) almost surely, then
\[
\sup_{\mathbf x\in\operatorname{supp}(\mathbf X)}
 \|\widehat{\mathbf D}_z(\mathbf x)-\mathbf D_z^\ast(\mathbf x)\|
=O_p(n^{-1/2}),
\]
with the analogous conditional bound for \(\widehat{\mathbf D}_z^{\ast,b}-\widehat{\mathbf D}_z\).
\end{lemma}

\begin{proof}
The moment assumptions and Lemma~\ref{lem:ordinary-weighted-first-step} give \(n^{-1}\sum_i b_z(O_i)\to_p q_z>0\) and the expansions
\[
\begin{aligned}
\sqrt n(\widehat\ell_{zk}-\ell_{zk})
&=\frac{n^{-1/2}\sum_i b_z(O_i)
              \{r_{zk}(\mathbf X_i;\widehat{\boldsymbol \theta}_A)-\ell_{zk}\}}
        {n^{-1}\sum_i b_z(O_i)}\\
&=\frac1{q_z\sqrt n}\sum_i b_z(O_i)
              \{r_{zk}(\mathbf X_i;\boldsymbol\theta_A^\ast)-\ell_{zk}\}
 +\frac{E_s\{b_z\nabla_{\boldsymbol \theta} r_{zk}(\mathbf X;\boldsymbol\theta_A^\ast)\}^\top}
        {q_z}\sqrt n(\widehat{\boldsymbol \theta}_A-\boldsymbol\theta_A^\ast)+o_p(1)\\
&=\frac1{\sqrt n}\sum_i\mathcal I_{\ell,zk}(O_i)+o_p(1).
\end{aligned}
\]
Centering at the population location gives
\[
\widehat v_{zk}
=\frac{\sum_i b_z(O_i)
       \{r_{zk}(\mathbf X_i;\widehat{\boldsymbol \theta}_A)-\ell_{zk}\}^2}
       {\sum_i b_z(O_i)}-(\widehat\ell_{zk}-\ell_{zk})^2.
\]
Hence Taylor expansion and \(\sqrt n(\widehat\ell_{zk}-\ell_{zk})^2=o_p(1)\) give
\[
\begin{aligned}
\sqrt n(\widehat v_{zk}-v_{zk})
&=\frac1{q_z\sqrt n}\sum_i b_z(O_i)
       [\{r_{zk}(\mathbf X_i;\boldsymbol\theta_A^\ast)-\ell_{zk}\}^2-v_{zk}]\\
&\quad+\frac{2E_s[b_z
       \{r_{zk}(\mathbf X;\boldsymbol\theta_A^\ast)-\ell_{zk}\}
       \nabla_{\boldsymbol \theta} r_{zk}(\mathbf X;\boldsymbol\theta_A^\ast)]^\top}{q_z}
       \sqrt n(\widehat{\boldsymbol \theta}_A-\boldsymbol\theta_A^\ast)+o_p(1)\\
&=\frac1{\sqrt n}\sum_i\mathcal I_{v,zk}(O_i)+o_p(1),\\
\sqrt n(\widehat s_{zk}-s_{zk})
&=\frac{\sqrt n(\widehat v_{zk}-v_{zk})}{\widehat s_{zk}+s_{zk}}\\
&=\frac1{\sqrt n}\sum_i\mathcal I_{s,zk}(O_i)+o_p(1).
\end{aligned}
\]
The raw-moment coordinates are square-integrable by \eqref{eq:primitive-score-root-moments}, so the expansions hold jointly by the finite-dimensional delta method. The bootstrap delta method in Chapter~23 of van der Vaart,\citep{vanderVaart2000} applied to the same maps and the joint expansion in Lemma~\ref{lem:primitive-score-root-bootstrap}, gives \eqref{eq:score-standardization-bootstrap}.

When the covariates are bounded, the derivatives of the raw coordinates are uniformly bounded near \(\boldsymbol\theta_A^\ast\), so
\[
\begin{aligned}
\sup_{\mathbf x\in\operatorname{supp}(\mathbf X)}
 \|\widehat{\mathbf D}_z(\mathbf x)
                         -\mathbf D_z^\ast(\mathbf x)\|
&\le C\left[\|\widehat{\boldsymbol \theta}_A-\boldsymbol\theta_A^\ast\|
 +\sum_{k=1}^2\{|\widehat\ell_{zk}-\ell_{zk}|
                    +|\widehat s_{zk}-s_{zk}|\}\right]\\
&=O_p(n^{-1/2}).
\end{aligned}
\]
Applying this bound to the replicate-minus-original differences gives \(O_{p^\ast}(n^{-1/2})\) conditionally in probability.
\end{proof}

\subsubsection{Fixed-score joint limit}

\begin{lemma}[Joint limit at fixed scores]
\label{lem:baseline-oracle-firststep-joint}
Fix a sampling regime \(A\in\{R,P\}\) and an estimand, PATE or PATT. We work under the fixed selected-sample law \(P_s\) at the limiting standardized scores \(\mathbf D_z^*\). Assume the corresponding population reduced-score identification and survey-recovery identities, Assumptions \ref{ass:residual-centering-survey-normalized}, \ref{ass:residual-qv-stabilization}, and \ref{ass:ordinary-weighted-m-estimation}. The unnormalized heterogeneity summand in the corresponding oracle decomposition has a finite \(2+\delta\) moment. Reduce \(\delta>0\) if necessary to use the same exponent for this summand, the observed residual, and \(\mathbf U_A(O;\boldsymbol\theta_A^\ast)\).

Let \(H_{n,A}\), \(E_{n,A}\), and \(B_{n,A}\) be the corresponding oracle decomposition terms. Then
\[
\binom{\sqrt n\{H_{n,A}+E_{n,A}\}}
      {\sqrt n\,\mathbb U_{A,n}(\boldsymbol\theta_A^\ast)}
\Rightarrow
N\left(\binom00,
\begin{bmatrix}
V_{A,\mathrm{known\text{-}score}}&\boldsymbol\gamma_{A,1}^\top\\
\boldsymbol\gamma_{A,1}&\boldsymbol\Sigma_{A,U}
\end{bmatrix}\right),
\]
where \(V_{A,\mathrm{known\text{-}score}}=V_{H,A}+V_{E,A}\). The covariance vector \(\boldsymbol\gamma_{A,1}\) is given in \eqref{eq:baseline-oracle-firststep-gamma}. The expectations and matching limits below define these deterministic covariance blocks without requiring closed-form expressions. We write \(G_{\tau,A}\) for the scalar Gaussian limit of the first coordinate and \(\mathbf G_{\theta,A}\) for the jointly Gaussian limit of \(\sqrt n(\widehat{\boldsymbol \theta}_A-\boldsymbol\theta_A^*)\), obtained by applying \(-\boldsymbol\Gamma_A^{-1}\) to the second coordinate. In particular,
\[
\begin{aligned}
\operatorname{Var}(\mathbf G_{\theta,A})
&=\boldsymbol\Gamma_A^{-1}\boldsymbol\Sigma_{A,U}\boldsymbol\Gamma_A^{-T}
=\boldsymbol\Sigma_{A,\theta},\\
\operatorname{Cov}(\mathbf G_{\theta,A},G_{\tau,A})
&=-\boldsymbol\Gamma_A^{-1}\boldsymbol\gamma_{A,1}.
\end{aligned}
\]
The minus sign follows from the estimating-equation expansion in Lemma~\ref{lem:ordinary-weighted-first-step}.

If the correction on these same oracle scores and matches satisfies
\begin{equation}\label{eq:baseline-oracle-firststep-bc-remainder}
\sqrt n\{\widehat B_{n,A}-B_{n,A}\}\to_p0,
\end{equation}
the same joint limit holds with \(\sqrt n\{\widehat\tau_A^{\mathrm{BC}}(\boldsymbol\theta_A^\ast)-\tau_A\}\) as its first coordinate.
\end{lemma}
\begin{proof}
For \(A\in\{R,P\}\), the residual coefficients in \eqref{eq:qv-four-scenario-coefficients} are
\[
c_{n,A,i}=
\begin{cases}
(2Z_i-1)\{w_{A,Z_i}(\mathbf X_i)+K_{n,Z_i}^{A}(i)\},
 &\text{PATE},\\
Z_iw_{A,1}(\mathbf X_i)-(1-Z_i)K_{n,0}^{A}(i),
 &\text{PATT}.
\end{cases}
\]
The decomposition for each estimand, with the stated count convention, gives
\[
\sqrt n(H_{n,A}+E_{n,A})
=\frac1{\sqrt n\,\bar w_{A,n}}\sum_i
 (h_{A,i}+c_{n,A,i}\varepsilon_{A,i})+o_p(1),
\]
where
\[
h_{A,i}:=
\begin{cases}
w_{A,Z_i}(\mathbf X_i)\{\tau_A^D(\mathbf X_i)-\tau_{\mathrm{PATE}}\},
 &\text{PATE},\\
Z_iw_{A,1}(\mathbf X_i)
 \{\tau_{\mathrm{PATT}}(\mathbf D_0^*(\mathbf X_i))-\tau_{\mathrm{PATT}}\},
 &\text{PATT}.
\end{cases}
\]
Population recovery gives \(E_s h_A=0\). By the iid CLT and the residual variance limit,
\[
\begin{aligned}
\left(\bar w_{A,n}^{-1}-\bar w_A^{-1}\right)
 \frac1{\sqrt n}\sum_i(h_{A,i}+c_{n,A,i}\varepsilon_{A,i})
&=o_p(1)O_p(1)\\
&=o_p(1).
\end{aligned}
\]

To obtain joint convergence, we reveal the matching variables, the remaining covariates, and the outcomes in succession, following the proof of Proposition~2 of Abadie and Imbens.\citep{AbadieImbens2016} Let \(\mathcal X_i=\sigma(Z_i,\mathbf X_i)\) and \(\mathcal X_{1:n}=\bigvee_{i=1}^n\mathcal X_i\). Since \(\mathcal G_{A,i}^{\mathrm{row}}\subseteq\mathcal X_i\), we define
\[
\begin{aligned}
u_{A,i}&=E_s\{\mathbf U_A(O_i;\boldsymbol\theta_A^\ast)\mid\mathcal G_{A,i}^{\mathrm{row}}\},\\
r_{A,i}&=\mathbf U_A(O_i;\boldsymbol\theta_A^\ast)-u_{A,i},\\
k_{A,i}&=E_s(\varepsilon_{A,i}r_{A,i}\mid\mathcal G_{A,i}^{\mathrm{row}}),\\
R_{A,i}&=E_s(r_{A,i}r_{A,i}^{\top}\mid\mathcal G_{A,i}^{\mathrm{row}}).
\end{aligned}
\]
The pairs \((\varepsilon_{A,i},r_{A,i})\) are independent and centered conditionally on \(\mathcal G_{A,n}^{\mathrm{match}}=\bigvee_i\mathcal G_{A,i}^{\mathrm{row}}\). For PATE, the three vector increments associated with row \(i\) are
\[
\boldsymbol\xi_{A,n,i}
=\frac1{\sqrt n}
\begin{pmatrix}
\displaystyle
\frac{w_{A,Z_i}(\mathbf X_i)
       \{\tau_A^D(\mathbf X_i)-\tau_{\mathrm{PATE}}\}}{\bar w_A}\\[3mm]
u_{A,i}
\end{pmatrix},
\]
\[
\boldsymbol\xi_{A,n,n+i}
=\frac1{\sqrt n}
\begin{pmatrix}
\displaystyle
\frac{(2Z_i-1)\{w_{A,Z_i}(\mathbf X_i)+K_{n,Z_i}^{A}(i)\}}
     {\bar w_A}
 E_s(\varepsilon_{A,i}\mid\mathcal X_i)\\[3mm]
E_s\{\mathbf U_A(O_i;\boldsymbol\theta_A^*)\mid\mathcal X_i\}-u_{A,i}
\end{pmatrix},
\]
and
\[
\boldsymbol\xi_{A,n,2n+i}
=\frac1{\sqrt n}
\begin{pmatrix}
\displaystyle
\frac{(2Z_i-1)\{w_{A,Z_i}(\mathbf X_i)+K_{n,Z_i}^{A}(i)\}}
     {\bar w_A}
 \{\varepsilon_{A,i}-E_s(\varepsilon_{A,i}\mid\mathcal X_i)\}\\[3mm]
\mathbf U_A(O_i;\boldsymbol\theta_A^*)-E_s\{\mathbf U_A(O_i;\boldsymbol\theta_A^*)\mid\mathcal X_i\}
\end{pmatrix},
\qquad 1\le i\le n.
\]
For PATT, the first matching coordinate uses the treated heterogeneity summand \(h_{A,i}/\bar w_A\) defined above. In the second and third blocks we replace the signed PATE factor by
\[
Z_iw_{A,1}(\mathbf X_i)-(1-Z_i)K_{n,0}^{A}(i).
\]
The nuisance coordinates are unchanged. Both arrays are adapted to
\[
\mathcal F_{A,n,k}=
\begin{cases}
 \bigvee_{i=1}^k\mathcal G_{A,i}^{\mathrm{row}},&0\le k\le n,\\
 \mathcal G_{A,n}^{\mathrm{match}}\vee\bigvee_{i=1}^{k-n}\mathcal X_i,
 &n<k\le2n,\\
 \mathcal X_{1:n}\vee\sigma(Y_1,\ldots,Y_{k-2n}),&2n<k\le3n.
\end{cases}
\]
The empty join denotes the trivial field. The tower property centers the second-stage increment:
\[
\begin{aligned}
E_s\{E_s(\varepsilon_{A,i}\mid\mathcal X_i)
                         \mid\mathcal G_{A,i}^{\mathrm{row}}\}
&=E_s(\varepsilon_{A,i}\mid\mathcal G_{A,i}^{\mathrm{row}})\\
&=0,\\
E_s\{E_s(\mathbf U_A(O_i;\boldsymbol\theta_A^\ast)\mid\mathcal X_i)-u_{A,i}
                         \mid\mathcal G_{A,i}^{\mathrm{row}}\}
&=u_{A,i}-u_{A,i}\\
&=0.
\end{aligned}
\]
The third-stage increments are centered given \(\mathcal X_i\). Row independence and measurability of \(c_{n,A,i}\) with respect to \(\mathcal G_{A,n}^{\mathrm{match}}\) therefore give
\[
E_s(\boldsymbol\xi_{A,n,k}\mid\mathcal F_{A,n,k-1})=0,
\qquad 1\le k\le3n.
\]
For \(l<k\), adaptedness and iterated expectation also give
\[
E_s(\boldsymbol\xi_{A,n,k}\boldsymbol\xi_{A,n,l}^{\top})
=E_s\{E_s(\boldsymbol\xi_{A,n,k}\mid\mathcal F_{A,n,k-1})
                         \boldsymbol\xi_{A,n,l}^{\top}\}=0.
\]
Different increments are uncorrelated, although the two coordinates within an increment may be correlated. Summing the three increments for observation \(i\) gives
\[
\boldsymbol\xi_{A,n,i}+\boldsymbol\xi_{A,n,n+i}+\boldsymbol\xi_{A,n,2n+i}
=\frac1{\sqrt n}
 \binom{\{h_{A,i}+c_{n,A,i}\varepsilon_{A,i}\}/\bar w_A}
       {\mathbf U_A(O_i;\boldsymbol\theta_A^\ast)}.
\]
We write the first-stage sum and the combined remaining sum as
\[
\mathcal M_{A,n}^{(1)}=\frac1{\sqrt n}\sum_i
 \binom{h_{A,i}/\bar w_A}{u_{A,i}},\qquad
\mathcal M_{A,n}^{(2:3)}=\frac1{\sqrt n}\sum_i
 \binom{c_{n,A,i}\varepsilon_{A,i}/\bar w_A}{r_{A,i}}.
\]
The first is an iid centered sum with covariance limit
\[
\Omega_{H,A}
=\operatorname{Var}_s\binom{h_A/\bar w_A}{u_A}.
\]
Conditional covariance decomposition separates the covariate and outcome contributions:
\[
\begin{aligned}
k_{A,i}
&=E_s\{E_s(\varepsilon_{A,i}r_{A,i}\mid\mathcal X_i)
                                          \mid\mathcal G_{A,i}^{\mathrm{row}}\}\\
&=E_s\!\left[
 E_s(\varepsilon_{A,i}\mid\mathcal X_i)
 \{E_s(\mathbf U_A(O_i;\boldsymbol\theta_A^\ast)\mid\mathcal X_i)-u_{A,i}\}
                                          \mid\mathcal G_{A,i}^{\mathrm{row}}\right]\\
&\quad+E_s\!\left[
 \operatorname{Cov}_s\{\mathbf U_A(O_i;\boldsymbol\theta_A^\ast),\varepsilon_{A,i}
                         \mid\mathcal X_i\}
                                          \mid\mathcal G_{A,i}^{\mathrm{row}}\right].
\end{aligned}
\]

Conditional Jensen and Cauchy--Schwarz, with \(p=1+\delta/2\), give
\[
\begin{aligned}
E_s(\|u_A\|^{2p}+\|r_A\|^{2p})
&\le C_pE_s\|\mathbf U_A\|^{2p}<\infty,\\
E_s\|k_A\|^p
&\le E_s\{E_s(\varepsilon_A^2\mid\mathcal G_{A,1}^{\mathrm{row}})
                   E_s(\|r_A\|^2\mid\mathcal G_{A,1}^{\mathrm{row}})\}^{p/2}\\
&\le\{E_s|\varepsilon_A|^{2p}\}^{1/2}
       \{E_s\|r_A\|^{2p}\}^{1/2}<\infty,\\
E_s\|R_A\|^p&\le E_s\|r_A\|^{2p}<\infty.
\end{aligned}
\]

For regime \(A\in\{R,P\}\), the joint conditional distribution of the weight, residual variance and residual--estimating-function covariance is
\begin{equation}\label{eq:baseline-oracle-firststep-mark-kernel}
\mathsf K_{A,a\mid z}^{U}(d,\cdot)
:=\mathcal L_s\{(w_{A,a}(\mathbf X),v_A,k_A)
             \mid Z=a,\mathbf D_z^*(\mathbf X)=d\}.
\end{equation}
\begin{equation}\label{eq:baseline-oracle-firststep-cross-constant}
\begin{aligned}
\mathcal C_{A,\mathrm{PATE}}
&:=\sum_{z=0}^1(2z-1)\int_{\mathcal S_z}h_{z\mid z}(d)\\
&\qquad{}\times
 \int E[\{w+K_{A,z}^\infty(d;w,v)\}k]\,
         \mathsf K_{A,z\mid z}^{U}(d,dw,dv,dk)\,dd,\\
\mathcal C_{A,\mathrm{PATT}}
&:=E_s\{Z w_{A,1}(\mathbf X)k_A\}\\
&\quad-\int_{\mathcal S_0}h_{0\mid0}(d)
 \int E[K_{A,0}^\infty(d;w,v)k]\,
         \mathsf K_{A,0\mid0}^{U}(d,dw,dv,dk)\,dd.
\end{aligned}
\end{equation}
The inner expectations condition on \((w,v,k)\) and use the coefficient in \eqref{eq:qv-poisson-exact-coefficient}. We write \(\mathcal C_A\) for the expression corresponding to the fixed estimand. For each matched arm and \(p=1+\delta/2\), the coefficient bound and \eqref{eq:qv-uniform-conditional-reuse} give
\[
\begin{aligned}
E_s\{\mathbf1(Z_i=z)\|c_{n,A,i}k_{A,i}\|^p\}
&\le C E_s\{\mathbf1(Z_i=z)(1+K_{M,z}(i))^p\|k_{A,i}\|^p\}\\
&=C E_s\!\left[\mathbf1(Z_i=z)\|k_{A,i}\|^p
 E_s\{(1+K_{M,z}(i))^p\mid Z_i,\mathbf D_z^*(\mathbf X_i)\}\right]\\
&\le C'E_s\{\mathbf1(Z_i=z)\|k_{A,i}\|^p\}<\infty.
\end{aligned}
\]
This moment bound permits the marked-limit and truncation argument in Proposition~\ref{prop:primitive-residual-qv-limit} to be applied coordinatewise to \eqref{eq:baseline-oracle-firststep-mark-kernel}. Combining it with the law of large numbers for the direct treated PATT term gives
\begin{equation}\label{eq:baseline-oracle-firststep-cross-limit}
\frac1n\sum_{i=1}^n c_{n,A,i}k_{A,i}\longrightarrow_{L^1}\mathcal C_A.
\end{equation}

Conditional independence, the residual variance limit, the cross-covariance limit, and the law of large numbers for \(R_A\) give
\begin{equation}\label{eq:baseline-oracle-firststep-conditional-qv}
\begin{aligned}
&\operatorname{Var}_s(\mathcal M_{A,n}^{(2:3)}\mid\mathcal G_{A,n}^{\mathrm{match}})\\
&=
\begin{bmatrix}
\displaystyle\frac1{n\bar w_A^2}\sum_i c_{n,A,i}^2v_{A,i}&
\displaystyle\frac1{n\bar w_A}\sum_i c_{n,A,i}k_{A,i}^\top\\
\displaystyle\frac1{n\bar w_A}\sum_i c_{n,A,i}k_{A,i}&
\displaystyle\frac1n\sum_iR_{A,i}
\end{bmatrix}\\
&\to_p
\Omega_{E,A}:=
\begin{bmatrix}
V_{E,A}&\mathcal C_A^\top/\bar w_A\\
\mathcal C_A/\bar w_A&E_s(R_{A,1})
\end{bmatrix}.
\end{aligned}
\end{equation}
To obtain the predictable covariance for the refined filtration, let
\[
\mathcal T_{A,n,i}
:=\operatorname{Var}_s\!\left\{
 \binom{c_{n,A,i}\varepsilon_{A,i}/\bar w_A}
       {\mathbf U_A(O_i;\boldsymbol\theta_A^\ast)}\,\middle|\,\mathcal X_{1:n}\right\}.
\]
For each \(i\), the preceding information at the second and third stages is, respectively,
\[
\begin{aligned}
\mathcal F_{A,n,n+i-1}
&=\mathcal G_{A,n}^{\mathrm{match}}\vee
                   \bigvee_{j<i}\mathcal X_j,\\
\mathcal F_{A,n,2n+i-1}
&=\mathcal X_{1:n}\vee\sigma(Y_1,\ldots,Y_{i-1}).
\end{aligned}
\]
Because the limiting score maps depend on individual observations, row independence yields conditional independence given \(\mathcal G_{A,n}^{\mathrm{match}}\). Revealing other observations' covariates leaves the conditional law of row \(i\) unchanged. Likewise, earlier outcomes do not change the law of \(Y_i\) given \(\mathcal X_{1:n}\). The matching coefficients are measurable with respect to \(\mathcal G_{A,n}^{\mathrm{match}}\), so
\[
\begin{aligned}
&nE_s(\boldsymbol\xi_{A,n,n+i}\boldsymbol\xi_{A,n,n+i}^{\top}
                              \mid\mathcal F_{A,n,n+i-1})\\
&=\operatorname{Var}_s\!\left[
 E_s\!\left\{\binom{c_{n,A,i}\varepsilon_{A,i}/\bar w_A}
                         {\mathbf U_A(O_i;\boldsymbol\theta_A^*)}
                    \,\middle|\,\mathcal X_{1:n}\right\}
                    \,\middle|\,\mathcal G_{A,n}^{\mathrm{match}}\right]\\
&=\operatorname{Var}_s\!\left\{
 \binom{c_{n,A,i}\varepsilon_{A,i}/\bar w_A}{\mathbf U_A(O_i;\boldsymbol\theta_A^*)}
                    \,\middle|\,\mathcal G_{A,n}^{\mathrm{match}}\right\}
 -E_s(\mathcal T_{A,n,i}\mid\mathcal G_{A,n}^{\mathrm{match}}).
\end{aligned}
\]
The last equality is conditional total covariance. At the third stage, subtracting the full-covariate conditional mean does not change conditional covariance, so
\[
nE_s(\boldsymbol\xi_{A,n,2n+i}\boldsymbol\xi_{A,n,2n+i}^{\top}
                              \mid\mathcal F_{A,n,2n+i-1})
=\mathcal T_{A,n,i}.
\]
Summing these identities over \(i\) and using conditional row independence gives
\begin{equation}\label{eq:joint-three-stage-qv-comparison}
\begin{aligned}
&\sum_{k=n+1}^{3n}
 E_s(\boldsymbol\xi_{A,n,k}\boldsymbol\xi_{A,n,k}^\top\mid\mathcal F_{A,n,k-1})\\
&=\operatorname{Var}_s(\mathcal M_{A,n}^{(2:3)}\mid\mathcal G_{A,n}^{\mathrm{match}})
 +\frac1n\sum_i\{\mathcal T_{A,n,i}
             -E_s(\mathcal T_{A,n,i}\mid\mathcal G_{A,n}^{\mathrm{match}})\}.
\end{aligned}
\end{equation}
For \(p=1+\delta/2>1\), conditional Jensen and the reuse bound \eqref{eq:qv-uniform-conditional-reuse} yield
\[
\begin{aligned}
\sup_n\frac1n\sum_i E_s\|\mathcal T_{A,n,i}\|^p
&\le C\sup_n\frac1n\sum_i
 E_s\{|c_{n,A,i}\varepsilon_{A,i}|^{2p}
                    +\|\mathbf U_A(O_i;\boldsymbol\theta_A^\ast)\|^{2p}\}\\
&\le C' E_s\{|\varepsilon_A|^{2p}+\|\mathbf U_A\|^{2p}\}
<\infty.
\end{aligned}
\]
Given \(\mathcal G_{A,n}^{\mathrm{match}}\), the matching coefficients are fixed and the remaining covariates are independent across rows. Hence the matrices \(\mathcal T_{A,n,i}\) are conditionally independent. We control their centered average by truncating each scalar entry \(T_{n,i}\) at a fixed level: \(T_{n,i}^{[L]}=\max\{-L,\min(T_{n,i},L)\}\). Conditional independence, Cauchy--Schwarz and the preceding moment bound give
\[
\begin{aligned}
&E_s\left|\frac1n\sum_i
 \{T_{n,i}-E_s(T_{n,i}\mid\mathcal G_{A,n}^{\mathrm{match}})\}\right|\\
&\le E_s\left|\frac1n\sum_i
 \{T_{n,i}^{[L]}-E_s(T_{n,i}^{[L]}\mid\mathcal G_{A,n}^{\mathrm{match}})\}\right|
 +\frac2n\sum_iE_s|T_{n,i}-T_{n,i}^{[L]}|\\
&\le\left\{\frac1{n^2}\sum_i
 E_s\operatorname{Var}_s(T_{n,i}^{[L]}\mid\mathcal G_{A,n}^{\mathrm{match}})\right\}^{1/2}
 +\frac{2L^{1-p}}n\sum_iE_s|T_{n,i}|^p\\
&\le\frac L{\sqrt n}+CL^{1-p}.
\end{aligned}
\]
Since \(C\) is independent of \(n\) and \(L\), first letting \(n\to\infty\) and then \(L\to\infty\) gives convergence to zero in \(L^1\), and hence in probability, because \(p>1\). Combining this conclusion with \eqref{eq:baseline-oracle-firststep-conditional-qv} and the first-stage covariance gives
\[
\sum_{k=1}^{3n}
 E_s(\boldsymbol\xi_{A,n,k}\boldsymbol\xi_{A,n,k}^\top\mid\mathcal F_{A,n,k-1})
\longrightarrow_p\Omega_{H,A}+\Omega_{E,A}.
\]

For a fixed scalar projection \(t=(t_0,t_U^\top)^\top\), the first-stage moment bound is
\[
\begin{aligned}
\sum_{k=1}^n E_s|t^\top\boldsymbol\xi_{A,n,k}|^{2+\delta}
&\le C_t n^{-\delta/2}
       E_s\{|h_A|^{2+\delta}+\|u_A\|^{2+\delta}\}\\
&\longrightarrow0.
\end{aligned}
\]
For the remaining stages, conditional Jensen and \eqref{eq:qv-uniform-conditional-reuse} give
\[
\begin{aligned}
\sum_{k=n+1}^{3n}E_s|t^\top\boldsymbol\xi_{A,n,k}|^{2+\delta}
&\le\frac{C_t}{n^{1+\delta/2}}\sum_i
 E_s\{|c_{n,A,i}\varepsilon_{A,i}|^{2+\delta}
                +\|\mathbf U_A(O_i;\boldsymbol\theta_A^\ast)\|^{2+\delta}\}\\
&\le C_t'n^{-\delta/2}
 E_s\{|\varepsilon_A|^{2+\delta}+\|\mathbf U_A\|^{2+\delta}\}\\
&\longrightarrow0.
\end{aligned}
\]
Markov's inequality converts these unconditional moment bounds into the conditional Lyapunov bound
\[
\sum_{k=1}^{3n}
E_s\{|t^\top\boldsymbol\xi_{A,n,k}|^{2+\delta}\mid\mathcal F_{A,n,k-1}\}
\longrightarrow_p0.
\]
The inequality \(x^2\mathbf1\{|x|>\epsilon\}\le \epsilon^{-\delta}|x|^{2+\delta}\) then verifies conditional Lindeberg for every fixed scalar projection. Together with the predictable covariance convergence and the martingale-difference property, the martingale central limit theorem of Hall and Heyde,\citep{HallHeyde1980} followed by the Cram\'er--Wold device, gives
\[
\sum_{k=1}^{3n}\boldsymbol\xi_{A,n,k}
\Rightarrow N(0,\Omega_{H,A}+\Omega_{E,A}).
\]
Zero-variance projections follow by the \(L^1\) covariance limits and Chebyshev's inequality. The lower covariance block is
\[
\begin{aligned}
E_s(u_Au_A^\top)+E_sR_A
&=E_s\{E_s(U_A\mathbf U_A^\top\mid\mathcal G_{A,1}^{\mathrm{row}})\}\\
&=\boldsymbol\Sigma_{A,U}.
\end{aligned}
\]
The other blocks are
\begin{equation}\label{eq:baseline-oracle-firststep-gamma}
\begin{aligned}
V_{A,\mathrm{known\text{-}score}}
&=E_s(h_{A,1}^2)/\bar w_A^2+V_{E,A}\\
&=V_{H,A}+V_{E,A},\\
\boldsymbol\gamma_{A,1}
&:=\bar w_A^{-1}
       \{E_s(h_{A,1}u_{A,1})+\mathcal C_A\}.
\end{aligned}
\end{equation}
These covariance blocks establish the joint limit. Applying \(-\boldsymbol\Gamma_A^{-1}\) to the estimating-function coordinate gives \(\operatorname{Cov}(\mathbf G_{\theta,A},G_{\tau,A}) =-\boldsymbol\Gamma_A^{-1}\boldsymbol\gamma_{A,1}\). The bias-correction remainder on the same matching graph, \eqref{eq:baseline-oracle-firststep-bc-remainder}, then gives the bias-corrected limit by Slutsky's theorem.
\end{proof}

\subsubsection{Local likelihood and discretization}\label{le-cam-and-discretization-lemmas-used-in-the-estimated-score-layer}

For \(P^{\boldsymbol\theta^\ast}\ll P^{\boldsymbol\theta_n}\), put \(L_n=\log(dP^{\boldsymbol\theta^\ast}/dP^{\boldsymbol\theta_n})\). The Gaussian form of Le Cam's third lemma, in Chapter~6 of van der Vaart\citep{vanderVaart2000} and Appendix~A of Andreou and Werker,\citep{AndreouWerker2012} gives
\begin{equation}\label{lem:gaussian-likelihood-shift}
\begin{gathered}
(Z_n,L_n)\Rightarrow
N\!\left(
 \binom0{-\mathbf h^\top Ih/2},
 \begin{bmatrix}\boldsymbol\Sigma_Z&Ch\\h^\top C^\top&\mathbf h^\top Ih\end{bmatrix}
 \right)
 \quad\text{under }P^{\boldsymbol\theta_n}\\
\Longrightarrow\quad
Z_n\Rightarrow N(Ch,\boldsymbol\Sigma_Z)
 \quad\text{under }P^{\boldsymbol\theta^\ast}.
\end{gathered}
\end{equation}

The following lemma gives a joint form of the finite-grid argument for Theorem~2 in Appendix~C of Andreou and Werker,\citep{AndreouWerker2012} with the linear local response allowed by their Remark~B1.

\begin{lemma}[Local discretization]
\label{lem:joint-local-grid-selection}
Let \(T_n(\mathbf h)\) be a scalar statistic indexed by the local coordinate \(\mathbf h\in\mathbb R^q\), and let \(\widehat{\mathbf h}_n\in\mathbb R^q\). The symbols \(T_n\), \(Z\), and \(H\) are local to this lemma. Suppose that there is a joint random vector \((Z,H)\), with \(H\) having a density, and a fixed vector \(b\in\mathbb R^q\), such that for every deterministic sequence \(\mathbf h_n\to \mathbf h\),
\[
\{T_n(\mathbf h_n),\widehat{\mathbf h}_n\}\Rightarrow\{Z+b^\top \mathbf h,H\}.
\]
For \(\eta>0\), take a deterministic grid offset \(a_{n,\eta}\in[-\eta/2,\eta/2]^q\), and let \(\bar{\mathbf h}_{n,\eta}\) be the midpoint of the side-\(\eta\) cube containing \(\widehat{\mathbf h}_n\) in the grid with midpoints \(a_{n,\eta}+\eta\mathbb Z^q\). Fix a boundary convention. The grid is in the local-parameter space: in the applications, \(\widehat{\mathbf h}_n=\sqrt n(\widehat{\boldsymbol \theta}_A-\boldsymbol\theta_A^*)\), so a cell has side length \(\eta/\sqrt n\) in the original parameter space. Then, for every bounded Lipschitz function \(\Psi:\mathbb R^{1+q}\to\mathbb R\),
\begin{equation}\label{eq:joint-grid-pair-limit}
\lim_{\eta\downarrow0}\limsup_{n\to\infty}
\left|E\{\Psi(T_n(\bar{\mathbf h}_{n,\eta}),\widehat{\mathbf h}_n)\}
      -E\{\Psi(Z+b^\top H,H)\}\right|=0.
\end{equation}
The same conclusion holds with \(\bar{\mathbf h}_{n,\eta}\) in the second coordinate, and for the first coordinate alone. If, in addition, for each \(\epsilon>0\),
\[
\lim_{\eta\downarrow0}\limsup_{n\to\infty}
P\{|T_n(\widehat{\mathbf h}_n)-T_n(\bar{\mathbf h}_{n,\eta})|>\epsilon\}=0,
\]
the same double-limit conclusion holds for \(T_n(\widehat{\mathbf h}_n)\). Since that statistic does not depend on \(\eta\), \(T_n(\widehat{\mathbf h}_n)\Rightarrow Z+b^\top H\).
\end{lemma}

\begin{proof}
For fixed \(\eta\), take a subsequence with \(a_{n,\eta}\to a_\eta\), and let \(r_{a_\eta,\eta}\) round to the limiting grid. At each midpoint \(\mathbf h_n^{(j)}=a_{n,\eta}+\eta j\), \(j\in\mathbb Z^q\), joint convergence determines the statistic together with selection of its grid cell, because the limiting boundary is null for \(H\). Tightness first restricts the local root to a compact set, where only finitely many cells occur. Summing over those cells and then enlarging the compact set gives
\[
E\{\Psi(T_n(\bar{\mathbf h}_{n,\eta}),\widehat{\mathbf h}_n)\}
\longrightarrow E\{\Psi(Z+b^\top r_{a_\eta,\eta}(H),H)\}.
\]
Since \(\|r_{a_\eta,\eta}(H)-H\|\le\sqrt q\,\eta/2\),
\[
\begin{aligned}
&\left|E\{\Psi(Z+b^\top r_{a_\eta,\eta}(H),H)\}
            -E\{\Psi(Z+b^\top H,H)\}\right|\\
&\le L_\Psi E|b^\top\{r_{a_\eta,\eta}(H)-H\}|\\
&\le L_\Psi\|b\|\sqrt q\,\eta/2\longrightarrow0.
\end{aligned}
\]
The bound is uniform in the offset and proves \eqref{eq:joint-grid-pair-limit}. Rounding the second coordinate adds at most \(L_\Psi\sqrt q\,\eta/2\). For a bounded Lipschitz function \(\varphi\) and \(\epsilon>0\),
\[
\begin{aligned}
&\left|E\{\varphi(T_n(\widehat{\mathbf h}_n))\}
       -E\{\varphi(T_n(\bar{\mathbf h}_{n,\eta}))\}\right|\\
&\le L_\varphi\epsilon
 +2\|\varphi\|_\infty
 P\{|T_n(\widehat{\mathbf h}_n)-T_n(\bar{\mathbf h}_{n,\eta})|>\epsilon\}.
\end{aligned}
\]
The assumed comparison makes the right side vanish after taking \(\lim_{\epsilon\downarrow0}\lim_{\eta\downarrow0}\limsup_n\).
\end{proof}

The compact restriction is in parameter space. The comparison between the fitted and discretized estimators below also localizes matching neighborhoods in score space.

\subsubsection{Local score expansions}
\label{primitive-local-asymptotic-expansion-for-the-first-step-estimators}

We use the design-specific estimating function of Assumption~\ref{ass:ordinary-weighted-m-estimation}: retrospective PS estimation is weighted, while prospective PS and arm-specific PG estimation are unweighted. For a fixed \(\mathbf h\in\mathbb R^{p_A}\), let \(\boldsymbol\theta_{A,n}=\boldsymbol\theta_A^\ast+\mathbf h/\sqrt n\). The normalized estimating-function sum is \(\sqrt n\,\mathbb U_{A,n}(\boldsymbol\theta)=n^{-1/2}\sum_i\mathbf U_A(O_i;\boldsymbol\theta)\). The local Jacobian convergence used in Lemma~\ref{lem:ordinary-weighted-first-step} gives, under \(P_s^{\otimes n}\),
\[
\sqrt n\,\mathbb U_{A,n}(\boldsymbol\theta_{A,n})
=\sqrt n\,\mathbb U_{A,n}(\boldsymbol\theta_A^\ast)+\boldsymbol\Gamma_A\mathbf h+o_p(1).
\]
Consequently,
\begin{equation}\label{lem:fixed-law-local-first-step}
\sqrt n(\widehat{\boldsymbol \theta}_A-\boldsymbol\theta_{A,n})
=-\boldsymbol\Gamma_A^{-1}\sqrt n\,\mathbb U_{A,n}(\boldsymbol\theta_{A,n})+o_p(1)
\Rightarrow N(-\mathbf h,\boldsymbol\Sigma_{A,\theta}).
\end{equation}
The auxiliary experiment below additionally requires nonsingular \(\boldsymbol\Sigma_{A,U}\).

We construct an auxiliary family whose population estimating-equation root has the same local coordinates as the matching parameter. If \(Q_{A,0}=P_s\) and \(E_{Q_{A,\mathbf t}}\{\mathbf U_A(O;\boldsymbol\theta_A(Q_{A,\mathbf t}))\}=0\), differentiation at zero gives
\[
0=\boldsymbol\Gamma_A
 \left.\frac{\partial\boldsymbol\theta_A(Q_{A,\mathbf t})}{\partial \mathbf t^\top}\right|_0
 +E_s\{\mathbf U_A(O;\boldsymbol\theta_A^*)\mathbf s_A(O)^\top\}.
\]
Thus a unit root derivative requires \(E_s\{\mathbf U_A(O;\boldsymbol\theta_A^*)\mathbf s_A(O)^\top\}=-\boldsymbol\Gamma_A\). To meet this requirement, take \(\mathbf s_A(O)=L_A\mathbf U_A(O;\boldsymbol\theta_A^*)\). The resulting matrix equation gives
\[
\begin{aligned}
\boldsymbol\Sigma_{A,U}L_A^\top&=-\boldsymbol\Gamma_A,\\
L_A^\top&=-\boldsymbol\Sigma_{A,U}^{-1}\boldsymbol\Gamma_A,\\
\mathbf s_A(O)&=-\boldsymbol\Gamma_A^\top\boldsymbol\Sigma_{A,U}^{-1}\mathbf U_A(O;\boldsymbol\theta_A^*).
\end{aligned}
\]
The following lemma constructs a probability family with this score and establishes the corresponding root and likelihood expansions.

\begin{lemma}[Auxiliary local likelihood]
\label{lem:auxiliary-godambe-experiment}
Let \(P_{A,0}\) be the selected-sample law of one observation at \(\boldsymbol\theta_A^\ast\). Assume Assumption~\ref{ass:ordinary-weighted-m-estimation} and suppose \(\boldsymbol\Sigma_{A,U}\) is nonsingular. Use the fixed-law expansion \eqref{lem:fixed-law-local-first-step}. Define the auxiliary likelihood score
\[
\mathbf s_A(O):=-\boldsymbol\Gamma_A^\top\boldsymbol\Sigma_{A,U}^{-1}\mathbf U_A(O;\boldsymbol\theta_A^\ast),
\qquad
\mathcal I_{A,G}:=\boldsymbol\Gamma_A^\top\boldsymbol\Sigma_{A,U}^{-1}\boldsymbol\Gamma_A.
\]
For \(\mathbf t\) in a neighborhood of zero, define a one-observation probability law by
\[
\begin{aligned}
g_{\mathrm{tilt}}(x)&:=1+\frac{x}{\sqrt{1+x^2}},\\
\frac{dQ_{A,\mathbf t}}{dP_{A,0}}(O)
&:=\frac{g_{\mathrm{tilt}}\{\mathbf t^\top \mathbf s_A(O)\}}
{E_s[g_{\mathrm{tilt}}\{\mathbf t^\top \mathbf s_A(O)\}]}.
\end{aligned}
\]
Write \(\boldsymbol\theta_{A,n}=\boldsymbol\theta_A^\ast+\mathbf h/\sqrt n\) for fixed \(\mathbf h\). Under \(Q_{A,\mathbf h/\sqrt n}^{\otimes n}\),
\[
\sqrt n\,\mathbb U_{A,n}(\boldsymbol\theta_{A,n})
\Rightarrow N(0,\boldsymbol\Sigma_{A,U}),
\]
and the reverse log-likelihood ratio satisfies
\[
\begin{aligned}
&\Lambda_{A,n}^{\mathrm{aux}}(\boldsymbol\theta_A^\ast\mid\boldsymbol\theta_{A,n})
:=\log\frac{dP_{A,0}^{\otimes n}}{dQ_{A,\mathbf h/\sqrt n}^{\otimes n}}\\
&=\mathbf h^\top\boldsymbol\Gamma_A^\top\boldsymbol\Sigma_{A,U}^{-1}
       \sqrt n\,\mathbb U_{A,n}(\boldsymbol\theta_{A,n})
-\frac12 \mathbf h^\top \mathcal I_{A,G}\mathbf h
+o_{Q_{A,\mathbf h/\sqrt n}^{\otimes n}}(1).
\end{aligned}
\]
The limiting covariance of \(\sqrt n(\widehat{\boldsymbol \theta}_A-\boldsymbol\theta_{A,n})\) with the reverse log-likelihood ratio is \(-\mathbf h\).

\end{lemma}
\begin{proof}
The column vector \(\mathbf s_A(O)\) satisfies \(E_s\{\mathbf s_A(O)\}=0\) and \(E_s\{\mathbf s_A(O)\mathbf s_A(O)^\top\}=\mathcal I_{A,G}\). Since \(0<g_{\mathrm{tilt}}<2\), the normalizing expectation is finite and positive. The density ratio integrates to one and defines a probability law on the observation space, with \(Q_{A,0}=P_{A,0}=P_s\). Moreover, \(g_{\mathrm{tilt}}(0)=g_{\mathrm{tilt}}'(0)=1\), so
\[
\begin{aligned}
\left.\nabla_{\mathbf t}\log\frac{dQ_{A,\mathbf t}}{dP_s}(O)\right|_{\mathbf t=0}
&=\mathbf s_A(O)-E_s\{\mathbf s_A(O)\}\\
&=\mathbf s_A(O).
\end{aligned}
\]
Thus the auxiliary family has likelihood score \(\mathbf s_A(O)\) at zero. Survey weights enter through the estimating equations in \(\mathbf s_A\), while \(g_{\mathrm{tilt}}\) supplies a positive, normalized perturbation with the specified derivative. The bounds \(|g_{\mathrm{tilt}}(x)-1|\le |x|\) and \(|\sqrt{g_{\mathrm{tilt}}(x)}-1|\le C|x|\) also hold. The quotient \(\{g_{\mathrm{tilt}}(x)-1-x\}/x^2\) is bounded and tends to zero as \(x\to0\). Dominated convergence with \(E_s\|\mathbf s_A(O)\|^2<\infty\) gives
\[
\begin{aligned}
E_s[g_{\mathrm{tilt}}\{\mathbf t^\top \mathbf s_A(O)\}]-1
&=E_s[g_{\mathrm{tilt}}\{\mathbf t^\top \mathbf s_A(O)\}-1-\mathbf t^\top \mathbf s_A(O)]\\
&=o(\|\mathbf t\|^2),\\
E_s\!\left[
 \left\{\sqrt{g_{\mathrm{tilt}}\{\mathbf t^\top \mathbf s_A(O)\}}
             -1-\tfrac12\mathbf t^\top \mathbf s_A(O)\right\}^2\right]
&=o(\|\mathbf t\|^2).
\end{aligned}
\]
Normalizing by \(E_s[g_{\mathrm{tilt}}\{\mathbf t^\top \mathbf s_A(O)\}]\) gives
\[
\left\|\sqrt{\frac{dQ_{A,\mathbf t}}{dP_s}(O)}
 -1-\tfrac12\mathbf t^\top \mathbf s_A(O)\right\|_{L^2(P_s)}
=o(\|\mathbf t\|).
\]
Thus the normalized family is differentiable in quadratic mean with score \(\mathbf s_A(O)\) and nonsingular information \(\mathcal I_{A,G}\). The iid local asymptotic normality theorem in Chapter~7 of van der Vaart\citep{vanderVaart2000} gives
\[
\log\frac{dQ_{A,\mathbf h/\sqrt n}^{\otimes n}}{dP_s^{\otimes n}}
=-\mathbf h^\top\boldsymbol\Gamma_A^\top\boldsymbol\Sigma_{A,U}^{-1}
 \sqrt n\,\mathbb U_{A,n}(\boldsymbol\theta_A^\ast)
-\tfrac12\mathbf h^\top \mathcal I_{A,G}\mathbf h+o_p(1).
\]
The quadratic-mean remainder is uniform as \(\mathbf t\to0\), so the likelihood expansion also holds for every deterministic \(\mathbf h_n\to \mathbf h\). The local Jacobian expansion is uniform for bounded \(\mathbf h_n\). The product laws describe \(n\) independent observations. Under \(P_s^{\otimes n}\), the forward log-likelihood ratio \(\log(dQ^{\otimes n}/dP_s^{\otimes n})\) has a normal limit with variance \(\mathbf h^\top\mathcal I_{A,G}\mathbf h\) and mean \(-\mathbf h^\top\mathcal I_{A,G}\mathbf h/2\). Its limiting exponential has expectation one. Joint convergence with the estimating-function sum therefore permits the following change of measure. Le Cam's first and third lemmas then give contiguity and the following limits under the local law:
\[
\begin{aligned}
\sqrt n\,\mathbb U_{A,n}(\boldsymbol\theta_A^\ast)
&\Rightarrow N(-\boldsymbol\Gamma_A\mathbf h,\boldsymbol\Sigma_{A,U}),\\
\sqrt n\,\mathbb U_{A,n}(\boldsymbol\theta_{A,n})
&=\sqrt n\,\mathbb U_{A,n}(\boldsymbol\theta_A^\ast)+\boldsymbol\Gamma_A\mathbf h
                  +o_{Q_{A,\mathbf h/\sqrt n}^{\otimes n}}(1)\\
&\Rightarrow N(0,\boldsymbol\Sigma_{A,U}),\\
\Lambda_{A,n}^{\mathrm{aux}}(\boldsymbol\theta_A^\ast\mid\boldsymbol\theta_{A,n})
&=\mathbf h^\top\boldsymbol\Gamma_A^\top\boldsymbol\Sigma_{A,U}^{-1}
                 \sqrt n\,\mathbb U_{A,n}(\boldsymbol\theta_A^\ast)
                 +\tfrac12\mathbf h^\top \mathcal I_{A,G}\mathbf h
                  +o_{Q_{A,\mathbf h/\sqrt n}^{\otimes n}}(1)\\
&=\mathbf h^\top\boldsymbol\Gamma_A^\top\boldsymbol\Sigma_{A,U}^{-1}
                 \{\sqrt n\,\mathbb U_{A,n}(\boldsymbol\theta_{A,n})-\boldsymbol\Gamma_A\mathbf h\}
                 +\tfrac12\mathbf h^\top \mathcal I_{A,G}\mathbf h
                  +o_{Q_{A,\mathbf h/\sqrt n}^{\otimes n}}(1)\\
&=\mathbf h^\top\boldsymbol\Gamma_A^\top\boldsymbol\Sigma_{A,U}^{-1}
                 \sqrt n\,\mathbb U_{A,n}(\boldsymbol\theta_{A,n})
                 -\tfrac12\mathbf h^\top \mathcal I_{A,G}\mathbf h
                  +o_{Q_{A,\mathbf h/\sqrt n}^{\otimes n}}(1).
\end{aligned}
\]
Contiguity also transfers \eqref{lem:fixed-law-local-first-step}. The limiting cross-covariance with the reverse likelihood is
\[
-\boldsymbol\Gamma_A^{-1}\boldsymbol\Sigma_{A,U}\boldsymbol\Sigma_{A,U}^{-1}\boldsymbol\Gamma_A\mathbf h=-\mathbf h.
\]

Finally the uniform score moment and integrable derivative envelope, together with bounded \(g_{\mathrm{tilt}},g_{\mathrm{tilt}}'\), make \((\mathbf t,\boldsymbol\theta)\mapsto E_{Q_{A,\mathbf t}}\mathbf U_A(O;\boldsymbol\theta)\) continuously differentiable near \((0,\boldsymbol\theta_A^\ast)\). The implicit-function theorem gives a local root \(\boldsymbol\theta_A(Q_{A,\mathbf t})\) satisfying \(E_{Q_{A,\mathbf t}}\{\mathbf U_A(O;\boldsymbol\theta_A(Q_{A,\mathbf t}))\}=0\). Differentiation of this expectation gives
\[
\begin{aligned}
0
&=\boldsymbol\Gamma_A
\left.\frac{\partial\boldsymbol\theta_A(Q_{A,\mathbf t})}{\partial \mathbf t^\top}\right|_{0}
+E_s\{\mathbf U_A(O;\boldsymbol\theta_A^\ast)\mathbf s_A(O)^\top\}\\
&=\boldsymbol\Gamma_A
\left.\frac{\partial\boldsymbol\theta_A(Q_{A,\mathbf t})}{\partial \mathbf t^\top}\right|_{0}
-\boldsymbol\Sigma_{A,U}\boldsymbol\Sigma_{A,U}^{-1}\boldsymbol\Gamma_A\\
&=\boldsymbol\Gamma_A
\left.\frac{\partial\boldsymbol\theta_A(Q_{A,\mathbf t})}{\partial \mathbf t^\top}\right|_{0}
-\boldsymbol\Gamma_A.
\end{aligned}
\]
Hence the root derivative is \(I_{p_A}\), and \(\boldsymbol\theta_A(Q_{A,\mathbf t})=\boldsymbol\theta_A^\ast+\mathbf t+o(\|\mathbf t\|)\). The auxiliary law therefore calibrates the parameter perturbation used for matching, with the estimating function and survey weight functions fixed along the path.
\end{proof}

\begin{lemma}[Conditional moments under the auxiliary law]
\label{lem:local-tilt-row-projections}
Use the auxiliary law of Lemma~\ref{lem:auxiliary-godambe-experiment}. Let \(\mathbf h_n\) be a bounded deterministic sequence, put \(Q_n=Q_{A,\mathbf h_n/\sqrt n}\), and suppose \(E_s\|\mathbf s_A(O)\|^{2+\delta}<\infty\) for some \(\delta>0\). For any sequence of deterministic conditioning fields \(\mathcal G_n\) generated by one observation, define
\[
g_n(O)=g_{\mathrm{tilt}}\{\mathbf h_n^\top \mathbf s_A(O)/\sqrt n\},\qquad
d_n=E_s(g_n\mid\mathcal G_n),\qquad
B_n=\{d_n<1/2\}.
\]
The fields need not converge. With \(B_{n,i}\) denoting this event evaluated at row \(i\),
\begin{equation}\label{eq:local-tilt-sampled-denominators}
P_s^{\otimes n}\!\left(\bigcup_{i=1}^nB_{n,i}\right)
+Q_n^{\otimes n}\!\left(\bigcup_{i=1}^nB_{n,i}\right)
=O(n^{-\delta/2}).
\end{equation}
For every fixed scalar function \(F(O)\in L^r(P_s)\), \(1\le r<\infty\),
\begin{equation}\label{eq:local-tilt-row-projection-limit}
\mathbf1_{B_n^c}E_{Q_n}(F\mid\mathcal G_n)
-E_s(F\mid\mathcal G_n)\longrightarrow_{L^r(P_s)}0.
\end{equation}
In particular, for a fixed finite-dimensional vector \(T=T(O)\) with \(E_s\|T\|^2<\infty\),
\begin{equation}\label{eq:local-tilt-row-covariance-limit}
E_s\left\|
\mathbf1_{B_n^c}\operatorname{Cov}_{Q_n}(T\mid\mathcal G_n)
-\operatorname{Cov}_s(T\mid\mathcal G_n)\right\|
\longrightarrow0.
\end{equation}
\end{lemma}

\begin{proof}
We first bound the probability of a small conditional normalizer. Conditional Jensen and Markov's inequalities give
\[
\begin{aligned}
|d_n-1|
&\le n^{-1/2}\|\mathbf h_n\|E_s(\|\mathbf s_A(O)\|\mid\mathcal G_n),\\
P_s(B_n)
&\le P_s(|d_n-1|>1/2)\\
&\le 2^{2+\delta}E_s|d_n-1|^{2+\delta}\\
&\le Cn^{-1-\delta/2}E_s\|\mathbf s_A(O)\|^{2+\delta}.
\end{aligned}
\]
Since \(dQ_n/dP_s\le4\) eventually,
\[
\begin{aligned}
P_s^{\otimes n}\!\left(\bigcup_i B_{n,i}\right)
+Q_n^{\otimes n}\!\left(\bigcup_i B_{n,i}\right)
&\le n\{P_s(B_n)+Q_n(B_n)\}\\
&\le5nP_s(B_n)=O(n^{-\delta/2}).
\end{aligned}
\]

Conditional change of measure gives
\[
E_{Q_n}(F\mid\mathcal G_n)
=\frac{E_s(g_nF\mid\mathcal G_n)}{d_n},\qquad
|\mathbf1_{B_n^c}E_{Q_n}(F\mid\mathcal G_n)|
\le4E_s(|F|\mid\mathcal G_n).
\]
For bounded \(F\), subtracting its baseline conditional mean gives
\[
\begin{aligned}
&\mathbf1_{B_n^c}E_{Q_n}(F\mid\mathcal G_n)-E_s(F\mid\mathcal G_n)\\
&=\frac{\mathbf1_{B_n^c}}{d_n}
 E_s[(g_n-1)\{F-E_s(F\mid\mathcal G_n)\}\mid\mathcal G_n]
 -\mathbf1_{B_n}E_s(F\mid\mathcal G_n).
\end{aligned}
\]
Therefore
\[
\begin{aligned}
&\|\mathbf1_{B_n^c}E_{Q_n}(F\mid\mathcal G_n)
                         -E_s(F\mid\mathcal G_n)\|_1\\
&\le4\|F\|_\infty E_s|g_n-1|+\|F\|_\infty P_s(B_n)
\longrightarrow0.
\end{aligned}
\]
Boundedness of the difference by \(5\|F\|_\infty\) strengthens this convergence to every finite \(L^r\) norm. For general \(F\in L^r\), put \(F_L=\max(-L,\min(F,L))\). Conditional contraction gives
\[
\begin{aligned}
&\limsup_n\|\mathbf1_{B_n^c}E_{Q_n}(F\mid\mathcal G_n)
                              -E_s(F\mid\mathcal G_n)\|_r\\
&\le4\|F-F_L\|_r+\|F-F_L\|_r\\
&\longrightarrow0\qquad(L\to\infty).
\end{aligned}
\]
Apply this projection limit with \(r=2\) to coordinates of \(T\) and \(r=1\) to their products. Cauchy--Schwarz controls products of conditional means. The identity \(\mathbf1_{B_n^c}^2=\mathbf1_{B_n^c}\) then gives \eqref{eq:local-tilt-row-covariance-limit}.
\end{proof}

The next lemma extends the inverse-coordinate calculation in Supplementary Lemma S.12 of Abadie and Imbens\citep{AbadieImbens2016} to the two score coordinates and observed survey weight.

\begin{lemma}[PATT conditional distributions]
\label{lem:patt-two-coordinate-conditional-law}
For PATT, fix \(A\in\{R,P\}\), with the two-block nuisance vector \(\boldsymbol\theta_A=(\alpha_A^\top,\beta_{A,0}^\top)^\top\). Use the auxiliary law \(Q_{A,t}\), tilt \(g_{\mathrm{tilt}}\), and score \(\mathbf s_A(O)\) of Lemma~\ref{lem:auxiliary-godambe-experiment}, under all its conditions. Assume the following coordinate and density regularity.

Suppose the covariate vector is bounded and can be partitioned as \(\mathbf X=(V^\top,L^\top)^\top\), \(V\in\mathbb R^2\), so that the raw logit/control-prognostic pair is
\[
S_{\boldsymbol \theta}(\mathbf X)=A_{\boldsymbol \theta} V+B_{\boldsymbol \theta} L,
\qquad
v_{\boldsymbol \theta}(s,l)=A_{\boldsymbol \theta}^{-1}(s-B_{\boldsymbol \theta} l),
\]
where \(A_{\boldsymbol\theta_A^\ast}\) is nonsingular and \(A_{\boldsymbol \theta},B_{\boldsymbol \theta}\) are continuous at \(\boldsymbol\theta_A^\ast\). An intercept, when present, belongs to \(L\). The observed weight is a fixed measurable function \(w_{A,z}(L)\), bounded above and bounded away from zero. As usual, \(w_{R,z}=w_z\) and \(w_{P,z}=w\). For each deterministic \(\boldsymbol\theta\), the double score \(\mathbf D_{0,\boldsymbol\theta}\) is obtained by applying expit to the first raw coordinate and then the specified deterministic componentwise centering and positive scaling. In particular,
\[
\sigma\{Z,\mathbf D_{0,\boldsymbol\theta}(\mathbf X),w_{A,Z}(L)\}
=\sigma\{Z,S_{\boldsymbol \theta}(\mathbf X),w_{A,Z}(L)\}.
\]

For each arm \(z\) of positive selected-sample probability, let
\[
\mathsf P_z^{L,Y}:=\mathcal L_s(L,Y\mid Z=z),
\qquad
\mathsf P_{z,w}^{L,Y}:=\mathcal L_s\{L,Y\mid Z=z,w_{A,z}(L)=w\}.
\]
Suppose \(V\mid L=l,Y=y,Z=z\) has a jointly measurable density \(f_z(v\mid l,y)\) whose zero extension is finite and continuous at Lebesgue-almost every \(v\), for \(\mathsf P_z^{L,Y}\)-almost every \((l,y)\). For some \(\delta>0\), suppose \(E_s|Y|^{2+\delta}<\infty\). For almost every baseline pair \((s,w)\) under \(\mathcal L_s\{S_{\boldsymbol\theta_A^\ast},w_{A,z}\mid Z=z\}\), assume there are a neighborhood \(N_{z,s,w}\) of \((\boldsymbol\theta_A^\ast,s)\) and an \(\mathsf P_{z,w}^{L,Y}\)-integrable \(J_{z,s,w}(l,y)\) such that
\begin{equation}
\sup_{(\boldsymbol\theta,s')\in N_{z,s,w}}
 f_z\{v_{\boldsymbol \theta}(s',l)\mid l,y\}(1+|y|)^{2+\delta}
\leq J_{z,s,w}(l,y)
\label{eq:patt-chart-local-density-envelope}
\end{equation}
for \(\mathsf P_{z,w}^{L,Y}\)-almost every \((l,y)\). Let \(m_0^A\) be Lipschitz on the compact baseline control-score support, with a fixed Lipschitz extension when an off-support value is written. Define
\[
\begin{aligned}
\zeta_A(O)&:=Y-m_0^A\{\mathbf D_0^\ast(\mathbf X)\},\\
\Phi_A(O)&:=
\bigl(\zeta_A(O),\mathbf U_A(O;\boldsymbol\theta_A^\ast)^\top\bigr)^\top.
\end{aligned}
\]

For every deterministic \(\boldsymbol\theta_n\to\boldsymbol\theta_A^\ast\) and \(t_n\to0\), a version of the conditional distribution of \(\Phi_A\) under \(Q_{A,t_n}\), given
\[
\mathcal G_n:=\sigma\{Z,\mathbf D_{0,\boldsymbol\theta_n}(\mathbf X),
                         w_{A,Z}(L)\},
\]
converges, evaluated at the current root observation, to its baseline conditional distribution given \(\mathcal G_\ast:=\sigma\{Z,\mathbf D_0^\ast(\mathbf X),w_{A,Z}(L)\}\). The convergence is in the bounded-Lipschitz metric, almost surely under \(P_s\), and hence in probability under \(Q_{A,t_n}\). Its first and second conditional moments also converge almost surely under \(P_s\) and in \(L^1(Q_{A,t_n})\):
\begin{equation}
\begin{aligned}
E_{Q_{A,t_n}}(\Phi_A\mid\mathcal G_n)
&\longrightarrow E_s(\Phi_A\mid\mathcal G_\ast),\\
E_{Q_{A,t_n}}(\Phi_A\Phi_A^\top\mid\mathcal G_n)
&\longrightarrow E_s(\Phi_A\Phi_A^\top\mid\mathcal G_\ast).
\end{aligned}
\label{eq:patt-chart-conditional-moment-continuity}
\end{equation}
The same conclusions hold for the conditional covariance matrix.
\end{lemma}

\begin{proof}
We express the conditional law in the raw-score coordinates. Condition on the observed arm and weight, and set \(v_{\boldsymbol \theta}(s,l)=A_{\boldsymbol \theta}^{-1}(s-B_{\boldsymbol \theta} l)\). Define
\[
r_{z,\boldsymbol\theta,t}(s,l,y)
=\frac{f_z\{v_{\boldsymbol \theta}(s,l)\mid l,y\}}{|\det A_{\boldsymbol \theta}|}
 g_{\mathrm{tilt}}\{t^\top \mathbf s_A(v_{\boldsymbol \theta}(s,l),l,y,z)\},\qquad
q_{z,\boldsymbol\theta,t}(s,w)=\int r_{z,\boldsymbol\theta,t}\,d\mathsf P_{z,w}^{L,Y}.
\]
Disintegration and change of variables give
\begin{equation}
\label{eq:patt-chart-weight-conditioned-fiber-law}
\mathcal L_{Q_{A,t}}\{L,Y\mid Z=z,S_{\boldsymbol \theta}=s,w_{A,z}(L)=w\}
=\frac{r_{z,\boldsymbol\theta,t}(s,l,y)}{q_{z,\boldsymbol\theta,t}(s,w)}\mathsf P_{z,w}^{L,Y}(dl,dy).
\end{equation}
Put \(s_n=S_{\boldsymbol\theta_n}(\mathbf X)\) and \(s_\ast=S_{\boldsymbol\theta_A^\ast}(\mathbf X)\) at the conditioning observation. By invertibility and Tonelli's theorem, the density discontinuity sets have measure zero for almost every conditioning observation and \(\mathsf P_{z,w}^{L,Y}\)-almost every \((l,y)\). The envelope in \eqref{eq:patt-chart-local-density-envelope} and \(0<g_{\mathrm{tilt}}<2\) therefore give
\[
\begin{aligned}
v_{\boldsymbol\theta_n}(s_n,l)&\longrightarrow v_{\boldsymbol\theta_A^\ast}(s_\ast,l),\\
\int|r_{z,\boldsymbol\theta_n,t_n}(s_n,l,y)
       -r_{z,\boldsymbol\theta_A^\ast,0}(s_\ast,l,y)|
             (1+|y|)^2\,d\mathsf P_{z,w}^{L,Y}&\longrightarrow0,\\
q_{z,\boldsymbol\theta_n,t_n}(s_n,w)
&\longrightarrow q_{z,\boldsymbol\theta_A^\ast,0}(s_\ast,w)\in(0,\infty).
\end{aligned}
\]
For bounded continuous \(f\), and also for coordinate and coordinate-product functions of \(\Phi_A\), dominated convergence gives
\[
\begin{aligned}
E_{Q_{A,t_n}}\{f(\Phi_A)\mid\mathcal G_n\}
&=\int f\{\Phi_A(v_{\boldsymbol\theta_n}(s_n,l),l,y,z)\}
 \frac{r_{z,\boldsymbol\theta_n,t_n}(s_n,l,y)}
      {q_{z,\boldsymbol\theta_n,t_n}(s_n,w)}\,d\mathsf P_{z,w}^{L,Y}\\
&\longrightarrow\int f\{\Phi_A(v_{\boldsymbol\theta_A^\ast}(s_\ast,l),l,y,z)\}
 \frac{r_{z,\boldsymbol\theta_A^\ast,0}(s_\ast,l,y)}
      {q_{z,\boldsymbol\theta_A^\ast,0}(s_\ast,w)}\,d\mathsf P_{z,w}^{L,Y}\\
&=E_s\{f(\Phi_A)\mid\mathcal G_\ast\}.
\end{aligned}
\]
Here \(\|\Phi_A(v_{\boldsymbol \theta}(s,l),l,y,z)\|\le C(1+|y|)\). A countable convergence-determining class makes the conditional-law convergence simultaneous on a set of probability one.

With \(r=1+\delta/2>1\), conditional Jensen and \(dQ_{A,t_n}/dP_s\le3\) eventually give
\[
\sup_n E_{Q_{A,t_n}}
 \{E_{Q_{A,t_n}}(\|\Phi_A\|^2\mid\mathcal G_n)\}^{r}
\le C E_s\|\Phi_A\|^{2+\delta}<\infty.
\]
Together with the corresponding baseline bound, this proves uniform integrability and hence convergence of the moments in \(L^1(Q_{A,t_n})\). Subtracting the conditional-mean products, controlled by Cauchy--Schwarz, gives the covariance limit.
\end{proof}

\subsubsection{Local PATT joint limit}

Let
\[
\boldsymbol\theta_n=\boldsymbol\theta_A^\ast+\frac{\mathbf h_n}{\sqrt n},\qquad
\mathbf h_n\longrightarrow \mathbf h,\qquad
Q_n=Q_{A,\mathbf h_n/\sqrt n}.
\]

We write \(\mathbf D_{0,n}=\mathbf D_{0,\boldsymbol\theta_n}\) for the corresponding deterministically normalized control score.

Write \(n_z=\sum_i\mathbf1(Z_i=z)\) and \(\mathcal A_n=\{n_0\ge M,n_1\ge1\}\) for the PATT count event. On this event, the weighted reuse coefficient is
\begin{equation}\label{eq:patt-local-qv-exact-reuse}
K_{n,0}^{A}(j;\mathbf D_{0,n})
=\sum_{i:Z_i=1}w_{A,1}(\mathbf X_i)
 \frac{w_{A,0}(\mathbf X_j)
       \mathbf1\{j\in\mathcal J_{\mathbf D_{0,n}}(i)\}}
      {\sum_{k\in\mathcal J_{\mathbf D_{0,n}}(i)}
                       w_{A,0}(\mathbf X_k)}.
\end{equation}
We set it to zero off \(\mathcal A_n\) and define
\[
\begin{aligned}
c_{n,A,i}
&=\mathbf1_{\mathcal A_n}
 \{Z_iw_{A,1}(\mathbf X_i)
 -(1-Z_i)K_{n,0}^{A}(i;\mathbf D_{0,n})\},\\
\bar w_{A,n}&=\frac1n\sum_i Z_iw_{A,1}(\mathbf X_i),\\
\bar w_A&=E_s\{Zw_{A,1}(\mathbf X)\}>0,\\
\zeta_{A,i}&=Y_i-m_0^A\{\mathbf D_0^\ast(\mathbf X_i)\}.
\end{aligned}
\]
We also set the reciprocal denominator to zero when \(n_1=0\) and the first coordinate below to zero off \(\mathcal A_n\).

The deterministic own-row and array conditioning fields are
\[
\begin{aligned}
\mathcal G_{A,n,i}^{\mathrm{row}}
&=\sigma\{Z_i,\mathbf D_{0,n}(\mathbf X_i),
                       w_{A,Z_i}(\mathbf X_i)\},\\
\mathcal G_{A,n}^{\mathrm{match}}&=\bigvee_{i=1}^n\mathcal G_{A,n,i}^{\mathrm{row}},\\
\mathcal G_{A,i}^{\mathrm{row},\ast}
&=\sigma\{Z_i,\mathbf D_0^\ast(\mathbf X_i),
                       w_{A,Z_i}(\mathbf X_i)\}.
\end{aligned}
\]
The conditional moments under the current law are
\begin{equation}\label{eq:patt-local-qv-conditional-marks}
\begin{aligned}
\eta_{n,A,i}^{Q}
&=E_{Q_n}(\zeta_{A,i}\mid\mathcal G_{A,n,i}^{\mathrm{row}}),\\
u_{n,A,i}^{Q}
&=E_{Q_n}\{\mathbf U_A(O_i;\boldsymbol\theta_A^\ast)\mid\mathcal G_{A,n,i}^{\mathrm{row}}\},\\
v_{n,A,i}^{Q}
&=\operatorname{Var}_{Q_n}(\zeta_{A,i}\mid\mathcal G_{A,n,i}^{\mathrm{row}}),\\
k_{n,A,i}^{Q}
&=\operatorname{Cov}_{Q_n}\{\mathbf U_A(O_i;\boldsymbol\theta_A^\ast),\zeta_{A,i}
                                  \mid\mathcal G_{A,n,i}^{\mathrm{row}}\},\\
R_{n,A,i}^{Q}
&=\operatorname{Cov}_{Q_n}\{\mathbf U_A(O_i;\boldsymbol\theta_A^\ast)
                                  \mid\mathcal G_{A,n,i}^{\mathrm{row}}\}.
\end{aligned}
\end{equation}
The column vector \(k_{n,A,i}^{Q}\) has dimension \(p_A=\dim\boldsymbol\theta_A\). Superscript \(P\) denotes moments under \(P_s\) conditional on the same perturbed-score field, and superscript \(\ast\) denotes moments under \(P_s\) conditional on \(\mathcal G_{A,i}^{\mathrm{row},\ast}\).

For any finite deterministic centering \(\tau_n\), put
\begin{equation}\label{eq:patt-local-qv-exact-pair}
\begin{aligned}
T_{A,n}(\tau_n)
&=\frac{\sum_i c_{n,A,i}\zeta_{A,i}
 -\mathbf1_{\mathcal A_n}\tau_n
                 \sum_iZ_iw_{A,1}(\mathbf X_i)}
 {\sqrt n\,\bar w_{A,n}},\\
\mathbb U_{A,n}^{0}
&=\mathbb U_{A,n}(\boldsymbol\theta_A^\ast)
 -E_{Q_n}\{\mathbf U_A(O;\boldsymbol\theta_A^\ast)\}.
\end{aligned}
\end{equation}
The first coordinate is \(H+E\) on the current matching graph, centered at \(\tau_n\) and evaluated using the limiting-score means.

Define the centered contribution after revealing the matching field by
\[
\mathcal E_{A,n}^{\mathrm{loc}}
=\binom{T_{A,n}(\tau_n)}{\sqrt n\,\mathbb U_{A,n}^{0}}
 -E_{Q_n}\!\left\{
 \binom{T_{A,n}(\tau_n)}{\sqrt n\,\mathbb U_{A,n}^{0}}
 \,\middle|\,\mathcal G_{A,n}^{\mathrm{match}}\right\}.
\]

\begin{proposition}[Local PATT conditional covariance]
\label{prop:patt-deterministic-local-conditional-qv}
For each deterministic sequence \(\mathbf h_n\to \mathbf h\), fix \(A\in\{R,P\}\), with \(w_{R,z}=w_z\) and \(w_{P,z}=w\). Suppose Assumption~\ref{ass:residual-qv-stabilization} and all the conditions of Lemma~\ref{lem:patt-two-coordinate-conditional-law} hold.

We use the fixed deterministic population normalization, or a continuously differentiable deterministic population normalization with positive component scales, in \(\mathbf D_{0,n}=\mathbf D_{0,\boldsymbol\theta_n}\).

The conditional mean of \(\{T_{A,n}(\tau_n),\sqrt n\,\mathbb U_{A,n}^{0}\}^{\top}\) is
\begin{equation}\label{eq:patt-local-qv-predictable-vector}
\begin{pmatrix}
\displaystyle
 \frac{\sum_i c_{n,A,i}\eta_{n,A,i}^{Q}
 -\mathbf1_{\mathcal A_n}\tau_n\sum_iZ_iw_{A,1}(\mathbf X_i)}
 {\sqrt n\,\bar w_{A,n}}\\[2mm]
\displaystyle
 \frac1{\sqrt n}\sum_i
 \{u_{n,A,i}^{Q}-E_{Q_n}\mathbf U_A(O;\boldsymbol\theta_A^\ast)\}
\end{pmatrix}.
\end{equation}
Its conditional centered-row covariance is
\begin{equation}\label{eq:patt-local-qv-matrix-limit}
\begin{aligned}
&\operatorname{Cov}_{Q_n}\!\left\{
 \binom{T_{A,n}(\tau_n)}{\sqrt n\,\mathbb U_{A,n}^{0}}
 \,\middle|\,\mathcal G_{A,n}^{\mathrm{match}}\right\}\\
&=\begin{bmatrix}
 \displaystyle\frac{\sum_i c_{n,A,i}^{2}v_{n,A,i}^{Q}}
                         {n\bar w_{A,n}^{2}}
 &\displaystyle\frac{\sum_i c_{n,A,i}(k_{n,A,i}^{Q})^\top}
                         {n\bar w_{A,n}}\\[2mm]
 \displaystyle\frac{\sum_i c_{n,A,i}k_{n,A,i}^{Q}}
                         {n\bar w_{A,n}}
 &\displaystyle\frac1n\sum_i R_{n,A,i}^{Q}
 \end{bmatrix}\\
&\longrightarrow_{Q_n^{\otimes n}\text{-prob.}}
 \Omega_{E,A}
 =\begin{bmatrix}
 V_{E,A}&\mathcal C_A^\top/\bar w_A\\
 \mathcal C_A/\bar w_A&E_sR_{A,1}^\ast
 \end{bmatrix}.
\end{aligned}
\end{equation}
Here \(V_{E,A}\), \(\mathcal C_A\), and \(\Omega_{E,A}\) are the baseline marked-matching constants in \eqref{eq:qv-poisson-mean-limits}, \eqref{eq:baseline-oracle-firststep-cross-constant}, and \eqref{eq:baseline-oracle-firststep-conditional-qv}. The centered-row limit \(\Omega_{E,A}\) may be singular. For every \(t\in\mathbb R^{1+p_A}\), the centered contribution \(\mathcal E_{A,n}^{\mathrm{loc}}\) satisfies
\begin{equation}\label{eq:patt-local-qv-centered-conditional-cf}
E_{Q_n}\{e^{it^\top\mathcal E_{A,n}^{\mathrm{loc}}}
                         \mid\mathcal G_{A,n}^{\mathrm{match}}\}
\longrightarrow_{Q_n^{\otimes n}\text{-prob.}}
\exp(-t^\top\Omega_{E,A}t/2).
\end{equation}
In particular \(\mathcal E_{A,n}^{\mathrm{loc}} \Rightarrow N(0,\Omega_{E,A})\) under the local product laws.

\end{proposition}

\begin{proof}
The \(H/B/E\) decomposition gives \eqref{eq:patt-local-qv-exact-pair}: the treated contrast plus residual is \(\zeta_A-\tau_n\), and the control residual is \(\zeta_A\). Conditional on \(\mathcal G_{A,n}^{\mathrm{match}}\), the rows are independent and the coefficients are measurable. Taking conditional means and covariances gives \eqref{eq:patt-local-qv-predictable-vector} and the covariance identity in \eqref{eq:patt-local-qv-matrix-limit}.

Let \(\mathcal X_i=\sigma(Z_i,\mathbf X_i)\), \(\mathcal X_{1:n}=\bigvee_i\mathcal X_i\), and define the full-covariate projections under the current law by
\[
\eta_{n,A,i}^{Q,X}=E_{Q_n}(\zeta_{A,i}\mid\mathcal X_i),
\qquad
u_{n,A,i}^{Q,X}=E_{Q_n}(\mathbf U_A(O_i;\boldsymbol\theta_A^*)\mid\mathcal X_i).
\]
With the population normalizer, the covariate and outcome increments are
\[
\begin{aligned}
D_{n,A,i}^{X}
&=\frac1{\sqrt n}
 \begin{pmatrix}
 c_{n,A,i}(\eta_{n,A,i}^{Q,X}-\eta_{n,A,i}^{Q})/\bar w_A\\
 (u_{n,A,i}^{Q,X}-u_{n,A,i}^{Q})
 \end{pmatrix},\\
D_{n,A,i}^{Y}
&=\frac1{\sqrt n}
 \begin{pmatrix}
 c_{n,A,i}(\zeta_{A,i}-\eta_{n,A,i}^{Q,X})/\bar w_A\\
 (\mathbf U_A(O_i;\boldsymbol\theta_A^*)-u_{n,A,i}^{Q,X})
 \end{pmatrix}.
\end{aligned}
\]
Since \(\mathcal G_{A,n,i}^{\mathrm{row}}\subseteq\mathcal X_i\), conditional independence and the tower property give
\[
\begin{aligned}
E_{Q_n}\!\left(D_{n,A,i}^{X}\,\middle|\,
 \mathcal G_{A,n}^{\mathrm{match}}\vee\bigvee_{r<i}\mathcal X_r\right)&=0,\\
E_{Q_n}\!\left(D_{n,A,i}^{Y}\,\middle|\,
 \mathcal X_{1:n}\vee\sigma(Y_1,\ldots,Y_{i-1})\right)&=0,\\
\mathcal E_{A,n}^{\mathrm{loc}}
&=\operatorname{diag}(\bar w_A/\bar w_{A,n},I_{p_A})
       \sum_i(D_{n,A,i}^{X}+D_{n,A,i}^{Y}).
\end{aligned}
\]

To establish the covariance limit, first compare the moving conditional moments with their baseline values. Let \(\Phi_A=(\zeta_A,\mathbf U_A^\top)^\top\). The moment assumptions give \(E_s\|\Phi_A\|^{2+\delta}<\infty\). At \(t_n=0\), Lemma~\ref{lem:patt-two-coordinate-conditional-law} and conditional contraction with continuous clipping give convergence of first conditional moments in \(L^2(P_s)\) and of second moments in \(L^1(P_s)\). Subtracting the conditional-mean products gives
\begin{equation}\label{eq:patt-local-qv-fixed-law-mark-limit}
E_s\left\{
 |v_{n,A,1}^{P}-v_{A,1}^\ast|
 +\|k_{n,A,1}^{P}-k_{A,1}^\ast\|
 +\|R_{n,A,1}^{P}-R_{A,1}^\ast\|\right\}
\longrightarrow0.
\end{equation}
For \(\mathcal B_{n,A,i}=\{E_s[g_{\mathrm{tilt}}(\mathbf h_n^\top \mathbf s_A(O)/\sqrt n) \mid\mathcal G_{A,n,i}^{\mathrm{row}}]<1/2\}\), Lemma~\ref{lem:local-tilt-row-projections} consequently yields
\begin{equation}\label{eq:patt-local-qv-tilted-mark-limit}
\begin{aligned}
E_s\bigl\{
 &|\mathbf1_{\mathcal B_{n,A,1}^c}v_{n,A,1}^{Q}-v_{A,1}^\ast|\\
 &+\|\mathbf1_{\mathcal B_{n,A,1}^c}k_{n,A,1}^{Q}-k_{A,1}^\ast\|\\
 &+\|\mathbf1_{\mathcal B_{n,A,1}^c}R_{n,A,1}^{Q}-R_{A,1}^\ast\|
 \bigr\}\longrightarrow0.
\end{aligned}
\end{equation}
The probability that any sampled conditional denominator is below one-half is \(O(n^{-\delta/2})\) under both product laws.

Under \(P_s\), smoothness of the bounded raw scores and positive deterministic normalization gives
\begin{equation}\label{eq:patt-local-qv-score-expansion}
\mathbf D_{0,n}(\mathbf X)
=\mathbf D_0^\ast(\mathbf X)
 +n^{-1/2}S_0(\mathbf X)\mathbf h_n+n^{-1/2}r_n(\mathbf X),
\end{equation}
with bounded \(S_0\) and \(\sup_{\mathbf x}\|r_n(\mathbf x)\|\to0\). Every selected local donor therefore lies within the baseline \(M\)-th radius plus \(a/\sqrt n\), for a fixed \(a\). The inflated reuse count in \eqref{eq:stabilizer-inflated-reuse} bounds \(K_{n,0}^{A}\le w_{\max} K_{M,0}^{+,a}\) and \(|c_{n,A,i}|\le C(1+K_{M,0}^{+,a}(i))\), with
\begin{equation}\label{eq:patt-local-qv-inflated-count-moment}
\sup_n\mathop{\rm ess\,sup}_{d}
E_s\!\left[\{1+K_{M,0}^{+,a}(j)\}^{q}
       \mid Z_j=0,\mathbf D_0^\ast(\mathbf X_j)=d\right]<\infty.
\end{equation}
Conditional on the baseline score and arm, this count is independent of the remaining root marks.

Next apply the marked-limit argument of Proposition~\ref{prop:primitive-residual-qv-limit} with \(\eta=(S_0,w_{A,Z},v_A^\ast,k_A^\ast)\). Writing \(\eta_0\) for the root variables, the translations \(\{S_0(\eta)-S_0(\eta_0)\}\mathbf h\) preserve the homogeneous marked intensity. The localization argument applies to \eqref{eq:patt-local-qv-score-expansion}: displacements are bounded, the remainder vanishes uniformly, and limiting distance ties have probability zero. Conditional Jensen gives \(p=1+\delta/2\) moments of \(v_A^\ast,k_A^\ast,R_A^\ast\). The inflated-count bound supplies the corresponding uniform \(p\)-moments of the weighted summands. The one-root limit and two-root concentration argument, combined with the ordinary laws of large numbers for the treated and nuisance terms, yield
\begin{equation}\label{eq:patt-local-qv-baseline-mark-llns}
\begin{aligned}
\frac1n\sum_i c_{n,A,i}^2v_{A,i}^\ast
&\longrightarrow_{L^1(P_s^{\otimes n})}
 \Lambda_{A,1}^{\mathrm{PATT}}+\Lambda_{A,0}^{\mathrm{PATT}},\\
\frac1n\sum_i c_{n,A,i}k_{A,i}^\ast
&\longrightarrow_{L^1(P_s^{\otimes n})}\mathcal C_A,\\
\frac1n\sum_iR_{A,i}^\ast
&\longrightarrow_{L^1(P_s^{\otimes n})}E_sR_{A,1}^\ast.
\end{aligned}
\end{equation}

For every nonnegative integrable own-row mark \(F_n\), the same conditional-root independence gives
\[
E_s\{\mathbf1(Z=0)(K_{M,0}^{+,a})^qF_n(O)\}
\le C_q E_s\{\mathbf1(Z=0)F_n(O)\}.
\]
Taking \(q=2\) for the variance difference and \(q=1\) for the cross difference in \eqref{eq:patt-local-qv-tilted-mark-limit}, we obtain
\begin{equation}\label{eq:patt-local-qv-weighted-mark-transfer}
\begin{aligned}
E_s\!\left[\frac1n\sum_i c_{n,A,i}^2
 |\mathbf1_{\mathcal B_{n,A,i}^c}v_{n,A,i}^{Q}-v_{A,i}^\ast|\right]
&\longrightarrow0,\\
E_s\!\left[\frac1n\sum_i |c_{n,A,i}|
 \|\mathbf1_{\mathcal B_{n,A,i}^c}k_{n,A,i}^{Q}-k_{A,i}^\ast\|\right]
&\longrightarrow0,\\
E_s\!\left[\frac1n\sum_i
 \|\mathbf1_{\mathcal B_{n,A,i}^c}R_{n,A,i}^{Q}-R_{A,i}^\ast\|\right]
&\longrightarrow0.
\end{aligned}
\end{equation}
The conditional-denominator bound, the three average limits, and \(\bar w_{A,n}\to_p\bar w_A>0\) give the matrix limit in \(P_s^{\otimes n}\)-probability. Contiguity transfers this limit to the local law.

To compare the refined predictable covariance with this limit, put \(p=1+\delta/2\) and
\[
\begin{aligned}
M_{n,A}
&=\frac1n\sum_i(1+|c_{n,A,i}|^{2p})
 E_{Q_n}(\|\Phi_A\|^{2p}\mid\mathcal G_{A,n,i}^{\mathrm{row}}),\\
T_{n,A,i}
&=nE_{Q_n}(D_{n,A,i}^{Y}D_{n,A,i}^{Y\top}\mid\mathcal X_{1:n}).
\end{aligned}
\]
On \(\bigcap_i\mathcal B_{n,A,i}^c\), the bounded tilt and its conditional denominator give
\[
\begin{aligned}
M_{n,A}
&\le\frac Cn\sum_i(1+|c_{n,A,i}|^{2p})
 E_s(\|\Phi_A\|^{2p}\mid\mathcal G_{A,n,i}^{\mathrm{row}}),\\
E_s\!\left[\frac1n\sum_i(1+|c_{n,A,i}|^{2p})
 E_s(\|\Phi_A\|^{2p}\mid\mathcal G_{A,n,i}^{\mathrm{row}})\right]
&\le C E_s\|\Phi_A\|^{2p}<\infty.
\end{aligned}
\]
The second inequality follows by conditioning the reuse count on its root score. Markov's inequality, the conditional-denominator bound, and contiguity give \(M_{n,A}=O_{Q_n^{\otimes n}}(1)\). Conditional total covariance, as in \eqref{eq:joint-three-stage-qv-comparison}, gives
\[
\begin{aligned}
&\sum_i E_{Q_n}(D_{n,A,i}^{X}D_{n,A,i}^{X\top}\mid\mathcal G_{A,n}^{\mathrm{match}})
 +\sum_i E_{Q_n}(D_{n,A,i}^{Y}D_{n,A,i}^{Y\top}\mid\mathcal X_{1:n})\\
&\quad-\sum_i E_{Q_n}\{(D_{n,A,i}^{X}+D_{n,A,i}^{Y})
                 (D_{n,A,i}^{X}+D_{n,A,i}^{Y})^\top\mid\mathcal G_{A,n}^{\mathrm{match}}\}\\
&=\frac1n\sum_i\{T_{n,A,i}-E_{Q_n}(T_{n,A,i}\mid\mathcal G_{A,n}^{\mathrm{match}})\}.
\end{aligned}
\]
For any scalar entry \(T_{n,A,i}^{ab}\), conditional Jensen bounds its average conditional \(p\)-moment by \(CM_{n,A}\). Clipping at height \(L\), conditional independence gives
\[
\begin{aligned}
&E_{Q_n}\!\left[\left|\frac1n\sum_i
 \{T_{n,A,i}^{ab}-E_{Q_n}(T_{n,A,i}^{ab}\mid\mathcal G_{A,n}^{\mathrm{match}})\}\right|
 \,\middle|\,\mathcal G_{A,n}^{\mathrm{match}}\right]\\
&\le \frac L{\sqrt n}+CL^{1-p}M_{n,A}.
\end{aligned}
\]
Letting \(n\to\infty\) and then \(L\to\infty\) makes this covariance difference vanish in local-law probability. Together with \(\bar w_{A,n}\to\bar w_A\), this gives \(\Omega_{E,A}\) for the combined covariate and outcome stages. Conditional Jensen also yields
\[
\sum_i E_{Q_n}\{\|D_{n,A,i}^{X}\|^{2p}
                   +\|D_{n,A,i}^{Y}\|^{2p}\mid\mathcal G_{A,n}^{\mathrm{match}}\}
\le Cn^{1-p}M_{n,A}\longrightarrow_p0.
\]
Conditional Markov's inequality gives the corresponding predictable Lyapunov bound for the outcome stage. The combined rows \(D_{n,A,i}^{X}+D_{n,A,i}^{Y}\) are independent and centered given \(\mathcal G_{A,n}^{\mathrm{match}}\). Their conditional Lyapunov bound, the covariance limit and Lemma~\ref{lem:conditional-residual-stable-clt} give \eqref{eq:patt-local-qv-centered-conditional-cf} after restoring the empirical normalizer. Conditional Chebyshev covers a zero-variance projection. Bounded characteristic functions give the unconditional vector limit.
\end{proof}

At the limiting scores, the first conditional-mean coordinate is \(\sqrt n H_{n,A}\): \(\eta_A^\ast=0\) for controls and \(\eta_A^\ast=\tau_{\mathrm{PATT}}(\mathbf D_0^\ast)\) for treated observations. We next derive the change in the conditional mean of \(Y-m_0^A(\mathbf D_0^*)\) when the conditioning score is perturbed along a fixed direction \(\mathbf h\). The residual itself remains defined at \(\mathbf D_0^*\). We denote this conditional-mean derivative by \(\dot m^\varepsilon_{A,\mathbf h}\), with scalar perturbation index \(t\).

\begin{lemma}[PATT control drift]
\label{lem:patt-signed-density-control-drift}
For PATT, fix \(A\in\{R,P\}\) under all the conditions of Proposition~\ref{prop:patt-deterministic-local-conditional-qv} and Lemma~\ref{lem:baseline-oracle-firststep-joint}. Use the coefficient in \eqref{eq:patt-local-qv-exact-reuse} and assume the following additional conditions.

Suppose that the bounded-covariate raw chart of Lemma~\ref{lem:patt-two-coordinate-conditional-law} applies, and that \(A_{\boldsymbol \theta},B_{\boldsymbol \theta}\) are twice continuously differentiable near \(\boldsymbol\theta_A^\ast\). Write \(A_\ast=A_{\boldsymbol\theta_A^\ast}\), \(B_\ast=B_{\boldsymbol\theta_A^\ast}\), and \(S_\ast=S_{\boldsymbol\theta_A^\ast}(\mathbf X)\). For almost every observed control weight \(w\), let
\[
\nu_w:=\mathcal L_s\{L\mid Z=0,w_{A,0}(L)=w\}.
\]
Let \(p_w(s,l)\) be a jointly measurable conditional density of \(S_\ast\), given \(L=l,Z=0,w_{A,0}(L)=w\), with support indicators included. Define the signed first-moment density and its ordinary marginal by
\begin{equation}\label{eq:patt-drift-signed-density-definition}
\begin{aligned}
a_w(s,l)
&:=p_w(s,l)
 E_s(\zeta_A\mid Z=0,S_\ast=s,L=l,w_{A,0}=w),\\
q_w(s)&:=\int p_w(s,l)\,\nu_w(dl),
\end{aligned}
\end{equation}
where \(\zeta_A=Y-m_0^A(\mathbf D_0^\ast)\). All the conditions below hold with constants uniform over almost every such \(w\).
\begin{enumerate}
\item[(i)] There are a compact set \(J\subset\mathbb R^2\) and a bounded open set \(N\supset J\), with \(\operatorname{dist}(J,N^c)>0\), such that, for \(\nu_w\)-almost every \(l\), \(a_w(\cdot,l)\) has a \(C^2\) extension to \(\mathbb R^2\) supported in \(J\). For multi-indices \(\alpha\),
\[
\int\sup_{s\in\mathbb R^2}
 \sum_{|\alpha|\le2}|\partial^\alpha a_w(s,l)|
 \,\nu_w(dl)\le C.
\]
\item[(ii)] On \(N\), \(p_w(\cdot,l)\) is \(C^1\) for \(\nu_w\)-almost every \(l\), and
\[
\begin{aligned}
\int\sup_{s\in N}
 \{|p_w(s,l)|+\|\nabla p_w(s,l)\|\}\,\nu_w(dl)&\le C,\\
\inf_{s\in N}q_w(s)&\ge c>0.
\end{aligned}
\]
\end{enumerate}

For a fixed direction \(\mathbf h\), set \(\boldsymbol\theta_t=\boldsymbol\theta_A^\ast+t\mathbf h\), and define
\begin{equation}\label{eq:patt-drift-chart-direction}
\begin{aligned}
C_t&:=A_\ast A_{\boldsymbol\theta_t}^{-1}B_{\boldsymbol\theta_t}-B_\ast,\\
\dot A_{\mathbf h}&:=\left.\frac{d}{dt}A_{\boldsymbol\theta_A^\ast+t\mathbf h}\right|_{t=0},
\qquad
\dot B_{\mathbf h}:=\left.\frac{d}{dt}B_{\boldsymbol\theta_A^\ast+t\mathbf h}\right|_{t=0},\\
\dot C_{\mathbf h}&:=\dot B_{\mathbf h}-\dot A_{\mathbf h} A_\ast^{-1}B_\ast,\\
\mathbf F^{\varepsilon}_{\mathbf h,w}(s)&:=\int a_w(s,l)\dot C_{\mathbf h}l\,\nu_w(dl),\\
\dot m^{\varepsilon}_{A,\mathbf h}(s,w)&:=-\frac{\operatorname{div}\mathbf F^{\varepsilon}_{\mathbf h,w}(s)}{q_w(s)}
 \quad(s\in N).
\end{aligned}
\end{equation}
Set \(\dot m^{\varepsilon}_{A,\mathbf h}=0\) outside \(N\). This function is uniformly bounded and Lipschitz in \(s\), and linear in \(\mathbf h\). For \(\boldsymbol\theta_n=\boldsymbol\theta_A^\ast+\mathbf h_n/\sqrt n\), \(\mathbf h_n\to \mathbf h\), the current-field baseline-law projection in \eqref{eq:patt-local-qv-conditional-marks} satisfies, on control rows outside a fixed null set,
\begin{equation}\label{eq:patt-drift-fixed-law-root-expansion}
\eta_{n,A}^{P}
=n^{-1/2}\dot m^{\varepsilon}_{A,\mathbf h_n}(S_\ast,w_{A,0})+O(n^{-1}),
\end{equation}
with a uniform remainder.

Let \(\kappa_{A,0}(d,w)\) be the expectation of the limiting weighted reuse coefficient \(K_{A,0}^{\infty}\) in \eqref{eq:qv-poisson-exact-coefficient}, conditional on the control root score \(d\) and observed weight \(w\). The superscript \(\infty\) denotes the limiting random local matching configuration. Then
\begin{equation}\label{eq:patt-drift-fixed-law-geometric-limit}
\begin{aligned}
&\frac1{\sqrt n}\sum_{i:Z_i=0}
 K_{n,0}^{A}(i;\mathbf D_{0,n})\eta_{n,A,i}^{P}\\
&\longrightarrow_{P_s^{\otimes n}\text{-prob.}}
 \Delta_A(\mathbf h),\\
\Delta_A(\mathbf h)
&:=E_s\{(1-Z)\kappa_{A,0}(\mathbf D_0^\ast,w_{A,0})
                    \dot m^{\varepsilon}_{A,\mathbf h}(S_\ast,w_{A,0})\}.
\end{aligned}
\end{equation}
The finite constant \(\Delta_A(\mathbf h)\) is linear in \(\mathbf h\).
\end{lemma}

\begin{proof}
An invertible transformation of the current raw score expresses the moving conditioning field as a perturbation of the baseline field:
\[
\begin{aligned}
A_\ast A_{\boldsymbol\theta_t}^{-1}S_{\boldsymbol\theta_t}
&=A_\ast V+A_\ast A_{\boldsymbol\theta_t}^{-1}B_{\boldsymbol\theta_t}L\\
&=S_\ast+C_tL=:T_t,\\
C_t&=t\{\dot B_{\mathbf h}-\dot A_hA_\ast^{-1}B_\ast\}+O(t^2)\\
&=t\dot C_{\mathbf h}+O(t^2).
\end{aligned}
\]
The inverse-matrix derivative gives the expansion of the perturbation, uniformly for bounded \(\mathbf h\). The current row field is \(\sigma(Z,T_t,w_{A,Z})\). Translation has Jacobian one, so
\begin{equation}\label{eq:patt-drift-exact-density-ratio}
\begin{aligned}
q_{t,w}(u)&=\int p_w(u-C_tl,l)\,\nu_w(dl),\\
N_{t,w}(u)&=\int a_w(u-C_tl,l)\,\nu_w(dl),\\
E_s(\zeta_A\mid Z=0,T_t=u,w_{A,0}=w)
&=\frac{N_{t,w}(u)}{q_{t,w}(u)}.
\end{aligned}
\end{equation}
Baseline control centering and continuity imply \(\int a_w(s,l)\nu_w(dl)=0\) everywhere. Expanding the translated signed density therefore gives
\[
\begin{aligned}
N_{t,w}(u)
&=\int\{a_w(u,l)-\nabla a_w(u,l)^\top C_tl\}\,\nu_w(dl)+O(t^2)\\
&=-\int\nabla a_w(u,l)^\top C_tl\,\nu_w(dl)+O(t^2)\\
&=-t\,\operatorname{div}\mathbf F^{\varepsilon}_{\mathbf h,w}(u)+O(t^2).
\end{aligned}
\]
On a fixed neighborhood of \(J\), bounded \(L\) and the ordinary density derivative bound give \(q_{t,w}=q_w+O(|t|)\ge c/2\). The support and derivative bounds make the zero extension of \(\dot m^{\varepsilon}_{A,\mathbf h}\) bounded and Lipschitz. Dividing by the marginal density gives
\[
\begin{aligned}
E_s(\zeta_A\mid Z=0,T_t,w_{A,0})
&=\frac{-t\operatorname{div}\mathbf F^{\varepsilon}_{\mathbf h,w_{A,0}}(T_t)+O(t^2)}
             {q_{w_{A,0}}(T_t)+O(|t|)}\\
&=t \dot m^{\varepsilon}_{A,\mathbf h}(T_t,w_{A,0})+O(t^2)\\
&=t \dot m^{\varepsilon}_{A,\mathbf h}(S_\ast,w_{A,0})+O(t^2).
\end{aligned}
\]
Outside that neighborhood the translated signed density is zero for small \(t\), giving the uniform expansion. Taking \(t=n^{-1/2}\), \(\mathbf h=\mathbf h_n\) proves \eqref{eq:patt-drift-fixed-law-root-expansion}.

Summing the normalized matching fractions gives
\begin{equation}\label{eq:patt-drift-reuse-mass-conservation}
\sum_{j:Z_j=0}K_{n,0}^{A}(j;\mathbf D_{0,n})
=\sum_{i:Z_i=1}w_{A,1}(\mathbf X_i)\le nw_{\max}.
\end{equation}
Multiplying the uniform projection expansion by the reuse coefficients and using this identity gives
\[
\frac1{\sqrt n}\sum_{i:Z_i=0}K_{n,0}^{A}(i)\eta_{n,A,i}^P
=\frac1n\sum_{i:Z_i=0}K_{n,0}^{A}(i)
 \dot m^{\varepsilon}_{A,\mathbf h}(S_{\ast,i},w_{A,0,i})+o_p(1).
\]
Linearity and boundedness permit \(\mathbf h_n\to \mathbf h\). The marked law of large numbers \eqref{eq:patt-local-qv-baseline-mark-llns}, applied to the bounded mark \(\dot m^{\varepsilon}_{A,\mathbf h}\), then gives \(\Delta_A(\mathbf h)\) and proves \eqref{eq:patt-drift-fixed-law-geometric-limit}.
\end{proof}

\begin{proposition}[Local PATT joint limit]
\label{prop:patt-signed-density-local-joint-law}
Under all the conditions of Lemma~\ref{lem:patt-signed-density-control-drift}, use \(Q_n=Q_{A,\mathbf h_n/\sqrt n}\), \(\mathbf h_n\to \mathbf h\), the deterministic population normalization, and the pair \((T_{A,n}(\tau_n),\sqrt n\,\mathbb U_{A,n}^{0})\) in \eqref{eq:patt-local-qv-exact-pair}. Let \(\tau_A\) be the identified baseline PATT target. For one generic row, write
\[
\begin{aligned}
\eta_A^\ast&=E_s(\zeta_A\mid\mathcal G_A^\ast),\\
u_A^\ast&=E_s\{\mathbf U_A(O;\boldsymbol\theta_A^\ast)\mid\mathcal G_A^\ast\},\\
k_A^\ast&=\operatorname{Cov}_s
 \{\mathbf U_A(O;\boldsymbol\theta_A^\ast),\zeta_A\mid\mathcal G_A^\ast\},\\
\chi_{A,\mathbf h}^\ast
&=\operatorname{Cov}_s(\zeta_A,\mathbf h^\top \mathbf s_A(O)\mid\mathcal G_A^\ast),\\
\mathcal L_{A,\mathbf h}
&=E_s\{(1-Z)\kappa_{A,0}(\mathbf D_0^\ast,w_{A,0})
                         \chi_{A,\mathbf h}^\ast\}.
\end{aligned}
\]
Under \(Q_n^{\otimes n}\),
\begin{equation}\label{eq:patt-drift-auxiliary-control-limit}
\frac1{\sqrt n}\sum_{i:Z_i=0}
 K_{n,0}^{A}(i;\mathbf D_{0,n})\eta_{n,A,i}^{Q}
\longrightarrow_p\Delta_A(\mathbf h)+\mathcal L_{A,\mathbf h}.
\end{equation}
For every finite deterministic centering sequence \(\tau_n\),
\begin{equation}\label{eq:patt-drift-local-joint-limit}
\begin{pmatrix}
T_{A,n}(\tau_n)+\sqrt n(\tau_n-\tau_A)\\
\sqrt n\,\mathbb U_{A,n}^{0}
\end{pmatrix}
\Rightarrow
N\!\left(
 \binom{\mu_A(\mathbf h)}{0},
 \begin{bmatrix}
 V_{A,\mathrm{known\text{-}score}}&\boldsymbol\gamma_{A,1}^{\top}\\
 \boldsymbol\gamma_{A,1}&\boldsymbol\Sigma_{A,U}
 \end{bmatrix}\right),
\end{equation}
where
\begin{equation}\label{eq:patt-drift-local-mean}
\begin{aligned}
\mu_A(\mathbf h)
&=\frac{E_s\{Zw_{A,1}(\zeta_A-\tau_A)\mathbf h^\top \mathbf s_A(O)\}
              -\Delta_A(\mathbf h)-\mathcal L_{A,\mathbf h}}{\bar w_A}\\
&=-\boldsymbol\gamma_{A,1}^{\top}\boldsymbol\Sigma_{A,U}^{-1}\boldsymbol\Gamma_A\mathbf h
             -\frac{\Delta_A(\mathbf h)}{\bar w_A}.
\end{aligned}
\end{equation}
The covariance is the baseline Gaussian covariance in Lemma~\ref{lem:baseline-oracle-firststep-joint}. The lower coordinate may equivalently be \(\sqrt n\,\mathbb U_{A,n}(\boldsymbol\theta_n)=n^{-1/2}\sum_i\mathbf U_A(O_i;\boldsymbol\theta_n)\).
\end{proposition}

\begin{proof}
Write \(g_n=g_{\mathrm{tilt}}(\mathbf h_n^\top \mathbf s_A(O)/\sqrt n)\), \(a_n=\sqrt n(g_n-1)\), \(d_n=E_s(g_n\mid\mathcal G_{A,n,1}^{\mathrm{row}})\), and \(\mathcal B_n=\{d_n<1/2\}\). We first calculate the additional control drift induced by the local law. The likelihood normalizer cancels from the conditional ratio:
\begin{equation}\label{eq:patt-drift-exact-tilt-covariance}
\sqrt n(\eta_{n,A}^{Q}-\eta_{n,A}^{P})
=\frac{\operatorname{Cov}_s(\zeta_A,a_n\mid\mathcal G_{A,n,1}^{\mathrm{row}})}{d_n}.
\end{equation}
Since \(|a_n|\le C\|\mathbf s_A(O)\|\), dominated convergence gives \(a_n\to \mathbf h^\top \mathbf s_A(O)\) in \(L^2(P_s)\). Conditional contraction and Cauchy--Schwarz, followed by \eqref{eq:patt-local-qv-fixed-law-mark-limit}, show that the numerator converges in \(L^1(P_s)\) to \(\chi_{A,\mathbf h}^\ast\). The factor \(\mathbf1_{\mathcal B_n^c}/d_n\) is bounded by two and converges in probability to one. Consequently
\begin{equation}\label{eq:patt-drift-good-row-correction}
r_n:=\mathbf1_{\mathcal B_n^c}\sqrt n
 (\eta_{n,A}^{Q}-\eta_{n,A}^{P})-\chi_{A,\mathbf h}^\ast,
\qquad E_s|r_n|\longrightarrow0.
\end{equation}
The definition of the auxiliary score gives
\begin{equation}\label{eq:patt-drift-tilt-mark-sign}
\chi_{A,\mathbf h}^\ast
=-(k_A^\ast)^\top\boldsymbol\Sigma_{A,U}^{-1}\boldsymbol\Gamma_A\mathbf h.
\end{equation}
This mark has a finite \(1+\delta/2\) moment. The marked law of large numbers for the incoming coefficient in \eqref{eq:patt-local-qv-baseline-mark-llns} gives the limit \(\mathcal L_{A,\mathbf h}\) for \(n^{-1}\sum_{Z_i=0}K_{n,0}^{A} \chi_{A,\mathbf h,i}^\ast\). The conditional inflated-count bound gives
\[
E_s\!\left\{\frac1n\sum_{Z_i=0}K_{n,0}^{A}(i)|r_n(O_i)|\right\}
\le C E_s\{(1-Z)|r_n(O)|\}\longrightarrow0.
\]
Together with the fixed-law drift and conditional-denominator bounds, this proves \eqref{eq:patt-drift-auxiliary-control-limit} in \(P_s^{\otimes n}\)-probability and, by contiguity, under the local law.

Baseline centering gives \(Z\eta_A^\ast=Z\tau_{\mathrm{PATT}}(\mathbf D_0^\ast)\), \((1-Z)\eta_A^\ast=0\), and \(E_su_A^\ast=0\). Lemma~\ref{lem:patt-two-coordinate-conditional-law} gives convergence of the moving projections in probability. Conditional Jensen, the fixed \(2+\delta\) moments and \(dQ_n/dP_s\le3\) give uniform integrability of their squares, hence
\begin{equation}\label{eq:patt-drift-treated-projection-ltwo}
E_{Q_n}\{ |\eta_{n,A}^{Q}-\eta_A^\ast|^2
                    +\|u_{n,A}^{Q}-u_A^\ast\|^2\}\longrightarrow0.
\end{equation}
Using the increments from Proposition~\ref{prop:patt-deterministic-local-conditional-qv}, define the three-stage vector array, for \(1\le i\le n\), by
\[
\begin{aligned}
\xi_{n,i}
&=\frac1{\sqrt n}\left[
 \begin{pmatrix}
 Z_iw_{A,1}(\mathbf X_i)(\eta_{n,A,i}^{Q}-\tau_A)/\bar w_A\\
 u_{n,A,i}^{Q}
 \end{pmatrix}
 -E_{Q_n}\begin{pmatrix}
 Zw_{A,1}(\eta_{n,A}^{Q}-\tau_A)/\bar w_A\\u_{n,A}^{Q}
 \end{pmatrix}\right],\\
\xi_{n,n+i}&=D_{n,A,i}^{X},\\
\xi_{n,2n+i}&=D_{n,A,i}^{Y}.
\end{aligned}
\]
Its filtration is
\[
\mathcal F_{n,k}^{\mathrm{loc}}=
\begin{cases}
 \displaystyle\bigvee_{i=1}^k\mathcal G_{A,n,i}^{\mathrm{row}},&0\le k\le n,\\
 \displaystyle\mathcal G_{A,n}^{\mathrm{match}}\vee\bigvee_{i=1}^{k-n}\mathcal X_i,
       &n<k\le2n,\\
 \mathcal X_{1:n}\vee\sigma(Y_1,\ldots,Y_{k-2n}),&2n<k\le3n.
\end{cases}
\]
Independence and the conditional projections give
\[
E_{Q_n}(\xi_{n,k}\mid\mathcal F_{n,k-1}^{\mathrm{loc}})=0,
\qquad 1\le k\le3n.
\]
For the centered projection differences, independence gives
\[
\begin{aligned}
&E_{Q_n}\left\|\frac1{\sqrt n}\sum_i
 \left[\binom{Z_iw_{A,1}(\mathbf X_i)(\eta_{n,A,i}^{Q}-\eta_{A,i}^\ast)/\bar w_A}
                  {(u_{n,A,i}^{Q}-u_{A,i}^\ast)}
 -E_{Q_n}\binom{Zw_{A,1}(\eta_{n,A}^{Q}-\eta_A^\ast)/\bar w_A}
                  {(u_{n,A}^{Q}-u_A^\ast)}\right]\right\|^2\\
&\le C E_{Q_n}\{|\eta_{n,A}^{Q}-\eta_A^\ast|^2
                        +\|u_{n,A}^{Q}-u_A^\ast\|^2\}\\
&\longrightarrow0.
\end{aligned}
\]
The uniform \(2+\delta\) moment bound gives the triangular-array CLT for the baseline projections. The covariance limit is
\[
\Omega_{H,A}
=\operatorname{Var}_s
 \binom{Zw_{A,1}\{\tau_{\mathrm{PATT}}(\mathbf D_0^\ast)-\tau_A\}/\bar w_A}
       {u_A^\ast}.
\]
The law of large numbers gives \(\bar w_{A,n}\to\bar w_A>0\), so restoring the empirical normalizer preserves the centered limit.

The tower property gives
\[
\begin{aligned}
E_{Q_n}[Zw_{A,1}(\eta_{n,A}^Q-\tau_A)]
&=E_{Q_n}\{E_{Q_n}[Zw_{A,1}(\zeta_A-\tau_A)
                         \mid\mathcal G_{A,n,1}^{\mathrm{row}}]\}\\
&=E_{Q_n}[Zw_{A,1}(\zeta_A-\tau_A)],\\
E_{Q_n}u_{n,A}^Q&=E_{Q_n}\mathbf U_A.
\end{aligned}
\]
Baseline recovery and \(a_n\to \mathbf h^\top \mathbf s_A(O)\) in \(L^2\) yield
\begin{equation}\label{eq:patt-drift-treated-likelihood-mean}
\sqrt n E_{Q_n}\{Zw_{A,1}(\zeta_A-\tau_A)\}
=\frac{E_s\{Zw_{A,1}(\zeta_A-\tau_A)a_n\}}{E_sg_n}
\longrightarrow E_s\{Zw_{A,1}(\zeta_A-\tau_A)\mathbf h^\top \mathbf s_A(O)\}.
\end{equation}
Subtracting the control drift gives, for the predictable vector at \(\tau_n=\tau_A\),
\begin{equation}\label{eq:patt-drift-predictable-local-limit}
\mathcal P_{A,n}
\Rightarrow N\!\left(\binom{\mu_A(\mathbf h)}0,\Omega_{H,A}\right).
\end{equation}
This vector is \(\mathcal G_{A,n}^{\mathrm{match}}\)-measurable. The conditional Gaussian limit \eqref{eq:patt-local-qv-centered-conditional-cf} and Lemma~\ref{lem:conditional-residual-stable-clt} give independent limits for the predictable vector and the combined covariate and outcome contribution. Their sum has the covariance \(\Omega_{H,A}+\Omega_{E,A}\) identified in \eqref{eq:baseline-oracle-firststep-conditional-qv} and \eqref{eq:baseline-oracle-firststep-gamma}. For arbitrary finite deterministic \(\tau_n\), the exact identity
\[
T_{A,n}(\tau_n)
=T_{A,n}(\tau_A)-\mathbf1_{\mathcal A_n}\sqrt n(\tau_n-\tau_A)
\]
gives the recentering on an event whose probability tends to one.

To replace the centered estimating function by its value at the local parameter, differentiate the estimating equation and the bounded tilt:
\[
\frac1{\sqrt n}\sum_i\{\mathbf U_A(O_i;\boldsymbol\theta_n)-\mathbf U_A(O_i;\boldsymbol\theta_A^\ast)\}
=\boldsymbol\Gamma_A\mathbf h+o_p(1),\qquad
\sqrt n E_{Q_n}\mathbf U_A(O;\boldsymbol\theta_A^\ast)=-\boldsymbol\Gamma_A\mathbf h+o(1).
\]
Hence \(\sqrt n\,\mathbb U_{A,n}(\boldsymbol\theta_n)=\sqrt n\,\mathbb U_{A,n}^0+o_p(1)\). For the mean formula, conditional covariance decomposition gives
\[
E_s\{Zw_{A,1}(\zeta_A-\tau_A)\mathbf U_A^\top\}
=E_s[Zw_{A,1}\{\tau_{\mathrm{PATT}}(\mathbf D_0^\ast)-\tau_A\}
 (u_A^\ast)^\top]+E_s\{Zw_{A,1}(k_A^\ast)^\top\}.
\]
Combining this identity with \(\mathcal C_A=E_s(Zw_{A,1}k_A^\ast) -E_s\{(1-Z)\kappa_{A,0}k_A^\ast\}\), \eqref{eq:patt-drift-tilt-mark-sign}, and the cross-covariance with \(\mathbf U_A\) in \eqref{eq:baseline-oracle-firststep-gamma}, we obtain
\[
\frac{E_s\{Zw_{A,1}(\zeta_A-\tau_A)\mathbf h^\top \mathbf s_A(O)\}
-\mathcal L_{A,\mathbf h}}{\bar w_A}
=-\boldsymbol\gamma_{A,1}^\top\boldsymbol\Sigma_{A,U}^{-1}\boldsymbol\Gamma_A\mathbf h.
\]
The remaining drift is \(-\Delta_A(\mathbf h)/\bar w_A\), proving \eqref{eq:patt-drift-local-mean}.
\end{proof}

For a deterministic local sequence, we write
\[
\boldsymbol\theta_n=\boldsymbol\theta_A^\ast+\frac{\mathbf h_n}{\sqrt n},
\qquad \mathbf h_n\longrightarrow \mathbf h,
\qquad Q_n=Q_{A,\mathbf h_n/\sqrt n}.
\]
For this deterministic \(\boldsymbol\theta_n\), use the raw coordinates
\[
r_{01}(\mathbf X;\boldsymbol\theta)=e(\mathbf X;\boldsymbol\alpha),
\qquad
r_{02}(\mathbf X;\boldsymbol\theta)=\mathbf X^\top\boldsymbol\beta_0.
\]

Let \(b_0(O)\) be the moment weight and write \(q_0=E_sb_0(O)\). For \(\boldsymbol\theta\) near \(\boldsymbol\theta_A^\ast\), define the corresponding population moments by
\[
\begin{aligned}
\ell_{0k}(\boldsymbol\theta)
&:=\frac{E_s\{b_0(O)r_{0k}(\mathbf X;\boldsymbol\theta)\}}{q_0},\\
s_{0k}(\boldsymbol\theta)^2
&:=\frac{E_s[b_0(O)
       \{r_{0k}(\mathbf X;\boldsymbol\theta)-\ell_{0k}(\boldsymbol\theta)\}^2]}{q_0},
\qquad k=1,2.
\end{aligned}
\]
Let \((\ell_{n,0k},s_{n,0k})\) be either the fixed baseline moments \((\ell_{0k}(\boldsymbol\theta_A^\ast),s_{0k}(\boldsymbol\theta_A^\ast))\), or these population moments at \(\boldsymbol\theta_n\).

The empirical moments and the two normalized score maps, evaluated on the same deterministic raw-score array, are
\begin{equation}
\label{eq:patt-empirical-standardization-moments}
\begin{aligned}
\widehat\ell_{n,0k}
&:=\frac{\sum_i b_0(O_i)r_{0k}(\mathbf X_i;\boldsymbol\theta_n)}
         {\sum_i b_0(O_i)},\\
\widehat s_{n,0k}^{\,2}
&:=\frac{\sum_i b_0(O_i)
  \{r_{0k}(\mathbf X_i;\boldsymbol\theta_n)-\widehat\ell_{n,0k}\}^2}
         {\sum_i b_0(O_i)},\\
D_{0,n,k}^{\mathrm p}(\mathbf x)
&:=\frac{r_{0k}(\mathbf x;\boldsymbol\theta_n)-\ell_{n,0k}}{s_{n,0k}},\\
D_{0,n,k}^{\mathrm e}(\mathbf x)
&:=\frac{r_{0k}(\mathbf x;\boldsymbol\theta_n)-\widehat\ell_{n,0k}}
         {\widehat s_{n,0k}}.
\end{aligned}
\end{equation}
On exceptional samples, we use the moment definitions of Lemma~\ref{lem:primitive-score-standardization}.

Let \(T_{A,n}^{\mathrm p}(\tau_n)\) be the \(H+E\) coordinate in \eqref{eq:patt-local-qv-exact-pair} on \(\mathbf D_{0,n}^{\mathrm p}\), and define \(T_{A,n}^{\mathrm e}(\tau_n)\) by changing only its matching graph to \(\mathbf D_{0,n}^{\mathrm e}\). Both use
\[
\zeta_{A,i}
=Y_i-m_0^A\{\mathbf D_0^\ast(\mathbf X_i)\},
\qquad
\bar w_{A,n}=\frac1n\sum_iZ_iw_{A,1}(\mathbf X_i),
\]
the same donor weights, and the same valid-count completion.

\begin{proposition}[Empirical normalization for local PATT scores]
\label{prop:patt-deterministic-local-empirical-standardization}
Under all the conditions of Proposition~\ref{prop:patt-signed-density-local-joint-law}, consider the deterministic local sequences and normalizations defined above.

Let \(b_0(O)\) be the known moment weight in Lemma~\ref{lem:primitive-score-standardization}, with
\[
0\le b_0(O)\le C_b,
\qquad q_0:=E_s b_0(O)>0.
\]
The limiting normalization \(\mathbf D_0^\ast\) in the preceding propositions corresponds to this same rule, with positive limiting coordinate variances. For every \(n\), we assume
\begin{equation}
\label{eq:patt-empirical-standardization-row-field}
\begin{aligned}
\mathcal G_{A,n,i}^{\mathrm{row}}
&:=\sigma\{Z_i,r_{01}(\mathbf X_i;\boldsymbol\theta_n),
                r_{02}(\mathbf X_i;\boldsymbol\theta_n),
                w_{A,Z_i}(\mathbf X_i)\},\\
b_0(O_i)&\text{ is }\mathcal G_{A,n,i}^{\mathrm{row}}\text{-measurable},\\
\mathcal G_{A,n}^{\mathrm{match}}&:=\bigvee_{i=1}^n\mathcal G_{A,n,i}^{\mathrm{row}}.
\end{aligned}
\end{equation}
We assume that tied selections under both normalizations are \(\mathcal G_{A,n}^{\mathrm{match}}\)-measurable, as with fixed row-label priority.

The preceding local joint limit applies under this deterministic normalization. For every finite deterministic centering sequence \(\tau_n\),
\begin{equation}
\label{eq:patt-empirical-standardization-hbe-transfer}
T_{A,n}^{\mathrm e}(\tau_n)-T_{A,n}^{\mathrm p}(\tau_n)
\longrightarrow_{Q_n^{\otimes n}\text{-prob.}}0.
\end{equation}
The corresponding comparison also holds under \(P_s^{\otimes n}\). Consequently, the joint limit \eqref{eq:patt-drift-local-joint-limit} also holds with \(T_{A,n}^{\mathrm e}\) in place of \(T_{A,n}\).
\end{proposition}

\begin{proof}
Empirical centering cancels from pairwise differences, as in Lemma~\ref{lem:primitive-score-standardization}:
\begin{equation}\label{eq:patt-empirical-standardization-distance-cancellation}
D_{0,n,k}^{e}(\mathbf X_i)-D_{0,n,k}^{e}(\mathbf X_j)
=\frac{s_{n,0k}}{\widehat s_{n,0k}}
 \{D_{0,n,k}^{p}(\mathbf X_i)-D_{0,n,k}^{p}(\mathbf X_j)\}.
\end{equation}
Thus the empirical metric applies \(A_n=\operatorname{diag}(s_{n,01}/\widehat s_{n,01}, s_{n,02}/\widehat s_{n,02})\) to deterministic-score differences. For the bounded triangular raw-moment rows, the iid variance bound is \(O(n^{-1})\). The ratio and square-root maps from that lemma and smooth population moments give, for either stated normalization,
\begin{equation}\label{eq:patt-empirical-standardization-scale-localization}
\max_{k=1,2}|\widehat s_{n,0k}-s_{n,0k}|=O_{P_s}(n^{-1/2}),
\qquad \|A_n-I_2\|_{\rm op}=O_{P_s}(n^{-1/2}).
\end{equation}
Let \(\mathcal E_{n,C}\) be the event of valid moments and scales with \(\|A_n-I_2\|_{\rm op}\le C/\sqrt n\). Then \(\lim_{C\to\infty}\limsup_nP_s(\mathcal E_{n,C}^c)=0\). By the assumed measurability of \(b_0\) and tied selections, both graphs and coefficient arrays are measurable with respect to the same \(\mathcal G_{A,n}^{\mathrm{match}}\). The rows remain conditionally independent.

For controls, write \(\Delta K_{n,A,i}=K_{n,0}^{A}(i;\mathbf D_{0,n}^e) -K_{n,0}^{A}(i;\mathbf D_{0,n}^p)\). The treated terms and target centering cancel:
\begin{equation}\label{eq:patt-empirical-standardization-exact-difference}
T_{A,n}^{e}(\tau_n)-T_{A,n}^{p}(\tau_n)
=-\frac1{\sqrt n\,\bar w_{A,n}}\sum_{i:Z_i=0}\Delta K_{n,A,i}\zeta_{A,i}.
\end{equation}
For fixed \(C\), write \(\epsilon_n=C/\sqrt n<1/2\). The bounds \((1-\epsilon_n)\|v\|\le\|A_nv\|\le(1+\epsilon_n)\|v\|\) show that a changed query set necessarily satisfies
\begin{equation}\label{eq:patt-empirical-standardization-query-gap}
(1-\epsilon_n)R_{n,M+1}^{p}(j)
\le(1+\epsilon_n)R_{n,M}^{p}(j).
\end{equation}
With exactly \(M\) controls the selected sets coincide.

The query-root local marked-process limit in Section~\ref{app:local-score-marked-limits} gives \((\sqrt nR_{n,M}^{p},\sqrt nR_{n,M+1}^{p}) \Rightarrow(R_M^\infty,R_{M+1}^\infty)\), where \(0<R_M^\infty<R_{M+1}^\infty<\infty\) almost surely. The limiting strict gap makes the probability of a changed query vanish. Bounded convergence and exchangeability therefore give, for the number \(N_{n,\mathrm{gap}}(C)\) of queries satisfying the gap event,
\begin{equation}\label{eq:patt-empirical-standardization-gap-fraction}
E_s\{N_{n,\mathrm{gap}}(C)/n\}\longrightarrow0.
\end{equation}
Set the gap indicator to zero when fewer than \(M+1\) controls are available.

On \(\mathcal E_{n,C}\), both point maps are within \(a(C)/\sqrt n\) of the baseline map, after omitting a common translation. The baseline inflated count consequently gives
\begin{equation}\label{eq:patt-empirical-standardization-reuse-envelope}
K_{n,0}^{A}(i;\mathbf D_{0,n}^p)
+K_{n,0}^{A}(i;\mathbf D_{0,n}^e)
\le2w_{\max} K_{M,0}^{+,a(C)}(i),
\qquad
|\Delta K_{n,A,i}|\le2w_{\max} K_{M,0}^{+,a(C)}(i).
\end{equation}
The two vectors of normalized fractions at a changed query differ in \(\ell^1\)-norm by at most two. Hence
\begin{equation}\label{eq:patt-empirical-standardization-coefficient-lone}
\sum_{i:Z_i=0}|\Delta K_{n,A,i}|
\le2w_{\max} N_{n,\mathrm{gap}}(C).
\end{equation}
At most \(2M N_{n,\mathrm{gap}}(C)\) donor rows are affected. Apply Cauchy--Schwarz first across these rows and then under the sample law. The uniform fourth moment of the inflated count gives
\begin{equation}\label{eq:patt-empirical-standardization-coefficient-ltwo}
\begin{aligned}
E_s\!\left\{\mathbf1_{\mathcal E_{n,C}}\frac1n
 \sum_{i:Z_i=0}(\Delta K_{n,A,i})^2\right\}
&\le C\left\{E_s\frac{2M N_{n,\mathrm{gap}}(C)}n\right\}^{1/2}
 \left\{E_s\frac1n\sum_{i:Z_i=0}
                  (K_{M,0}^{+,a(C)}(i))^4\right\}^{1/2}\\
&\le C_C\{E_s(N_{n,\mathrm{gap}}(C)/n)\}^{1/2}\\
&\longrightarrow0.
\end{aligned}
\end{equation}
For any nonnegative deterministic own-row mark \(F_n\), conditional independence of the unweighted inflated count from the remaining root marks also gives, for \(j=1,2\),
\begin{equation}\label{eq:patt-empirical-standardization-weighted-mark-bound}
E_s\!\left\{\mathbf1_{\mathcal E_{n,C}}\frac1n
 \sum_{i:Z_i=0}|\Delta K_{n,A,i}|^jF_n(O_i)\right\}
\le C_{C,j}E_s\{(1-Z)F_n(O)\}.
\end{equation}
This bound follows by dropping the event indicator before conditioning on each root's baseline score and treatment.

We now bound the centered and conditional-mean parts of the control sum. Use the conditional marks on the common current field and write \(\mathcal B_n=\{E_s(g_n\mid\mathcal G_{A,n,1}^{\mathrm{row}})<1/2\}\). Equation~\eqref{eq:patt-local-qv-tilted-mark-limit} gives \(E_s|\mathbf1_{\mathcal B_n^c}v_{n,A}^Q-v_A^\ast|\to0\). For fixed \(C,T\), the weighted-mark bound gives
\[
\begin{aligned}
&E_s\!\left[\mathbf1_{\mathcal E_{n,C}}\frac1n
 \sum_{i:Z_i=0}(\Delta K_{n,A,i})^2
                       \mathbf1_{\mathcal B_n^c}(O_i)v_{n,A,i}^Q\right]\\
&\le C_C E_s\{(1-Z)|\mathbf1_{\mathcal B_n^c}v_{n,A}^Q-v_A^\ast|\}
 +E_s\!\left[\mathbf1_{\mathcal E_{n,C}}\frac1n
              \sum_{i:Z_i=0}(\Delta K_{n,A,i})^2v_{A,i}^\ast\right]\\
&\le o(1)+T E_s\!\left[\mathbf1_{\mathcal E_{n,C}}\frac1n
              \sum_{i:Z_i=0}(\Delta K_{n,A,i})^2\right]
 +C_CE_s\{(1-Z)v_A^\ast\mathbf1(v_A^\ast>T)\}.
\end{aligned}
\]
The iterated limit \(n\to\infty\), then \(T\to\infty\), is zero. The conditional-denominator bound, followed by \(C\to\infty\) and contiguity, gives
\begin{equation}\label{eq:patt-empirical-standardization-conditional-variance}
\operatorname{Var}_{Q_n}\!\left[
 \frac1{\sqrt n}\sum_{i:Z_i=0}\Delta K_{n,A,i}
 (\zeta_{A,i}-\eta_{n,A,i}^Q)\,\middle|\,\mathcal G_{A,n}^{\mathrm{match}}\right]
=\frac1n\sum_{i:Z_i=0}(\Delta K_{n,A,i})^2v_{n,A,i}^Q
\longrightarrow_{Q_n^{\otimes n}\text{-prob.}}0.
\end{equation}
The sum has conditional mean zero, so conditional Chebyshev gives an \(o_p(1)\) centered difference.

On controls define \(\widetilde r_{n,A}=\mathbf1_{\mathcal B_n^c}\sqrt n\,\eta_{n,A}^Q -\dot m^{\varepsilon}_{A,\mathbf h}(S_\ast,w_{A,0})-\chi_{A,\mathbf h}^\ast\), and set it to zero on treated rows. The uniform signed-density expansion and \eqref{eq:patt-drift-good-row-correction}, together with \(\mathbf h_n\to \mathbf h\), give
\begin{equation}\label{eq:patt-empirical-standardization-mean-rate}
E_s|\widetilde r_{n,A}|\longrightarrow0.
\end{equation}
For the integrable mark \(\dot m^{\varepsilon}_{A,\mathbf h}+\chi_{A,\mathbf h}^\ast\), \eqref{eq:patt-empirical-standardization-coefficient-lone} and the \(j=1\) weighted-mark bound give
\[
\begin{aligned}
&E_s\!\left[\mathbf1_{\mathcal E_{n,C}}\frac1n\sum_{i:Z_i=0}
 |\Delta K_{n,A,i}|\,
 |\dot m^{\varepsilon}_{A,\mathbf h}(S_{*,i},w_{A,0,i})+\chi_{A,\mathbf h,i}^\ast
                         +\widetilde r_{n,A,i}|\right]\\
&\le T E_s\!\left[\mathbf1_{\mathcal E_{n,C}}\frac1n
                         \sum_{i:Z_i=0}|\Delta K_{n,A,i}|\right]
       +C_C E_s|\widetilde r_{n,A}|\\
&\quad+C_C E_s\!\left[(1-Z)|\dot m^{\varepsilon}_{A,\mathbf h}+\chi_{A,\mathbf h}^\ast|
                    \mathbf1\{|\dot m^{\varepsilon}_{A,\mathbf h}+\chi_{A,\mathbf h}^\ast|>T\}\right].
\end{aligned}
\]
Taking \(n\to\infty\), then \(T\to\infty\), gives zero. The conditional-denominator and localization bounds, followed by contiguity, give
\begin{equation}\label{eq:patt-empirical-standardization-predictable-transfer}
\frac1{\sqrt n}\sum_{i:Z_i=0}\Delta K_{n,A,i}\eta_{n,A,i}^Q
\longrightarrow_{Q_n^{\otimes n}\text{-prob.}}0.
\end{equation}
Combining the centered and conditional-mean comparisons in \eqref{eq:patt-empirical-standardization-exact-difference}, with \(\bar w_{A,n}\to_p\bar w_A>0\), proves \eqref{eq:patt-empirical-standardization-hbe-transfer}. The nuisance coordinate is identical under both normalizations, so the joint limit follows. The same proof under \(P_s^{\otimes n}\) uses the fixed-law marks and signed-density expansion, without the likelihood correction.
\end{proof}

\subsubsection{Fitted-score WDSM consistency}
\label{app:primary-fitted-score-consistency}

Let \(A\in\{R,P\}\), with weights \(w_{R,z}=w_z\), \(w_{P,z}=w\), population reduced means \(m_z^A\), and residuals
\[
\varepsilon_{z,i}^A
=Y_i-m_z^A\{\mathbf D_z^\ast(\mathbf X_i)\},\qquad Z_i=z.
\]
For PATT, we use
\[
\tau_{\mathrm{PATT}}(d)=m_{1\mid1}(d)-m_0^A(d),
\qquad
\varepsilon_{1\mid1,i}
=Y_i-m_{1\mid1}\{\mathbf D_0^\ast(\mathbf X_i)\},\quad Z_i=1.
\]
The matching graph is constructed from the fitted and empirically standardized scores \(\widehat{\mathbf D}_z\).

\begin{proposition}[Consistency with estimated scores]
\label{cor:primary-fitted-score-consistency}
Assume the following conditions for the specified estimand:
\begin{enumerate}
\item[(i)] Population recovery and mean balancing hold, together with Assumption~\ref{ass:residual-centering-survey-normalized}.
\item[(ii)] Observations are independent and identically distributed under \(P_s\), with \(p_a:=P_s(Z=a)\in(0,1)\), and \(M\) is fixed. Each limiting score has a compact convex two-dimensional support with nonempty interior. The donor densities are bounded above and away from zero, and the query densities are bounded above on these supports.
\item[(iii)] Observed survey weights are bounded and positive. The outcome and each required observed-arm residual have finite selected-sample \(p=2+\delta_0>2\) moments.
\item[(iv)] The conditions on bounded regressors, finite roots, rank, and outcome moments in Lemma~\ref{lem:primitive-logistic-linear-roots} hold. Standardization uses the fixed bounded weights and positive scales in Lemma~\ref{lem:primitive-score-standardization}.
\end{enumerate}
The resulting Taylor expansion gives the bounded local score expansion in Section~\ref{app:local-score-reuse}.

Then, for either estimand and sampling design, the residual term evaluated with fitted-score matching coefficients and limiting-score residuals converges to zero in probability. If the bias-correction remainder at the fitted graph satisfies
\[
\widehat B_{n,A}^{D,\mathcal T}
   (\widehat{\mathbf D})
-B_{n,A}^{D,\mathcal T}
   (\widehat{\mathbf D};\mathbf D^\ast,m)
\to_p0,
\qquad
\mathcal T\in\{\mathrm{PATE},\mathrm{PATT}\},
\]
then the bias-corrected WDSM estimator is consistent:
\[
\widehat\tau_{\mathcal T}^{A,\mathrm{BC}}
   (\widehat{\mathbf D})
\to_p\tau_{\mathcal T}.
\]
For PATT only \(\widehat{\mathbf D}_0\) is used.
\end{proposition}

\begin{proof}
For PATE, iterated expectation reduces the joint-score centering assumption to \(E_s(\varepsilon_z^A\mid Z=z,\mathbf D_z^*,w_{A,z})=0\). For PATT, the two corresponding centering conditions are assumed directly. We first consider a deterministic local index and the weighted query functional
\[
\overline r_{n,A,z,i}(\mathbf h)
=\mathbf1\{Z_i=1-z\}w_{A,1-z}(\mathbf X_i)
 \frac{\sum_{j\in\mathcal J_{z,n,\mathbf h}(i)}
       w_{A,z}(\mathbf X_j)\varepsilon_{z,j}^A}
      {\sum_{j\in\mathcal J_{z,n,\mathbf h}(i)}w_{A,z}(\mathbf X_j)} .
\]
By the local marked law in Section~\ref{app:local-score-marked-limits}, applied with \((S_z,w_{A,Z},\varepsilon_z^A)\), the limiting selected donor marks \((W_j,E_j)\) at location \(d\) satisfy
\[
E\left(\left.
 W_q\frac{\sum_{j=1}^M W_jE_j}{\sum_{j=1}^M W_j}
 \right|d,\text{query mark},W_1,\ldots,W_M\right)
=\frac{W_q\sum_{j=1}^M W_jE(E_j\mid d,W_j)}
       {\sum_{j=1}^M W_j}=0.
\]
The denominator is bounded below by \(Mw_{\min}\).

For any donor set \(J\) of size \(M\), Jensen's inequality gives
\[
\left|w_{A,1-z}(\mathbf X_i)
 \frac{\sum_{j\in J}w_{A,z,j}\varepsilon_{z,j}^A}
      {\sum_{j\in J}w_{A,z,j}}\right|^p
\le\frac{w_{\max}^{p+1}}{Mw_{\min}}
       \sum_{j\in J}|\varepsilon_{z,j}^A|^p.
\]
For a uniformly sampled root \(I_n\), reindexing through \eqref{eq:stabilizer-inflated-reuse} gives
\[
\begin{aligned}
E_s|\overline r_{n,A,z,I_n}(\mathbf h)|^p
&=\frac1n\sum_i E_s|\overline r_{n,A,z,i}(\mathbf h)|^p\\
&\le\frac Cn\sum_{j=1}^n
 E_s[\mathbf1(Z_j=z)K_{M,z}^{+,a}(j)|\varepsilon_{z,j}^A|^p]\\
&\le C'E_s[\mathbf1(Z=z)|\varepsilon_z^A|^p]<\infty.
\end{aligned}
\]
The last inequality conditions the inflated count on the root score and treatment, under which it is independent of the remaining root marks. Cauchy--Schwarz then gives uniform integrability of the two-root products. For a uniform ordered pair \((I_n,J_n)\) of distinct indices, independent of the data, the two-root limit gives
\[
\begin{aligned}
E_s\!\left[\left\{\frac1n\sum_i\overline r_{n,A,z,i}(\mathbf h)\right\}^2\right]
&=\frac1n E_s\!\left[\{\overline r_{n,A,z,I_n}(\mathbf h)\}^2\right]\\
&\quad+\frac{n-1}{n}E_s[
 \overline r_{n,A,z,I_n}(\mathbf h)\overline r_{n,A,z,J_n}(\mathbf h)]\\
&\longrightarrow0.
\end{aligned}
\]

To evaluate the average at the fitted local index, we control the effect of changing a match set. If donor sets \(J,J'\) coincide, their weighted ratios coincide. Otherwise, the difference is bounded by
\[
\frac{w_{\max}}{Mw_{\min}}
 \left(\sum_{j\in J}|\varepsilon_{z,j}^A|
       +\sum_{j\in J'}|\varepsilon_{z,j}^A|\right).
\]
At a fixed grid mesh, the nearest-neighbor-gap law of large numbers applies simultaneously at the finitely many grid points. It bounds the fraction of queries that change within a grid cell by \(C\delta+o_p(1)\). For two bounded local indices, H\"older's inequality with three factors gives
\[
\begin{aligned}
&\frac1n\sum_i
 |\overline r_{n,A,z,i}(\mathbf h)-\overline r_{n,A,z,i}(\mathbf h')|\\
&\le C\left\{\frac1n\#\{i:\mathcal J_{z,n,\mathbf h}(i)
                          \ne\mathcal J_{z,n,\mathbf h'}(i)\}\right\}^{1/2-1/p}\\
&\quad\times\left\{\frac1n\sum_{j:Z_j=z}
                    \{K_{M,z}^{+,a}(j)\}^2\right\}^{1/2}
 \left\{\frac1n\sum_{j:Z_j=z}|\varepsilon_{z,j}^A|^p\right\}^{1/p}.
\end{aligned}
\]
At most \(2M\) donor rows occur per changed match set. The uniform score-expansion remainder is controlled by the same gap bound. Since \(1/2-1/p>0\), the moment bounds and tightness of the fitted local index give, as \(n\to\infty\), then \(\delta\downarrow0\), and finally the index radius tends to infinity,
\begin{equation}\label{eq:primary-actual-residual-query-lln}
\frac1n\sum_{i:Z_i=1-z}w_{A,1-z}(\mathbf X_i)
 \frac{\sum_{j\in\mathcal J_{\widehat{\mathbf D}_z}(i)}
       w_{A,z}(\mathbf X_j)\varepsilon_{z,j}^A}
      {\sum_{j\in\mathcal J_{\widehat{\mathbf D}_z}(i)}
       w_{A,z}(\mathbf X_j)}
\longrightarrow_p0.
\end{equation}

Reindexing this query sum gives \(n^{-1}\sum_{j:Z_j=z}K_{n,z}^{A} (j;\widehat{\mathbf D}_z)\varepsilon_{z,j}^A\). The direct weighted residual averages converge to zero by the iid law of large numbers and centering conditional on the score and weight. Thus
\[
\begin{aligned}
E_{n,A}^{D,\mathrm{PATE}}
&=\frac1{n\bar w_{A,n}}\sum_{z=0}^1(2z-1)
 \sum_{j:Z_j=z}(w_{A,z}+K_{n,z}^{A})
                   \varepsilon_{z,j}^A\longrightarrow_p0,\\
E_{n,A}^{D,\mathrm{PATT}}
&=\frac1{n\bar w_{A,1,n}}\left[
 \sum_{i:Z_i=1}w_{A,1}\varepsilon_{1\mid1,i}
 -\sum_{j:Z_j=0}K_{n,0}^{A,\mathrm{PATT}}
                   \varepsilon_{0,j}^A\right]\longrightarrow_p0.
\end{aligned}
\]
These coefficients are evaluated on the fitted graphs. The total and treated weight denominators converge to positive limits.

Population recovery and identification center the heterogeneity numerators. Conditional Jensen and bounded weights give integrability within the observed arm. The query upper and donor lower density bounds transfer reduced-mean moments to the opposite arm. The treated PATT conditional mean is integrable under its own law. The iid LLNs consequently give \(H_{n,A}^{D,\mathcal T}\to_p0\) for both estimands. Finally, on the fitted graph,
\[
\widehat\tau_{\mathcal T}^{A,\mathrm{BC}}-\tau_{\mathcal T}
=H_{n,A}^{D,\mathcal T}+E_{n,A}^{D,\mathcal T}
+B_{n,A}^{D,\mathcal T}-\widehat B_{n,A}^{D,\mathcal T}.
\]
The bias-correction condition on the fitted graph makes the last difference \(o_p(1)\), proving consistency.
\end{proof}

\subsubsection{Estimated-score limits by discretization}\label{estimated-score-exact-wdsm-via-le-cams-third-lemma-and-discretization}
\label{joint-local-gaussian-limit-for-the-known-score-exact-wdsm-nuisance-estimator-and-likelihood-ratio}

\begin{proposition}[Discretized-score limit]
\label{prop:local-mean-corrected-grid-variance}
\label{prop:joint-local-gaussian-with-bias}
Assume all conditions of Lemma~\ref{lem:auxiliary-godambe-experiment} and use its auxiliary laws \(Q_{A,t}\), with \(Q_{A,0}=P_s\).

For every deterministic sequence \(\mathbf h_n\to \mathbf h\), put \(\boldsymbol\theta_{A,n}=\boldsymbol\theta_A^\ast+\mathbf h_n/\sqrt n\). Suppose that, under \(Q_{A,\mathbf h_n/\sqrt n}^{\otimes n}\),
\begin{equation}\label{eq:moving-law-joint-input-with-bias}
\begin{pmatrix}
\sqrt n\{\widehat\tau_A^{\mathrm{BC}}(\boldsymbol\theta_{A,n})
              -\tau_A(P_s)\}\\
\sqrt n\,\mathbb U_{A,n}(\boldsymbol\theta_{A,n})
\end{pmatrix}
\Rightarrow
N\left[
\begin{pmatrix}\widetilde{\boldsymbol \gamma}_{A,2}^\top \mathbf h\\0\end{pmatrix},
\begin{pmatrix}
V_{A,\mathrm{known-score}}&\boldsymbol\gamma_{A,1}^\top\\
\boldsymbol\gamma_{A,1}&\boldsymbol\Sigma_{A,U}
\end{pmatrix}
\right].
\end{equation}
Here \(\widetilde{\boldsymbol \gamma}_{A,2}\) is the coefficient of the local limiting mean when the statistic is centered at the fixed target \(\tau_A(P_s)\).

The limiting constants are common to all such sequences.

Under \(P_s^{\otimes n}\), for every \(\mathbf h_n\to \mathbf h\),
\begin{equation}\label{eq:fixed-law-local-shift-with-bias}
\begin{pmatrix}
\sqrt n\{\widehat\tau_A^{\mathrm{BC}}(\boldsymbol\theta_A^\ast+\mathbf h_n/\sqrt n)
                  -\tau_A(P_s)\}\\
\sqrt n(\widehat{\boldsymbol \theta}_A-\boldsymbol\theta_A^\ast)
\end{pmatrix}
\Rightarrow
\begin{pmatrix}
G_{\tau,A}+
\{\widetilde{\boldsymbol \gamma}_{A,2}
       +\boldsymbol\Gamma_A^\top\boldsymbol\Sigma_{A,U}^{-1}\boldsymbol\gamma_{A,1}\}^\top \mathbf h\\
\mathbf G_{\theta,A}
\end{pmatrix},
\end{equation}
where \((G_{\tau,A},\mathbf G_{\theta,A})\) is centered Gaussian with covariance
\[
\begin{pmatrix}
V_{A,\mathrm{known-score}}&-\boldsymbol\gamma_{A,1}^\top\boldsymbol\Gamma_A^{-T}\\
-\boldsymbol\Gamma_A^{-1}\boldsymbol\gamma_{A,1}&\boldsymbol\Sigma_{A,\theta}
\end{pmatrix}.
\]
Let \(\bar{\boldsymbol \theta}_{A,n,\eta}\) be the midpoint of the side-\(\eta/\sqrt n\) grid cell containing \(\widehat{\boldsymbol \theta}_A\), with deterministic grid offset. Define the Gaussian projection residual variance
\[
V_{A,\mathrm{proj}}
:=V_{A,\mathrm{known-score}}
       -\boldsymbol\gamma_{A,1}^\top\boldsymbol\Sigma_{A,U}^{-1}\boldsymbol\gamma_{A,1}
\]
and
\begin{equation}\label{eq:local-mean-corrected-grid-variance}
V_{A,\mathrm{grid}}
:=V_{A,\mathrm{proj}}
 +\widetilde{\boldsymbol \gamma}_{A,2}^\top
       \boldsymbol\Sigma_{A,\theta}\widetilde{\boldsymbol \gamma}_{A,2}.
\end{equation}
For every bounded Lipschitz function \(\varphi\),
\[
\begin{aligned}
\lim_{\eta\downarrow0}\limsup_{n\to\infty}
\Big|&
E_s\!\left[
\varphi\!\left(\sqrt n\{
\widehat\tau_A^{\mathrm{BC}}(\bar{\boldsymbol \theta}_{A,n,\eta})
-\tau_A(P_s)\}\right)\right]\\
&-E\{\varphi(N_A)\}\Big|=0,
\qquad N_A\sim N(0,V_{A,\mathrm{grid}}).
\end{aligned}
\]
Standardization requires \(V_{A,\mathrm{grid}}>0\).
\end{proposition}
\begin{proof}
\par\smallskip\noindent\emph{(i) Joint local experiment.}\quad
We apply the local-likelihood argument of Andreou and Werker,\citep{AndreouWerker2012} with a possibly nonzero local mean as in their Remark~B1. By Lemma~\ref{lem:auxiliary-godambe-experiment} and Slutsky's theorem, the triple consisting of the first coordinate in \eqref{eq:moving-law-joint-input-with-bias}, \(\sqrt n(\widehat{\boldsymbol \theta}_A-\boldsymbol\theta_{A,n})\), and \(\Lambda_{A,n}^{\mathrm{aux}}(\boldsymbol\theta_A^\ast\mid\boldsymbol\theta_{A,n})\) converges under these local laws to a Gaussian vector with mean
\[
\begin{pmatrix}
\widetilde{\boldsymbol \gamma}_{A,2}^\top \mathbf h\\0\\
-\tfrac12\mathbf h^\top\boldsymbol\Gamma_A^\top\boldsymbol\Sigma_{A,U}^{-1}\boldsymbol\Gamma_A\mathbf h
\end{pmatrix}
\]
and covariance matrix
\[
\begin{pmatrix}
V_{A,\mathrm{known-score}}
 &-\boldsymbol\gamma_{A,1}^\top\boldsymbol\Gamma_A^{-T}
 &\boldsymbol\gamma_{A,1}^\top\boldsymbol\Sigma_{A,U}^{-1}\boldsymbol\Gamma_A\mathbf h\\
-\boldsymbol\Gamma_A^{-1}\boldsymbol\gamma_{A,1}
 &\boldsymbol\Sigma_{A,\theta}&-\mathbf h\\
 \mathbf h^\top\boldsymbol\Gamma_A^\top\boldsymbol\Sigma_{A,U}^{-1}\boldsymbol\gamma_{A,1}
 &-\mathbf h^\top&\mathbf h^\top\boldsymbol\Gamma_A^\top\boldsymbol\Sigma_{A,U}^{-1}\boldsymbol\Gamma_A\mathbf h
\end{pmatrix}.
\]
The mean and covariance follow from the affine transformation
\[
(x,v)\longmapsto
\left(x,-\boldsymbol\Gamma_A^{-1}v,
\mathbf h^\top\boldsymbol\Gamma_A^\top\boldsymbol\Sigma_{A,U}^{-1}v-\tfrac12\mathbf h^\top \mathcal I_{A,G}\mathbf h\right).
\]
\par\smallskip\noindent\emph{(ii) Change of law and discretization.}\quad
Le Cam's third lemma, applied to the centered triple through \eqref{lem:gaussian-likelihood-shift}, shifts the statistic and nuisance means by \(\boldsymbol\gamma_{A,1}^\top\boldsymbol\Sigma_{A,U}^{-1}\boldsymbol\Gamma_A\mathbf h\) and \(-\mathbf h\). Restoring \(\widetilde{\boldsymbol \gamma}_{A,2}^\top \mathbf h\) and adding \(\mathbf h\) to the limiting nuisance coordinate gives \eqref{eq:fixed-law-local-shift-with-bias}.

The sandwich covariance satisfies
\begin{equation}\label{eq:patt-joint-projection-matrix-product}
\begin{aligned}
\boldsymbol\Sigma_{A,\theta}\boldsymbol\Gamma_A^\top\boldsymbol\Sigma_{A,U}^{-1}\boldsymbol\gamma_{A,1}
&=\boldsymbol\Gamma_A^{-1}\boldsymbol\Sigma_{A,U}\boldsymbol\Gamma_A^{-T}
  \boldsymbol\Gamma_A^\top\boldsymbol\Sigma_{A,U}^{-1}\boldsymbol\gamma_{A,1}\\
&=\boldsymbol\Gamma_A^{-1}\boldsymbol\Sigma_{A,U}\boldsymbol\Sigma_{A,U}^{-1}\boldsymbol\gamma_{A,1}\\
&=\boldsymbol\Gamma_A^{-1}\boldsymbol\gamma_{A,1}.
\end{aligned}
\end{equation}
Consequently,
\begin{equation}\label{eq:patt-joint-projection-quadratic-product}
\begin{aligned}
&(\boldsymbol\Gamma_A^\top\boldsymbol\Sigma_{A,U}^{-1}\boldsymbol\gamma_{A,1})^\top
 \boldsymbol\Sigma_{A,\theta}(\boldsymbol\Gamma_A^\top\boldsymbol\Sigma_{A,U}^{-1}\boldsymbol\gamma_{A,1})\\
&=(\boldsymbol\Gamma_A^\top\boldsymbol\Sigma_{A,U}^{-1}\boldsymbol\gamma_{A,1})^\top
 \boldsymbol\Gamma_A^{-1}\boldsymbol\gamma_{A,1}\\
&=\boldsymbol\gamma_{A,1}^\top\boldsymbol\Sigma_{A,U}^{-1}
  \boldsymbol\Gamma_A\boldsymbol\Gamma_A^{-1}\boldsymbol\gamma_{A,1}\\
&=\boldsymbol\gamma_{A,1}^\top\boldsymbol\Sigma_{A,U}^{-1}\boldsymbol\gamma_{A,1}.
\end{aligned}
\end{equation}
For the centered Gaussian pair in \eqref{eq:fixed-law-local-shift-with-bias}, Gaussian projection therefore gives
\[
G_{\tau,A}^\perp=G_{\tau,A}+\boldsymbol\gamma_{A,1}^\top\boldsymbol\Sigma_{A,U}^{-1}\boldsymbol\Gamma_A\mathbf G_{\theta,A},
\qquad G_{\tau,A}^\perp\perp \mathbf G_{\theta,A},\qquad
\operatorname{Var}(G_{\tau,A}^\perp)=V_{A,\mathrm{proj}}\ge0.
\]
The nonsingular nuisance covariance gives \(\mathbf G_{\theta,A}\) a density. Lemma~\ref{lem:joint-local-grid-selection} therefore yields the joint grid limit \(G_{\tau,A}^\perp+\widetilde{\boldsymbol \gamma}_{A,2}^\top \mathbf G_{\theta,A}\). Its variance is \eqref{eq:local-mean-corrected-grid-variance}. The likelihood argument gives the fixed-law limit, and joint discretization permits substitution of the discretized parameter estimate. Gaussian projection expresses the variance of this limit.
\end{proof}

\subsubsection{Continuity of the local matching process}

The discretization argument above gives the limit on a local grid. To compare the fitted score with its grid approximation, we control changes in the weighted imputation numerator as the coefficients and scales vary together. The next two propositions establish this continuity, first for a Poisson sample and then for a fixed sample size. Their proofs appear in Section~\ref{app:local-matching-process-proofs}.

The proofs first restrict candidate donors to enlarged neighborhoods in the rescaled score space. Within each neighborhood, a small parameter change can alter the selected matches only near a distance tie. Combining these tie bounds with moment bounds for the centered sum, while retaining the normalized survey weights, gives the required uniform comparison after removing the neighborhood restriction. The grid comparison uses the stated iterated limit.

For PATT, fix \(A\in\{R,P\}\). Use the common control score \(\mathbf D_0^\ast\), the residual reference
\[
\zeta_A=Y-m_0^A\{\mathbf D_0^\ast(\mathbf X)\},
\]
and observed weights \(w_{A,Z}\), where \(w_{R,z}=w_z\) and \(w_{P,z}=w\). Let \(p_A=\dim\boldsymbol\theta_A\), and let \(K\subset\mathbb R^{p_A+2}\) be a fixed compact box. The points of \(K\) are the pairs \((\mathbf h,l)\), with \(\mathbf h\in\mathbb R^{p_A}\) and \(l=(l_1,l_2)^\top\in\mathbb R^2\). Define the deterministic local score family
\begin{equation}
\label{eq:patt-local-process-score-family}
\mathcal D_{n,\mathbf h,l}(\mathbf X)
=
\operatorname{diag}\left(1+\frac{l_1}{\sqrt n},
                       1+\frac{l_2}{\sqrt n}\right)
\mathbf D_{0,\boldsymbol\theta_A^\ast+\mathbf h/\sqrt n}^{\,p}(\mathbf X).
\end{equation}
Here \(\mathbf D_{0,\boldsymbol\theta}^{\,p}\) is the specified deterministic population-normalized pair of the probability \(e(\mathbf X;\boldsymbol\alpha)\) and the raw control prognostic score \(\mathbf X^\top\boldsymbol\beta_0\). Its baseline value is \(\mathbf D_0^\ast\).

For \(m\) observations, let \(\mathcal J_{\mathcal D_{n,\mathbf h,l}}^{(m)}(i)\) denote the set of \(M\) matched controls for individual \(i\). The local parameter and scale perturbations remain divided by \(\sqrt n\), even when \(m\ne n\). On \(m_0:=\sum_{i=1}^m(1-Z_i)\ge M\), define the control-imputation numerator by
\begin{equation}
\label{eq:patt-local-process-imputation-numerator}
\begin{aligned}
\mathcal I_{A,m;n}(\mathbf h,l)
&:=\sum_{\substack{i\le m\\Z_i=1}}w_{A,1}(\mathbf X_i)
 \sum_{j\in\mathcal J_{\mathcal D_{n,\mathbf h,l}}^{(m)}(i)}
 \frac{w_{A,0}(\mathbf X_j)\zeta_{A,j}}
 {\sum_{k\in\mathcal J_{\mathcal D_{n,\mathbf h,l}}^{(m)}(i)}
                       w_{A,0}(\mathbf X_k)}\\
&=\sum_{\substack{i\le m\\Z_i=1}}
 \sum_{\substack{j\le m\\Z_j=0}}
 \frac{w_{A,1}(\mathbf X_i)w_{A,0}(\mathbf X_j)
       \mathbf1\{j\in\mathcal J_{\mathcal D_{n,\mathbf h,l}}^{(m)}(i)\}}
 {\sum_{k\in\mathcal J_{\mathcal D_{n,\mathbf h,l}}^{(m)}(i)}
                       w_{A,0}(\mathbf X_k)}
 \zeta_{A,j}\\
&=\sum_{\substack{j\le m\\Z_j=0}}
 \left\{\sum_{\substack{i\le m\\Z_i=1}}
 \frac{w_{A,1}(\mathbf X_i)w_{A,0}(\mathbf X_j)
       \mathbf1\{j\in\mathcal J_{\mathcal D_{n,\mathbf h,l}}^{(m)}(i)\}}
 {\sum_{k\in\mathcal J_{\mathcal D_{n,\mathbf h,l}}^{(m)}(i)}
                       w_{A,0}(\mathbf X_k)}\right\}
 \zeta_{A,j}\\
&=\sum_{\substack{j\le m\\Z_j=0}}
 K_{m,0}^{A}(j;\mathcal D_{n,\mathbf h,l})\zeta_{A,j}.
\end{aligned}
\end{equation}
Set the numerator and coefficients to zero when \(m_0<M\). The case \(m=n\) gives the fixed-size numerator.

Let \(N_n\sim\operatorname{Poisson}(n)\), independently of an infinite iid sequence of observations. For \(m=N_n\), the observations form the Poisson process
\[
\mathcal P_n=\sum_{i=1}^{N_n}\delta_{O_i},
\qquad
E\mathcal P_n(dO)=nP_s(dO).
\]
Define
\begin{equation}
\label{eq:patt-poisson-centered-process}
\mathcal Z_{A,n}^{P}(\mathbf h,l)
:=
\frac{\mathcal I_{A,N_n;n}(\mathbf h,l)
      -E\mathcal I_{A,N_n;n}(\mathbf h,l)}{\sqrt n},
\end{equation}
where expectations include the count and the iid rows.

Here \(E^\ast\) denotes outer expectation.

\begin{proposition}[Stochastic equicontinuity under Poisson sampling]
\label{prop:patt-poisson-local-process-equicontinuity}
The population normalization is fixed at the baseline or continuously differentiable in \(\boldsymbol\theta\), with positive component scales.

Suppose the selected rows are iid with law \(P_s\) and both arms have positive probability. The baseline control score is supported on a fixed compact convex set \(S\subset\mathbb R^2\) with nonempty interior, and its density is bounded above and away from zero on \(S\). The baseline treated-score density is bounded above and supported on \(S\). Covariates are bounded, \(M\) is fixed, and
\[
0<w_{\min}\le w_{A,Z}(\mathbf X)\le w_{\max}<\infty,
\qquad
E_s(1+|\zeta_A|)^p<\infty
\quad\text{for some }p=2+\delta>2.
\]
Matching uses unweighted Euclidean distance on \(\mathcal D_{n,\mathbf h,l}\), with replacement, and a fixed measurable outcome-independent tie rule. The logistic/linear raw scores and the stated smooth population normalization give bounded local derivatives. In particular, for every \(K\), a finite \(C_K\) satisfies
\begin{equation}
\label{eq:patt-poisson-bounded-score-displacement}
\begin{aligned}
\sup_{(\mathbf h,l)\in K,\,\mathbf X}
 \|\mathcal D_{n,\mathbf h,l}(\mathbf X)-\mathbf D_0^\ast(\mathbf X)\|
&\le \frac{C_K}{\sqrt n},\\
\sup_{\mathbf X}
 \|\mathcal D_{n,\mathbf h,l}(\mathbf X)
       -\mathcal D_{n,\mathbf h',l'}(\mathbf X)\|
&\le \frac{C_K}{\sqrt n}\|(\mathbf h,l)-(\mathbf h',l')\|.
\end{aligned}
\end{equation}
Assume also the donor subdensity bound
\begin{equation}
\label{eq:patt-poisson-current-donor-density}
\sup_{n\ge n_K}\sup_{(\mathbf h,l)\in K}
\left\|
\frac{dP_s\{Z=0,\mathcal D_{n,\mathbf h,l}(\mathbf X)\in\cdot\}}{du}
\right\|_\infty<\infty.
\end{equation}

Then
\begin{equation}
\label{eq:patt-poisson-full-process-modulus}
\lim_{\eta\downarrow0}\limsup_{n\to\infty}
E^\ast
\sup_{\substack{(\mathbf h,l),(\mathbf h',l')\in K\\
                \|(\mathbf h,l)-(\mathbf h',l')\|\le\eta}}
\left|\mathcal Z_{A,n}^{P}(\mathbf h,l)
             -\mathcal Z_{A,n}^{P}(\mathbf h',l')\right|=0.
\end{equation}
\end{proposition}

Baseline control centering also makes the change in the numerator negligible when the Poisson sample size is replaced by its mean. This transfers the preceding continuity result to the fixed-size sample.

Define the fixed-size centered process by
\[
\mathcal Z^{B}_{A,n}(\mathbf h,l)
 :=\frac{\mathcal I_{A,n;n}(\mathbf h,l)
               -E\mathcal I_{A,n;n}(\mathbf h,l)}{\sqrt n}.
\]

\begin{proposition}[De-Poissonization of the local process]
\label{prop:patt-binomial-local-process-transfer}
Retain all the hypotheses of Proposition~\ref{prop:patt-poisson-local-process-equicontinuity}, the score family in \eqref{eq:patt-local-process-score-family}, and the normalized weighted numerator \(\mathcal I_{A,m;n}(\mathbf h,l)\) in \eqref{eq:patt-local-process-imputation-numerator}. The first index \(m\) is the number of observations. Every score perturbation remains scaled by \(\sqrt n\), also when \(m\ne n\). Assume baseline control centering:
\begin{equation}\label{eq:patt-binomial-baseline-control-centering}
E_s\{\zeta_A\mid Z=0,\mathbf D_0^\ast,w_{A,0}\}=0,
\qquad
\zeta_A=Y-m_0^A(\mathbf D_0^\ast).
\end{equation}

Let \(N_n\sim\operatorname{Poisson}(n)\) be independent of the entire iid row sequence. For every fixed compact local-index box \(K\subset\mathbb R^{p_A+2}\),
\begin{equation}\label{eq:patt-binomial-uniform-depoissonization}
E^*\sup_{(\mathbf h,l)\in K}
 \frac{|\mathcal I_{A,N_n;n}(\mathbf h,l)
             -\mathcal I_{A,n;n}(\mathbf h,l)|}{\sqrt n}
\longrightarrow0.
\end{equation}
Consequently,
\begin{equation}\label{eq:patt-binomial-expectation-transfer}
\sup_{(\mathbf h,l)\in K}
 \frac{|E\mathcal I_{A,N_n;n}(\mathbf h,l)
             -E\mathcal I_{A,n;n}(\mathbf h,l)|}{\sqrt n}
\longrightarrow0.
\end{equation}
The fixed-size centered process \(\mathcal Z^B_{A,n}\) satisfies
\begin{equation}\label{eq:patt-binomial-centered-equicontinuity}
\lim_{\eta\downarrow0}\limsup_{n\to\infty}
 E^*\sup_{\substack{(\mathbf h,l),(\mathbf h',l')\in K\\
                  \|(\mathbf h,l)-(\mathbf h',l')\|\le\eta}}
 \left|\mathcal Z^{B}_{A,n}(\mathbf h,l)
             -\mathcal Z^{B}_{A,n}(\mathbf h',l')\right|=0.
\end{equation}
The conclusions hold for the specified measurable, outcome-independent tie rule.
\end{proposition}

The proof is given in Section~\ref{app:local-matching-process-proofs}.

\subsubsection{PATT with estimated scores}

We combine the local limit, discretization, and process continuity to derive the PATT limit with estimated scores. Part (i) gives the local mean and variance on the grid. Part (ii) accounts for empirical standardization and replaces the grid parameter by the estimated parameter, jointly with the first-stage estimator.

\begin{theorem}[Local and fitted-score PATT limits]
\label{cor:patt-verified-local-grid-law}
\label{thm:patt-sufficient-branch-actual-score-law}
\label{lem:patt-empirical-local-process-index}
\label{lem:patt-local-process-mean}
\label{cor:patt-joint-actual-score-law}
For PATT, fix \(A\in\{R,P\}\).
\begin{enumerate}
\item[(i)] Retain the hypotheses of Proposition~\ref{prop:patt-signed-density-local-joint-law}, Lemma~\ref{lem:patt-auxiliary-scientific-target}, and Proposition~\ref{prop:patt-deterministic-local-empirical-standardization}. Use the estimator's normalization and the first-step conditions in Proposition~\ref{prop:local-mean-corrected-grid-variance}.

For every deterministic \(\mathbf h_n\to \mathbf h\), let \(\boldsymbol\theta_{A,n}=\boldsymbol\theta_A^\ast+\mathbf h_n/\sqrt n\) and \(\mathbf D_{0,n}^{e}\) be the empirically standardized probability-propensity and raw prognostic scores evaluated at these deterministic coefficients. The empirical moments use the fixed, field-measurable moment weight specified in the standardization result. For this family, assume the bias-correction remainder satisfies
\begin{equation}\label{eq:patt-local-grid-accepted-bc-remainder}
\sqrt n\left[
\widehat B_{n,A}^{D,\mathrm{PATT}}(\mathbf D_{0,n}^{e})
-B_{n,A}^{D,\mathrm{PATT}}
 (\mathbf D_{0,n}^{e};\mathbf D_0^\ast,m_0^A)
\right]\longrightarrow_{P_s^{\otimes n}\text{-prob.}}0.
\end{equation}
Both terms use the same empirical graph. Here \(B\) uses the baseline reduced mean and \(\widehat B\) uses the prescribed regression refit. The remainder is required for every deterministic local sequence.

Write \(\mathbf e_r\) for the \(r\)-th coordinate vector in \(\mathbb R^{p_A}\). The finite linear functional \(\Delta_A(\mathbf h)\) is the one derived in \eqref{eq:patt-drift-fixed-law-geometric-limit}. The resulting local response coefficient is
\begin{equation}\label{eq:patt-verified-local-mean-coefficient}
\widetilde{\boldsymbol \gamma}_{A,2}
=-\boldsymbol\Gamma_A^\top\boldsymbol\Sigma_{A,U}^{-1}\boldsymbol\gamma_{A,1}
 -\left(\frac{\Delta_A(\mathbf e_r)}{\bar w_A}\right)_{r=1}^{p_A},
\end{equation}
where the parenthesized array is a column vector. Then the moving-law joint limit \eqref{eq:moving-law-joint-input-with-bias} holds for this PATT family, with its stated baseline Gaussian covariance and the coefficient in \eqref{eq:patt-verified-local-mean-coefficient}.

Consequently the double-limit grid conclusion of Proposition~\ref{prop:local-mean-corrected-grid-variance} applies to this same family. Its variance has the nonnegative form
\begin{equation}\label{eq:patt-verified-geometric-grid-variance}
\begin{aligned}
V_{A,\mathrm{grid}}
&=V_{A,\mathrm{proj}}\\
&\quad{}+
 \left\{-\boldsymbol\Gamma_A^\top\boldsymbol\Sigma_{A,U}^{-1}\boldsymbol\gamma_{A,1}
 -\left(\frac{\Delta_A(\mathbf e_r)}{\bar w_A}\right)_{r=1}^{p_A}\right\}^{\!\top}
 \boldsymbol\Sigma_{A,\theta}\\
&\qquad\quad{}\times
 \left\{-\boldsymbol\Gamma_A^\top\boldsymbol\Sigma_{A,U}^{-1}\boldsymbol\gamma_{A,1}
 -\left(\frac{\Delta_A(\mathbf e_r)}{\bar w_A}\right)_{r=1}^{p_A}\right\}.
\end{aligned}
\end{equation}
The variance is finite and is positive if and only if \(V_{A,\mathrm{proj}}>0\) or the displayed vector is nonzero.

\item[(ii)] Retain the hypotheses of part (i) and Proposition~\ref{prop:patt-binomial-local-process-transfer}, except that the tie-measurability condition of Proposition~\ref{prop:patt-deterministic-local-empirical-standardization} is imposed only on comparison graphs with fixed priority based on row labels. The estimator uses its specified outcome-independent tie rule. Use the same normalization, baseline means, fixed-\(M\) survey-normalized graph and prescribed BC fit throughout. In addition to the deterministic-local-family remainder in \eqref{eq:patt-local-grid-accepted-bc-remainder}, assume the corresponding remainder on the fitted graph satisfies
\begin{equation}
\label{eq:patt-actual-process-accepted-bc}
\sqrt n\left[
\widehat B_{n,A}^{D,\mathrm{PATT}}(\widehat{\mathbf D}_0)
-B_{n,A}^{D,\mathrm{PATT}}
(\widehat{\mathbf D}_0;\mathbf D_0^\ast,m_0^A)
\right]\longrightarrow_p0.
\end{equation}

Then the estimator evaluated at the fitted score root satisfies
\begin{equation}
\label{eq:patt-sufficient-branch-actual-score-law}
\sqrt n\{
\widehat\tau_A^{\mathrm{BC}}(\widehat{\boldsymbol \theta}_A)-\tau_A
\}
\Rightarrow N(0,V_{A,\mathrm{grid}}),
\end{equation}
where \(V_{A,\mathrm{grid}}\) is the finite nonnegative variance in \eqref{eq:patt-verified-geometric-grid-variance}. Its positivity criterion is given in part (i). Let \(p_A=\dim\boldsymbol\theta_A\), and let \(\mathbf e_r\) denote its \(r\)-th coordinate vector. Define the column vector
\begin{equation}
\label{eq:patt-joint-geometric-coefficient}
\mathbf j_A:=\left(\frac{\Delta_A(\mathbf e_r)}{\bar w_A}\right)_{r=1}^{p_A},
\qquad
\frac{\Delta_A(\mathbf h)}{\bar w_A}=\mathbf j_A^\top \mathbf h.
\end{equation}

Write the fitted and grid parameters in local coordinates as
\[
\begin{aligned}
\widehat{\mathbf h}_n
&=\sqrt n(\widehat{\boldsymbol \theta}_A-\boldsymbol\theta_A^\ast),\\
\bar{\mathbf h}_{n,\eta}
&=\sqrt n(\bar{\boldsymbol \theta}_{A,n,\eta}-\boldsymbol\theta_A^\ast).
\end{aligned}
\]
Then
\begin{equation}
\label{eq:patt-joint-actual-score-law}
\begin{pmatrix}
\sqrt n\{\widehat\tau_A^{\mathrm{BC}}(\widehat{\boldsymbol \theta}_A)-\tau_A\}\\
\sqrt n(\widehat{\boldsymbol \theta}_A-\boldsymbol\theta_A^\ast)
\end{pmatrix}
\Rightarrow
\begin{pmatrix}
G_{\tau,A}-\mathbf j_A^\top \mathbf G_{\theta,A}\\
\mathbf G_{\theta,A}
\end{pmatrix},
\end{equation}
where the centered Gaussian pair \((G_{\tau,A},\mathbf G_{\theta,A})\) is the one in \eqref{eq:fixed-law-local-shift-with-bias}, with covariance
\begin{equation}
\label{eq:patt-joint-baseline-gaussian-covariance}
\operatorname{Var}
\begin{pmatrix}G_{\tau,A}\\\mathbf G_{\theta,A}\end{pmatrix}
=
\begin{pmatrix}
V_{A,\mathrm{known-score}}
 &-\boldsymbol\gamma_{A,1}^\top\boldsymbol\Gamma_A^{-T}\\
-\boldsymbol\Gamma_A^{-1}\boldsymbol\gamma_{A,1}
 &\boldsymbol\Sigma_{A,\theta}
\end{pmatrix}.
\end{equation}
In particular, the variance and covariance column of the limiting pair satisfy
\begin{equation}
\label{eq:patt-joint-actual-limit-covariances}
\begin{aligned}
\operatorname{Var}(G_{\tau,A}-\mathbf j_A^\top \mathbf G_{\theta,A})
&=V_{A,\mathrm{known-score}}
 -\boldsymbol\gamma_{A,1}^\top\boldsymbol\Sigma_{A,U}^{-1}\boldsymbol\gamma_{A,1}\\
&\quad{}+
 (-\boldsymbol\Gamma_A^\top\boldsymbol\Sigma_{A,U}^{-1}\boldsymbol\gamma_{A,1}-\mathbf j_A)^\top
 \boldsymbol\Sigma_{A,\theta}
 (-\boldsymbol\Gamma_A^\top\boldsymbol\Sigma_{A,U}^{-1}\boldsymbol\gamma_{A,1}-\mathbf j_A)\\
&=V_{A,\mathrm{grid}},\\
\operatorname{Cov}(\mathbf G_{\theta,A},G_{\tau,A}-\mathbf j_A^\top \mathbf G_{\theta,A})
&=-\boldsymbol\Gamma_A^{-1}\boldsymbol\gamma_{A,1}-\boldsymbol\Sigma_{A,\theta}\mathbf j_A.
\end{aligned}
\end{equation}
\end{enumerate}
\end{theorem}
\begin{proof}
\par\smallskip\noindent\emph{Step 1: limits at deterministic local and grid parameters.}\quad
We first place the bias-corrected statistic under the local law. Contiguity transfers \eqref{eq:patt-local-grid-accepted-bc-remainder} to \(Q_n^{\otimes n}\), where \(Q_n=Q_{A,\mathbf h_n/\sqrt n}\). Removing the bias-correction remainder from the \(H/B/E\) identity and applying the empirical-normalization result give
\[
\begin{aligned}
\sqrt n\{\widehat\tau_A^{\mathrm{BC}}(\boldsymbol\theta_{A,n})-\tau_A(P_s)\}
&=T_{A,n}^{e}\{\tau_A(P_s)\}+o_p(1)\\
&=T_{A,n}^{p}\{\tau_A(P_s)\}+o_p(1).
\end{aligned}
\]
The population-normalized statistic has the established local joint law. Slutsky's theorem therefore gives
\begin{equation}\label{eq:patt-verified-moving-law-input}
\begin{pmatrix}
\sqrt n\{\widehat\tau_A^{\mathrm{BC}}(\boldsymbol\theta_{A,n})-\tau_A(P_s)\}\\
\sqrt n\,\mathbb U_{A,n}(\boldsymbol\theta_{A,n})
\end{pmatrix}
\Rightarrow
N\!\left[
\binom{\widetilde{\boldsymbol \gamma}_{A,2}^\top \mathbf h}{0},
\begin{pmatrix}
V_{A,\mathrm{known-score}}&\boldsymbol\gamma_{A,1}^\top\\
\boldsymbol\gamma_{A,1}&\boldsymbol\Sigma_{A,U}
\end{pmatrix}\right]
\end{equation}
for every deterministic \(\mathbf h_n\to \mathbf h\), with \(\widetilde{\boldsymbol \gamma}_{A,2}^\top \mathbf h =-\boldsymbol\gamma_{A,1}^\top\boldsymbol\Sigma_{A,U}^{-1}\boldsymbol\Gamma_A\mathbf h -\Delta_A(\mathbf h)/\bar w_A\). Linearity gives \eqref{eq:patt-verified-local-mean-coefficient}. Proposition~\ref{prop:local-mean-corrected-grid-variance} then yields \eqref{eq:patt-verified-geometric-grid-variance}. The nonnegative Schur complement and positive-definite \(\boldsymbol\Sigma_{A,\theta}\) give the positivity criterion. This proves part (i). To prove part (ii), we next express the empirical scales as local process indices and compare the fitted and grid values.
\par\smallskip\noindent\emph{Step 2: empirical normalization.}\quad
For arbitrary \(\boldsymbol\theta\) near \(\boldsymbol\theta_A^\ast\), let \(\widehat\ell_{n,0k}(\boldsymbol\theta)\) and \(\widehat s_{n,0k}(\boldsymbol\theta)\), \(k=1,2\), be the specified empirical moments evaluated at \(r_{0k}(\mathbf X;\boldsymbol\theta)\). Let \(s_{0k}^{\mathrm p}(\boldsymbol\theta)\) denote the corresponding deterministic scale used in the local process: either the fixed baseline scale, or the population scale at \(\boldsymbol\theta\). It is positive and continuously differentiable near \(\boldsymbol\theta_A^\ast\). Define
\begin{equation}
\label{eq:patt-actual-empirical-scale-index}
L_{n,k}(\mathbf h)
:=\sqrt n\left\{
\frac{s_{0k}^{\mathrm p}(\boldsymbol\theta_A^\ast+\mathbf h/\sqrt n)}
{\widehat s_{n,0k}(\boldsymbol\theta_A^\ast+\mathbf h/\sqrt n)}-1
\right\},
\qquad k=1,2.
\end{equation}
Then \(L_n(0)=O_p(1)\). For every fixed \(H<\infty\), there is \(C_H<\infty\) such that, with probability tending to one,
\begin{equation}
\label{eq:patt-actual-scale-index-lipschitz}
\|L_n(\mathbf h)-L_n(\mathbf h')\|\le C_H\|\mathbf h-\mathbf h'\|,
\qquad \|\mathbf h\|,\|\mathbf h'\|\le H.
\end{equation}
Consequently \(\sup_{\|\mathbf h\|\le H}\|L_n(\mathbf h)\|=O_p(1)\). On the valid moment event, the empirical matching distance array at \(\boldsymbol\theta_A^\ast+\mathbf h/\sqrt n\) equals that of \(\mathcal D_{n,\mathbf h,L_n(\mathbf h)}\). To establish these claims, Lemma~\ref{lem:primitive-score-standardization} first gives \(L_n(0)=O_p(1)\). Since \(E_sb_0>0\), bounded covariates and \(\operatorname{expit}'\le1/4\) give, on the valid moment event,
\[
\begin{aligned}
\nabla_{\boldsymbol \theta}\widehat\ell_{n,0k}(\boldsymbol\theta)
&=\frac{\sum_i b_0(O_i)\nabla_{\boldsymbol \theta} r_{0k}(\mathbf X_i;\boldsymbol\theta)}
              {\sum_i b_0(O_i)},\\
\nabla_{\boldsymbol \theta}\widehat s_{n,0k}(\boldsymbol\theta)^2
&=\frac{2\sum_i b_0(O_i)
       \{r_{0k}(\mathbf X_i;\boldsymbol\theta)-\widehat\ell_{n,0k}(\boldsymbol\theta)\}
                   \nabla_{\boldsymbol \theta} r_{0k}(\mathbf X_i;\boldsymbol\theta)}
              {\sum_i b_0(O_i)},\\
\sup_{\boldsymbol\theta\text{ near }\boldsymbol\theta_A^\ast,\ k=1,2}
 \|\nabla_{\boldsymbol \theta}\widehat s_{n,0k}(\boldsymbol\theta)^2\|&\le C.
\end{aligned}
\]
The baseline variances are positive. Consistency and the derivative bound therefore keep the empirical scales away from zero throughout the local neighborhood:
\begin{equation}
\label{eq:patt-actual-scale-positive-neighborhood}
P_s\!\left\{
\inf_{\|\mathbf h\|\le H,\ k=1,2}
\widehat s_{n,0k}(\boldsymbol\theta_A^\ast+\mathbf h/\sqrt n)>c_H
\right\}\longrightarrow1.
\end{equation}
On this event, square-root and quotient differentiation give
\[
\begin{aligned}
\nabla_hL_{n,k}(\mathbf h)
&=\left.\nabla_{\boldsymbol \theta}\{s_{0k}^{\mathrm p}(\boldsymbol\theta)/
                         \widehat s_{n,0k}(\boldsymbol\theta)\}
\right|_{\boldsymbol\theta=\boldsymbol\theta_A^\ast+\mathbf h/\sqrt n}\\
&=\left.\left\{
 \frac{\nabla_{\boldsymbol \theta} s_{0k}^{\mathrm p}(\boldsymbol\theta)}{\widehat s_{n,0k}(\boldsymbol\theta)}
 -\frac{s_{0k}^{\mathrm p}(\boldsymbol\theta)
          \nabla_{\boldsymbol \theta}\widehat s_{n,0k}(\boldsymbol\theta)^2}
              {2\widehat s_{n,0k}(\boldsymbol\theta)^3}\right\}
\right|_{\boldsymbol\theta=\boldsymbol\theta_A^\ast+\mathbf h/\sqrt n}\\
&=O_p(1)
\end{aligned}
\]
uniformly on the \(\mathbf h\)-ball. The mean-value theorem proves \eqref{eq:patt-actual-scale-index-lipschitz} and \(\sup_{\|\mathbf h\|\le H}\|L_n(\mathbf h)\|\le\|L_n(0)\|+C_HH=O_p(1)\).

Finally, empirical centers cancel in each pair distance, leaving only the scale ratio:
\[
D_{0,\boldsymbol\theta,k}^{\mathrm e}(\mathbf X_i)
-D_{0,\boldsymbol\theta,k}^{\mathrm e}(\mathbf X_j)
=\frac{s_{0k}^{\mathrm p}(\boldsymbol\theta)}{\widehat s_{n,0k}(\boldsymbol\theta)}
 \{D_{0,\boldsymbol\theta,k}^{\mathrm p}(\mathbf X_i)
                  -D_{0,\boldsymbol\theta,k}^{\mathrm p}(\mathbf X_j)\}.
\]
At \(\boldsymbol\theta=\boldsymbol\theta_A^\ast+\mathbf h/\sqrt n\), this is the distance array of \(\mathcal D_{n,\mathbf h,L_n(\mathbf h)}\). Under the hypotheses of Proposition~\ref{prop:patt-binomial-local-process-transfer}, \eqref{eq:patt-binomial-tie-rule-transfer} and tightness of \(L_n\) give an \(o_p(1)\) difference between the corresponding \(n^{-1/2}\)-normalized numerators for the specified tie rule.
\par\smallskip\noindent\emph{Step 3: local mean.}\quad
We next control the mean of the numerator, so that centered process continuity also controls the uncentered increments. Let \(\mathcal D_{n,\mathbf h,l}\) and \(\mathcal I_{A,m;n}(\mathbf h,l)\) be the deterministic score family and the incoming numerator in \eqref{eq:patt-local-process-score-family} and \eqref{eq:patt-local-process-imputation-numerator}. For every fixed compact box \(K\subset\mathbb R^{p_A+2}\),
\begin{equation}
\label{eq:patt-local-process-uniform-mean}
\sup_{(\mathbf h,l)\in K}
\left|
\frac{E_s\{\mathcal I_{A,n;n}(\mathbf h,l)\}}{\sqrt n}
-\Delta_A(\mathbf h)
\right|\longrightarrow0,
\end{equation}
where the finite linear functional \(\Delta_A(\mathbf h)\) is the one proved in \eqref{eq:patt-drift-fixed-law-geometric-limit}. In particular its limit does not depend on \(l\). To prove uniform mean convergence, fix a deterministic \((\mathbf h_n,l_n)\to(\mathbf h,l)\). Positive deterministic diagonal rescaling leaves the own-row score/arm/weight field unchanged, so \eqref{eq:patt-drift-fixed-law-root-expansion} gives
\begin{equation}
\label{eq:patt-local-process-mean-projection}
\eta_{n,A}^{P}
=n^{-1/2}\dot m^{\varepsilon}_{A,\mathbf h_n}(S_\ast,w_{A,0})+O(n^{-1})
\end{equation}
uniformly on valid control rows and bounded \(\mathbf h_n\), independently of \(l_n\). The normalized fractions satisfy
\begin{equation}\label{eq:patt-local-process-mean-mass}
\sum_{j:Z_j=0}K_{n,0}^{A}(j;\mathcal D_{n,\mathbf h_n,l_n})
=\mathbf1_{\{n_0\ge M\}}\sum_{i:Z_i=1}w_{A,1}(\mathbf X_i)
\le nw_{\max} .
\end{equation}
At a deterministic index, the current donor density makes ties null. Conditional independence permits projection of each residual:
\[
\frac{E_s\mathcal I_{A,n;n}(\mathbf h_n,l_n)}{\sqrt n}
=E_s\!\left[\frac1n\sum_{j:Z_j=0}
 K_{n,0}^{A}(j;\mathcal D_{n,\mathbf h_n,l_n})
 \dot m^{\varepsilon}_{A,\mathbf h_n}(S_{\ast,j},w_{A,0,j})\right]+O(n^{-1/2}).
\]
The inflated reuse moments justify the conditional expectation, and mass conservation bounds the remainder by \(Cw_{\max}/\sqrt n\).

The score displacement is
\[
\sqrt n\{\mathcal D_{n,\mathbf h_n,l_n}(\mathbf X)-\mathbf D_0^\ast(\mathbf X)\}
=S_0(\mathbf X)\mathbf h+\operatorname{diag}(l)\mathbf D_0^\ast(\mathbf X)+o(1)
\]
uniformly on the covariate support. In a bounded rescaled window about \(d\), the diagonal term is the common translation \(\operatorname{diag}(l)d+o(1)\). Proposition~\ref{prop:patt-deterministic-local-conditional-qv} therefore gives the one-root limit. The inflated-count envelope, bounded \(\dot m^{\varepsilon}_{A,\mathbf h}\) and linearity in \(\mathbf h\) yield
\[
\frac{E_s\mathcal I_{A,n;n}(\mathbf h_n,l_n)}{\sqrt n}
\longrightarrow
E_s[(1-Z)\kappa_{A,0}(\mathbf D_0^\ast,w_{A,0})
                 \dot m^{\varepsilon}_{A,\mathbf h}(S_\ast,w_{A,0})]
=\Delta_A(\mathbf h).
\]
Every sequence in the compact index set has a convergent subsequence. The preceding limit and continuity of \(\Delta_A(\mathbf h)\) therefore give \eqref{eq:patt-local-process-uniform-mean}. At \(\mathbf h=0\), baseline control centering gives \(E_s\mathcal I_{A,n;n}(0,l)=0\) for every deterministic \(l\).
\par\smallskip\noindent\emph{Step 4: comparison of estimated and grid parameters.}\quad
For fixed row-label priority, Proposition~\ref{prop:patt-deterministic-local-empirical-standardization} gives the deterministic-local \(H+E\) limit. Equation~\eqref{eq:patt-binomial-tie-rule-transfer} and the empirical-index bounds in Step 2 extend this limit to the estimator's tie rule on compact local families. Contiguity gives the same extension under \(Q_{A,\mathbf h_n/\sqrt n}^{\otimes n}\). Combining it with \eqref{eq:patt-local-grid-accepted-bc-remainder} on the estimator's graph gives \eqref{eq:patt-verified-moving-law-input} and the local and grid conclusions of part (i) for this rule.

\emph{Changes in the matching statistic.} Only the imputation numerator changes with the local parameter. For \(m=n\), the \(H+E\) statistic based on the common control score therefore satisfies
\begin{equation}
\label{eq:patt-poisson-hbe-increment-identity}
\begin{aligned}
&T_{A,n}(\mathbf h,l;\tau_n)-T_{A,n}(\mathbf h',l';\tau_n)\\
&=-\frac{\mathcal I_{A,n;n}(\mathbf h,l)
             -\mathcal I_{A,n;n}(\mathbf h',l')}
             {\sqrt n\,\bar w_{A,n}},
\qquad
\bar w_{A,n}=\frac1n\sum_{i=1}^nZ_iw_{A,1}(\mathbf X_i).
\end{aligned}
\end{equation}
Here \(T_{A,n}\) is defined by \eqref{eq:patt-local-qv-exact-pair} with baseline residual \(\zeta_A\) and common target sequence. The direct treated term and target cancel.

The first-step result gives \(\widehat{\mathbf h}_n=\sqrt n(\widehat{\boldsymbol \theta}_A-\boldsymbol\theta_A^\ast) =O_p(1)\). Its prescribed grid midpoint \(\bar{\mathbf h}_{n,\eta}\) satisfies \(\|\widehat{\mathbf h}_n-\bar{\mathbf h}_{n,\eta}\|\le\sqrt{p_A}\,\eta/2\), uniformly in the deterministic offset. On \(\|\widehat{\mathbf h}_n\|\le H\) and sufficiently small \(\eta\), Step 2 places \((\widehat{\mathbf h}_n,L_n(\widehat{\mathbf h}_n))\) and \((\bar{\mathbf h}_{n,\eta},L_n(\bar{\mathbf h}_{n,\eta}))\) in a compact box with arbitrarily high probability, at distance at most \(\sqrt{1+C_{H+1}^2}\sqrt{p_A}\,\eta/2\). On this compact box, stochastic equicontinuity in \eqref{eq:patt-binomial-centered-equicontinuity} and Markov's inequality control the centered numerator difference. Uniform mean convergence \eqref{eq:patt-local-process-uniform-mean} and linearity of \(\Delta_A(\mathbf h)\) bound its deterministic mean difference by \(o(1)+O(\eta)\). Removing compact truncation yields
\begin{equation}
\label{eq:patt-actual-grid-incoming-comparison}
\begin{aligned}
\lim_{\eta\downarrow0}\limsup_n P_s\!\left\{
\frac1{\sqrt n}\left|
\mathcal I_{A,n;n}(\widehat{\mathbf h}_n,L_n(\widehat{\mathbf h}_n))
\right.\right.\\[-2pt]
\left.\left.
-\mathcal I_{A,n;n}(\bar{\mathbf h}_{n,\eta},L_n(\bar{\mathbf h}_{n,\eta}))
\right|>\epsilon\right\}=0.
\end{aligned}
\end{equation}
The exact identity \eqref{eq:patt-poisson-hbe-increment-identity}, \(\bar w_{A,n}\to_p\bar w_A>0\), and the empirical-index distance and uniform tie comparisons give the same limit for the empirical \(H+E\) coordinate.

It remains to compare the bias-correction remainders. On the same empirical graph, with the baseline reduced mean, define
\begin{equation}
\label{eq:patt-actual-local-bc-remainder-definition}
\begin{aligned}
R_{A,n}(\mathbf h)
:=\sqrt n\big[
&\widehat B_{n,A}^{D,\mathrm{PATT}}
(\mathbf D_{0,\boldsymbol\theta_A^\ast+\mathbf h/\sqrt n}^{\mathrm e})\\
&-B_{n,A}^{D,\mathrm{PATT}}
(\mathbf D_{0,\boldsymbol\theta_A^\ast+\mathbf h/\sqrt n}^{\mathrm e};
 \mathbf D_0^\ast,m_0^A)
\big].
\end{aligned}
\end{equation}
If the following bound failed on a compact set, a sequence of indices where it fails would have a convergent subsequence, contradicting the deterministic-local BC condition. Thus
\begin{equation}
\label{eq:patt-actual-bc-uniform-pointwise-probability}
\sup_{\|\mathbf h\|\le H}P_s\{|R_{A,n}(\mathbf h)|>\epsilon\}\longrightarrow0.
\end{equation}
For each fixed \(\eta\), a bounded root event involves at most \(C_H(1+\eta^{-1})^{p_A}\) deterministic midpoints, uniformly in the offset. A finite union bound and root tightness imply \(R_{A,n}(\bar{\mathbf h}_{n,\eta})\to_p0\). Equation~\eqref{eq:patt-actual-process-accepted-bc} gives \(R_{A,n}(\widehat{\mathbf h}_n)\to_p0\). The \(H/B/E\) identity gives
\[
\sqrt n\{\widehat\tau_A^{\mathrm{BC}}
       (\boldsymbol\theta_A^\ast+\mathbf h/\sqrt n)-\tau_A\}
=T_{A,n}^{\mathrm e}(\mathbf h;\tau_A)-R_{A,n}(\mathbf h).
\]
Subtracting at the fitted and grid indices proves
\begin{equation}
\label{eq:patt-actual-bc-grid-comparison}
\lim_{\eta\downarrow0}\limsup_n
P_s\!\left\{\sqrt n\left|
\widehat\tau_A^{\mathrm{BC}}(\widehat{\boldsymbol \theta}_A)
-\widehat\tau_A^{\mathrm{BC}}(\bar{\boldsymbol \theta}_{A,n,\eta})
\right|>\epsilon\right\}=0.
\end{equation}

\par\smallskip\noindent\emph{Step 5: joint limit and covariance.}\quad
Equations~\eqref{eq:patt-verified-local-mean-coefficient} and \eqref{eq:patt-joint-geometric-coefficient} give
\[
\widetilde{\boldsymbol \gamma}_{A,2}+\boldsymbol\Gamma_A^\top\boldsymbol\Sigma_{A,U}^{-1}\boldsymbol\gamma_{A,1}=-\mathbf j_A.
\]
Consequently the empirical-score family in part (i) satisfies, for every deterministic \(\mathbf h_n\to \mathbf h\),
\begin{equation}
\label{eq:patt-joint-deterministic-local-pair}
\begin{pmatrix}
\sqrt n\{
\widehat\tau_A^{\mathrm{BC}}(\boldsymbol\theta_A^\ast+\mathbf h_n/\sqrt n)
-\tau_A\}\\
\widehat{\mathbf h}_n
\end{pmatrix}
\Rightarrow
\begin{pmatrix}
G_{\tau,A}-\mathbf j_A^\top \mathbf h\\
\mathbf G_{\theta,A}
\end{pmatrix}.
\end{equation}
Positive definite \(\boldsymbol\Sigma_{A,\theta}\) makes grid-cell boundaries null. For this proof, denote the centered bias-corrected statistic by
\[
T_{A,n}^{\mathrm{BC}}(\mathbf h)
:=\sqrt n\{\widehat\tau_A^{\mathrm{BC}}
                  (\boldsymbol\theta_A^\ast+\mathbf h/\sqrt n)-\tau_A\}.
\]
Lemma~\ref{lem:joint-local-grid-selection}, including its deterministic offset-subsequence argument, applies with coefficient \(-\mathbf j_A\) and gives, for every bounded Lipschitz \(\Psi\),
\begin{equation}
\label{eq:patt-joint-grid-selection-limit}
\begin{aligned}
\lim_{\eta\downarrow0}\limsup_{n\to\infty}
\Big|&E_s\{\Psi(T_{A,n}^{\mathrm{BC}}(\bar{\mathbf h}_{n,\eta}),\widehat{\mathbf h}_n)\}\\
&-E\{\Psi(G_{\tau,A}-\mathbf j_A^\top \mathbf G_{\theta,A},\mathbf G_{\theta,A})\}\Big|=0.
\end{aligned}
\end{equation}

Step 4 gives the actual-versus-grid comparison
\begin{equation}
\label{eq:patt-joint-actual-grid-probability}
\lim_{\eta\downarrow0}\limsup_{n\to\infty}
P_s\{|T_{A,n}^{\mathrm{BC}}(\widehat{\mathbf h}_n)
       -T_{A,n}^{\mathrm{BC}}(\bar{\mathbf h}_{n,\eta})|>\epsilon\}=0.
\end{equation}
The nuisance coordinate is identical in the fitted and grid pairs. The bounded-Lipschitz criterion, with \(n\to\infty\) and then \(\eta\downarrow0\), proves \eqref{eq:patt-joint-actual-score-law}. Projection onto its first coordinate gives \eqref{eq:patt-sufficient-branch-actual-score-law}.

Using \eqref{eq:patt-joint-projection-matrix-product}-- \eqref{eq:patt-joint-projection-quadratic-product}, we obtain
\[
\begin{aligned}
\operatorname{Var}(G_{\tau,A}-\mathbf j_A^\top \mathbf G_{\theta,A})
&=\operatorname{Var}(G_{\tau,A})-2\mathbf j_A^\top\operatorname{Cov}(\mathbf G_{\theta,A},G_{\tau,A})
  +\mathbf j_A^\top\operatorname{Var}(\mathbf G_{\theta,A})\mathbf j_A\\
&=V_{A,\mathrm{known-score}}+2\mathbf j_A^\top\boldsymbol\Gamma_A^{-1}\boldsymbol\gamma_{A,1}
  +\mathbf j_A^\top\boldsymbol\Sigma_{A,\theta}\mathbf j_A\\
&=V_{A,\mathrm{known-score}}
  -\boldsymbol\gamma_{A,1}^\top\boldsymbol\Sigma_{A,U}^{-1}\boldsymbol\gamma_{A,1}\\
&\quad{}+(-\boldsymbol\Gamma_A^\top\boldsymbol\Sigma_{A,U}^{-1}\boldsymbol\gamma_{A,1}-\mathbf j_A)^\top
  \boldsymbol\Sigma_{A,\theta}
  (-\boldsymbol\Gamma_A^\top\boldsymbol\Sigma_{A,U}^{-1}\boldsymbol\gamma_{A,1}-\mathbf j_A)\\
&=V_{A,\mathrm{grid}},
\end{aligned}
\]
where the last equality follows from \eqref{eq:patt-verified-geometric-grid-variance}. Bilinearity gives the covariance column in \eqref{eq:patt-joint-actual-limit-covariances}. The positivity criterion follows from the nonnegative projection variance and positive-definite \(\boldsymbol\Sigma_{A,\theta}\).
\end{proof}

\subsubsection{Local PATE joint limit}
\label{app:pate-two-arm-joint-martingale}

PATE combines the two arm-specific imputed means, which share the propensity score estimator. We derive their joint limit with the stacked first-stage estimator before taking the difference between arms. We fix the design \(A\in\{R,P\}\), the baseline selected law \(P_s\), the number of matches \(M\), and the score stack \(\boldsymbol\theta_A=(\alpha_A^\top,\beta_{A,0}^\top,\beta_{A,1}^\top)^\top\). Here \(p_A=\dim\boldsymbol\theta_A\). Throughout this subsection the design label \(A\) refers to the corresponding PATE nuisance stack. We write \(w_{R,z}=w_z\) and \(w_{P,z}=w\), holding these functions fixed under score perturbations. We assume the population identification of the reduced means and the corresponding survey-recovery identity in Lemma~\ref{lem:retrospective-population-recovery}(i)--(ii). The finite-sample decomposition is given by Proposition~\ref{prop:exact-pate-score-space-decomposition}. We also work under Assumptions~\ref{ass:residual-centering-survey-normalized}, \ref{ass:residual-qv-stabilization}, and \ref{ass:ordinary-weighted-m-estimation}, and the moment conditions of Lemma~\ref{lem:baseline-oracle-firststep-joint}. In particular, the selected observations are iid and both treatment probabilities are positive. For each matching direction, the baseline donor density is bounded above and away from zero on its compact convex two-dimensional support, and the density of target individuals is bounded above on the same support. The observed weights satisfy \(0<w_{\min}\le w_{A,Z}\le w_{\max}<\infty\). The residuals and first-stage estimating-function contributions below have a common finite \(2+\delta\) moment, for some \(\delta>0\).

We write \(\tau_{A,z}=E_\Omega Y(z)=E_\Omega m_z^A(\mathbf D_z^*)\) for the identified arm means, whose difference is the PATE target. Define the individual heterogeneity contributions at the limiting scores by

\[
h_{A,z}(O)
=w_{A,Z}(\mathbf X)
 \{m_z^A(\mathbf D_z^*(\mathbf X))-\tau_{A,z}\},\qquad z=0,1.
\]

Survey recovery gives \(E_sh_{A,z}=0\), and
\[
h_{A,1}-h_{A,0}
=w_{A,Z}\{\tau_A^D(\mathbf X)-\tau_{\mathrm{PATE}}\}.
\]

For any deterministic \(\mathbf h_n\to \mathbf h\), let \(\boldsymbol\theta_n=\boldsymbol\theta_A^*+\mathbf h_n/\sqrt n\), and initially use fixed baseline population normalization or continuously differentiable deterministic population centering and positive scales near the baseline. Both arms use the specified probability-score Euclidean metric. Let \(n_z=\sum_i\mathbf1(Z_i=z)\) and \(\mathcal A_n=\{n_0\ge M,n_1\ge M\}\). On this event, we define

\[
\begin{aligned}
K_{n,z}(j)
&=\sum_{i:Z_i=1-z}w_{A,1-z}(\mathbf X_i)
 \frac{w_{A,z}(\mathbf X_j)
 \mathbf1\{j\in\mathcal J_{\mathbf D_{z,\boldsymbol\theta_n}^{\mathrm p}}(i)\}}
 {\sum_{k\in\mathcal J_{\mathbf D_{z,\boldsymbol\theta_n}^{\mathrm p}}(i)}
                              w_{A,z}(\mathbf X_k)},\\
c_{n,z,j}
&=\mathbf1(Z_j=z)\mathbf1_{\mathcal A_n}\{w_{A,z}(\mathbf X_j)+K_{n,z}(j)\},\\
\varepsilon_{A,j}
&=Y_j-m_{Z_j}^A(\mathbf D_{Z_j}^*(\mathbf X_j)),\\
\bar w_{A,n}
&=n^{-1}\sum_jw_{A,Z_j}(\mathbf X_j),\qquad
\bar w_A=E_s w_{A,Z}(\mathbf X)>0.
\end{aligned}
\]

Here \(K_{n,z}(j)=K_{n,z}^{A}(j;\mathbf D_{z,\boldsymbol\theta_n}^{\mathrm p})\), and \(\varepsilon_{A,j}=\varepsilon_{z,j}^A\) when \(Z_j=z\). The superscript \(\mathrm p\) denotes the specified deterministic population normalization with baseline \(\mathbf D_z^*\). We set both arm-specific statistics and their reuse coefficients to zero outside \(\mathcal A_n\), leaving the estimating-function sum unchanged. The following identities hold on \(\mathcal A_n\), whose complement has exponentially small probability.

Separate the two armwise contributions as

\[
\mathcal T_{n,z}
=\frac{\mathbf1_{\mathcal A_n}}{\sqrt n\bar w_{A,n}}
 \sum_i\{h_{A,z}(O_i)+c_{n,z,i}\varepsilon_{A,i}\}.
\]

The finite-sample decomposition expresses the PATE \(\sqrt n(H+E)\) term as the difference between these two arm-specific statistics. At a deterministic local parameter \(\mathbf h\), it is
\begin{equation}\label{eq:pate-two-arm-joint-pate-coordinate}
\begin{aligned}
&T_{A,n}^{\mathrm p}(\mathbf h;\tau_{\mathrm{PATE}})\\
&:=\frac{\mathbf1_{\mathcal A_n}}{\sqrt n\bar w_{A,n}}
 \sum_iw_{A,Z_i}(\mathbf X_i)
       \{\tau_A^D(\mathbf X_i)-\tau_{\mathrm{PATE}}\}\\
&\quad+\frac{\mathbf1_{\mathcal A_n}}{\sqrt n\bar w_{A,n}}
 \sum_{z=0}^1(2z-1)\sum_{i:Z_i=z}
 \{w_{A,z}(\mathbf X_i)
   +K_{n,z}^{A}
     (i;\mathbf D_{z,\boldsymbol\theta_A^*+\mathbf h/\sqrt n}^{\mathrm p})\}
       \varepsilon_{z,i}^{A}.
\end{aligned}
\end{equation}
Thus \(\mathcal T_{n,1}-\mathcal T_{n,0} =T_{A,n}^{\mathrm p}(\mathbf h_n;\tau_{\mathrm{PATE}})\).

\Needspace{12\baselineskip}\paragraph{Conditional covariance.}

Let

\[
\begin{aligned}
\mathcal F_{n,i}
&=\sigma\{Z_i,\mathbf D_{0,\boldsymbol\theta_n}^{\mathrm p}(\mathbf X_i),
                 \mathbf D_{1,\boldsymbol\theta_n}^{\mathrm p}(\mathbf X_i),
                 w_{A,Z_i}(\mathbf X_i)\},\\
\mathcal F_n&=\bigvee_i\mathcal F_{n,i},\\
\mathcal F_{*,i}
&=\sigma\{Z_i,\mathbf D_0^*(\mathbf X_i),
                 \mathbf D_1^*(\mathbf X_i),w_{A,Z_i}(\mathbf X_i)\}.
\end{aligned}
\]

We use the fixed tie convention measurable with respect to the coefficient field. Under the density conditions below, deterministic-index distance ties are null, so all valid tie choices agree almost surely. Both coefficient arrays and the common denominator are \(\mathcal F_n\)-measurable. Conditional on this field, the match sets and weights are fixed. Because the field is generated by deterministic functions applied separately to each independent observation, the remaining observation-specific quantities are conditionally independent.

To compute the conditional covariance, define the following conditional moments for observation \(i\):

\[
\begin{aligned}
\bar h_{n,z,i}&=E_s(h_{A,z}\mid\mathcal F_{n,i}),&
\eta_{n,i}&=E_s(\varepsilon_A\mid\mathcal F_{n,i}),\\
u_{n,i}&=E_s(\mathbf U_A\mid\mathcal F_{n,i}),&
v_{n,i}&=\operatorname{Var}_s(\varepsilon_A\mid\mathcal F_{n,i}),\\
k_{n,i}&=\operatorname{Cov}_s(\mathbf U_A,\varepsilon_A\mid\mathcal F_{n,i}),&
R_{n,i}&=\operatorname{Cov}_s(\mathbf U_A\mid\mathcal F_{n,i}),\\
a_{n,zz',i}&=\operatorname{Cov}_s(h_{A,z},h_{A,z'}\mid\mathcal F_{n,i}),&
b_{n,z,i}&=\operatorname{Cov}_s(h_{A,z},\varepsilon_A\mid\mathcal F_{n,i}),\\
d_{n,z,i}&=\operatorname{Cov}_s(\mathbf U_A,h_{A,z}\mid\mathcal F_{n,i}).
\end{aligned}
\]

Here \(\mathbf U_A=\mathbf U_A(O;\boldsymbol\theta_A^*)\), and \(k,d\) are column vectors. The quantities \(v_*,k_*,R_*\) are defined in the same way using the limiting-score information \(\mathcal F_*\), and \(u_*=E_s(\mathbf U_A\mid\mathcal F_*)\). The nuisance coordinate uses \(\mathbf U_A\). The quantity \(\bar h_{n,z,i}\) is the conditional projection of \(h_{A,z}\) onto the current field.

\begin{theorem}[Local PATE joint limit]
\label{thm:pate-two-arm-population-joint-law}
\label{lem:pate-two-arm-joint-fiber}
\label{lem:pate-two-arm-joint-local-mean}
Suppose that the baseline conditions stated at the start of this subsection hold, together with the following conditions.
\begin{enumerate}
\item[(i)] Partition bounded covariates as \(\mathbf X=(V^\top,L^\top)^\top\), \(V\in\mathbb R^3\), placing the intercept in \(L\). Assume the raw three-coordinate score stack has the form

\[
S_{\boldsymbol \theta}(\mathbf X)
=\begin{pmatrix}\mathbf X^\top\alpha\\
                 \mathbf X^\top\beta_0\\
                 \mathbf X^\top\beta_1\end{pmatrix}
=A_{\boldsymbol \theta} V+B_{\boldsymbol \theta} L,
\qquad \det A_{\boldsymbol\theta_A^*}\ne0.
\]

Assume \(A_{\boldsymbol \theta},B_{\boldsymbol \theta}\) are \(C^2\) near the baseline and \(w_{A,z}=w_{A,z}(L)\). For each arm let \(H_z^w=\mathcal L_s(L,Y\mid Z=z,w_{A,z}=w)\). Require that \(V\mid L=l,Y=y,Z=z\) has a measurable density \(f_z(v\mid l,y)\), with zero extension finite and continuous at Lebesgue-almost every \(v\). At almost every baseline \((s,w)\), require a neighborhood of \((\boldsymbol\theta_A^*,s)\) and an \(H_z^w\)-integrable envelope for

\[
\sup_{(\boldsymbol\theta,s')\text{ in that neighborhood}}
 f_z\{A_{\boldsymbol \theta}^{-1}(s'-B_{\boldsymbol \theta} l)\mid l,y\}
 (1+|y|)^{2+\delta}.
\]

Assume Lipschitz baseline reduced means on their compact supports, with fixed Lipschitz extensions. The shared propensity coordinate, expit transformation and positive normalization imply \(\sigma(\mathbf D_{0,\boldsymbol\theta},\mathbf D_{1,\boldsymbol\theta})=\sigma(S_{\boldsymbol \theta})\).

\item[(ii)] For each arm \(z\), conditional on its observed weight \(w\), put \(\nu_{z,w}=\mathcal L_s(L\mid Z=z,w_{A,z}=w)\). Let \(p_{z,w}(s,l)\) be the conditional density of \(S_*\) given \((L=l,Z=z,w_{A,z}=w)\), with support indicators retained, and define

\[
\begin{aligned}
a_{z,w}(s,l)
&=p_{z,w}(s,l)
 E_s(\varepsilon_A\mid Z=z,S_*=s,L=l,w_{A,z}=w),\\
q_{z,w}(s)&=\int p_{z,w}(s,l)\,\nu_{z,w}(dl).
\end{aligned}
\]

Suppose there are a compact set \(J\subset\mathbb R^3\) and a bounded open set \(N\supset J\), with \(\operatorname{dist}(J,N^c)>0\), such that the following conditions hold for both arms, uniformly over almost every observed weight \(w\). For \(\nu_{z,w}\)-almost every \(l\), the signed density \(a_{z,w}(\cdot,l)\) has a \(C^2\) extension to \(\mathbb R^3\) supported in \(J\), and \(p_{z,w}(\cdot,l)\) is \(C^1\) on \(N\). For multi-indices \(\alpha\),
\begin{equation}\label{eq:pate-two-arm-joint-signed-density-bounds}
\begin{aligned}
\int\sup_{s\in\mathbb R^3}
 \sum_{|\alpha|\le2}|\partial^\alpha a_{z,w}(s,l)|
               \,\nu_{z,w}(dl)&\le C,\\
\int\sup_{s\in N}
 \{|p_{z,w}(s,l)|+\|\nabla p_{z,w}(s,l)\|\}
               \,\nu_{z,w}(dl)&\le C,\\
\inf_{s\in N}q_{z,w}(s)&\ge c>0.
\end{aligned}
\end{equation}

\end{enumerate}
Under condition (i), the conditional-moment limits in \eqref{eq:pate-two-arm-joint-projection-limits} hold. Under conditions (i) and (ii), the predictable residual means and their weighted sums satisfy \eqref{eq:pate-two-arm-joint-local-mean-expansion} and \eqref{eq:pate-two-arm-joint-drift-limit}, respectively. For every deterministic \(\mathbf h_n\to \mathbf h\), use the two matching graphs at \(\boldsymbol\theta_n=\boldsymbol\theta_A^*+\mathbf h_n/\sqrt n\), with the specified deterministic population normalization. Then the pair consisting of \(T_{A,n}^{\mathrm p}(\mathbf h_n;\tau_{\mathrm{PATE}})\) and \(\sqrt n\,\mathbb U_{A,n}(\boldsymbol\theta_A^*)=n^{-1/2}\sum_i\mathbf U_A(O_i;\boldsymbol\theta_A^*)\) has the joint Gaussian limit \eqref{eq:pate-two-arm-joint-population-pair-limit}, with mean and covariance defined in \eqref{eq:pate-two-arm-joint-pate-constants}.
\end{theorem}
\begin{proof}
\par\smallskip\noindent\emph{(i) Conditional projections.}\quad
We first show that conditioning on the perturbed score field approaches conditioning on the baseline field. This identifies the conditional moments needed in the martingale calculation. Write
\[
\Phi=(h_{A,1},h_{A,0},\varepsilon_A,\mathbf U_A^\top)^\top.
\]
The change-of-variables argument in Lemma~\ref{lem:patt-two-coordinate-conditional-law} applies with three raw coordinates, \(t=0\), and the joined mark \(\Phi\). For \(v_{\boldsymbol \theta}(s,l)=A_{\boldsymbol \theta}^{-1}(s-B_{\boldsymbol \theta} l)\), it gives
\begin{equation}\label{eq:pate-two-arm-joint-fiber-densities}
\begin{aligned}
r_{z,\boldsymbol\theta}(s,l,y)
&=\frac{f_z\{v_{\boldsymbol \theta}(s,l)\mid l,y\}}{|\det A_{\boldsymbol \theta}|},\\
q_{z,\boldsymbol\theta}(s,w)
&=\int r_{z,\boldsymbol\theta}(s,l,y)\,H_z^w(dl,dy).
\end{aligned}
\end{equation}
The conditional kernel is \(r_{z,\boldsymbol\theta}H_z^w/q_{z,\boldsymbol\theta}\). The envelope in condition (i) permits dominated convergence of this kernel. Conditional contraction then gives the moment convergence in Lemma~\ref{lem:patt-two-coordinate-conditional-law}: first moments converge in \(L^2\), and products in \(L^1\). Since \(h_{A,z}\in\mathcal F_*\) and the residual is centered on the joined field, subtraction of the conditional-mean products gives
\begin{equation}\label{eq:pate-two-arm-joint-projection-limits}
\begin{gathered}
\bar h_{n,z}\longrightarrow h_{A,z}\text{ in }L^2,
\quad u_n\longrightarrow u_*:=E_s(\mathbf U_A\mid\mathcal F_* )\text{ in }L^2,
\quad \eta_n\longrightarrow0\text{ in }L^2,\\
v_n\longrightarrow v_*,\quad k_n\longrightarrow k_*,
\quad R_n\longrightarrow R_*\qquad\text{in }L^1,\\
a_{n,zz'}\longrightarrow0,\quad b_{n,z}\longrightarrow0,
\quad d_{n,z}\longrightarrow0\qquad\text{in }L^1.
\end{gathered}
\end{equation}
\par\smallskip\noindent\emph{(ii) Martingale array and conditional variance.}\quad
Let \(\mathcal X_i=\sigma(Z_i,\mathbf X_i)\), \(\mathcal X_{1:n}=\bigvee_i\mathcal X_i\), and write
\[
\eta_{A,i}^{X}=E_s(\varepsilon_{A,i}\mid\mathcal X_i),
\qquad
u_{A,i}^{X}=E_s(U_{A,i}\mid\mathcal X_i).
\]
Using the population normalization \(\bar w_A\), define the centered contribution of observation \(i\) as

\begin{equation}\label{eq:pate-two-arm-joint-centered-row}
D_{n,i}=\frac1{\sqrt n}
\begin{pmatrix}
\{h_{A,1,i}-\bar h_{n,1,i}
                   +c_{n,1,i}(\varepsilon_{A,i}-\eta_{n,i})\}/\bar w_A\\
\{h_{A,0,i}-\bar h_{n,0,i}
                   +c_{n,0,i}(\varepsilon_{A,i}-\eta_{n,i})\}/\bar w_A\\
(U_{A,i}-u_{n,i})
\end{pmatrix}.
\end{equation}

To reveal covariates before outcomes, write \(D_{n,i}=D_{n,i}^{X}+D_{n,i}^{Y}\), where
\[
\begin{aligned}
D_{n,i}^{X}
&=\frac1{\sqrt n}
\begin{pmatrix}
\{h_{A,1,i}-\bar h_{n,1,i}
             +c_{n,1,i}(\eta_{A,i}^{X}-\eta_{n,i})\}/\bar w_A\\
\{h_{A,0,i}-\bar h_{n,0,i}
             +c_{n,0,i}(\eta_{A,i}^{X}-\eta_{n,i})\}/\bar w_A\\
(u_{A,i}^{X}-u_{n,i})
\end{pmatrix},\\
D_{n,i}^{Y}
&=\frac1{\sqrt n}
\begin{pmatrix}
c_{n,1,i}(\varepsilon_{A,i}-\eta_{A,i}^{X})/\bar w_A\\
c_{n,0,i}(\varepsilon_{A,i}-\eta_{A,i}^{X})/\bar w_A\\
(U_{A,i}-u_{A,i}^{X})
\end{pmatrix}.
\end{aligned}
\]

With \(z\) indexing the arm, the top covariance block is

\[
\sum_iE_s(D_{n,i,z}D_{n,i,z'}\mid\mathcal F_n)
=\frac1{n\bar w_A^2}\sum_i
\{a_{n,zz',i}+c_{n,z',i}b_{n,z,i}
 +c_{n,z,i}b_{n,z',i}+c_{n,z,i}c_{n,z',i}v_{n,i}\}.
\]

For \(z\ne z'\), \(c_{n,1,i}c_{n,0,i}=0\), since a row can be a donor in only its observed arm. The top/lower and lower/lower blocks are

\[
\begin{aligned}
\sum_i E_s\{D_{n,i,U}D_{n,i,z}\mid\mathcal F_n\}
&=\frac1{n\bar w_A}\sum_i\{d_{n,z,i}+c_{n,z,i}k_{n,i}\},\\
\sum_i E_s\{D_{n,i,U}D_{n,i,U}^\top\mid\mathcal F_n\}
&=\frac1n\sum_iR_{n,i}.
\end{aligned}
\]

We reveal the score fields first, then the remaining covariates, and finally the outcomes. For a fixed projection \(t\in\mathbb R^{2+p_A}\), put
\[
\begin{aligned}
\xi_{n,i}&=\frac{t^\top}{\sqrt n}
 \begin{pmatrix}\bar h_{n,1,i}/\bar w_A\\
                 \bar h_{n,0,i}/\bar w_A\\u_{n,i}\end{pmatrix},\\
\xi_{n,n+i}&=t^\top D_{n,i}^{X},\\
\xi_{n,2n+i}&=t^\top D_{n,i}^{Y},\qquad 1\le i\le n,
\end{aligned}
\]
and use the filtration
\[
\mathcal F_{n,k}^{\mathrm{seq}}=
\begin{cases}
 \bigvee_{i=1}^k\mathcal F_{n,i},&0\le k\le n,\\
 \mathcal F_n\vee\bigvee_{i=1}^{k-n}\mathcal X_i,&n<k\le2n,\\
 \mathcal X_{1:n}\vee\sigma(Y_1,\ldots,Y_{k-2n}),&2n<k\le3n.
\end{cases}
\]
The empty join is the trivial field. Population recovery and the estimating equations give zero mean for the first-stage contributions. The next two stages are centered by subtracting their conditional means. Thus
\[
E_s(\xi_{n,k}\mid\mathcal F_{n,k-1}^{\mathrm{seq}})=0,
\quad 1\le k\le3n,
\qquad
\sum_{k=1}^{3n}\xi_{n,k}
=t^\top\left(\mathcal H_n+\sum_iD_{n,i}\right).
\]
Here \(\mathcal H_n\) is the sum of the displayed first-stage rows. The predictable residual contribution is treated in part (iii).

For each direction, bounded derivatives give the uniform expansion
\[
\mathbf D_{z,\boldsymbol\theta_n}^{\mathrm p}(\mathbf X)
=\mathbf D_z^*(\mathbf X)+n^{-1/2}S_z(\mathbf X)\mathbf h_n+o(n^{-1/2}).
\]
The baseline \(M\)-donor competitor therefore bounds the current incoming load by \(K_{M,z}^{+,a}\) for some fixed \(a\). This bound controls the reuse coefficients while the conditioning field changes. Conditional on treatment and baseline own score, this unweighted count is independent of the remaining root marks. Equation~\eqref{eq:stabilizer-inflated-reuse} gives, for every integrable deterministic own-row mark \(F_n\) and fixed \(q>0\),
\[
E_s[\mathbf1(Z=z)(1+K_{M,z}^{+,a})^q|F_n|]
\le C_q E_s[\mathbf1(Z=z)|F_n|].
\]
Since \(c_{n,z}\le C\mathbf1(Z=z)(1+K_{M,z}^{+,a})\), the conditional-moment convergence in part (i) remains valid after multiplication by the reuse coefficients:
\[
\begin{aligned}
E_s\!\left[\frac1n\sum_i c_{n,z,i}^2|v_{n,i}-v_{*,i}|\right]
&\le C E_s\{\mathbf1(Z=z)|v_n-v_*|\}\longrightarrow0,\\
E_s\!\left[\frac1n\sum_i |c_{n,z,i}|
                  (\|k_{n,i}-k_{*,i}\|+|b_{n,z',i}|)\right]
&\le C E_s\{\mathbf1(Z=z)(\|k_n-k_*\|+|b_{n,z'}|)\}
                                                        \longrightarrow0,\\
E_s\!\left[\frac1n\sum_i
 (|a_{n,zz',i}|+\|d_{n,z,i}\|+\|R_{n,i}-R_{*,i}\|)\right]
&=E_s(|a_{n,zz'}|+\|d_{n,z}\|+\|R_n-R_*\|)\longrightarrow0.
\end{aligned}
\]

It remains to average the baseline covariance marks over the local graphs. Apply Proposition~\ref{prop:primitive-residual-qv-limit} and \eqref{eq:patt-local-qv-baseline-mark-llns} armwise, with mark \((\mathbf D_0^*,\mathbf D_1^*,S_z,w_{A,Z},v_*,k_*)\) conditional on \((Z,\mathbf D_z^*)\). The same support, density and displacement bounds apply. For \(r=1+\delta/2>1\), conditional Jensen and H\"older give \(E_s(v_*^r+\|k_*\|^r+\|R_*\|^r)<\infty\), and the count bound gives
\[
E_s[\mathbf1(Z=z)c_{n,z}^{2r}v_*^r]
\le C E_s[\mathbf1(Z=z)v_*^r]<\infty.
\]
The analogous bound treats \(c_{n,z}k_*\). These moment bounds permit removal of the mark truncation from the concentration argument in \eqref{eq:qv-truncated-concentration}--\eqref{eq:qv-armwise-lln}, giving
\begin{equation}\label{eq:pate-two-arm-joint-baseline-mark-limits}
\begin{aligned}
n^{-1}\sum_i c_{n,z,i}^2v_{*,i}&\longrightarrow_{L^1}\Lambda_{A,z},\\
n^{-1}\sum_i c_{n,z,i}k_{*,i}&\longrightarrow_{L^1}\mathcal C_{A,z},\\
\Lambda_{A,z}
&=E_s[\mathbf1(Z=z)E\{(w_{A,z}+K_{A,z}^\infty)^2v_*\mid O\}],\\
\mathcal C_{A,z}
&=E_s[\mathbf1(Z=z)E\{(w_{A,z}+K_{A,z}^\infty)k_*\mid O\}].
\end{aligned}
\end{equation}
The inner expectation fixes all root marks and integrates the auxiliary neighborhood. Its coefficient is
\begin{equation}\label{eq:pate-two-arm-joint-poisson-coefficient}
K_{A,z}^{\infty}(d;w)
=\sum_{y\text{ a query}}W_y
 \frac{w\,\mathbf1\{0\in\mathcal J_z^\infty(y)\}}
      {\sum_{k\in\mathcal J_z^\infty(y)}W_k},
\qquad W_0=w.
\end{equation}
The donor and query processes are independent, with intensities \(P_s(Z=z)f_{z\mid z}(d)\) and \(P_s(Z=1-z)f_{1-z\mid z}(d)\), respectively. We use their baseline conditional weight laws and fixed root weight \(w\). In particular,
\begin{equation}\label{eq:pate-two-arm-joint-incoming-mean-definition}
\kappa_{A,z}(d,w):=E\{K_{A,z}^{\infty}(d;w)\}.
\end{equation}
These constants are the unsigned arm contributions to \eqref{eq:qv-poisson-mean-limits} and \eqref{eq:baseline-oracle-firststep-cross-constant}. The signed PATE cross constant is \(\mathcal C_{A,1}-\mathcal C_{A,0}\). Combining the conditional covariance formulas with these limits gives
\begin{equation}\label{eq:pate-two-arm-joint-conditional-qv}
\sum_iE_s(D_{n,i}D_{n,i}^\top\mid\mathcal F_n)
\longrightarrow_p
\Omega_E=
\begin{pmatrix}
\Lambda_{A,1}/\bar w_A^2&0&\mathcal C_{A,1}^\top/\bar w_A\\
0&\Lambda_{A,0}/\bar w_A^2&\mathcal C_{A,0}^\top/\bar w_A\\
\mathcal C_{A,1}/\bar w_A&\mathcal C_{A,0}/\bar w_A&E_sR_*
\end{pmatrix}.
\end{equation}
For the outcome-stage covariance, write \(T_{n,i}=nE_s(D_{n,i}^{Y}D_{n,i}^{Y\top}\mid\mathcal X_{1:n})\). The same own-row bound and conditional Jensen give, for \(r=1+\delta/2>1\),
\[
\begin{aligned}
\frac1n\sum_i E_s\|T_{n,i}\|^r
&\le\frac Cn\sum_i E_s\!\left[
 (1+|c_{n,1,i}|^{2r}+|c_{n,0,i}|^{2r})
 E_s(\|\Phi\|^{2r}\mid\mathcal F_{n,i})\right]\\
&\le C E_s\|\Phi\|^{2r}<\infty.
\end{aligned}
\]
To compare the covariance before and after revealing the covariates, clip each matrix entry and use conditional independence. As in \eqref{eq:joint-three-stage-qv-comparison}, the bounded part and the moment-controlled tail yield
\[
E_s\left|\frac1n\sum_i
 \{T_{n,i}^{ab}-E_s(T_{n,i}^{ab}\mid\mathcal F_n)\}\right|
\le\frac L{\sqrt n}+CL^{1-r}.
\]
Letting \(n\to\infty\) and then \(L\to\infty\), conditional total covariance gives the predictable covariance of the refined array:
\[
\begin{aligned}
&\sum_iE_s(D_{n,i}^{X}D_{n,i}^{X\top}\mid\mathcal F_n)
 +\sum_iE_s(D_{n,i}^{Y}D_{n,i}^{Y\top}\mid\mathcal X_{1:n})\\
&=\sum_iE_s(D_{n,i}D_{n,i}^{\top}\mid\mathcal F_n)
 +\frac1n\sum_i\{T_{n,i}-E_s(T_{n,i}\mid\mathcal F_n)\}\\
&\longrightarrow_p\Omega_E.
\end{aligned}
\]
With \(p=2+\delta\), the same count bound and conditional Jensen give
\[
E_s\sum_i\{\|D_{n,i}\|^p+\|D_{n,i}^{X}\|^p+\|D_{n,i}^{Y}\|^p\}
\le Cn^{1-p/2}E_s\|\Phi\|^p\longrightarrow0.
\]
Markov's inequality gives the predictable Lyapunov condition for both remaining stages. In particular, conditional row independence permits the conditioning fields in their predictable second moments to be reduced to \(\mathcal F_n\) and \(\mathcal X_{1:n}\), respectively. The preceding covariance calculation therefore gives, for the same scalar projection used in the three-stage array,
\begin{equation}\label{eq:pate-two-arm-joint-predictable-covariance}
\sum_{k=n+1}^{3n}
 E_s(\xi_{n,k}^{2}\mid\mathcal F_{n,k-1}^{\mathrm{seq}})
\longrightarrow_p t^\top\Omega_Et.
\end{equation}

The remaining first-stage vector is
\[
\mathcal H_n=\frac1{\sqrt n}\sum_i
 \begin{pmatrix}\bar h_{n,1,i}/\bar w_A\\
                 \bar h_{n,0,i}/\bar w_A\\u_{n,i}\end{pmatrix}.
\]
Its summands are iid and centered. Independence and part (i) give
\[
\begin{aligned}
&E_s\left\|\mathcal H_n-\frac1{\sqrt n}\sum_i
 \begin{pmatrix}h_{A,1,i}/\bar w_A\\h_{A,0,i}/\bar w_A\\u_{*,i}\end{pmatrix}
                                                   \right\|^2\\
&=E_s\left\|
 \begin{pmatrix}(\bar h_{n,1}-h_{A,1})/\bar w_A\\
                (\bar h_{n,0}-h_{A,0})/\bar w_A\\(u_n-u_*)\end{pmatrix}
                                                   \right\|^2\\
&\longrightarrow0.
\end{aligned}
\]
The vector CLT gives
\begin{equation}\label{eq:pate-two-arm-joint-firststage-limit}
\mathcal H_n\Rightarrow N(0,\Omega_H),\qquad
\Omega_H=\operatorname{Var}_s
 \begin{pmatrix}h_{A,1}/\bar w_A\\h_{A,0}/\bar w_A\\u_*\end{pmatrix}.
\end{equation}
For the first stage, independence of the row fields makes each predictable second moment equal to its unconditional value. The same \(L^2\) convergence of the projected row gives
\[
\begin{aligned}
\sum_{i=1}^n
 E_s(\xi_{n,i}^{2}\mid\mathcal F_{n,i-1}^{\mathrm{seq}})
&=t^\top\operatorname{Var}_s
 \begin{pmatrix}\bar h_{n,1}/\bar w_A\\
                 \bar h_{n,0}/\bar w_A\\u_n\end{pmatrix}t\\
&\longrightarrow t^\top\Omega_Ht.
\end{aligned}
\]
Conditional Jensen also controls the first-stage moments. For \(p=2+\delta\),
\[
\begin{aligned}
\sum_{i=1}^n E_s|\xi_{n,i}|^p
&\le C_t n^{1-p/2}
 E_s\{|\bar h_{n,1}|^p+|\bar h_{n,0}|^p+\|u_n\|^p\}\\
&\le C_t' n^{1-p/2}E_s\|\Phi\|^p
\longrightarrow0.
\end{aligned}
\]
Together with the bounds for the other two stages, Markov's inequality and \(x^2\mathbf1\{|x|>\epsilon\}\le \epsilon^{-\delta}|x|^{2+\delta}\) establish conditional Lindeberg for the full \(3n\)-step array. Its predictable variance converges to \(t^\top(\Omega_H+\Omega_E)t\). The martingale part of Lemma~\ref{lem:conditional-residual-stable-clt}, followed by the Cram\'er--Wold device, consequently gives
\[
\mathcal H_n+\sum_iD_{n,i}
\Rightarrow N(0,\Omega_H+\Omega_E).
\]
If a projected variance is zero, the preceding \(L^1\) covariance limits and the \(L^2\) convergence of the first-stage rows give that projection's convergence to zero by Chebyshev's inequality.

\par\smallskip\noindent\emph{(iii) Predictable residual contribution.}\quad
The centered array accounts for the random fluctuation. We now compute the residual means induced by perturbing the score field. Joint-score centering and the integrated \(C^2\) envelope give \(\int a_{z,w}(s,l)\nu_{z,w}(dl)=0\) everywhere by continuity. For \(\boldsymbol\theta_t=\boldsymbol\theta_A^*+t\mathbf h\), set
\[
\begin{aligned}
C_t&=A_*A_{\boldsymbol\theta_t}^{-1}B_{\boldsymbol\theta_t}-B_*,\\
\dot C_{\mathbf h}&=\dot B_{\mathbf h}-\dot A_hA_*^{-1}B_*,\\
\mathbf F^{\varepsilon,\mathrm J}_{A,z,\mathbf h,w}(s)&=\int a_{z,w}(s,l)\dot C_{\mathbf h}l\,\nu_{z,w}(dl),\\
\dot m^{\varepsilon,\mathrm J}_{A,z,\mathbf h}(s,w)
&=-\operatorname{div}\mathbf F^{\varepsilon,\mathrm J}_{A,z,\mathbf h,w}(s)/q_{z,w}(s)
\quad(s\in N),
\end{aligned}
\]
and extend \(\dot m^{\varepsilon,\mathrm J}\) by zero outside \(N\). The identity \(A_*A_{\boldsymbol\theta_t}^{-1}S_{\boldsymbol\theta_t}=S_*+C_tL=:T_t\) preserves the joined field, with \(C_t=t\dot C_{\mathbf h}+O(t^2)\) uniformly for bounded \(\mathbf h\). Translation has Jacobian one, so
\[
E_s(\varepsilon_A\mid Z=z,T_t=u,w_{A,z}=w)
=\frac{\int a_{z,w}(u-C_tl,l)\nu_{z,w}(dl)}
       {\int p_{z,w}(u-C_tl,l)\nu_{z,w}(dl)}.
\]
The compact signed support and derivative envelopes in condition (ii) control the numerator expansion uniformly, including near its boundary. The denominator stays positive on the relevant neighborhood. The Taylor argument in Lemma~\ref{lem:patt-signed-density-control-drift} therefore applies in three coordinates. Taking \(t=n^{-1/2}\) gives, uniformly,
\begin{equation}\label{eq:pate-two-arm-joint-local-mean-expansion}
\eta_{n,i}
=n^{-1/2}\dot m^{\varepsilon,\mathrm J}_{A,Z_i,\mathbf h_n}(S_{*,i},w_{A,Z_i})
       +O(n^{-1})
\end{equation}

The direct term has exact expectation zero by the tower property, \(E_s[\mathbf1(Z=z)w_{A,z}\eta_n] =E_s[\mathbf1(Z=z)w_{A,z}\varepsilon_A]=0\). Since it is an iid, bounded \(O(n^{-1/2})\) row function,
\[
E_s\left|\frac1{\sqrt n}\sum_i
 \mathbf1(Z_i=z)w_{A,z}(\mathbf X_i)\eta_{n,i}\right|^2
=E_s[\mathbf1(Z=z)w_{A,z}^2\eta_n^2]\le C/n.
\]
Valid-count completion preserves this conclusion. The normalized fractions give \(\sum_{j:Z_j=z}K_{n,z}(j) =\sum_{i:Z_i=1-z}w_{A,1-z}(\mathbf X_i)\le nw_{\max}\) on \(\mathcal A_n\). Hence
\begin{equation}\label{eq:pate-two-arm-joint-drift-limit}
\begin{aligned}
\frac1{\sqrt n}\sum_{j:Z_j=z}K_{n,z}(j)\eta_{n,j}
&=\frac1n\sum_{j:Z_j=z}K_{n,z}(j)
                  \dot m^{\varepsilon,\mathrm J}_{A,z,\mathbf h_n}(S_{*,j},w_{A,z,j})+O(n^{-1/2})\\
&=\frac1n\sum_{j:Z_j=z}K_{n,z}(j)
                  \dot m^{\varepsilon,\mathrm J}_{A,z,\mathbf h}(S_{*,j},w_{A,z,j})+o_p(1)\\
&\longrightarrow_p\Delta_{A,z}(\mathbf h),\\
\Delta_{A,z}(\mathbf h)
&=E_s[\mathbf1(Z=z)\kappa_{A,z}(\mathbf D_z^*,w_{A,z})
                          \dot m^{\varepsilon,\mathrm J}_{A,z,\mathbf h}(S_*,w_{A,z})].
\end{aligned}
\end{equation}
Thus the local mean comes from incoming reuse. The last step uses the marked \(L^1\) law from part (ii) with bounded mark \(\dot m^{\varepsilon,\mathrm J}_{A,z,\mathbf h}\). Conservation and the uniform remainder also give \(L^1\) convergence of the displayed sum.

If the own-arm signed-density conditions additionally apply, let \(\mathcal F_n^{(z)}\) contain only treatment, the current arm-\(z\) score, and observed weights, and write \(\eta_{n,j}^{(z)}=E_s(\varepsilon_{A,j}\mid\mathcal F_{n,j}^{(z)})\). The own-arm and joined fields make the same coefficient measurable. Projecting the residual onto either field therefore leaves the expectation of the weighted sum unchanged:
\[
\begin{aligned}
E_s\!\left[\sum_{j:Z_j=z}K_{n,z}(j)\varepsilon_{A,j}\right]
&=E_s\sum_{j:Z_j=z}K_{n,z}(j)\eta_{n,j}\\
&=E_s\sum_{j:Z_j=z}K_{n,z}(j)\eta_{n,j}^{(z)}.
\end{aligned}
\]
The two \(L^1\) limits identify the same integrated \(\Delta_{A,z}(\mathbf h)\).
\par\smallskip\noindent\emph{(iv) Joint Gaussian limit.}\quad
Combining the centered fluctuation and predictable mean, on the valid-count event we have
\[
\begin{aligned}
&\begin{pmatrix}
(\sqrt n\bar w_A)^{-1}\sum_i(h_{A,1,i}+c_{n,1,i}\varepsilon_{A,i})\\
(\sqrt n\bar w_A)^{-1}\sum_i(h_{A,0,i}+c_{n,0,i}\varepsilon_{A,i})\\
n^{-1/2}\sum_iU_{A,i}
\end{pmatrix}\\
&=\mathcal H_n+\sum_iD_{n,i}
+\begin{pmatrix}
(\sqrt n\bar w_A)^{-1}\sum_ic_{n,1,i}\eta_{n,i}\\
(\sqrt n\bar w_A)^{-1}\sum_ic_{n,0,i}\eta_{n,i}\\0
\end{pmatrix}.
\end{aligned}
\]
Parts (ii) and (iii) give mean \((\Delta_{A,1}(\mathbf h)/\bar w_A,\Delta_{A,0}(\mathbf h)/\bar w_A,0)^\top\) and covariance \(\Omega_H+\Omega_E\). Its lower block is \(E_s(u_*u_*^\top)+E_sR_*=\boldsymbol\Sigma_{A,U}\), by conditional variance decomposition. Its cross covariance between each arm and the nuisance block is \(\boldsymbol\gamma_{A,1}^{(z)} =\bar w_A^{-1}\{E_s(h_{A,z}u_*)+\mathcal C_{A,z}\}\). Slutsky's theorem restores the common denominator \(\bar w_{A,n}\to_p\bar w_A>0\). Taking the first coordinate minus the second gives
\begin{equation}\label{eq:pate-two-arm-joint-pate-constants}
\begin{aligned}
\delta_A(\mathbf h)&=(\Delta_{A,1}(\mathbf h)-\Delta_{A,0}(\mathbf h))/\bar w_A,\\
V_{A,\mathrm{known\text{-}score}}
&=\{E_s[(h_{A,1}-h_{A,0})^2]+\Lambda_{A,1}+\Lambda_{A,0}\}/\bar w_A^2,\\
\boldsymbol\gamma_{A,1}
&=\boldsymbol\gamma_{A,1}^{(1)}-\boldsymbol\gamma_{A,1}^{(0)}\\
&=\bar w_A^{-1}\{E_s[(h_{A,1}-h_{A,0})u_*]
                         +\mathcal C_{A,1}-\mathcal C_{A,0}\}.
\end{aligned}
\end{equation}
For every deterministic \(\mathbf h_n\to \mathbf h\), we therefore have
\begin{equation}\label{eq:pate-two-arm-joint-population-pair-limit}
\begin{pmatrix}
T_{A,n}^{\mathrm p}(\mathbf h_n;\tau_{\mathrm{PATE}})\\
\sqrt n\,\mathbb U_{A,n}(\boldsymbol\theta_A^*)
\end{pmatrix}
\Rightarrow
N\left(\begin{pmatrix}\delta_A(\mathbf h)\\0\end{pmatrix},
\begin{pmatrix}V_{A,\mathrm{known\text{-}score}}&\boldsymbol\gamma_{A,1}^\top\\
\boldsymbol\gamma_{A,1}&\boldsymbol\Sigma_{A,U}\end{pmatrix}\right).
\end{equation}
The root expansion gives
\begin{equation}\label{eq:pate-two-arm-joint-root-pair-limit}
\begin{pmatrix}
T_{A,n}^{\mathrm p}(\mathbf h_n;\tau_{\mathrm{PATE}})\\
\sqrt n(\widehat{\boldsymbol \theta}_A-\boldsymbol\theta_A^*)
\end{pmatrix}
\Rightarrow
N\!\left[
 \begin{pmatrix}\delta_A(\mathbf h)\\0\end{pmatrix},
 \begin{pmatrix}
 V_{A,\mathrm{known\text{-}score}}&
              -\boldsymbol\gamma_{A,1}^{\top}\boldsymbol\Gamma_A^{-T}\\
 -\boldsymbol\Gamma_A^{-1}\boldsymbol\gamma_{A,1}&\boldsymbol\Sigma_{A,\theta}
 \end{pmatrix}
 \right],
\end{equation}
where \(\boldsymbol\Sigma_{A,\theta}=\boldsymbol\Gamma_A^{-1}\boldsymbol\Sigma_{A,U}\boldsymbol\Gamma_A^{-T}\). Equivalently, \(\delta_A(\mathbf h)=\boldsymbol b_{A,\mathrm{loc}}^\top \mathbf h\), where \(b_{A,\mathrm{loc},r}=(\Delta_{A,1}(\mathbf e_r)-\Delta_{A,0}(\mathbf e_r))/\bar w_A\). Here \(\mathbf e_r\) is the \(r\)th unit vector in \(\mathbb R^{p_A}\), and \(\boldsymbol b_{A,\mathrm{loc}}\) gives the change in the statistic's limiting mean per unit local parameter perturbation. The marked count bounds give finiteness. Nonnegativity follows from the covariance representation. Positive baseline observed-field residual variance on a positive-probability arm subset gives \(\Lambda_{A,z}>0\) because \(w_{A,z}+K_{A,z}^\infty\ge w_{\min}\).

\par\smallskip\noindent\emph{(v) Likelihood shift.}\quad

The fixed-law joint limit also determines the local likelihood shift. Under the conditions of Lemma~\ref{lem:auxiliary-godambe-experiment}, the normalized likelihood and Le Cam's third lemma shift the PATE coordinate by \(-\boldsymbol\gamma_{A,1}^\top\boldsymbol\Sigma_{A,U}^{-1}\boldsymbol\Gamma_A\mathbf h\) and the baseline estimating-function coordinate by \(-\boldsymbol\Gamma_A\mathbf h\). The actual local equation adds \(\boldsymbol\Gamma_A\mathbf h\) to the latter. If the target along this same normalized family additionally satisfies the identification, differentiability and integrability conditions of Proposition~\ref{prop:scientific-target-path-derivative}, with derivative \(\boldsymbol\gamma_{A,2}\), its target-centered moving-law mean is
\[
\widetilde{\boldsymbol \gamma}_{A,2}^\top \mathbf h-\boldsymbol\gamma_{A,2}^\top \mathbf h
=\delta_A(\mathbf h)-\boldsymbol\gamma_{A,1}^\top\boldsymbol\Sigma_{A,U}^{-1}\boldsymbol\Gamma_A\mathbf h
                         -\boldsymbol\gamma_{A,2}^\top \mathbf h
\]
for a statistic sharing the displayed local \(H+E\) limit.
\end{proof}
\subsubsection{PATE with estimated scores and empirical standardization}
\label{app:pate-two-arm-actual-grid}

We next transfer the joint local limit to empirical normalization and fitted scores. The two directed imputation processes share the nuisance parameter, but their increments can be bounded separately and combined through the PATE contrast.

Fix \(A\in\{R,P\}\), with

\[
w_{R,z}(\mathbf X)=w_z(\mathbf X),\qquad
w_{P,z}(\mathbf X)=w(\mathbf X),\qquad z\in\{0,1\}.
\]

The nuisance vector is

\[
\boldsymbol\theta_A=(\boldsymbol\alpha_A^\top,
          \boldsymbol\beta_{A,0}^\top,
          \boldsymbol\beta_{A,1}^\top)^\top,
\qquad p_A=\dim\boldsymbol\theta_A.
\]

The propensity equation is weighted for \(A=R\) and unweighted for \(A=P\). The two armwise prognostic equations are unweighted.

For each donor arm \(z\), use the propensity score probability and linear prognostic score

\[
r_{z1}(\mathbf X;\boldsymbol\theta)=e(\mathbf X;\boldsymbol\alpha),\qquad
r_{z2}(\mathbf X;\boldsymbol\theta)=\mathbf X^\top\boldsymbol\beta_z.
\]

All derivatives are with respect to \(\boldsymbol\theta_A\). The derivatives with respect to the prognostic coefficients for the opposite arm are zero.

We use the local score family
\begin{equation}\label{eq:pate-two-arm-grid-score-family}
\mathcal D_{z,n,\mathbf h,l_z}(\mathbf X)
=\operatorname{diag}(1+l_{z1}/\sqrt n,1+l_{z2}/\sqrt n)
 \mathbf D_{z,\boldsymbol\theta_A^*+\mathbf h/\sqrt n}^{\mathrm p}(\mathbf X),
\qquad l_z\in\mathbb R^2.
\end{equation}

For direction \(z\), we denote the two arm densities by \(f_{z\mid z}\) and \(f_{1-z\mid z}\). The supports \(\mathcal S_0\) and \(\mathcal S_1\) need not coincide. We write
\[
\zeta_{A,z}=Y-m_z^A\{\mathbf D_z^*(\mathbf X)\},\qquad z=0,1,
\]
where \(m_z^A\) is the baseline population reduced mean. On donor rows, \(\zeta_{A,z}=\varepsilon_z^A\). The joint-score centering in Assumption~\ref{ass:residual-centering-survey-normalized} implies the baseline centering below by iterated expectation.

\begin{assumption}[PATE local processes]
\label{ass:pate-two-arm-grid-process}
\begin{enumerate}
\item[(i)] The observed rows \(O_i=(\mathbf X_i,Z_i,Y_i)\) are iid under the selected law \(P_s\). Both treatment probabilities are positive. The fixed number of matches is \(M\ge1\). Covariates are bounded and

\begin{equation}\label{eq:pate-two-arm-grid-weight-bounds}
0<w_{\min}\le w_{A,Z}(\mathbf X)\le w_{\max}<\infty.
\end{equation}

\item[(ii)] Assume root tightness, as given by Lemma~\ref{lem:primitive-logistic-linear-roots} under its conditions:

\begin{equation}\label{eq:pate-two-arm-grid-root-tightness}
\widehat{\mathbf h}_n:=\sqrt n(\widehat{\boldsymbol \theta}_A-\boldsymbol\theta_A^*)=O_p(1).
\end{equation}

\item[(iii)] Use the specified deterministic normalization \(\mathbf D_{z,\boldsymbol\theta}^{\mathrm p}\), with baseline \(\mathbf D_z^*\), and positive component scales. The normalization may be fixed at baseline or continuously differentiable in \(\boldsymbol\theta\).

For the empirical-normalization application in Theorem~\ref{lem:pate-two-arm-empirical-index}, use the specified fixed measurable moment weights \(b_z(O)\), with \(0\le b_z\le C_b<\infty\) and \(E_s b_z>0\). The baseline scales must agree with the population limits of those same moment rules:
\[
\{s_{zk}^{\mathrm p}(\boldsymbol\theta_A^*)\}^2
=\frac{E_s\{b_zr_{zk}(\mathbf X;\boldsymbol\theta_A^*)^2\}}{E_sb_z}
 -\left[\frac{E_s\{b_zr_{zk}(\mathbf X;\boldsymbol\theta_A^*)\}}{E_sb_z}\right]^2>0.
\]
Here \(b_z\) is the rule in \eqref{eq:pate-two-arm-grid-empirical-moments}.

Matching uses unweighted Euclidean distance on the probability-PG pair, with replacement and the specified fixed, measurable, outcome-independent tie rule.

\item[(iv)] For each \(z=0,1\), given \(Z=z\), \(\mathbf D_z^*\) is supported on a compact convex set \(\mathcal S_z\subset\mathbb R^2\) with nonempty interior, and its density is bounded above and away from zero on \(\mathcal S_z\). Given \(Z=1-z\), its density is bounded above and supported on the same \(\mathcal S_z\).
\item[(v)] For each \(z=0,1\), \(m_z^A\) is Lipschitz on \(\mathcal S_z\), with a fixed Lipschitz extension wherever needed. For some \(p>2\), \(E_s(1+|\zeta_{A,z}|)^p<\infty\), and

   \begin{equation}\label{eq:pate-two-arm-grid-baseline-centering}
E_s(\zeta_{A,z}\mid Z=z,\mathbf D_z^*,w_{A,z})=0.
\end{equation}

\item[(vi)] For each \(z=0,1\), the deterministic-current donor subdensity satisfies, for every compact \(K_z\subset\mathbb R^{p_A+2}\),

   \begin{equation}\label{eq:pate-two-arm-grid-current-donor-density}
\sup_{n\ge n_{K_z}}\sup_{(\mathbf h,l_z)\in K_z}
   \left\|\frac{dP_s\{Z=z,
            \mathcal D_{z,n,\mathbf h,l_z}(\mathbf X)\in\cdot\}}{du}
   \right\|_\infty<\infty.
\end{equation}

\end{enumerate}
\end{assumption}

For each donor arm \(z\), write the covariate partition \(\mathbf X=(V_z^\top,L_z^\top)^\top\), with \(V_z\in\mathbb R^2\), and write the raw logit/PG pair as

\begin{equation}\label{eq:pate-two-arm-grid-raw-chart}
\begin{aligned}
S_{z,\boldsymbol\theta}
&:=\begin{pmatrix}\mathbf X^\top\boldsymbol\alpha\\
                  \mathbf X^\top\boldsymbol\beta_z\end{pmatrix}\\
&=A_{z,\boldsymbol\theta}V_z+B_{z,\boldsymbol\theta}L_z.
\end{aligned}
\end{equation}

An intercept belongs to \(L_z\). The partitions for the two arms may differ. For almost every donor weight \(w\), we write
\[
\nu_{z,w}=\mathcal L_s(L_z\mid Z=z,w_{A,z}=w).
\]

When the conditional density of \(S_{z,*}\) given \(L_z=l,Z=z,w_{A,z}=w\) exists, we denote it by \(p_{z,w}(s,l)\) and define
\begin{equation}\label{eq:pate-two-arm-grid-signed-densities}
\begin{aligned}
a_{z,w}(s,l)
&=p_{z,w}(s,l)
 E_s(\zeta_{A,z}\mid Z=z,S_{z,*}=s,L_z=l,w_{A,z}=w),\\
q_{z,w}(s)&=\int p_{z,w}(s,l)\,\nu_{z,w}(dl).
\end{aligned}
\end{equation}

\begin{assumption}[Armwise signed-density conditions]
\label{ass:pate-two-arm-grid-signed-density}
\begin{enumerate}
\item[(i)] For each donor arm \(z\in\{0,1\}\), covariates are bounded, \(A_{z,*}\) is nonsingular, and both coefficient matrices are \(C^2\) near \(\boldsymbol\theta_A^*\). On donor rows the observed weight is a fixed measurable function \(w_{A,z}(L_z)\).

\item[(ii)] For each donor arm \(z\in\{0,1\}\) and almost every donor weight \(w\), there is a jointly measurable conditional density \(p_{z,w}(s,l)\) of \(S_{z,*}\) given \(L_z=l,Z=z,w_{A,z}=w\), including support indicators.

\item[(iii)] For each arm there are a compact \(J_z\subset\mathbb R^2\) and bounded open \(N_z\supset J_z\), with positive separation from \(N_z^c\), such that the following bounds hold uniformly over almost every \(w\):

\begin{equation}\label{eq:pate-two-arm-grid-signed-envelopes}
\begin{aligned}
&a_{z,w}(\cdot,l)\text{ has a }C^2\text{ extension supported in }J_z,\\
&\int\sup_{s\in\mathbb R^2}
  \sum_{|\alpha|\le2}|\partial^\alpha a_{z,w}(s,l)|\,
                     \nu_{z,w}(dl)\le C_z,\\
&p_{z,w}(\cdot,l)\text{ is }C^1\text{ on }N_z,\\
&\int\sup_{s\in N_z}
 (|p_{z,w}(s,l)|+\|\nabla p_{z,w}(s,l)\|)\,
                      \nu_{z,w}(dl)\le C_z,\\
&\inf_{s\in N_z}q_{z,w}(s)\ge c_z>0.
\end{aligned}
\end{equation}
\end{enumerate}
\end{assumption}

For \(m\) observations, denote the \(M\)-donor set by \(\mathcal J_{z,n,\mathbf h,l_z}^{(m)}(i)\) for query \(i\) in arm \(1-z\), with all local displacements still divided by \(\sqrt n\). Write \(m_z=\sum_{i=1}^m\mathbf1\{Z_i=z\}\). On \(m_z\ge M\),

\begin{equation}\label{eq:pate-two-arm-grid-incoming-numerator}
\begin{aligned}
\mathcal I_{A,z,m;n}(\mathbf h,l_z)
&:=\sum_{\substack{i\le m\\Z_i=1-z}}w_{A,1-z}(\mathbf X_i)
 \sum_{j\in\mathcal J_{z,n,\mathbf h,l_z}^{(m)}(i)}
 \frac{w_{A,z}(\mathbf X_j)\varepsilon_{z,j}^A}
      {\sum_{k\in\mathcal J_{z,n,\mathbf h,l_z}^{(m)}(i)}w_{A,z}(\mathbf X_k)}\\
&=\sum_{\substack{j\le m\\Z_j=z}}
 K_{m,z}^{A}(j;\mathcal D_{z,n,\mathbf h,l_z})
 \varepsilon_{z,j}^A.
\end{aligned}
\end{equation}

We set this directional numerator to zero when \(m_z<M\). The PATE identities below hold on \(\mathcal A_n=\{n_0\ge M,n_1\ge M\}\), whose complement has exponentially small probability.

Both directions use the common denominator

\begin{equation}\label{eq:pate-two-arm-grid-denominator}
\bar w_{A,n}=\frac1n\sum_iw_{A,Z_i}(\mathbf X_i),\qquad
\bar w_A=E_sw_{A,Z}(\mathbf X)>0.
\end{equation}

For \(l=(l_0^\top,l_1^\top)^\top\), separate the direct observations from the imputed contributions in the baseline-mean H+E coordinate:

\begin{equation}\label{eq:pate-two-arm-grid-hbe-coordinate}
\begin{aligned}
T_{A,n}(\mathbf h,l;\tau_{\mathrm{PATE}})
&=\frac1{\sqrt n\bar w_{A,n}}
 \sum_i w_{A,Z_i}(\mathbf X_i)
       \{\tau_A^D(\mathbf X_i)-\tau_{\mathrm{PATE}}\}\\
&\quad+\frac1{\sqrt n\bar w_{A,n}}
 \sum_{z=0}^1(2z-1)\sum_{j:Z_j=z}
 \{w_{A,z}(\mathbf X_j)+K_{n,z}^{A}
                    (j;\mathcal D_{z,n,\mathbf h,l_z})\}\varepsilon_{z,j}^A\\
&=\frac1{\sqrt n\bar w_{A,n}}
 \sum_i w_{A,Z_i}(\mathbf X_i)
       \{\tau_A^D(\mathbf X_i)-\tau_{\mathrm{PATE}}\}\\
&\quad+\frac1{\sqrt n\bar w_{A,n}}
 \sum_{z=0}^1(2z-1)\sum_{j:Z_j=z}
                 w_{A,z}(\mathbf X_j)\varepsilon_{z,j}^A\\
&\quad+\frac1{\sqrt n\bar w_{A,n}}
 \sum_{z=0}^1(2z-1)\sum_{j:Z_j=z}
 K_{n,z}^{A}(j;\mathcal D_{z,n,\mathbf h,l_z})\varepsilon_{z,j}^A\\
&=\frac1{\sqrt n\bar w_{A,n}}
 \sum_i w_{A,Z_i}(\mathbf X_i)
       \{\tau_A^D(\mathbf X_i)-\tau_{\mathrm{PATE}}\}\\
&\quad+\frac1{\sqrt n\bar w_{A,n}}
 \sum_{z=0}^1(2z-1)\sum_{j:Z_j=z}
                 w_{A,z}(\mathbf X_j)\varepsilon_{z,j}^A\\
&\quad+\frac{\mathcal I_{A,1,n;n}(\mathbf h,l_1)
                 -\mathcal I_{A,0,n;n}(\mathbf h,l_0)}{\sqrt n\bar w_{A,n}}.
\end{aligned}
\end{equation}

\Needspace{8\baselineskip}
The heterogeneity and direct residual terms do not depend on the local index. By \eqref{eq:pate-two-arm-grid-incoming-numerator}, their cancellation gives

\begin{equation}\label{eq:pate-two-arm-grid-exact-difference}
\begin{aligned}
&T_{A,n}(\mathbf h,l;\tau_{\mathrm{PATE}})
 -T_{A,n}(\mathbf h',l';\tau_{\mathrm{PATE}})\\
&=\frac{\mathcal I_{A,1,n;n}(\mathbf h,l_1)
              -\mathcal I_{A,0,n;n}(\mathbf h,l_0)}{\sqrt n\bar w_{A,n}}
 -\frac{\mathcal I_{A,1,n;n}(\mathbf h',l'_1)
              -\mathcal I_{A,0,n;n}(\mathbf h',l'_0)}{\sqrt n\bar w_{A,n}}\\
&=\frac{\mathcal I_{A,1,n;n}(\mathbf h,l_1)
              -\mathcal I_{A,1,n;n}(\mathbf h',l'_1)}{\sqrt n\bar w_{A,n}}
 -\frac{\mathcal I_{A,0,n;n}(\mathbf h,l_0)
              -\mathcal I_{A,0,n;n}(\mathbf h',l'_0)}{\sqrt n\bar w_{A,n}}.
\end{aligned}
\end{equation}

For \(k=1,2\), the empirical moments at \(\boldsymbol\theta\) are

\begin{equation}\label{eq:pate-two-arm-grid-empirical-moments}
\begin{aligned}
\widehat\ell_{n,zk}(\boldsymbol\theta)
&=\frac{\sum_i b_z(O_i)r_{zk}(\mathbf X_i;\boldsymbol\theta)}{\sum_i b_z(O_i)},\\
\widehat s_{n,zk}(\boldsymbol\theta)^2
&=\frac{\sum_i b_z(O_i)r_{zk}(\mathbf X_i;\boldsymbol\theta)^2}{\sum_i b_z(O_i)}
 -\widehat\ell_{n,zk}(\boldsymbol\theta)^2.
\end{aligned}
\end{equation}

On exceptional samples, we use the moment definitions of Lemma~\ref{lem:primitive-score-standardization}. With the same deterministic scales \(s_{zk}^{\mathrm p}(\boldsymbol\theta)\) used in \eqref{eq:pate-two-arm-grid-score-family}, define

\begin{equation}\label{eq:pate-two-arm-grid-scale-indices}
\begin{aligned}
L_{n,z,k}(\mathbf h)&=\sqrt n\left\{
 \frac{s_{zk}^{\mathrm p}(\boldsymbol\theta_A^*+\mathbf h/\sqrt n)}
      {\widehat s_{n,zk}(\boldsymbol\theta_A^*+\mathbf h/\sqrt n)}-1\right\},\\
L_n(\mathbf h)&=(L_{n,0}(\mathbf h)^\top,L_{n,1}(\mathbf h)^\top)^\top.
\end{aligned}
\end{equation}

Let \(T_{A,n}^{\mathrm p}(\mathbf h;\tau_{\mathrm{PATE}}) :=T_{A,n}(\mathbf h,0;\tau_{\mathrm{PATE}})\). Let \(T_{A,n}^{\mathrm e}(\mathbf h;\tau_{\mathrm{PATE}})\) denote the same baseline-reference \(H+E\) coordinate in \eqref{eq:pate-two-arm-grid-hbe-coordinate}, evaluated on the two empirically normalized graphs at \(\boldsymbol\theta_A^*+\mathbf h/\sqrt n\), including the specified selections at ties. The residuals, heterogeneity, target and common denominator are unchanged.

Let \(\bar{\mathbf h}_{n,\eta}\) be the midpoint on a deterministic-offset side-\(\eta\) grid containing \(\widehat{\mathbf h}_n\). Its parameter version is \(\bar{\boldsymbol \theta}_{A,n,\eta}=\boldsymbol\theta_A^*+\bar{\mathbf h}_{n,\eta}/\sqrt n\). For every allowed offset and boundary convention,

\begin{equation}\label{eq:pate-two-arm-grid-grid-distance}
\|\widehat{\mathbf h}_n-\bar{\mathbf h}_{n,\eta}\|\le\sqrt{p_A}\,\eta/2.
\end{equation}

For the current pair of empirical graphs, the baseline-mean discrepancy is

\begin{equation}\label{eq:pate-two-arm-grid-bc-discrepancy}
\begin{aligned}
&B_{n,A}^{D,\mathrm{PATE}}
 ((\mathbf D_{z,\boldsymbol\theta}^{\mathrm e})_{z=0}^1;
                       (\mathbf D_z^*,m_z^A)_{z=0}^1)\\
&=\frac1{n\bar w_{A,n}}\sum_i w_{A,Z_i}(\mathbf X_i)(2Z_i-1)
 \sum_j\lambda_{i\to j}^{A,(1-Z_i)}
              (\mathbf D_{1-Z_i,\boldsymbol\theta}^{\mathrm e})
 [m_{1-Z_i}^A\{\mathbf D_{1-Z_i}^*(\mathbf X_i)\}
  -m_{1-Z_i}^A\{\mathbf D_{1-Z_i}^*(\mathbf X_j)\}].
\end{aligned}
\end{equation}

The two baseline functions do not move with \(\boldsymbol\theta\). The estimated discrepancy \(\widehat B\) uses the prescribed BC fits on these same current graphs. Define the total remainder

\begin{equation}\label{eq:pate-two-arm-grid-bc-remainder}
\begin{aligned}
R_{A,n}(\mathbf h)=\sqrt n[&\widehat B_{n,A}^{D,\mathrm{PATE}}
                   ((\mathbf D_{z,\boldsymbol\theta_A^*+\mathbf h/\sqrt n}^{\mathrm e})_z)\\
&-B_{n,A}^{D,\mathrm{PATE}}
 ((\mathbf D_{z,\boldsymbol\theta_A^*+\mathbf h/\sqrt n}^{\mathrm e})_z;
                               (\mathbf D_z^*,m_z^A)_z)].
\end{aligned}
\end{equation}

\begin{assumption}[Empirical-graph bias-correction remainders]
\label{ass:pate-two-arm-grid-bc}
We assume that the bias-correction remainder satisfies

\begin{equation}\label{eq:pate-two-arm-grid-bc-quantifiers}
\begin{aligned}
R_{A,n}(\mathbf h_n)&\to_p0
 &&\text{for every convergent deterministic sequence }\mathbf h_n,\\
R_{A,n}(\widehat{\mathbf h}_n)&\to_p0
 &&\text{for the actual fitted graph.}
\end{aligned}
\end{equation}

\end{assumption}

\begin{theorem}[Empirical-normalization and fitted-score transfer for PATE]
\label{thm:pate-two-arm-actual-grid-comparison}
\label{prop:pate-two-arm-grid-centered-processes}
\label{lem:pate-two-arm-empirical-index}
\label{prop:pate-two-arm-process-comparison}
\label{prop:pate-two-arm-empirical-joint-transfer}
\label{app:pate-two-arm-empirical-joint-bridge}
\label{lem:pate-two-arm-grid-local-means}
We use the empirical moment rules in Assumption~\ref{ass:pate-two-arm-grid-process} and the preceding notation.
\begin{enumerate}
\item[(i)] Under Assumption~\ref{ass:pate-two-arm-grid-process}, each centered directed process is stochastically equicontinuous as in \eqref{eq:pate-two-arm-grid-centered-modulus}. The same holds for their signed sum in \eqref{eq:pate-two-arm-grid-signed-modulus}. The empirical normalization indices are tight and locally Lipschitz as in \eqref{eq:pate-two-arm-grid-tight-indices}, and reproduce the empirical distances through \eqref{eq:pate-two-arm-grid-empirical-distances}. If Assumption~\ref{ass:pate-two-arm-grid-signed-density} also holds, the conditional projection \eqref{eq:pate-two-arm-grid-root-projection} and the expectation limit \eqref{eq:pate-two-arm-grid-local-mean} hold, uniformly on compact local-index sets as in \eqref{eq:pate-two-arm-grid-uniform-mean}. The limiting expectations are finite and linear in the nuisance direction.
\item[(ii)] Retain the conditions of Theorem~\ref{thm:pate-two-arm-population-joint-law} and Assumptions~\ref{ass:pate-two-arm-grid-process} and~\ref{ass:pate-two-arm-grid-signed-density}. Suppose each deterministic baseline scale equals the population limit of its specified fixed bounded empirical moment rule. Then, for every deterministic sequence \(\mathbf h_n\to \mathbf h\),
\begin{equation}\label{eq:pate-two-arm-empirical-joint-transfer}
T_{A,n}^{\mathrm e}(\mathbf h_n;\tau_{\mathrm{PATE}})
-T_{A,n}^{\mathrm p}(\mathbf h_n;\tau_{\mathrm{PATE}})
\longrightarrow_p0.
\end{equation}
The same comparison holds jointly with the unchanged nuisance root.
\item[(iii)] Under Assumptions~\ref{ass:pate-two-arm-grid-process} and~\ref{ass:pate-two-arm-grid-signed-density}, the two-arm \(H+E\) coordinates satisfy \eqref{eq:pate-two-arm-grid-hbe-comparison}.
\item[(iv)] If the conditions of part (iii) and Assumption~\ref{ass:pate-two-arm-grid-bc} hold, the fixed-\(M\), survey-normalized, bias-corrected PATE estimator satisfies \eqref{eq:pate-two-arm-grid-actual-comparison} and the joint comparison \eqref{eq:pate-two-arm-grid-joint-bl-comparison}.
\end{enumerate}
\end{theorem}
\begin{proof}
\par\smallskip\noindent\emph{(i) Centered process and empirical normalization.}\quad
Each matching direction has the form of the one-sided imputation process. For each \(z\), apply Proposition~\ref{prop:patt-binomial-local-process-transfer} with donor/query indicators \((\mathbf1\{Z=z\},\mathbf1\{Z=1-z\})\), score \(\mathcal D_{z,n,\mathbf h,l_z}\), weights \((w_{A,z},w_{A,1-z})\), and donor mark \(\varepsilon_z^A\). The geometry, moments, baseline centering and donor density required there are included in Assumption~\ref{ass:pate-two-arm-grid-process}. The unused nuisance coordinates leave the directed matching rule unchanged. Hence
\[
\mathcal Z_{A,z,n}^B(\mathbf h,l_z)
=\frac{\mathcal I_{A,z,n;n}(\mathbf h,l_z)
       -E_s\mathcal I_{A,z,n;n}(\mathbf h,l_z)}{\sqrt n}
\]
satisfies, for every compact \(K_z\),
\begin{equation}\label{eq:pate-two-arm-grid-centered-modulus}
\lim_{\eta\downarrow0}\limsup_n E^*
\sup_{\substack{(\mathbf h,l_z),(\mathbf h',l'_z)\in K_z\\
                \|(\mathbf h,l_z)-(\mathbf h',l'_z)\|\le\eta}}
|\mathcal Z_{A,z,n}^B(\mathbf h,l_z)-\mathcal Z_{A,z,n}^B(\mathbf h',l'_z)|=0.
\end{equation}
The same mapping gives the tie-rule bound \eqref{eq:patt-binomial-tie-rule-transfer}. Taking \(K_z\) to contain the corresponding projection of \(K\), the triangle inequality combines the two directional bounds in \eqref{eq:pate-two-arm-grid-centered-modulus}:
\begin{equation}\label{eq:pate-two-arm-grid-signed-modulus}
\lim_{\eta\downarrow0}\limsup_n
E^*\sup_{\substack{(\mathbf h,l),(\mathbf h',l')\in K\\
                    \|(\mathbf h,l)-(\mathbf h',l')\|\le\eta}}
\left|\sum_{z=0}^1(2z-1)
[\mathcal Z_{A,z,n}^B(\mathbf h,l_z)-\mathcal Z_{A,z,n}^B(\mathbf h',l'_z)]\right|=0.
\end{equation}
We now include the empirical scales in the process index. Apply \eqref{eq:patt-actual-scale-index-lipschitz} to all four \((r_{zk},b_z)\) coordinates. The same bounded derivatives and positive moment denominators and variances give \(L_n(0)=O_p(1)\), with scales uniformly positive on compact local-index sets with probability tending to one. The ratio derivative is
\begin{equation}\label{eq:pate-two-arm-grid-scale-derivative}
\nabla_hL_{n,z,k}(\mathbf h)=
\left.\nabla_{\boldsymbol \theta}\{s_{zk}^{\mathrm p}(\boldsymbol\theta)/
                     \widehat s_{n,zk}(\boldsymbol\theta)\}
\right|_{\boldsymbol\theta=\boldsymbol\theta_A^*+\mathbf h/\sqrt n}.
\end{equation}
The derivative bound gives
\begin{equation}\label{eq:pate-two-arm-grid-tight-indices}
\begin{aligned}
\|L_n(\mathbf h)-L_n(\mathbf h')\|&\le C_H\|\mathbf h-\mathbf h'\|,
                 &&\|\mathbf h\|,\|\mathbf h'\|\le H,\\
\sup_{\|\mathbf h\|\le H}\|L_n(\mathbf h)\|
&\le\|L_n(0)\|+C_HH=O_p(1)
\end{aligned}
\end{equation}
Empirical centers cancel, giving
\begin{equation}\label{eq:pate-two-arm-grid-empirical-distances}
\begin{aligned}
D_{z,\boldsymbol\theta,k}^{\mathrm e}(\mathbf X_i)
 -D_{z,\boldsymbol\theta,k}^{\mathrm e}(\mathbf X_j)
&=\frac{r_{zk}(\mathbf X_i;\boldsymbol\theta)-r_{zk}(\mathbf X_j;\boldsymbol\theta)}
             {\widehat s_{n,zk}(\boldsymbol\theta)}\\
&=\frac{s_{zk}^{\mathrm p}(\boldsymbol\theta)}{\widehat s_{n,zk}(\boldsymbol\theta)}
 [D_{z,\boldsymbol\theta,k}^{\mathrm p}(\mathbf X_i)
                   -D_{z,\boldsymbol\theta,k}^{\mathrm p}(\mathbf X_j)].
\end{aligned}
\end{equation}
This identifies the empirical distance arrays with those of \(\mathcal D_{z,n,\mathbf h,L_{n,z}(\mathbf h)}\). Equation~\eqref{eq:patt-binomial-tie-rule-transfer} handles the specified tie rule.
\par\smallskip\noindent\emph{Local expectations.}\quad
To recover the uncentered processes, we compute their mean increments. Suppose in addition that Assumption~\ref{ass:pate-two-arm-grid-signed-density} holds. For each arm \(z\), we apply the signed-density calculation in the proof of Lemma~\ref{lem:patt-signed-density-control-drift} to the raw score \(S_{z,*}\), observed weight \(w_{A,z}\), and residual \(\zeta_{A,z}\), with arm \(z\) providing matches for observations in arm \(1-z\). The chart, support and density conditions are those in Assumption~\ref{ass:pate-two-arm-grid-signed-density}. Baseline centering and continuity give
\begin{equation}\label{eq:pate-two-arm-grid-signed-cancellation}
\int a_{z,w}(s,l)\,\nu_{z,w}(dl)=0
\quad\text{for every }s.
\end{equation}
For \(\boldsymbol\theta_t=\boldsymbol\theta_A^*+t\mathbf h\), the transformed chart and derivative are
\begin{equation}\label{eq:pate-two-arm-grid-inverse-chart}
T_{z,t}:=A_{z,*}A_{z,\boldsymbol\theta_t}^{-1}S_{z,\boldsymbol\theta_t}
       =S_{z,*}+C_{z,t}L_z,\qquad
C_{z,t}:=A_{z,*}A_{z,\boldsymbol\theta_t}^{-1}B_{z,\boldsymbol\theta_t}-B_{z,*}.
\end{equation}
\begin{equation}\label{eq:pate-two-arm-grid-chart-direction}
\dot C_{z,\mathbf h}=-\dot A_{z,\mathbf h}A_{z,*}^{-1}B_{z,*}+\dot B_{z,\mathbf h},
\qquad C_{z,t}=t\dot C_{z,\mathbf h}+O(t^2).
\end{equation}
Set
\begin{equation}\label{eq:pate-two-arm-grid-conditional-drift}
\mathbf F^{\varepsilon}_{A,z,\mathbf h,w}(s)=\int a_{z,w}(s,l)\dot C_{z,\mathbf h}l\,\nu_{z,w}(dl),
\qquad
\dot m^{\varepsilon}_{A,z,\mathbf h}(s,w)=-\frac{\operatorname{div}\mathbf F^{\varepsilon}_{A,z,\mathbf h,w}(s)}{q_{z,w}(s)}
\quad(s\in N_z),
\end{equation}
and extend by zero outside \(N_z\). The signed-density expansion, uniform also near the boundary of its support, gives
\begin{equation}\label{eq:pate-two-arm-grid-root-projection}
E_s(\zeta_{A,z}\mid Z=z,T_{z,t},w_{A,z})
=t \dot m^{\varepsilon}_{A,z,\mathbf h}(S_{z,*},w_{A,z})+O(t^2).
\end{equation}
Here \(\dot m^{\varepsilon}_{A,z,\mathbf h}\) is uniformly bounded and Lipschitz in \(s\), uniformly over almost every \(w\) for \(\mathbf h\) in a compact set, and is linear in \(\mathbf h\). Positive deterministic diagonal scaling preserves this conditioning field.

Let \((\mathbf h_n,l_{z,n})\to(\mathbf h,l_z)\) be a deterministic sequence in a fixed compact local-index set. The exact reuse identity gives
\begin{equation}\label{eq:pate-two-arm-grid-reuse-mass}
\sum_{j:Z_j=z}K_{n,z}^{A}(j;\mathcal D_{z,n,\mathbf h_n,l_{z,n}})
=\mathbf1_{\{n_z\ge M\}}\sum_{i:Z_i=1-z}w_{A,1-z}(\mathbf X_i)
\le nw_{\max}.
\end{equation}
\begin{equation}\label{eq:pate-two-arm-grid-projected-expectation}
\begin{aligned}
\frac{E_s\mathcal I_{A,z,n;n}(\mathbf h_n,l_{z,n})}{\sqrt n}
&=E_s\!\left[\frac1{\sqrt n}\sum_{j:Z_j=z}
 K_{n,z}^{A}(j;\mathcal D_{z,n,\mathbf h_n,l_{z,n}})
 E_s(\varepsilon_{z,j}^A\mid
 Z_j,\mathcal D_{z,n,\mathbf h_n,l_{z,n}}(\mathbf X_j),w_{A,Z_j})\right]\\
&=E_s\!\left[\frac1n\sum_{j:Z_j=z}
 K_{n,z}^{A}(j;\mathcal D_{z,n,\mathbf h_n,l_{z,n}})
 \dot m^{\varepsilon}_{A,z,\mathbf h}(S_{z,*,j},w_{A,z,j})\right]+o(1).
\end{aligned}
\end{equation}
Mass conservation bounds the summed projection error by \(O(n^{-1/2})\). Linearity then permits \(\mathbf h_n\to \mathbf h\). To evaluate the remaining weighted average, use the coefficient \(K_{A,z}^\infty(d;w)\) in \eqref{eq:pate-two-arm-joint-poisson-coefficient}, with first moment \(\kappa_{A,z}(d,w)\). The score expansion is
\begin{equation}\label{eq:pate-two-arm-grid-score-expansion}
\begin{aligned}
\mathcal D_{z,n,\mathbf h_n,l_{z,n}}(\mathbf X)
&=\mathbf D_z^*(\mathbf X)
 +n^{-1/2}\dot{\mathbf D}_z^*(\mathbf X)\mathbf h_n
 +n^{-1/2}\operatorname{diag}(l_{z,n})\mathbf D_z^*(\mathbf X)
 +o(n^{-1/2}),
\end{aligned}
\end{equation}
where \(\dot{\mathbf D}_z^*=\nabla_{\boldsymbol \theta}\mathbf D_{z,\boldsymbol\theta}^{\mathrm p} |_{\boldsymbol\theta=\boldsymbol\theta_A^*}\). The diagonal term cancels in relative local positions. The one-root limit with these derivative and drift marks, and the inflated-count moment bound, give
\begin{equation}\label{eq:pate-two-arm-grid-local-mean}
\frac{E_s\mathcal I_{A,z,n;n}(\mathbf h_n,l_{z,n})}{\sqrt n}
\longrightarrow
E_s[\mathbf1\{Z=z\}\kappa_{A,z}(\mathbf D_z^*,w_{A,z})
                         \dot m^{\varepsilon}_{A,z,\mathbf h}(S_{z,*},w_{A,z})]
=:\Delta_{A,z}(\mathbf h).
\end{equation}
The same limit holds along every convergent index sequence. Since the limiting mean is continuous, sequential compactness gives
\begin{equation}\label{eq:pate-two-arm-grid-uniform-mean}
\sup_{(\mathbf h,l_z)\in K_z}
\left|\frac{E_s\mathcal I_{A,z,n;n}(\mathbf h,l_z)}{\sqrt n}
                  -\Delta_{A,z}(\mathbf h)\right|\longrightarrow0.
\end{equation}
Thus
\begin{equation}\label{eq:pate-two-arm-grid-mean-modulus}
\begin{aligned}
&\frac{|E_s\mathcal I_{A,z,n;n}(\mathbf h,l_z)
         -E_s\mathcal I_{A,z,n;n}(\mathbf h',l'_z)|}{\sqrt n}\\
&\le2\sup_{(u,v)\in K_z}
 \left|n^{-1/2}E_s\mathcal I_{A,z,n;n}(u,v)-\Delta_{A,z}(u)\right|
 +\|(\Delta_{A,z}(\mathbf e_r))_{r=1}^{p_A}\|\|\mathbf h-\mathbf h'\|.
\end{aligned}
\end{equation}
The signed displacement is
\begin{equation}\label{eq:pate-two-arm-grid-signed-mean}
\frac{\Delta_{A,1}(\mathbf h)-\Delta_{A,0}(\mathbf h)}{\bar w_A}.
\end{equation}
Let \(\mathbf e_r\) be the \(r\)th coordinate vector in \(\mathbb R^{p_A}\), where \(p_A=\dim\boldsymbol\theta_A\), and define
\[
\boldsymbol b_{A,\mathrm{loc}}=\left(
\frac{\Delta_{A,1}(\mathbf e_r)-\Delta_{A,0}(\mathbf e_r)}{\bar w_A}
\right)_{r=1}^{p_A}.
\]
The change in the limiting mean in \eqref{eq:pate-two-arm-grid-signed-mean} is therefore \(\boldsymbol b_{A,\mathrm{loc}}^\top \mathbf h\). When both own-arm and joined-field mean limits apply, their integrated constants agree by the exact conditional-expectation identity in Theorem~\ref{lem:pate-two-arm-joint-local-mean}.

\par\smallskip\noindent\emph{(ii) Transfer to empirical normalization.}\quad
Fix a donor arm \(z\), deterministic \(\mathbf h_n\to \mathbf h\), and \(l_z\in\mathbb R^2\). Set \(\boldsymbol\theta_n=\boldsymbol\theta_A^\ast+\mathbf h_n/\sqrt n\) and
\begin{equation}\label{eq:pate-two-arm-bridge-coefficient-difference}
\begin{aligned}
K_{n,z}^{l_z}(j)
&:=K_{n,z}^{A}
       (j;\mathcal D_{z,n,\mathbf h_n,l_z}),\\
K_{n,z}^{0}(j)
&:=K_{n,z}^{A}
       (j;\mathcal D_{z,n,\mathbf h_n,0}),\\
\Delta K_{n,z}(j)
&:=K_{n,z}^{l_z}(j)-K_{n,z}^{0}(j).
\end{aligned}
\end{equation}
Both coefficients use the same observations and donor weights. Only the relative scale of the two matching coordinates changes. At root \(j\), put \(x_{n,i}=\sqrt n\{\mathbf D_{z,\boldsymbol\theta_n}^{\mathrm p}(\mathbf X_i) -\mathbf D_{z,\boldsymbol\theta_n}^{\mathrm p}(\mathbf X_j)\}\). The other metric has
\[
x_{n,i}^{l_z}=\operatorname{diag}(1+l_z/\sqrt n)x_{n,i},
\qquad
\sup_{\|x_{n,i}\|\le R}\|x_{n,i}^{l_z}-x_{n,i}\|
\le R\|\operatorname{diag}(l_z)\|/\sqrt n\to0.
\]
On a fixed local window, the scale perturbation vanishes. The \hyperref[app:qv-localization]{localization argument} with strict cone margins therefore gives the same marked limit for both metrics, without ties or cutoff-boundary points. Removing the window gives \(\Delta K_{n,z}(j)\to_p0\) at an integrated random donor root.

The common inflated count has all fixed conditional moments bounded and is independent of remaining root marks given baseline score and treatment. Hence
\[
E_s[\mathbf1(Z_j=z)|\Delta K_{n,z}(j)|^p|\varepsilon_{z,j}^A|^p]
\le C E_s[\mathbf1(Z=z)|\varepsilon_z^A|^p]<\infty.
\]
For \(p>2\), this gives uniform integrability of the squared product. Combining convergence at a random donor root with this bound, and using the unmarked count bound for the first moment, yields
\begin{equation}\label{eq:pate-two-arm-bridge-weighted-ui}
\begin{aligned}
E_s[\mathbf1\{Z_j=z\}\Delta K_{n,z}(j)^2
                                (\varepsilon_{z,j}^A)^2]
&\longrightarrow0,\\
E_s[\mathbf1\{Z_j=z\}|\Delta K_{n,z}(j)|]
&\longrightarrow0.
\end{aligned}
\end{equation}

Both deterministic metrics have the same own-row field
\[
\mathcal F_{n,j}^z=\sigma\{Z_j,\mathbf D_{z,\boldsymbol\theta_n}^{\mathrm p}(\mathbf X_j),
 w_{A,Z_j}(\mathbf X_j)\},\qquad
\mathcal F_n^z=\bigvee_j\mathcal F_{n,j}^z .
\]
The invertible diagonal map preserves this field, so both coefficient arrays can be held fixed in the same conditional calculation. Conditional product disintegration centers each residual at \(\eta_{n,j}^z=E_s(\varepsilon_{z,j}^A\mid\mathcal F_{n,j}^z)\), giving
\[
\begin{aligned}
&E_s\left[\left\{\frac1{\sqrt n}\sum_{j:Z_j=z}
 \Delta K_{n,z}(j)(\varepsilon_{z,j}^A-\eta_{n,j}^z)\right\}^2\right]\\
&=\frac1nE_s\sum_{j:Z_j=z}\Delta K_{n,z}(j)^2
       \operatorname{Var}_s(\varepsilon_{z,j}^A\mid\mathcal F_{n,j}^z)\\
&\le\frac1nE_s\sum_{j:Z_j=z}\Delta K_{n,z}(j)^2
       E_s\{(\varepsilon_{z,j}^A)^2\mid\mathcal F_{n,j}^z\}\\
&=\frac1n E_s\sum_{j:Z_j=z}
 \Delta K_{n,z}(j)^2(\varepsilon_{z,j}^A)^2\\
&\longrightarrow0.
\end{aligned}
\]
Conditional independence eliminates the cross-row products, and the tower property gives the last equality. For the predictable part, the own-arm signed-density expansion gives \(\eta_{n,j}^z=n^{-1/2}\dot m^{\varepsilon}_{A,z,\mathbf h_n}(S_{z,\ast,j},w_{A,z,j})+O(n^{-1})\) uniformly, with bounded \(\dot m^{\varepsilon}_{A,z,\mathbf h_n}\). The first-moment convergence in \eqref{eq:pate-two-arm-bridge-weighted-ui}, together with reuse-mass conservation, yields
\[
\begin{aligned}
E_s\left|\frac1{\sqrt n}\sum_{j:Z_j=z}
                  \Delta K_{n,z}(j)\eta_{n,j}^z\right|
&\le\frac Cn E_s\sum_{j:Z_j=z}|\Delta K_{n,z}(j)|+O(n^{-1/2})\\
&=C E_s[\mathbf1\{Z_1=z\}|\Delta K_{n,z}(1)|]+O(n^{-1/2})\\
&\longrightarrow0.
\end{aligned}
\]
Here exchangeability gives the equality, and \(\sum_{j:Z_j=z}|\Delta K_{n,z}(j)|\le2nw_{\max}\) bounds the contribution of the uniform \(O(n^{-1})\) projection remainder. Combining centered and predictable terms yields
\begin{equation}\label{eq:pate-two-arm-bridge-fixed-scale}
\frac{\mathcal I_{A,z,n;n}(\mathbf h_n,l_z)
             -\mathcal I_{A,z,n;n}(\mathbf h_n,0)}{\sqrt n}
\longrightarrow_p0
\end{equation}

On each compact scale box \(L\), \eqref{eq:pate-two-arm-grid-centered-modulus} and the \(l_z\)-independent mean limit \eqref{eq:pate-two-arm-grid-uniform-mean} give
\[
\begin{aligned}
&\lim_{\eta\downarrow0}\limsup_n P_s^*\!\left\{
 \sup_{\substack{l_z,l'_z\in L\\\|l_z-l'_z\|\le\eta}}
 \frac{|\mathcal I_{A,z,n;n}(\mathbf h_n,l_z)
             -\mathcal I_{A,z,n;n}(\mathbf h_n,l'_z)|}{\sqrt n}>\epsilon\right\}
=0.
\end{aligned}
\]
The fixed-scale comparison \eqref{eq:pate-two-arm-bridge-fixed-scale} controls a finite \(\eta\)-net of \(L\), and the preceding modulus controls the gaps between its points. Letting \(n\to\infty\) and \(\eta\downarrow0\) gives
\begin{equation}\label{eq:pate-two-arm-bridge-uniform-scale}
\sup_{l_z\in L}
 \frac{|\mathcal I_{A,z,n;n}(\mathbf h_n,l_z)
            -\mathcal I_{A,z,n;n}(\mathbf h_n,0)|}{\sqrt n}
\longrightarrow_{P_s}^\ast0.
\end{equation}
Tightness in part (i) permits \(l_z=L_{n,z}(\mathbf h_n)\), by expanding \(L\).

The pair-distance identity \eqref{eq:pate-two-arm-grid-empirical-distances} now identifies the empirical graphs. Applying the uniform tie comparison and cancelling the common \(H+E\) terms give
\[
\begin{aligned}
&T_{A,n}^{\mathrm e}(\mathbf h_n;\tau_{\mathrm{PATE}})
-T_{A,n}^{\mathrm p}(\mathbf h_n;\tau_{\mathrm{PATE}})\\
&=\frac1{\bar w_{A,n}}\sum_{z=0}^1(2z-1)
 \frac{\mathcal I_{A,z,n;n}(\mathbf h_n,L_{n,z}(\mathbf h_n))
              -\mathcal I_{A,z,n;n}(\mathbf h_n,0)}{\sqrt n}+o_p(1)\\
&\longrightarrow_p0.
\end{aligned}
\]
The error accounts for the specified tie rule. The exceptional count and scale events have vanishing probability, and \(\bar w_{A,n}\to_p\bar w_A>0\). The nuisance coordinate is identical, so Slutsky gives the joint transfer.

\par\smallskip\noindent\emph{(iii) Matching-coordinate comparison.}\quad
We apply the same modulus to the fitted nuisance root and its grid midpoint. The midpoint distance is at most \(\sqrt{p_A}\,\eta/2\), and the full-index distance on a compact root/scale event is at most \(\sqrt{1+C_{H+1}^2}\sqrt{p_A}\,\eta/2\) by \eqref{eq:pate-two-arm-grid-tight-indices}. Apply the centered signed modulus, Markov's inequality and the \(o(1)+O(\eta)\) mean bound in \eqref{eq:pate-two-arm-grid-mean-modulus}. Root and scale tightness then give
\begin{equation}\label{eq:pate-two-arm-grid-incoming-comparison}
\lim_{\eta\downarrow0}\limsup_n P_s\left\{
\frac1{\sqrt n}\left|
\sum_{z=0}^1(2z-1)
 [\mathcal I_{A,z,n;n}(\widehat{\mathbf h}_n,L_{n,z}(\widehat{\mathbf h}_n))
 -\mathcal I_{A,z,n;n}(\bar{\mathbf h}_{n,\eta},L_{n,z}(\bar{\mathbf h}_{n,\eta}))]
\right|>\epsilon\right\}=0.
\end{equation}
The difference identity reduces the comparison to the imputation numerators. The distance identity and tie bounds account for the empirical match sets, and the exceptional event has vanishing probability. Since \(\bar w_{A,n}\ge w_{\min}\), division gives
\begin{equation}\label{eq:pate-two-arm-grid-hbe-comparison}
\lim_{\eta\downarrow0}\limsup_n P_s\{
 |T_{A,n}^{\mathrm e}(\widehat{\mathbf h}_n;\tau_{\mathrm{PATE}})
  -T_{A,n}^{\mathrm e}(\bar{\mathbf h}_{n,\eta};\tau_{\mathrm{PATE}})|>\epsilon\}=0.
\end{equation}
\par\smallskip\noindent\emph{(iv) Bias correction and joint comparison.}\quad
We finally transfer the comparison from the matching coordinate to the bias-corrected estimator. Applying the sequential compactness argument in the proof of Theorem~\ref{thm:patt-sufficient-branch-actual-score-law} to the bias-correction condition for deterministic local sequences gives
\begin{equation}\label{eq:pate-two-arm-grid-bc-uniform-probability}
\sup_{\|\mathbf h\|\le H}P_s\{|R_{A,n}(\mathbf h)|>\epsilon\}\longrightarrow0.
\end{equation}
There are at most \(C_H(1+\eta^{-1})^{p_A}\) midpoints in a bounded ball, uniformly in the prescribed offset. Hence
\begin{equation}\label{eq:pate-two-arm-grid-bc-finite-cells}
\begin{aligned}
P_s\{|R_{A,n}(\bar{\mathbf h}_{n,\eta})|>\epsilon\}
&\le P_s(\|\widehat{\mathbf h}_n\|>H)
 +\sum_{\substack{\mathbf h\text{ a grid midpoint}\\\|\mathbf h\|\le H+1}}
                         P_s\{|R_{A,n}(\mathbf h)|>\epsilon\}\\
&\le P_s(\|\widehat{\mathbf h}_n\|>H)
 +C_H(1+\eta^{-1})^{p_A}
         \sup_{\|\mathbf h\|\le H+1}P_s\{|R_{A,n}(\mathbf h)|>\epsilon\}.
\end{aligned}
\end{equation}
For fixed \(\eta\), let \(n\to\infty\) and then \(H\to\infty\). This makes the grid remainder \(o_p(1)\). The second BC condition controls the fitted-root remainder. The finite-sample decomposition on the valid event is
\begin{equation}\label{eq:pate-two-arm-grid-bc-hbe-identity}
\begin{aligned}
&\sqrt n\{\widehat\tau_{\mathrm{PATE}}^{A,\mathrm{BC}}
(\boldsymbol\theta_A^*+\mathbf h/\sqrt n)-\tau_{\mathrm{PATE}}\}\\
&=T_{A,n}^{\mathrm e}(\mathbf h;\tau_{\mathrm{PATE}})-R_{A,n}(\mathbf h).
\end{aligned}
\end{equation}
Subtract at the two indices and apply \eqref{eq:pate-two-arm-grid-hbe-comparison} and the two BC remainders to obtain
\begin{equation}\label{eq:pate-two-arm-grid-actual-comparison}
\lim_{\eta\downarrow0}\limsup_n P_s\left\{
\sqrt n\left|
\widehat\tau_{\mathrm{PATE}}^{A,\mathrm{BC}}(\widehat{\boldsymbol \theta}_A)
-\widehat\tau_{\mathrm{PATE}}^{A,\mathrm{BC}}(\bar{\boldsymbol \theta}_{A,n,\eta})
\right|>\epsilon\right\}=0.
\end{equation}
For the joint comparison, denote the centered statistic within this proof by \(T_{A,n}^{\mathrm{BC}}(\mathbf h)=\sqrt n\{\widehat\tau_{\mathrm{PATE}}^{A,\mathrm{BC}} (\boldsymbol\theta_A^*+\mathbf h/\sqrt n)-\tau_{\mathrm{PATE}}\}\). Both pairs retain the same nuisance coordinate, so for bounded Lipschitz \(\Psi\),
\begin{equation}\label{eq:pate-two-arm-grid-joint-bl-comparison}
\begin{aligned}
&|E_s\Psi(T_{A,n}^{\mathrm{BC}}(\widehat{\mathbf h}_n),\widehat{\mathbf h}_n)
       -E_s\Psi(T_{A,n}^{\mathrm{BC}}(\bar{\mathbf h}_{n,\eta}),\widehat{\mathbf h}_n)|\\
&\le\operatorname{Lip}(\Psi)\epsilon+
2\|\Psi\|_\infty P_s\{|T_{A,n}^{\mathrm{BC}}(\widehat{\mathbf h}_n)
                         -T_{A,n}^{\mathrm{BC}}(\bar{\mathbf h}_{n,\eta})|>\epsilon\}.
\end{aligned}
\end{equation}

\end{proof}

\begin{theorem}[Fitted-score PATE limit under joined-score conditions]
\label{thm:pate-two-arm-actual-joint-law}
\label{cor:pate-two-arm-actual-positive-variance}
Retain the conditions for part (ii) of Theorem~\ref{prop:pate-two-arm-empirical-joint-transfer} and the two specified BC quantifiers in Assumption~\ref{ass:pate-two-arm-grid-bc}. Use the stated estimating equations and their ordinary first-step expansion, with nonsingular \(\boldsymbol\Gamma_A\) and positive definite \(\boldsymbol\Sigma_{A,U}\). The column vector \(\boldsymbol b_{A,\mathrm{loc}}\) records the local limiting mean change caused by perturbing the matching parameter while keeping the selected-sample law fixed. Its coordinates are
\begin{equation}\label{eq:pate-two-arm-actual-drift}
b_{A,\mathrm{loc},r}
:=\frac{\Delta_{A,1}(\mathbf e_r)-\Delta_{A,0}(\mathbf e_r)}{\bar w_A},
\qquad r=1,\ldots,p_A .
\end{equation}
The joined-field and own-arm expectation limits identify the same integrated incoming means \(\Delta_{A,z}(\mathbf h)\). Let
\begin{equation}\label{eq:pate-two-arm-actual-baseline-pair}
\begin{pmatrix}G_{\tau,A}\\\mathbf G_{\theta,A}\end{pmatrix}
\sim N\left(0,
\begin{pmatrix}
V_{A,\mathrm{known\text{-}score}}
       &-\boldsymbol\gamma_{A,1}^{\top}\boldsymbol\Gamma_A^{-T}\\
-\boldsymbol\Gamma_A^{-1}\boldsymbol\gamma_{A,1}&\boldsymbol\Sigma_{A,\theta}
\end{pmatrix}\right),
\qquad
\boldsymbol\Sigma_{A,\theta}=\boldsymbol\Gamma_A^{-1}\boldsymbol\Sigma_{A,U}\boldsymbol\Gamma_A^{-T}.
\end{equation}
With the covariance constants from the joint martingale proof, the fitted-score estimator satisfies
\begin{equation}\label{eq:pate-two-arm-actual-joint-law}
\begin{pmatrix}
\sqrt n\{\widehat\tau_{\mathrm{PATE}}^{A,\mathrm{BC}}
                   (\widehat{\boldsymbol \theta}_A)-\tau_{\mathrm{PATE}}\}\\
\sqrt n(\widehat{\boldsymbol \theta}_A-\boldsymbol\theta_A^\ast)
\end{pmatrix}
\Rightarrow
\begin{pmatrix}G_{\tau,A}+\boldsymbol b_{A,\mathrm{loc}}^\top \mathbf G_{\theta,A}\\\mathbf G_{\theta,A}\end{pmatrix}.
\end{equation}
Its first coordinate is centered Gaussian with finite variance
\begin{equation}\label{eq:pate-two-arm-actual-variance}
\begin{aligned}
V_{A,\mathrm{actual}}
&=V_{A,\mathrm{known\text{-}score}}
 -\boldsymbol\gamma_{A,1}^\top\boldsymbol\Sigma_{A,U}^{-1}\boldsymbol\gamma_{A,1}\\
&\quad{}+(\boldsymbol b_{A,\mathrm{loc}}-\boldsymbol\Gamma_A^\top\boldsymbol\Sigma_{A,U}^{-1}\boldsymbol\gamma_{A,1})^\top
 \boldsymbol\Sigma_{A,\theta}
 (\boldsymbol b_{A,\mathrm{loc}}-\boldsymbol\Gamma_A^\top\boldsymbol\Sigma_{A,U}^{-1}\boldsymbol\gamma_{A,1}).
\end{aligned}
\end{equation}
Its covariance with the nuisance root is
\begin{equation}\label{eq:pate-two-arm-actual-root-covariance}
\operatorname{Cov}(\mathbf G_{\theta,A},G_{\tau,A}+\boldsymbol b_{A,\mathrm{loc}}^\top \mathbf G_{\theta,A})
=-\boldsymbol\Gamma_A^{-1}\boldsymbol\gamma_{A,1}+\boldsymbol\Sigma_{A,\theta}\boldsymbol b_{A,\mathrm{loc}} .
\end{equation}
For positivity, let \(\mathcal F_\ast=\sigma(Z,\mathbf D_0^\ast,\mathbf D_1^\ast, w_{A,Z})\), \(\varepsilon_A=Y-m_Z^A(\mathbf D_Z^\ast)\), and \(v_\ast=\operatorname{Var}_s(\varepsilon_A\mid\mathcal F_\ast)\). If \(v_\ast>0\) on a positive-probability observed-arm set where the opposing query density is positive, then \(0<V_{A,\mathrm{actual}}<\infty\).
\end{theorem}
\begin{proof}
\par\smallskip\noindent\emph{(i) Fitted-score limit.}\quad
The deterministic joint limit identifies the grid law, and the preceding comparison transfers that law to the fitted root. Within this proof, write
\[
T_{A,n}^{\mathrm{BC}}(\mathbf h)
=\sqrt n\{\widehat\tau_{\mathrm{PATE}}^{A,\mathrm{BC}}
(\boldsymbol\theta_A^*+\mathbf h/\sqrt n)-\tau_{\mathrm{PATE}}\},
\qquad
\widehat{\mathbf h}_n=\sqrt n(\widehat{\boldsymbol \theta}_A-\boldsymbol\theta_A^*).
\]
For every deterministic \(\mathbf h_n\to \mathbf h\), Theorem~\ref{prop:pate-two-arm-empirical-joint-transfer}(ii) and the first, deterministic-local BC condition in \eqref{eq:pate-two-arm-grid-bc-quantifiers} give
\[
T_{A,n}^{\mathrm{BC}}(\mathbf h_n)=T_{A,n}^{\mathrm e}(\mathbf h_n;\tau_{\mathrm{PATE}})+o_p(1)
\]
and hence
\begin{equation}\label{eq:pate-two-arm-grid-joint-law-interface}
\begin{pmatrix}
\sqrt n\{\widehat\tau_{\mathrm{PATE}}^{A,\mathrm{BC}}
(\boldsymbol\theta_A^*+\mathbf h_n/\sqrt n)-\tau_{\mathrm{PATE}}\}\\
\sqrt n(\widehat{\boldsymbol \theta}_A-\boldsymbol\theta_A^*)
\end{pmatrix}
\Rightarrow
\begin{pmatrix}G_{\tau,A}+\boldsymbol b_{A,\mathrm{loc}}^\top \mathbf h\\\mathbf G_{\theta,A}\end{pmatrix}.
\end{equation}
The nuisance covariance is positive definite, so Gaussian grid-cell boundaries have probability zero. Tightness confines \(\widehat{\mathbf h}_n\) to a compact set with arbitrarily high probability. For each fixed mesh width, that set contains finitely many cells. The joint limit above controls both the statistic at each midpoint and the event that the fitted root selects its cell. Lemma~\ref{lem:joint-local-grid-selection} therefore gives the joint grid limit \((G_{\tau,A}+\boldsymbol b_{A,\mathrm{loc}}^\top \mathbf G_{\theta,A},\mathbf G_{\theta,A})\). The comparison between estimated and grid parameters in \eqref{eq:pate-two-arm-grid-actual-comparison}, together with the bias-correction remainder for the fitted match sets, proves \eqref{eq:pate-two-arm-actual-joint-law}. Expanding the variance and completing the quadratic form, using \eqref{eq:patt-joint-projection-matrix-product}-- \eqref{eq:patt-joint-projection-quadratic-product} for the present covariance matrices, gives
\begin{equation}\label{eq:pate-two-arm-grid-variance-implication}
\begin{aligned}
\operatorname{Var}(G_{\tau,A}+\boldsymbol b_{A,\mathrm{loc}}^\top \mathbf G_{\theta,A})
&=\operatorname{Var}(G_{\tau,A})+2\boldsymbol b_{A,\mathrm{loc}}^\top\operatorname{Cov}(\mathbf G_{\theta,A},G_{\tau,A})
  +\boldsymbol b_{A,\mathrm{loc}}^\top\operatorname{Var}(\mathbf G_{\theta,A})\boldsymbol b_{A,\mathrm{loc}}\\
&=V_{A,\mathrm{known\text{-}score}}
 -2\boldsymbol b_{A,\mathrm{loc}}^\top\boldsymbol\Gamma_A^{-1}\boldsymbol\gamma_{A,1}
 +\boldsymbol b_{A,\mathrm{loc}}^\top\boldsymbol\Sigma_{A,\theta}\boldsymbol b_{A,\mathrm{loc}}\\
&=V_{A,\mathrm{known\text{-}score}}
 -\boldsymbol\gamma_{A,1}^\top\boldsymbol\Sigma_{A,U}^{-1}\boldsymbol\gamma_{A,1}\\
&\quad{}+(\boldsymbol b_{A,\mathrm{loc}}-\boldsymbol\Gamma_A^\top\boldsymbol\Sigma_{A,U}^{-1}\boldsymbol\gamma_{A,1})^\top
 \boldsymbol\Sigma_{A,\theta}
 (\boldsymbol b_{A,\mathrm{loc}}-\boldsymbol\Gamma_A^\top\boldsymbol\Sigma_{A,U}^{-1}\boldsymbol\gamma_{A,1})\\
&=V_{A,\mathrm{actual}},
\end{aligned}
\end{equation}
and
\[
\begin{aligned}
\operatorname{Cov}(\mathbf G_{\theta,A},G_{\tau,A}+\boldsymbol b_{A,\mathrm{loc}}^\top \mathbf G_{\theta,A})
&=\operatorname{Cov}(\mathbf G_{\theta,A},G_{\tau,A})+\operatorname{Var}(\mathbf G_{\theta,A})\boldsymbol b_{A,\mathrm{loc}}\\
&=-\boldsymbol\Gamma_A^{-1}\boldsymbol\gamma_{A,1}+\boldsymbol\Sigma_{A,\theta}\boldsymbol b_{A,\mathrm{loc}}.
\end{aligned}
\]
The stated moment conditions make these entries finite.
\par\smallskip\noindent\emph{(ii) Positive variance.}\quad
We bound the variance below by the randomness of the local reuse coefficient after the observed row is fixed. Conditionally on the baseline row field \(\mathcal F_*\), draw the auxiliary coefficient \(K_{A,Z}^\infty\) independently of the remaining root row with its law given \((Z,\mathbf D_Z^*,w_{A,Z})\). The joint marked calculation gives this conditional law and finite second moments. Put \(u_*=E_s(\mathbf U_A\mid\mathcal F_*)\) and
\[
R_A=\frac{(2Z-1)(w_{A,Z}+K_{A,Z}^\infty)\varepsilon_A}{\bar w_A}
-\boldsymbol b_{A,\mathrm{loc}}^\top\boldsymbol\Gamma_A^{-1}(\mathbf U_A-u_*).
\]
Baseline centering gives \(E(R_A\mid\mathcal F_*)=0\). The conditional-projection covariance decomposition and total variance give
\[
\begin{aligned}
V_{A,\mathrm{actual}}
&=\operatorname{Var}_s\!\left\{
\frac{h_{A,1}-h_{A,0}}{\bar w_A}-\boldsymbol b_{A,\mathrm{loc}}^\top\boldsymbol\Gamma_A^{-1}u_*\right\}
+E(R_A^2)\\
&=\operatorname{Var}_s\!\left\{
\frac{h_{A,1}-h_{A,0}}{\bar w_A}-\boldsymbol b_{A,\mathrm{loc}}^\top\boldsymbol\Gamma_A^{-1}u_*\right\}
 +E\{\operatorname{Var}(R_A\mid O)\}
 +E\{E(R_A\mid O)^2\}\\
&\ge E\{\operatorname{Var}(R_A\mid O)\}.
\end{aligned}
\]
Given \(O\), only the auxiliary coefficient is random. Thus
\[
\begin{aligned}
E\{\operatorname{Var}(R_A\mid O)\}
&=\bar w_A^{-2}E_s\!\left[
 \varepsilon_A^2\operatorname{Var}
       \{K_{A,Z}^\infty\mid Z,\mathbf D_Z^*,w_{A,Z}\}\right]\\
&=\bar w_A^{-2}E_s\!\left[
 v_*\operatorname{Var}
       \{K_{A,Z}^\infty\mid Z,\mathbf D_Z^*,w_{A,Z}\}\right].
\end{aligned}
\]
Consequently,
\begin{equation}\label{eq:pate-two-arm-actual-variance-lower-bound}
V_{A,\mathrm{actual}}
\ge\bar w_A^{-2}E_s\left[
 v_\ast
 \operatorname{Var}\{K_{A,Z}^{\infty}\mid Z,\mathbf D_Z^\ast,w_{A,Z}\}
 \right].
\end{equation}

At an interior root with positive query intensity \(\lambda_{1-z}(d)\), condition on the donor configuration and weight marks. The finite-cone shield makes its order-\(M\) catchment \(C\) bounded. A ball \(B\) of radius less than half the nearest other donor's distance is contained in \(C\), so \(0<|B|\le|C|<\infty\) almost surely. The independent query process has
\[
P\{N(C)=0\mid\text{donors}\}=e^{-\lambda_{1-z}(d)|C|}>0,\qquad
P\{N(B)\ge1\mid\text{donors}\}=1-e^{-\lambda_{1-z}(d)|B|}>0.
\]
The coefficient is zero on the first event and positive on the second, so its conditional law is nondegenerate. Integrating over donor configurations gives \(\operatorname{Var}(K_{A,z}^\infty\mid Z=z,\mathbf D_z^*=d,w_{A,z}=w)>0\) almost everywhere under the overlap condition. The residual condition then makes the lower bound positive.

\par\smallskip\noindent\emph{(iii) Local response coefficient.}\quad
Define the coefficient
\[
\widetilde{\boldsymbol \gamma}_{A,2}
=\boldsymbol b_{A,\mathrm{loc}}-\boldsymbol\Gamma_A^\top\boldsymbol\Sigma_{A,U}^{-1}\boldsymbol\gamma_{A,1}.
\]
Completing the quadratic form in \eqref{eq:pate-two-arm-grid-variance-implication} gives
\[
\begin{aligned}
&V_{A,\mathrm{known\text{-}score}}
-\boldsymbol\gamma_{A,1}^\top\boldsymbol\Sigma_{A,U}^{-1}\boldsymbol\gamma_{A,1}
+\widetilde{\boldsymbol \gamma}_{A,2}^\top\boldsymbol\Sigma_{A,\theta}\widetilde{\boldsymbol \gamma}_{A,2}\\
&=V_{A,\mathrm{actual}}.
\end{aligned}
\]
Under the auxiliary-law conditions, this coefficient also gives the local mean of the statistic centered at the original target. It equals the target derivative \(\boldsymbol\gamma_{A,2}\) only when the local limit centered at the moving target has mean zero, as specified in Section~\ref{score-indexed-target-maps}.
\end{proof}
\subsubsection{PATE with correctly specified prognostic means}
\label{app:pate-fullx-branch}

When the reduced means equal the full conditional means, the residuals have conditional mean zero for every deterministic local matching graph. We derive the joint limit under this condition and the stated bias-correction conditions.

\begin{theorem}[PATE under full-covariate centering]
\label{thm:pate-fullx-actual-joint-law}
Fix \(A\in\{R,P\}\) and consider the PATE estimator in Proposition~\ref{prop:exact-pate-score-space-decomposition}. Assume:
\begin{enumerate}
\item[(i)] Population recovery, reduced-mean identification, and Assumption~\ref{ass:pate-two-arm-grid-process} hold. The empirical moment weights are \(b_z(O)\), the baseline variances are positive, and the deterministic baseline scales equal their population limits.
\item[(ii)] The finite logistic root, positive Hessian, and positive prognostic Gram conditions in Lemma~\ref{lem:primitive-logistic-linear-roots} hold for both arms. For some \(p>2\), \(E_s|Y|^p<\infty\), and \(\boldsymbol\Sigma_{A,U}\) is positive definite. The propensity estimating equation is weighted for \(A=R\) and unweighted for \(A=P\). Both prognostic estimating equations are unweighted.
\item[(iii)] For \(z=0,1\),
\begin{equation}\label{eq:pate-fullx-mean-condition}
\begin{aligned}
m_z^A\{\mathbf D_z^*(\mathbf X)\}
&=\mu_z(\mathbf X)=\mathbf X^\top\boldsymbol\beta_{A,z}^*,\\
E_s(Y\mid\mathbf X,Z=z)&=\mu_z(\mathbf X).
\end{aligned}
\end{equation}
The second equality holds under treatment and outcome-selection ignorability when both prognostic means are correct. The propensity model may be misspecified.
\item[(iv)] Both bias-correction conditions in \eqref{eq:pate-two-arm-grid-bc-quantifiers} hold, with \(R_{A,n}\) evaluated on the specified empirical graphs:
\[
R_{A,n}(\mathbf h_n)\to_p0\quad\hbox{for every deterministic }\mathbf h_n\to \mathbf h,
\qquad
R_{A,n}(\widehat{\mathbf h}_n)\to_p0,
\quad \widehat{\mathbf h}_n=\sqrt n(\widehat{\boldsymbol \theta}_A-\boldsymbol\theta_A^*).
\]
\end{enumerate}
Then the estimator and nuisance root satisfy
\begin{equation}\label{eq:pate-fullx-actual-joint-law}
\begin{pmatrix}
\sqrt n\{\widehat\tau_{\mathrm{PATE}}^{A,\mathrm{BC}}
                     (\widehat{\boldsymbol \theta}_A)-\tau_{\mathrm{PATE}}\}\\
\sqrt n(\widehat{\boldsymbol \theta}_A-\boldsymbol\theta_A^*)
\end{pmatrix}
\Rightarrow
\begin{pmatrix}G_{\tau,A}\\\mathbf G_{\theta,A}\end{pmatrix},
\end{equation}
where the centered Gaussian pair satisfies
\begin{equation}\label{eq:pate-fullx-baseline-pair}
\begin{aligned}
\operatorname{Var}(G_{\tau,A})&=V_{A,\mathrm{known\text{-}score}},\\
\operatorname{Cov}(\mathbf G_{\theta,A},G_{\tau,A})&=-\boldsymbol\Gamma_A^{-1}\boldsymbol\gamma_{A,1},\\
\operatorname{Var}(\mathbf G_{\theta,A})&=\boldsymbol\Sigma_{A,\theta}
 =\boldsymbol\Gamma_A^{-1}\boldsymbol\Sigma_{A,U}\boldsymbol\Gamma_A^{-T}.
\end{aligned}
\end{equation}
With \(\boldsymbol\gamma_{A,1}\) defined using \(\mathbf U_A\), the limiting variance is finite and is strictly positive if \(V_{H,A}>0\) or
\[
\sum_z E_s[\mathbf1(Z=z)\operatorname{Var}_s(Y\mid\mathbf X,Z=z)]>0.
\]

\end{theorem}

\begin{proof}
\par\smallskip\noindent\emph{Martingale representation.}\quad
We first establish the joint limit at deterministic local scores. Let \(\mathcal G_n^X=\sigma\{(\mathbf X_i,Z_i):i\le n\}\). The corresponding graphs, observed weights and denominators are measurable with respect to this field, and the outcomes are conditionally independent. On \(Z_i=z\), \eqref{eq:pate-fullx-mean-condition} gives
\[
\begin{aligned}
E_s(\varepsilon_{z,i}^A\mid\mathcal G_n^X)
&=E_s(Y_i\mid\mathbf X_i,Z_i=z)-m_z^A(\mathbf D_{z,i}^*)\\
&=\mu_z(\mathbf X_i)-m_z^A(\mathbf D_{z,i}^*)\\
&=0.
\end{aligned}
\]
Consequently
\begin{equation}\label{eq:pate-fullx-zero-incoming-mean}
\begin{aligned}
E_s\mathcal I_{A,z,n;n}(\mathbf h,l_z)
&=E_s\sum_{j:Z_j=z}K_{n,z}^{A}(j;\mathcal D_{z,n,\mathbf h,l_z})
                  \varepsilon_{z,j}^A\\
&=E_s\sum_{j:Z_j=z}K_{n,z}^{A}(j;\mathcal D_{z,n,\mathbf h,l_z})
                  E_s(\varepsilon_{z,j}^A\mid\mathcal G_n^X)\\
&=0.
\end{aligned}
\end{equation}
The count moment bound justifies the conditional expectation, including for each sample size and radius truncation. By the tower property, both own-score/weight and joint-score/weight centering also hold.

Fix deterministic \(\mathbf h_n\to \mathbf h\), put \(\boldsymbol\theta_n=\boldsymbol\theta_A^*+\mathbf h_n/\sqrt n\), and write \(\sigma_z^2(\mathbf X)=\operatorname{Var}_s(Y\mid\mathbf X,Z=z)\). After revealing \(\mathcal G_n^X\), we define the outcome-row differences by
\begin{equation}\label{eq:pate-fullx-second-stage}
D_{n,i}^X=\frac1{\sqrt n}
\begin{pmatrix}
\mathbf1(Z_i=1)
 \{w_{A,1}(\mathbf X_i)+K_{n,1}^{A}
                  (i;\mathbf D_{1,\boldsymbol\theta_n}^{\mathrm p})\}
                         \varepsilon_{1,i}^A/\bar w_A\\
\mathbf1(Z_i=0)
 \{w_{A,0}(\mathbf X_i)+K_{n,0}^{A}
                  (i;\mathbf D_{0,\boldsymbol\theta_n}^{\mathrm p})\}
                         \varepsilon_{0,i}^A/\bar w_A\\
\{\mathbf U_A(O_i;\boldsymbol\theta_A^*)
       -E_s[\mathbf U_A(O_i;\boldsymbol\theta_A^*)\mid\mathbf X_i,Z_i]\}
\end{pmatrix}.
\end{equation}
For \(k=0,\ldots,2n\), define
\[
\mathcal F_{n,k}^X=
\begin{cases}
\sigma\{(\mathbf X_i,Z_i):1\le i\le k\},&0\le k\le n,\\
\mathcal G_n^X\vee\sigma(Y_1,\ldots,Y_{k-n}),&n<k\le2n,
\end{cases}
\]
and the complete vector array
\[
\boldsymbol\xi_{n,k}^X=
\begin{cases}
\displaystyle\frac1{\sqrt n}
\begin{pmatrix}
h_{A,1}(O_k)/\bar w_A\\
h_{A,0}(O_k)/\bar w_A\\
E_s\{\mathbf U_A(O_k;\boldsymbol\theta_A^*)\mid\mathbf X_k,Z_k\}
\end{pmatrix},&1\le k\le n,\\[3mm]
D_{n,k-n}^X,&n<k\le2n.
\end{cases}
\]
Population recovery and the estimating equations center the first-stage rows. Full-covariate residual centering handles the outcome stage, giving
\[
E_s(\boldsymbol\xi_{n,k}^X\mid\mathcal F_{n,k-1}^X)=0,
\qquad 1\le k\le2n.
\]
Thus \(\{\sum_{j=1}^k\boldsymbol\xi_{n,j}^X,\mathcal F_{n,k}^X\}\) is a martingale. Conditional independence gives
\[
\begin{aligned}
&\operatorname{Var}_s\!\left[
 \frac1{\bar w_A\sqrt n}\sum_{i:Z_i=z}
 \{w_{A,z}(\mathbf X_i)+K_{n,z}^{A}
                    (i;\mathbf D_{z,\boldsymbol\theta_n}^{\mathrm p})\}
 \varepsilon_{z,i}^A\,\middle|\,\mathcal G_n^X\right]\\
&=\frac1{n\bar w_A^2}\sum_{i:Z_i=z}
 \{w_{A,z}(\mathbf X_i)+K_{n,z}^{A}
                    (i;\mathbf D_{z,\boldsymbol\theta_n}^{\mathrm p})\}^2
 E_s[(\varepsilon_{z,i}^A)^2\mid\mathcal G_n^X]\\
&=\frac1{n\bar w_A^2}\sum_{i:Z_i=z}
 \{w_{A,z}(\mathbf X_i)+K_{n,z}^{A}
                    (i;\mathbf D_{z,\boldsymbol\theta_n}^{\mathrm p})\}^2
 \sigma_z^2(\mathbf X_i).
\end{aligned}
\]
The cross-arm conditional covariance is zero: off-diagonal row products vanish by conditional independence and centering, and same-row products contain \(\mathbf1(Z_i=1)\mathbf1(Z_i=0)=0\). The covariance of the estimating-function sum with the arm-\(z\) coordinate is
\[
\frac1{n\bar w_A}\sum_{i:Z_i=z}
 \{w_{A,z}(\mathbf X_i)+K_{n,z}^{A}
                    (i;\mathbf D_{z,\boldsymbol\theta_n}^{\mathrm p})\}
 E_s(\mathbf U_A\varepsilon_z^A\mid\mathbf X_i,Z_i=z),
\]
and its conditional variance is \(n^{-1}\sum_i\operatorname{Var}_s(\mathbf U_A\mid\mathbf X_i,Z_i)\). The unweighted PG equations and measurable PS equation give
\[
E_s(\varepsilon_z^A U_{A,\alpha}\mid\mathbf X,Z=z)=0,\qquad
E_s(\varepsilon_z^A U_{A,\beta_a}\mid\mathbf X,Z=z)
=\mathbf1(a=z)\mathbf X\sigma_z^2(\mathbf X).
\]

\par\smallskip\noindent\emph{Conditional covariance and asymptotic normality.}\quad
To identify the conditional covariance limit, apply the marked-limit argument for \eqref{eq:patt-local-qv-baseline-mark-llns} in each direction with mark \((\mathbf X,S_z,w_{A,Z},\sigma_z^2, E_s(\varepsilon_z^A \mathbf U_A\mid\mathbf X,Z=z))\). The own-score density bounds and bounded local displacements are unchanged. The limiting coefficient is \(K_{A,z}^\infty(d;w)\) in \eqref{eq:pate-two-arm-joint-poisson-coefficient}, with the same baseline intensities and conditional weight laws. For \(s=p/2>1\), conditional Jensen and H\"older give finite \(s\)-moments of the covariance marks. The inflated-count bound, conditional on the baseline own score and treatment, gives
\[
E_s[\mathbf1(Z=z)(w_{A,z}+K_{n,z})^{2s}
                         \{\sigma_z^2(\mathbf X)\}^s]
\le C E_s[\mathbf1(Z=z)|\varepsilon_z^A|^p]<\infty .
\]
The first-power bound applies to the cross mark. These bounds allow removal of the mark truncations and give the \(L^1\) limits
\begin{equation}\label{eq:pate-fullx-marked-constants}
\begin{aligned}
\Lambda_{A,z}
&=E_s\!\left[\mathbf1(Z=z)
 E\{(w_{A,z}+K_{A,z}^{\infty})^2\mid\mathbf X,Z\}
                                  \sigma_z^2(\mathbf X)\right],\\
\mathcal C_{A,z}
&=E_s\!\left[\mathbf1(Z=z)
 \{w_{A,z}+\kappa_{A,z}(\mathbf D_z^*,w_{A,z})\}
             E_s(\varepsilon_z^A \mathbf U_A\mid\mathbf X,Z=z)\right].
\end{aligned}
\end{equation}
The auxiliary expectation is conditional on the root score and weight. Each directional average converges to a constant. Together with the ordinary LLN for the nuisance variance, this gives
\begin{equation}\label{eq:pate-fullx-conditional-qv}
\Omega_E^X=
\begin{pmatrix}
\Lambda_{A,1}/\bar w_A^2&0&\mathcal C_{A,1}^\top/\bar w_A\\
0&\Lambda_{A,0}/\bar w_A^2&\mathcal C_{A,0}^\top/\bar w_A\\
\mathcal C_{A,1}/\bar w_A&\mathcal C_{A,0}/\bar w_A&
 E_s\operatorname{Var}_s(\mathbf U_A\mid\mathbf X,Z)
\end{pmatrix}.
\end{equation}
The same count bound gives, for every \(\epsilon>0\),
\[
\begin{aligned}
&E_s\sum_i E_s\!\left[
 \|D_{n,i}^X\|^2\mathbf1\{\|D_{n,i}^X\|>\epsilon\}
 \mid\mathcal G_n^X\right]\\
&\le\epsilon^{2-p}E_s\sum_iE_s(\|D_{n,i}^X\|^p\mid\mathcal G_n^X)\\
&\le C\epsilon^{2-p}n^{1-p/2}E_s(1+|Y|+\|\mathbf U_A\|)^p\\
&\longrightarrow0.
\end{aligned}
\]
The first-stage predictable covariance follows from iid sampling. Together with the second-stage covariance and Lindeberg bounds above, this verifies the hypotheses of Lemma~\ref{lem:conditional-residual-martingale-clt} for each scalar projection of the full two-stage array. The martingale CLT and the Cram\'er--Wold device give the joint Gaussian limit, including degenerate projections.

Let \(\mathcal F_*\subset\sigma(\mathbf X,Z)\) be the baseline joined score/weight/treatment field. Auxiliary coefficient moments are \(\mathcal F_*\)-measurable, so the tower property replaces \(\sigma_z^2(\mathbf X)\) by \(E_s((\varepsilon_z^A)^2\mid\mathcal F_*)\) inside \(\Lambda_{A,z}\), and similarly replaces the cross mark inside \(\mathcal C_{A,z}\). Combining the two stages therefore recovers
\begin{equation}\label{eq:pate-fullx-u-covariance}
\boldsymbol\gamma_{A,1}
=\frac{E_s\{(h_{A,1}-h_{A,0})\mathbf U_A\}}{\bar w_A}
 +\frac{\mathcal C_{A,1}-\mathcal C_{A,0}}{\bar w_A}.
\end{equation}
The total nuisance covariance is
\[
\begin{aligned}
&\operatorname{Var}_s\{E_s(\mathbf U_A\mid\mathbf X,Z)\}
       +E_s\operatorname{Var}_s(\mathbf U_A\mid\mathbf X,Z)\\
&=\operatorname{Var}_s(\mathbf U_A)\\
&=\boldsymbol\Sigma_{A,U}.
\end{aligned}
\]
The root expansion, arm contrast and \(\bar w_{A,n}\to_p\bar w_A>0\) give
\begin{equation}\label{eq:pate-fullx-deterministic-joint-law}
\{T_{A,n}^{\mathrm p}(\mathbf h_n;\tau_{\mathrm{PATE}}),\widehat{\mathbf h}_n\}
\Rightarrow (G_{\tau,A},\mathbf G_{\theta,A}),
\end{equation}
with \eqref{eq:pate-fullx-baseline-pair} and zero local drift.

\par\smallskip\noindent\emph{Estimated-score substitution.}\quad
Theorem~\ref{lem:pate-two-arm-empirical-index}(i) gives tight Lipschitz scale indices for the specified \(b_z\), including outcome-dependent weights. For fixed \(l_z\), the coefficient comparison in the proof of Theorem~\ref{prop:pate-two-arm-empirical-joint-transfer}(ii) and the preceding moment bound apply to the full-covariate mark. Conditional centering gives
\[
E_s\left[\left\{\frac1{\sqrt n}\sum_{j:Z_j=z}
 (K_{n,z}^{l_z}(j)-K_{n,z}^{0}(j))\varepsilon_{z,j}^A\right\}^2\right]
=\frac1n E_s\sum_{j:Z_j=z}
 (K_{n,z}^{l_z}(j)-K_{n,z}^{0}(j))^2\sigma_z^2(\mathbf X_j)\to0.
\]
Here the notation is that of \eqref{eq:pate-two-arm-bridge-coefficient-difference}. Theorem~\ref{prop:pate-two-arm-grid-centered-processes}(i) applies by own-score centering and the current-donor bound. Its modulus and uniform-tie bound extend the comparison over finite scale nets. Tightness then permits the empirical indices. Cancelling the common \(H+E\) terms gives
\begin{equation}\label{eq:pate-fullx-empirical-joint-transfer}
T_{A,n}^{\mathrm e}(\mathbf h_n;\tau_{\mathrm{PATE}})
-T_{A,n}^{\mathrm p}(\mathbf h_n;\tau_{\mathrm{PATE}})\to_p0.
\end{equation}

For fitted and midpoint indices, the empirical-index Lipschitz bound makes their full-index distance \(O(\eta)\) on a high-probability compact set. The full incoming modulus controls the numerator difference, the uniform tie bound identifies the estimator's graph, and \(\bar w_{A,n}\ge w_{\min}\) controls the denominator in the cancellation. Thus
\begin{equation}\label{eq:pate-fullx-hbe-grid-transfer}
\lim_{\eta\downarrow0}\limsup_n P_s\{
 |T_{A,n}^{\mathrm e}(\widehat{\mathbf h}_n;\tau_{\mathrm{PATE}})
 -T_{A,n}^{\mathrm e}(\bar{\mathbf h}_{n,\eta};\tau_{\mathrm{PATE}})|>\epsilon\}=0.
\end{equation}
The finite-cell argument for \eqref{eq:pate-two-arm-grid-bc-uniform-probability} and \eqref{eq:pate-two-arm-grid-bc-finite-cells}, the bias-correction condition for deterministic local sequences, and tightness of the root-$n$-scaled parameter estimation error control the grid remainder. Condition (iv) controls the remainder at the estimated parameter. The estimator satisfies
\[
\begin{aligned}
&\sqrt n\{\widehat\tau_{\mathrm{PATE}}^{A,\mathrm{BC}}
(\boldsymbol\theta_A^*+\mathbf h/\sqrt n)-\tau_{\mathrm{PATE}}\}\\
&=T_{A,n}^{\mathrm e}(\mathbf h;\tau_{\mathrm{PATE}})-R_{A,n}(\mathbf h).
\end{aligned}
\]
Subtracting this identity at the fitted and grid indices transfers the comparison to the estimator. Equations~\eqref{eq:pate-fullx-deterministic-joint-law} and \eqref{eq:pate-fullx-empirical-joint-transfer} supply the joint local law. Positive definite \(\boldsymbol\Sigma_{A,U}\) and nonsingular \(\boldsymbol\Gamma_A\) give the nuisance limit a density, so Lemma~\ref{lem:joint-local-grid-selection} with zero drift proves \eqref{eq:pate-fullx-actual-joint-law}.

Finiteness follows from the marked limits. Since \(K_{A,z}^\infty\ge0\),
\begin{equation}\label{eq:pate-fullx-variance-positive}
\begin{aligned}
V_{A,\mathrm{actual}}
&=V_{A,\mathrm{known\text{-}score}}\\
&=V_{H,A}+\bar w_A^{-2}(\Lambda_{A,1}+\Lambda_{A,0})\\
&\ge\bar w_A^{-2}(\Lambda_{A,1}+\Lambda_{A,0})\\
&\ge w_{\min}^2\bar w_A^{-2}
 \sum_{z=0}^1 E_s[\mathbf1(Z=z)\sigma_z^2(\mathbf X)].
\end{aligned}
\end{equation}
This proves either stated positivity criterion.

\end{proof}

If the separate same-path differentiability and integrability conditions of Proposition~\ref{prop:scientific-target-path-derivative} hold, the local response identity becomes
\[
\widetilde{\boldsymbol \gamma}_{A,2}=-\boldsymbol\Gamma_A^\top\boldsymbol\Sigma_{A,U}^{-1}\boldsymbol\gamma_{A,1}.
\]

\subsubsection{PATT with a correctly specified control prognostic mean}
\label{app:patt-grouped-fullx}

Following the one-sided ATT construction of Section~6 of Yang and Zhang,\citep{yang_multiply_2023} we combine the treated heterogeneity and residual into one observed-outcome contribution. Under the stated sampling assumptions, correct specification of the control prognostic mean gives conditional mean zero for the imputed control residual. We use the conditional decomposition of Abadie and Imbens\citep{Abadie2006} to obtain its joint limit with the estimating equations, following the joint local-limit construction of Yang and Zhang.\citep{yang_multiply_2023} The discretization and process comparison above then transfer this limit to fitted scores.

\Needspace{9\baselineskip}
\begin{theorem}[PATT under full-control-mean centering]
\label{thm:patt-grouped-fullx-actual}
Fix \(A\in\{R,P\}\), with \(w_{R,z}=w_z\), \(w_{P,z}=w\), \(\tau_A=\tau_{\mathrm{PATT}}\), and nuisance vector \(\boldsymbol\theta_A=(\alpha_A^\top,\beta_{A,0}^\top)^\top\). Assume:
\begin{enumerate}
\item[(i)] Population recovery, one-sided identification, and the common-\(\mathbf D_0^*\) reduced-mean identities hold, together with the geometric and moment conditions of Proposition~\ref{prop:patt-poisson-local-process-equicontinuity}. Matching uses the specified probability/control-PG Euclidean metric and measurable, outcome-independent tie rule.
\item[(ii)] The finite logistic root, positive Hessian, and positive control Gram conditions of Lemma~\ref{lem:primitive-logistic-linear-roots} hold. The matrix \(\boldsymbol\Sigma_{A,U}\) is positive definite. The propensity estimating equation is weighted for \(A=R\) and unweighted for \(A=P\). The control prognostic estimating equation is unweighted.
\item[(iii)] For some \(p>2\), \(E_s|Y|^p<\infty\), and
\begin{equation}\label{eq:patt-grouped-fullx-control-mean}
\begin{aligned}
m_0^A\{\mathbf D_0^*(\mathbf X)\}
&=\mu_0(\mathbf X)=\mathbf X^\top\beta_{A,0}^*,\\
E_s(Y\mid\mathbf X,Z=0)&=\mu_0(\mathbf X).
\end{aligned}
\end{equation}
Outcome-selection and one-sided treatment ignorability give the second equality when the control prognostic mean is correct. The propensity model may be misspecified.
\item[(iv)] The specified empirical standardization uses fixed measurable \(0\le b_0(O)\le C_b\), \(E_sb_0>0\), and positive baseline variances equal to the population limits of that same moment rule. The deterministic normalization in \eqref{eq:patt-local-process-score-family} is fixed at those baseline moments or is the specified positive smooth population normalization. The fixed row function \(b_0\) may depend on \(Y\).
\item[(v)] The bias-correction remainders at the corresponding empirical graphs satisfy
\begin{equation}\label{eq:patt-grouped-fullx-bc}
\begin{aligned}
R_{A,n}(\mathbf h_n)&\to_p0\quad\text{for every deterministic }\mathbf h_n\to \mathbf h,\\
R_{A,n}(\widehat{\mathbf h}_n)&\to_p0,
\qquad \widehat{\mathbf h}_n=\sqrt n(\widehat{\boldsymbol \theta}_A-\boldsymbol\theta_A^*),
\end{aligned}
\end{equation}
where \(R_{A,n}\) is defined in \eqref{eq:patt-actual-local-bc-remainder-definition}.
\end{enumerate}
Let \(\bar w_A=E_s[Zw_{A,1}(\mathbf X)]>0\), and let \(K_{A,0}^{\infty}\) and \(\kappa_{A,0}(d,w)\) denote the weighted Poisson incoming coefficient and its first moment. Define the finite constants
\begin{equation}\label{eq:patt-grouped-fullx-constants}
\begin{aligned}
\widetilde V_{A,\mathrm{PATT}}
={}&\frac{E_s[Zw_{A,1}^2\{Y-m_0^A(\mathbf D_0^*)-\tau_A\}^2]}
            {\bar w_A^2}\\
&+\frac{E_s[(1-Z)E\{(K_{A,0}^{\infty})^2\mid
                    \mathbf D_0^*,w_{A,0},Z=0\}
                  \operatorname{Var}_s(Y\mid\mathbf X,Z=0)]}
             {\bar w_A^2},\\
\widetilde{\boldsymbol \gamma}_{A,1}
={}&\frac{E_s[Zw_{A,1}\{Y-m_0^A(\mathbf D_0^*)-\tau_A\}\mathbf U_A]}
             {\bar w_A}\\
&-\frac{E_s[(1-Z)\kappa_{A,0}(\mathbf D_0^*,w_{A,0})
                    \{Y-m_0^A(\mathbf D_0^*)\}\mathbf U_A]}{\bar w_A}.
\end{aligned}
\end{equation}
Here \(\mathbf U_A=\mathbf U_A(O;\boldsymbol\theta_A^*)\). Arm-specific expressions are used only on their observed arm. Then
\begin{equation}\label{eq:patt-grouped-fullx-joint-law}
\begin{pmatrix}
\sqrt n\{\widehat\tau_{\mathrm{PATT}}^{A,\mathrm{BC}}
                      (\widehat{\boldsymbol \theta}_A)-\tau_A\}\\
\sqrt n(\widehat{\boldsymbol \theta}_A-\boldsymbol\theta_A^*)
\end{pmatrix}
\Rightarrow\begin{pmatrix}\widetilde G_{\tau,A}\\\mathbf G_{\theta,A}\end{pmatrix},
\end{equation}
where the centered Gaussian pair satisfies
\begin{equation}\label{eq:patt-grouped-fullx-gaussian-pair}
\begin{aligned}
\operatorname{Var}(\widetilde G_{\tau,A})&=\widetilde V_{A,\mathrm{PATT}},\\
\operatorname{Cov}(\mathbf G_{\theta,A},\widetilde G_{\tau,A})
 &=-\boldsymbol\Gamma_A^{-1}\widetilde{\boldsymbol \gamma}_{A,1},\\
\operatorname{Var}(\mathbf G_{\theta,A})&=\boldsymbol\Sigma_{A,\theta}
 =\boldsymbol\Gamma_A^{-1}\boldsymbol\Sigma_{A,U}\boldsymbol\Gamma_A^{-T}.
\end{aligned}
\end{equation}
The first variance is positive if either displayed contribution is positive.

\end{theorem}

\begin{proof}
\par\smallskip\noindent\emph{Martingale representation.}\quad
We begin by combining the treated contributions. On the valid-count event, the \(H+E\) coordinate at any deterministic graph has the exact form
\begin{equation}\label{eq:patt-grouped-fullx-hbe}
\begin{aligned}
T_{A,n}^{\mathrm p}(\mathbf h;\tau_A)
={}&\frac1{\bar w_{A,n}\sqrt n}\sum_i Z_iw_{A,1}(\mathbf X_i)
 [\tau_{\mathrm{PATT}}(\mathbf D_{0,i}^*)-\tau_A
                                      +\varepsilon_{1\mid1,i}]\\
&-\frac1{\bar w_{A,n}\sqrt n}\sum_{j:Z_j=0}
 K_{n,0}^{A,\mathrm{PATT}}
        (j;\mathbf D_{0,\boldsymbol\theta_A^*+\mathbf h/\sqrt n}^{\mathrm p})
                                      \varepsilon_{0,j}^A\\
={}&\frac1{\bar w_{A,n}\sqrt n}\sum_i Z_iw_{A,1}(\mathbf X_i)
                 [Y_i-m_0^A(\mathbf D_{0,i}^*)-\tau_A]\\
&-\frac1{\bar w_{A,n}\sqrt n}\sum_{j:Z_j=0}
 K_{n,0}^{A,\mathrm{PATT}}
        (j;\mathbf D_{0,\boldsymbol\theta_A^*+\mathbf h/\sqrt n}^{\mathrm p})
                                      \varepsilon_{0,j}^A.
\end{aligned}
\end{equation}
The treated reduced mean cancels between the common-score contrast and treated residual, giving {\hypersetup{linkcolor=black}\eqref{eq:patt-grouped-fullx-hbe}}. Population recovery and one-sided identification give
\begin{align*}
&E_s[Zw_{A,1}\{Y-m_0^A(\mathbf D_0^*)-\tau_A\}]\\
&=\bar w_A E_\Omega[Y(1)-\mu_0(\mathbf X)-\tau_A\mid Z=1]\\
&=\bar w_A\{E_\Omega[Y(1)-Y(0)\mid Z=1]-\tau_A\}\\
&=0.
\end{align*}
This centers the treated contribution without imposing a treated outcome model. For deterministic \(\mathbf h_n\to \mathbf h\), we first reveal \(\mathcal G_n^T=\sigma\{\mathbf X_i,Z_i,Z_iY_i:1\le i\le n\}\) and then the control outcomes. Every deterministic local graph, complete donor denominator and count event is measurable with respect to this field. By product disintegration, the control outcomes are conditionally independent, and
\begin{align*}
E_s(\varepsilon_{0,i}^A\mid\mathcal G_n^T,Z_i=0)
&=E_s[Y_i-m_0^A(\mathbf D_{0,i}^*)\mid\mathbf X_i,Z_i=0]\\
&=\mu_0(\mathbf X_i)-m_0^A(\mathbf D_{0,i}^*)\\
&=0.
\end{align*}
Only the control prognostic equation still depends on the unrevealed control outcomes. The estimating function therefore satisfies
\[
E_s(\mathbf U_A\mid\mathbf X,Z,ZY)
 =\begin{pmatrix}U_{A,\alpha}(\mathbf X,Z)\\0\end{pmatrix},
\qquad
\mathbf U_A-E_s(\mathbf U_A\mid\mathbf X,Z,ZY)
 =\begin{pmatrix}0\\(1-Z)\mathbf X\varepsilon_0^A\end{pmatrix}.
\]
Combining the revealed rows with these centered control contributions gives the martingale array
\begin{equation}\label{eq:patt-grouped-fullx-martingale}
\boldsymbol\xi_{n,k}^T=
\begin{cases}
\displaystyle\frac1{\sqrt n}
\begin{pmatrix}
Z_kw_{A,1}[Y_k-m_0^A(\mathbf D_{0,k}^*)-\tau_A]/\bar w_A\\
E_s\{\mathbf U_A(O_k;\boldsymbol\theta_A^*)\mid\mathbf X_k,Z_k,Z_kY_k\}
\end{pmatrix},&1\le k\le n,\\[3mm]
\displaystyle\frac1{\sqrt n}
\begin{pmatrix}
-(1-Z_i)K_{n,0}^{A,\mathrm{PATT}}
 (i;\mathbf D_{0,\boldsymbol\theta_A^*+\mathbf h_n/\sqrt n}^{\mathrm p})
 \varepsilon_{0,i}^A/\bar w_A\\
\{\mathbf U_A(O_i;\boldsymbol\theta_A^*)-E_s(\mathbf U_A\mid\mathbf X_i,Z_i,Z_iY_i)\}
\end{pmatrix},&k=n+i,\ 1\le i\le n.
\end{cases}
\end{equation}
The corresponding filtration is
\[
\mathcal F_{n,k}^T=
\begin{cases}
\sigma\{(\mathbf X_i,Z_i,Z_iY_i):1\le i\le k\},&0\le k\le n,\\
\mathcal G_n^T\vee\sigma(Y_1,\ldots,Y_{k-n}),&n<k\le2n.
\end{cases}
\]
The first-stage rows are iid with zero mean and finite covariance. For the second stage the preceding conditional-mean identities give
\[
E_s(\boldsymbol\xi_{n,k}^T\mid\mathcal F_{n,k-1}^T)=0,
\qquad1\le k\le2n.
\]
Therefore \(\{\sum_{j=1}^k\boldsymbol\xi_{n,j}^T,\mathcal F_{n,k}^T\}\) is a martingale.

\par\smallskip\noindent\emph{Conditional covariance.}\quad
Put \(\sigma_0^2(\mathbf X)=\operatorname{Var}_s(Y\mid\mathbf X,Z=0)\). The conditional variance of the first coordinate and its covariance with the nuisance coordinate are
\begin{align*}
&\frac1{n\bar w_A^2}\sum_{i:Z_i=0}
 [K_{n,0}^{A,\mathrm{PATT}}
             (i;\mathbf D_{0,\boldsymbol\theta_A^*+\mathbf h_n/\sqrt n}^{\mathrm p})]^2
                                      \sigma_0^2(\mathbf X_i),\\
&-\frac1{n\bar w_A}\sum_{i:Z_i=0}
 K_{n,0}^{A,\mathrm{PATT}}
             (i;\mathbf D_{0,\boldsymbol\theta_A^*+\mathbf h_n/\sqrt n}^{\mathrm p})
 E_s(\varepsilon_{0,i}^A \mathbf U_A\mid\mathbf X_i,Z_i=0).
\end{align*}
The nuisance variance is the average of \(\operatorname{Var}_s(\mathbf U_A\mid\mathbf X,Z,ZY)\).

Apply the marked-limit argument for \eqref{eq:patt-local-qv-baseline-mark-llns} with control mark \((\mathbf X,\sigma_0^2(\mathbf X), E_s(\varepsilon_0^A \mathbf U_A\mid\mathbf X,Z=0))\). The control density bounds and bounded smooth score map give its geometric and displacement conditions. For \(s=p/2>1\), conditional Jensen and the count bound \eqref{eq:patt-local-qv-inflated-count-moment} give
\[
\begin{aligned}
&E_s\!\left[(1-Z_i)
 \left\{K_{n,0}^{A,\mathrm{PATT}}
 (i;\mathbf D_{0,\boldsymbol\theta_A^*+\mathbf h_n/\sqrt n}^{\mathrm p})\right\}^{2s}
 \{\sigma_0^2(\mathbf X_i)\}^s\right]\\
&\le C E_s[(1-Z)\{\sigma_0^2(\mathbf X)\}^s]\\
&\le C E_s[(1-Z)|\varepsilon_0^A|^p]<\infty,\\
&E_s\!\left[(1-Z)
 \|E_s(\varepsilon_0^A \mathbf U_A\mid\mathbf X,Z=0)\|^s\right]\\
&\le E_s[(1-Z)|\varepsilon_0^A|^s\|\mathbf U_A\|^s]\\
&\le\{E_s[(1-Z)|\varepsilon_0^A|^p]
             E_s[(1-Z)\|\mathbf U_A\|^p]\}^{1/2}<\infty.
\end{aligned}
\]
These bounds permit removal of the covariance-mark truncations. The marked averages therefore converge to the control contributions in \eqref{eq:patt-grouped-fullx-constants}, with coefficient \(K_{n,0}^{A,\mathrm{PATT}}\).

The same reuse bound verifies the conditional Lindeberg condition. The expected sum of second-stage \(p\)-moments in \eqref{eq:patt-grouped-fullx-martingale} is bounded by
\[
\begin{aligned}
&\frac C{n^{p/2}}\sum_i E_s\Big[(1-Z_i)
 \{K_{n,0}^{A,\mathrm{PATT}}
 (i;\mathbf D_{0,\boldsymbol\theta_A^*+\mathbf h_n/\sqrt n}^{\mathrm p})\}^{p}
 |\varepsilon_{0,i}^A|^p\\
&\qquad+\|\mathbf U_A(O_i;\boldsymbol\theta_A^*)
       -E_s(\mathbf U_A\mid\mathbf X_i,Z_i,Z_iY_i)\|^p\Big]\\
&\le C'n^{1-p/2}E_s(1+|Y|+\|\mathbf U_A\|)^p\\
&\longrightarrow0.
\end{aligned}
\]
The first-stage predictable covariance follows from iid sampling. The displayed covariance and moment bounds verify the martingale CLT conditions for every scalar projection of the full two-stage array. Lemma~\ref{lem:conditional-residual-martingale-clt} and the Cram\'er--Wold device therefore give the joint Gaussian limit. The covariance of the first-stage estimating functions is the treated term of \(\widetilde{\boldsymbol \gamma}_{A,1}\). The second-stage covariance is its negative control term. For the nuisance block,
\[
\begin{aligned}
&\operatorname{Var}_s\{E_s(\mathbf U_A\mid\mathbf X,Z,ZY)\}
 +E_s\operatorname{Var}_s(\mathbf U_A\mid\mathbf X,Z,ZY)\\
&=\operatorname{Var}_s(\mathbf U_A)\\
&=\boldsymbol\Sigma_{A,U}.
\end{aligned}
\]
The ordinary root expansion multiplies the estimating-function sum by \(-\boldsymbol\Gamma_A^{-1}\). Hence, for every deterministic \(\mathbf h_n\to \mathbf h\),
\begin{equation}\label{eq:patt-grouped-fullx-deterministic-law}
\{T_{A,n}^{\mathrm p}(\mathbf h_n;\tau_A),\widehat{\mathbf h}_n\}
 \Rightarrow(\widetilde G_{\tau,A},\mathbf G_{\theta,A}).
\end{equation}
Slutsky applies to the treated denominator \(\bar w_{A,n}\to_p\bar w_A>0\).

\par\smallskip\noindent\emph{Estimated-score substitution.}\quad
For every deterministic \((\mathbf h,l)\), at every sample size \(m\), the control incoming numerator satisfies
\begin{equation}\label{eq:patt-grouped-fullx-zero-mean}
\begin{aligned}
E_s\mathcal I_{A,m;n}(\mathbf h,l)
&=E_s\sum_{j:Z_j=0}
 K_{m,0}^{A,\mathrm{PATT}}(j;\mathcal D_{n,\mathbf h,l})
                                      \varepsilon_{0,j}^A\\
&=E_s\sum_{j:Z_j=0}
 K_{m,0}^{A,\mathrm{PATT}}(j;\mathcal D_{n,\mathbf h,l})
             E_s(\varepsilon_{0,j}^A\mid\mathcal G_m^T)\\
&=0.
\end{aligned}
\end{equation}
Full control-mean centering also gives this identity after radius truncation. Thus Proposition~\ref{prop:patt-binomial-local-process-transfer} applies with its current-donor bound, and its centered modulus is the full numerator modulus.

The calculation for the empirical scale ratios in \eqref{eq:patt-actual-empirical-scale-index}-- \eqref{eq:patt-actual-scale-positive-neighborhood} applies to the specified bounded \(b_0(O)\), bounded raw-score derivatives and positive baseline variances. Thus \(L_n(0)=O_p(1)\), the scale indices are Lipschitz on each compact local set, and the empirical distance array is that of \(\mathcal D_{n,\mathbf h,L_n(\mathbf h)}\).

For fixed \(l\) and deterministic \(\mathbf h_n\), the relative scale perturbation is \(\operatorname{diag}(l)x/\sqrt n\). The fixed-window coefficient comparison and the preceding count bound apply as in the proof of \eqref{eq:pate-fullx-empirical-joint-transfer}. Conditional centering therefore gives
\begin{align*}
&E_s\left[\left\{\frac1{\sqrt n}\sum_{j:Z_j=0}
 [K_{n,0}^{A,\mathrm{PATT}}(j;\mathcal D_{n,\mathbf h_n,l})
 -K_{n,0}^{A,\mathrm{PATT}}(j;\mathcal D_{n,\mathbf h_n,0})]
                                      \varepsilon_{0,j}^A\right\}^2\right]\\
&=\frac1n E_s\sum_{j:Z_j=0}
 [K_{n,0}^{A,\mathrm{PATT}}(j;\mathcal D_{n,\mathbf h_n,l})
 -K_{n,0}^{A,\mathrm{PATT}}(j;\mathcal D_{n,\mathbf h_n,0})]^2
                                      \sigma_0^2(\mathbf X_j)\\
&\longrightarrow0.
\end{align*}
The full numerator modulus controls the gaps in finite nets of compact \(l\)-boxes. Tightness of \(L_n(\mathbf h_n)\) and the uniform-tie bound therefore give \(T_{A,n}^{\mathrm e}(\mathbf h_n;\tau_A) -T_{A,n}^{\mathrm p}(\mathbf h_n;\tau_A)\to_p0\).

The fitted root and its grid midpoint have full-index distance \(O(\eta)\) on compact events. Cancelling the treated term in \eqref{eq:patt-grouped-fullx-hbe} and using the positive denominator limit gives a vanishing \(H+E\) difference as \(n\to\infty\), then \(\eta\downarrow0\). Compactness and deterministic-local BC control the grid remainders, while \eqref{eq:patt-grouped-fullx-bc} controls the fitted-index remainder.

The bias-corrected estimator and matching coordinate are related by
\[
\sqrt n\{\widehat\tau_A^{\mathrm{BC}}
                    (\boldsymbol\theta_A^*+\mathbf h/\sqrt n)-\tau_A\}
 =T_{A,n}^{\mathrm e}(\mathbf h;\tau_A)-R_{A,n}(\mathbf h).
\]
The empirical-scale comparison and the two bias-correction remainders now connect the deterministic joint law \eqref{eq:patt-grouped-fullx-deterministic-law} to the fitted estimator. Lemma~\ref{lem:joint-local-grid-selection} applies with zero drift. Because the nuisance covariance is nonsingular, the Gaussian limit assigns probability zero to grid-cell boundaries. This proves \eqref{eq:patt-grouped-fullx-joint-law}. The marked bounds and iid treated moment prove finiteness. The two nonnegative terms in \eqref{eq:patt-grouped-fullx-constants} give positivity.

\end{proof}

Under the treated-centering condition in Assumption~\ref{ass:residual-centering-survey-normalized}, iterated expectation gives \(\widetilde V_{A,\mathrm{PATT}}=V_{A,\mathrm{known\text{-}score}}\) and \(\widetilde{\boldsymbol \gamma}_{A,1}=\boldsymbol\gamma_{A,1}\).

\Needspace{8\baselineskip}
\subsubsection{Target derivatives}
\label{fixed-target-target-invariance-in-all-scenario-specifics}
\label{score-indexed-target-maps}

With the population law fixed, the targets are constant in the matching-score parameter:
\begin{equation}\label{prop:fixed-law-scientific-target-invariance}
\begin{aligned}
\frac{\partial}{\partial\boldsymbol\theta_A}E_\Omega\{Y(1)-Y(0)\}&=0,\\
\frac{\partial}{\partial\boldsymbol\theta_A}E_\Omega\{Y(1)-Y(0)\mid Z=1\}&=0.
\end{aligned}
\end{equation}

Along a likelihood path through \(P_s\), the selected-observation law and the population law recovered from it can change. We differentiate the identified target through its conditional means and its weighted averaging distribution, holding the survey weight functions fixed. This derivative describes the target under neighboring laws, whereas \(\boldsymbol b_{A,\mathrm{loc}}\) and \(-\mathbf j_A\) describe the PATE and PATT statistics under local changes in the matching inputs with \(P_s\) fixed.

\begin{proposition}[Pathwise derivatives of PATE and PATT]
\label{prop:scientific-target-path-derivative}
Fix an estimand, PATE or PATT, and a sampling regime \(A\in\{R,P\}\). Let \(Q_{A,\boldsymbol\theta}\) be a dominated selected-observation likelihood family for \(O=(Y,\mathbf X,Z)\), with \(Q_{A,\boldsymbol\theta_A^\ast}=P_s\). Write \(E_{s,\boldsymbol\theta}\) for its expectation and define its column likelihood score by
\[
\mathbf s_A^{\mathrm{path}}(O)
:=
\left.
\frac{\partial}{\partial\boldsymbol\theta}\log q_{A,\boldsymbol\theta}(O)
\right|_{\boldsymbol\theta=\boldsymbol\theta_A^\ast},
\qquad
E_s\{\mathbf s_A^{\mathrm{path}}(O)\}=0,
\qquad
E_s\|\mathbf s_A^{\mathrm{path}}(O)\|^2<\infty.
\]
Assume differentiability of the joint, marginal, and conditional densities on their locally common relevant supports, and domination sufficient to differentiate every expectation and conditional mean displayed below. In particular, the displayed derivatives and products are integrable. The survey weight functions are positive, finite, known, and held fixed along the path:
\[
w_{A,z}(\mathbf X):=
\begin{cases}
w_z(\mathbf X), & A=R,\\
w(\mathbf X), & A=P.
\end{cases}
\]
Assume the required weighted denominators are positive and finite. For PATE assume both arms have positive selected probability on the covariate support. For PATT, assume a positive treated population and positive control probability wherever the treated covariate law has mass.

The population observed law recovered along this path is
\[
dQ_{\Omega,\boldsymbol\theta}^{\mathrm{obs}}(O)
=
\frac{w_{A,Z}(\mathbf X)}
     {E_{s,\boldsymbol\theta}\{w_{A,Z}(\mathbf X)\}}\,
dQ_{A,\boldsymbol\theta}(O).
\]
At \(\boldsymbol\theta=\boldsymbol\theta_A^*\), population recovery gives \(Q_{\Omega,\boldsymbol\theta_A^*}^{\mathrm{obs}}=P_\Omega^{\mathrm{obs}}\). Away from this point the recovered law is generally different. The parameter \(\boldsymbol\theta\) here indexes the observation law, rather than changing a fitted score while holding that law fixed. Assume that the sampling and potential-outcome identification conditions hold throughout a neighborhood of \(\boldsymbol\theta_A^\ast\): consistency, correct fixed inverse-inclusion weights, sampling ignorability for the relevant potential-outcome means, and treatment mean exchangeability for both potential outcomes for PATE, or for \(Y(0)\) alone for PATT. Equivalently, the population recovery and arm-specific mean identification identities used below must hold along the path.

Define the true selected and recovered population propensities by
\[
e_{s,\boldsymbol\theta}(\mathbf X)
:=Q_{A,\boldsymbol\theta}(Z=1\mid\mathbf X),
\]
\[
e_{\Omega,\boldsymbol\theta}(\mathbf X)
:=
\frac{w_{A,1}(\mathbf X)e_{s,\boldsymbol\theta}(\mathbf X)}
     {w_{A,1}(\mathbf X)e_{s,\boldsymbol\theta}(\mathbf X)
      +w_{A,0}(\mathbf X)\{1-e_{s,\boldsymbol\theta}(\mathbf X)\}}.
\]
For the prospective scenarios, \(e_{\Omega,\boldsymbol\theta}=e_{s,\boldsymbol\theta}\). Define
\[
\mu_{z,\boldsymbol\theta}(\mathbf X)
:=E_{s,\boldsymbol\theta}[Y\mid\mathbf X,Z=z],
\qquad
\mu_{1\mid1,\boldsymbol\theta}(\mathbf X)
:=E_{s,\boldsymbol\theta}[Y\mid\mathbf X,Z=1].
\]
Retain the scenario weights
\[
\omega_A(\mathbf X,Z):=
\begin{cases}
w_{A,Z}(\mathbf X), & \text{PATE},\\
Zw_{A,1}(\mathbf X), & \text{PATT},
\end{cases}
\]
and use the conditional mean contrast
\[
\eta_{A,\boldsymbol\theta}(\mathbf X):=
\begin{cases}
\mu_{1,\boldsymbol\theta}(\mathbf X)-\mu_{0,\boldsymbol\theta}(\mathbf X),
 & \text{PATE},\\
\mu_{1\mid1,\boldsymbol\theta}(\mathbf X)-\mu_{0,\boldsymbol\theta}(\mathbf X),
 & \text{PATT}.
\end{cases}
\]
Then the identified target is
\[
\tau_A(\boldsymbol\theta)
:=
\frac{E_{s,\boldsymbol\theta}[\omega_A(\mathbf X,Z)
                         \eta_{A,\boldsymbol\theta}(\mathbf X)]}
     {E_{s,\boldsymbol\theta}[\omega_A(\mathbf X,Z)]},
\]
and its derivative along this likelihood family is
\[
\boldsymbol\gamma_{A,2}
:=
\left.
\frac{\partial}{\partial\boldsymbol\theta}\tau_A(\boldsymbol\theta)
\right|_{\boldsymbol\theta=\boldsymbol\theta_A^\ast}
=
E_s\{\phi_A(O)\mathbf s_A^{\mathrm{path}}(O)\}.
\]
All unindexed functions below are evaluated at the baseline law. The four representers are obtained from the following two formulas using \(w_{A,z}=w_z\) retrospectively and \(w_{A,z}=w\) prospectively:
\[
\begin{aligned}
E_s(w_{A,Z})\phi_A(O)
&=w_{A,Z}\{\mu_1(\mathbf X)-\mu_0(\mathbf X)-\tau_{\mathrm{PATE}}\}\\
&\quad+\frac{Zw_{A,1}}{e_\Omega(\mathbf X)}
                   \{Y-\mu_1(\mathbf X)\}\\
&\quad-\frac{(1-Z)w_{A,0}}{1-e_\Omega(\mathbf X)}
                   \{Y-\mu_0(\mathbf X)\},
\qquad\text{for PATE},
\end{aligned}
\]
and
\[
\begin{aligned}
E_s(Zw_{A,1})\phi_A(O)
&=Zw_{A,1}\{\mu_{1\mid1}(\mathbf X)-\mu_0(\mathbf X)
                         -\tau_{\mathrm{PATT}}\}\\
&\quad+Zw_{A,1}\{Y-\mu_{1\mid1}(\mathbf X)\}\\
&\quad-(1-Z)w_{A,0}\frac{e_\Omega(\mathbf X)}{1-e_\Omega(\mathbf X)}
                    \{Y-\mu_0(\mathbf X)\},
\qquad\text{for PATT}.
\end{aligned}
\]
Each representer has mean zero. If \(E_s\{\phi_A(O)^2\}<\infty\), these representers are \(L^2(P_s)\) gradients.
\end{proposition}

\begin{proof}
Population recovery and arm-specific identification give the target quotient. Its numerator changes through both the conditional contrast and the likelihood. Its denominator changes through the likelihood. Write \(\dot\eta_A(\mathbf X)\) and \(\dot\mu_z(\mathbf X)\) for the column derivatives at \(\boldsymbol\theta_A^\ast\). At this parameter,
\[
\left.\frac{dQ_{A,\boldsymbol\theta}}{dP_s}(O)\right|_{\boldsymbol\theta=\boldsymbol\theta_A^*}=1,
\qquad
\left.\nabla_{\boldsymbol \theta}\frac{dQ_{A,\boldsymbol\theta}}{dP_s}(O)
 \right|_{\boldsymbol\theta=\boldsymbol\theta_A^*}=\mathbf s_A^{\mathrm{path}}(O).
\]
Changing measure and applying the product rule under the stated domination give
\[
\begin{aligned}
&\left.\nabla_{\boldsymbol \theta} E_{s,\boldsymbol\theta}
 [\omega_A(\mathbf X,Z)\eta_{A,\boldsymbol\theta}(\mathbf X)]
 \right|_{\boldsymbol\theta=\boldsymbol\theta_A^\ast}\\
&=\left.\nabla_{\boldsymbol \theta} E_s\!\left[
 \omega_A(\mathbf X,Z)\eta_{A,\boldsymbol\theta}(\mathbf X)
 \frac{dQ_{A,\boldsymbol\theta}}{dP_s}(O)\right]
 \right|_{\boldsymbol\theta=\boldsymbol\theta_A^\ast}\\
&=E_s\{\omega_A(\mathbf X,Z)\dot\eta_A(\mathbf X)\}
 +E_s\{\omega_A(\mathbf X,Z)\eta_A(\mathbf X)\mathbf s_A^{\mathrm{path}}(O)\},\\
&\left.\nabla_{\boldsymbol \theta} E_{s,\boldsymbol\theta}
 [\omega_A(\mathbf X,Z)]\right|_{\boldsymbol\theta=\boldsymbol\theta_A^\ast}
\\
&=\left.\nabla_{\boldsymbol \theta} E_s\!\left[
 \omega_A(\mathbf X,Z)\frac{dQ_{A,\boldsymbol\theta}}{dP_s}(O)\right]
 \right|_{\boldsymbol\theta=\boldsymbol\theta_A^\ast}\\
&=E_s\{\omega_A(\mathbf X,Z)\mathbf s_A^{\mathrm{path}}(O)\}.
\end{aligned}
\]
The weight functions remain fixed, but their expectations need not: the latter change through the probability law. The quotient rule and \(E_s(\omega_A\eta_A)=\tau_AE_s\omega_A\) therefore give
\[
\begin{aligned}
\boldsymbol\gamma_{A,2}
&=\frac{E_s(\omega_A\dot\eta_A)+E_s\{\omega_A\eta_A\mathbf s_A^{\mathrm{path}}(O)\}}
         {E_s\omega_A}
 -\frac{E_s(\omega_A\eta_A)E_s\{\omega_A\mathbf s_A^{\mathrm{path}}(O)\}}{(E_s\omega_A)^2}\\
&=\frac{E_s(\omega_A\dot\eta_A)+E_s\{\omega_A\eta_A\mathbf s_A^{\mathrm{path}}(O)\}}
         {E_s\omega_A}
 -\frac{\tau_AE_s\{\omega_A\mathbf s_A^{\mathrm{path}}(O)\}}{E_s\omega_A}\\
&=\frac{E_s(\omega_A\dot\eta_A)
       +E_s\{\omega_A(\eta_A-\tau_A)\mathbf s_A^{\mathrm{path}}(O)\}}
       {E_s\omega_A}.
\end{aligned}
\]
For the conditional means, conditional change of measure gives
\[
\mu_{z,\boldsymbol\theta}(\mathbf X)
=\frac{E_s\!\left[Y\,\dfrac{dQ_{A,\boldsymbol\theta}}{dP_s}(O)
                    \,\middle|\,\mathbf X,Z=z\right]}
       {E_s\!\left[\dfrac{dQ_{A,\boldsymbol\theta}}{dP_s}(O)
                    \,\middle|\,\mathbf X,Z=z\right]}.
\]
The denominator is one at \(\boldsymbol\theta_A^*\). Differentiating numerator and denominator separately therefore gives
\[
\begin{aligned}
\dot\mu_z(\mathbf X)
&=\frac{E_s\{Y\mathbf s_A^{\mathrm{path}}(O)\mid\mathbf X,Z=z\}}{1}
 -\frac{E_s(Y\mid\mathbf X,Z=z)}{1^2}
       E_s\{\mathbf s_A^{\mathrm{path}}(O)\mid\mathbf X,Z=z\}\\
&=E_s\{Y\mathbf s_A^{\mathrm{path}}(O)\mid\mathbf X,Z=z\}
-\mu_z(\mathbf X)E_s\{\mathbf s_A^{\mathrm{path}}(O)\mid\mathbf X,Z=z\}\\
&=E_s[\{Y-\mu_z(\mathbf X)\}\mathbf s_A^{\mathrm{path}}(O)\mid\mathbf X,Z=z],
\end{aligned}
\]
also for the observed treated mean \(\mu_{1\mid1}\). For PATE, let \(p_{\Omega,1}=e_\Omega\), \(p_{\Omega,0}=1-e_\Omega\). Conditioning converts the weighted conditional-mean derivative to an observed-arm residual term:
\[
\begin{aligned}
E_s(w_{A,Z}\dot\mu_z)
&=E_s[\{w_{A,1}e_s+w_{A,0}(1-e_s)\}\dot\mu_z]\\
&=E_s\!\left[
 \frac{\mathbf1(Z=z)w_{A,z}}{p_{\Omega,z}}\dot\mu_z\right]\\
&=E_s\!\left[
 \frac{\mathbf1(Z=z)w_{A,z}}{p_{\Omega,z}}
 (Y-\mu_z)\mathbf s_A^{\mathrm{path}}(O)\right],\qquad z=0,1.
\end{aligned}
\]
For PATT, the outer expectation is over treated rows. The treated derivative is already on its observed arm. The control derivative uses the recovered treatment odds. Conditioning gives
\[
\begin{aligned}
E_s(Zw_{A,1}\dot\mu_{1\mid1})
&=E_s\{Zw_{A,1}(Y-\mu_{1\mid1})\mathbf s_A^{\mathrm{path}}(O)\},\\
E_s(Zw_{A,1}\dot\mu_0)
&=E_s\{w_{A,1}e_s\dot\mu_0\}\\
&=E_s\!\left[(1-Z)w_{A,0}
              \frac{e_\Omega}{1-e_\Omega}\dot\mu_0\right]\\
&=E_s\!\left[
 (1-Z)w_{A,0}\frac{e_\Omega}{1-e_\Omega}(Y-\mu_0)\mathbf s_A^{\mathrm{path}}(O)\right],
\end{aligned}
\]
where \(w_{A,1}e_s/(1-e_s)=w_{A,0}e_\Omega/(1-e_\Omega)\). Substitution into the quotient derivative gives \(\boldsymbol\gamma_{A,2}=E_s\{\phi_A(O)\mathbf s_A^{\mathrm{path}}(O)\}\). Conditional centering given \((\mathbf X,Z)\) and the target identity give \(E_s\{\phi_A(O)\}=0\). Square integrability gives the \(L^2(P_s)\) gradient property.
\end{proof}

The inner product of \(\phi_A(O)\) with the likelihood score represents the target derivative \(\boldsymbol\gamma_{A,2}\). This does not identify \(\phi_A\) as the influence function of the matching estimator. For PATE, the three terms correspond to changes in the weighted averaging distribution and normalization, the treated conditional mean, and the control conditional mean. For PATT, the normalization is \(E_s\{Zw_{A,1}(\mathbf X)\}\).

For either estimand and sampling regime, we use
\begin{equation}\label{eq:pate-target-design-weights}
w_{A,z}(\mathbf X)=
\begin{cases}
w_z(\mathbf X),&A=R,\\
w(\mathbf X),&A=P.
\end{cases}
\end{equation}
For PATE, the nuisance vector is
\[
\boldsymbol\theta_A=(\boldsymbol\alpha^\top,
          \boldsymbol\beta_0^\top,
          \boldsymbol\beta_1^\top)^\top,
\]
with estimating function
\begin{equation}\label{eq:pate-target-estimating-function}
\begin{aligned}
\mathbf U_A(O;\boldsymbol\theta_A^\ast)&=
\begin{pmatrix}
a_A(O)\mathbf X
 \{Z-e(\mathbf X;\boldsymbol\alpha_A^\ast)\}\\
(1-Z)\mathbf X\{Y-\mathbf X^\top\boldsymbol\beta_{A,0}^\ast\}\\
Z\mathbf X\{Y-\mathbf X^\top\boldsymbol\beta_{A,1}^\ast\}
\end{pmatrix},
\\
a_A(O)&=
\begin{cases}
w_Z(\mathbf X),&A=R,\\
1,&A=P.
\end{cases}
\end{aligned}
\end{equation}
For PATT, we omit the last prognostic block, so that \(\boldsymbol\theta_A=(\boldsymbol\alpha^\top,\boldsymbol\beta_0^\top)^\top\).

The auxiliary law is
\begin{equation}\label{eq:pate-target-auxiliary-law}
\begin{aligned}
\mathbf s_A(O)&=-\boldsymbol\Gamma_A^\top\boldsymbol\Sigma_{A,U}^{-1}
                          \mathbf U_A(O;\boldsymbol\theta_A^\ast),\\
g_{\mathrm{tilt}}(v)&=1+\frac{v}{\sqrt{1+v^2}},\\
\frac{dQ_{A,\mathbf t}}{dP_s}(O)
&=\frac{g_{\mathrm{tilt}}\{\mathbf t^\top \mathbf s_A(O)\}}
        {E_s[g_{\mathrm{tilt}}\{\mathbf t^\top \mathbf s_A(O)\}]}.
\end{aligned}
\end{equation}

We define the conditional moments and normalization by
\begin{equation}\label{eq:pate-target-auxiliary-definitions}
\begin{aligned}
d_{\mathbf t,z}(\mathbf X)
&=E_s[g_{\mathrm{tilt}}\{\mathbf t^\top \mathbf s_A(O)\}\mid\mathbf X,Z=z],\\
\mu_{z,\mathbf t}(\mathbf X)
&=E_{Q_{A,\mathbf t}}(Y\mid\mathbf X,Z=z),\\
\mu_z(\mathbf X)&=E_s(Y\mid\mathbf X,Z=z),\\
\bar w_A&=
\begin{cases}
E_s\{w_{A,Z}(\mathbf X)\},&\text{PATE},\\
E_s\{Zw_{A,1}(\mathbf X)\},&\text{PATT}.
\end{cases}
\end{aligned}
\end{equation}
and the recovered population propensity
\begin{equation}\label{eq:pate-target-baseline-propensity-recovery}
e_\Omega(\mathbf X)=
\frac{w_{A,1}(\mathbf X)e_s(\mathbf X)}
 {w_{A,1}(\mathbf X)e_s(\mathbf X)
    +w_{A,0}(\mathbf X)\{1-e_s(\mathbf X)\}}.
\end{equation}
For prospective sampling, \(e_\Omega=e_s\).

\begin{lemma}[PATE and PATT expansions under the auxiliary law]
\label{lem:patt-auxiliary-scientific-target}
\label{lem:pate-auxiliary-scientific-target}
Fix an estimand, PATE or PATT, and a sampling regime \(A\in\{R,P\}\). Retain the baseline identification conditions for that estimand, the true fixed inverse-inclusion weight functions in Proposition~\ref{prop:scientific-target-path-derivative}, and the conditions of Lemma~\ref{lem:auxiliary-godambe-experiment}.

Assume also:
\begin{enumerate}
\item[(i)] \(\mathbf X\) is bounded.
\item[(ii)] \(0<w_{\min}\le w_{A,z}(\mathbf X)\le w_{\max}<\infty\), and \(0<\pi_{A,z}(\mathbf X)=1/w_{A,z}(\mathbf X)\le1\).
\item[(iii)] For a fixed \(\epsilon>0\), the selected propensity satisfies
\begin{equation}\label{eq:pate-target-selected-overlap}
\epsilon\le e_s(\mathbf X):=P_s(Z=1\mid\mathbf X)
             \le1-\epsilon\quad\text{a.s.}
\end{equation}
\item[(iv)] For some \(\delta>0\),
\begin{equation}\label{eq:pate-target-uniform-conditional-moment}
\mathop{\rm ess\,sup}_{\mathbf x,z}
E_s\{|Y|^{2+\delta}\mid\mathbf X=\mathbf x,Z=z\}<\infty.
\end{equation}
\end{enumerate}
The auxiliary law defined above has column likelihood score \(\mathbf s_A(O)\) at zero.

\emph{(i) PATE.} The target functional is
\begin{equation}\label{eq:pate-target-auxiliary-observed-functional}
\tau_A(Q_{A,\mathbf t})
=\frac{E_{Q_{A,\mathbf t}}[
 w_{A,Z}(\mathbf X)\{\mu_{1,\mathbf t}(\mathbf X)-\mu_{0,\mathbf t}(\mathbf X)\}]}
 {E_{Q_{A,\mathbf t}}\{w_{A,Z}(\mathbf X)\}}.
\end{equation}
Its gradient representer is
\begin{equation}\label{eq:pate-target-representer}
\begin{aligned}
\bar w_A\phi_A(O)
&=w_{A,Z}(\mathbf X)
 \{\mu_1(\mathbf X)-\mu_0(\mathbf X)-\tau_A\}\\
&\quad+\frac{Zw_{A,1}(\mathbf X)}{e_\Omega(\mathbf X)}
                        \{Y-\mu_1(\mathbf X)\}\\
&\quad-\frac{(1-Z)w_{A,0}(\mathbf X)}
                          {1-e_\Omega(\mathbf X)}
                        \{Y-\mu_0(\mathbf X)\},
\end{aligned}
\end{equation}

\emph{(ii) PATT.} The target functional is
\begin{equation}\label{eq:patt-target-auxiliary-observed-functional}
\begin{aligned}
\tau_A(Q_{A,\mathbf t})
&:=\frac{E_{Q_{A,\mathbf t}}\!\left[
 Zw_{A,1}(\mathbf X)
 \{\mu_{1,\mathbf t}(\mathbf X)-\mu_{0,\mathbf t}(\mathbf X)\}\right]}
 {E_{Q_{A,\mathbf t}}\{Zw_{A,1}(\mathbf X)\}}\\
&=\frac{E_{Q_{A,\mathbf t}}\!\left[
 Zw_{A,1}(\mathbf X)\{Y-\mu_{0,\mathbf t}(\mathbf X)\}\right]}
 {E_{Q_{A,\mathbf t}}\{Zw_{A,1}(\mathbf X)\}}.
\end{aligned}
\end{equation}
Here \(\mu_{1,\mathbf t}=\mu_{1\mid1,\mathbf t}\) is the observed treated mean. The PATT gradient representer can be written as
\begin{equation}\label{eq:patt-target-collapsed-representer}
\begin{aligned}
\bar w_A\phi_A(O)
&=Zw_{A,1}(\mathbf X)
                \{Y-\mu_0(\mathbf X)-\tau_A\}\\
&\quad{}-(1-Z)w_{A,0}(\mathbf X)
 \frac{e_\Omega(\mathbf X)}{1-e_\Omega(\mathbf X)}
                \{Y-\mu_0(\mathbf X)\}.
\end{aligned}
\end{equation}

For either estimand, \(\tau_A(Q_{A,0})=\tau_A\), and \(\phi_A\) is mean zero and square integrable. The recovered observed laws admit potential-outcome laws with the same inverse-inclusion functions and the corresponding identified target. In a sufficiently small neighborhood of zero, the conditional and target denominators are uniformly positive, and
\begin{equation}\label{eq:pate-target-auxiliary-derivative}
\begin{aligned}
\tau_A(Q_{A,\mathbf t})-\tau_A
&=\mathbf t^\top\boldsymbol\gamma_{A,2}
     +O\{\|\mathbf t\|^{1+\min(\delta,1)}\},\\
\boldsymbol\gamma_{A,2}
&=E_s\{\phi_A(O)\mathbf s_A(O)\}\\
&=-\boldsymbol\Gamma_A^\top\boldsymbol\Sigma_{A,U}^{-1}
        E_s\{\mathbf U_A(O;\boldsymbol\theta_A^\ast)\phi_A(O)\},
\end{aligned}
\end{equation}
Consequently, for every deterministic \(\mathbf h_n\to \mathbf h\), the local target shift is given by \eqref{eq:pate-target-auxiliary-local-shift}.
\end{lemma}

\begin{proof}
We first show that the auxiliary perturbation preserves positive conditional denominators and overlap locally. The estimating equations give \(\|\mathbf s_A(O)\|\le C(1+|Y|)\). With \(e_{s,\mathbf t}(\mathbf X)=Q_{A,\mathbf t}(Z=1\mid\mathbf X)\), the bounds \(0<g_{\mathrm{tilt}}<2\), \(|g_{\mathrm{tilt}}'|\le1\) and \(|g_{\mathrm{tilt}}(v)-1|\le|v|\) imply
\[
\begin{aligned}
\mathop{\rm ess\,sup}_{\mathbf x,z}|d_{\mathbf t,z}(\mathbf x)-1|
&\le\|\mathbf t\|\mathop{\rm ess\,sup}_{\mathbf x,z}
                        E_s(\|\mathbf s_A(O)\|\mid\mathbf X=\mathbf x,Z=z)\\
&\le C\|\mathbf t\|,\\
e_{s,\mathbf t}(\mathbf X)
&=\frac{e_s(\mathbf X)d_{\mathbf t,1}(\mathbf X)}
 {e_s(\mathbf X)d_{\mathbf t,1}(\mathbf X)
       +\{1-e_s(\mathbf X)\}d_{\mathbf t,0}(\mathbf X)}\\
&=e_s(\mathbf X)+O(\|\mathbf t\|).
\end{aligned}
\]
The conditional denominators and selected-arm probabilities are therefore bounded away from zero locally. The uniform conditional moment bound also permits differentiation of the conditional means. The recovered population observed law is
\begin{equation}\label{eq:patt-target-auxiliary-population-law}
dQ_{\Omega,\mathbf t}^{\mathrm{obs}}
=\frac{w_{A,Z}}{E_{Q_{A,\mathbf t}}w_{A,Z}}\,dQ_{A,\mathbf t}.
\end{equation}
At \(\mathbf t=0\), this recovers \(P_\Omega^{\mathrm{obs}}\). For nonzero \(\mathbf t\), the target is defined under the neighboring recovered population law, expressed through \(Q_{A,\mathbf t}\). To verify identification along the path, assign each potential outcome its observed arm law independently of \(Z\) given \(\mathbf X\), and sample with probability \(1/w_{A,Z}\). This recovers \(Q_{A,\mathbf t}\) and the corresponding target quotient. Proposition~\ref{prop:scientific-target-path-derivative} with \(\mathbf s_A^{\mathrm{path}}(O)=\mathbf s_A(O)\) gives \(\boldsymbol\gamma_{A,2}=E_s\{\phi_A(O)\mathbf s_A(O)\}\), with the PATE representer \eqref{eq:pate-target-representer}. Combining the treated terms for PATT gives \eqref{eq:patt-target-collapsed-representer}. It remains to bound the Taylor remainder. Using \(|g_{\mathrm{tilt}}(v)-1-v|\le C|v|^{1+\min(\delta,1)}\), uniformly in \((\mathbf X,Z)\),
\[
\begin{aligned}
&E_s\!\left[(1+|Y|)
 |g_{\mathrm{tilt}}\{\mathbf t^\top \mathbf s_A(O)\}-1-\mathbf t^\top \mathbf s_A(O)|
                                        \mid\mathbf X,Z\right]\\
&\le C\|\mathbf t\|^{1+\min(\delta,1)}
 E_s[(1+|Y|)^{2+\min(\delta,1)}\mid\mathbf X,Z]\\
&\le C\|\mathbf t\|^{1+\min(\delta,1)},\\
\mu_{z,\mathbf t}(\mathbf X)-\mu_z(\mathbf X)
&=\frac{E_s[\{Y-\mu_z(\mathbf X)\}g_{\mathrm{tilt}}\{\mathbf t^\top \mathbf s_A(O)\}
                                      \mid\mathbf X,Z=z]}{d_{\mathbf t,z}(\mathbf X)}\\
&=\mathbf t^\top E_s[\{Y-\mu_z(\mathbf X)\}\mathbf s_A(O)\mid\mathbf X,Z=z]
                   +O(\|\mathbf t\|^{1+\min(\delta,1)}).
\end{aligned}
\]
For either estimand, use \(\omega_A(\mathbf X,Z)\) from Proposition~\ref{prop:scientific-target-path-derivative}. The denominator expansion is
\[
\begin{aligned}
E_{Q_{A,\mathbf t}}\{\omega_A(\mathbf X,Z)\}
&=\frac{E_s[\omega_A(\mathbf X,Z)
                  g_{\mathrm{tilt}}\{\mathbf t^\top \mathbf s_A(O)\}]}
        {E_s[g_{\mathrm{tilt}}\{\mathbf t^\top \mathbf s_A(O)\}]}\\
&=\bar w_A+\mathbf t^\top E_s\{\omega_A(\mathbf X,Z)\mathbf s_A(O)\}
  +O(\|\mathbf t\|^2).
\end{aligned}
\]
The conditional-mean and denominator expansions can now be substituted into the corresponding target quotient, giving \eqref{eq:pate-target-auxiliary-derivative}. Its denominator is bounded away from zero and \(\|\mathbf t\|^2=O(\|\mathbf t\|^{1+\min(\delta,1)})\). Bounded true inverse propensities and weights give \(E_s\phi_A^2<\infty\). Centering follows from identification and arm-specific conditioning. For every deterministic \(\mathbf h_n\to \mathbf h\), we obtain
\begin{equation}\label{eq:pate-target-auxiliary-local-shift}
\begin{aligned}
\sqrt n\{\tau_A(Q_{A,\mathbf h_n/\sqrt n})-\tau_A\}
&=\mathbf h_n^\top\boldsymbol\gamma_{A,2}
 +O\{n^{-\min(\delta,1)/2}\|\mathbf h_n\|^{1+\min(\delta,1)}\}\\
&\longrightarrow \mathbf h^\top\boldsymbol\gamma_{A,2}.
\end{aligned}
\end{equation}

Finally, the target expansion specifies the centering in the local matching limits. Under the conditions of this lemma and Proposition~\ref{prop:patt-signed-density-local-joint-law}, take \(\tau_n=\tau_A(Q_{A,\mathbf h_n/\sqrt n})\) in \eqref{eq:patt-drift-local-joint-limit}. By \eqref{eq:pate-target-auxiliary-local-shift} and Slutsky's theorem, the limiting mean centered at this target is
\begin{equation}\label{eq:patt-target-scientifically-centered-local-mean}
\begin{aligned}
\mu_A(\mathbf h)-\mathbf h^\top\boldsymbol\gamma_{A,2}
&=-\boldsymbol\gamma_{A,1}^\top\boldsymbol\Sigma_{A,U}^{-1}\boldsymbol\Gamma_A\mathbf h
       -\Delta_A(\mathbf h)/\bar w_A-\mathbf h^\top\boldsymbol\gamma_{A,2}.
\end{aligned}
\end{equation}
The bias-corrected conclusion additionally requires the BC remainder on the same graph and local law.

Under the conditions of this lemma, Theorem~\ref{thm:pate-two-arm-population-joint-law}, part (ii) of Theorem~\ref{prop:pate-two-arm-empirical-joint-transfer}, and the deterministic-local BC remainder on the empirical graphs, let \(\boldsymbol b_{A,\mathrm{loc}}\) be defined by \eqref{eq:pate-two-arm-actual-drift}. The likelihood shift gives the local response coefficient in \eqref{eq:moving-law-joint-input-with-bias}, with centering at the fixed target \(\tau_A(P_s)\):
\begin{equation}\label{eq:pate-target-scientifically-centered-local-mean}
\begin{aligned}
\widetilde{\boldsymbol \gamma}_{A,2}^\top \mathbf h
&=\boldsymbol b_{A,\mathrm{loc}}^\top \mathbf h-
  \mathbf h^\top\boldsymbol\Gamma_A^\top\boldsymbol\Sigma_{A,U}^{-1}\boldsymbol\gamma_{A,1}\\
&=\mathbf h^\top\{\boldsymbol b_{A,\mathrm{loc}}-\boldsymbol\Gamma_A^\top\boldsymbol\Sigma_{A,U}^{-1}\boldsymbol\gamma_{A,1}\}.
\end{aligned}
\end{equation}
Centering instead at \(\tau_A(Q_{A,\mathbf h/\sqrt n})\) subtracts \(\boldsymbol\gamma_{A,2}^\top \mathbf h\) from this limiting mean. Hence
\[
\widetilde{\boldsymbol \gamma}_{A,2}=\boldsymbol\gamma_{A,2}
\quad\Longleftrightarrow\quad
(\widetilde{\boldsymbol \gamma}_{A,2}-\boldsymbol\gamma_{A,2})^\top \mathbf h=0
\quad\text{for every }\mathbf h\in\mathbb R^{p_A}.
\]
The equality requires a mean-zero local limit centered at the moving target under the same path, score construction, and bias correction. The identification, residual-centering, and remainder conditions must support this centered local limit along the path, not only the result at \(P_s\). Without this additional result, the variance expression retains \(\widetilde{\boldsymbol\gamma}_{A,2}\). The estimated-score CLT under \(P_s\) remains centered at \(\tau_A(P_s)\).
\end{proof}
\subsection{Replication and variance estimation}

\paragraph{Ratio expansion and multinomial moments.}
Let \(\mathcal F_n=\sigma(O_1,\ldots,O_n)\). Conditionally on \(\mathcal F_n\), we draw \((m_1^\ast,\ldots,m_n^\ast)\sim \operatorname{Multinomial}(n;1/n,\ldots,1/n)\). For fixed rows \(q_i,d_i\), let \(\widehat t=\sum_iq_i/\sum_id_i\), \(\phi_i=q_i-d_i\widehat t\), and \(\bar d^\ast=n^{-1}\sum_i m_i^\ast d_i\). If the replicated numerator is \(\sum_i m_i^\ast(q_i+r_i)\), then \(\sum_i\phi_i=0\), and on the positive-denominator event
\begin{equation}\label{eq:fixed-match-centered-ratio-algebra}
\begin{aligned}
n\bar d^\ast(\widehat t^\ast-\widehat t)
&=\sum_i m_i^\ast\phi_i+\sum_i m_i^\ast r_i\\
&=\sum_i(m_i^\ast-1)\phi_i+\sum_i m_i^\ast r_i.
\end{aligned}
\end{equation}
The refit differences \(r_i\) may depend on the entire count vector. Under the full multinomial law, \(\operatorname{Cov}^\ast(m_i^\ast,m_j^\ast\mid\mathcal F_n) =\mathbf1(i=j)-1/n\). Thus, with \(\bar d=n^{-1}\sum_i d_i\),
\begin{equation}\label{eq:fixed-match-multinomial-moments}
\begin{aligned}
\operatorname{Var}^\ast\left(n^{-1/2}\sum_i(m_i^\ast-1)\phi_i
                      \mid\mathcal F_n\right)
&=\frac1n\sum_{i,j}\phi_i\phi_j\{\mathbf1(i=j)-1/n\}\\
&=\frac1n\sum_i\phi_i^2-\left(\frac1n\sum_i\phi_i\right)^2\\
&=\frac1n\sum_i\phi_i^2,\\
E^\ast(\bar d^\ast\mid\mathcal F_n)&=\bar d,\\
\operatorname{Var}^\ast(\bar d^\ast\mid\mathcal F_n)
&=\frac1n\left\{\frac1n\sum_i d_i^2-\bar d^2\right\}.
\end{aligned}
\end{equation}
We take each reuse coefficient to be zero outside its matching arm. To express the refit contribution along matching edges, write
\[
\Delta_{z,i}^\ast
=\widehat m_z^\ast\{\widehat{\mathbf D}_z^\ast(\mathbf X_i)\}
 -\widehat m_z\{\widehat{\mathbf D}_z(\mathbf X_i)\}.
\]

For PATE, \(d_i=w_{A,Z_i}(\mathbf X_i)\). For PATT, \(d_i=Z_iw_{A,1}(\mathbf X_i)\). Bounded positive weights and the iid law give \(\bar d\to_p E_s(d_i)>0\), and \eqref{eq:fixed-match-multinomial-moments} gives \(\bar d^\ast-\bar d=O_{p^\ast}(n^{-1/2})\). In the PATT case, with \(n_1=\sum_i Z_i\),
\begin{equation}\label{eq:fixed-match-treated-denominator-zero}
P^\ast(\bar d^\ast=0\mid\mathcal F_n)
=(1-n_1/n)^n\le e^{-n_1}.
\end{equation}

\subsubsection{PATT prediction refitting}
\label{app:patt-full-replication-interface}

Fix \(A\in\{R,P\}\), with \(w_{R,z}=w_z\), \(w_{P,z}=w\), and \(\tau_A=\tau_{\mathrm{PATT}}\). We suppress the replicate number. Each replicate refits the propensity model using the weights specified by the sampling design, the control prognostic model using multinomial counts without survey weights, and the control bias-correction regression using multinomial counts multiplied by survey weights. Empirical standardization follows the specified rule. We hold the fitted, empirically standardized control graph fixed throughout replication, including each donor denominator.

We write \(\mathcal J_0(i)\) for the original selected control set. Its coefficients satisfy
\begin{equation}\label{eq:patt-frozen-reuse-mass}
\begin{aligned}
\lambda_{i\to j}^{A,(0)}
&=\frac{w_{A,0}(\mathbf X_j)
        \mathbf1\{j\in\mathcal J_0(i)\}}
       {\sum_{\ell\in\mathcal J_0(i)}w_{A,0}(\mathbf X_\ell)},\\
K_{n,0}^{A,\mathrm{PATT}}(j)
&=\sum_{i:Z_i=1}w_{A,1}(\mathbf X_i)
                    \lambda_{i\to j}^{A,(0)},\\
\sum_{j:Z_j=0}\lambda_{i\to j}^{A,(0)}
&=1,\\
\sum_{j:Z_j=0}K_{n,0}^{A,\mathrm{PATT}}(j)
&=\sum_{i:Z_i=1}w_{A,1}(\mathbf X_i).
\end{aligned}
\end{equation}
We extend the incoming coefficient by zero on treated rows and denote the treated denominators by
\[
\bar w_{A,n}=\frac1n\sum_iZ_iw_{A,1}(\mathbf X_i),
\qquad
\bar w_{A,n}^\ast=\frac1n\sum_i m_i^\ast Z_iw_{A,1}(\mathbf X_i),
\qquad
\bar w_A=E_s[Zw_{A,1}(\mathbf X)]>0.
\]
The ratios below hold when donor counts are sufficient, the prescribed fits exist, and scales and denominators are positive. The specified finite outputs apply otherwise. Reindexing the bias-corrected matching estimator on the fixed graph gives
\begin{equation}\label{eq:patt-fixed-prediction-ratios}
\begin{aligned}
n\bar w_{A,n}\widehat\tau_A^{\mathrm{BC}}
&=\sum_{i:Z_i=1}w_{A,1}(\mathbf X_i)
  [Y_i-\widehat m_0\{\widehat{\mathbf D}_0(\mathbf X_i)\}]\\
&\quad-\sum_{j:Z_j=0}K_{n,0}^{A,\mathrm{PATT}}(j)
  [Y_j-\widehat m_0\{\widehat{\mathbf D}_0(\mathbf X_j)\}],\\
n\bar w_{A,n}^\ast\widehat\tau_A^\ast
&=\sum_{i:Z_i=1}m_i^\ast w_{A,1}(\mathbf X_i)
  [Y_i-\widehat m_0^\ast\{\widehat{\mathbf D}_0^\ast(\mathbf X_i)\}]\\
&\quad-\sum_{j:Z_j=0}m_j^\ast K_{n,0}^{A,\mathrm{PATT}}(j)
  [Y_j-\widehat m_0^\ast\{\widehat{\mathbf D}_0^\ast(\mathbf X_j)\}].
\end{aligned}
\end{equation}

We center the original numerator row in \eqref{eq:patt-fixed-prediction-ratios} by \(Z_iw_{A,1}(\mathbf X_i)\widehat\tau_A^{\mathrm{BC}}\), obtaining
\begin{equation}\label{eq:patt-common-fitted-centered-row}
\begin{aligned}
\widehat\varphi_{A,\mathrm{PATT},i}
={}&[Z_iw_{A,1}(\mathbf X_i)
 -(1-Z_i)K_{n,0}^{A,\mathrm{PATT}}(i)]
 [Y_i-\widehat m_0\{\widehat{\mathbf D}_0(\mathbf X_i)\}]\\
&-Z_iw_{A,1}(\mathbf X_i)\widehat\tau_A^{\mathrm{BC}}.
\end{aligned}
\end{equation}
Thus \(\sum_i\widehat\varphi_{A,\mathrm{PATT},i}=0\). Replacing the original prediction by its complete refit changes the numerator by
\begin{equation}\label{eq:patt-full-prediction-exact-refit-remainder}
\begin{aligned}
\mathcal R_{A,\mathrm{PATT},n}^\ast
={}&-\sum_i m_i^\ast
 [Z_iw_{A,1}(\mathbf X_i)
 -(1-Z_i)K_{n,0}^{A,\mathrm{PATT}}(i)]\\
&\qquad{}\times
 [\widehat m_0^\ast\{\widehat{\mathbf D}_0^\ast(\mathbf X_i)\}
 -\widehat m_0\{\widehat{\mathbf D}_0(\mathbf X_i)\}].
\end{aligned}
\end{equation}
Reindexing the frozen coefficients in \eqref{eq:patt-frozen-reuse-mass} gives, for \(A=R,P\),
\begin{equation}\label{eq:patt-common-refit-edge-count}
\begin{aligned}
\mathcal R_{A,\mathrm{PATT},n}^\ast
&=-\sum_{i:Z_i=1}w_{A,1}(\mathbf X_i)
 \sum_{j:Z_j=0}\lambda_{i\to j}^{A,(0)}
 (m_i^\ast\Delta_{0,i}^\ast-m_j^\ast\Delta_{0,j}^\ast)\\
&=-\sum_{i:Z_i=1}w_{A,1}(\mathbf X_i)
 \sum_{j:Z_j=0}\lambda_{i\to j}^{A,(0)}
 m_i^\ast(\Delta_{0,i}^\ast-\Delta_{0,j}^\ast)\\
&\quad-\sum_{i:Z_i=1}w_{A,1}(\mathbf X_i)
 \sum_{j:Z_j=0}\lambda_{i\to j}^{A,(0)}
 (m_i^\ast-m_j^\ast)\Delta_{0,j}^\ast.
\end{aligned}
\end{equation}
Applying \eqref{eq:fixed-match-centered-ratio-algebra} gives
\begin{equation}\label{eq:patt-common-centered-replication-ratio}
\begin{aligned}
\sqrt n(\widehat\tau_A^\ast-\widehat\tau_A^{\mathrm{BC}})
&=\frac{\sum_i(m_i^\ast-1)\widehat\varphi_{A,\mathrm{PATT},i}}
        {\bar w_{A,n}^\ast\sqrt n}
 +\frac{\mathcal R_{A,\mathrm{PATT},n}^\ast}
        {\bar w_{A,n}^\ast\sqrt n}.
\end{aligned}
\end{equation}

Combining the limiting score and reduced mean with the incoming coefficients of the fitted graph, define
\begin{equation}\label{eq:patt-frozen-baseline-row}
\varphi_{A,\mathrm{PATT},i}
=
\begin{cases}
w_{A,1}(\mathbf X_i)
 [Y_i-m_0^A\{\mathbf D_0^\ast(\mathbf X_i)\}-\tau_A],
 &Z_i=1,\\
-K_{n,0}^{A,\mathrm{PATT}}(i)
 [Y_i-m_0^A\{\mathbf D_0^\ast(\mathbf X_i)\}],
 &Z_i=0.
\end{cases}
\end{equation}
For a prediction function \(v\), define
\begin{equation}\label{eq:patt-frozen-regression-functionals}
\begin{aligned}
\mathfrak B_{A,n}^{\mathrm{fr}}(v)
={}&\frac1{n\bar w_{A,n}}
\Big[\sum_{i:Z_i=1}w_{A,1}(\mathbf X_i)
 \{v(\mathbf X_i)-m_0^A(\mathbf D_0^\ast(\mathbf X_i))\}\\
&\hspace{8mm}
-\sum_{j:Z_j=0}K_{n,0}^{A,\mathrm{PATT}}(j)
 \{v(\mathbf X_j)-m_0^A(\mathbf D_0^\ast(\mathbf X_j))\}\Big],\\
\mathfrak B_{A,n}^{\mathrm{fr},\ast}(v)
={}&\frac1{n\bar w_{A,n}^\ast}
\Big[\sum_{i:Z_i=1}m_i^\ast w_{A,1}(\mathbf X_i)
 \{v(\mathbf X_i)-m_0^A(\mathbf D_0^\ast(\mathbf X_i))\}\\
&\hspace{8mm}
-\sum_{j:Z_j=0}m_j^\ast K_{n,0}^{A,\mathrm{PATT}}(j)
 \{v(\mathbf X_j)-m_0^A(\mathbf D_0^\ast(\mathbf X_j))\}\Big].
\end{aligned}
\end{equation}
At the original fitted prediction, \(\mathfrak B_{A,n}^{\mathrm{fr}} (\widehat m_0\circ\widehat{\mathbf D}_0)=\widehat B-B\) on the same graph.

\paragraph{Resampling counts.}
\label{par:patt-frozen-edge-count-term}
Write \(\mathbf D_{0,i}^\ast=\mathbf D_0^\ast(\mathbf X_i)\) for the limiting score and \(\widehat{\mathbf D}_{0,i}^\ast =\widehat{\mathbf D}_0^\ast(\mathbf X_i)\) for the replicate score. Reindexing by \eqref{eq:patt-frozen-reuse-mass} gives
\begin{equation}\label{eq:patt-frozen-regression-edge-count}
\begin{aligned}
&n\bar w_{A,n}^\ast\mathfrak B_{A,n}^{\mathrm{fr},\ast}
             (\widehat m_0^\ast\circ\widehat{\mathbf D}_0^\ast)\\
&=\sum_{i:Z_i=1}w_{A,1}(\mathbf X_i)
 \sum_{j:Z_j=0}\lambda_{i\to j}^{A,(0)}
 \Big[m_i^\ast
   \{\widehat m_0^\ast(\widehat{\mathbf D}_{0,i}^\ast)
                         -m_0^A(\mathbf D_{0,i}^\ast)\}\\
&\qquad
 -m_j^\ast
   \{\widehat m_0^\ast(\widehat{\mathbf D}_{0,j}^\ast)
                         -m_0^A(\mathbf D_{0,j}^\ast)\}\Big]\\
&=\sum_{i:Z_i=1}m_i^\ast w_{A,1}(\mathbf X_i)
 \sum_{j:Z_j=0}\lambda_{i\to j}^{A,(0)}
 \Big[\widehat m_0^\ast(\widehat{\mathbf D}_{0,i}^\ast)
                         -m_0^A(\mathbf D_{0,i}^\ast)\\
&\qquad
 -\widehat m_0^\ast(\widehat{\mathbf D}_{0,j}^\ast)
                         +m_0^A(\mathbf D_{0,j}^\ast)\Big]\\
&\quad+\sum_{i:Z_i=1}w_{A,1}(\mathbf X_i)
 \sum_{j:Z_j=0}\lambda_{i\to j}^{A,(0)}
 (m_i^\ast-m_j^\ast)
 \{\widehat m_0^\ast(\widehat{\mathbf D}_{0,j}^\ast)
                         -m_0^A(\mathbf D_{0,j}^\ast)\}.
\end{aligned}
\end{equation}
The second equality adds and subtracts the donor prediction error multiplied by \(m_i^\ast\). Both sums use the same frozen graph.

\begin{proposition}[PATT row moments and joint multipliers]
\label{prop:patt-actual-graph-row-multiplier-law}
Assume the following conditions for PATT:
\begin{enumerate}
\item[(i)] Population recovery, reduced-mean balancing, residual centering, geometry, moments, score roots, and normalization satisfy Proposition~\ref{cor:primary-fitted-score-consistency}.
\item[(ii)] For the joint nuisance conclusion, the first-step conditions hold and \(E_s\|\mathbf U_A(O;\boldsymbol\theta_A^\ast)\|^p<\infty\).
\end{enumerate}
The resulting bounded local score expansion is given in Section~\ref{app:local-score-reuse}.

Using the fitted-match reuse coefficient in \eqref{eq:patt-frozen-baseline-row},
\begin{equation}\label{eq:patt-actual-graph-row-moments}
\begin{aligned}
\frac1n\sum_i\varphi_{A,\mathrm{PATT},i}
&\longrightarrow_p0,\\
\frac1{n\bar w_A^2}\sum_i\varphi_{A,\mathrm{PATT},i}^2
&\longrightarrow_p V_{A,\mathrm{known\text{-}score}},\\
\frac1{n\bar w_A}\sum_i
 \varphi_{A,\mathrm{PATT},i}\{\mathbf U_A(O_i;\boldsymbol\theta_A^\ast)\}
&\longrightarrow_p\boldsymbol\gamma_{A,1}.
\end{aligned}
\end{equation}
Here \(V_{A,\mathrm{known\text{-}score}}=V_H+V_E\) is the known-score variance, and \(\boldsymbol\gamma_{A,1}\) is defined in \eqref{eq:baseline-oracle-firststep-gamma}. These moment limits imply the conditional law
\begin{equation}\label{eq:patt-actual-graph-joint-multiplier-law}
\begin{aligned}
&\mathcal L^\ast\left(
\begin{bmatrix}
\displaystyle\frac1{\bar w_A\sqrt n}
       \sum_i(m_i^\ast-1)\varphi_{A,\mathrm{PATT},i}\\
\displaystyle\frac1{\sqrt n}
       \sum_i(m_i^\ast-1)\{\mathbf U_A(O_i;\boldsymbol\theta_A^\ast)\}
\end{bmatrix}
\middle|\mathcal F_n\right)\\
&\qquad\Longrightarrow_{P_s}
N\left(0,
\begin{bmatrix}
V_{A,\mathrm{known\text{-}score}}&\boldsymbol\gamma_{A,1}^{\top}\\
\boldsymbol\gamma_{A,1}&\boldsymbol\Sigma_{A,U}
\end{bmatrix}\right).
\end{aligned}
\end{equation}

\end{proposition}

\begin{proof}
\par\smallskip\noindent\emph{Row moments.}\quad
Let \(\mathcal E_{n,C}\) be the event that the fitted local nuisance and scale indices have norm at most \(C\) and the uniform score-expansion remainder is at most \(n^{-1/2}\). On \(\mathcal E_{n,C}\), the bound in Section~\ref{app:local-score-reuse} gives \(K_{n,0}^{A,\mathrm{PATT}}\le w_{\max} K_{M,0}^{+,a}\) for a fixed \(a=a(C)\). Conditional on \((Z_j,\mathbf D_{0,j}^\ast)\), the count depends only on the other rows. Hence \eqref{eq:stabilizer-inflated-reuse} gives, for every integrable nonnegative own-row mark \(F(O_j)\) and \(r>0\),
\begin{equation}\label{eq:patt-actual-row-marked-count-bound}
\begin{aligned}
&E_s[(1-Z_j)\{K_{M,0}^{+,a}(j)\}^{r}F(O_j)]\\
&=E_s\left[(1-Z_j)F(O_j)
 E_s\{\{K_{M,0}^{+,a}(j)\}^{r}\mid O_j\}\right]\\
&\le C_{r,a}E_s[(1-Z)F(O)].
\end{aligned}
\end{equation}

First fix a convergent deterministic local-index sequence. Apply the marked-law and \hyperref[app:qv-localization]{localization argument} of Proposition~\ref{prop:primitive-residual-qv-limit} with mark \((\varepsilon_0,\mathbf U_A,S_0,w_{A,Z})\) under the stated conditions. The donor/query bounds and bounded score displacement give the fixed-\(P_s\) limit in \eqref{eq:patt-local-qv-baseline-mark-llns}. The relative scale perturbation vanishes by \eqref{eq:pate-two-arm-grid-score-expansion}. Thus the complete weighted coefficient converges to its baseline incoming law, with independent limits at two distinct roots. The limiting spatial ranks are tie-free.

For \(1<s<p/2\), \eqref{eq:patt-actual-row-marked-count-bound} gives
\begin{equation}\label{eq:patt-actual-row-square-uniform-integrability}
\begin{aligned}
&E_s[\mathbf1_{\mathcal E_{n,C}}(1-Z_j)
 \{K_{n,0}^{A,\mathrm{PATT}}(j)^2
                           \varepsilon_{0,j}^2\}^{s}]\\
&\le C E_s[(1-Z_j)
       \{K_{M,0}^{+,a}(j)\}^{2s}|\varepsilon_{0,j}|^{2s}]\\
&\le C'E_s[(1-Z)|\varepsilon_0|^{2s}]<\infty .
\end{aligned}
\end{equation}
For deterministic local indices in the same bounded set, this bound holds without the indicator for all sufficiently large \(n\). The cross mark has a corresponding bound since
\[
E_s|\varepsilon_0\mathbf U_A|^s
\le \{E_s|\varepsilon_0|^{2s}\}^{1/2}
     \{E_s\|\mathbf U_A\|^{2s}\}^{1/2}<\infty.
\]
Let \(C_{n,i}\) be a scalar root functional clipped at height \(T\), and let \(I_n,J_n\) be a uniform ordered pair of distinct indices, independent of the data. Then
\begin{equation}\label{eq:patt-actual-row-random-root-averaging}
\begin{aligned}
E_s\left(\frac1n\sum_iC_{n,i}\right)
&=E_s C_{n,I_n},\\
E_s\left[\left\{\frac1n\sum_iC_{n,i}\right\}^{2}\right]
&=\frac1n E_s C_{n,I_n}^{2}
 +\frac{n-1}{n}E_s(C_{n,I_n}C_{n,J_n}).
\end{aligned}
\end{equation}
Conditional on the labels, the two-root limit makes the variance of this average tend to zero. The clipping error is bounded by \(CT^{1-s}\) by \eqref{eq:patt-actual-row-square-uniform-integrability}. Letting \(n\to\infty\) and then \(T\to\infty\) proves the deterministic-index moment limits.

\par\smallskip\noindent\emph{Evaluation at the estimated parameter.}\quad
To pass to the fitted index, take a finite lattice of mesh \(\delta\) on a bounded local-index set. Between an index and its lattice representative, let \(F_n\) be the changed-query fraction and \(K_j,\widetilde K_j\) the two coefficient arrays. At most \(2MnF_n\) coefficients change, including retained donors whose denominator changes. For \(r>1\) with \(1/r+2/p<1\), H\"older's inequality gives
\begin{equation}\label{eq:patt-actual-row-lattice-square-bound}
\begin{aligned}
&\frac1n\sum_{j:Z_j=0}
 |K_j^2-\widetilde K_j^2|\varepsilon_{0,j}^{2}\\
&\le C F_n^{\,1-1/r-2/p}
 \left\{\frac1n\sum_{j:Z_j=0}
           \{K_{M,0}^{+,a}(j)\}^{2r}\right\}^{1/r}
 \left\{\frac1n\sum_{j:Z_j=0}
           |\varepsilon_{0,j}|^p\right\}^{2/p}.
\end{aligned}
\end{equation}
Both empirical moment factors are \(O_p(1)\), uniformly in the mesh, by the count bound and the iid mark moment. For the cross functional the corresponding bound is
\begin{align*}
&\frac1n\sum_{j:Z_j=0}
 |K_j-\widetilde K_j|\,|\varepsilon_{0,j}\mathbf U_A(O_j;\boldsymbol\theta_A^\ast)|\\
&\le C F_n^{\,1-1/r-2/p}
 \left\{\frac1n\sum_{j:Z_j=0}
           \{K_{M,0}^{+,a}(j)\}^{r}\right\}^{1/r}\\
&\qquad{}\times
 \left\{\frac1n\sum_{j:Z_j=0}
       |\varepsilon_{0,j}\mathbf U_A(O_j;\boldsymbol\theta_A^\ast)|^{p/2}
 \right\}^{2/p}.
\end{align*}
The last factor is \(O_p(1)\) by H\"older and the global \(p\)-moments. The gap-indicator argument in Section~\ref{app:local-score-marked-limits} gives \(F_n\le C\delta+o_p(1)\), including the uniform Taylor error and the original valid tie choices. Letting \(n\to\infty\), then the lattice mesh \(\delta\downarrow0\), and finally the bounded-index radius \(C\uparrow\infty\) proves convergence of the unscaled empirical moments at the data-dependent index.

It remains to identify the limits. The residual LLN and the iid treated contrast law give the first line of \eqref{eq:patt-actual-graph-row-moments}. Conditional on the root score and weight, the squared control-root limit is the observed-field residual variance times the incoming second moment from the oracle variance proof. The treated square is an iid average whose contrast/residual cross term vanishes by treated observed-field centering. Their sum, divided by \(\bar w_A^2\), is \(V_H+V_E\). For the cross moment, the incoming first moment is exactly \(\kappa_{A,0}(\mathbf D_0^\ast,w_{A,0})\). The control mark is
\[
k_A^\ast
=\operatorname{Cov}_s(\zeta_A,\mathbf U_A\mid\mathcal G_A^\ast),
\qquad
\zeta_A=Y-m_0^A(\mathbf D_0^\ast),
\]
because its control conditional residual mean is zero. The treated tower calculation and the sign of \(\mathbf U_A\) give \(\boldsymbol\gamma_{A,1}\) in \eqref{eq:baseline-oracle-firststep-gamma}.

\par\smallskip\noindent\emph{Conditional multiplier CLT.}\quad
On each bounded-index event, \eqref{eq:patt-actual-row-marked-count-bound} also gives \(n^{-1}\sum_i|\varphi_{A,\mathrm{PATT},i}|^p=O_p(1)\). For every \(\epsilon>0\),
\begin{equation}\label{eq:patt-actual-row-lindeberg}
\begin{aligned}
&\frac1n\sum_i\varphi_{A,\mathrm{PATT},i}^2
 \mathbf1\{|\varphi_{A,\mathrm{PATT},i}|>\epsilon\sqrt n\}\\
&\le(\epsilon\sqrt n)^{-(p-2)}
       \frac1n\sum_i|\varphi_{A,\mathrm{PATT},i}|^p\\
&\longrightarrow_p0.
\end{aligned}
\end{equation}
Localization removes the index restriction in probability. The same bound holds for every fixed linear combination with \(\mathbf U_A(O_i;\boldsymbol\theta_A^\ast)\). Conditionally on \(\mathcal F_n\), multinomial resampling is equivalent to \(n\) iid draws from the empirical row distribution. For fixed scalar \(t\) and vector \(u\), the conditional variance is
\[
\begin{aligned}
&\operatorname{Var}^*\!\left[
 \frac1{\sqrt n}\sum_i(m_i^*-1)
 \left\{\frac{t\varphi_{A,\mathrm{PATT},i}}{\bar w_A}
        +u^\top \mathbf U_A(O_i;\boldsymbol\theta_A^*)\right\}
 \,\middle|\,\mathcal F_n\right]\\
&=\frac1n\sum_i
 \left\{\frac{t\varphi_{A,\mathrm{PATT},i}}{\bar w_A}
        +u^\top \mathbf U_A(O_i;\boldsymbol\theta_A^*)\right\}^2\\
&\quad-\left[\frac1n\sum_i
 \left\{\frac{t\varphi_{A,\mathrm{PATT},i}}{\bar w_A}
        +u^\top \mathbf U_A(O_i;\boldsymbol\theta_A^*)\right\}\right]^2\\
&\longrightarrow_p t^2 V_{A,\mathrm{known\text{-}score}}
       +2t u^\top\boldsymbol\gamma_{A,1}+u^\top\boldsymbol\Sigma_{A,U}u,
\end{aligned}
\]
by \eqref{eq:patt-actual-graph-row-moments} and the iid laws for \(\mathbf U_A\). Thus the Lindeberg--Feller theorem (Theorem~27.2 of Billingsley\citep{Billingsley1995}), applied conditionally along almost-surely convergent further subsequences, and Cram\'er--Wold give \eqref{eq:patt-actual-graph-joint-multiplier-law}. Zero-variance projections follow from conditional Chebyshev.
\end{proof}

For the prediction derivative \(\dot m_{A,0}^{\mathrm{raw}}\) in the expansion below, we define the frozen-prediction derivative
\begin{equation}\label{eq:patt-frozen-prediction-derivative}
\begin{aligned}
\ell_{A,0}
={}&E_s[Zw_{A,1}(\mathbf X)\dot m_{A,0}^{\mathrm{raw}}(\mathbf X)]\\
&-E_s[(1-Z)\kappa_{A,0}(\mathbf D_0^\ast,w_{A,0})
                                      \dot m_{A,0}^{\mathrm{raw}}(\mathbf X)].
\end{aligned}
\end{equation}
Here \(\kappa_{A,0}\) is the limiting mean of the reuse coefficient for fixed \(M\).

\begin{proposition}[PATT replication limit]
\label{prop:patt-full-prediction-replication-law}\label{prop:patt-frozen-baseline-replication-identity}\label{cor:patt-full-prediction-joint-comparison}
Assume the following conditions:
\begin{enumerate}
\item[(i)] Proposition~\ref{prop:patt-actual-graph-row-multiplier-law} holds, the fitted estimator is consistent, and the first-step bootstrap expansion of Lemma~\ref{lem:primitive-score-root-bootstrap} holds. The fit and denominator failure probabilities tend to zero.
\item[(ii)] The prescribed prediction refit admits
\begin{equation}\label{eq:patt-full-prediction-refit-expansion}
\begin{aligned}
&\widehat m_0^\ast\{\widehat{\mathbf D}_0^\ast(\mathbf X_i)\}
 -\widehat m_0\{\widehat{\mathbf D}_0(\mathbf X_i)\}\\
&=\dot m_{A,0}^{\mathrm{raw}}(\mathbf X_i)^\top
                       (\widehat{\boldsymbol \theta}_A^\ast-\widehat{\boldsymbol \theta}_A)
 +q_{A,0}\{\mathbf D_0^\ast(\mathbf X_i)\}^{\top}t_{A,n}^\ast
 +r_{A,n,i}^\ast .
\end{aligned}
\end{equation}
Here \(\dot m_{A,0}^{\mathrm{raw}}\) is a fixed bounded vector of dimension \(p_A=\dim\boldsymbol\theta_A\). The fixed finite vector \(q_{A,0}\) consists of bounded Lipschitz functions of that score, and \(\sqrt n\,t_{A,n}^\ast=O_{p^\ast}(1)\).
\item[(iii)] The weighted remainder satisfies
\begin{equation}\label{eq:patt-full-prediction-weighted-remainder}
\begin{aligned}
\frac1{\sqrt n}\sum_i m_i^\ast
 &[Z_iw_{A,1}(\mathbf X_i)
 +(1-Z_i)K_{n,0}^{A,\mathrm{PATT}}(i)]\\
&\hspace{8mm}\times |r_{A,n,i}^\ast|
=o_{p^\ast}(1)
\quad\text{in }P_s\text{-probability}.
\end{aligned}
\end{equation}
\item[(iv)] The original prediction is consistent in the weighted empirical norm:
\begin{equation}\label{eq:patt-weighted-prediction-consistency}
\begin{aligned}
\frac1n\sum_i
 &[Z_iw_{A,1}(\mathbf X_i)^2
 +(1-Z_i)K_{n,0}^{A,\mathrm{PATT}}(i)^2]\\
&\times
 [\widehat m_0\{\widehat{\mathbf D}_0(\mathbf X_i)\}
         -m_0^A\{\mathbf D_0^\ast(\mathbf X_i)\}]^2
\longrightarrow_p0.
\end{aligned}
\end{equation}
\end{enumerate}
Then
\begin{equation}\label{eq:patt-full-prediction-replication-law}
\mathcal L^\ast\left\{
\sqrt n(\widehat\tau_A^\ast-\widehat\tau_A^{\mathrm{BC}})
\mid\mathcal F_n\right\}
\Longrightarrow_{P_s}N(0,V_{A,\mathrm{rep}}),
\end{equation}
where
\begin{equation}\label{eq:patt-full-prediction-replication-variance}
\begin{aligned}
V_{A,\mathrm{rep}}
={}&V_{A,\mathrm{known\text{-}score}}
 +\frac{2}{\bar w_A}
       \ell_{A,0}^{\top}\boldsymbol\Gamma_A^{-1}\boldsymbol\gamma_{A,1}\\
&+\frac1{\bar w_A^2}
       \ell_{A,0}^{\top}\boldsymbol\Sigma_{A,\theta}\ell_{A,0}.
\end{aligned}
\end{equation}
The variance is nonnegative. A zero variance gives a point-mass limit.

\par\smallskip\noindent\emph{Sampling and replication.}\quad
The joint conditional law with \(\sqrt n(\widehat{\boldsymbol \theta}_A^\ast-\widehat{\boldsymbol \theta}_A)\) has first-coordinate variance \eqref{eq:patt-full-prediction-replication-variance}, nuisance covariance \(\boldsymbol\Sigma_{A,\theta}\), and cross covariance
\begin{equation}\label{eq:patt-full-prediction-joint-cross-covariance}
-\boldsymbol\Gamma_A^{-1}\boldsymbol\gamma_{A,1}
       -\boldsymbol\Sigma_{A,\theta}\ell_{A,0}/\bar w_A.
\end{equation}
If, in addition, the hypotheses of Theorem~\ref{cor:patt-joint-actual-score-law} hold, its sampling law is \((G_{\tau,A}-\mathbf j_A^\top \mathbf G_{\theta,A},\mathbf G_{\theta,A})\), where
\[
\operatorname{Var}(G_{\tau,A})=V_{A,\mathrm{known\text{-}score}},
\qquad
\operatorname{Cov}(\mathbf G_{\theta,A},G_{\tau,A})=-\boldsymbol\Gamma_A^{-1}\boldsymbol\gamma_{A,1},
\qquad
\operatorname{Var}(\mathbf G_{\theta,A})=\boldsymbol\Sigma_{A,\theta}.
\]

For positive definite \(\boldsymbol\Sigma_{A,\theta}\), equality of these \emph{joint} centered Gaussian laws is equivalent to
\begin{equation}\label{eq:patt-full-prediction-joint-compatibility}
\ell_{A,0}/\bar w_A=\mathbf j_A.
\end{equation}
The scalar laws agree if and only if their limiting variances agree.
\end{proposition}

\begin{proof}
Adding and subtracting the limiting reduced mean in \eqref{eq:patt-fixed-prediction-ratios} gives
\begin{equation}\label{eq:patt-frozen-baseline-numerators}
\begin{aligned}
n\bar w_{A,n}^\ast(\widehat\tau_A^\ast-\tau_A)
&=\sum_i m_i^\ast\varphi_{A,\mathrm{PATT},i}
 -n\bar w_{A,n}^\ast\mathfrak B_{A,n}^{\mathrm{fr},\ast}
       (\widehat m_0^\ast\circ\widehat{\mathbf D}_0^\ast),\\
n\bar w_{A,n}(\widehat\tau_A^{\mathrm{BC}}-\tau_A)
&=\sum_i\varphi_{A,\mathrm{PATT},i}
 -n\bar w_{A,n}\mathfrak B_{A,n}^{\mathrm{fr}}
       (\widehat m_0\circ\widehat{\mathbf D}_0).
\end{aligned}
\end{equation}
Subtracting these numerators and using \(n(\bar w_{A,n}^\ast-\bar w_{A,n}) =\sum_i(m_i^\ast-1)Z_iw_{A,1}(\mathbf X_i)\), we obtain
\begin{equation}\label{eq:patt-frozen-baseline-exact-ratio}
\begin{aligned}
&\sqrt n(\widehat\tau_A^\ast-\widehat\tau_A^{\mathrm{BC}})\\
&=\frac1{\bar w_{A,n}^\ast\sqrt n}
       \sum_i(m_i^\ast-1)\varphi_{A,\mathrm{PATT},i}\\
&\quad-\frac{\widehat\tau_A^{\mathrm{BC}}-\tau_A}
                {\bar w_{A,n}^\ast\sqrt n}
       \sum_i(m_i^\ast-1)Z_iw_{A,1}(\mathbf X_i)\\
&\quad+\frac{\bar w_{A,n}}{\bar w_{A,n}^\ast}
       \sqrt n\,\mathfrak B_{A,n}^{\mathrm{fr}}
                (\widehat m_0\circ\widehat{\mathbf D}_0)\\
&\quad-\sqrt n\,\mathfrak B_{A,n}^{\mathrm{fr},\ast}
                (\widehat m_0^\ast\circ\widehat{\mathbf D}_0^\ast).
\end{aligned}
\end{equation}

\par\smallskip\noindent\emph{Prediction derivative.}\quad
The one-root and two-root proof in Proposition~\ref{prop:patt-actual-graph-row-multiplier-law}, applied to the fixed bounded mark \(\dot m_{A,0}^{\mathrm{raw}}\), gives
\begin{equation}\label{eq:patt-frozen-prediction-derivative-lln}
\begin{aligned}
&\frac1n\Big[
 \sum_{i:Z_i=1}w_{A,1}(\mathbf X_i)\dot m_{A,0}^{\mathrm{raw}}(\mathbf X_i)
 -\sum_{j:Z_j=0}K_{n,0}^{A,\mathrm{PATT}}(j)
                                      \dot m_{A,0}^{\mathrm{raw}}(\mathbf X_j)\Big]\\
&\hspace{15mm}\longrightarrow_p\ell_{A,0}.
\end{aligned}
\end{equation}
For the baseline-score function \(q_{A,0}\), mass reindexing gives
\begin{align*}
&\frac1n\Big[
 \sum_{i:Z_i=1}w_{A,1}(\mathbf X_i)
                         q_{A,0}(\mathbf D_{0,i}^\ast)
 -\sum_{j:Z_j=0}K_{n,0}^{A,\mathrm{PATT}}(j)
                         q_{A,0}(\mathbf D_{0,j}^\ast)\Big]\\
&=\frac1n\sum_{i:Z_i=1}w_{A,1}(\mathbf X_i)
 \sum_{j:Z_j=0}\lambda_{i\to j}^{A,(0)}
 [q_{A,0}(\mathbf D_{0,i}^\ast)-q_{A,0}(\mathbf D_{0,j}^\ast)]\\
&\longrightarrow_p0.
\end{align*}
Indeed, Lipschitz continuity gives
\[
\begin{aligned}
&\left\|\frac1n\sum_{i:Z_i=1}w_{A,1}(\mathbf X_i)
 \sum_{j:Z_j=0}\lambda_{i\to j}^{A,(0)}
 [q_{A,0}(\mathbf D_{0,i}^\ast)-q_{A,0}(\mathbf D_{0,j}^\ast)]\right\|\\
&\le\frac{\operatorname{Lip}(q_{A,0})}{n}
 \sum_{i:Z_i=1}w_{A,1}(\mathbf X_i)
 \sum_{j:Z_j=0}\lambda_{i\to j}^{A,(0)}
 \|\mathbf D_{0,i}^\ast-\mathbf D_{0,j}^\ast\|\\
&=O_p(n^{-1/2})\longrightarrow_p0,
\end{aligned}
\]
by Lemma~\ref{lem:oracle-score-radius} and the uniform \(O_p(n^{-1/2})\) score perturbation.

Inserting \(m_i^\ast\) in either bounded-mark average preserves its limit. For any fixed bounded scalar component \(b_i\), the multinomial identity bounds the variance of its increment by
\begin{align*}
&\operatorname{Var}^\ast\left[
 \frac1n\sum_i(m_i^\ast-1)
 [Z_iw_{A,1}(\mathbf X_i)
       -(1-Z_i)K_{n,0}^{A,\mathrm{PATT}}(i)]b_i
 \,\middle|\,\mathcal F_n\right]\\
&\le\frac{C}{n^2}\sum_i
 [Z_iw_{A,1}(\mathbf X_i)^2
       +(1-Z_i)K_{n,0}^{A,\mathrm{PATT}}(i)^2]\\
&=O_p(n^{-1}).
\end{align*}
Here the squared-reuse bound follows from the preceding proposition with the mark set to one.

\par\smallskip\noindent\emph{Remainder after refitting.}\quad
Substitution of \eqref{eq:patt-full-prediction-refit-expansion} into \eqref{eq:patt-full-prediction-exact-refit-remainder}, together with \eqref{eq:patt-full-prediction-weighted-remainder}, gives
\begin{equation}\label{eq:patt-full-prediction-refit-first-order}
\begin{aligned}
\frac{\mathcal R_{A,\mathrm{PATT},n}^\ast}{\sqrt n}
={}&-\Big\{\frac1n\sum_i m_i^\ast
 [Z_iw_{A,1}(\mathbf X_i)
       -(1-Z_i)K_{n,0}^{A,\mathrm{PATT}}(i)]
                              \dot m_{A,0}^{\mathrm{raw}}(\mathbf X_i)\Big\}^{\top}\\
&\hspace{15mm}\times
             \sqrt n(\widehat{\boldsymbol \theta}_A^\ast-\widehat{\boldsymbol \theta}_A)\\
&-\Big\{\frac1n\sum_i m_i^\ast
 [Z_iw_{A,1}(\mathbf X_i)
       -(1-Z_i)K_{n,0}^{A,\mathrm{PATT}}(i)]
                       q_{A,0}(\mathbf D_{0,i}^\ast)\Big\}^{\top}
             \sqrt n\,t_{A,n}^\ast+o_{p^\ast}(1)\\
={}&-\ell_{A,0}^{\top}
             \sqrt n(\widehat{\boldsymbol \theta}_A^\ast-\widehat{\boldsymbol \theta}_A)
             +o_{p^\ast}(1)\\
={}&\ell_{A,0}^{\top}\boldsymbol\Gamma_A^{-1}
       \frac1{\sqrt n}\sum_i(m_i^\ast-1)
                   \{\mathbf U_A(O_i;\boldsymbol\theta_A^\ast)\}+o_{p^\ast}(1).
\end{aligned}
\end{equation}
The bounded-mark limits give the second equality, and the score-root bootstrap expansion gives the third.

\par\smallskip\noindent\emph{Centered multiplier substitution.}\quad
The fitted row \eqref{eq:patt-common-fitted-centered-row} and the baseline row \eqref{eq:patt-frozen-baseline-row} satisfy
\begin{align*}
&\frac1n\sum_i
 |\widehat\varphi_{A,\mathrm{PATT},i}
                 -\varphi_{A,\mathrm{PATT},i}|^2\\
&\le\frac2n\sum_i
 [Z_iw_{A,1}(\mathbf X_i)^2
       +(1-Z_i)K_{n,0}^{A,\mathrm{PATT}}(i)^2]\\
&\qquad{}\times
 [\widehat m_0\{\widehat{\mathbf D}_0(\mathbf X_i)\}
                      -m_0^A\{\mathbf D_0^\ast(\mathbf X_i)\}]^2\\
&\quad+2(\widehat\tau_A^{\mathrm{BC}}-\tau_A)^2
                       \frac1n\sum_iZ_iw_{A,1}(\mathbf X_i)^2\\
&\longrightarrow_p0.
\end{align*}
Weighted prediction consistency \eqref{eq:patt-weighted-prediction-consistency} makes the first term vanish. Estimator consistency and bounded weights handle the second. The multinomial covariance identity then gives
\[
\begin{aligned}
&\operatorname{Var}^*\!\left[
 \frac1{\sqrt n}\sum_i(m_i^*-1)
 (\widehat\varphi_{A,\mathrm{PATT},i}-\varphi_{A,\mathrm{PATT},i})
 \,\middle|\,\mathcal F_n\right]\\
&\le\frac1n\sum_i
 |\widehat\varphi_{A,\mathrm{PATT},i}-\varphi_{A,\mathrm{PATT},i}|^2\\
&\longrightarrow_p0.
\end{aligned}
\]
Conditional Chebyshev therefore gives
\[
\frac1{\sqrt n}\sum_i(m_i^\ast-1)
 [\widehat\varphi_{A,\mathrm{PATT},i}
                   -\varphi_{A,\mathrm{PATT},i}]
=o_{p^\ast}(1).
\]
Since the fitted row sums to zero, the ratio identity and \eqref{eq:patt-full-prediction-refit-first-order} now imply
\begin{equation}\label{eq:patt-full-prediction-replicate-linearization}
\begin{aligned}
&\sqrt n(\widehat\tau_A^\ast-\widehat\tau_A^{\mathrm{BC}})\\
&=\frac1{\bar w_{A,n}^\ast\sqrt n}
       \sum_i(m_i^\ast-1)\widehat\varphi_{A,\mathrm{PATT},i}
       +\frac{\mathcal R_{A,\mathrm{PATT},n}^\ast}
              {\bar w_{A,n}^\ast\sqrt n}\\
&=\frac1{\bar w_A\sqrt n}
       \sum_i(m_i^\ast-1)\varphi_{A,\mathrm{PATT},i}\\
&\quad+\frac1{\bar w_A}\ell_{A,0}^{\top}\boldsymbol\Gamma_A^{-1}
       \frac1{\sqrt n}\sum_i(m_i^\ast-1)
                    \{\mathbf U_A(O_i;\boldsymbol\theta_A^\ast)\}
       +o_{p^\ast}(1).
\end{aligned}
\end{equation}
The denominator has the positive limit established above. Combining the joint multiplier law \eqref{eq:patt-actual-graph-joint-multiplier-law} and conditional Slutsky therefore yields \eqref{eq:patt-full-prediction-replication-law}. Applying the first-stage expansion to the joint Gaussian limit gives \((G_{\tau,A},\mathbf G_{\theta,A})\). The replication variance is
\[
\begin{aligned}
\operatorname{Var}(G_{\tau,A}-\ell_{A,0}^\top \mathbf G_{\theta,A}/\bar w_A)
&=\operatorname{Var}(G_{\tau,A})
 -\frac{2\ell_{A,0}^\top\operatorname{Cov}(\mathbf G_{\theta,A},G_{\tau,A})}{\bar w_A}
 +\frac{\ell_{A,0}^\top\operatorname{Var}(\mathbf G_{\theta,A})\ell_{A,0}}{\bar w_A^2}\\
&=V_{A,\mathrm{known\text{-}score}}
 +\frac{2\ell_{A,0}^\top\boldsymbol\Gamma_A^{-1}\boldsymbol\gamma_{A,1}}{\bar w_A}
 +\frac{\ell_{A,0}^\top\boldsymbol\Sigma_{A,\theta}\ell_{A,0}}{\bar w_A^2},
\end{aligned}
\]
where \(\boldsymbol\Sigma_{A,\theta}=\boldsymbol\Gamma_A^{-1}\boldsymbol\Sigma_{A,U}\boldsymbol\Gamma_A^{-\top}\).

\par\smallskip\noindent\emph{Sampling and replication.}\quad
Applying the bootstrap-root expansion and \eqref{eq:patt-full-prediction-replicate-linearization} to the joint multiplier limit \eqref{eq:patt-actual-graph-joint-multiplier-law} gives \((G_{\tau,A}-\ell_{A,0}^\top \mathbf G_{\theta,A}/\bar w_A,\mathbf G_{\theta,A})\). The sandwich identity \(\boldsymbol\Sigma_{A,\theta}=\boldsymbol\Gamma_A^{-1}\boldsymbol\Sigma_{A,U}\boldsymbol\Gamma_A^{-\top}\) gives \eqref{eq:patt-full-prediction-joint-cross-covariance}. The difference from the sampling cross block is
\[
\begin{aligned}
&\operatorname{Cov}(\mathbf G_{\theta,A},G_{\tau,A}-\ell_{A,0}^\top \mathbf G_{\theta,A}/\bar w_A)
 -\operatorname{Cov}(\mathbf G_{\theta,A},G_{\tau,A}-\mathbf j_A^\top \mathbf G_{\theta,A})\\
&=-\boldsymbol\Gamma_A^{-1}\boldsymbol\gamma_{A,1}-\boldsymbol\Sigma_{A,\theta}\ell_{A,0}/\bar w_A
 -\{-\boldsymbol\Gamma_A^{-1}\boldsymbol\gamma_{A,1}-\boldsymbol\Sigma_{A,\theta}\mathbf j_A\}\\
&=\boldsymbol\Sigma_{A,\theta}(\mathbf j_A-\ell_{A,0}/\bar w_A).
\end{aligned}
\]
Since \(\boldsymbol\Sigma_{A,\theta}\) is positive definite, this is zero exactly under \eqref{eq:patt-full-prediction-joint-compatibility}. Conversely, that equality makes the Gaussian transformations identical. Scalar agreement follows by comparing the variances.
\end{proof}

\subsubsection{PATE prediction refitting}
\label{app:pate-full-prediction-replication}
Fix \(A\in\{R,P\}\), with \(w_{R,z}=w_z\) and \(w_{P,z}=w\). We use the full PATE parameter stack \(\boldsymbol\theta_{A,\mathrm{PATE}}\). The coefficients \(K_{n,z}^{A,\mathrm{PATE}}\) use the estimated, empirically standardized scores. The match sets and donor-weight denominators are held fixed across replicates. With
\[
\bar w_{A,\mathrm{PATE},n}^\ast
 =\frac1n\sum_i m_i^\ast w_{A,Z_i}(\mathbf X_i),
\]
the fixed-match replicate is
\begin{equation}\label{eq:pate-fixed-prediction-ratio}
\begin{aligned}
n\bar w_{A,\mathrm{PATE},n}^\ast
 \widehat\tau_{\mathrm{PATE}}^{A,\ast}
&=\sum_i m_i^\ast w_{A,Z_i}(\mathbf X_i)
 \left[\widehat m_1^\ast\{\widehat{\mathbf D}_1^\ast(\mathbf X_i)\}
       -\widehat m_0^\ast\{\widehat{\mathbf D}_0^\ast(\mathbf X_i)\}\right]\\
&\quad+\sum_i m_i^\ast(2Z_i-1)
 \{w_{A,Z_i}(\mathbf X_i)+K_{n,Z_i}^{A,\mathrm{PATE}}(i)\}\\
&\qquad{}\times
 \left[Y_i-\widehat m_{Z_i}^\ast
                 \{\widehat{\mathbf D}_{Z_i}^\ast(\mathbf X_i)\}\right].
\end{aligned}
\end{equation}
Taking every count to be one and using the original fitted predictions gives \(n\bar w_{A,\mathrm{PATE},n} \widehat\tau_{\mathrm{PATE}}^{A,\mathrm{BC}}\). Center its \(i\)th summand by \(w_{A,Z_i}(\mathbf X_i)\widehat\tau_{\mathrm{PATE}}^{A,\mathrm{BC}}\) to obtain \(\widehat\varphi_{A,\mathrm{PATE},i}\). Cancellation of the observed-arm prediction terms leaves the refit contribution
\begin{equation}\label{eq:pate-common-fixed-refit-remainder}
\begin{aligned}
\mathcal R_{A,\mathrm{PATE},n}^\ast
&=\sum_{z=0}^1(2z-1)\sum_i m_i^\ast
 \left[\mathbf1\{Z_i=1-z\}w_{A,1-z}(\mathbf X_i)
       -\mathbf1\{Z_i=z\}K_{n,z}^{A,\mathrm{PATE}}(i)\right]\\
&\qquad{}\times
 \left[\widehat m_z^\ast\{\widehat{\mathbf D}_z^\ast(\mathbf X_i)\}
       -\widehat m_z\{\widehat{\mathbf D}_z(\mathbf X_i)\}\right].
\end{aligned}
\end{equation}
For \(A=R,P\), the corresponding edge identity is
\begin{equation}\label{eq:pate-common-refit-edge-count}
\begin{aligned}
\mathcal R_{A,\mathrm{PATE},n}^\ast
&=\sum_{z=0}^1(2z-1)\sum_{i:Z_i=1-z}w_{A,1-z}(\mathbf X_i)
 \sum_{j:Z_j=z}\lambda_{i\to j}^{A,(z)}
 (m_i^\ast\Delta_{z,i}^\ast-m_j^\ast\Delta_{z,j}^\ast)\\
&=\sum_{z=0}^1(2z-1)\sum_{i:Z_i=1-z}w_{A,1-z}(\mathbf X_i)
 \sum_{j:Z_j=z}\lambda_{i\to j}^{A,(z)}
 m_i^\ast(\Delta_{z,i}^\ast-\Delta_{z,j}^\ast)\\
&\quad+\sum_{z=0}^1(2z-1)\sum_{i:Z_i=1-z}w_{A,1-z}(\mathbf X_i)
 \sum_{j:Z_j=z}\lambda_{i\to j}^{A,(z)}
 (m_i^\ast-m_j^\ast)\Delta_{z,j}^\ast.
\end{aligned}
\end{equation}
Equation~\eqref{eq:fixed-match-centered-ratio-algebra} therefore gives
\begin{equation}\label{eq:pate-common-centered-replication-ratio}
\begin{aligned}
&\sqrt n(\widehat\tau_{\mathrm{PATE}}^{A,\ast}
       -\widehat\tau_{\mathrm{PATE}}^{A,\mathrm{BC}})\\
&=\frac{\sum_i(m_i^\ast-1)\widehat\varphi_{A,\mathrm{PATE},i}}
        {\bar w_{A,\mathrm{PATE},n}^\ast\sqrt n}
 +\frac{\mathcal R_{A,\mathrm{PATE},n}^\ast}
        {\bar w_{A,\mathrm{PATE},n}^\ast\sqrt n}.
\end{aligned}
\end{equation}

On these graphs, we define the unnormalized baseline-mean row by
\begin{equation}\label{eq:pate-actual-graph-baseline-row}
\begin{aligned}
\varphi_{A,\mathrm{PATE},i}
:={}&w_{A,Z_i}(\mathbf X_i)
       \{\tau_A^D(\mathbf X_i)-\tau_{\mathrm{PATE}}\}\\
&+(2Z_i-1)
 \{w_{A,Z_i}(\mathbf X_i)
       +K_{n,Z_i}^{A,\mathrm{PATE}}(i)\}
       \varepsilon_{Z_i,i}^A,\\
\tau_A^D(\mathbf X)
:={}&m_1^A\{\mathbf D_1^\ast(\mathbf X)\}
       -m_0^A\{\mathbf D_0^\ast(\mathbf X)\},\\
\varepsilon_{z,i}^A
:={}&Y_i-m_z^A\{\mathbf D_z^\ast(\mathbf X_i)\},
       \qquad Z_i=z.
\end{aligned}
\end{equation}
As in the scenario definitions, put \(\bar w_{A,\mathrm{PATE}}=E_s\{w_{A,Z}(\mathbf X)\}>0\).

\Needspace{8\baselineskip}
\begin{proposition}[PATE row moments and joint multipliers]
\label{prop:pate-actual-graph-row-multiplier-law}
Assume the following conditions for PATE:
\begin{enumerate}
\item[(i)] Population recovery and the reduced-mean identities hold. The residuals satisfy the joint-score/observed-weight centering in Assumption~\ref{ass:residual-centering-survey-normalized}.
\item[(ii)] Both directional graphs satisfy conditions (i)--(iv) of Proposition~\ref{cor:primary-fitted-score-consistency}. In particular, observations are selected-iid and \(M\) is fixed. Each baseline support is compact, convex, and two-dimensional, with nonempty interior. The donor density is bounded above and away from zero, the query density is bounded above, and the survey weights are bounded and positive.
\item[(iii)] The bounded-regressor finite-root/rank conditions in Lemma~\ref{lem:primitive-logistic-linear-roots} hold. Standardization uses the fixed bounded weights and positive scales in Lemma~\ref{lem:primitive-score-standardization}.
\item[(iv)] For some \(p>2\), the observed residuals, the heterogeneity summand in \eqref{eq:pate-actual-graph-baseline-row}, and \(U_{A,\mathrm{PATE}}(O;\boldsymbol\theta_{A,\mathrm{PATE}}^\ast)\) have finite global \(p\)-moments.
\end{enumerate}

Then
\begin{equation}\label{eq:pate-actual-graph-row-moments}
\begin{aligned}
\frac1n\sum_i\varphi_{A,\mathrm{PATE},i}
&\longrightarrow_p0,\\
\frac1{n\bar w_{A,\mathrm{PATE}}^2}
       \sum_i\varphi_{A,\mathrm{PATE},i}^2
&\longrightarrow_p V_{A,\mathrm{PATE},\mathrm{known\text{-}score}},\\
\frac1{n\bar w_{A,\mathrm{PATE}}}
       \sum_i\varphi_{A,\mathrm{PATE},i}
       \{U_{A,\mathrm{PATE}}(O_i;\boldsymbol\theta_{A,\mathrm{PATE}}^\ast)\}
&\longrightarrow_p\boldsymbol\gamma_{A,\mathrm{PATE},1}.
\end{aligned}
\end{equation}
The constants are defined for PATE in \eqref{eq:baseline-oracle-firststep-gamma}. Consequently,
\begin{equation}\label{eq:pate-actual-graph-joint-multiplier-law}
\begin{aligned}
&\mathcal L^\ast\left(
\begin{bmatrix}
\displaystyle\frac1{\bar w_{A,\mathrm{PATE}}\sqrt n}
       \sum_i(m_i^\ast-1)\varphi_{A,\mathrm{PATE},i}\\
\displaystyle\frac1{\sqrt n}\sum_i(m_i^\ast-1)
       \{U_{A,\mathrm{PATE}}(O_i;\boldsymbol\theta_{A,\mathrm{PATE}}^\ast)\}
\end{bmatrix}\middle|\mathcal F_n\right)\\
&\qquad\Longrightarrow_{P_s}
N\left(0,
\begin{bmatrix}
V_{A,\mathrm{PATE},\mathrm{known\text{-}score}}
       &\boldsymbol\gamma_{A,\mathrm{PATE},1}^{\top}\\
\boldsymbol\gamma_{A,\mathrm{PATE},1}&\boldsymbol\Sigma_{A,\mathrm{PATE},U}
\end{bmatrix}\right).
\end{aligned}
\end{equation}

\end{proposition}

\begin{proof}
For each matching direction, augment the mark in the proof of Proposition~\ref{prop:patt-actual-graph-row-multiplier-law} by both scores, heterogeneity and the full estimating function. Its conditional count bound then gives, for every integrable nonnegative own-row mark \(F\) and \(r>0\),
\begin{equation}\label{eq:pate-actual-row-marked-count-bound}
E_s[\mathbf1\{Z_j=z\}\{1+K_{M,z}^{+,a}(j)\}^{r}F(O_j)]
\le C_{r,a}E_s[\mathbf1\{Z=z\}F(O)].
\end{equation}
Expanding \eqref{eq:pate-actual-graph-baseline-row} gives
\begin{equation}\label{eq:pate-actual-row-square-decomposition}
\begin{aligned}
\varphi_{A,\mathrm{PATE},i}^2
={}&w_{A,Z_i}^2\{\tau_A^D(\mathbf X_i)-\tau_{\mathrm{PATE}}\}^2\\
&+2(2Z_i-1)w_{A,Z_i}\{\tau_A^D(\mathbf X_i)-\tau_{\mathrm{PATE}}\}
 \{w_{A,Z_i}+K_{n,Z_i}^{A,\mathrm{PATE}}(i)\}
 \varepsilon_{Z_i,i}^A\\
&+\{w_{A,Z_i}+K_{n,Z_i}^{A,\mathrm{PATE}}(i)\}^2
 (\varepsilon_{Z_i,i}^A)^2.
\end{aligned}
\end{equation}
The first term is iid. The remaining terms involve only one arm's coefficient, to degree at most two. For \(1<s<p/2\), the count bound and H\"older's inequality give uniform \(s\)-moments for all square and cross terms. Equation~\eqref{eq:patt-actual-row-random-root-averaging} therefore applies, with clipping error bounded by \(CT^{1-s}\).

For the fitted indices, apply \eqref{eq:patt-actual-row-lattice-square-bound} and its cross-term version on a finite lattice for the full score/scale stack. Each arm has \(F_{n,z}\le C\delta+o_p(1)\) and at most \(2MnF_{n,z}\) affected coefficients. For \(r>1\) with \(1/r+2/p<1\), the error is \(F_{n,z}^{1-1/r-2/p}O_p(1)\). Letting \(n\to\infty\), then \(\delta\downarrow0\), then the index radius tend to infinity gives both fitted-index limits.

To identify the limits, we condition on the root arm, both baseline scores and its observed weight. Heterogeneity is measurable and the residual is centered by the joint-score condition. The limiting incoming process is independent of remaining root marks conditional on the own score and weight. In particular, the arm-\(z\) limit of the middle term is zero because
\[
\begin{aligned}
&E_s\!\left[\mathbf1(Z=z)w_{A,z}
 \{\tau_A^D(\mathbf X)-\tau_{\mathrm{PATE}}\}
 \{w_{A,z}+\kappa_{A,z}(\mathbf D_z^*,w_{A,z})\}
 \varepsilon_z^A\right]\\
&=E_s\!\left[\mathbf1(Z=z)w_{A,z}
 \{\tau_A^D(\mathbf X)-\tau_{\mathrm{PATE}}\}
 \{w_{A,z}+\kappa_{A,z}(\mathbf D_z^*,w_{A,z})\}
 E_s(\varepsilon_z^A\mid Z,\mathbf D_0^*,\mathbf D_1^*,w_{A,Z})\right]\\
&=0.
\end{aligned}
\]
The empirical average of the remaining square terms, divided by \(\bar w_{A,\mathrm{PATE}}^2\), gives \(V_H+V_E\). The full residual/\(U\) and heterogeneity/\(U\) cross products, averaged and divided by \(\bar w_{A,\mathrm{PATE}}\), give \(\boldsymbol\gamma_{A,\mathrm{PATE},1}\). The residual LLNs and population identification give the zero row mean, completing \eqref{eq:pate-actual-graph-row-moments}.

Finally, \(n^{-1}\sum_i|\varphi_{A,\mathrm{PATE},i}|^p=O_p(1)\) and the \(p\)-moment of \(U\) give \eqref{eq:patt-actual-row-lindeberg} for every projection. The conditional empirical-row CLT used for PATT proves \eqref{eq:pate-actual-graph-joint-multiplier-law}.
\end{proof}

For the prediction derivatives \(\dot m_{A,z}^{\mathrm{raw}}\) in the expansion below, we define the two prediction-imbalance derivatives by
\begin{equation}\label{eq:pate-frozen-prediction-derivatives}
\begin{aligned}
\ell_{A,z}
:={}&E_s[\mathbf1\{Z=1-z\}w_{A,1-z}(\mathbf X)\dot m_{A,z}^{\mathrm{raw}}(\mathbf X)]\\
&-E_s\left[\mathbf1\{Z=z\}
 E\{K_{A,z}^{\infty}\mid\mathbf D_z^\ast,w_{A,z},Z=z\}
       \dot m_{A,z}^{\mathrm{raw}}(\mathbf X)\right],\qquad z=0,1.
\end{aligned}
\end{equation}
Here \(K_{A,z}^{\infty}\) is the incoming marked-Poisson coefficient defined in the oracle proof.

\begin{proposition}[PATE replication limit]
\label{prop:pate-full-prediction-replication-law}\label{prop:pate-sampling-full-refit-comparison}\label{app:pate-sampling-replication-comparisons}
Assume the following conditions:
\begin{enumerate}
\item[(i)] Proposition~\ref{prop:pate-actual-graph-row-multiplier-law} holds, the fitted estimator is consistent, and the PATE first-step bootstrap expansion in Lemma~\ref{lem:primitive-score-root-bootstrap} holds.
\item[(ii)] For each arm, the prescribed prediction refit admits
\begin{equation}\label{eq:pate-full-prediction-refit-expansion}
\begin{aligned}
&\widehat m_z^\ast\{\widehat{\mathbf D}_z^\ast(\mathbf X_i)\}
       -\widehat m_z\{\widehat{\mathbf D}_z(\mathbf X_i)\}\\
&=\dot m_{A,z}^{\mathrm{raw}}(\mathbf X_i)^\top
       (\widehat{\boldsymbol \theta}_{A,\mathrm{PATE}}^\ast
                   -\widehat{\boldsymbol \theta}_{A,\mathrm{PATE}})
 +q_{A,z}\{\mathbf D_z^\ast(\mathbf X_i)\}^{\top}t_{A,z,n}^\ast
 +r_{A,z,n,i}^\ast,\qquad z=0,1.
\end{aligned}
\end{equation}
Each \(\dot m_{A,z}^{\mathrm{raw}}\) is a fixed bounded vector in the common PATE parameter dimension. Each \(q_{A,z}\) is a fixed finite vector of bounded Lipschitz functions of its own baseline score, and \(\sqrt n\,t_{A,z,n}^\ast=O_{p^\ast}(1)\).
\item[(iii)] For both arms, the weighted remainder satisfies
\begin{equation}\label{eq:pate-full-prediction-weighted-remainder}
\begin{aligned}
\frac1{\sqrt n}\sum_i m_i^\ast
 &[\mathbf1\{Z_i=1-z\}w_{A,1-z}(\mathbf X_i)
   +\mathbf1\{Z_i=z\}K_{n,z}^{A,\mathrm{PATE}}(i)]\\
&\hspace{12mm}{}\times|r_{A,z,n,i}^\ast|=o_{p^\ast}(1),
\end{aligned}
\end{equation}
\item[(iv)] For both arms, prediction consistency holds in the form
\begin{equation}\label{eq:pate-weighted-prediction-consistency}
\begin{aligned}
\frac1n\sum_i
 &[\mathbf1\{Z_i=1-z\}w_{A,1-z}(\mathbf X_i)^2
   +\mathbf1\{Z_i=z\}K_{n,z}^{A,\mathrm{PATE}}(i)^2]\\
&\quad{}\times
 [\widehat m_z\{\widehat{\mathbf D}_z(\mathbf X_i)\}
       -m_z^A\{\mathbf D_z^\ast(\mathbf X_i)\}]^2
       \longrightarrow_p0.
\end{aligned}
\end{equation}
\item[(v)] The prescribed finite outputs are used on unsuccessful fits or insufficient treatment-arm counts. The original and conditional fit-failure probabilities tend to zero. Any alternative finite extension satisfies the disagreement condition in Proposition~\ref{prop:bootstrap-monte-carlo-iterated-variance}.
\end{enumerate}
All conditional statements are in \(P_s\)-probability. Then
\begin{equation}\label{eq:pate-full-prediction-replication-law}
\mathcal L^\ast\left\{
\sqrt n(\widehat\tau_{\mathrm{PATE}}^{A,\ast}
       -\widehat\tau_{\mathrm{PATE}}^{A,\mathrm{BC}})
       \mid\mathcal F_n\right\}
\Longrightarrow_{P_s}N(0,V_{A,\mathrm{PATE},\mathrm{rep}}),
\end{equation}
where
\begin{equation}\label{eq:pate-full-prediction-replication-variance}
\begin{aligned}
V_{A,\mathrm{PATE},\mathrm{rep}}
={}&V_{A,\mathrm{PATE},\mathrm{known\text{-}score}}\\
&-\frac2{\bar w_{A,\mathrm{PATE}}}
 (\ell_{A,1}-\ell_{A,0})^\top\boldsymbol\Gamma_{A,\mathrm{PATE}}^{-1}
       \boldsymbol\gamma_{A,\mathrm{PATE},1}\\
&+\frac1{\bar w_{A,\mathrm{PATE}}^2}
 (\ell_{A,1}-\ell_{A,0})^\top\boldsymbol\Sigma_{A,\mathrm{PATE},\boldsymbol\theta}
       (\ell_{A,1}-\ell_{A,0}).
\end{aligned}
\end{equation}

\par\smallskip\noindent\emph{Sampling and replication.}\quad
Suppose additionally that the hypotheses of Theorem~\ref{thm:pate-two-arm-actual-joint-law} hold. Every parameter and covariance below refers to the same full PATE nuisance stack, and \(\bar w_A\) is its total survey-weight denominator limit. The two limiting scalar laws agree if and only if
\begin{equation}\label{eq:pate-sampling-replication-agreement}
\begin{aligned}
&\left(\boldsymbol b_{A,\mathrm{loc}}-\frac{\ell_{A,1}-\ell_{A,0}}{\bar w_A}\right)^\top\\
&\quad{}\times
 \left\{-2\boldsymbol\Gamma_A^{-1}\boldsymbol\gamma_{A,1}
 +\boldsymbol\Sigma_{A,\theta}
       \left(\boldsymbol b_{A,\mathrm{loc}}+\frac{\ell_{A,1}-\ell_{A,0}}{\bar w_A}\right)\right\}
 =0.
\end{aligned}
\end{equation}
The joint laws, including the full nuisance root, agree if and only if \(\boldsymbol b_{A,\mathrm{loc}}=(\ell_{A,1}-\ell_{A,0})/\bar w_A\). Wald inference requires agreement between the replication and sampling variances, together with the Monte Carlo conditions in Proposition~\ref{prop:bootstrap-monte-carlo-iterated-variance}.
\end{proposition}

\begin{proof}
For each direction \(z\), the one-arm calculation in the proof of Proposition~\ref{prop:patt-full-prediction-replication-law} applies with control and treated rows replaced by \(Z=z\) and \(Z=1-z\). The bounded-mark limits follow from Proposition~\ref{prop:pate-actual-graph-row-multiplier-law}:
\begin{equation}\label{eq:pate-frozen-prediction-derivative-lln}
\begin{aligned}
&\frac1n\left[
 \sum_{i:Z_i=1-z}w_{A,1-z}(\mathbf X_i)\dot m_{A,z}^{\mathrm{raw}}(\mathbf X_i)
 -\sum_{j:Z_j=z}K_{n,z}^{A,\mathrm{PATE}}(j)
       \dot m_{A,z}^{\mathrm{raw}}(\mathbf X_j)\right]\\
&\hspace{15mm}\longrightarrow_p\ell_{A,z}.
\end{aligned}
\end{equation}
For the own-score terms, reindexing gives
\begin{equation}\label{eq:pate-score-prediction-balance}
\begin{aligned}
&\frac1n\left[
 \sum_{i:Z_i=1-z}w_{A,1-z}(\mathbf X_i)q_{A,z}(\mathbf D_{z,i}^\ast)
 -\sum_{j:Z_j=z}K_{n,z}^{A,\mathrm{PATE}}(j)
       q_{A,z}(\mathbf D_{z,j}^\ast)\right]\\
&=\frac1n\sum_{i:Z_i=1-z}w_{A,1-z}(\mathbf X_i)
 \sum_{j:Z_j=z}\lambda_{i\to j}^{A,(z)}
 [q_{A,z}(\mathbf D_{z,i}^\ast)-q_{A,z}(\mathbf D_{z,j}^\ast)]\\
&=O_p(n^{-1/2}).
\end{aligned}
\end{equation}
The last bound follows by applying the PATT Lipschitz and radius argument to each graph. For either fixed bounded mark, inserting the multinomial counts changes the average by a conditionally mean-zero quantity with variance at most
\[
\frac C{n^2}\sum_i
 [\mathbf1\{Z_i=1-z\}w_{A,1-z}(\mathbf X_i)^2
  +\mathbf1\{Z_i=z\}K_{n,z}^{A,\mathrm{PATE}}(i)^2]
=O_p(n^{-1}).
\]
Consequently, both counted averages have the same limits.

Substituting the prediction expansion into \eqref{eq:pate-common-fixed-refit-remainder} gives
\begin{equation}\label{eq:pate-full-prediction-refit-first-order}
\begin{aligned}
\frac{\mathcal R_{A,\mathrm{PATE},n}^\ast}{\sqrt n}
={}&\sum_{z=0}^1(2z-1)
 \Bigg\{\frac1n\sum_i m_i^\ast
 [\mathbf1\{Z_i=1-z\}w_{A,1-z}(\mathbf X_i)\\
&\hspace{28mm}
 -\mathbf1\{Z_i=z\}K_{n,z}^{A,\mathrm{PATE}}(i)]
       \dot m_{A,z}^{\mathrm{raw}}(\mathbf X_i)\Bigg\}^{\top}
       \sqrt n(\widehat{\boldsymbol \theta}_{A,\mathrm{PATE}}^\ast
                   -\widehat{\boldsymbol \theta}_{A,\mathrm{PATE}})\\
&+\sum_{z=0}^1(2z-1)
 \Bigg\{\frac1n\sum_i m_i^\ast
 [\mathbf1\{Z_i=1-z\}w_{A,1-z}(\mathbf X_i)\\
&\hspace{28mm}
 -\mathbf1\{Z_i=z\}K_{n,z}^{A,\mathrm{PATE}}(i)]
       q_{A,z}(\mathbf D_{z,i}^\ast)\Bigg\}^{\top}
       \sqrt n\,t_{A,z,n}^\ast+o_{p^\ast}(1)\\
={}&\sum_{z=0}^1(2z-1)\ell_{A,z}^{\top}
       \sqrt n(\widehat{\boldsymbol \theta}_{A,\mathrm{PATE}}^\ast
                   -\widehat{\boldsymbol \theta}_{A,\mathrm{PATE}})+o_{p^\ast}(1)\\
={}&(\ell_{A,1}-\ell_{A,0})^{\top}
       \sqrt n(\widehat{\boldsymbol \theta}_{A,\mathrm{PATE}}^\ast
                   -\widehat{\boldsymbol \theta}_{A,\mathrm{PATE}})+o_{p^\ast}(1)\\
={}&-(\ell_{A,1}-\ell_{A,0})^{\top}\boldsymbol\Gamma_{A,\mathrm{PATE}}^{-1}
       \frac1{\sqrt n}\sum_i(m_i^\ast-1)
       \{U_{A,\mathrm{PATE}}(O_i;\boldsymbol\theta_{A,\mathrm{PATE}}^\ast)\}
       +o_{p^\ast}(1).
\end{aligned}
\end{equation}
The weighted remainder condition controls the first expansion. The counted averages converge to the derivative limits, after which summing over the two arms and applying the bootstrap score-root expansion give the last two equalities.

The fitted and baseline rows differ by
\begin{equation}\label{eq:pate-full-prediction-row-difference}
\begin{aligned}
&\widehat\varphi_{A,\mathrm{PATE},i}-\varphi_{A,\mathrm{PATE},i}\\
&=\sum_{z=0}^1(2z-1)
 [\mathbf1\{Z_i=1-z\}w_{A,1-z}(\mathbf X_i)
 -\mathbf1\{Z_i=z\}K_{n,z}^{A,\mathrm{PATE}}(i)]\\
&\quad{}\times
 [\widehat m_z\{\widehat{\mathbf D}_z(\mathbf X_i)\}
       -m_z^A\{\mathbf D_z^\ast(\mathbf X_i)\}]\\
&\quad-w_{A,Z_i}(\mathbf X_i)
       (\widehat\tau_{\mathrm{PATE}}^{A,\mathrm{BC}}
                  -\tau_{\mathrm{PATE}}).
\end{aligned}
\end{equation}
The own-arm prediction weight cancels its direct residual weight. Thus
\[
\begin{aligned}
&\frac1n\sum_i
       |\widehat\varphi_{A,\mathrm{PATE},i}-\varphi_{A,\mathrm{PATE},i}|^2\\
&\le\frac3n\sum_{z=0}^1\sum_i
 [\mathbf1\{Z_i=1-z\}w_{A,1-z}(\mathbf X_i)^2
 +\mathbf1\{Z_i=z\}K_{n,z}^{A,\mathrm{PATE}}(i)^2]\\
&\qquad{}\times
 [\widehat m_z\{\widehat{\mathbf D}_z(\mathbf X_i)\}
       -m_z^A\{\mathbf D_z^\ast(\mathbf X_i)\}]^2\\
&\quad+3(\widehat\tau_{\mathrm{PATE}}^{A,\mathrm{BC}}
                  -\tau_{\mathrm{PATE}})^2
       \frac1n\sum_iw_{A,Z_i}(\mathbf X_i)^2\\
&\longrightarrow_p0,
\end{aligned}
\]
by weighted prediction consistency and estimator consistency. Equation~\eqref{eq:fixed-match-multinomial-moments} and conditional Chebyshev therefore permit replacing the fitted row by its baseline row in the centered multiplier sum. The same multinomial moments and the iid weight law give \(\bar w_{A,\mathrm{PATE},n}^\ast\to_{p^\ast}\bar w_{A,\mathrm{PATE}}>0\). Hence the exact ratio becomes
\begin{equation}\label{eq:pate-full-prediction-replicate-linearization}
\begin{aligned}
&\sqrt n(\widehat\tau_{\mathrm{PATE}}^{A,\ast}
       -\widehat\tau_{\mathrm{PATE}}^{A,\mathrm{BC}})\\
&=\frac1{\bar w_{A,\mathrm{PATE},n}^\ast\sqrt n}
       \sum_i(m_i^\ast-1)\widehat\varphi_{A,\mathrm{PATE},i}
 +\frac{\mathcal R_{A,\mathrm{PATE},n}^\ast}
       {\bar w_{A,\mathrm{PATE},n}^\ast\sqrt n}\\
&=\frac1{\bar w_{A,\mathrm{PATE}}\sqrt n}
       \sum_i(m_i^\ast-1)\varphi_{A,\mathrm{PATE},i}\\
&\quad-\frac{(\ell_{A,1}-\ell_{A,0})^\top}{\bar w_{A,\mathrm{PATE}}}
       \boldsymbol\Gamma_{A,\mathrm{PATE}}^{-1}
       \frac1{\sqrt n}\sum_i(m_i^\ast-1)
       \{U_{A,\mathrm{PATE}}(O_i;\boldsymbol\theta_{A,\mathrm{PATE}}^\ast)\}
       +o_{p^\ast}(1).
\end{aligned}
\end{equation}
The joint multiplier law and conditional Slutsky give
\begin{equation}\label{eq:pate-sampling-full-refit-joint-limit}
\begin{aligned}
&\mathcal L^*\left(
 \begin{pmatrix}
 \sqrt n(\widehat\tau_{\mathrm{PATE}}^{A,*}
              -\widehat\tau_{\mathrm{PATE}}^{A,\mathrm{BC}})\\
 \sqrt n(\widehat{\boldsymbol \theta}_{A,\mathrm{PATE}}^*
              -\widehat{\boldsymbol \theta}_{A,\mathrm{PATE}})
 \end{pmatrix}\middle|\mathcal F_n\right)\\
&\quad\Longrightarrow_{P_s}
 \mathcal L\left(
 \begin{pmatrix}
 G_{\tau,A}+\dfrac{(\ell_{A,1}-\ell_{A,0})^\top}{\bar w_A}\mathbf G_{\theta,A}\\\mathbf G_{\theta,A}
 \end{pmatrix}\right).
\end{aligned}
\end{equation}
Here \(\operatorname{Cov}(\mathbf G_{\theta,A},G_{\tau,A})=-\boldsymbol\Gamma_A^{-1}\boldsymbol\gamma_{A,1}\) and \(\operatorname{Var}(\mathbf G_{\theta,A})=\boldsymbol\Sigma_{A,\theta}\). The first-coordinate variance is \eqref{eq:pate-full-prediction-replication-variance}. Its covariance with the nuisance root is
\begin{equation}\label{eq:pate-full-prediction-nuisance-cross-covariance}
-\boldsymbol\Gamma_{A,\mathrm{PATE}}^{-1}\boldsymbol\gamma_{A,\mathrm{PATE},1}
 +\boldsymbol\Sigma_{A,\mathrm{PATE},\boldsymbol\theta}
       (\ell_{A,1}-\ell_{A,0})/\bar w_{A,\mathrm{PATE}}.
\end{equation}

Under the additional sampling hypotheses, the sampling pair is \((G_{\tau,A}+\boldsymbol b_{A,\mathrm{loc}}^\top \mathbf G_{\theta,A},\mathbf G_{\theta,A})\). Since \(\boldsymbol\Sigma_{A,\theta}\) is symmetric,
\[
\begin{aligned}
&\operatorname{Var}(G_{\tau,A}+\boldsymbol b_{A,\mathrm{loc}}^\top \mathbf G_{\theta,A})
 -\operatorname{Var}\!\left(G_{\tau,A}+
       \frac{(\ell_{A,1}-\ell_{A,0})^\top \mathbf G_{\theta,A}}{\bar w_A}\right)\\
&=-2\left(\boldsymbol b_{A,\mathrm{loc}}-\frac{\ell_{A,1}-\ell_{A,0}}{\bar w_A}\right)^\top
             \boldsymbol\Gamma_A^{-1}\boldsymbol\gamma_{A,1}
 +\boldsymbol b_{A,\mathrm{loc}}^\top\boldsymbol\Sigma_{A,\theta}\boldsymbol b_{A,\mathrm{loc}}
 -\frac{(\ell_{A,1}-\ell_{A,0})^\top
       \boldsymbol\Sigma_{A,\theta}(\ell_{A,1}-\ell_{A,0})}{\bar w_A^2}\\
&=\left(\boldsymbol b_{A,\mathrm{loc}}-\frac{\ell_{A,1}-\ell_{A,0}}{\bar w_A}\right)^\top
 \left\{-2\boldsymbol\Gamma_A^{-1}\boldsymbol\gamma_{A,1}
 +\boldsymbol\Sigma_{A,\theta}
 \left(\boldsymbol b_{A,\mathrm{loc}}+\frac{\ell_{A,1}-\ell_{A,0}}{\bar w_A}\right)\right\}.
\end{aligned}
\]
Equality of the centered scalar Gaussian laws is therefore equivalent to \eqref{eq:pate-sampling-replication-agreement}. Their nuisance cross-covariance difference is
\[
\boldsymbol\Sigma_{A,\theta}
 \left(\boldsymbol b_{A,\mathrm{loc}}-\frac{\ell_{A,1}-\ell_{A,0}}{\bar w_A}\right).
\]
By positive definiteness, this difference vanishes exactly when the two loading vectors agree. That equality also makes the joint Gaussian transformations identical.

\label{par:pate-full-prediction-variance-scope}
The joint covariance matrix gives \(V_{A,\mathrm{PATE},\mathrm{known\text{-}score}} -\boldsymbol\gamma_{A,\mathrm{PATE},1}^{\top} \boldsymbol\Sigma_{A,\mathrm{PATE},U}^{-1}\boldsymbol\gamma_{A,\mathrm{PATE},1}\ge0\) when \(\boldsymbol\Sigma_{A,\mathrm{PATE},U}>0\). The sandwich identity yields
\begin{equation}\label{eq:pate-full-prediction-positive-decomposition}
\begin{aligned}
V_{A,\mathrm{PATE},\mathrm{rep}}
={}&V_{A,\mathrm{PATE},\mathrm{known\text{-}score}}
 -\boldsymbol\gamma_{A,\mathrm{PATE},1}^{\top}
  \boldsymbol\Sigma_{A,\mathrm{PATE},U}^{-1}\boldsymbol\gamma_{A,\mathrm{PATE},1}\\
&+\left(-\boldsymbol\Gamma_{A,\mathrm{PATE}}^\top
       \boldsymbol\Sigma_{A,\mathrm{PATE},U}^{-1}\boldsymbol\gamma_{A,\mathrm{PATE},1}
       +\frac{\ell_{A,1}-\ell_{A,0}}{\bar w_{A,\mathrm{PATE}}}\right)^\top
       \boldsymbol\Sigma_{A,\mathrm{PATE},\boldsymbol\theta}\\
&\quad{}\times\left(-\boldsymbol\Gamma_{A,\mathrm{PATE}}^\top
       \boldsymbol\Sigma_{A,\mathrm{PATE},U}^{-1}\boldsymbol\gamma_{A,\mathrm{PATE},1}
       +\frac{\ell_{A,1}-\ell_{A,0}}{\bar w_{A,\mathrm{PATE}}}\right).
\end{aligned}
\end{equation}
The variance is finite and nonnegative, and positive if either the projection variance is positive or the displayed vector is nonzero.
\end{proof}

\Needspace{8\baselineskip}
\subsubsection{Series regression refits and bias correction}
\label{app:growing-series-bc}

The following results give sufficient conditions for survey-weighted series refitting and bias correction with estimated scores, following Abadie and Imbens.\citep{AbadieImbens2011} The empirical implementation uses the fixed quadratic basis
\[
q(d)=(1,d_1,d_2,d_1^2,d_1d_2,d_2^2)^\top.
\]
Corollary~\ref{cor:oracle-regression-bc} applies on fixed scores under its derivative consistency condition. The following proof includes fixed bases that represent the reduced mean exactly. The results for estimated scores also require the stated approximation, graph, and bias correction conditions.

\begin{proposition}[Survey-weighted series refits]
\label{prop:full-refit-growing-series-realization}
Assume the following conditions for each fitted arm:
\begin{enumerate}
\item[(i)] The sampling-row, consistency, score-root, finite-output, and empirical-normalization hypotheses of the relevant PATE or PATT replication proposition hold. Its prediction-expansion, prediction-remainder, and prediction-consistency conditions are replaced by (ii)--(iv) below.
\item[(ii)] The prescribed deterministic \(C^1\) basis \(q_{z,n}\) is unchanged within each replicate, on a fixed compact convex neighborhood of the baseline and fitted scores. Its dimension may grow, and
\[
\sup_d\|q_{z,n}(d)\|\le B_{z,n},\qquad
\|q_{z,n}(d)-q_{z,n}(d')\|\le L_{z,n}\|d-d'\|,\qquad B_{z,n},L_{z,n}\ge1.
\]
The weights are bounded and positive, \(E_sY^2<\infty\), and
\begin{equation}\label{eq:full-refit-series-growth}
 \frac{(B_{z,n}+L_{z,n})
       \{B_{z,n}+(1+B_{z,n})L_{z,n}\}}{\sqrt n}
 \longrightarrow0.
\end{equation}
\item[(iii)] The population weighted least-squares projection \(c_{z,n}\) satisfies
\[
 E_s[\mathbf1\{Z=z\}w_{A,z}q_{z,n}(\mathbf D_z^*)
       \{Y-q_{z,n}(\mathbf D_z^*)^\top c_{z,n}\}]=0.
\]
Also, \(\sup_n\|c_{z,n}\|<\infty\), and the smallest eigenvalue of \(E_s[\mathbf1\{Z=z\}w_{A,z}q_{z,n}(\mathbf D_z^*) q_{z,n}(\mathbf D_z^*)^\top]\) is bounded away from zero uniformly in \(n\).
\item[(iv)] On the fixed neighborhood,
\[
 \|q_{z,n}^\top c_{z,n}-m_z^A\|_{C^1}\longrightarrow0,
\]
where \(m_z^A\) is \(C^2\) with bounded derivatives.
\end{enumerate}

We fit the series by survey-weighted least squares on \(\widehat{\mathbf D}_z\) and refit it with weights \(m_i^*w_{A,z}\) on \(\widehat{\mathbf D}_z^*\). We recompute the scores and their standardization while retaining the original reuse coefficients. The corresponding full-prediction replication proposition gives the variance and covariance with the parameter estimator, using the derivative
\begin{equation}\label{eq:full-refit-growing-series-tangent}
 \dot m_{A,z}^{\mathrm{raw}}(\mathbf X)^\top \mathbf h
 =\nabla m_z^A(\mathbf D_z^*)^\top
   \dot{\mathbf D}_{z,\mathbf h}^{\,\mathrm{raw}}(\mathbf X).
\end{equation}
The derivative is with respect to the propensity and prognostic score parameters in direction \(\mathbf h\), with the limiting centers and scales held fixed. Each replicate recomputes standardization and uses the same deterministic basis rule.
\end{proposition}

\Needspace{10\baselineskip}
\begin{proof}
Fix an arm and put \(r_{z,n}=\{B_{z,n}+(1+B_{z,n})L_{z,n}\}/\sqrt n\). We first bound the Gram matrices and coefficient errors under conditions (ii)--(iv). The score displacement is uniformly \(O_p(n^{-1/2})\). For the Gram matrix evaluated at the limiting scores, the i.i.d.\ selected observations give
\[
\begin{aligned}
&E_s\left\|\frac1n\sum_i\mathbf1(Z_i=z)w_{A,z}(\mathbf X_i)
 q_{z,n}(\mathbf D_{z,i}^*)q_{z,n}(\mathbf D_{z,i}^*)^\top
 -E_s[\mathbf1(Z=z)w_{A,z}q_{z,n}(\mathbf D_z^*)
                         q_{z,n}(\mathbf D_z^*)^\top]\right\|_F^2\\
&\le\frac1n E_s\!\left[
 \mathbf1(Z=z)w_{A,z}^2\|q_{z,n}(\mathbf D_z^*)\|^4\right]\\
&\le\frac{CB_{z,n}^4}{n}.
\end{aligned}
\]
Thus the Frobenius-norm fluctuation is \(O_p(B_{z,n}^2/\sqrt n)\). Uniform score displacement \(O_p(n^{-1/2})\) and
\[
\|q(d+\Delta)q(d+\Delta)^\top-q(d)q(d)^\top\|
\le2B_{z,n}L_{z,n}\|\Delta\|+L_{z,n}^2\|\Delta\|^2
\]
bound the generated-score Gram change by \(O_p(B_{z,n}L_{z,n}/\sqrt n+L_{z,n}^2/n)\). Both vanish by \eqref{eq:full-refit-series-growth}. We write the original and replicate Gram matrices as
\[
\begin{aligned}
G_n
&=\frac1n\sum_i\mathbf1(Z_i=z)w_{A,z}(\mathbf X_i)
 q_{z,n}(\widehat{\mathbf D}_{z,i})
 q_{z,n}(\widehat{\mathbf D}_{z,i})^\top,\\
G_n^\ast
&=\frac1n\sum_i m_i^\ast\mathbf1(Z_i=z)w_{A,z}(\mathbf X_i)
 q_{z,n}(\widehat{\mathbf D}_{z,i}^\ast)
 q_{z,n}(\widehat{\mathbf D}_{z,i}^\ast)^\top.
\end{aligned}
\]
The multinomial covariance identity and the same score-displacement inequality give
\[
\begin{aligned}
&E^\ast\left\|\frac1n\sum_i(m_i^\ast-1)\mathbf1(Z_i=z)w_{A,z}
 q_{z,n}(\widehat{\mathbf D}_{z,i})
 q_{z,n}(\widehat{\mathbf D}_{z,i})^\top\right\|_F^2
 \le \frac{CB_{z,n}^4}{n},\\
&\|G_n^\ast-G_n\|
 =O_{p^\ast}\!\left\{
 \frac{B_{z,n}^2+B_{z,n}L_{z,n}}{\sqrt n}
 +\frac{L_{z,n}^2}{n}\right\}
 =o_{p^\ast}(1),
\end{aligned}
\]
because \(\max_i\|\widehat{\mathbf D}_{z,i}^\ast -\widehat{\mathbf D}_{z,i}\| =O_{p^\ast}(n^{-1/2})\). If \(c>0\) is the stipulated population eigenvalue bound, then
\[
\begin{aligned}
\lambda_{\min}(G_n)
&\ge c-\left\|G_n-
 E_s[\mathbf1(Z=z)w_{A,z}q_{z,n}(\mathbf D_z^\ast)
                          q_{z,n}(\mathbf D_z^\ast)^\top]\right\|\\
&=c-o_p(1),\\
\lambda_{\min}(G_n^\ast)
&\ge\lambda_{\min}(G_n)-\|G_n^\ast-G_n\|\\
&\ge c-o_p(1)-o_{p^\ast}(1).
\end{aligned}
\]
Thus \(\|G_n^{-1}\|=O_p(1)\) and \(\|(G_n^\ast)^{-1}\|=O_{p^\ast}(1)\) on events whose respective probabilities tend to one.

The population normal equation and uniformly bounded projection predictions give
\[
\begin{aligned}
&E_s\left\|\frac1n\sum_i\mathbf1(Z_i=z)w_{A,z}(\mathbf X_i)
 q_{z,n}(\mathbf D_{z,i}^*)
 \{Y_i-q_{z,n}(\mathbf D_{z,i}^*)^\top c_{z,n}\}\right\|^2\\
&=\frac1n E_s\!\left[
 \mathbf1(Z=z)w_{A,z}^2\|q_{z,n}(\mathbf D_z^*)\|^2
 \{Y-q_{z,n}(\mathbf D_z^*)^\top c_{z,n}\}^2\right]\\
&\le\frac{CB_{z,n}^2}{n}E_s(1+Y^2).
\end{aligned}
\]
Replacing the baseline score by the fitted score changes each vector summand by
\[
\begin{aligned}
&q_{z,n}(\widehat{\mathbf D}_{z,i})
 \{Y_i-q_{z,n}(\widehat{\mathbf D}_{z,i})^\top c_{z,n}\}
 -q_{z,n}(\mathbf D_{z,i}^*)
 \{Y_i-q_{z,n}(\mathbf D_{z,i}^*)^\top c_{z,n}\}\\
&=\{q_{z,n}(\widehat{\mathbf D}_{z,i})-q_{z,n}(\mathbf D_{z,i}^*)\}
       \{Y_i-q_{z,n}(\mathbf D_{z,i}^*)^\top c_{z,n}\}\\
&\quad-q_{z,n}(\widehat{\mathbf D}_{z,i})
 \{q_{z,n}(\widehat{\mathbf D}_{z,i})-q_{z,n}(\mathbf D_{z,i}^*)\}^\top c_{z,n}.
\end{aligned}
\]
Its weighted average norm is bounded by
\[
\begin{aligned}
&L_{z,n}\max_i\|\widehat{\mathbf D}_{z,i}-\mathbf D_{z,i}^*\|
 \left\{\frac1n\sum_i\mathbf1(Z_i=z)w_{A,z}
 |Y_i-q_{z,n}(\mathbf D_{z,i}^*)^\top c_{z,n}|
       +w_{\max} B_{z,n}\|c_{z,n}\|\right\}\\
&=O_p\{(1+B_{z,n})L_{z,n}/\sqrt n\}.
\end{aligned}
\]
The sample normal equation is
\[
G_n(\widehat c_{z,n}-c_{z,n})
=\frac1n\sum_i\mathbf1\{Z_i=z\}w_{A,z}
 q_{z,n}(\widehat{\mathbf D}_{z,i})
 \{Y_i-q_{z,n}(\widehat{\mathbf D}_{z,i})^\top c_{z,n}\}.
\]
Combining the inverse and score-vector bounds gives
\begin{equation}\label{eq:full-refit-growing-series-coefficient-rate}
\begin{aligned}
\|\widehat c_{z,n}-c_{z,n}\|
&\le\|G_n^{-1}\|
 \left\|\frac1n\sum_i\mathbf1(Z_i=z)w_{A,z}
 q_{z,n}(\widehat{\mathbf D}_{z,i})
 \{Y_i-q_{z,n}(\widehat{\mathbf D}_{z,i})^\top c_{z,n}\}\right\|\\
&=O_p(1)\left[O_p(B_{z,n}/\sqrt n)
            +O_p\{(1+B_{z,n})L_{z,n}/\sqrt n\}\right]\\
&=O_p(r_{z,n}).
\end{aligned}
\end{equation}

For the replicate coefficient error, first hold the fitted scores fixed. The multinomial covariance identity gives
\[
\begin{aligned}
&E^*\left\|\frac1n\sum_i(m_i^*-1)\mathbf1(Z_i=z)w_{A,z}
 q_{z,n}(\widehat{\mathbf D}_{z,i})
 \{Y_i-q_{z,n}(\widehat{\mathbf D}_{z,i})^\top\widehat c_{z,n}\}
 \right\|^2\\
&\le\frac{B_{z,n}^2}{n^2}\sum_i\mathbf1(Z_i=z)w_{A,z}^2
 \{Y_i-q_{z,n}(\widehat{\mathbf D}_{z,i})^\top\widehat c_{z,n}\}^2\\
&=O_p(B_{z,n}^2/n).
\end{aligned}
\]
The empirical squared residual average is \(O_p(1)\) by the coefficient bound and \((B_{z,n}+L_{z,n})r_{z,n}\to0\). Conditional Markov's inequality bounds the residual averages weighted by the multinomial counts. Subtracting the original normal equation gives
\[
\begin{aligned}
&\left\{\frac1n\sum_i m_i^*\mathbf1(Z_i=z)w_{A,z}
 q_{z,n}(\widehat{\mathbf D}_{z,i}^*)
 q_{z,n}(\widehat{\mathbf D}_{z,i}^*)^\top\right\}
       (\widehat c_{z,n}^*-\widehat c_{z,n})\\
&=\frac1n\sum_i m_i^*\mathbf1(Z_i=z)w_{A,z}
 q_{z,n}(\widehat{\mathbf D}_{z,i}^*)
 \{Y_i-q_{z,n}(\widehat{\mathbf D}_{z,i}^*)^\top\widehat c_{z,n}\}.
\end{aligned}
\]
The centered-count and score-displacement bounds give
\begin{equation}\label{eq:full-refit-growing-series-bootstrap-rate}
\begin{aligned}
\|\widehat c_{z,n}^*-\widehat c_{z,n}\|
&\le O_{p^*}(1)\left[
 O_{p^*}(B_{z,n}/\sqrt n)
 +O_{p^*}\{(1+B_{z,n})L_{z,n}/\sqrt n\}\right]\\
&=O_{p^*}(r_{z,n}).
\end{aligned}
\end{equation}
These bounds also show that the original and replicate Gram matrices satisfy the required invertibility condition with probability tending to one.

Writing \(\widehat m_z=q_{z,n}^\top\widehat c_{z,n}\),
\[
\begin{aligned}
\|\widehat m_z-m_z^A\|_{C^1}
&\le(B_{z,n}+L_{z,n})\|\widehat c_{z,n}-c_{z,n}\|
       +\|q_{z,n}^\top c_{z,n}-m_z^A\|_{C^1}\\
&=O_p\{(B_{z,n}+L_{z,n})r_{z,n}\}+o(1)\\
&=o_p(1),\\
\|\widehat m_z^*-m_z^A\|_{C^1}
&\le(B_{z,n}+L_{z,n})\|\widehat c_{z,n}^*-\widehat c_{z,n}\|
       +\|\widehat m_z-m_z^A\|_{C^1}\\
&=o_{p^*}(1).
\end{aligned}
\]
For a fixed basis that represents \(m_z^A\) exactly, \(B_{z,n}+L_{z,n}=O(1)\) and the approximation error is zero. Equation~\eqref{eq:full-refit-growing-series-coefficient-rate} and the preceding inequality then give
\[
\begin{aligned}
\|\widehat c_{z,n}-c_{z,n}\|&=O_p(n^{-1/2}),\\
\|\widehat m_z-m_z^A\|_{C^1}
&\le (B_{z,n}+L_{z,n})\|\widehat c_{z,n}-c_{z,n}\|
=O_p(n^{-1/2}).
\end{aligned}
\]
This verifies the derivative condition in Corollary~\ref{cor:oracle-regression-bc}. In particular, when the prognostic coordinate is the correct conditional mean, its fixed affine standardization makes \(m_z^A\) affine in the second score coordinate, which is contained in the quadratic basis. The separate conditions for bias correction with estimated scores are addressed in the next two propositions.

The fundamental theorem of calculus and the Lipschitz gradient of \(m_z^A\) yield uniformly over observed points
\[
\begin{aligned}
\widehat m_z^*(\widehat{\mathbf D}_{z,i}^*)
 -\widehat m_z(\widehat{\mathbf D}_{z,i})
&=\widehat m_z^*(\widehat{\mathbf D}_{z,i}^*)
 -\widehat m_z^*(\widehat{\mathbf D}_{z,i})
 +q_{z,n}(\widehat{\mathbf D}_{z,i})^\top
       (\widehat c_{z,n}^*-\widehat c_{z,n})\\
&=\int_0^1\nabla\widehat m_z^*\!
 \left\{\widehat{\mathbf D}_{z,i}
       +t(\widehat{\mathbf D}_{z,i}^*-\widehat{\mathbf D}_{z,i})\right\}^\top
       (\widehat{\mathbf D}_{z,i}^*-\widehat{\mathbf D}_{z,i})\,dt\\
&\quad+q_{z,n}(\widehat{\mathbf D}_{z,i})^\top
       (\widehat c_{z,n}^*-\widehat c_{z,n})\\
&=\nabla m_z^A(\mathbf D_{z,i}^*)^\top
       (\widehat{\mathbf D}_{z,i}^*-\widehat{\mathbf D}_{z,i})\\
&\quad+q_{z,n}(\widehat{\mathbf D}_{z,i})^\top
       (\widehat c_{z,n}^*-\widehat c_{z,n})
       +o_{p^*}(n^{-1/2})\\
&=\nabla m_z^A(\mathbf D_{z,i}^*)^\top
       (\widehat{\mathbf D}_{z,i}^*-\widehat{\mathbf D}_{z,i})\\
&\quad+q_{z,n}(\mathbf D_{z,i}^*)^\top
       (\widehat c_{z,n}^*-\widehat c_{z,n})
       +o_{p^*}(n^{-1/2}).
\end{aligned}
\]
The error from evaluating the basis at the baseline rather than the fitted score is bounded by \(L_{z,n}O_p(n^{-1/2})O_{p^*}(r_{z,n})=o_{p^*}(n^{-1/2})\). The score Taylor expansion separates the derivative in \eqref{eq:full-refit-growing-series-tangent} from the empirical centering and scaling terms. The latter are fixed bounded Lipschitz functions of the corresponding score multiplied by increments that are tight at the root-\(n\) scale.

Let \(a_{z,n,i}=\mathbf1\{Z_i=1-z\}w_{A,1-z}(\mathbf X_i) -\mathbf1\{Z_i=z\}K_{n,z}(i)\), where \(K_{n,z}\) is the reuse coefficient for the relevant estimand, held fixed at the original estimated-score matches. Since the within-match weights sum to one, reindexing gives
\[
\begin{aligned}
&\left\|\frac1{\sqrt n}\sum_i a_{z,n,i}q_{z,n}(\mathbf D_{z,i}^*)\right\|\\
&=\left\|\frac1{\sqrt n}\sum_{i:Z_i=1-z}w_{A,1-z}(\mathbf X_i)
 \sum_{j:Z_j=z}\lambda_{i\to j}^{A,(z)}
 \{q_{z,n}(\mathbf D_{z,i}^*)-q_{z,n}(\mathbf D_{z,j}^*)\}\right\|\\
&\le\frac{L_{z,n}}{\sqrt n}\sum_{i:Z_i=1-z}w_{A,1-z}(\mathbf X_i)
 \sum_{j:Z_j=z}\lambda_{i\to j}^{A,(z)}
 \|\mathbf D_{z,i}^*-\mathbf D_{z,j}^*\|\\
&=O_p(L_{z,n}).
\end{aligned}
\]
The multinomial covariance identity and squared-reuse bound give
\[
E^*\left\|n^{-1/2}\sum_i(m_i^*-1)a_{z,n,i}
 q_{z,n}(\mathbf D_{z,i}^*)\right\|^2
\le B_{z,n}^2\,n^{-1}\sum_i a_{z,n,i}^2=O_p(B_{z,n}^2).
\]
Consequently,
\[
\begin{aligned}
&\left|\left\{\frac1{\sqrt n}\sum_i m_i^*a_{z,n,i}
            q_{z,n}(\mathbf D_{z,i}^*)\right\}^\top
       (\widehat c_{z,n}^*-\widehat c_{z,n})\right|\\
&\le\left\|\frac1{\sqrt n}\sum_i m_i^*a_{z,n,i}
            q_{z,n}(\mathbf D_{z,i}^*)\right\|
       \|\widehat c_{z,n}^*-\widehat c_{z,n}\|\\
&=O_{p^*}(B_{z,n}+L_{z,n})O_{p^*}(r_{z,n})\\
&=o_{p^*}(1).
\end{aligned}
\]
The matched-distance bound also makes the standardization terms negligible. The normalized sum of absolute coefficients weighted by the multinomial counts is tight and controls the uniform Taylor remainder. Uniform prediction consistency and the squared-reuse bound establish the required consistency under squared coefficients. The full-prediction expansions therefore apply with \(\ell_{A,1}-\ell_{A,0}\) for PATE and \(-\ell_{A,0}\) for PATT, using their respective normalizers and covariances with the parameter estimator.
\end{proof}

Under the corresponding sampling hypotheses, this joint Gaussian refit law gives the comparisons in Proposition~\ref{prop:pate-sampling-full-refit-comparison} and Proposition~\ref{cor:patt-full-prediction-joint-comparison}.

\begin{proposition}[PATE series bias correction]
\label{prop:pate-growing-bc-compatibility}\label{cor:pate-growing-bc-residual-agreement}
Assume the following conditions:
\begin{enumerate}
\item[(i)] The unscaled score, normalization, and graph conditions of Proposition~\ref{prop:pate-actual-graph-row-multiplier-law} for deterministic local and actual fitted graphs hold.
\item[(ii)] Each arm satisfies the population projection, positive Gram, \(C^1\) approximation, bounded-coefficient and growth conditions in Proposition~\ref{prop:full-refit-growing-series-realization}, with its fixed \(C^2\) target \(m_z^A\). The deterministic \(C^1\) basis is defined on a fixed compact convex neighborhood of the baseline and current scores, with the stated vector sup and Lipschitz bounds.
\item[(iii)] Covariates and positive survey weights are bounded, \(E_sY^2<\infty\), and the residual and first-step moments satisfy the cited marked-graph conditions.
\end{enumerate}
We fit survey-weighted least squares on the current scores, using the prescribed finite output when a fit does not exist.

Let \(\dot m_{A,z}^{\mathrm{raw}}\) be the raw derivative in \eqref{eq:full-refit-growing-series-tangent}, and define \(\ell_{A,z}\) by \eqref{eq:pate-frozen-prediction-derivatives}. For every deterministic full direction \(\mathbf h_n\to \mathbf h\), with its own specified empirical normalization and regression fit,
\begin{equation}\label{eq:pate-growing-bc-local-limit}
\begin{aligned}
&\sqrt n\{B_{n,A}^{D,\mathrm{PATE}}
 (\mathbf D_n^e(\mathbf h_n);\mathbf D^*,m^A)
 -\widehat B_{n,A}^{D,\mathrm{PATE}}(\mathbf D_n^e(\mathbf h_n))\}\\
&\hspace{12mm}\longrightarrow_p
 \frac{(\ell_{A,1}-\ell_{A,0})^\top \mathbf h}{\bar w_{A,\mathrm{PATE}}}.
\end{aligned}
\end{equation}
At the fitted scores, using the corresponding regression,
\begin{equation}\label{eq:pate-growing-bc-actual-expansion}
\begin{aligned}
&\sqrt n\{B_{n,A}^{D,\mathrm{PATE}}
 (\widehat{\mathbf D};\mathbf D^*,m^A)
 -\widehat B_{n,A}^{D,\mathrm{PATE}}(\widehat{\mathbf D})\}\\
&=\frac{(\ell_{A,1}-\ell_{A,0})^\top
 \sqrt n(\widehat{\boldsymbol \theta}_{A,\mathrm{PATE}}
                -\boldsymbol\theta_{A,\mathrm{PATE}}^*)}
 {\bar w_{A,\mathrm{PATE}}}+o_p(1).
\end{aligned}
\end{equation}
Under these regression conditions and the full-dimensional interior parameter space, the BC remainder condition at every deterministic local parameter is equivalent to \(\ell_{A,1}-\ell_{A,0}=0\). This equality also gives the BC remainder condition at the fitted parameter.

\par\smallskip\noindent\emph{Replication validity.}\quad
Assume additionally the first, deterministic-local clause of \eqref{eq:pate-two-arm-grid-bc-quantifiers}, all other hypotheses of Theorem~\ref{thm:pate-two-arm-actual-joint-law}, the first-step bootstrap and finite-completion conditions, and the full-refit series-regression conditions in Proposition~\ref{prop:full-refit-growing-series-realization}. For each arm, suppose that the signed-density conditions in Assumption~\ref{ass:pate-two-arm-grid-signed-density} hold and that
\begin{equation}\label{eq:pate-weighted-drift-residual-moment}
E_s(\varepsilon_z^A\mathbf X\mid S_{z,*},w_{A,z},Z=z)=0.
\end{equation}
Then the sampling and replication joint laws agree:
\begin{equation}\label{eq:pate-growing-bc-joint-agreement}
\begin{aligned}
\ell_{A,1}-\ell_{A,0}&=0,\qquad \boldsymbol b_{A,\mathrm{loc}}=0,\\
V_{A,\mathrm{actual}}
 &=V_{A,\mathrm{PATE},\mathrm{rep}}
  =V_{A,\mathrm{known\text{-}score}},\\
\operatorname{Cov}(\mathbf G_{\theta,A},G_{\tau,A})&=-\boldsymbol\Gamma_A^{-1}\boldsymbol\gamma_{A,1}.
\end{aligned}
\end{equation}
The common variance is finite, and is positive if \(E_s[(\varepsilon_Z^A)^2]>0\).
\end{proposition}

\begin{proof}
We prove the two expansions by applying the same calculation separately to \(\boldsymbol\theta_n=\boldsymbol\theta_A^*+\mathbf h_n/\sqrt n\) and \(\boldsymbol\theta_n=\widehat{\boldsymbol \theta}_A\), in each case using the corresponding empirical scores \(\mathbf D_{z,n,i}\), graph and regression. The root and normalization results give \(\sqrt n(\boldsymbol\theta_n-\boldsymbol\theta_A^*)=O_p(1)\) and uniform score displacement \(O_p(n^{-1/2})\). The baseline nearest donors remain candidates for the current matches. Hence the weighted averages of current-edge lengths, measured at either the current or baseline scores, are \(O_p(n^{-1/2})\).

The Gram and normal-equation calculation leading to \eqref{eq:full-refit-growing-series-coefficient-rate} applies under the stated regression conditions to these score sequences. With \(r_{z,n}=\{B_{z,n}+(1+B_{z,n})L_{z,n}\}/\sqrt n\), it gives
\begin{equation}\label{eq:pate-growing-bc-coefficient-rate}
\|\widehat c_{z,n}-c_{z,n}\|=O_p(r_{z,n}),\qquad
(B_{z,n}+L_{z,n})r_{z,n}\longrightarrow0.
\end{equation}
Consequently \(\|\widehat m_{z,n}-m_z^A\|_{C^1}=o_p(1)\). The Gram bound gives vanishing fit-failure probabilities.

For the current graph, we write
\[
a_{z,n,i}=\mathbf1\{Z_i=1-z\}w_{A,1-z}(\mathbf X_i)
-\mathbf1\{Z_i=z\}K_{n,z}^{A,\mathrm{PATE}}
                         (i;\mathbf D_{z,n}).
\]
Because the positive donor fractions have unit sum, for any row values \(g_i\),
\begin{equation}\label{eq:pate-growing-bc-edge-load}
\begin{aligned}
\sum_i a_{z,n,i}&=0,\qquad
\sum_i|a_{z,n,i}|=2\sum_{i:Z_i=1-z}w_{A,1-z}(\mathbf X_i)=O_p(n),\\
\frac1{\sqrt n}\sum_i a_{z,n,i}g_i
&=\frac1{\sqrt n}\sum_{i:Z_i=1-z}w_{A,1-z}(\mathbf X_i)
 \sum_{j:Z_j=z}\lambda_{i\to j}^{A,(z)}(g_i-g_j).
\end{aligned}
\end{equation}
For \(f_{z,n}=\widehat m_{z,n}-m_z^A\), this identity gives
\[
\left|\frac1{\sqrt n}\sum_i a_{z,n,i}f_{z,n}(\mathbf D_{z,n,i})\right|
\le \|f_{z,n}\|_{C^1}\frac1{\sqrt n}
 \sum_{i:Z_i=1-z}w_{A,1-z}(\mathbf X_i)
 \sum_{j:Z_j=z}\lambda_{i\to j}^{A,(z)}
 \|\mathbf D_{z,n,i}-\mathbf D_{z,n,j}\|
=o_p(1).
\]

\par\smallskip\noindent\emph{Score expansion.}\quad
Let \(\Delta\boldsymbol\theta_n=\boldsymbol\theta_n-\boldsymbol\theta_A^*\), and denote empirical location and scale errors by \(\Delta\ell_{zk,n}\) and \(\Delta s_{zk,n}\). Bounded covariates, positive limiting scales and bounded second derivatives give uniformly over observed points
\[
\begin{aligned}
m_z^A(\mathbf D_{z,n,i})-m_z^A(\mathbf D_{z,i}^*)
&=\dot m_{A,z}^{\mathrm{raw}}(\mathbf X_i)^\top\Delta\boldsymbol\theta_n
-\sum_{k=1}^2\frac{\partial_km_z^A(\mathbf D_{z,i}^*)}{s_{zk}}
                     \Delta\ell_{zk,n}\\
&\quad-\sum_{k=1}^2
 \frac{D_{z,i,k}^*\partial_km_z^A(\mathbf D_{z,i}^*)}{s_{zk}}
                     \Delta s_{zk,n}+O_p(n^{-1}).
\end{aligned}
\]
Each normalization derivative is a fixed bounded Lipschitz own-score function. For its location term, \eqref{eq:pate-growing-bc-edge-load} gives
\[
\begin{aligned}
&\left|\frac{\Delta\ell_{zk,n}}{\sqrt n}\sum_i a_{z,n,i}
       \frac{\partial_km_z^A(\mathbf D_{z,i}^*)}{s_{zk}}\right|\\
&\le C|\Delta\ell_{zk,n}|\frac1{\sqrt n}
 \sum_{i:Z_i=1-z}w_{A,1-z}(\mathbf X_i)
 \sum_{j:Z_j=z}\lambda_{i\to j}^{A,(z)}
 \|\mathbf D_{z,i}^*-\mathbf D_{z,j}^*\|\\
&=O_p(n^{-1/2})O_p(1)\\
&=o_p(1).
\end{aligned}
\]
The same bound applies to the scale term with \(\Delta s_{zk,n}\) and \(d_k\partial_km_z^A(d)/s_{zk}\). The uniform Taylor remainder contributes at most
\[
\frac1{\sqrt n}\sum_i|a_{z,n,i}|\,O_p(n^{-1})
=O_p(\sqrt n)O_p(n^{-1})
=o_p(1).
\]

The deterministic- and fitted-index versions of \eqref{eq:pate-frozen-prediction-derivative-lln} give \(n^{-1}\sum_i a_{z,n,i}\dot m_{A,z}^{\mathrm{raw}}(\mathbf X_i)\to_p\ell_{A,z}\). Combining this limit with \(\widehat m_{z,n}(\mathbf D_{z,n,i})-m_z^A(\mathbf D_{z,i}^*) =f_{z,n}(\mathbf D_{z,n,i}) +m_z^A(\mathbf D_{z,n,i})-m_z^A(\mathbf D_{z,i}^*)\) yields
\begin{equation}\label{eq:pate-growing-bc-signed-expansion}
\begin{aligned}
&\frac1{\sqrt n}\sum_i a_{z,n,i}
 [\widehat m_{z,n}(\mathbf D_{z,n,i})-m_z^A(\mathbf D_{z,i}^*)]\\
&=\frac1{\sqrt n}\sum_i a_{z,n,i}f_{z,n}(\mathbf D_{z,n,i})
 +\frac1{\sqrt n}\sum_i a_{z,n,i}
 [m_z^A(\mathbf D_{z,n,i})-m_z^A(\mathbf D_{z,i}^*)]\\
&=\left\{\frac1n\sum_i a_{z,n,i}\dot m_{A,z}^{\mathrm{raw}}(\mathbf X_i)\right\}^\top
       \sqrt n\Delta\boldsymbol\theta_n+o_p(1)\\
&=\ell_{A,z}^\top\sqrt n\Delta\boldsymbol\theta_n+o_p(1).
\end{aligned}
\end{equation}
The law-of-large-numbers error is \(o_p(1)\), and its product with the tight root-\(n\)-scaled parameter error is negligible. Substitution into the \(H+B+E\) decomposition gives
\[
\begin{aligned}
\sqrt n(B_{n,A}^{D,\mathrm{PATE}}-\widehat B_{n,A}^{D,\mathrm{PATE}})
&=\frac1{\bar w_{A,\mathrm{PATE},n}}
 \sum_{z=0}^1(2z-1)\frac1{\sqrt n}\sum_i a_{z,n,i}
 [\widehat m_{z,n}(\mathbf D_{z,n,i})-m_z^A(\mathbf D_{z,i}^*)]\\
&=\frac{(\ell_{A,1}-\ell_{A,0})^\top\sqrt n\Delta\boldsymbol\theta_n}
        {\bar w_{A,\mathrm{PATE}}}+o_p(1).
\end{aligned}
\]
The matching discrepancy and its estimate use the same current matches. The law of large numbers for the survey weights gives a positive denominator limit, completing both expansions. Since \(\bar w_{A,\mathrm{PATE}}>0\),
\[
\frac{(\ell_{A,1}-\ell_{A,0})^\top \mathbf h}{\bar w_{A,\mathrm{PATE}}}=0
 \quad\text{for every }\mathbf h\in\mathbb R^{p_A}
\quad\Longleftrightarrow\quad\ell_{A,1}-\ell_{A,0}=0.
\]
Substitution into the expansion at the fitted parameter gives its BC remainder condition.

\par\smallskip\noindent\emph{Replication validity.}\quad
The BC condition for deterministic local parameters and the two expansions above give a zero aggregate derivative and a negligible BC remainder at the fitted parameter. The sampling theorem then gives consistency, allowing application of the conditional replication limit with all regressions refitted. Substitution into \eqref{eq:pate-full-prediction-replication-variance} gives the known-score replication variance. For each arm, \(L_z\) is a subvector of the bounded covariate vector. The flux definition in \eqref{eq:pate-two-arm-grid-conditional-drift} therefore gives
\[
\mathbf F^{\varepsilon}_{A,z,\mathbf h,w}(s)
=q_{z,w}(s)E_s(\varepsilon_z^A\dot C_{z,\mathbf h}L_z
                    \mid Z=z,S_{z,*}=s,w_{A,z}=w)=0
\]
almost everywhere by \eqref{eq:pate-weighted-drift-residual-moment}. The continuously differentiable signed-flux version is zero everywhere. The definitions in \eqref{eq:pate-two-arm-grid-conditional-drift} give
\[
\mathbf F^{\varepsilon}_{A,z,\mathbf h,w}=0
\quad\Longrightarrow\quad\nabla\!\cdot \mathbf F^{\varepsilon}_{A,z,\mathbf h,w}=0
\quad\Longrightarrow\quad \dot m^{\varepsilon}_{A,z,\mathbf h}=0
\quad\Longrightarrow\quad\Delta_{A,z}(\mathbf h)=0,
\qquad \boldsymbol b_{A,\mathrm{loc}}=0.
\]
Both joint laws reduce to \((G_{\tau,A},\mathbf G_{\theta,A})\).

Joint conditional centering makes the heterogeneity--residual cross term zero in the marked-Poisson variance representation. Nonnegative incoming coefficients and weights bounded below by \(w_{\min}>0\) give
\begin{align*}
V_{A,\mathrm{known\text{-}score}}
&\ge\frac{E[(w_{A,Z}+K_{A,Z}^{\infty})^2(\varepsilon_Z^A)^2]}
 {\bar w_{A,\mathrm{PATE}}^2}\\
&\ge\frac{w_{\min}^2E_s[(\varepsilon_Z^A)^2]}
 {\bar w_{A,\mathrm{PATE}}^2}\\
&>0.
\end{align*}
The incoming and row-moment bounds give finiteness.
\end{proof}

\begin{proposition}[PATT series bias correction]
\label{prop:patt-growing-bc-compatibility}\label{cor:patt-growing-bc-residual-agreement}
Fix \(A\in\{R,P\}\), with \(w_{R,z}=w_z\), \(w_{P,z}=w\), and the PATT nuisance vector \(\boldsymbol\theta_A=(\alpha_A^\top,\beta_{A,0}^\top)^\top\). Assume the following conditions:
\begin{enumerate}
\item[(i)] The unscaled conditions of Proposition~\ref{prop:patt-actual-graph-row-multiplier-law} hold, including residual centering conditional on the common control score and observed survey weight, the original score parameter estimators, and empirical standardization.
\item[(ii)] The control regression satisfies the conditions of Proposition~\ref{prop:full-refit-growing-series-realization} on the deterministic basis, {population projection}, positive Gram matrix, bounded coefficients, \(C^1\) approximation to the fixed \(C^2\) function \(m_0^A\), finite \(E_sY^2\), and basis growth. The basis is defined on the same fixed compact convex neighborhood of the limiting and current scores.
\end{enumerate}
We use the prescribed survey-weighted least-squares fit on each current score, with its specified finite output on unsuccessful fits. We write
\[
\bar w_{A,n}=\frac1n\sum_iZ_iw_{A,1}(\mathbf X_i)
 \longrightarrow_p\bar w_A=E_s[Zw_{A,1}(\mathbf X)]>0,
\]
and use \(\dot m_{A,0}^{\mathrm{raw}}\) from \eqref{eq:full-refit-growing-series-tangent} and \(\ell_{A,0}\) from \eqref{eq:patt-frozen-prediction-derivative}. For every deterministic full direction \(\mathbf h_n\to \mathbf h\), the regression is refitted at its own empirical score \(\mathbf D_{0,n}^e(\mathbf h_n)\). Then
\begin{equation}\label{eq:patt-growing-bc-local-limit}
\begin{aligned}
&\sqrt n\{B_{n,A}^{D,\mathrm{PATT}}
 (\mathbf D_{0,n}^e(\mathbf h_n);\mathbf D_0^*,m_0^A)
 -\widehat B_{n,A}^{D,\mathrm{PATT}}(\mathbf D_{0,n}^e(\mathbf h_n))\}\\
&\hspace{12mm}\longrightarrow_p-\ell_{A,0}^\top \mathbf h/\bar w_A.
\end{aligned}
\end{equation}
Separately, for the actual nuisance fit and its own BC regression,
\begin{equation}\label{eq:patt-growing-bc-actual-expansion}
\begin{aligned}
&\sqrt n\{B_{n,A}^{D,\mathrm{PATT}}
 (\widehat{\mathbf D}_0;\mathbf D_0^*,m_0^A)
 -\widehat B_{n,A}^{D,\mathrm{PATT}}(\widehat{\mathbf D}_0)\}\\
&=-\frac{\ell_{A,0}^\top\sqrt n(\widehat{\boldsymbol \theta}_A-\boldsymbol\theta_A^*)}
              {\bar w_A}+o_p(1).
\end{aligned}
\end{equation}
Under these regression conditions and the full-dimensional interior parameter space, the BC remainder condition at every deterministic local parameter holds exactly when \(\ell_{A,0}=0\). This equality also gives the BC remainder condition at the fitted parameter.

\par\smallskip\noindent\emph{Replication validity.}\quad
Assume additionally the deterministic-local BC condition \eqref{eq:patt-local-grid-accepted-bc-remainder} and all other hypotheses of Theorem~\ref{cor:patt-joint-actual-score-law}, including the signed-density drift conditions, and the first-step bootstrap and completion conditions for Proposition~\ref{prop:full-refit-growing-series-realization}. Suppose also that the control residual-covariate moment satisfies
\begin{equation}\label{eq:patt-growing-bc-residual-x-moment}
E_s[\{Y-m_0^A(\mathbf D_0^*)\}\mathbf X
       \mid Z=0,\mathbf D_0^*,w_{A,0}]=0.
\end{equation}
Then \(\ell_{A,0}=0\) and \(\mathbf j_A=0\). The sampling and replication joint laws both equal the centered Gaussian law of \((G_{\tau,A},\mathbf G_{\theta,A})\), with
\begin{equation}\label{eq:patt-growing-bc-joint-agreement}
\begin{aligned}
V_{A,\mathrm{actual}}&=V_{A,\mathrm{rep}}\\
 &=V_{A,\mathrm{known\text{-}score}},\\
\operatorname{Cov}(\mathbf G_{\theta,A},G_{\tau,A})&=-\boldsymbol\Gamma_A^{-1}\boldsymbol\gamma_{A,1},\qquad
\operatorname{Var}(\mathbf G_{\theta,A})=\boldsymbol\Sigma_{A,\theta}.
\end{aligned}
\end{equation}
The common variance is finite and strictly positive if \(E_s[Zw_{A,1}^2\varepsilon_{1\mid1}^2]>0\). Here \(\varepsilon_{1\mid1}=Y-m_{1\mid1}(\mathbf D_0^*)\) on treated rows. We also impose the stated treated-centering assumption.
\end{proposition}

\begin{proof}
We apply the one-arm calculation in Proposition~\ref{prop:pate-growing-bc-compatibility} separately at \(\boldsymbol\theta_n=\boldsymbol\theta_A^*+\mathbf h_n/\sqrt n\) and \(\boldsymbol\theta_n=\widehat{\boldsymbol \theta}_A\), using the corresponding control graph, empirical normalization and regression. Here
\[
a_{0,n,i}=Z_iw_{A,1}(\mathbf X_i)
 -(1-Z_i)K_{n,0}^{A,\mathrm{PATT}}(i;\mathbf D_{0,n}).
\]
The complete donor fractions have unit sum, so
\[
\sum_i a_{0,n,i}=0,\qquad
\sum_i|a_{0,n,i}|=2\sum_iZ_iw_{A,1}(\mathbf X_i)=O_p(n).
\]
For a fixed bounded Lipschitz function \(q\),
\begin{equation}\label{eq:patt-local-score-function-balance}
\begin{aligned}
\left\|\frac1n\sum_i a_{0,n,i}q(\mathbf D_{0,i}^*)\right\|
&=\left\|\frac1n\sum_{i:Z_i=1}w_{A,1}(\mathbf X_i)
 \sum_{j:Z_j=0}\lambda_{i\to j,n}^{A,(0)}
 [q(\mathbf D_{0,i}^*)-q(\mathbf D_{0,j}^*)]\right\|\\
&\le\frac{\operatorname{Lip}(q)}n\sum_{i:Z_i=1}w_{A,1}(\mathbf X_i)
 \sum_{j:Z_j=0}\lambda_{i\to j,n}^{A,(0)}
 \|\mathbf D_{0,i}^*-\mathbf D_{0,j}^*\|\\
&=O_p(n^{-1/2}).
\end{aligned}
\end{equation}
The radius bound also holds for current-score edge lengths. The control version of \eqref{eq:full-refit-growing-series-coefficient-rate} gives
\[
\begin{aligned}
\|\widehat c_{0,n}-c_{0,n}\|
&=O_p\!\left(\frac{B_{0,n}+(1+B_{0,n})L_{0,n}}{\sqrt n}\right),\\
\|\widehat m_{0,n}-m_0^A\|_{C^1}&=o_p(1).
\end{aligned}
\]
We can therefore apply the cancellation of the current regression error and the score expansion in \eqref{eq:pate-growing-bc-signed-expansion}:
\[
\begin{aligned}
&\frac1{\sqrt n}\sum_i a_{0,n,i}
 [\widehat m_{0,n}(\mathbf D_{0,n,i})-m_0^A(\mathbf D_{0,i}^*)]\\
&=\left\{\frac1n\sum_i a_{0,n,i}\dot m_{A,0}^{\mathrm{raw}}(\mathbf X_i)\right\}^{\!\top}
 \sqrt n(\boldsymbol\theta_n-\boldsymbol\theta_A^*)+o_p(1).
\end{aligned}
\]
Here \eqref{eq:patt-local-score-function-balance} controls both normalization derivatives, while the \(C^1\) error is multiplied by the current edge-length bound. The Gram bound makes the regression-failure probability tend to zero.

The bounded-mark limits in Proposition~\ref{prop:patt-actual-graph-row-multiplier-law} hold separately for deterministic local indices and the actual fitted index:
\begin{equation}\label{eq:patt-local-prediction-mark-balance}
\begin{aligned}
\frac1n\sum_i a_{0,n,i}\dot m_{A,0}^{\mathrm{raw}}(\mathbf X_i)
&\longrightarrow_p\ell_{A,0}\\
&=E_s[Zw_{A,1}\dot m_{A,0}^{\mathrm{raw}}]
 -E_s[(1-Z)\kappa_{A,0}(\mathbf D_0^*,w_{A,0})\dot m_{A,0}^{\mathrm{raw}}].
\end{aligned}
\end{equation}
For \(e_{n,i}=\widehat m_{0,n}(\mathbf D_{0,n,i}) -m_0^A(\mathbf D_{0,i}^*)\), the exact BC identity is
\begin{equation}\label{eq:patt-local-bc-exact-prediction-error}
\begin{aligned}
\sqrt n(\widehat B_{n,A}^{D,\mathrm{PATT}}-B_{n,A}^{D,\mathrm{PATT}})
&=\frac1{\bar w_{A,n}\sqrt n}
 \sum_{i:Z_i=1}w_{A,1}(\mathbf X_i)
 \sum_{j:Z_j=0}\lambda_{i\to j,n}^{A,(0)}(e_{n,i}-e_{n,j})\\
&=\frac1{\bar w_{A,n}\sqrt n}\sum_i a_{0,n,i}e_{n,i}.
\end{aligned}
\end{equation}
Since \(\bar w_{A,n}\to_p\bar w_A>0\) and \(\sqrt n(\boldsymbol\theta_n-\boldsymbol\theta_A^*)=O_p(1)\), the two applications give
\[
\sqrt n(B_{n,A}^{D,\mathrm{PATT}}-\widehat B_{n,A}^{D,\mathrm{PATT}})
=-\frac{\ell_{A,0}^\top\sqrt n(\boldsymbol\theta_n-\boldsymbol\theta_A^*)}{\bar w_A}
 +o_p(1).
\]
This establishes both expansions. The BC remainder condition for all deterministic local parameters is characterized by
\[
\ell_{A,0}^\top \mathbf h=0\ \text{for every }\mathbf h\in\mathbb R^{p_A}
\quad\Longleftrightarrow\quad\ell_{A,0}=0,
\]
which also gives the BC remainder condition at the fitted parameter.

Under the additional inference hypotheses, deterministic-local BC gives \(\ell_{A,0}=0\). To identify the sampling loading, we use the signed flux in \eqref{eq:patt-drift-signed-density-definition} and \eqref{eq:patt-drift-chart-direction}. Because \(L\) is a subvector of \(\mathbf X\), the residual-covariate moment gives
\[
\begin{aligned}
\mathbf F^{\varepsilon}_{\mathbf h,w}(s)
&=\int a_w(s,l)\dot C_{\mathbf h}l\,\nu_w(dl)\\
&=q_w(s)E_s[\{Y-m_0^A(\mathbf D_0^*)\}\dot C_hL
                         \mid Z=0,S_*=s,w_{A,0}=w]\\
&=0.
\end{aligned}
\]
The continuously differentiable signed-flux version is therefore zero, and
\[
\nabla\!\cdot \mathbf F^{\varepsilon}_{\mathbf h,w}=0
\quad\Longrightarrow\quad \dot m^{\varepsilon}_{A,\mathbf h}=0
\quad\Longrightarrow\quad\Delta_A(\mathbf h)=0,
\qquad \mathbf j_A=0.
\]
The sampling law \((G_{\tau,A}-\mathbf j_A^\top \mathbf G_{\theta,A},\mathbf G_{\theta,A})\) and replication law \((G_{\tau,A}-\ell_{A,0}^\top \mathbf G_{\theta,A}/\bar w_A,\mathbf G_{\theta,A})\) both reduce to \((G_{\tau,A},\mathbf G_{\theta,A})\). Finally, under the stated treated-centering condition,
\[
\begin{aligned}
V_{A,\mathrm{known\text{-}score}}
&=V_H+V_E\\
&\ge V_E\\
&\ge E_s[Zw_{A,1}^2\varepsilon_{1\mid1}^2]/\bar w_A^2\\
&>0.
\end{aligned}
\]
Finiteness follows from the incoming-coefficient and row-moment bounds.
\end{proof}

\subsubsection{Replication under full-mean centering}

\begin{corollary}[Replication under full-covariate centering]
\label{cor:pate-fullx-full-refit-comparison}
Suppose the conditions of Theorem~\ref{thm:pate-fullx-actual-joint-law} hold. Assume additionally the prediction-expansion, squared-load consistency, weighted-remainder, bootstrap-root, and finite-output conditions of Proposition~\ref{prop:pate-full-prediction-replication-law}. The growing-series conditions in Proposition~\ref{prop:full-refit-growing-series-realization} may replace its prediction conditions. Then the replicate with score and prediction refits and frozen reuse satisfies
\begin{equation}\label{eq:pate-fullx-replicate-joint-law}
\begin{aligned}
&\mathcal L^*\!\left(
\begin{pmatrix}
\sqrt n(\widehat\tau_{\mathrm{PATE}}^{A,*}
               -\widehat\tau_{\mathrm{PATE}}^{A,\mathrm{BC}})\\
\sqrt n(\widehat{\boldsymbol \theta}_{A,\mathrm{PATE}}^*
               -\widehat{\boldsymbol \theta}_{A,\mathrm{PATE}})
\end{pmatrix}\middle|\mathcal F_n\right)\\
&\quad\Longrightarrow_{P_s}
\mathcal L\!\left(
\begin{pmatrix}
G_{\tau,A}+\dfrac{(\ell_{A,1}-\ell_{A,0})^\top}{\bar w_A}\mathbf G_{\theta,A}\\\mathbf G_{\theta,A}
\end{pmatrix}\right).
\end{aligned}
\end{equation}
Here \(\ell_{A,z}\) is defined in \eqref{eq:pate-frozen-prediction-derivatives}. Joint sampling and replication agreement holds exactly when \(\ell_{A,1}-\ell_{A,0}=0\). Scalar agreement is equivalent to
\begin{equation}\label{eq:pate-fullx-scalar-agreement}
\frac{(\ell_{A,1}-\ell_{A,0})^\top}{\bar w_A}
\left\{-2\boldsymbol\Gamma_A^{-1}\boldsymbol\gamma_{A,1}
 +\boldsymbol\Sigma_{A,\theta}\frac{\ell_{A,1}-\ell_{A,0}}{\bar w_A}\right\}=0.
\end{equation}
Under the conditions of Proposition~\ref{prop:pate-growing-bc-compatibility}, the local BC family forces \(\ell_{A,1}-\ell_{A,0}=0\), and hence gives joint agreement for those series regressions.
\end{corollary}

\begin{proof}
Theorem~\ref{thm:pate-fullx-actual-joint-law} gives consistency, and iterated expectation reduces full-field centering to baseline joint-score/weight centering. Together with the own-score, moment, root and normalization conditions, this establishes the actual-row joint multiplier law in Proposition~\ref{prop:pate-actual-graph-row-multiplier-law}. Applying the full-prediction and bootstrap-root expansions to this joint limit gives \eqref{eq:pate-fullx-replicate-joint-law} by conditional Slutsky.

By \eqref{eq:pate-fullx-baseline-pair},
\[
\begin{aligned}
&\operatorname{Cov}\!\left(\mathbf G_{\theta,A},G_{\tau,A}+
 \frac{(\ell_{A,1}-\ell_{A,0})^\top \mathbf G_{\theta,A}}{\bar w_A}\right)
       -\operatorname{Cov}(\mathbf G_{\theta,A},G_{\tau,A})\\
&=\frac{\operatorname{Var}(\mathbf G_{\theta,A})(\ell_{A,1}-\ell_{A,0})}{\bar w_A}\\
&=\frac{\boldsymbol\Sigma_{A,\theta}(\ell_{A,1}-\ell_{A,0})}{\bar w_A},\\[1mm]
&\operatorname{Var}\!\left(G_{\tau,A}+
 \frac{(\ell_{A,1}-\ell_{A,0})^\top \mathbf G_{\theta,A}}{\bar w_A}\right)
       -\operatorname{Var}(G_{\tau,A})\\
&=\frac{2(\ell_{A,1}-\ell_{A,0})^\top\operatorname{Cov}(\mathbf G_{\theta,A},G_{\tau,A})}{\bar w_A}
 +\frac{(\ell_{A,1}-\ell_{A,0})^\top\boldsymbol\Sigma_{A,\theta}
                    (\ell_{A,1}-\ell_{A,0})}{\bar w_A^2}\\
&=\frac{(\ell_{A,1}-\ell_{A,0})^\top}{\bar w_A}
 \left\{-2\boldsymbol\Gamma_A^{-1}\boldsymbol\gamma_{A,1}
       +\boldsymbol\Sigma_{A,\theta}\frac{\ell_{A,1}-\ell_{A,0}}{\bar w_A}\right\}.
\end{aligned}
\]
Positive definiteness gives joint agreement exactly when \(\ell_{A,1}-\ell_{A,0}=0\). Equality of the scalar variances gives \eqref{eq:pate-fullx-scalar-agreement}, including degenerate Gaussian laws.

Proposition~\ref{prop:pate-growing-bc-compatibility} gives the coefficient \((\ell_{A,1}-\ell_{A,0})/\bar w_A\) in the local BC remainder. The BC condition for deterministic local parameters makes this coefficient zero and also gives a negligible remainder at the fitted parameter, establishing the final assertion.
\end{proof}

\begin{corollary}[Replication under full-control-mean centering]
\label{cor:patt-grouped-fullx-refits}
Suppose the conditions of Theorem~\ref{thm:patt-grouped-fullx-actual} hold. Assume additionally the prediction expansion, weighted remainder and squared-load prediction consistency in \eqref{eq:patt-full-prediction-refit-expansion}--\eqref{eq:patt-weighted-prediction-consistency}, the original bootstrap-root conditions and vanishing completion probabilities. Alternatively, the stated control growing-series conditions of Proposition~\ref{prop:full-refit-growing-series-realization} may replace the prediction conditions. The bootstrap-root and completion conditions are required in either case.

Then the replicate with prediction refits and frozen reuse satisfies
\begin{equation}\label{eq:patt-grouped-fullx-refit-pair}
\begin{aligned}
&\mathcal L^*\left(
\begin{pmatrix}
\sqrt n(\widehat\tau_A^*-\widehat\tau_A^{\mathrm{BC}})\\
\sqrt n(\widehat{\boldsymbol \theta}_A^*-\widehat{\boldsymbol \theta}_A)
\end{pmatrix}\middle|\mathcal F_n\right)\\
&\qquad\Longrightarrow_{P_s}
\mathcal L\left(
\begin{pmatrix}\widetilde G_{\tau,A}-\ell_{A,0}^\top \mathbf G_{\theta,A}/\bar w_A\\\mathbf G_{\theta,A}\end{pmatrix}
\right),
\end{aligned}
\end{equation}
with \(\ell_{A,0}\) defined in \eqref{eq:patt-frozen-prediction-derivative}. Joint sampling/replication agreement holds if and only if \(\ell_{A,0}=0\). Scalar Gaussian-law agreement requires only
\begin{equation}\label{eq:patt-grouped-fullx-scalar-agreement}
\frac{\ell_{A,0}^\top}{\bar w_A}
 \left\{\boldsymbol\Sigma_{A,\theta}\frac{\ell_{A,0}}{\bar w_A}
                   +2\boldsymbol\Gamma_A^{-1}\widetilde{\boldsymbol \gamma}_{A,1}\right\}=0.
\end{equation}
Under the growing-series conditions, the deterministic-local BC family forces \(\ell_{A,0}=0\). The resulting full-refit joint law therefore agrees with the sampling law.
\end{corollary}

\begin{proof}
Apply the control-row argument of Proposition~\ref{prop:patt-actual-graph-row-multiplier-law} under the stated geometry, centering, moment and empirical-index conditions. The treated contribution in \eqref{eq:patt-frozen-baseline-row} is iid, centered and has a finite \(p\)-moment. Thus
\[
\begin{aligned}
\frac1n\sum_i\varphi_{A,\mathrm{PATT},i}&\to_p0,\\
\frac1{n\bar w_A^2}\sum_i\varphi_{A,\mathrm{PATT},i}^2
 &\to_p\widetilde V_{A,\mathrm{PATT}},\\
\frac1{n\bar w_A}\sum_i\varphi_{A,\mathrm{PATT},i}(\mathbf U_A)
 &\to_p\widetilde{\boldsymbol \gamma}_{A,1}.
\end{aligned}
\]
The treated/control product vanishes because the treatment indicators are disjoint. Control centering and iterated expectation identify the control variance. The same \(p\)-moment bound gives the joint multinomial Lindeberg condition. Thus \eqref{eq:patt-actual-graph-joint-multiplier-law} holds with \(V_{A,\mathrm{known\text{-}score}}\) and \(\boldsymbol\gamma_{A,1}\) replaced by \(\widetilde V_{A,\mathrm{PATT}}\) and \(\widetilde{\boldsymbol \gamma}_{A,1}\), respectively.

In the ratio \eqref{eq:patt-frozen-baseline-exact-ratio}, apply the prediction expansion as in Proposition~\ref{prop:patt-full-prediction-replication-law}. That argument requires only control residual centering and gives the signed derivative \(\ell_{A,0}\), while the own-score terms and weighted remainder vanish. Sampling consistency controls the denominator term, so the bootstrap-root expansion gives \eqref{eq:patt-grouped-fullx-refit-pair} with its negative PATT sign. The Gaussian covariance differences are
\[
\begin{aligned}
&\operatorname{Cov}(\mathbf G_{\theta,A},\widetilde G_{\tau,A}-\ell_{A,0}^\top \mathbf G_{\theta,A}/\bar w_A)
       -\operatorname{Cov}(\mathbf G_{\theta,A},\widetilde G_{\tau,A})\\
&=-\boldsymbol\Sigma_{A,\theta}\ell_{A,0}/\bar w_A,\\
&\operatorname{Var}(\widetilde G_{\tau,A}-\ell_{A,0}^\top \mathbf G_{\theta,A}/\bar w_A)
       -\operatorname{Var}(\widetilde G_{\tau,A})\\
&=2\ell_{A,0}^\top\boldsymbol\Gamma_A^{-1}\widetilde{\boldsymbol \gamma}_{A,1}/\bar w_A
       +\ell_{A,0}^\top\boldsymbol\Sigma_{A,\theta}\ell_{A,0}/\bar w_A^2\\
&=\frac{\ell_{A,0}^\top}{\bar w_A}
 \left\{\boldsymbol\Sigma_{A,\theta}\frac{\ell_{A,0}}{\bar w_A}
                +2\boldsymbol\Gamma_A^{-1}\widetilde{\boldsymbol \gamma}_{A,1}\right\}.
\end{aligned}
\]
Positive definiteness gives agreement of the joint laws exactly when \(\ell_{A,0}=0\). The variance difference gives \eqref{eq:patt-grouped-fullx-scalar-agreement}.

For series fits, apply the control-only normal-equation, Gram, approximation and current-edge arguments of Proposition~\ref{prop:patt-growing-bc-compatibility}. These give \eqref{eq:patt-growing-bc-local-limit} and \eqref{eq:patt-growing-bc-actual-expansion} separately, before BC or consistency is invoked. The full-dimensional local BC family forces \(\ell_{A,0}=0\). The actual sampling and complete-refit laws then apply, with the growing signed basis load controlled by Proposition~\ref{prop:full-refit-growing-series-realization}.
\end{proof}

\subsubsection{Monte Carlo variance and Wald inference}
\label{app:monte-carlo-inference}

Fix the sampling design \(A\in\{R,P\}\) and estimand \(T\in\{\mathrm{PATE},\mathrm{PATT}\}\). Let \(\widehat\tau_T^{A,\mathrm{BC}}\) be the bias-corrected WDSM estimator using estimated scores, and let \(\widehat\tau_T^{A,\ast,b}\) be its fixed-match replicate, with scores, standardization and bias-correction regressions refitted according to the prescribed procedure.

Conditionally on the original data, fits and graph \(\mathcal F_n\), we draw the \(B\) complete replicates independently and define
\begin{equation}\label{eq:bootstrap-mc-original-variance}
\begin{aligned}
\bar\tau_{T,B}^{A,\ast}
&=\frac1B\sum_{b=1}^B\widehat\tau_T^{A,\ast,b},\\
\widehat{\operatorname{Var}}_{n,B}
 \bigl(\widehat\tau_T^{A,\mathrm{BC}}\bigr)
&=\frac1B\sum_{b=1}^B
 \left(\widehat\tau_T^{A,\ast,b}
              -\bar\tau_{T,B}^{A,\ast}\right)^2.
\end{aligned}
\end{equation}

\begin{proposition}[Monte Carlo variance and Wald inference]
\label{prop:bootstrap-monte-carlo-iterated-variance}

For the conditionally independent complete replicates defined above, suppose the conditional law satisfies
\begin{equation}\label{eq:bootstrap-mc-proved-weak-law}
\begin{aligned}
&\mathcal L^\ast\!\left[
 \sqrt n\left\{
 \widehat\tau_T^{A,\ast}
 -\widehat\tau_T^{A,\mathrm{BC}}
 \right\}
 \,\middle|\,\mathcal F_n\right]\\
&\hspace{12mm}\Longrightarrow_{P_s}
 \mathcal N(0,V_{A,T,\mathrm{rep}}),
\qquad 0\le V_{A,T,\mathrm{rep}}<\infty.
\end{aligned}
\end{equation}
Each replicate uses the prescribed finite-valued output. If the weak limit is proved using an alternative finite extension, assume its conditional disagreement probability tends to zero.

For every fixed integer \(B\ge2\),
\begin{equation}\label{eq:bootstrap-mc-fixed-b-law}
\begin{aligned}
&\mathcal L^\ast\!\left[
 n\widehat{\operatorname{Var}}_{n,B}
  \bigl(\widehat\tau_T^{A,\mathrm{BC}}\bigr)
 \,\middle|\,\mathcal F_n\right]\\
&\hspace{12mm}\Longrightarrow_{P_s}
 \mathcal L\!\left\{
 \frac{V_{A,T,\mathrm{rep}}}{B}\chi^2_{B-1}
 \right\}.
\end{aligned}
\end{equation}
The same convergence holds unconditionally. The right-hand side is the point mass at zero when \(V_{A,T,\mathrm{rep}}=0\). Writing \(P\) for the joint probability over the original sample and the independent replication draws, one has, for every \(\epsilon>0\),
\begin{equation}\label{eq:bootstrap-mc-iterated-consistency}
\lim_{B\to\infty}\limsup_{n\to\infty}
P\!\left\{
 \left|
 n\widehat{\operatorname{Var}}_{n,B}
  \bigl(\widehat\tau_T^{A,\mathrm{BC}}\bigr)
 -V_{A,T,\mathrm{rep}}
 \right|>\epsilon
\right\}=0.
\end{equation}
There also exists a deterministic, sufficiently slowly increasing sequence of integers \(B_{\mathrm{rep},n}\to\infty\), possibly depending on the fixed data-generating law, such that
\begin{equation}\label{eq:bootstrap-mc-slow-diagonal-consistency}
n\widehat{\operatorname{Var}}_{n,B_{\mathrm{rep},n}}
 \bigl(\widehat\tau_T^{A,\mathrm{BC}}\bigr)
\longrightarrow_P V_{A,T,\mathrm{rep}}.
\end{equation}
The conditional error probability also tends to zero in \(P_s\)-probability along this sequence.

Finally, suppose a separate sampling result proves
\begin{equation}\label{eq:bootstrap-mc-matched-sampling-law}
\sqrt n\left\{
 \widehat\tau_T^{A,\mathrm{BC}}-\tau_T
 \right\}
\Longrightarrow\mathcal N(0,V_{A,T}),
\qquad
V_{A,T}=V_{A,T,\mathrm{rep}}>0.
\end{equation}
For \(0<\alpha<1\), the normal Wald interval based on \eqref{eq:bootstrap-mc-original-variance} has asymptotic coverage \(1-\alpha\) along a sequence in \eqref{eq:bootstrap-mc-slow-diagonal-consistency}, and also in the iterated sense with \(n\) first and \(B\) second.
\end{proposition}

\begin{proof}
Write \(T_{n,b}^*=\sqrt n(\widehat\tau_T^{A,*,b} -\widehat\tau_T^{A,\mathrm{BC}})\), \(S_{n,B}=B^{-1}\sum_b(T_{n,b}^*-\bar T_{n,B}^*)^2\), and \(V=V_{A,T,\mathrm{rep}}\). For fixed \(B\), conditional independence gives joint convergence of the replicates to \(\mathcal N(0,V)^{\otimes B}\). The continuous mapping theorem, applied to their empirical variance, and the normal quadratic-form identity (Lemma~17.1 of van der Vaart\citep{vanderVaart2000}) give
\[
n\widehat{\operatorname{Var}}_{n,B}
 (\widehat\tau_T^{A,\mathrm{BC}})
=S_{n,B}
\Rightarrow \frac VB\chi^2_{B-1},
\]
conditionally in probability and unconditionally, including \(V=0\). For \(L_B=(V/B)\chi^2_{B-1}\),
\[
\begin{aligned}
E\!\left[(L_B-V)^2\right]
&=\operatorname{Var}(L_B)+\{E(L_B)-V\}^2\\
&=\frac{2V^2(B-1)}{B^2}
       +\left\{\frac{V(B-1)}B-V\right\}^2\\
&=\frac{V^2(2B-1)}{B^2}.
\end{aligned}
\]
When \(V>0\), the limiting law has no atoms at the boundaries of \(\{|x-V|>\epsilon\}\), so
\[
\begin{aligned}
\limsup_nP(|S_{n,B}-V|>\epsilon)
&=P(|L_B-V|>\epsilon)\\
&\le\frac{V^2(2B-1)}{B^2\epsilon^2}\\
&\longrightarrow0.
\end{aligned}
\]
When \(V=0\), this probability tends to zero already at fixed \(B\).

To obtain a diagonal sequence, choose increasing \(B_k\ge\max(2,k)\) and \(N_k\uparrow\infty\) such that \(P(|S_{n,B_k}-V|>1/k)<1/k\) for every \(n\ge N_k\). We set \(B_{\mathrm{rep},n}=B_k\) on \(N_k\le n<N_{k+1}\). Then \(S_{n,B_{\mathrm{rep},n}}\to_P V\), and
\[
P_s\{P^*(|S_{n,B_{\mathrm{rep},n}}-V|>\epsilon\mid\mathcal F_n)>\eta\}
\le\eta^{-1}P(|S_{n,B_{\mathrm{rep},n}}-V|>\epsilon)\longrightarrow0.
\]
To transfer the result from an alternative finite extension, let \(q_n\to_{P_s}0\) be its conditional one-replicate disagreement probability. Since \(0\le q_n\le1\), \(E_sq_n\to0\). The union bound \(BE_sq_n\) transfers the fixed-\(B\) limits. Enlarging \(N_k\) so that \(B_kE_sq_n<1/k\) throughout the \(k\)th block transfers the diagonal limit to the prescribed finite outputs.

When the sampling and replication variances agree and are positive, Slutsky's theorem yields
\[
\frac{\widehat\tau_T^{A,\mathrm{BC}}-\tau_T}
 {\{\widehat{\operatorname{Var}}_{n,B_{\mathrm{rep},n}}
          (\widehat\tau_T^{A,\mathrm{BC}})\}^{1/2}}
=\frac{\sqrt n(\widehat\tau_T^{A,\mathrm{BC}}-\tau_T)}
       {\sqrt{S_{n,B_{\mathrm{rep},n}}}}
\Rightarrow\mathcal N(0,1).
\]
For the iterated assertion, on \(\{|S_{n,B}-V|\le\eta\}\), with \(0<\eta<V\), the Wald coverage event lies between the two events having deterministic critical values \(z_{1-\alpha/2}\sqrt{V-\eta}\) and \(z_{1-\alpha/2}\sqrt{V+\eta}\). The sampling CLT and the iterated variance limit therefore bound its iterated lower and upper coverage limits by
\[
2\Phi\!\left(z_{1-\alpha/2}\sqrt{1-\eta/V}\right)-1
\quad\text{and}\quad
2\Phi\!\left(z_{1-\alpha/2}\sqrt{1+\eta/V}\right)-1.
\]
Letting \(\eta\downarrow0\) proves the result.
\end{proof}

\subsection{Retrospective PATE}\label{app:retrospective-pate}
\subsubsection{Conditions}
Throughout this section, we take expectations under the selected-sample law unless explicitly stated otherwise. The retrospective inverse-inclusion weights are \(w_z(\mathbf X)=\pi_z(\mathbf X)^{-1}\), \(z=0,1\). The oracle graphs use the fixed limiting standardized scores \(\mathbf D_z^*\). We write \(m_z=m_z^R\) for the population reduced means from Section~\ref{oracle-score-space-decomposition-behind-the-double-robust-theorems}, and use the main-text definitions
\[
\tau_{\mathbf D}(\mathbf X)
=m_1\{\mathbf D_1^*(\mathbf X)\}-m_0\{\mathbf D_0^*(\mathbf X)\},
\qquad
\varepsilon_{z,i}=Y_i-m_z\{\mathbf D_z^*(\mathbf X_i)\},\quad Z_i=z.
\]
The common centering condition invoked below conditions residuals on both limiting scores and the observed survey weight.

\begin{assumption}[Retrospective PATE conditions]
\label{ass:retrospective-pate-oracle}
\leavevmode
\begin{enumerate}
\item[(i)] The retrospective sampling, treatment mean-balancing, and selected-i.i.d.\ conditions in Section~\ref{population-representation-identities-and-score-relations} hold for both treatment arms.
\item[(ii)] Assumptions~\ref{ass:residual-centering-survey-normalized} and~\ref{ass:residual-qv-stabilization} hold with \(A=R\) and estimand PATE.
\item[(iii)] Each matching support is compact, convex, and two-dimensional. On each support the density in the matched treatment arm is bounded above and away from zero, and the density in the opposite treatment arm is bounded above. Survey weights are bounded above and away from zero.
\item[(iv)] Each \(m_z\) is Lipschitz on its matching support. For some \(\delta>0\), the observed reduced-score residuals have finite \((2+\delta)\)-moments and
\[
E_s\!\left[
\left|w_Z(\mathbf X)\{\tau_{\mathbf D}(\mathbf X)-\tau_{\mathrm{PATE}}\}\right|^{2+\delta}
\right]<\infty.
\]
\end{enumerate}
\end{assumption}
\label{retrospective-pate-with-exact-within-match-survey-weight-normalized-imputation}

\subsubsection{Estimator and reuse coefficients}\label{exact-estimator-and-exact-survey-weighted-matched-set-coefficient}

For each individual \(i\) with \(Z_i=0\), let \(\mathcal J_{\mathbf D_1^*}(i)\) be the set of its \(M\) nearest treated individuals in Euclidean distance on \(\mathbf D_1^*\). For \(Z_i=1\), let \(\mathcal J_{\mathbf D_0^*}(i)\) be the set of its \(M\) nearest control individuals on \(\mathbf D_0^*\). Matching is with replacement across individuals and uses a fixed outcome-independent tie rule. We define the survey-weighted within-match coefficients by
\[\lambda_{i\to j}^{R,(1)}
:=
\frac{w_1(\mathbf X_j)}{\sum_{k\in\mathcal J_{\mathbf D_1^*}(i)} w_1(\mathbf X_k)}\,\mathbf 1\{j\in\mathcal J_{\mathbf D_1^*}(i)\},\]
for \(Z_i=0\), and \[\lambda_{i\to j}^{R,(0)}
:=
\frac{w_0(\mathbf X_j)}{\sum_{k\in\mathcal J_{\mathbf D_0^*}(i)} w_0(\mathbf X_k)}\,\mathbf 1\{j\in\mathcal J_{\mathbf D_0^*}(i)\},\]
for \(Z_i=1\).

The corresponding survey-weighted imputed counterfactuals are
\[\widehat Y_i^{R}(1)
:=
\begin{cases}
Y_i,& Z_i=1,\\[4pt]
\sum_{j=1}^n \lambda_{i\to j}^{R,(1)}Y_j,& Z_i=0,
\end{cases}\] and \[\widehat Y_i^{R}(0)
:=
\begin{cases}
\sum_{j=1}^n \lambda_{i\to j}^{R,(0)}Y_j,& Z_i=1,\\[4pt]
Y_i,& Z_i=0.
\end{cases}\] The retrospective PATE WDSM estimator is
\[\widehat\tau_{\mathrm{PATE}}^{R}
:=
\frac{\sum_{i=1}^n w_{Z_i}(\mathbf X_i)\{\widehat Y_i^{R}(1)-\widehat Y_i^{R}(0)\}}{\sum_{i=1}^n w_{Z_i}(\mathbf X_i)}.\]
We denote the retrospective PATE denominator average by
\[\bar w_{R,\mathrm{PATE},n}
:=
\frac1n\sum_{i=1}^n w_{Z_i}(\mathbf X_i).\]

Reindexing the matched contributions leads to the weighted reuse coefficients
\[K_{n,1}^{R,\mathrm{PATE}}(j)
:=
\sum_{i:Z_i=0} w_0(\mathbf X_i)
\frac{w_1(\mathbf X_j)}{\sum_{k\in\mathcal J_{\mathbf D_1^*}(i)}w_1(\mathbf X_k)}\,
\mathbf 1\{j\in\mathcal J_{\mathbf D_1^*}(i)\},\] and
\[K_{n,0}^{R,\mathrm{PATE}}(j)
:=
\sum_{i:Z_i=1} w_1(\mathbf X_i)
\frac{w_0(\mathbf X_j)}{\sum_{k\in\mathcal J_{\mathbf D_0^*}(i)}w_0(\mathbf X_k)}\,
\mathbf 1\{j\in\mathcal J_{\mathbf D_0^*}(i)\}.\]

For the estimator decomposition, we write
\[H_{n,R,\mathrm{PATE}}
:=
\frac{1}{n\bar w_{R,\mathrm{PATE},n}}
\sum_{i=1}^n w_{Z_i}(\mathbf X_i)\{\tau_{\mathbf D}(\mathbf X_i)-\tau_{\mathrm{PATE}}\},\]
\[\begin{aligned}
&B_{n,R,\mathrm{PATE}}\\
&:=
\frac{1}{n\bar w_{R,\mathrm{PATE},n}}
\sum_{i=1}^n w_{Z_i}(\mathbf X_i)(2Z_i-1)
\left[
m_{1-Z_i}\{\mathbf D_{1-Z_i}^*(\mathbf X_i)\}-\sum_{j=1}^n \lambda_{i\to j}^{R,(1-Z_i)}m_{1-Z_i}\{\mathbf D_{1-Z_i}^*(\mathbf X_j)\}
\right],
\end{aligned}\] and \[E_{n,R,\mathrm{PATE}}
:=
\frac{1}{n\bar w_{R,\mathrm{PATE},n}}
\sum_{j=1}^n(2Z_j-1)
\{w_{Z_j}(\mathbf X_j)+K_{n,Z_j}^{R,\mathrm{PATE}}(j)\}
\varepsilon_{Z_j,j}.\]

\subsubsection{Known-score asymptotics}
\label{exact-decomposition}
\label{exact-conditional-residual-variance}
\label{exact-known-score-asymptotics}

For a bias estimate \(\widehat B_{n,R,\mathrm{PATE}}\) on the same oracle match graph, define
\[
\widehat\tau_{\mathrm{PATE}}^{R,\mathrm{BC}}
:=\widehat\tau_{\mathrm{PATE}}^{R}
-\widehat B_{n,R,\mathrm{PATE}}.
\]

The heterogeneity variance is
\[V_{H,R,\mathrm{PATE}}
=
\frac{\operatorname{Var}_s\!\left(
 w_Z(\mathbf X)
 \bigl\{
 m_1\{\mathbf D_1^*(\mathbf X)\}
 -m_0\{\mathbf D_0^*(\mathbf X)\}
 -\tau_{\mathrm{PATE}}
 \bigr\}
\right)}{\left(E_s[w_Z(\mathbf X)]\right)^2}.
\]

\begin{theorem}[Retrospective PATE with known scores]
\label{thm:retrospective-pate-oracle}
\label{cor:retrospective-pate-score-variance}
Under Assumption~\ref{ass:retrospective-pate-oracle},
\[
\begin{pmatrix}
\sqrt n\,H_{n,R,\mathrm{PATE}}\\
\sqrt n\,E_{n,R,\mathrm{PATE}}
\end{pmatrix}
\Rightarrow
N\!\left[
\begin{pmatrix}0\\0\end{pmatrix},
\begin{pmatrix}
V_{H,R,\mathrm{PATE}}&0\\
0&V_{E,R,\mathrm{PATE}}
\end{pmatrix}
\right].
\]
The matching discrepancy satisfies \(B_{n,R,\mathrm{PATE}}=O_p(n^{-1/2})\), and the uncorrected estimator is consistent. If the correction on the same oracle scores and match graph additionally satisfies
\[\sqrt n\left(B_{n,R,\mathrm{PATE}}-\widehat B_{n,R,\mathrm{PATE}}\right)\to_p 0,\]
then
\[\sqrt n\left(\widehat\tau_{\mathrm{PATE}}^{R,\mathrm{BC}}-\tau_{\mathrm{PATE}}\right)
\Rightarrow
N\!\left(0,V_{H,R,\mathrm{PATE}}+V_{E,R,\mathrm{PATE}}\right).\]
Here the total variance is finite and nonnegative. It is strictly positive if \(V_{H,R,\mathrm{PATE}}>0\) or if either observed-arm residual variance is positive with positive selected-sample probability, by Proposition~\ref{prop:primitive-residual-qv-limit}. Otherwise the normal law is interpreted as degenerate when the total variance is zero. Corollary~\ref{cor:oracle-regression-bc} gives sufficient derivative-consistency conditions for this remainder.

\label{lem:retrospective-pate-hbe}
\label{lem:retrospective-pate-coefficient-rewrite}
\label{lem:retrospective-pate-conditional-variance}
\label{lem:retrospective-pate-heterogeneity-clt}
\label{lem:retrospective-pate-qv}
\end{theorem}
\begin{proof}
\emph{Decomposition.} Using Proposition~\ref{prop:exact-pate-score-space-decomposition} with \(A=R\) and the coefficients above, we separate the matching discrepancy from the two centered terms:
\[
\begin{aligned}
\widehat\tau_{\mathrm{PATE}}^{R}-\tau_{\mathrm{PATE}}
&=H_{n,R,\mathrm{PATE}}+B_{n,R,\mathrm{PATE}}+E_{n,R,\mathrm{PATE}},\\
\sqrt n\{\widehat\tau_{\mathrm{PATE}}^{R}
              -\tau_{\mathrm{PATE}}-B_{n,R,\mathrm{PATE}}\}
&=\sqrt n(H_{n,R,\mathrm{PATE}}+E_{n,R,\mathrm{PATE}}).
\end{aligned}
\]

\emph{Martingale representation.} We construct a two-stage martingale representation using the survey-weighted reuse coefficients, following Abadie and Imbens\citep{AbadieImbens2012}. Population recovery and Theorem~\ref{thm:four-scenario-score-identification} first give centering and a positive denominator limit:
\[
E_s[w_Z(\mathbf X)\{\tau_{\mathbf D}(\mathbf X)-\tau_{\mathrm{PATE}}\}]=0,\qquad
\bar w_{R,\mathrm{PATE},n}\longrightarrow_p E_s\{w_Z(\mathbf X)\}>0.
\]
We condition on
\[
\mathcal G_{R,n}^{\mathrm{match}}
=\sigma\{Z_i,\mathbf D_0^*(\mathbf X_i),\mathbf D_1^*(\mathbf X_i),
w_{Z_i}(\mathbf X_i):1\le i\le n\}.
\]
This field determines the oracle match sets and coefficients. For \(Z_j=z\), let
\[
v_{z,j}:=E_s\!\left[\varepsilon_{z,j}^2\mid
Z_j=z,\mathbf D_0^*(\mathbf X_j),\mathbf D_1^*(\mathbf X_j),
w_z(\mathbf X_j)\right].
\]
Then
\[E_s\!\left[E_{n,R,\mathrm{PATE}}\mid \mathcal G_{R,n}^{\mathrm{match}}\right]=0,\]
and
\[\operatorname{Var}_s\!\left(\sqrt n\,E_{n,R,\mathrm{PATE}}\mid \mathcal G_{R,n}^{\mathrm{match}}\right)
=
\frac{1}{n\bar w_{R,\mathrm{PATE},n}^2}
\sum_{z=0}^1\sum_{j:Z_j=z}
\{w_z(\mathbf X_j)+K_{n,z}^{R,\mathrm{PATE}}(j)\}^{2}v_{z,j}.\]
The observed residuals are conditionally independent and centered given \(\mathcal G_{R,n}^{\mathrm{match}}\). Since \(H_{n,R,\mathrm{PATE}}\) is measurable in this field,
\[
E_s(H_{n,R,\mathrm{PATE}}E_{n,R,\mathrm{PATE}})
=E_s\!\left[H_{n,R,\mathrm{PATE}}E_s(E_{n,R,\mathrm{PATE}}\mid\mathcal G_{R,n}^{\mathrm{match}})\right]=0.
\]
To establish the joint limit, for fixed \(a,b\in\mathbb R\) we define the \(2n\) increments
\[
\begin{aligned}
\xi_{n,i}(a,b)
&=\frac{a}{\sqrt n\,E_s\{w_Z(\mathbf X)\}}\,
 w_{Z_i}(\mathbf X_i)\{\tau_{\mathbf D}(\mathbf X_i)-\tau_{\mathrm{PATE}}\},\qquad 1\le i\le n,\\
\xi_{n,n+i}(a,b)
&=\frac{b}{\sqrt n\,E_s\{w_Z(\mathbf X)\}}
 (2Z_i-1)\{w_{Z_i}(\mathbf X_i)+K_{n,Z_i}^{R,\mathrm{PATE}}(i)\}
 \varepsilon_{Z_i,i},
 \qquad 1\le i\le n.
\end{aligned}
\]
We use \(\mathcal F_{n,0}=\{\varnothing,\Omega\}\) and
\[
\mathcal F_{n,k}=
\begin{cases}
\sigma\{Z_j,\mathbf D_0^*(\mathbf X_j),\mathbf D_1^*(\mathbf X_j),w_{Z_j}(\mathbf X_j):1\le j\le k\},
   &1\le k\le n,\\
\mathcal G_{R,n}^{\mathrm{match}}\vee\sigma(Y_1,\ldots,Y_{k-n}),
   &n<k\le2n.
\end{cases}
\]
The coefficients in the second block are \(\mathcal F_{n,n}\)-measurable. Conditional centering therefore gives
\[
\begin{aligned}
E_s\{\xi_{n,i}(a,b)\mid\mathcal F_{n,i-1}\}
&=\frac{a}{\sqrt n\,E_s\{w_Z(\mathbf X)\}}E_s[w_Z(\mathbf X)\{\tau_{\mathbf D}(\mathbf X)-\tau_{\mathrm{PATE}}\}]=0,\\
E_s\{\xi_{n,n+i}(a,b)\mid\mathcal F_{n,n+i-1}\}
&=0,\\
\sum_{k=1}^{2n}\xi_{n,k}(a,b)
&=\frac{\bar w_{R,\mathrm{PATE},n}}{E_s\{w_Z(\mathbf X)\}}
  \{a\sqrt n\,H_{n,R,\mathrm{PATE}}+b\sqrt n\,E_{n,R,\mathrm{PATE}}\}+o_p(1).
\end{aligned}
\]

\emph{Conditional variances and Lindeberg condition.} The first block consists of independent centered observations, so
\[
\sum_{i=1}^n E_s\{\xi_{n,i}(a,b)^2\mid\mathcal F_{n,i-1}\}
=a^2V_{H,R,\mathrm{PATE}}.
\]
Proposition~\ref{prop:primitive-residual-qv-limit}, with the coefficients and conditional variances displayed above, gives
\[
\begin{aligned}
\operatorname{Var}_s(\sqrt n\,E_{n,R,\mathrm{PATE}}\mid \mathcal G_{R,n}^{\mathrm{match}})
&\longrightarrow_{L^1}V_{E,R,\mathrm{PATE}}\\
&=\frac{\Lambda_{R,0}^{\mathrm{PATE}}
       +\Lambda_{R,1}^{\mathrm{PATE}}}{(E_s\{w_Z(\mathbf X)\})^2}<\infty,\\
\sum_{i=1}^n E_s\{\xi_{n,n+i}(a,b)^2\mid\mathcal F_{n,n+i-1}\}
&=b^2\frac{\bar w_{R,\mathrm{PATE},n}^{\,2}}{(E_s\{w_Z(\mathbf X)\})^2}
       \operatorname{Var}_s(\sqrt n\,E_{n,R,\mathrm{PATE}}\mid \mathcal G_{R,n}^{\mathrm{match}})\\
&\longrightarrow_p b^2V_{E,R,\mathrm{PATE}}.
\end{aligned}
\]
For every \(\epsilon>0\), the finite second moment of the heterogeneity numerator gives
\[
\begin{aligned}
&\sum_{i=1}^n
 E_s\!\left[\xi_{n,i}(a,b)^2
 \mathbf1\{|\xi_{n,i}(a,b)|>\epsilon\}
 \mid\mathcal F_{n,i-1}\right]\\
&=\frac{a^2}{(E_s\{w_Z(\mathbf X)\})^2}
 E_s\!\left[
 (w_Z(\mathbf X)\{\tau_{\mathbf D}(\mathbf X)-\tau_{\mathrm{PATE}}\})^2
 \mathbf1\{|aw_Z(\mathbf X)\{\tau_{\mathbf D}(\mathbf X)-\tau_{\mathrm{PATE}}\}|>\epsilon\sqrt n\,E_s\{w_Z(\mathbf X)\}\}
 \right]\\
&\longrightarrow0.
\end{aligned}
\]
For the residual block, \eqref{eq:qv-derived-lyapunov} and \(\bar w_{R,\mathrm{PATE},n}\to_p E_s\{w_Z(\mathbf X)\}\) imply
\[
\begin{aligned}
&\sum_{i=1}^n E_s\!\left[
 \xi_{n,n+i}(a,b)^2\mathbf1\{|\xi_{n,n+i}(a,b)|>\epsilon\}
 \mid\mathcal F_{n,n+i-1}\right]\\
&\le\epsilon^{-\delta}
 \sum_{i=1}^n E_s\{|\xi_{n,n+i}(a,b)|^{2+\delta}
                 \mid\mathcal F_{n,n+i-1}\}\\
&\longrightarrow_p0.
\end{aligned}
\]
The martingale central limit theorem (Corollary~3.1 and the constant-variance remark in Hall and Heyde,\citep{HallHeyde1980} pp.~58--59) and the Cram\'er--Wold device give
\[
\begin{pmatrix}\sqrt n\,H_{n,R,\mathrm{PATE}}\\\sqrt n\,E_{n,R,\mathrm{PATE}}\end{pmatrix}
\Rightarrow
N\!\left(0,
\begin{pmatrix}
V_{H,R,\mathrm{PATE}}&0\\
0&V_{E,R,\mathrm{PATE}}
\end{pmatrix}\right).
\]
Combining this joint limit, which allows zero-variance coordinates, with the estimator decomposition proves the bias-centered limit.

\emph{Matching discrepancy and bias correction.} Lemma~\ref{lem:oracle-score-radius}, applied to both matching directions, and the positive limit of the outer denominator give \(B_{n,R,\mathrm{PATE}}=O_p(n^{-1/2})=o_p(1)\). The joint limit above implies \(H_{n,R,\mathrm{PATE}}+E_{n,R,\mathrm{PATE}}=o_p(1)\). The exact decomposition therefore proves consistency of the uncorrected estimator. Under the stated bias-correction remainder,
\[
\begin{aligned}
\sqrt n\{\widehat\tau_{\mathrm{PATE}}^{R,\mathrm{BC}}
                  -\tau_{\mathrm{PATE}}\}
&=\sqrt n(H_{n,R,\mathrm{PATE}}+E_{n,R,\mathrm{PATE}})
 +\sqrt n(B_{n,R,\mathrm{PATE}}-\widehat B_{n,R,\mathrm{PATE}})\\
&=\sqrt n(H_{n,R,\mathrm{PATE}}+E_{n,R,\mathrm{PATE}})+o_p(1).
\end{aligned}
\]
Slutsky's theorem proves the bias-corrected limit. The second-moment assumptions and Proposition~\ref{prop:primitive-residual-qv-limit} give the stated finiteness and positivity conclusions.
\end{proof}

\subsubsection{Estimated-score asymptotics}\label{retrospective-pate-estimated-score-results}

Proposition~\ref{cor:primary-fitted-score-consistency} establishes consistency under its stated conditions, with the bias-correction remainder evaluated on the estimated-score match sets.

For the retrospective PATE scenario, we write the oracle variance as
\[
\begin{aligned}
V_{R,\mathrm{PATE}}^{\mathrm{known\text{-}score}}
&:=V_{H,R,\mathrm{PATE}}+V_{E,R,\mathrm{PATE}}.
\end{aligned}
\]
Let \(\widehat{\boldsymbol \theta}_{R,\mathrm{PATE}}\) collect the retrospective survey-weighted propensity-score estimator and the two unweighted arm-specific prognostic-score estimators. Their stacked estimating function is
\[
U_{R,\mathrm{PATE}}(O;\boldsymbol\theta)
=
\begin{pmatrix}
w_Z(\mathbf X)\mathbf X\{Z-e(\mathbf X;\boldsymbol\alpha)\}\\
(1-Z)\mathbf X\{Y-\mathbf X^\top\boldsymbol\beta_0\}\\
Z\mathbf X\{Y-\mathbf X^\top\boldsymbol\beta_1\}
\end{pmatrix},
\qquad
\boldsymbol\theta=
(\boldsymbol\alpha^\top,\boldsymbol\beta_0^\top,
 \boldsymbol\beta_1^\top)^\top.
\]
The probability limit may be a pseudo-true parameter. Let \(\bar{\boldsymbol \theta}_{R,\mathrm{PATE},n,\eta}\) denote the midpoint discretization on cubes of side length \(\eta/\sqrt n\), with deterministic grid offset.

The conditions in Lemma~\ref{lem:primitive-logistic-linear-roots} give the first-stage parameter expansion in Lemma~\ref{lem:ordinary-weighted-first-step}. The following theorem gives the resulting estimated-score limit under the matching conditions.

\begin{theorem}[Retrospective PATE with estimated scores]
\label{thm:retrospective-pate-estimated-scores}
\label{par:retrospective-pate-two-arm-sampling}
\emph{(i)} Under all conditions of Theorem~\ref{thm:pate-two-arm-actual-joint-law}, with \(A=R\), including empirical normalization and the distinct deterministic-local and actual-graph BC remainders,
\[
\begin{pmatrix}
\sqrt n\{\widehat\tau_{\mathrm{PATE}}^{R,\mathrm{BC}}
              (\widehat{\boldsymbol \theta}_R)-\tau_{\mathrm{PATE}}\}\\
\sqrt n(\widehat{\boldsymbol \theta}_R-\boldsymbol\theta_R^*)
\end{pmatrix}
\Rightarrow
\begin{pmatrix}G_{\tau,R}+\boldsymbol b_{R,\mathrm{loc}}^\top \mathbf G_{\theta,R}\\\mathbf G_{\theta,R}\end{pmatrix},
\]
where \(\boldsymbol b_{R,\mathrm{loc}}\) is defined by \eqref{eq:pate-two-arm-actual-drift}. Consequently the fitted-estimator variance is
\[
\begin{aligned}
V_{R,\mathrm{actual}}
&=V_{R,\mathrm{known\text{-}score}}
 -\boldsymbol\gamma_{R,1}^\top\boldsymbol\Sigma_{R,U}^{-1}\boldsymbol\gamma_{R,1}\\
&\quad{}+(\boldsymbol b_{R,\mathrm{loc}}-\boldsymbol\Gamma_R^\top\boldsymbol\Sigma_{R,U}^{-1}\boldsymbol\gamma_{R,1})^\top
 \boldsymbol\Sigma_{R,\theta}
 (\boldsymbol b_{R,\mathrm{loc}}-\boldsymbol\Gamma_R^\top\boldsymbol\Sigma_{R,U}^{-1}\boldsymbol\gamma_{R,1}).
\end{aligned}
\]
\emph{(ii)} Under the full-mean and all other conditions of Theorem~\ref{thm:pate-fullx-actual-joint-law}, the corresponding joint limit is instead \((G_{\tau,R},\mathbf G_{\theta,R})\), with \(\operatorname{Var}(G_{\tau,R})=V_{R,\mathrm{known\text{-}score}}\).
\end{theorem}
\begin{proof}
For part (i), set \(A=R\) in Theorem~\ref{thm:pate-two-arm-actual-joint-law}, using the estimating equations specified above. This gives the stated joint limit with \(\boldsymbol b_{R,\mathrm{loc}}\) from \eqref{eq:pate-two-arm-actual-drift}. Its covariance is
\[
\operatorname{Cov}(\mathbf G_{\theta,R},G_{\tau,R})
=-\boldsymbol\Gamma_R^{-1}\boldsymbol\gamma_{R,1},\qquad
\operatorname{Var}(\mathbf G_{\theta,R})=\boldsymbol\Sigma_{R,\theta}.
\]
Consequently,
\[
\begin{aligned}
\operatorname{Var}(G_{\tau,R}+\boldsymbol b_{R,\mathrm{loc}}^\top \mathbf G_{\theta,R})
&=V_{R,\mathrm{known\text{-}score}}
 -2\boldsymbol b_{R,\mathrm{loc}}^\top\boldsymbol\Gamma_R^{-1}\boldsymbol\gamma_{R,1}
 +\boldsymbol b_{R,\mathrm{loc}}^\top\boldsymbol\Sigma_{R,\theta}\boldsymbol b_{R,\mathrm{loc}}\\
&=V_{R,\mathrm{actual}},
\end{aligned}
\]
where the last equality follows by expanding the displayed projection form and using \(\boldsymbol\Sigma_{R,\theta}=\boldsymbol\Gamma_R^{-1}\boldsymbol\Sigma_{R,U}\boldsymbol\Gamma_R^{-T}\). For part (ii), apply Theorem~\ref{thm:pate-fullx-actual-joint-law} with \(A=R\) under its full set of conditions.
\end{proof}

Under the pathwise identification and differentiation conditions of Proposition~\ref{prop:scientific-target-path-derivative}, the scenario target and its derivative can be written as
\[
\tau_A(Q_{A,t})
=\frac{E_{Q_{A,t}}[w_Z(\mathbf X)\{\mu_{1,t}(\mathbf X)-\mu_{0,t}(\mathbf X)\}]}
       {E_{Q_{A,t}}[w_Z(\mathbf X)]},
\]
and
\[
\boldsymbol\gamma_{R,\mathrm{PATE},2}
=E_s\{\phi_R(O)\mathbf s_A(O)\},
\qquad \mathbf s_A(O)=-\boldsymbol\Gamma_A^\top\boldsymbol\Sigma_{A,U}^{-1}\mathbf U_A(O;\boldsymbol\theta_A^\ast),
\]
with the weight functions fixed along the path. Define
\[
V_{R,\mathrm{PATE}}^{\mathrm{fix}}
=V_{R,\mathrm{PATE}}^{\mathrm{known\text{-}score}}
-\boldsymbol\gamma_{R,\mathrm{PATE},1}^{\top}
 \boldsymbol\Sigma_{A,U}^{-1}\boldsymbol\gamma_{R,\mathrm{PATE},1},
\]
and
\[
V_{R,\mathrm{PATE}}^{\mathrm{grid}}
=V_{R,\mathrm{PATE}}^{\mathrm{fix}}
+\widetilde{\boldsymbol \gamma}_{R,\mathrm{PATE},2}^\top
 \boldsymbol\Sigma_{A,\theta}\widetilde{\boldsymbol \gamma}_{R,\mathrm{PATE},2}.
\]

\begin{proposition}[Discretized-score approximation]
\label{prop:retrospective-pate-conditional-score-calculation}
For PATE, put \(A=R\). Retain the retrospective PATE oracle conditions and the first-step and auxiliary-experiment conditions of Lemma~\ref{lem:auxiliary-godambe-experiment}, including \(\boldsymbol\Sigma_{A,U}>0\). Suppose that \eqref{eq:moving-law-joint-input-with-bias} holds for the specified standardization, matching and bias-correction family at every deterministic \(\mathbf h_n\to \mathbf h\), with
\[
V_{A,\mathrm{known\text{-}score}}
=V_{R,\mathrm{PATE}}^{\mathrm{known\text{-}score}}
=V_{H,R,\mathrm{PATE}}+V_{E,R,\mathrm{PATE}},\qquad
\boldsymbol\gamma_{A,1}=\boldsymbol\gamma_{R,\mathrm{PATE},1},\quad
\widetilde{\boldsymbol \gamma}_{A,2}=\widetilde{\boldsymbol \gamma}_{R,\mathrm{PATE},2}.
\]
Suppose also that \(\tau_A(Q_{A,t})\) is identified and differentiable along these same auxiliary laws, with
\[
\begin{aligned}
\tau_A(Q_{A,t})
&=\tau_{\mathrm{PATE}}+\boldsymbol\gamma_{R,\mathrm{PATE},2}^{\top}t+o(\|t\|),\\
\boldsymbol\gamma_{R,\mathrm{PATE},2}
&:=\left.\frac{\partial}{\partial t}\tau_A(Q_{A,t})\right|_{t=0}.
\end{aligned}
\]
Then, in the order \(n\to\infty\), \(\eta\downarrow0\),
\[
\sqrt n\{\widehat\tau_{\mathrm{PATE}}^{R,\mathrm{BC}}
(\bar{\boldsymbol \theta}_{A,n,\eta})-\tau_{\mathrm{PATE}}\}
\Rightarrow N(0,V_{R,\mathrm{PATE}}^{\mathrm{grid}}).
\]
The double limit is in the bounded-Lipschitz sense of Proposition~\ref{prop:local-mean-corrected-grid-variance}.
\end{proposition}
\begin{proof}
Applying Proposition~\ref{prop:local-mean-corrected-grid-variance} with the constants above gives the grid limit \(G_{\tau,A}^\perp+\widetilde{\boldsymbol \gamma}_{A,2}^\top \mathbf G_{\theta,A}\), where \(G_{\tau,A}^\perp\perp \mathbf G_{\theta,A}\). By Lemma~\ref{lem:ordinary-weighted-first-step}, \(\boldsymbol\Sigma_{A,\theta}=\boldsymbol\Gamma_A^{-1}\boldsymbol\Sigma_{A,U}\boldsymbol\Gamma_A^{-T}\). Independence therefore gives
\[
\begin{aligned}
&\operatorname{Var}\{G_{\tau,A}^\perp+\widetilde{\boldsymbol \gamma}_{A,2}^\top \mathbf G_{\theta,A}\}\\
&=\operatorname{Var}(G_{\tau,A}^\perp)
 +\widetilde{\boldsymbol \gamma}_{A,2}^\top\boldsymbol\Sigma_{A,\theta}\widetilde{\boldsymbol \gamma}_{A,2}\\
&=V_{R,\mathrm{PATE}}^{\mathrm{fix}}
 +\widetilde{\boldsymbol \gamma}_{R,\mathrm{PATE},2}^\top
 \boldsymbol\Sigma_{A,\theta}\widetilde{\boldsymbol \gamma}_{R,\mathrm{PATE},2}\\
&=V_{R,\mathrm{PATE}}^{\mathrm{grid}}.
\end{aligned}
\]
The target representer follows from Proposition~\ref{prop:scientific-target-path-derivative} with weight \(w_Z(\mathbf X)\), under its stated identification and differentiation conditions.
\end{proof}

\subsubsection{Fixed-match replication}\label{retrospective-pate-bootstrap-main-text}

For each fixed-match replicate \(b=1,\ldots,B\), we draw
\[
(m_1^{\ast,b},\ldots,m_n^{\ast,b})
\sim
\mathrm{Multinomial}(n;1/n,\ldots,1/n),
\]
and define the replicated retrospective survey weight
\[
w_i^{\ast,b}:=m_i^{\ast,b}w_{Z_i}(\mathbf X_i).
\]
We refit the retrospective propensity score with weights \(w_i^{\ast,b}\) and the arm-specific prognostic scores with weights \(m_i^{\ast,b}\), using the same estimating equations as in the original sample. These refits determine \(\widehat{\mathbf D}_z^{\ast,b}\), including standardization by the original rule. Within arm \(z\), we refit the bias-correction regression with weights \(w_i^{\ast,b}\) on these replicate-specific scores and denote the fitted reduced mean by \(\widehat m_z^{\ast,b}\). The original match sets, their normalized within-match weights \(\lambda_{i\to j}^{R,(z)}\), and hence \(K_{n,z}^{R,\mathrm{PATE}}(j)\), remain fixed at the original estimated-score graph.

The retrospective PATE fixed-match replicate is
\begin{equation}\label{eq:retrospective-pate-fixed-match-replicate}
\begin{aligned}
\widehat\tau_{\mathrm{PATE}}^{R,\ast,b}
=
\frac{1}{\sum_{i=1}^n w_i^{\ast,b}}
\sum_{i=1}^n
\Big[
& w_i^{\ast,b}
\Big\{
\widehat m_1^{\ast,b}
 \{\widehat{\mathbf D}_1^{\ast,b}(\mathbf X_i)\}
-
\widehat m_0^{\ast,b}
 \{\widehat{\mathbf D}_0^{\ast,b}(\mathbf X_i)\}
\Big\}\\
&+
m_i^{\ast,b}(2Z_i-1)
\Bigl\{
w_{Z_i}(\mathbf X_i)+K_{n,Z_i}^{R,\mathrm{PATE}}(i)
\Bigr\}\\
&\qquad{}\times
\Bigl[
Y_i-\widehat m_{Z_i}^{\ast,b}
 \{\widehat{\mathbf D}_{Z_i}^{\ast,b}(\mathbf X_i)\}
\Bigr]
\Big].
\end{aligned}
\end{equation}
The centering estimator \(\widehat\tau_{\mathrm{PATE}}^{R,\mathrm{BC}}\) uses the same graph and fitted means \(\widehat m_z\{\widehat{\mathbf D}_z(\mathbf X)\}\). The identities below hold when the original and replicated fits exist and the replicated denominator is positive.

We denote the original and replicated normalized denominators by
\[
\bar w_{R,\mathrm{PATE},n}
=\frac1n\sum_{i=1}^n w_{Z_i}(\mathbf X_i),
\qquad
\bar w_{R,\mathrm{PATE},n}^{\ast,b}
=\frac1n\sum_{i=1}^n m_i^{\ast,b}w_{Z_i}(\mathbf X_i).
\]
The centered contribution of individual \(i\) is
\begin{equation}\label{eq:retrospective-pate-fitted-centered-row}
\begin{aligned}
\widehat\varphi_{R,\mathrm{PATE},i}
:={}&w_{Z_i}(\mathbf X_i)
\Big[
\widehat m_1\{\widehat{\mathbf D}_1(\mathbf X_i)\}
-\widehat m_0\{\widehat{\mathbf D}_0(\mathbf X_i)\}
-\widehat\tau_{\mathrm{PATE}}^{R,\mathrm{BC}}
\Big]\\
&+(2Z_i-1)
\{w_{Z_i}(\mathbf X_i)+K_{n,Z_i}^{R,\mathrm{PATE}}(i)\}\\
&\qquad{}\times
\Big[Y_i-\widehat m_{Z_i}
 \{\widehat{\mathbf D}_{Z_i}(\mathbf X_i)\}\Big].
\end{aligned}
\end{equation}

The full refit contribution is
\begin{equation}\label{eq:retrospective-pate-bootstrap-refit-remainder}
\begin{aligned}
\mathcal R_{R,\mathrm{PATE},n}^{\ast,b}
={}&\sum_{z=0}^1(2z-1)
\Bigg[
\sum_{i:Z_i=1-z}m_i^{\ast,b}w_{1-z}(\mathbf X_i)\\
&\quad{}\times
\Big[
\widehat m_z^{\ast,b}
 \{\widehat{\mathbf D}_z^{\ast,b}(\mathbf X_i)\}
-\widehat m_z\{\widehat{\mathbf D}_z(\mathbf X_i)\}
\Big]\\
&-
\sum_{j:Z_j=z}m_j^{\ast,b}
 K_{n,z}^{R,\mathrm{PATE}}(j)\\
&\quad{}\times
\Big[
\widehat m_z^{\ast,b}
 \{\widehat{\mathbf D}_z^{\ast,b}(\mathbf X_j)\}
-\widehat m_z\{\widehat{\mathbf D}_z(\mathbf X_j)\}
\Big]
\Bigg].
\end{aligned}
\end{equation}

\begin{proposition}[Fixed-match replication for retrospective PATE]
\label{prop:retrospective-pate-fixed-match-identity}
Then \(\sum_i\widehat\varphi_{R,\mathrm{PATE},i}=0\) exactly, and
\begin{equation}\label{eq:retrospective-pate-bootstrap-exact-ratio}
\begin{aligned}
&\sqrt n\Big(
\widehat\tau_{\mathrm{PATE}}^{R,\ast,b}
-\widehat\tau_{\mathrm{PATE}}^{R,\mathrm{BC}}
\Big)\\
&=
\frac{1}{\bar w_{R,\mathrm{PATE},n}^{\ast,b}}
\frac1{\sqrt n}\sum_{i=1}^n
(m_i^{\ast,b}-1)\widehat\varphi_{R,\mathrm{PATE},i}
+
\frac{\mathcal R_{R,\mathrm{PATE},n}^{\ast,b}}
{\sqrt n\,\bar w_{R,\mathrm{PATE},n}^{\ast,b}},
\end{aligned}
\end{equation}

Let \(\mathcal F_n\) contain the original observations, fits, and graph. The conditional variance of the centered multiplier numerator is exactly
\begin{equation}\label{eq:retrospective-pate-bootstrap-numerator-variance}
\operatorname{Var}^{\ast}\left(
\frac1{\sqrt n}\sum_{i=1}^n
(m_i^{\ast,b}-1)\widehat\varphi_{R,\mathrm{PATE},i}
\,\middle|\,\mathcal F_n
\right)
=
\frac1n\sum_{i=1}^n\widehat\varphi_{R,\mathrm{PATE},i}^{\,2}.
\end{equation}

If, in addition, all conditions of Proposition~\ref{prop:pate-full-prediction-replication-law} hold with \(A=R\), then
\[
\mathcal L^\ast\!\left[
 \sqrt n(\widehat\tau_{\mathrm{PATE}}^{R,\ast,b}-\widehat\tau_{\mathrm{PATE}}^{R,\mathrm{BC}})
 \,\middle|\,\mathcal F_n\right]
 \Longrightarrow_{P_s}N(0,V_{R,\mathrm{PATE},\mathrm{rep}}),
\]
where
\[
\begin{aligned}
V_{R,\mathrm{PATE},\mathrm{rep}}
&=V_{R,\mathrm{PATE},\mathrm{known\text{-}score}}
 -\frac{2}{\bar w_{R,\mathrm{PATE}}}\,
 (\ell_{R,1}-\ell_{R,0})^\top\boldsymbol\Gamma_{R,\mathrm{PATE}}^{-1}\boldsymbol\gamma_{R,\mathrm{PATE},1}\\
&\quad+\frac1{\bar w_{R,\mathrm{PATE}}^2}\,
 (\ell_{R,1}-\ell_{R,0})^\top\boldsymbol\Sigma_{R,\mathrm{PATE},\boldsymbol\theta}(\ell_{R,1}-\ell_{R,0}).
\end{aligned}
\]
Suppose a sampling limit centered at \(\tau_{\mathrm{PATE}}\) has also been established, and
\[
\sqrt n(\widehat\tau_{\mathrm{PATE}}^{R,\mathrm{BC}}-\tau_{\mathrm{PATE}})
\Rightarrow N(0,V_{R,\mathrm{PATE}}),\qquad
V_{R,\mathrm{PATE}}=V_{R,\mathrm{PATE},\mathrm{rep}}>0.
\]
For conditionally independent replicates with the prescribed finite-valued outputs, the variance estimator in \eqref{eq:bootstrap-mc-original-variance} is consistent in the iterated limit \(n\to\infty\), then \(B\to\infty\), and along a sufficiently slowly increasing deterministic sequence \(B_n\). The associated normal Wald interval has asymptotic coverage \(1-\alpha\), for \(0<\alpha<1\), in the same regimes.
\label{par:retrospective-pate-bootstrap-law-comparison}
\end{proposition}

\begin{proof}
Specialize the PATE ratio representation \eqref{eq:pate-fixed-prediction-ratio} and its centered form \eqref{eq:pate-common-centered-replication-ratio} to \(A=R\) and \(w_{A,z}(\mathbf X_i)=w_z(\mathbf X_i)\). Centering the numerator gives \eqref{eq:retrospective-pate-fitted-centered-row}. Collecting the prediction changes in \eqref{eq:pate-common-fixed-refit-remainder} gives \eqref{eq:retrospective-pate-bootstrap-refit-remainder}. Since the centered rows sum to zero, the multinomial covariance identity \eqref{eq:fixed-match-multinomial-moments} yields \eqref{eq:retrospective-pate-bootstrap-numerator-variance}. Under the additional replication hypotheses, the refit expansion \eqref{eq:pate-full-prediction-refit-first-order} reduces the remainder to
\[
\frac{\mathcal R_{R,\mathrm{PATE},n}^{\ast,b}}{\sqrt n}
=(\ell_{R,1}-\ell_{R,0})^\top
 \sqrt n(\widehat{\boldsymbol \theta}_{R,\mathrm{PATE}}^{\ast,b}
              -\widehat{\boldsymbol \theta}_{R,\mathrm{PATE}})
 +o_{p^\ast}(1).
\]
The joint multiplier law and positive denominator limit give the stated conditional distribution. Under the stipulated positive sampling--replication variance agreement, Proposition~\ref{prop:bootstrap-monte-carlo-iterated-variance} gives variance consistency and Wald coverage in the stated regimes.
\end{proof}

\subsection{Retrospective PATT}\label{app:retrospective-patt}
\subsubsection{Conditions}
For retrospective PATT, the outer average uses the treated survey weight \(w_1(\mathbf X)=\pi_1(\mathbf X)^{-1}\). Matching uses the limiting standardized control double score \(\mathbf D_0^*(\mathbf X)\).

\Needspace{5\baselineskip}
We express both reduced means on the common control score as
\[
\begin{aligned}
m_0^R(\mathbf d)&=E_\Omega\{Y(0)\mid\mathbf D_0^*(\mathbf X)=\mathbf d\},\\
m_{1\mid1}(\mathbf d)&=E_\Omega\{Y(1)\mid Z=1,\mathbf D_0^*(\mathbf X)=\mathbf d\},\\
\tau_{\mathrm{PATT}}(\mathbf d)&=m_{1\mid1}(\mathbf d)-m_0^R(\mathbf d),\\
\varepsilon_{0,i}&=Y_i-m_0^R\{\mathbf D_0^*(\mathbf X_i)\},\qquad Z_i=0,\\
\varepsilon_{1\mid1,i}&=Y_i-m_{1\mid1}\{\mathbf D_0^*(\mathbf X_i)\},\qquad Z_i=1.
\end{aligned}
\]
Thus \(\varepsilon_{0,i}\) is the earlier \(\varepsilon_{0,i}^R\).

\begin{assumption}[Retrospective PATT conditions]
\label{ass:retro-patt-oracle-regularity}
\leavevmode
\begin{enumerate}
\item[(i)] The one-sided treatment condition \(Y(0)\perp Z\mid\mathbf X\) and the retrospective sampling condition in Section~\ref{population-representation-identities-and-score-relations} hold.
\item[(ii)] Either the limiting propensity model is correctly specified and \(\mathbf D_0^*(\mathbf X)\) determines the population propensity, or the limiting control prognostic mean is correctly specified and \(\mathbf D_0^*(\mathbf X)\) determines \(\mu_0(\mathbf X)=E_\Omega\{Y(0)\mid\mathbf X\}\). Standardization is fixed and nonsingular.
\item[(iii)] Assumption~\ref{ass:residual-qv-stabilization} holds for retrospective PATT.
\item[(iv)] The population reduced control mean \(m_0^R\) is Lipschitz on the control-score support, with constant \(L_0<\infty\), and
\[
E_s\!\left[
 Z w_1(\mathbf X)^2
 \left\{\tau_{\mathrm{PATT}}\{\mathbf D_0^*(\mathbf X)\}
                  -\tau_{\mathrm{PATT}}\right\}^{2}
\right]<\infty.
\]
\end{enumerate}
\end{assumption}
\label{retrospective-patt-with-exact-one-sided-within-match-survey-weight-normalized-imputation}

When the propensity coordinate is correct, the balancing argument in Lemma~\ref{lem:armwise-double-score-balancing} applies to \(\mathbf D_0^*\). When the control mean is correct, measurability of \(\mu_0(\mathbf X)\) in \(\mathbf D_0^*(\mathbf X)\) gives instead
\[
\begin{aligned}
E_\Omega\{Y(0)\mid Z,\mathbf D_0^*(\mathbf X)\}
&=E_\Omega\!\left[
 E_\Omega\{Y(0)\mid\mathbf X,Z\}
 \mid Z,\mathbf D_0^*(\mathbf X)\right]\\
&=E_\Omega\{\mu_0(\mathbf X)\mid Z,\mathbf D_0^*(\mathbf X)\}\\
&=\mu_0(\mathbf X)\\
&=m_0^R\{\mathbf D_0^*(\mathbf X)\}.
\end{aligned}
\]
The equalities use the tower property, one-sided ignorability and score measurability. Under either route, the identification calculation in Section~\ref{oracle-score-space-decomposition-behind-the-double-robust-theorems} therefore gives
\[
\tau_{\mathrm{PATT}}
=E_\Omega\!\left[
 \tau_{\mathrm{PATT}}\{\mathbf D_0^*(\mathbf X)\}\mid Z=1\right].
\]
\subsubsection{Estimator and reuse coefficients}\label{exact-estimator-and-exact-survey-weighted-matched-set-coefficient-1}

For each treated individual \(i\), let \(\mathcal J_{\mathbf D_0^*}(i)\) be the set of its \(M\) nearest control individuals in the fixed limiting score metric. The survey-weighted coefficients for these matches are
\[\lambda_{i\to j}^{R,(0)}
:=
\frac{w_0(\mathbf X_j)}{\sum_{k\in\mathcal J_{\mathbf D_0^*}(i)} w_0(\mathbf X_k)}\,\mathbf 1\{j\in\mathcal J_{\mathbf D_0^*}(i)\},
\qquad Z_i=1.\] We then define the survey-weighted imputed control outcome for treated
unit \(i\) by \[\widehat Y_i^{R}(0)
:=
\sum_{j=1}^n \lambda_{i\to j}^{R,(0)}Y_j,
\qquad Z_i=1.\] The retrospective PATT WDSM estimator is
\[\widehat\tau_{\mathrm{PATT}}^{R}
:=
\frac{\sum_{i:Z_i=1} w_1(\mathbf X_i)\{Y_i-\widehat Y_i^{R}(0)\}}{\sum_{i:Z_i=1} w_1(\mathbf X_i)}.\]
We denote the denominator average by \[\bar w_{R,\mathrm{PATT},n}
:=
\frac1n\sum_{i=1}^n Z_iw_1(\mathbf X_i).\] The weighted control reuse coefficient is
\[K_{n,0}^{R,\mathrm{PATT}}(j)
:=
\sum_{i:Z_i=1}w_1(\mathbf X_i)
\frac{w_0(\mathbf X_j)}{\sum_{k\in\mathcal J_{\mathbf D_0^*}(i)}w_0(\mathbf X_k)}\,
\mathbf 1\{j\in\mathcal J_{\mathbf D_0^*}(i)\}.\]

We define the three components by
\[H_{n,R,\mathrm{PATT}}
:=
\frac{1}{n\bar w_{R,\mathrm{PATT},n}}
\sum_{i:Z_i=1} w_1(\mathbf X_i)\{\tau_{\mathrm{PATT}}\{\mathbf D_0^*(\mathbf X_i)\}-\tau_{\mathrm{PATT}}\},\]
\[B_{n,R,\mathrm{PATT}}
:=
\frac{1}{n\bar w_{R,\mathrm{PATT},n}}
\sum_{i:Z_i=1} w_1(\mathbf X_i)
\left[
m_0^R\{\mathbf D_0^*(\mathbf X_i)\}-\sum_{j=1}^n \lambda_{i\to j}^{R,(0)}m_0^R\{\mathbf D_0^*(\mathbf X_j)\}
\right],\] and \[E_{n,R,\mathrm{PATT}}
:=
\frac{1}{n\bar w_{R,\mathrm{PATT},n}}
\left[
\sum_{i:Z_i=1} w_1(\mathbf X_i)\varepsilon_{1\mid1,i}
-
\sum_{j:Z_j=0}K_{n,0}^{R,\mathrm{PATT}}(j)\varepsilon_{0,j}
\right].\]

\subsubsection{Known-score asymptotics}
\label{exact-decomposition-and-exact-conditional-variance}
\label{exact-known-score-asymptotics-1}

For the bias-corrected estimator on the same oracle match graph, define
\[
\widehat\tau_{\mathrm{PATT}}^{R,\mathrm{BC}}
:=\widehat\tau_{\mathrm{PATT}}^{R}
                         -\widehat B_{n,R,\mathrm{PATT}}.
\]

The heterogeneity variance is
\[
V_{H,R,\mathrm{PATT}}
=
\frac{\operatorname{Var}_s\!\left(
 Zw_1(\mathbf X)
 \left\{m_{1\mid1}\{\mathbf D_0^*(\mathbf X)\}
        -m_0^R\{\mathbf D_0^*(\mathbf X)\}
        -\tau_{\mathrm{PATT}}\right\}
\right)}{(E_s[Zw_1(\mathbf X)])^2}.
\]
The one-sided bias discrepancy is
\[
\begin{aligned}
&m_0^R\{\mathbf D_0^*(\mathbf X_i)\}
 -\sum_j\lambda_{i\to j}^{R,(0)}
                         m_0^R\{\mathbf D_0^*(\mathbf X_j)\}\\
&=\sum_j\lambda_{i\to j}^{R,(0)}
 \left[m_0^R\{\mathbf D_0^*(\mathbf X_i)\}
                         -m_0^R\{\mathbf D_0^*(\mathbf X_j)\}\right].
\end{aligned}
\]

\begin{theorem}[Retrospective PATT with known scores]
\label{thm:retro-patt-oracle}
\label{cor:retro-patt-score-variance}
Under Assumption~\ref{ass:retro-patt-oracle-regularity},
\[
\begin{pmatrix}
\sqrt n\,H_{n,R,\mathrm{PATT}}\\
\sqrt n\,E_{n,R,\mathrm{PATT}}
\end{pmatrix}
\Rightarrow
N\!\left[
\begin{pmatrix}0\\0\end{pmatrix},
\begin{pmatrix}
V_{H,R,\mathrm{PATT}}&0\\
0&V_{E,R,\mathrm{PATT}}
\end{pmatrix}
\right].
\]
The matching discrepancy satisfies \(B_{n,R,\mathrm{PATT}}=O_p(n^{-1/2})\), and the uncorrected estimator is consistent. If the correction on the same oracle scores and match graph satisfies
\[
B_{n,R,\mathrm{PATT}}
-\widehat B_{n,R,\mathrm{PATT}}
=o_p(n^{-1/2}),
\]
then this bias-corrected estimator is consistent and
\[
\sqrt n\left(
 \widehat\tau_{\mathrm{PATT}}^{R,\mathrm{BC}}-\tau_{\mathrm{PATT}}\right)
\Rightarrow
N\!\left(0,V_{H,R,\mathrm{PATT}}+V_{E,R,\mathrm{PATT}}\right).
\]
The total variance is finite and nonnegative. It is strictly positive, for example, if the heterogeneity numerator has positive variance or if \(P_s(Z_i=1,v_{1,i}^{\mathrm{PATT}}>0)>0\). Corollary~\ref{cor:oracle-regression-bc} gives sufficient derivative-consistency conditions for this remainder.
\label{lem:retro-patt-decomposition}
\label{lem:retro-patt-reindex}
\label{lem:retro-patt-conditional-variance}
\label{lem:retro-patt-bias-order}
\label{lem:retro-patt-heterogeneity-clt}
\label{lem:retro-patt-qv}
\end{theorem}
\begin{proof}
\emph{Decomposition.} Proposition~\ref{prop:exact-pate-score-space-decomposition} with \(A=R\) and the coefficients above separates the matching discrepancy from the two centered terms:
\[
\begin{aligned}
\widehat\tau_{\mathrm{PATT}}^{R}-\tau_{\mathrm{PATT}}
&=H_{n,R,\mathrm{PATT}}+B_{n,R,\mathrm{PATT}}+E_{n,R,\mathrm{PATT}},\\
\sqrt n\{\widehat\tau_{\mathrm{PATT}}^{R}
              -\tau_{\mathrm{PATT}}-B_{n,R,\mathrm{PATT}}\}
&=\sqrt n(H_{n,R,\mathrm{PATT}}+E_{n,R,\mathrm{PATT}}).
\end{aligned}
\]

\emph{Martingale representation.} Population recovery and Theorem~\ref{thm:four-scenario-score-identification} give
\[
E_s[Zw_1(\mathbf X)\{\tau_{\mathrm{PATT}}\{\mathbf D_0^*(\mathbf X)\}-\tau_{\mathrm{PATT}}\}]=0,\qquad
\bar w_{R,\mathrm{PATT},n}\longrightarrow_p E_s\{Zw_1(\mathbf X)\}>0.
\]
For the residual stage, condition on the initial field
\[
\mathcal G_{R,n}^{\mathrm{match}}
=\sigma\{Z_i,\mathbf D_0^*(\mathbf X_i),w_{Z_i}(\mathbf X_i):1\le i\le n\},
\]
and, for \(Z_i=z\), define the conditional residual variance
\[
v_{z,i}^{\mathrm{PATT}}
:=\operatorname{Var}_s\{Y_i\mid Z_i=z,\mathbf D_0^*(\mathbf X_i),
                                      w_z(\mathbf X_i)\}.
\]
\[
E_s\!\left[E_{n,R,\mathrm{PATT}}
                       \mid\mathcal G_{R,n}^{\mathrm{match}}\right]=0,
\]
and
\[
\operatorname{Var}_s\!\left(\sqrt n E_{n,R,\mathrm{PATT}}
                                      \mid\mathcal G_{R,n}^{\mathrm{match}}\right)
=\frac{1}{n\bar w_{R,\mathrm{PATT},n}^{2}}
\left[
\sum_{i:Z_i=1}w_1(\mathbf X_i)^2v_{1,i}^{\mathrm{PATT}}
+\sum_{j:Z_j=0}K_{n,0}^{R,\mathrm{PATT}}(j)^2
                         v_{0,j}^{\mathrm{PATT}}
\right].
\]
The observed residuals are conditionally independent and centered given \(\mathcal G_{R,n}^{\mathrm{match}}\). Since \(H_{n,R,\mathrm{PATT}}\) is measurable in this field,
\[
E_s(H_{n,R,\mathrm{PATT}}E_{n,R,\mathrm{PATT}})
=E_s\!\left[H_{n,R,\mathrm{PATT}}E_s(E_{n,R,\mathrm{PATT}}\mid\mathcal G_{R,n}^{\mathrm{match}})\right]=0.
\]
To establish the joint limit, for fixed \(a,b\in\mathbb R\) we define the \(2n\) increments
\[
\begin{aligned}
\xi_{n,i}(a,b)
&=\frac{a}{\sqrt n\,E_s\{Zw_1(\mathbf X)\}}\,
 Z_iw_1(\mathbf X_i)\{\tau_{\mathrm{PATT}}\{\mathbf D_0^*(\mathbf X_i)\}-\tau_{\mathrm{PATT}}\},\qquad 1\le i\le n,\\
\xi_{n,n+i}(a,b)
&=\frac{b}{\sqrt n\,E_s\{Zw_1(\mathbf X)\}}
 \left[Z_iw_1(\mathbf X_i)\varepsilon_{1\mid1,i}
 -(1-Z_i)K_{n,0}^{R,\mathrm{PATT}}(i)\varepsilon_{0,i}\right],
 \qquad 1\le i\le n.
\end{aligned}
\]
We use \(\mathcal F_{n,0}=\{\varnothing,\Omega\}\) and
\[
\mathcal F_{n,k}=
\begin{cases}
\sigma\{Z_j,\mathbf D_0^*(\mathbf X_j),w_{Z_j}(\mathbf X_j):1\le j\le k\},
   &1\le k\le n,\\
\mathcal G_{R,n}^{\mathrm{match}}\vee\sigma(Y_1,\ldots,Y_{k-n}),
   &n<k\le2n.
\end{cases}
\]
The coefficients in the second block are \(\mathcal F_{n,n}\)-measurable. Conditional centering therefore gives
\[
\begin{aligned}
E_s\{\xi_{n,i}(a,b)\mid\mathcal F_{n,i-1}\}
&=\frac{a}{\sqrt n\,E_s\{Zw_1(\mathbf X)\}}E_s[Zw_1(\mathbf X)\{\tau_{\mathrm{PATT}}\{\mathbf D_0^*(\mathbf X)\}-\tau_{\mathrm{PATT}}\}]=0,\\
E_s\{\xi_{n,n+i}(a,b)\mid\mathcal F_{n,n+i-1}\}
&=0,\\
\sum_{k=1}^{2n}\xi_{n,k}(a,b)
&=\frac{\bar w_{R,\mathrm{PATT},n}}{E_s\{Zw_1(\mathbf X)\}}
  \{a\sqrt n\,H_{n,R,\mathrm{PATT}}+b\sqrt n\,E_{n,R,\mathrm{PATT}}\}+o_p(1).
\end{aligned}
\]

\emph{Conditional variances and Lindeberg condition.} The first block consists of independent centered observations, so
\[
\sum_{i=1}^n E_s\{\xi_{n,i}(a,b)^2\mid\mathcal F_{n,i-1}\}
=a^2V_{H,R,\mathrm{PATT}}.
\]
Proposition~\ref{prop:primitive-residual-qv-limit}, with the coefficients and conditional variances displayed above, gives
\[
\begin{aligned}
\operatorname{Var}_s(\sqrt n\,E_{n,R,\mathrm{PATT}}\mid \mathcal G_{R,n}^{\mathrm{match}})
&\longrightarrow_{L^1}V_{E,R,\mathrm{PATT}}\\
&=\frac{\Lambda_{R,0}^{\mathrm{PATT}}
       +\Lambda_{R,1}^{\mathrm{PATT}}}{(E_s\{Zw_1(\mathbf X)\})^2}<\infty,\\
\sum_{i=1}^n E_s\{\xi_{n,n+i}(a,b)^2\mid\mathcal F_{n,n+i-1}\}
&=b^2\frac{\bar w_{R,\mathrm{PATT},n}^{\,2}}{(E_s\{Zw_1(\mathbf X)\})^2}
       \operatorname{Var}_s(\sqrt n\,E_{n,R,\mathrm{PATT}}\mid \mathcal G_{R,n}^{\mathrm{match}})\\
&\longrightarrow_p b^2V_{E,R,\mathrm{PATT}}.
\end{aligned}
\]
The first block has i.i.d. summands with finite second moment. Dominated convergence therefore gives, for every \(\epsilon>0\),
\[
\sum_{i=1}^n
 E_s\!\left[\xi_{n,i}(a,b)^2
 \mathbf1\{|\xi_{n,i}(a,b)|>\epsilon\}
 \mid\mathcal F_{n,i-1}\right]
\longrightarrow0.
\]
For the residual block, \eqref{eq:qv-derived-lyapunov} and \(\bar w_{R,\mathrm{PATT},n}\to_p E_s\{Zw_1(\mathbf X)\}\) imply
\[
\begin{aligned}
&\sum_{i=1}^n E_s\!\left[
 \xi_{n,n+i}(a,b)^2\mathbf1\{|\xi_{n,n+i}(a,b)|>\epsilon\}
 \mid\mathcal F_{n,n+i-1}\right]\\
&\le\epsilon^{-\delta}
 \sum_{i=1}^n E_s\{|\xi_{n,n+i}(a,b)|^{2+\delta}
                 \mid\mathcal F_{n,n+i-1}\}\\
&\longrightarrow_p0.
\end{aligned}
\]
The martingale central limit theorem (Corollary~3.1 and the constant-variance remark in Hall and Heyde,\citep{HallHeyde1980} pp.~58--59) and the Cram\'er--Wold device give
\[
\begin{pmatrix}\sqrt n\,H_{n,R,\mathrm{PATT}}\\\sqrt n\,E_{n,R,\mathrm{PATT}}\end{pmatrix}
\Rightarrow
N\!\left(0,
\begin{pmatrix}
V_{H,R,\mathrm{PATT}}&0\\
0&V_{E,R,\mathrm{PATT}}
\end{pmatrix}\right).
\]
Combining this joint limit, including zero-variance coordinates, with the estimator decomposition proves the bias-centered limit.

\emph{Matching discrepancy and bias correction.} Lemma~\ref{lem:oracle-score-radius} with \(z=0\), together with the positive limit of the treated denominator, gives \(B_{n,R,\mathrm{PATT}}=O_p(n^{-1/2})=o_p(1)\). The joint limit above implies \(H_{n,R,\mathrm{PATT}}+E_{n,R,\mathrm{PATT}}=o_p(1)\). The exact decomposition therefore proves consistency of the uncorrected estimator. Under the stated bias-correction remainder,
\[
\begin{aligned}
\sqrt n\{\widehat\tau_{\mathrm{PATT}}^{R,\mathrm{BC}}
                  -\tau_{\mathrm{PATT}}\}
&=\sqrt n(H_{n,R,\mathrm{PATT}}+E_{n,R,\mathrm{PATT}})
 +\sqrt n(B_{n,R,\mathrm{PATT}}-\widehat B_{n,R,\mathrm{PATT}})\\
&=\sqrt n(H_{n,R,\mathrm{PATT}}+E_{n,R,\mathrm{PATT}})+o_p(1).
\end{aligned}
\]
Slutsky's theorem proves the bias-corrected limit. The second-moment assumptions and Proposition~\ref{prop:primitive-residual-qv-limit} give the stated finiteness and positivity conclusions.
\end{proof}

\subsubsection{Estimated-score asymptotics}\label{retrospective-patt-estimated-score-results}

Proposition~\ref{cor:primary-fitted-score-consistency} establishes consistency under its stated conditions, with the bias-correction remainder evaluated on the estimated-score match sets.

For the retrospective PATT scenario, let
\[
V_{R,\mathrm{PATT}}^{\mathrm{known\text{-}score}}
:=
V_{H,R,\mathrm{PATT}}+V_{E,R,\mathrm{PATT}}.
\]
Let \(\widehat{\boldsymbol \theta}_{R,\mathrm{PATT}}\) collect the retrospective survey-weighted propensity-score estimator together with the unweighted control prognostic-score estimator, and let \(\bar{\boldsymbol \theta}_{R,\mathrm{PATT},n,\eta}\) denote its midpoint discretization on cubes of side length \(\eta/\sqrt n\), with deterministic grid offset. The nuisance vector is \(\boldsymbol\theta=(\boldsymbol\alpha^\top,\boldsymbol\beta_0^\top)^\top\). The common-\(\mathbf D_0^\ast\) contrast and treated normalization from Theorem~\ref{thm:retro-patt-oracle} remain
\[
\begin{aligned}
\tau_{\mathrm{PATT}}(\mathbf D_0^\ast)
&=m_{1\mid1}(\mathbf D_0^\ast)-m_0^R(\mathbf D_0^\ast),\\
\bar w_{R,\mathrm{PATT},n}
&=\frac1n\sum_{i=1}^n Z_iw_1(\mathbf X_i).
\end{aligned}
\]
For retrospective PATT, Theorem~\ref{cor:patt-verified-local-grid-law}(i) gives the local Gaussian limit under its stated conditions, with mean \eqref{eq:patt-verified-local-mean-coefficient}. The control-side local matching drift coefficient is
\[
\mathbf j_R:=\left\{\frac{\Delta_R(\mathbf e_r)}
 {\bar w_{R,\mathrm{PATT}}}\right\}_{r=1}^{p_R},
\qquad \bar w_{R,\mathrm{PATT}}=E_s\{Zw_1(\mathbf X)\},
\]
where \(\Delta_R\) is defined in \eqref{eq:patt-drift-fixed-law-geometric-limit}. Thus \(\widetilde{\boldsymbol \gamma}_{R,\mathrm{PATT},2} =-\mathbf j_R-\boldsymbol\Gamma_R^\top\boldsymbol\Sigma_{R,U}^{-1}\boldsymbol\gamma_{R,\mathrm{PATT},1}\).

Under the pathwise identification and differentiation conditions of Proposition~\ref{prop:scientific-target-path-derivative}, the scenario target and its derivative can be written as
\[
\tau_A(Q_{A,t})
=\frac{E_{Q_{A,t}}[Zw_1(\mathbf X)\{\mu_{1\mid1,t}(\mathbf X)-\mu_{0,t}(\mathbf X)\}]}
       {E_{Q_{A,t}}[Zw_1(\mathbf X)]},
\]
and
\[
\boldsymbol\gamma_{R,\mathrm{PATT},2}
=E_s\{\phi_R(O)\mathbf s_A(O)\},
\qquad \mathbf s_A(O)=-\boldsymbol\Gamma_A^\top\boldsymbol\Sigma_{A,U}^{-1}\mathbf U_A(O;\boldsymbol\theta_A^\ast),
\]
with the weight functions fixed along the path. Define
\[
V_{R,\mathrm{PATT}}^{\mathrm{fix}}
=V_{R,\mathrm{PATT}}^{\mathrm{known\text{-}score}}
-\boldsymbol\gamma_{R,\mathrm{PATT},1}^{\top}
 \boldsymbol\Sigma_{A,U}^{-1}\boldsymbol\gamma_{R,\mathrm{PATT},1},
\]
and
\[
V_{R,\mathrm{PATT}}^{\mathrm{grid}}
=V_{R,\mathrm{PATT}}^{\mathrm{fix}}
+\widetilde{\boldsymbol \gamma}_{R,\mathrm{PATT},2}^\top
 \boldsymbol\Sigma_{A,\theta}\widetilde{\boldsymbol \gamma}_{R,\mathrm{PATT},2}.
\]

\begin{theorem}[Retrospective PATT with estimated scores]
\label{thm:retrospective-patt-estimated-scores}
\label{prop:retrospective-patt-estimated-score-conditional}
For PATT, put \(A=R\). Retain the first-step and auxiliary-experiment conditions of Lemma~\ref{lem:auxiliary-godambe-experiment}, including \(\boldsymbol\Sigma_{A,U}>0\). Suppose that \eqref{eq:moving-law-joint-input-with-bias} holds for the specified standardization, matching and bias-correction family at every deterministic \(\mathbf h_n\to \mathbf h\), with
\[
V_{A,\mathrm{known\text{-}score}}
=V_{R,\mathrm{PATT}}^{\mathrm{known\text{-}score}}
=V_{H,R,\mathrm{PATT}}+V_{E,R,\mathrm{PATT}},\qquad
\boldsymbol\gamma_{A,1}=\boldsymbol\gamma_{R,\mathrm{PATT},1},\quad
\widetilde{\boldsymbol \gamma}_{A,2}=\widetilde{\boldsymbol \gamma}_{R,\mathrm{PATT},2}.
\]
Suppose also that \(\tau_A(Q_{A,t})\) is identified and differentiable along these same auxiliary laws, with
\[
\begin{aligned}
\tau_A(Q_{A,t})
&=\tau_{\mathrm{PATT}}+\boldsymbol\gamma_{R,\mathrm{PATT},2}^{\top}t+o(\|t\|),\\
\boldsymbol\gamma_{R,\mathrm{PATT},2}
&:=\left.\frac{\partial}{\partial t}\tau_A(Q_{A,t})\right|_{t=0}.
\end{aligned}
\]
If, in addition, all hypotheses of Theorem~\ref{thm:patt-sufficient-branch-actual-score-law} hold with \(A=R\) for PATT, using the retrospective weights and the original treated normalization displayed above, then the original fixed-\(M\), survey-normalized estimator obeys
\[
\sqrt n\{\widehat\tau_{\mathrm{PATT}}^{R,\mathrm{BC}}
(\widehat{\boldsymbol \theta}_A)-\tau_{\mathrm{PATT}}\}
\Rightarrow N(0,V_{R,\mathrm{PATT}}^{\mathrm{grid}}).
\]
The scenario variance, local mean and strict-positivity alternatives are given by \eqref{eq:patt-verified-geometric-grid-variance}. The variance is finite and nonnegative.
\label{cor:retrospective-patt-actual-sufficient}

For the midpoint discretization, in the order \(n\to\infty\), \(\eta\downarrow0\),
\[
\sqrt n\{\widehat\tau_{\mathrm{PATT}}^{R,\mathrm{BC}}
(\bar{\boldsymbol \theta}_{A,n,\eta})-\tau_{\mathrm{PATT}}\}
\Rightarrow N(0,V_{R,\mathrm{PATT}}^{\mathrm{grid}}).
\]
The double limit is in the bounded-Lipschitz sense of Proposition~\ref{prop:local-mean-corrected-grid-variance}.

\end{theorem}
\begin{proof}
The estimating function is
\[
\mathbf U_A(O;\boldsymbol\theta)=
\begin{pmatrix}
w_Z(\mathbf X)\mathbf X\{Z-e(\mathbf X;\boldsymbol\alpha)\}\\
(1-Z)\mathbf X\{Y-\mathbf X^\top\boldsymbol\beta_0\}
\end{pmatrix},
\qquad
e(\mathbf X;\boldsymbol\alpha)=\operatorname{expit}
(\mathbf X^\top\boldsymbol\alpha).
\]
Its derivative is
\[
\boldsymbol\Gamma_A=-\operatorname{diag}\!\left(
E_s[w_Z(\mathbf X)e^\ast(1-e^\ast)\mathbf X\mathbf X^\top],
E_s[(1-Z)\mathbf X\mathbf X^\top]\right),
\]
where \(e^\ast=e(\mathbf X;\boldsymbol\alpha^\ast)\). Using the full two-block covariance matrix \(\boldsymbol\Sigma_{A,U}\), Proposition~\ref{prop:local-mean-corrected-grid-variance} gives the grid limit \(G_{\tau,A}^\perp+\widetilde{\boldsymbol \gamma}_{A,2}^\top \mathbf G_{\theta,A}\). Its nuisance covariance is \(\boldsymbol\Sigma_{A,\theta}=\boldsymbol\Gamma_A^{-1}\boldsymbol\Sigma_{A,U}\boldsymbol\Gamma_A^{-T}\) by Lemma~\ref{lem:ordinary-weighted-first-step}. The Gaussian projection calculation in the retrospective PATE proof then gives \(V_{R,\mathrm{PATT}}^{\mathrm{grid}}\). The target representer follows from Proposition~\ref{prop:scientific-target-path-derivative}, under its conditions, with weight \(Zw_1(\mathbf X)\).

Under the additional hypotheses for the undiscretized estimator, Theorem~\ref{thm:patt-sufficient-branch-actual-score-law} uses the same control score, normalized matching fractions and treated denominator. Its variance \(V_{A,\mathrm{grid}}\) is therefore \(V_{R,\mathrm{PATT}}^{\mathrm{grid}}\).
\end{proof}

\paragraph{Full-control-mean centering.}
\label{par:retrospective-patt-grouped-sampling}
Under all conditions of Theorem~\ref{thm:patt-grouped-fullx-actual}, with \(A=R\), including its control-mean equality, score geometry, estimating equations, empirical normalization and the two distinct BC remainders,
\[
\begin{pmatrix}
\sqrt n\{\widehat\tau_{\mathrm{PATT}}^{R,\mathrm{BC}}
              (\widehat{\boldsymbol \theta}_R)-\tau_{\mathrm{PATT}}\}\\
\sqrt n(\widehat{\boldsymbol \theta}_R-\boldsymbol\theta_R^*)
\end{pmatrix}
\Rightarrow
\begin{pmatrix}\widetilde G_{\tau,R}\\\mathbf G_{\theta,R}\end{pmatrix}.
\]
Here
\[
\operatorname{Var}(\widetilde G_{\tau,R})
 =\widetilde V_{R,\mathrm{PATT}},\qquad
\operatorname{Cov}(\mathbf G_{\theta,R},\widetilde G_{\tau,R})
 =-\boldsymbol\Gamma_R^{-1}\widetilde{\boldsymbol \gamma}_{R,1},
\]
with the whole treated-observation contribution retained in \eqref{eq:patt-grouped-fullx-constants}.

\subsubsection{Fixed-match replication}\label{retrospective-patt-bootstrap-main-text}

For each fixed-match replicate \(b=1,\ldots,B\), we draw
\[
(m_1^{\ast,b},\ldots,m_n^{\ast,b})
\sim
\mathrm{Multinomial}(n;1/n,\ldots,1/n),
\]
and define
\[
w_i^{\ast,b}:=m_i^{\ast,b}w_{Z_i}(\mathbf X_i).
\]
We refit the retrospective propensity score with weights \(w_i^{\ast,b}\) and the control prognostic score with weights \(m_i^{\ast,b}\) on \(Z_i=0\), using the original estimating equations. We recompute \(\widehat{\mathbf D}_0^{\ast,b}\), including its standardization by the original rule, and refit the control bias-correction regression on these scores with weights \(w_i^{\ast,b}\) to obtain \(\widehat m_0^{\ast,b}\). The control match sets, the donor-normalized coefficients \(\lambda_{i\to j}^{R,(0)}\), and \(K_{n,0}^{R,\mathrm{PATT}}(j)\) remain fixed at the original estimated-score graph.

For a treated observation, the common-score cancellation is
\[
\begin{aligned}
&\tau_{\mathrm{PATT}}\{\mathbf D_0^*(\mathbf X_i)\}
-\tau_{\mathrm{PATT}}+\varepsilon_{1\mid1,i}\\
&=
m_{1\mid1}\{\mathbf D_0^*(\mathbf X_i)\}
-m_0^R\{\mathbf D_0^*(\mathbf X_i)\}
-\tau_{\mathrm{PATT}}+\varepsilon_{1\mid1,i}\\
&=
Y_i-m_0^R\{\mathbf D_0^*(\mathbf X_i)\}
-\tau_{\mathrm{PATT}}.
\end{aligned}
\]
The one-sided fixed-match replicate is
\begin{equation}\label{eq:retrospective-patt-fixed-match-replicate}
\begin{aligned}
\widehat\tau_{\mathrm{PATT}}^{R,\ast,b}
={}&\frac1{\sum_{i:Z_i=1}w_i^{\ast,b}}
\Bigg[
\sum_{i:Z_i=1}w_i^{\ast,b}
\Big[
Y_i-\widehat m_0^{\ast,b}
 \{\widehat{\mathbf D}_0^{\ast,b}(\mathbf X_i)\}
\Big]\\
&\qquad-
\sum_{j:Z_j=0}m_j^{\ast,b}
 K_{n,0}^{R,\mathrm{PATT}}(j)
\Big[
Y_j-\widehat m_0^{\ast,b}
 \{\widehat{\mathbf D}_0^{\ast,b}(\mathbf X_j)\}
\Big]
\Bigg].
\end{aligned}
\end{equation}
The centering estimator \(\widehat\tau_{\mathrm{PATT}}^{R,\mathrm{BC}}\) uses the same graph and control fit \(\widehat m_0\{\widehat{\mathbf D}_0(\mathbf X)\}\). The identities below hold when the original and replicated fits exist and the replicated treated denominator is positive.

We use the treated denominator averages
\[
\begin{aligned}
\bar w_{R,\mathrm{PATT},n}
&=\frac1n\sum_i Z_iw_1(\mathbf X_i),\\
\bar w_{R,\mathrm{PATT},n}^{\ast,b}
&=\frac1n\sum_i m_i^{\ast,b}Z_iw_1(\mathbf X_i).
\end{aligned}
\]
The centered contribution of individual \(i\), evaluated at the original fitted values, is
\begin{equation}\label{eq:retrospective-patt-fitted-centered-row}
\widehat\varphi_{R,\mathrm{PATT},i}
:=
\begin{cases}
w_1(\mathbf X_i)
\bigl[
Y_i-\widehat m_0\{\widehat{\mathbf D}_0(\mathbf X_i)\}
-\widehat\tau_{\mathrm{PATT}}^{R,\mathrm{BC}}
\bigr],
& Z_i=1,\\[1mm]
-K_{n,0}^{R,\mathrm{PATT}}(i)
\bigl[
Y_i-\widehat m_0\{\widehat{\mathbf D}_0(\mathbf X_i)\}
\bigr],
& Z_i=0.
\end{cases}
\end{equation}

The full refit contribution is
\begin{equation}\label{eq:retrospective-patt-bootstrap-refit-remainder}
\begin{aligned}
\mathcal R_{R,\mathrm{PATT},n}^{\ast,b}
={}&-\sum_{i:Z_i=1}m_i^{\ast,b}w_1(\mathbf X_i)
\Big[
\widehat m_0^{\ast,b}
 \{\widehat{\mathbf D}_0^{\ast,b}(\mathbf X_i)\}
-\widehat m_0\{\widehat{\mathbf D}_0(\mathbf X_i)\}
\Big]\\
&+\sum_{j:Z_j=0}m_j^{\ast,b}
 K_{n,0}^{R,\mathrm{PATT}}(j)
\Big[
\widehat m_0^{\ast,b}
 \{\widehat{\mathbf D}_0^{\ast,b}(\mathbf X_j)\}
-\widehat m_0\{\widehat{\mathbf D}_0(\mathbf X_j)\}
\Big].
\end{aligned}
\end{equation}

\begin{proposition}[Fixed-match replication for retrospective PATT]
\label{prop:retrospective-patt-fixed-match-identity}
Then \(\sum_i\widehat\varphi_{R,\mathrm{PATT},i}=0\), and the exact ratio is
\begin{equation}\label{eq:retrospective-patt-bootstrap-exact-ratio}
\begin{aligned}
&\sqrt n\Big(
\widehat\tau_{\mathrm{PATT}}^{R,\ast,b}
-\widehat\tau_{\mathrm{PATT}}^{R,\mathrm{BC}}
\Big)\\
&=
\frac1{\bar w_{R,\mathrm{PATT},n}^{\ast,b}}
\frac1{\sqrt n}\sum_i
(m_i^{\ast,b}-1)\widehat\varphi_{R,\mathrm{PATT},i}
+
\frac{\mathcal R_{R,\mathrm{PATT},n}^{\ast,b}}
{\sqrt n\,\bar w_{R,\mathrm{PATT},n}^{\ast,b}},
\end{aligned}
\end{equation}

With \(\mathcal F_n\) containing the original observations, fits, and graph, the full multinomial law gives
\begin{equation}\label{eq:retrospective-patt-bootstrap-numerator-variance}
\operatorname{Var}^\ast\left(
\frac1{\sqrt n}\sum_i(m_i^{\ast,b}-1)
\widehat\varphi_{R,\mathrm{PATT},i}
\,\middle|\,\mathcal F_n\right)
=
\frac1n\sum_i\widehat\varphi_{R,\mathrm{PATT},i}^{\,2}.
\end{equation}

If, in addition, all conditions of Proposition~\ref{prop:patt-full-prediction-replication-law} hold with \(A=R\), then
\[
\mathcal L^\ast\!\left[
 \sqrt n(\widehat\tau_{\mathrm{PATT}}^{R,\ast,b}-\widehat\tau_{\mathrm{PATT}}^{R,\mathrm{BC}})
 \,\middle|\,\mathcal F_n\right]
 \Longrightarrow_{P_s}N(0,V_{R,\mathrm{rep}}),
\]
where
\[
\begin{aligned}
V_{R,\mathrm{rep}}
&=V_{R,\mathrm{known\text{-}score}}
 +\frac{2}{\bar w_{R}}\,
 \ell_{R,0}^\top\boldsymbol\Gamma_{R}^{-1}\boldsymbol\gamma_{R,1}\\
&\quad+\frac1{\bar w_{R}^2}\,
 \ell_{R,0}^\top\boldsymbol\Sigma_{R,\theta}\ell_{R,0}.
\end{aligned}
\]
Suppose a sampling limit centered at \(\tau_{\mathrm{PATT}}\) has also been established, and
\[
\sqrt n(\widehat\tau_{\mathrm{PATT}}^{R,\mathrm{BC}}-\tau_{\mathrm{PATT}})
\Rightarrow N(0,V_{R,\mathrm{PATT}}),\qquad
V_{R,\mathrm{PATT}}=V_{R,\mathrm{rep}}>0.
\]
For conditionally independent replicates with the prescribed finite-valued outputs, the variance estimator in \eqref{eq:bootstrap-mc-original-variance} is consistent in the iterated limit \(n\to\infty\), then \(B\to\infty\), and along a sufficiently slowly increasing deterministic sequence \(B_n\). The associated normal Wald interval has asymptotic coverage \(1-\alpha\), for \(0<\alpha<1\), in the same regimes.
\label{par:retrospective-patt-bootstrap-law-comparison}
\end{proposition}

\begin{proof}
Specialize the PATT ratio representation \eqref{eq:patt-fixed-prediction-ratios} and its centered form \eqref{eq:patt-common-centered-replication-ratio} to \(A=R\) and \(w_{A,z}(\mathbf X_i)=w_z(\mathbf X_i)\). The fitted row in \eqref{eq:patt-common-fitted-centered-row} is then \eqref{eq:retrospective-patt-fitted-centered-row}. Collecting the prediction changes in \eqref{eq:patt-full-prediction-exact-refit-remainder} gives \eqref{eq:retrospective-patt-bootstrap-refit-remainder}. The centered rows sum to zero, so \eqref{eq:fixed-match-multinomial-moments} gives the numerator variance. The zero-denominator bound \eqref{eq:fixed-match-treated-denominator-zero}, together with the iid denominator law, gives a positive denominator limit. Under the additional replication hypotheses, the refit expansion \eqref{eq:patt-full-prediction-refit-first-order} reduces the remainder to
\[
\frac{\mathcal R_{R,\mathrm{PATT},n}^{\ast,b}}{\sqrt n}
=-\ell_{R,0}^\top
 \sqrt n(\widehat{\boldsymbol \theta}_{R,\mathrm{PATT}}^{\ast,b}
              -\widehat{\boldsymbol \theta}_{R,\mathrm{PATT}})
 +o_{p^\ast}(1).
\]
The joint multiplier law gives the stated conditional distribution. Under the stipulated positive sampling--replication variance agreement, Proposition~\ref{prop:bootstrap-monte-carlo-iterated-variance} gives variance consistency and Wald coverage in the stated regimes.
\end{proof}

\subsection{Prospective PATE}
\label{app:prospective_pate}
\subsubsection{Conditions}

The prospective inverse-inclusion weight is \(w(\mathbf X)=\pi(\mathbf X)^{-1}\). For the oracle analysis, matching uses the fixed limiting standardized scores \(\mathbf D_z^*(\mathbf X)\). We write
\[
\begin{aligned}
m_z(\mathbf d)&:=E_\Omega\{Y(z)\mid\mathbf D_z^*(\mathbf X)=\mathbf d\},\\
\tau_{\mathbf D}(\mathbf X)
&:=m_1\{\mathbf D_1^*(\mathbf X)\}
  -m_0\{\mathbf D_0^*(\mathbf X)\},\\
\varepsilon_{z,i}
&:=Y_i-m_z\{\mathbf D_z^*(\mathbf X_i)\},
\qquad Z_i=z.
\end{aligned}
\]
Under prospective reduced-score identification, \(m_z=m_z^P\), \(\varepsilon_{z,i}=\varepsilon_{z,i}^P\), and \(\tau_{\mathbf D}=\tau_P^D\) in the preceding common notation. Under the identification conditions below, armwise mean balancing holds if the propensity coordinate is correct or the prognostic coordinate is the correct conditional outcome mean.

\begin{assumption}[Prospective PATE conditions]
\label{ass:prospective-pate-oracle}
\leavevmode
\begin{enumerate}
\item[(i)] Consistency and strong treatment overlap hold. Prospective survey selection satisfies \(\{Y(0),Y(1)\}\perp S\mid\mathbf X\), with \(\pi(\mathbf X)\) bounded away from zero. Treatment assignment satisfies \(Z\perp\{Y(0),Y(1),S\}\mid\mathbf X\).
\item[(ii)] For each treatment arm, the limiting double score satisfies population conditional-mean balancing.
\item[(iii)] The PATE conditions in Assumptions~\ref{ass:residual-centering-survey-normalized} and~\ref{ass:residual-qv-stabilization} hold for \(A=P\) and both matching directions. The selected observations are i.i.d. The two fixed score spaces have the stated compact convex supports and the stated within-arm and opposite-arm score densities. Weights are bounded above and away from zero.
\item[(iv)] The reduced-score residuals have finite \((2+\delta)\)-moments for some \(\delta>0\) and are centered given both scores and the observed weight. The functions \(m_0,m_1\) are Lipschitz on their respective score supports, and
\[
E_s\!\left[
 \left|w(\mathbf X)
 \{\tau_{\mathbf D}(\mathbf X)-\tau_{\mathrm{PATE}}\}\right|^{2+\delta}
 \right]<\infty.
\]
\end{enumerate}
\end{assumption}
\label{prospective-pate-with-exact-within-match-survey-weight-normalized-imputation}

\subsubsection{Estimator and reuse coefficients}\label{exact-estimator-exact-survey-weighted-matched-set-coefficients-and-exact-matched-set-coefficient-representation}

For each individual \(i\) in arm \(1-z\), let \(\mathcal J_{\mathbf D_z^*}(i)\) be the set of its \(M\) nearest individuals in arm \(z\) on \(\mathbf D_z^*(\mathbf X)\).

We define the survey-weighted within-match coefficient by
\[\lambda_{i\to j}^{P,(z)}
:=
\frac{w(\mathbf X_j)}{\sum_{k\in\mathcal J_{\mathbf D_z^*}(i)} w(\mathbf X_k)}\,\mathbf 1\{j\in\mathcal J_{\mathbf D_z^*}(i)\},
\qquad j=1,\dots,n,\] when \(Z_i=1-z\). Define it to equal \(0\) when
\(Z_i=z\).

The corresponding survey-weighted imputed counterfactuals are
\[\widehat Y_i^{P}(z)
:=
\begin{cases}
Y_i,& Z_i=z,\\[4pt]
\sum_{j=1}^n \lambda_{i\to j}^{P,(z)}Y_j,& Z_i=1-z.
\end{cases}\] Then the prospective PATE WDSM estimator is
\[\widehat\tau_{\mathrm{PATE}}^{P}
:=
\frac{\sum_{i=1}^n w(\mathbf X_i)\bigl\{\widehat Y_i^{P}(1)-\widehat Y_i^{P}(0)\bigr\}}
{\sum_{i=1}^n w(\mathbf X_i)}.\] We denote the denominator mean by
\[\bar w_{P,\mathrm{PATE},n}:=\frac{1}{n}\sum_{i=1}^n w(\mathbf X_i).\] The weighted reuse coefficients are \[K_{n,1}^{P,\mathrm{PATE}}(j)
:=
\sum_{i:Z_i=0}
 w(\mathbf X_i)
 \frac{w(\mathbf X_j)\mathbf 1\{j\in\mathcal J_{\mathbf D_1^*}(i)\}}
 {\sum_{k\in\mathcal J_{\mathbf D_1^*}(i)} w(\mathbf X_k)},
\qquad Z_j=1,\] and \[K_{n,0}^{P,\mathrm{PATE}}(j)
:=
\sum_{i:Z_i=1}
 w(\mathbf X_i)
 \frac{w(\mathbf X_j)\mathbf 1\{j\in\mathcal J_{\mathbf D_0^*}(i)\}}
 {\sum_{k\in\mathcal J_{\mathbf D_0^*}(i)} w(\mathbf X_k)},
\qquad Z_j=0.\]

We define the three components by \[H_{n,P,\mathrm{PATE}}
:=
\frac{1}{n\bar w_{P,\mathrm{PATE},n}}
\sum_{i=1}^n w(\mathbf X_i)\{\tau_{\mathbf D}(\mathbf X_i)-\tau_{\mathrm{PATE}}\},\]
\[\begin{aligned}
&B_{n,P,\mathrm{PATE}}\\
&:=
\frac{1}{n\bar w_{P,\mathrm{PATE},n}}
\sum_{i=1}^n w(\mathbf X_i)(2Z_i-1)
\left[
m_{1-Z_i}\{\mathbf D_{1-Z_i}^*(\mathbf X_i)\}-\sum_{j=1}^n \lambda_{i\to j}^{P,(1-Z_i)}m_{1-Z_i}\{\mathbf D_{1-Z_i}^*(\mathbf X_j)\}
\right],
\end{aligned}\] and \[E_{n,P,\mathrm{PATE}}
:=
\frac{1}{n\bar w_{P,\mathrm{PATE},n}}
\sum_{j=1}^n(2Z_j-1)
\{w(\mathbf X_j)+K_{n,Z_j}^{P,\mathrm{PATE}}(j)\}
\varepsilon_{Z_j,j}.\]

\subsubsection{Known-score asymptotics}
\label{exact-decomposition-and-exact-conditional-residual-variance}
\label{exact-known-score-asymptotics-2}

Define the oracle bias-corrected estimator by
\[
\widehat\tau_{\mathrm{PATE}}^{P,\mathrm{BC}}
=\widehat\tau_{\mathrm{PATE}}^{P}
-\widehat B_{n,P,\mathrm{PATE}}.
\]

The heterogeneity variance is
\[
V_{H,P,\mathrm{PATE}}
=
\frac{\operatorname{Var}_s\!\left(
w(\mathbf X)
\left[
m_1\{\mathbf D_1^*(\mathbf X)\}
-m_0\{\mathbf D_0^*(\mathbf X)\}
-\tau_{\mathrm{PATE}}
\right]\right)}
{\{E_s[w(\mathbf X)]\}^2}.
\]
The matching discrepancy is
\[
m_{1-Z_i}\{\mathbf D_{1-Z_i}^*(\mathbf X_i)\}
-\sum_{j=1}^n\lambda_{i\to j}^{P,(1-Z_i)}
m_{1-Z_i}\{\mathbf D_{1-Z_i}^*(\mathbf X_j)\}.
\]

\begin{theorem}[Prospective PATE with known scores]
\label{thm:prospective-pate-oracle}
\label{cor:prospective-pate-score-variance}
Under Assumption~\ref{ass:prospective-pate-oracle},
\[
\begin{pmatrix}
\sqrt n\,H_{n,P,\mathrm{PATE}}\\
\sqrt n\,E_{n,P,\mathrm{PATE}}
\end{pmatrix}
\Rightarrow
N\!\left[
\begin{pmatrix}0\\0\end{pmatrix},
\begin{pmatrix}
V_{H,P,\mathrm{PATE}}&0\\
0&V_{E,P,\mathrm{PATE}}
\end{pmatrix}
\right].
\]
The matching discrepancy satisfies \(B_{n,P,\mathrm{PATE}}=O_p(n^{-1/2})\), and the uncorrected estimator is consistent. If the auxiliary-regression correction satisfies the specified remainder condition at these same oracle scores, matches and coefficients,
\begin{equation}\label{eq:prospective-pate-oracle-bc-remainder}
\sqrt n\left(
\widehat B_{n,P,\mathrm{PATE}}
-B_{n,P,\mathrm{PATE}}
\right)\to_p0,
\end{equation}
then
\[
\sqrt n\left(
\widehat\tau_{\mathrm{PATE}}^{P,\mathrm{BC}}
-\tau_{\mathrm{PATE}}
\right)
\Rightarrow
N\!\left(0,V_{H,P,\mathrm{PATE}}+
             V_{E,P,\mathrm{PATE}}\right).
\]
Under this additional remainder condition, the bias-corrected estimator is also consistent. Its Gaussian limit is nondegenerate when \(V_{H,P,\mathrm{PATE}}+V_{E,P,\mathrm{PATE}}>0\). It suffices that \(V_{H,P,\mathrm{PATE}}>0\) or \(P_s\{v_{Z_i,i}>0\}>0\), the latter by Proposition~\ref{prop:primitive-residual-qv-limit}. Corollary~\ref{cor:oracle-regression-bc} gives sufficient derivative-consistency conditions for this remainder.
\label{lem:prospective-pate-decomposition}
\label{prop:prospective-pate-coefficients}
\label{lem:prospective-pate-residual-variance}
\label{prop:prospective-pate-bias-order}
\label{lem:prospective-pate-heterogeneity-clt}
\label{lem:prospective-pate-qv}
\end{theorem}
\begin{proof}
The proof of Theorem~\ref{thm:retrospective-pate-oracle} applies with \(w_0=w_1=w\), since the prospective treatment assignment condition implies \(\pi_0=\pi_1=\pi\). This substitution leaves the fixed scores and match sets unchanged and gives the normalized coefficients, reuse coefficients and denominator defined above. Proposition~\ref{prop:exact-pate-score-space-decomposition} therefore gives
\[
\widehat\tau_{\mathrm{PATE}}^{P}-\tau_{\mathrm{PATE}}
=H_{n,P,\mathrm{PATE}}
 +B_{n,P,\mathrm{PATE}}
 +E_{n,P,\mathrm{PATE}}.
\]
Population recovery and mean balancing imply
\[
E_s[w(\mathbf X)\{\tau_{\mathbf D}(\mathbf X)-\tau_{\mathrm{PATE}}\}]=0,
\qquad
\bar w_{P,\mathrm{PATE},n}\to_p E_s\{w(\mathbf X)\}>0.
\]

The initial field in the residual stage is
\[
\mathcal G_{P,n}^{\mathrm{match}}
=\sigma\!\left\{
Z_i,\mathbf D_0^*(\mathbf X_i),\mathbf D_1^*(\mathbf X_i),
w(\mathbf X_i):1\le i\le n
\right\}.
\]
For \(Z_j=z\), let
\[
v_{z,j}
=E_s\!\left[\varepsilon_{z,j}^2\mid
Z_j=z,\mathbf D_0^*(\mathbf X_j),\mathbf D_1^*(\mathbf X_j),
w(\mathbf X_j)\right].
\]
Assumption~\ref{ass:prospective-pate-oracle}(iii)--(iv) gives conditional independence and centering of the residuals in this field. Consequently,
\[
\begin{aligned}
\operatorname{Var}_s\!\left(
 \sqrt n E_{n,P,\mathrm{PATE}}\mid\mathcal G_{P,n}^{\mathrm{match}}\right)
&=\frac{1}{n\bar w_{P,\mathrm{PATE},n}^{\,2}}
 \sum_{z=0}^1\sum_{j:Z_j=z}
 \{w(\mathbf X_j)+K_{n,z}^{P,\mathrm{PATE}}(j)\}^2v_{z,j}\\
&\longrightarrow_{L^1}
 \frac{\Lambda_{P,0}^{\mathrm{PATE}}+\Lambda_{P,1}^{\mathrm{PATE}}}
      {\{E_s w(\mathbf X)\}^2}
 =V_{E,P,\mathrm{PATE}},
\end{aligned}
\]
by Proposition~\ref{prop:primitive-residual-qv-limit}. The heterogeneity summands have finite second moments, and \eqref{eq:qv-derived-lyapunov} supplies the residual-stage Lindeberg condition. Thus the two-stage martingale argument in Theorem~\ref{thm:retrospective-pate-oracle}, with the substitutions above, gives, for every \(a,b\in\mathbb R\),
\[
a\sqrt n H_{n,P,\mathrm{PATE}}
+b\sqrt n E_{n,P,\mathrm{PATE}}
\Rightarrow
N\!\left(0,a^2V_{H,P,\mathrm{PATE}}
          +b^2V_{E,P,\mathrm{PATE}}\right).
\]
The Cram\'er--Wold device proves the stated joint limit, including zero-variance coordinates.

Lemma~\ref{lem:oracle-score-radius}, applied to both matching directions, and the positive limit of the outer denominator give \(B_{n,P,\mathrm{PATE}}=O_p(n^{-1/2})=o_p(1)\). Together with the joint limit above and the exact decomposition, this proves consistency of the uncorrected estimator. Under \eqref{eq:prospective-pate-oracle-bc-remainder},
\[
\begin{aligned}
\sqrt n\{\widehat\tau_{\mathrm{PATE}}^{P,\mathrm{BC}}
                 -\tau_{\mathrm{PATE}}\}
&=\sqrt n(H_{n,P,\mathrm{PATE}}
          +E_{n,P,\mathrm{PATE}})
  +\sqrt n(B_{n,P,\mathrm{PATE}}
           -\widehat B_{n,P,\mathrm{PATE}})\\
&=\sqrt n(H_{n,P,\mathrm{PATE}}
          +E_{n,P,\mathrm{PATE}})+o_p(1).
\end{aligned}
\]
Slutsky's theorem gives the bias-corrected limit and consistency. Finiteness and the stated sufficient conditions for positivity follow from Proposition~\ref{prop:primitive-residual-qv-limit} and the heterogeneity variance.
\end{proof}

\subsubsection{Estimated-score asymptotics}\label{prospective-pate-estimated-score-results}

Proposition~\ref{cor:primary-fitted-score-consistency} establishes consistency under its stated conditions, with the bias-correction remainder evaluated on the estimated-score match sets.

For prospective PATE, put \(A=P\) and retain the deterministic limiting-score oracle variance
\[
V_{P,\mathrm{PATE}}^{\mathrm{known\text{-}score}}
:=
V_{H,P,\mathrm{PATE}}+V_{E,P,\mathrm{PATE}}
\]
from Theorem~\ref{thm:prospective-pate-oracle}, with its stated bias-correction remainder. Let \(\widehat{\boldsymbol \theta}_A\) collect the prospective propensity-score estimator and both prognostic-score estimators, and let \(\bar{\boldsymbol \theta}_{A,n,\eta}\) be its midpoint discretization on cubes of side length \(\eta/\sqrt n\), with deterministic grid offset.

The selected-sample estimating function for the logistic propensity and linear prognostic models in the main text is
\begin{equation}\label{eq:prospective-pate-selected-first-step}
\mathbf U_A(O;\boldsymbol\theta)
=
\begin{pmatrix}
\mathbf X\{Z-e(\mathbf X;\boldsymbol\alpha)\}\\
(1-Z)\mathbf X\{Y-\Psi_0(\mathbf X;\boldsymbol\beta_0)\}\\
Z\mathbf X\{Y-\Psi_1(\mathbf X;\boldsymbol\beta_1)\}
\end{pmatrix},
\end{equation}
where \(\boldsymbol\theta=(\boldsymbol\alpha^\top,\boldsymbol\beta_0^\top, \boldsymbol\beta_1^\top)^\top\), \(e(\mathbf X;\boldsymbol\alpha)=\operatorname{expit} (\mathbf X^\top\boldsymbol\alpha)\), and \(\Psi_z(\mathbf X;\boldsymbol\beta_z)=\mathbf X^\top\boldsymbol\beta_z\). The moment and rank conditions and existence of a finite solution to the population estimating equations in Lemma~\ref{lem:primitive-logistic-linear-roots} imply Assumption~\ref{ass:ordinary-weighted-m-estimation}. Lemma~\ref{lem:ordinary-weighted-first-step} then gives the first-stage parameter expansion.

\begin{theorem}[Prospective PATE with estimated scores]
\label{thm:prospective-pate-estimated-scores}
\label{par:prospective-pate-two-arm-sampling}
\emph{(i)} Under all conditions of Theorem~\ref{thm:pate-two-arm-actual-joint-law}, with \(A=P\), including empirical normalization and the distinct deterministic-local and actual-graph BC remainders,
\[
\begin{pmatrix}
\sqrt n\{\widehat\tau_{\mathrm{PATE}}^{P,\mathrm{BC}}
              (\widehat{\boldsymbol \theta}_P)-\tau_{\mathrm{PATE}}\}\\
\sqrt n(\widehat{\boldsymbol \theta}_P-\boldsymbol\theta_P^*)
\end{pmatrix}
\Rightarrow
\begin{pmatrix}G_{\tau,P}+\boldsymbol b_{P,\mathrm{loc}}^\top \mathbf G_{\theta,P}\\\mathbf G_{\theta,P}\end{pmatrix},
\]
where \(\boldsymbol b_{P,\mathrm{loc}}\) is defined by \eqref{eq:pate-two-arm-actual-drift}. Consequently the fitted-estimator variance is
\[
\begin{aligned}
V_{P,\mathrm{actual}}
&=V_{P,\mathrm{known\text{-}score}}
 -\boldsymbol\gamma_{P,1}^\top\boldsymbol\Sigma_{P,U}^{-1}\boldsymbol\gamma_{P,1}\\
&\quad{}+(\boldsymbol b_{P,\mathrm{loc}}-\boldsymbol\Gamma_P^\top\boldsymbol\Sigma_{P,U}^{-1}\boldsymbol\gamma_{P,1})^\top
 \boldsymbol\Sigma_{P,\theta}
 (\boldsymbol b_{P,\mathrm{loc}}-\boldsymbol\Gamma_P^\top\boldsymbol\Sigma_{P,U}^{-1}\boldsymbol\gamma_{P,1}).
\end{aligned}
\]
\emph{(ii)} Under the full-mean and all other conditions of Theorem~\ref{thm:pate-fullx-actual-joint-law}, the corresponding joint limit is instead \((G_{\tau,P},\mathbf G_{\theta,P})\), with \(\operatorname{Var}(G_{\tau,P})=V_{P,\mathrm{known\text{-}score}}\).
\end{theorem}
\begin{proof}
For part (i), set \(A=P\) in Theorem~\ref{thm:pate-two-arm-actual-joint-law}, using the estimating equations specified above. This gives the stated joint limit with \(\boldsymbol b_{P,\mathrm{loc}}\) from \eqref{eq:pate-two-arm-actual-drift}. Its covariance is
\[
\operatorname{Cov}(\mathbf G_{\theta,P},G_{\tau,P})
=-\boldsymbol\Gamma_P^{-1}\boldsymbol\gamma_{P,1},\qquad
\operatorname{Var}(\mathbf G_{\theta,P})=\boldsymbol\Sigma_{P,\theta}.
\]
Consequently,
\[
\begin{aligned}
\operatorname{Var}(G_{\tau,P}+\boldsymbol b_{P,\mathrm{loc}}^\top \mathbf G_{\theta,P})
&=V_{P,\mathrm{known\text{-}score}}
 -2\boldsymbol b_{P,\mathrm{loc}}^\top\boldsymbol\Gamma_P^{-1}\boldsymbol\gamma_{P,1}
 +\boldsymbol b_{P,\mathrm{loc}}^\top\boldsymbol\Sigma_{P,\theta}\boldsymbol b_{P,\mathrm{loc}}\\
&=V_{P,\mathrm{actual}},
\end{aligned}
\]
where the last equality follows by expanding the displayed projection form and using \(\boldsymbol\Sigma_{P,\theta}=\boldsymbol\Gamma_P^{-1}\boldsymbol\Sigma_{P,U}\boldsymbol\Gamma_P^{-T}\). For part (ii), apply Theorem~\ref{thm:pate-fullx-actual-joint-law} with \(A=P\) under its full set of conditions.
\end{proof}

Under the pathwise identification and differentiation conditions of Proposition~\ref{prop:scientific-target-path-derivative}, the scenario target and its derivative can be written as
\[
\tau_A(Q_{A,t})
=\frac{E_{Q_{A,t}}[w(\mathbf X)\{\mu_{1,t}(\mathbf X)-\mu_{0,t}(\mathbf X)\}]}
       {E_{Q_{A,t}}[w(\mathbf X)]},
\]
and
\[
\boldsymbol\gamma_{P,\mathrm{PATE},2}
=E_s\{\phi_P(O)\mathbf s_A(O)\},
\qquad \mathbf s_A(O)=-\boldsymbol\Gamma_A^\top\boldsymbol\Sigma_{A,U}^{-1}\mathbf U_A(O;\boldsymbol\theta_A^\ast),
\]
with the weight functions fixed along the path. Define
\begin{equation}\label{eq:prospective-pate-projection-variance}
V_{P,\mathrm{PATE}}^{\mathrm{fix}}
=V_{P,\mathrm{PATE}}^{\mathrm{known\text{-}score}}
-\boldsymbol\gamma_{P,\mathrm{PATE},1}^{\top}
 \boldsymbol\Sigma_{A,U}^{-1}\boldsymbol\gamma_{P,\mathrm{PATE},1},
\end{equation}
and
\begin{equation}\label{eq:prospective-pate-grid-variance}
V_{P,\mathrm{PATE}}^{\mathrm{grid}}
=V_{P,\mathrm{PATE}}^{\mathrm{fix}}
+\widetilde{\boldsymbol \gamma}_{P,\mathrm{PATE},2}^\top
 \boldsymbol\Sigma_{A,\theta}\widetilde{\boldsymbol \gamma}_{P,\mathrm{PATE},2}.
\end{equation}

\begin{proposition}[Discretized-score approximation]
\label{prop:prospective-pate-estimated-conditional}
For PATE, put \(A=P\). Retain the first-step and auxiliary-experiment conditions of Lemma~\ref{lem:auxiliary-godambe-experiment}, including \(\boldsymbol\Sigma_{A,U}>0\). Suppose that \eqref{eq:moving-law-joint-input-with-bias} holds for the specified standardization, matching and bias-correction family at every deterministic \(\mathbf h_n\to \mathbf h\), with
\[
V_{A,\mathrm{known\text{-}score}}
=V_{P,\mathrm{PATE}}^{\mathrm{known\text{-}score}}
=V_{H,P,\mathrm{PATE}}+V_{E,P,\mathrm{PATE}},\qquad
\boldsymbol\gamma_{A,1}=\boldsymbol\gamma_{P,\mathrm{PATE},1},\quad
\widetilde{\boldsymbol \gamma}_{A,2}=\widetilde{\boldsymbol \gamma}_{P,\mathrm{PATE},2}.
\]
Suppose also that \(\tau_A(Q_{A,t})\) is identified and differentiable along these same auxiliary laws, with
\begin{equation}\label{eq:prospective-pate-aux-target-derivative}
\begin{aligned}
\tau_A(Q_{A,t})
&=\tau_{\mathrm{PATE}}+\boldsymbol\gamma_{P,\mathrm{PATE},2}^{\top}t+o(\|t\|),\\
\boldsymbol\gamma_{P,\mathrm{PATE},2}
&:=\left.\frac{\partial}{\partial t}\tau_A(Q_{A,t})\right|_{t=0}.
\end{aligned}
\end{equation}
Then, in the order \(n\to\infty\), \(\eta\downarrow0\),
\[
\sqrt n\{\widehat\tau_{\mathrm{PATE}}^{P,\mathrm{BC}}
(\bar{\boldsymbol \theta}_{A,n,\eta})-\tau_{\mathrm{PATE}}\}
\Rightarrow N(0,V_{P,\mathrm{PATE}}^{\mathrm{grid}}).
\]
The double limit is in the bounded-Lipschitz sense of Proposition~\ref{prop:local-mean-corrected-grid-variance}.
\end{proposition}
\begin{proof}
Applying Proposition~\ref{prop:local-mean-corrected-grid-variance} with the constants above gives the grid limit \(G_{\tau,A}^\perp+\widetilde{\boldsymbol \gamma}_{A,2}^\top \mathbf G_{\theta,A}\). By Lemma~\ref{lem:ordinary-weighted-first-step}, \(\boldsymbol\Sigma_{A,\theta}=\boldsymbol\Gamma_A^{-1}\boldsymbol\Sigma_{A,U}\boldsymbol\Gamma_A^{-T}\). The Gaussian projection calculation in the retrospective PATE proof therefore gives \eqref{eq:prospective-pate-projection-variance} and \eqref{eq:prospective-pate-grid-variance}. Under the conditions of Proposition~\ref{prop:scientific-target-path-derivative}, the target representer uses weight \(w(\mathbf X)\).
\end{proof}

\subsubsection{Fixed-match replication}\label{prospective-pate-bootstrap-main-text}

For each fixed-match replicate \(b=1,\ldots,B\), we draw
\[
(m_1^{\ast,b},\ldots,m_n^{\ast,b})
\sim
\mathrm{Multinomial}(n;1/n,\ldots,1/n),
\]
and define the replicated prospective survey weight
\[
w_i^{\ast,b}:=m_i^{\ast,b}w(\mathbf X_i).
\]
We refit the prospective propensity score and both arm-specific prognostic scores with the multinomial weights \(m_i^{\ast,b}\), using the same unweighted score estimating equations as in the original sample. These refits determine \(\widehat{\mathbf D}_z^{\ast,b}\), including standardization by the original rule. Within arm \(z\), we refit the bias-correction regression with weights \(w_i^{\ast,b}\) on these replicate-specific scores and denote the fitted reduced mean by \(\widehat m_z^{\ast,b}\). The original match sets, their normalized within-match weights \(\lambda_{i\to j}^{P,(z)}\), and hence \(K_{n,z}^{P,\mathrm{PATE}}(j)\), remain fixed at the original estimated-score graph.

The prospective PATE fixed-match replicate is
\begin{equation}\label{eq:prospective-pate-fixed-match-replicate}
\begin{aligned}
\widehat\tau_{\mathrm{PATE}}^{P,\ast,b}
=
\frac{1}{\sum_{i=1}^n w_i^{\ast,b}}
\sum_{i=1}^n
\Big[
& w_i^{\ast,b}
\Big\{
\widehat m_1^{\ast,b}
 \{\widehat{\mathbf D}_1^{\ast,b}(\mathbf X_i)\}
-
\widehat m_0^{\ast,b}
 \{\widehat{\mathbf D}_0^{\ast,b}(\mathbf X_i)\}
\Big\}\\
&+
m_i^{\ast,b}(2Z_i-1)
\Bigl\{
w(\mathbf X_i)+K_{n,Z_i}^{P,\mathrm{PATE}}(i)
\Bigr\}\\
&\qquad{}\times
\Bigl[
Y_i-\widehat m_{Z_i}^{\ast,b}
 \{\widehat{\mathbf D}_{Z_i}^{\ast,b}(\mathbf X_i)\}
\Bigr]
\Big].
\end{aligned}
\end{equation}
The centering estimator \(\widehat\tau_{\mathrm{PATE}}^{P,\mathrm{BC}}\) uses the same graph and fitted means \(\widehat m_z\{\widehat{\mathbf D}_z(\mathbf X)\}\). The identities below hold when the original and replicated fits exist and the replicated denominator is positive.

We denote the original and replicated normalized denominators by
\[
\bar w_{P,\mathrm{PATE},n}
=\frac1n\sum_{i=1}^n w(\mathbf X_i),
\qquad
\bar w_{P,\mathrm{PATE},n}^{\ast,b}
=\frac1n\sum_{i=1}^n m_i^{\ast,b}w(\mathbf X_i).
\]
The centered contribution of individual \(i\) is
\begin{equation}\label{eq:prospective-pate-fitted-centered-row}
\begin{aligned}
\widehat\varphi_{P,\mathrm{PATE},i}
:={}&w(\mathbf X_i)
\Big[
\widehat m_1\{\widehat{\mathbf D}_1(\mathbf X_i)\}
-\widehat m_0\{\widehat{\mathbf D}_0(\mathbf X_i)\}
-\widehat\tau_{\mathrm{PATE}}^{P,\mathrm{BC}}
\Big]\\
&+(2Z_i-1)
\{w(\mathbf X_i)+K_{n,Z_i}^{P,\mathrm{PATE}}(i)\}\\
&\qquad{}\times
\Big[Y_i-\widehat m_{Z_i}
 \{\widehat{\mathbf D}_{Z_i}(\mathbf X_i)\}\Big].
\end{aligned}
\end{equation}

The full refit contribution is
\begin{equation}\label{eq:prospective-pate-bootstrap-refit-remainder}
\begin{aligned}
\mathcal R_{P,\mathrm{PATE},n}^{\ast,b}
={}&\sum_{z=0}^1(2z-1)
\Bigg[
\sum_{i:Z_i=1-z}m_i^{\ast,b}w(\mathbf X_i)\\
&\quad{}\times
\Big[
\widehat m_z^{\ast,b}
 \{\widehat{\mathbf D}_z^{\ast,b}(\mathbf X_i)\}
-\widehat m_z\{\widehat{\mathbf D}_z(\mathbf X_i)\}
\Big]\\
&-
\sum_{j:Z_j=z}m_j^{\ast,b}
 K_{n,z}^{P,\mathrm{PATE}}(j)\\
&\quad{}\times
\Big[
\widehat m_z^{\ast,b}
 \{\widehat{\mathbf D}_z^{\ast,b}(\mathbf X_j)\}
-\widehat m_z\{\widehat{\mathbf D}_z(\mathbf X_j)\}
\Big]
\Bigg].
\end{aligned}
\end{equation}

\begin{proposition}[Fixed-match replication for prospective PATE]
\label{prop:prospective-pate-fixed-match-identity}
Then \(\sum_i\widehat\varphi_{P,\mathrm{PATE},i}=0\) exactly, and
\begin{equation}\label{eq:prospective-pate-bootstrap-exact-ratio}
\begin{aligned}
&\sqrt n\Big(
\widehat\tau_{\mathrm{PATE}}^{P,\ast,b}
-\widehat\tau_{\mathrm{PATE}}^{P,\mathrm{BC}}
\Big)\\
&=
\frac{1}{\bar w_{P,\mathrm{PATE},n}^{\ast,b}}
\frac1{\sqrt n}\sum_{i=1}^n
(m_i^{\ast,b}-1)\widehat\varphi_{P,\mathrm{PATE},i}
+
\frac{\mathcal R_{P,\mathrm{PATE},n}^{\ast,b}}
{\sqrt n\,\bar w_{P,\mathrm{PATE},n}^{\ast,b}},
\end{aligned}
\end{equation}

Let \(\mathcal F_n\) contain the original observations, fits, and graph. The conditional variance of the centered multiplier numerator is exactly
\begin{equation}\label{eq:prospective-pate-bootstrap-numerator-variance}
\operatorname{Var}^{\ast}\left(
\frac1{\sqrt n}\sum_{i=1}^n
(m_i^{\ast,b}-1)\widehat\varphi_{P,\mathrm{PATE},i}
\,\middle|\,\mathcal F_n
\right)
=
\frac1n\sum_{i=1}^n\widehat\varphi_{P,\mathrm{PATE},i}^{\,2}.
\end{equation}

If, in addition, all conditions of Proposition~\ref{prop:pate-full-prediction-replication-law} hold with \(A=P\), then
\[
\mathcal L^\ast\!\left[
 \sqrt n(\widehat\tau_{\mathrm{PATE}}^{P,\ast,b}-\widehat\tau_{\mathrm{PATE}}^{P,\mathrm{BC}})
 \,\middle|\,\mathcal F_n\right]
 \Longrightarrow_{P_s}N(0,V_{P,\mathrm{PATE},\mathrm{rep}}),
\]
where
\[
\begin{aligned}
V_{P,\mathrm{PATE},\mathrm{rep}}
&=V_{P,\mathrm{PATE},\mathrm{known\text{-}score}}
 -\frac{2}{\bar w_{P,\mathrm{PATE}}}\,
 (\ell_{P,1}-\ell_{P,0})^\top\boldsymbol\Gamma_{P,\mathrm{PATE}}^{-1}\boldsymbol\gamma_{P,\mathrm{PATE},1}\\
&\quad+\frac1{\bar w_{P,\mathrm{PATE}}^2}\,
 (\ell_{P,1}-\ell_{P,0})^\top\boldsymbol\Sigma_{P,\mathrm{PATE},\boldsymbol\theta}(\ell_{P,1}-\ell_{P,0}).
\end{aligned}
\]
Suppose a sampling limit centered at \(\tau_{\mathrm{PATE}}\) has also been established, and
\[
\sqrt n(\widehat\tau_{\mathrm{PATE}}^{P,\mathrm{BC}}-\tau_{\mathrm{PATE}})
\Rightarrow N(0,V_{P,\mathrm{PATE}}),\qquad
V_{P,\mathrm{PATE}}=V_{P,\mathrm{PATE},\mathrm{rep}}>0.
\]
For conditionally independent replicates with the prescribed finite-valued outputs, the variance estimator in \eqref{eq:bootstrap-mc-original-variance} is consistent in the iterated limit \(n\to\infty\), then \(B\to\infty\), and along a sufficiently slowly increasing deterministic sequence \(B_n\). The associated normal Wald interval has asymptotic coverage \(1-\alpha\), for \(0<\alpha<1\), in the same regimes.
\label{par:prospective-pate-bootstrap-law-comparison}
\end{proposition}

\begin{proof}
Specialize the PATE ratio representation \eqref{eq:pate-fixed-prediction-ratio} and its centered form \eqref{eq:pate-common-centered-replication-ratio} to \(A=P\) and \(w_{A,z}(\mathbf X_i)=w(\mathbf X_i)\). Centering the numerator gives \eqref{eq:prospective-pate-fitted-centered-row}. Collecting the prediction changes in \eqref{eq:pate-common-fixed-refit-remainder} gives \eqref{eq:prospective-pate-bootstrap-refit-remainder}. Since the centered rows sum to zero, the multinomial covariance identity \eqref{eq:fixed-match-multinomial-moments} yields \eqref{eq:prospective-pate-bootstrap-numerator-variance}. Under the additional replication hypotheses, the refit expansion \eqref{eq:pate-full-prediction-refit-first-order} reduces the remainder to
\[
\frac{\mathcal R_{P,\mathrm{PATE},n}^{\ast,b}}{\sqrt n}
=(\ell_{P,1}-\ell_{P,0})^\top
 \sqrt n(\widehat{\boldsymbol \theta}_{P,\mathrm{PATE}}^{\ast,b}
              -\widehat{\boldsymbol \theta}_{P,\mathrm{PATE}})
 +o_{p^\ast}(1).
\]
The joint multiplier law and positive denominator limit give the stated conditional distribution. Under the stipulated positive sampling--replication variance agreement, Proposition~\ref{prop:bootstrap-monte-carlo-iterated-variance} gives variance consistency and Wald coverage in the stated regimes.
\end{proof}

\subsection{Prospective PATT}
\label{app:prospective_patt}
\subsubsection{Conditions}
For prospective PATT, both the outer treated average and the imputed control outcome use the inverse-inclusion weight \(w(\mathbf X)=\pi(\mathbf X)^{-1}\). Matching uses Euclidean distance on the fixed standardized control score \(\mathbf D_0^*(\mathbf X)\). The nonsingular standardization preserves the information in the propensity and control prognostic scores, so either correctly specified coordinate suffices for the stated one-sided identification.

We use the common control-score means
\[
m_{1\mid1}(\mathbf d)
:=E_\Omega\{Y(1)\mid Z=1,\mathbf D_0^*(\mathbf X)=\mathbf d\},
\qquad
m_0(\mathbf d):=m_0^P(\mathbf d)
=E_\Omega\{Y(0)\mid\mathbf D_0^*(\mathbf X)=\mathbf d\},
\]
and
\[
\tau_{\mathrm{PATT}}(\mathbf d):=m_{1\mid1}(\mathbf d)-m_0(\mathbf d).
\]
The observed-arm residuals are
\[
\varepsilon_{1\mid1,i}
:=Y_i-m_{1\mid1}\{\mathbf D_0^*(\mathbf X_i)\},\quad Z_i=1,
\qquad
\varepsilon_{0,j}
:=Y_j-m_0\{\mathbf D_0^*(\mathbf X_j)\},\quad Z_j=0.
\]
When an arm-specific residual or variance is multiplied by its treatment indicator, we take the product to be zero outside that observed arm.

The oracle initial field and observed-arm conditional variances are
\begin{equation}\label{eq:prospective-patt-oracle-field}
\mathcal G_{P,n}^{\mathrm{match},\mathrm{PATT}}
:=\sigma\{Z_i,\mathbf D_0^*(\mathbf X_i),w(\mathbf X_i):1\le i\le n\},
\end{equation}
\begin{equation}\label{eq:prospective-patt-variance-marks}
v_{z,i}^{\mathrm{PATT}}
:=\operatorname{Var}_s\{Y_i\mid
 Z_i=z,\mathbf D_0^*(\mathbf X_i),w(\mathbf X_i)\},\qquad Z_i=z.
\end{equation}
This is the prospective instance of \(\mathcal G_{A,n}^{\mathrm{match}}\) in the common centering lemma.

One-sided ignorability and overlap, together with a correct propensity score or a valid control prognostic score, imply the one-sided population identification condition below.

\begin{assumption}[Prospective PATT conditions]
\label{ass:prospective-patt-oracle}
\leavevmode
\begin{enumerate}
\item[(i)] The prospective recovery identity in Section~\ref{prospective-population-representation-identities} and the one-sided population identification
\[
E_\Omega\{Y(0)\mid Z=1,\mathbf D_0^*(\mathbf X)\}
=m_0\{\mathbf D_0^*(\mathbf X)\}
\]
hold.
\item[(ii)] Assumption~\ref{ass:residual-qv-stabilization} holds for prospective PATT.
\item[(iii)] The control-score density is bounded above and away from zero on a compact convex full-dimensional support. The treated-score density is bounded above and supported there. The common observed weight is bounded above and away from zero.
\item[(iv)] Both PATT centering conditions in Assumption~\ref{ass:residual-centering-survey-normalized} hold. The function \(m_0\) is Lipschitz on the control-score support. For some \(\delta>0\),
\[
E_s\!\left[Z|\varepsilon_{1\mid1}|^{2+\delta}
 +(1-Z)|\varepsilon_0|^{2+\delta}\right]<\infty,
\]
and
\[
E_s\!\left[
 \left|Zw(\mathbf X)
 \left[\tau_{\mathrm{PATT}}\{\mathbf D_0^*(\mathbf X)\}
       -\tau_{\mathrm{PATT}}\right]\right|^{2+\delta}
\right]<\infty.
\]
\end{enumerate}
\end{assumption}
\label{prospective-patt-with-exact-one-sided-within-match-survey-weight-normalized-imputation}

\subsubsection{Estimator and reuse coefficients}\label{exact-estimator-exact-survey-weighted-matched-set-coefficient-and-exact-matched-set-coefficient-representation}

For each treated individual \(i\), we define the within-match coefficient
\[\lambda_{i\to j}^{P,\mathrm{PATT}}
:=
\frac{w(\mathbf X_j)}{\sum_{k\in\mathcal J_{\mathbf D_0^*}(i)} w(\mathbf X_k)}\,\mathbf 1\{j\in\mathcal J_{\mathbf D_0^*}(i)\},
\qquad j=1,\dots,n,\] where \(\mathcal J_{\mathbf D_0^*}(i)\) is the set of the \(M\)
nearest control observations to treated unit \(i\) on \(\mathbf D_0^*(\mathbf X)\). Define
the survey-weighted imputed control outcome \[\widehat Y_i^{P}(0)
:=
\sum_{j=1}^n\lambda_{i\to j}^{P,\mathrm{PATT}}Y_j,
\qquad Z_i=1.\] The prospective PATT WDSM estimator is
\[\widehat\tau_{\mathrm{PATT}}^{P}
:=
\frac{\sum_{i:Z_i=1} w(\mathbf X_i)\{Y_i-\widehat Y_i^{P}(0)\}}{\sum_{i:Z_i=1} w(\mathbf X_i)}.\]
We denote the denominator mean by
\[\bar w_{P,\mathrm{PATT},n}:=\frac1n\sum_{i=1}^n Z_i w(\mathbf X_i).\] The one-sided weighted reuse coefficient is
\[K_{n,0}^{P,\mathrm{PATT}}(j)
:=
\sum_{i:Z_i=1}
 w(\mathbf X_i)
 \frac{w(\mathbf X_j)\mathbf 1\{j\in\mathcal J_{\mathbf D_0^*}(i)\}}{\sum_{k\in\mathcal J_{\mathbf D_0^*}(i)} w(\mathbf X_k)},
\qquad Z_j=0.\]

We define the three components by \[H_{n,P,\mathrm{PATT}}
:=
\frac{1}{n\bar w_{P,\mathrm{PATT},n}}
\sum_{i:Z_i=1}w(\mathbf X_i)\{\tau_{\mathrm{PATT}}\{\mathbf D_0^*(\mathbf X_i)\}-\tau_{\mathrm{PATT}}\},\]
\[B_{n,P,\mathrm{PATT}}
:=
\frac{1}{n\bar w_{P,\mathrm{PATT},n}}
\sum_{i:Z_i=1}w(\mathbf X_i)
\left[m_0\{\mathbf D_0^*(\mathbf X_i)\}-\sum_{j=1}^n\lambda_{i\to j}^{P,\mathrm{PATT}}m_0\{\mathbf D_0^*(\mathbf X_j)\}\right],\]
and \[E_{n,P,\mathrm{PATT}}
:=
\frac{1}{n\bar w_{P,\mathrm{PATT},n}}
\left[
\sum_{i:Z_i=1}w(\mathbf X_i)\varepsilon_{1\mid1,i}
-
\sum_{j:Z_j=0}K_{n,0}^{P,\mathrm{PATT}}(j)\varepsilon_{0,j}
\right].\]

For the known-score bias correction, we keep the same matched sets and normalized coefficients and define
\begin{equation}\label{eq:prospective-patt-oracle-bias-correction}
\widehat B_{n,P,\mathrm{PATT}}
:=
\frac{1}{n\bar w_{P,\mathrm{PATT},n}}
\sum_{i:Z_i=1}w(\mathbf X_i)
\left[
\widehat m_0\{\mathbf D_0^*(\mathbf X_i)\}
-\sum_j\lambda_{i\to j}^{P,\mathrm{PATT}}
        \widehat m_0\{\mathbf D_0^*(\mathbf X_j)\}
\right].
\end{equation}
The bias-correction remainder condition is
\begin{equation}\label{eq:prospective-patt-approved-bc-remainder}
\sqrt n\left(B_{n,P,\mathrm{PATT}}
            -\widehat B_{n,P,\mathrm{PATT}}\right)
\to_p0.
\end{equation}
Its error contrast is
\[
\begin{aligned}
B_{n,P,\mathrm{PATT}}
-\widehat B_{n,P,\mathrm{PATT}}
&=
\frac{1}{n\bar w_{P,\mathrm{PATT},n}}
\sum_{i:Z_i=1}w(\mathbf X_i)
\sum_j\lambda_{i\to j}^{P,\mathrm{PATT}}\\
&\quad\times
\left[
(m_0-\widehat m_0)\{\mathbf D_0^*(\mathbf X_i)\}
-(m_0-\widehat m_0)\{\mathbf D_0^*(\mathbf X_j)\}
\right].
\end{aligned}
\]

\subsubsection{Known-score asymptotics}
\label{exact-decomposition-and-exact-conditional-residual-variance-1}
\label{exact-known-score-asymptotics-3}

The heterogeneity variance is
\begin{equation}\label{eq:prospective-patt-h-variance}
V_{H,P,\mathrm{PATT}}
=
\frac{\operatorname{Var}_s\!\left(
 Zw(\mathbf X)
 \left[
 m_{1\mid1}\{\mathbf D_0^*(\mathbf X)\}
 -m_0\{\mathbf D_0^*(\mathbf X)\}-\tau_{\mathrm{PATT}}
 \right]\right)}
     {\left[E_s\{Zw(\mathbf X)\}\right]^2}.
\end{equation}
The one-sided prospective bias discrepancy is exactly
\[
\begin{aligned}
&m_0\{\mathbf D_0^*(\mathbf X_i)\}
-\sum_j\lambda_{i\to j}^{P,\mathrm{PATT}}
              m_0\{\mathbf D_0^*(\mathbf X_j)\}\\
&=\sum_j\lambda_{i\to j}^{P,\mathrm{PATT}}
 \left[
 m_0\{\mathbf D_0^*(\mathbf X_i)\}
 -m_0\{\mathbf D_0^*(\mathbf X_j)\}
 \right].
\end{aligned}
\]

For the same oracle match graph, we define \(\widehat\tau_{\mathrm{PATT}}^{P,\mathrm{BC}} :=\widehat\tau_{\mathrm{PATT}}^{P} -\widehat B_{n,P,\mathrm{PATT}}\).

\begin{theorem}[Prospective PATT with known scores]
\label{thm:prospective-patt-known-score}
\label{cor:prospective-patt-score-variance}
Under Assumption~\ref{ass:prospective-patt-oracle}, the two components have the joint limit
\[
\left(\sqrt n\,H_{n,P,\mathrm{PATT}},
      \sqrt n\,E_{n,P,\mathrm{PATT}}\right)
\Rightarrow (G_H,G_E),
\]
where \(G_H\sim N(0,V_{H,P,\mathrm{PATT}})\) and \(G_E\sim N(0,V_{E,P,\mathrm{PATT}})\) are independent. The matching discrepancy satisfies \(B_{n,P,\mathrm{PATT}}=O_p(n^{-1/2})\), and the uncorrected oracle estimator is consistent. If the bias correction in \eqref{eq:prospective-patt-oracle-bias-correction} satisfies \eqref{eq:prospective-patt-approved-bc-remainder}, then
\[
\sqrt n\left(
\widehat\tau_{\mathrm{PATT}}^{P,\mathrm{BC}}-\tau_{\mathrm{PATT}}
\right)
\Rightarrow N\!\left(
0,V_{H,P,\mathrm{PATT}}+V_{E,P,\mathrm{PATT}}
\right).
\]
The bias-corrected estimator is therefore consistent under the stated remainder condition. Corollary~\ref{cor:oracle-regression-bc} gives sufficient derivative-consistency conditions for this remainder.

The total variance is finite and nonnegative. It is strictly positive, for example, if the heterogeneity summand \(Zw(\mathbf X)[\tau_{\mathrm{PATT}}\{\mathbf D_0^*(\mathbf X)\} -\tau_{\mathrm{PATT}}]\) has positive selected-sample variance, or if \(P_s(Z=1,v_1^{\mathrm{PATT}}>0)>0\). The control-only sufficient condition in Proposition~\ref{prop:primitive-residual-qv-limit} also applies.

\label{lem:prospective-patt-decomposition}
\label{prop:prospective-patt-representation}
\label{lem:prospective-patt-residual-variance}
\label{prop:prospective-patt-bias-order}
\label{lem:prospective-patt-h-clt}
\label{lem:prospective-patt-qv}
\end{theorem}
\begin{proof}
The martingale argument in Theorem~\ref{thm:retro-patt-oracle} applies with \(w_0=w_1=w\). The prospective recovery identity and the one-sided mean-balancing identity in Assumption~\ref{ass:prospective-patt-oracle}(i) give
\[
E_s\!\left[Zw(\mathbf X)
 \{\tau_{\mathrm{PATT}}(\mathbf D_0^*(\mathbf X))
                 -\tau_{\mathrm{PATT}}\}\right]=0,
\qquad
\bar w_{P,\mathrm{PATT},n}\to_p E_s\{Zw(\mathbf X)\}>0.
\]
The normalized matching coefficients and their reindexing are the corresponding retrospective quantities with this common weight. Hence Proposition~\ref{prop:exact-pate-score-space-decomposition} gives
\[
\widehat\tau_{\mathrm{PATT}}^{P}-\tau_{\mathrm{PATT}}
=H_{n,P,\mathrm{PATT}}
 +B_{n,P,\mathrm{PATT}}
 +E_{n,P,\mathrm{PATT}}.
\]

Conditional on \eqref{eq:prospective-patt-oracle-field}, the observed-arm residuals are independent and centered by Assumption~\ref{ass:prospective-patt-oracle}(ii)--(iv). With the variances in \eqref{eq:prospective-patt-variance-marks},
\begin{equation}\label{eq:prospective-patt-exact-qv}
\begin{aligned}
&\operatorname{Var}_s\!\left(
 \sqrt n E_{n,P,\mathrm{PATT}}
 \mid\mathcal G_{P,n}^{\mathrm{match},\mathrm{PATT}}\right)\\
&\quad=
\frac{1}{n\bar w_{P,\mathrm{PATT},n}^{\,2}}
\left[
 \sum_{i:Z_i=1}w(\mathbf X_i)^2v_{1,i}^{\mathrm{PATT}}
 +\sum_{j:Z_j=0}
   K_{n,0}^{P,\mathrm{PATT}}(j)^2v_{0,j}^{\mathrm{PATT}}
\right].
\end{aligned}
\end{equation}
Proposition~\ref{prop:primitive-residual-qv-limit} yields
\[
\operatorname{Var}_s\!\left(
 \sqrt n E_{n,P,\mathrm{PATT}}
 \mid\mathcal G_{P,n}^{\mathrm{match},\mathrm{PATT}}\right)
\longrightarrow_{L^1}
\frac{\Lambda_{P,0}^{\mathrm{PATT}}+\Lambda_{P,1}^{\mathrm{PATT}}}
     {\{E_s[Zw(\mathbf X)]\}^2}
=V_{E,P,\mathrm{PATT}}.
\]
The heterogeneity summands have finite second moments under Assumption~\ref{ass:prospective-patt-oracle}(iv), and \eqref{eq:qv-derived-lyapunov} gives the residual-stage Lindeberg condition. Applying the two-stage argument of Theorem~\ref{thm:retro-patt-oracle} with this initial field and normalizer gives, for every \(a,b\in\mathbb R\),
\[
a\sqrt n H_{n,P,\mathrm{PATT}}
+b\sqrt n E_{n,P,\mathrm{PATT}}
\Rightarrow
N\!\left(0,a^2V_{H,P,\mathrm{PATT}}
          +b^2V_{E,P,\mathrm{PATT}}\right).
\]
The Cram\'er--Wold device gives the asserted joint limit and independence of its Gaussian coordinates, including the degenerate cases.

Lemma~\ref{lem:oracle-score-radius} with \(z=0\), together with the positive limit of the treated denominator, gives \(B_{n,P,\mathrm{PATT}}=O_p(n^{-1/2})=o_p(1)\). The joint limit above and the exact decomposition therefore prove consistency of the uncorrected estimator. Under \eqref{eq:prospective-patt-approved-bc-remainder},
\[
\begin{aligned}
\sqrt n\{\widehat\tau_{\mathrm{PATT}}^{P,\mathrm{BC}}
                 -\tau_{\mathrm{PATT}}\}
&=\sqrt n(H_{n,P,\mathrm{PATT}}
          +E_{n,P,\mathrm{PATT}})
  +\sqrt n(B_{n,P,\mathrm{PATT}}
           -\widehat B_{n,P,\mathrm{PATT}})\\
&=\sqrt n(H_{n,P,\mathrm{PATT}}
          +E_{n,P,\mathrm{PATT}})+o_p(1).
\end{aligned}
\]
Slutsky's theorem proves the corrected limit and consistency. The heterogeneity variance and Proposition~\ref{prop:primitive-residual-qv-limit} give the stated finiteness and positivity conclusions.
\end{proof}

\subsubsection{Estimated-score asymptotics}\label{prospective-patt-estimated-score-results}

Proposition~\ref{cor:primary-fitted-score-consistency} establishes consistency under its stated conditions, with the bias-correction remainder evaluated on the estimated-score match sets.

For the prospective PATT scenario, let
\[
V_{P,\mathrm{PATT}}^{\mathrm{known\text{-}score}}
:=
V_{H,P,\mathrm{PATT}}+V_{E,P,\mathrm{PATT}}.
\]
Let \(\widehat{\boldsymbol \theta}_{P,\mathrm{PATT}}\) collect the prospective unweighted propensity-score estimator together with the unweighted control prognostic-score estimator, and let \(\bar{\boldsymbol \theta}_{P,\mathrm{PATT},n,\eta}\) denote its midpoint discretization on cubes of side length \(\eta/\sqrt n\), with deterministic grid offset. The nuisance vector is \(\boldsymbol\theta=(\boldsymbol\alpha^\top,\boldsymbol\beta_0^\top)^\top\). The common-\(\mathbf D_0^\ast\) contrast and treated normalization from Theorem~\ref{thm:prospective-patt-known-score} remain
\[
\begin{aligned}
\tau_{\mathrm{PATT}}(\mathbf D_0^\ast)
&=m_{1\mid1}(\mathbf D_0^\ast)-m_0(\mathbf D_0^\ast),\\
\bar w_{P,\mathrm{PATT},n}
&=\frac1n\sum_{i=1}^n Z_iw(\mathbf X_i).
\end{aligned}
\]
For prospective PATT, Theorem~\ref{cor:patt-verified-local-grid-law}(i) gives the local Gaussian limit under its stated conditions, with mean \eqref{eq:patt-verified-local-mean-coefficient}.

Under the pathwise identification and differentiation conditions of Proposition~\ref{prop:scientific-target-path-derivative}, the scenario target and its derivative can be written as
\[
\tau_A(Q_{A,t})
=\frac{E_{Q_{A,t}}[Zw(\mathbf X)\{\mu_{1\mid1,t}(\mathbf X)-\mu_{0,t}(\mathbf X)\}]}
       {E_{Q_{A,t}}[Zw(\mathbf X)]},
\]
and
\[
\boldsymbol\gamma_{P,\mathrm{PATT},2}
=E_s\{\phi_P(O)\mathbf s_A(O)\},
\qquad \mathbf s_A(O)=-\boldsymbol\Gamma_A^\top\boldsymbol\Sigma_{A,U}^{-1}\mathbf U_A(O;\boldsymbol\theta_A^\ast),
\]
with the weight functions fixed along the path. Define
\[
V_{P,\mathrm{PATT}}^{\mathrm{fix}}
=V_{P,\mathrm{PATT}}^{\mathrm{known\text{-}score}}
-\boldsymbol\gamma_{P,\mathrm{PATT},1}^{\top}
 \boldsymbol\Sigma_{A,U}^{-1}\boldsymbol\gamma_{P,\mathrm{PATT},1},
\]
and
\[
V_{P,\mathrm{PATT}}^{\mathrm{grid}}
=V_{P,\mathrm{PATT}}^{\mathrm{fix}}
+\widetilde{\boldsymbol \gamma}_{P,\mathrm{PATT},2}^\top
 \boldsymbol\Sigma_{A,\theta}\widetilde{\boldsymbol \gamma}_{P,\mathrm{PATT},2}.
\]

\begin{theorem}[Prospective PATT with estimated scores]
\label{thm:prospective-patt-estimated-scores}
\label{prop:prospective-patt-estimated-score-conditional}
For PATT, put \(A=P\). Retain the first-step and auxiliary-experiment conditions of Lemma~\ref{lem:auxiliary-godambe-experiment}, including \(\boldsymbol\Sigma_{A,U}>0\). Suppose that \eqref{eq:moving-law-joint-input-with-bias} holds for the specified standardization, matching and bias-correction family at every deterministic \(\mathbf h_n\to \mathbf h\), with
\[
V_{A,\mathrm{known\text{-}score}}
=V_{P,\mathrm{PATT}}^{\mathrm{known\text{-}score}}
=V_{H,P,\mathrm{PATT}}+V_{E,P,\mathrm{PATT}},\qquad
\boldsymbol\gamma_{A,1}=\boldsymbol\gamma_{P,\mathrm{PATT},1},\quad
\widetilde{\boldsymbol \gamma}_{A,2}=\widetilde{\boldsymbol \gamma}_{P,\mathrm{PATT},2}.
\]
Suppose also that \(\tau_A(Q_{A,t})\) is identified and differentiable along these same auxiliary laws, with
\[
\begin{aligned}
\tau_A(Q_{A,t})
&=\tau_{\mathrm{PATT}}+\boldsymbol\gamma_{P,\mathrm{PATT},2}^{\top}t+o(\|t\|),\\
\boldsymbol\gamma_{P,\mathrm{PATT},2}
&:=\left.\frac{\partial}{\partial t}\tau_A(Q_{A,t})\right|_{t=0}.
\end{aligned}
\]
If, in addition, all hypotheses of Theorem~\ref{thm:patt-sufficient-branch-actual-score-law} hold with \(A=P\) for PATT, using the prospective weights and the original treated normalization displayed above, then the original fixed-\(M\), survey-normalized estimator obeys
\[
\sqrt n\{\widehat\tau_{\mathrm{PATT}}^{P,\mathrm{BC}}
(\widehat{\boldsymbol \theta}_A)-\tau_{\mathrm{PATT}}\}
\Rightarrow N(0,V_{P,\mathrm{PATT}}^{\mathrm{grid}}).
\]
The scenario variance, local mean and strict-positivity alternatives are given by \eqref{eq:patt-verified-geometric-grid-variance}. The variance is finite and nonnegative.
\label{cor:prospective-patt-actual-sufficient}

For the midpoint discretization, in the order \(n\to\infty\), \(\eta\downarrow0\),
\[
\sqrt n\{\widehat\tau_{\mathrm{PATT}}^{P,\mathrm{BC}}
(\bar{\boldsymbol \theta}_{A,n,\eta})-\tau_{\mathrm{PATT}}\}
\Rightarrow N(0,V_{P,\mathrm{PATT}}^{\mathrm{grid}}).
\]
The double limit is in the bounded-Lipschitz sense of Proposition~\ref{prop:local-mean-corrected-grid-variance}.

\end{theorem}
\begin{proof}
The estimating function is
\[
\mathbf U_A(O;\boldsymbol\theta)=
\begin{pmatrix}
\mathbf X\{Z-e(\mathbf X;\boldsymbol\alpha)\}\\
(1-Z)\mathbf X\{Y-\mathbf X^\top\boldsymbol\beta_0\}
\end{pmatrix},
\qquad
e(\mathbf X;\boldsymbol\alpha)=\operatorname{expit}
(\mathbf X^\top\boldsymbol\alpha).
\]
Its derivative is
\[
\boldsymbol\Gamma_A=-\operatorname{diag}\!\left(
E_s[e^\ast(1-e^\ast)\mathbf X\mathbf X^\top],
E_s[(1-Z)\mathbf X\mathbf X^\top]\right),
\]
where \(e^\ast=e(\mathbf X;\boldsymbol\alpha^\ast)\). Using the full two-block covariance matrix \(\boldsymbol\Sigma_{A,U}\), Proposition~\ref{prop:local-mean-corrected-grid-variance} gives the grid limit \(G_{\tau,A}^\perp+\widetilde{\boldsymbol \gamma}_{A,2}^\top \mathbf G_{\theta,A}\). Its nuisance covariance is \(\boldsymbol\Sigma_{A,\theta}=\boldsymbol\Gamma_A^{-1}\boldsymbol\Sigma_{A,U}\boldsymbol\Gamma_A^{-T}\) by Lemma~\ref{lem:ordinary-weighted-first-step}. The Gaussian projection calculation in the retrospective PATE proof then gives \(V_{P,\mathrm{PATT}}^{\mathrm{grid}}\). The target representer follows from Proposition~\ref{prop:scientific-target-path-derivative}, under its conditions, with weight \(Zw(\mathbf X)\).

Under the additional hypotheses for the undiscretized estimator, Theorem~\ref{thm:patt-sufficient-branch-actual-score-law} uses the same control score, normalized matching fractions and treated denominator. Its variance \(V_{A,\mathrm{grid}}\) is therefore \(V_{P,\mathrm{PATT}}^{\mathrm{grid}}\).
\end{proof}

\paragraph{Full-control-mean centering.}
\label{par:prospective-patt-grouped-sampling}
Under all conditions of Theorem~\ref{thm:patt-grouped-fullx-actual}, with \(A=P\), including its control-mean equality, score geometry, estimating equations, empirical normalization and the two distinct BC remainders,
\[
\begin{pmatrix}
\sqrt n\{\widehat\tau_{\mathrm{PATT}}^{P,\mathrm{BC}}
              (\widehat{\boldsymbol \theta}_P)-\tau_{\mathrm{PATT}}\}\\
\sqrt n(\widehat{\boldsymbol \theta}_P-\boldsymbol\theta_P^*)
\end{pmatrix}
\Rightarrow
\begin{pmatrix}\widetilde G_{\tau,P}\\\mathbf G_{\theta,P}\end{pmatrix}.
\]
Here
\[
\operatorname{Var}(\widetilde G_{\tau,P})
 =\widetilde V_{P,\mathrm{PATT}},\qquad
\operatorname{Cov}(\mathbf G_{\theta,P},\widetilde G_{\tau,P})
 =-\boldsymbol\Gamma_P^{-1}\widetilde{\boldsymbol \gamma}_{P,1},
\]
with the whole treated-observation contribution retained in \eqref{eq:patt-grouped-fullx-constants}.

\subsubsection{Fixed-match replication}\label{prospective-patt-bootstrap-main-text}

For each fixed-match replicate \(b=1,\ldots,B\), we draw
\[
(m_1^{\ast,b},\ldots,m_n^{\ast,b})
\sim
\mathrm{Multinomial}(n;1/n,\ldots,1/n),
\]
and define
\[
w_i^{\ast,b}:=m_i^{\ast,b}w(\mathbf X_i).
\]
We refit the prospective propensity score with multinomial weights \(m_i^{\ast,b}\) and the control prognostic score with the same multinomial weights on \(Z_i=0\), using the original unweighted score equations. We recompute \(\widehat{\mathbf D}_0^{\ast,b}\), including its standardization by the original rule, and refit the control bias-correction regression on these scores with weights \(w_i^{\ast,b}\) to obtain \(\widehat m_0^{\ast,b}\). The control match sets, the donor-normalized coefficients \(\lambda_{i\to j}^{P,(0)}\), and \(K_{n,0}^{P,\mathrm{PATT}}(j)\) remain fixed at the original estimated-score graph.

For a treated observation, the common-score cancellation is
\[
\begin{aligned}
&\tau_{\mathrm{PATT}}\{\mathbf D_0^*(\mathbf X_i)\}
-\tau_{\mathrm{PATT}}+\varepsilon_{1\mid1,i}\\
&=
m_{1\mid1}\{\mathbf D_0^*(\mathbf X_i)\}
-m_0^P\{\mathbf D_0^*(\mathbf X_i)\}
-\tau_{\mathrm{PATT}}+\varepsilon_{1\mid1,i}\\
&=
Y_i-m_0^P\{\mathbf D_0^*(\mathbf X_i)\}
-\tau_{\mathrm{PATT}}.
\end{aligned}
\]
The one-sided fixed-match replicate is
\begin{equation}\label{eq:prospective-patt-fixed-match-replicate}
\begin{aligned}
\widehat\tau_{\mathrm{PATT}}^{P,\ast,b}
={}&\frac1{\sum_{i:Z_i=1}w_i^{\ast,b}}
\Bigg[
\sum_{i:Z_i=1}w_i^{\ast,b}
\Big[
Y_i-\widehat m_0^{\ast,b}
 \{\widehat{\mathbf D}_0^{\ast,b}(\mathbf X_i)\}
\Big]\\
&\qquad-
\sum_{j:Z_j=0}m_j^{\ast,b}
 K_{n,0}^{P,\mathrm{PATT}}(j)
\Big[
Y_j-\widehat m_0^{\ast,b}
 \{\widehat{\mathbf D}_0^{\ast,b}(\mathbf X_j)\}
\Big]
\Bigg].
\end{aligned}
\end{equation}
The centering estimator \(\widehat\tau_{\mathrm{PATT}}^{P,\mathrm{BC}}\) uses the same graph and control fit \(\widehat m_0\{\widehat{\mathbf D}_0(\mathbf X)\}\). The identities below hold when the original and replicated fits exist and the replicated treated denominator is positive.

We use the treated denominator averages
\[
\begin{aligned}
\bar w_{P,\mathrm{PATT},n}
&=\frac1n\sum_i Z_iw(\mathbf X_i),\\
\bar w_{P,\mathrm{PATT},n}^{\ast,b}
&=\frac1n\sum_i m_i^{\ast,b}Z_iw(\mathbf X_i).
\end{aligned}
\]
The centered contribution of individual \(i\), evaluated at the original fitted values, is
\begin{equation}\label{eq:prospective-patt-fitted-centered-row}
\widehat\varphi_{P,\mathrm{PATT},i}
:=
\begin{cases}
w(\mathbf X_i)
\bigl[
Y_i-\widehat m_0\{\widehat{\mathbf D}_0(\mathbf X_i)\}
-\widehat\tau_{\mathrm{PATT}}^{P,\mathrm{BC}}
\bigr],
& Z_i=1,\\[1mm]
-K_{n,0}^{P,\mathrm{PATT}}(i)
\bigl[
Y_i-\widehat m_0\{\widehat{\mathbf D}_0(\mathbf X_i)\}
\bigr],
& Z_i=0.
\end{cases}
\end{equation}

The full refit contribution is
\begin{equation}\label{eq:prospective-patt-bootstrap-refit-remainder}
\begin{aligned}
\mathcal R_{P,\mathrm{PATT},n}^{\ast,b}
={}&-\sum_{i:Z_i=1}m_i^{\ast,b}w(\mathbf X_i)
\Big[
\widehat m_0^{\ast,b}
 \{\widehat{\mathbf D}_0^{\ast,b}(\mathbf X_i)\}
-\widehat m_0\{\widehat{\mathbf D}_0(\mathbf X_i)\}
\Big]\\
&+\sum_{j:Z_j=0}m_j^{\ast,b}
 K_{n,0}^{P,\mathrm{PATT}}(j)
\Big[
\widehat m_0^{\ast,b}
 \{\widehat{\mathbf D}_0^{\ast,b}(\mathbf X_j)\}
-\widehat m_0\{\widehat{\mathbf D}_0(\mathbf X_j)\}
\Big].
\end{aligned}
\end{equation}

\begin{proposition}[Fixed-match replication for prospective PATT]
\label{prop:prospective-patt-fixed-match-identity}
Then \(\sum_i\widehat\varphi_{P,\mathrm{PATT},i}=0\), and the exact ratio is
\begin{equation}\label{eq:prospective-patt-bootstrap-exact-ratio}
\begin{aligned}
&\sqrt n\Big(
\widehat\tau_{\mathrm{PATT}}^{P,\ast,b}
-\widehat\tau_{\mathrm{PATT}}^{P,\mathrm{BC}}
\Big)\\
&=
\frac1{\bar w_{P,\mathrm{PATT},n}^{\ast,b}}
\frac1{\sqrt n}\sum_i
(m_i^{\ast,b}-1)\widehat\varphi_{P,\mathrm{PATT},i}
+
\frac{\mathcal R_{P,\mathrm{PATT},n}^{\ast,b}}
{\sqrt n\,\bar w_{P,\mathrm{PATT},n}^{\ast,b}},
\end{aligned}
\end{equation}

With \(\mathcal F_n\) containing the original observations, fits, and graph, the full multinomial law gives
\begin{equation}\label{eq:prospective-patt-bootstrap-numerator-variance}
\operatorname{Var}^\ast\left(
\frac1{\sqrt n}\sum_i(m_i^{\ast,b}-1)
\widehat\varphi_{P,\mathrm{PATT},i}
\,\middle|\,\mathcal F_n\right)
=
\frac1n\sum_i\widehat\varphi_{P,\mathrm{PATT},i}^{\,2}.
\end{equation}

If, in addition, all conditions of Proposition~\ref{prop:patt-full-prediction-replication-law} hold with \(A=P\), then
\[
\mathcal L^\ast\!\left[
 \sqrt n(\widehat\tau_{\mathrm{PATT}}^{P,\ast,b}-\widehat\tau_{\mathrm{PATT}}^{P,\mathrm{BC}})
 \,\middle|\,\mathcal F_n\right]
 \Longrightarrow_{P_s}N(0,V_{P,\mathrm{rep}}),
\]
where
\[
\begin{aligned}
V_{P,\mathrm{rep}}
&=V_{P,\mathrm{known\text{-}score}}
 +\frac{2}{\bar w_{P}}\,
 \ell_{P,0}^\top\boldsymbol\Gamma_{P}^{-1}\boldsymbol\gamma_{P,1}\\
&\quad+\frac1{\bar w_{P}^2}\,
 \ell_{P,0}^\top\boldsymbol\Sigma_{P,\theta}\ell_{P,0}.
\end{aligned}
\]
Suppose a sampling limit centered at \(\tau_{\mathrm{PATT}}\) has also been established, and
\[
\sqrt n(\widehat\tau_{\mathrm{PATT}}^{P,\mathrm{BC}}-\tau_{\mathrm{PATT}})
\Rightarrow N(0,V_{P,\mathrm{PATT}}),\qquad
V_{P,\mathrm{PATT}}=V_{P,\mathrm{rep}}>0.
\]
For conditionally independent replicates with the prescribed finite-valued outputs, the variance estimator in \eqref{eq:bootstrap-mc-original-variance} is consistent in the iterated limit \(n\to\infty\), then \(B\to\infty\), and along a sufficiently slowly increasing deterministic sequence \(B_n\). The associated normal Wald interval has asymptotic coverage \(1-\alpha\), for \(0<\alpha<1\), in the same regimes.
\label{par:prospective-patt-bootstrap-law-comparison}
\end{proposition}

\begin{proof}
Specialize the PATT ratio representation \eqref{eq:patt-fixed-prediction-ratios} and its centered form \eqref{eq:patt-common-centered-replication-ratio} to \(A=P\) and \(w_{A,z}(\mathbf X_i)=w(\mathbf X_i)\). The fitted row in \eqref{eq:patt-common-fitted-centered-row} is then \eqref{eq:prospective-patt-fitted-centered-row}. Collecting the prediction changes in \eqref{eq:patt-full-prediction-exact-refit-remainder} gives \eqref{eq:prospective-patt-bootstrap-refit-remainder}. The centered rows sum to zero, so \eqref{eq:fixed-match-multinomial-moments} gives the numerator variance. The zero-denominator bound \eqref{eq:fixed-match-treated-denominator-zero}, together with the iid denominator law, gives a positive denominator limit. Under the additional replication hypotheses, the refit expansion \eqref{eq:patt-full-prediction-refit-first-order} reduces the remainder to
\[
\frac{\mathcal R_{P,\mathrm{PATT},n}^{\ast,b}}{\sqrt n}
=-\ell_{P,0}^\top
 \sqrt n(\widehat{\boldsymbol \theta}_{P,\mathrm{PATT}}^{\ast,b}
              -\widehat{\boldsymbol \theta}_{P,\mathrm{PATT}})
 +o_{p^\ast}(1).
\]
The joint multiplier law gives the stated conditional distribution. Under the stipulated positive sampling--replication variance agreement, Proposition~\ref{prop:bootstrap-monte-carlo-iterated-variance} gives variance consistency and Wald coverage in the stated regimes.
\end{proof}

\subsection{Conditionally stabilized WDSM}
\label{app:conditional-stabilization-extension}

\subsubsection{Definition and identification}
\label{app:stabilization-definition}

The conditionally stabilized WDSM estimator averages the adjusted matched residuals with coefficient \(1/M\). We establish consistency for this alternative estimator below. The preceding asymptotic normality and replication results concern the original WDSM estimator. {The simulations and case studies in the main manuscript evaluate the original estimator.}

We use the same \(w_{R,z}=w_z\), \(w_{P,z}=w\), and reduced population means \(m_z^A\), for \(A\in\{R,P\}\). Let
\[
\rho_{A,z}(d)
:=E_s\{w_{A,z}(\mathbf X)\mid Z=z,\mathbf D_z^\ast(\mathbf X)=d\}.
\]
The weighted reduced-mean recovery identity of Theorem~\ref{lem:weighted-score-population-mean}(i) is
\[
m_z^A(d)
=
\frac{E_s\{w_{A,z}(\mathbf X)Y\mid Z=z,\mathbf D_z^\ast=d\}}
     {\rho_{A,z}(d)}.
\]
We assume population mean-balancing identification: \(m_z^A(\mathbf D_z^\ast)\) identifies the relevant potential-outcome mean in both arms for PATE, and for \(Y(0)\) on the common control score \(\mathbf D_0^\ast\) for PATT.

Suppose the two conditional expectations in the preceding ratio are finite and \(0<\rho_{A,z}(d)<\infty\) on the relevant score support. For \(Z_i\ne z\), define the stabilized oracle imputation using the fixed-\(M\) nearest-neighbor set \(\mathcal J_{\mathbf D_z^\ast}(i)\) on deterministic population-standardized scores:
\[
\begin{aligned}
\widetilde Y_i^{A,\mathrm{BC,or}}(z)
&:=
m_z^A\{\mathbf D_z^\ast(\mathbf X_i)\}\\
&\quad+
\frac1M\sum_{j\in\mathcal J_{\mathbf D_z^\ast}(i)}
\frac{w_{A,z}(\mathbf X_j)}
{\rho_{A,z}\{\mathbf D_z^\ast(\mathbf X_j)\}}
\left[
Y_j-m_z^A\{\mathbf D_z^\ast(\mathbf X_j)\}
\right],
\end{aligned}
\]
and set \(\widetilde Y_i^{A,\mathrm{BC,or}}(Z_i)=Y_i\). The resulting estimators use the original outer averages:
\[
\widetilde\tau_{\mathrm{PATE}}^{A,\mathrm{BC,or}}
=
\frac{\sum_iw_{A,Z_i}(\mathbf X_i)
\{\widetilde Y_i^{A,\mathrm{BC,or}}(1)
-\widetilde Y_i^{A,\mathrm{BC,or}}(0)\}}
{\sum_iw_{A,Z_i}(\mathbf X_i)},
\]
\[
\widetilde\tau_{\mathrm{PATT}}^{A,\mathrm{BC,or}}
=
\frac{\sum_iZ_iw_{A,1}(\mathbf X_i)
\{Y_i-\widetilde Y_i^{A,\mathrm{BC,or}}(0)\}}
{\sum_iZ_iw_{A,1}(\mathbf X_i)}.
\]
\subsubsection{Oracle consistency}
\label{app:stabilization-oracle}

Assume the following conditions.
\begin{enumerate}
\item[(i)] The selected observations are i.i.d., with \(P_s(Z=a)\in(0,1)\), \(a=0,1\). Matching uses fixed \(M\) and deterministic two-dimensional Euclidean score metrics.
\item[(ii)] For each matching direction, the score support \(\mathcal S_z\) in arm \(z\) is compact and convex with nonempty interior. The density in arm \(z\) is bounded above and away from zero. The density in arm \(1-z\) is bounded above and supported on \(\mathcal S_z\). Write \(f_{a\mid z}\) for the density of \(\mathbf D_z^\ast\) conditional on \(Z=a\).
\item[(iii)] For some \(\delta_0>0\),
\[
0<w_{\min}\le w_{A,Z}(\mathbf X)\le w_{\max}<\infty,
\qquad E_s|Y|^{2+\delta_0}<\infty.
\]
In particular \(w_{\min}\le\rho_{A,z}\le w_{\max}\).
\item[(iv)] The relevant population-recovery conditions in Lemma~\ref{lem:retrospective-population-recovery} and the preceding reduced-mean and potential-outcome identification conditions hold.
\end{enumerate}

\begin{proposition}[Oracle consistency]
\label{prop:stabilizer-oracle-consistency}
\label{prop:stabilizer-identification}
Under these conditions, the two stabilized oracle estimators are consistent for their corresponding identified PATE and PATT.
\end{proposition}
\begin{proof}
The weighted regression ratio centers the adjusted donor residual:
\[
\begin{aligned}
E_s\!\left[\left.
\frac{w_{A,z}}{\rho_{A,z}(d)}\{Y-m_z^A(d)\}
\right|Z=z,\mathbf D_z^\ast=d\right]
&=\frac{E_s(w_{A,z}Y\mid Z=z,\mathbf D_z^\ast=d)}
       {\rho_{A,z}(d)}\\
&\quad-m_z^A(d)
 \frac{E_s(w_{A,z}\mid Z=z,\mathbf D_z^\ast=d)}{\rho_{A,z}(d)}\\
&=\frac{E_s(w_{A,z}Y\mid Z=z,\mathbf D_z^\ast=d)}
       {\rho_{A,z}(d)}-m_z^A(d)\\
&=0.
\end{aligned}
\]

For one treatment arm, write
\[
r_{A,z,j}
:=
\frac{w_{A,z}(\mathbf X_j)}
{\rho_{A,z}\{\mathbf D_z^\ast(\mathbf X_j)\}}
\left[Y_j-m_z^A\{\mathbf D_z^\ast(\mathbf X_j)\}\right],
\qquad Z_j=z.
\]
The centering calculation gives \(E_s(r_{A,z,j}\mid Z_j=z,\mathbf D_z^*(\mathbf X_j))=0\). Applying conditional Jensen's inequality under the weighted conditional law gives
\[
\begin{aligned}
|m_z^A(d)|^{2+\delta_0}
&\le\frac{E_s(w_{A,z}|Y|^{2+\delta_0}\mid Z=z,\mathbf D_z^\ast=d)}
              {\rho_{A,z}(d)}\\
&\le\frac{w_{\max}}{w_{\min}}
 E_s(|Y|^{2+\delta_0}\mid Z=z,\mathbf D_z^\ast=d).
\end{aligned}
\]
The donor-density lower and query-density upper bounds therefore imply, for \(a=0,1\),
\[
\begin{aligned}
E_s\{\mathbf1(Z=a)|m_z^A(\mathbf D_z^\ast)|^{2+\delta_0}\}
&\le C E_s\{\mathbf1(Z=z)|Y|^{2+\delta_0}\}<\infty,\\
E_s\{\mathbf1(Z=z)|r_{A,z}|^{2+\delta_0}\}
&\le C E_s\!\left[\mathbf1(Z=z)
 \{|Y|^{2+\delta_0}+|m_z^A(\mathbf D_z^\ast)|^{2+\delta_0}\}\right]
<\infty.
\end{aligned}
\]

We condition on the arm-specific field
\[
\mathcal H_{n,z}
:=
\sigma\!\left\{
(Z_i,\mathbf D_z^\ast(\mathbf X_i))_{i=1}^n,\,
\bigl(\mathbf1\{Z_i=1-z\}w_{A,1-z}(\mathbf X_i)\bigr)_{i=1}^n
\right\}.
\]
Under i.i.d.\ sampling from the selected-sample law, the residuals in arm \(z\) are conditionally independent given \(\mathcal H_{n,z}\), and
\[
E_s(r_{A,z,j}\mid\mathcal H_{n,z})=0,\qquad Z_j=z.
\]
Reindexing the matched residuals gives the coefficient
\[
C_{n,A,z}(j)
:=
\frac1M\sum_{i:Z_i=1-z}
w_{A,1-z}(\mathbf X_i)
\mathbf1\{j\in\mathcal J_{\mathbf D_z^\ast}(i)\}.
\]
It is \(\mathcal H_{n,z}\)-measurable. With \(K_{M,z}(j)=\sum_{i:Z_i=1-z} \mathbf1\{j\in\mathcal J_{\mathbf D_z^\ast}(i)\}\),
\[
0\le C_{n,A,z}(j)\le\frac{w_{\max}}{M}K_{M,z}(j).
\]
Reindexing the stabilized donor average by donor gives
\[
\begin{aligned}
R_{n,A,z}^{\mathrm{or}}
&:=
\frac1{nM}\sum_{i:Z_i=1-z}w_{A,1-z}(\mathbf X_i)
\sum_{j\in\mathcal J_{\mathbf D_z^\ast}(i)}r_{A,z,j}\\
&=
\frac1n\sum_{j:Z_j=z}C_{n,A,z}(j)r_{A,z,j}.
\end{aligned}
\]
Conditional centering and independence remove the cross terms, giving
\[
\begin{aligned}
E_s\{(R_{n,A,z}^{\mathrm{or}})^2\}
&=
\frac1{n^2}E_s\!\left[
\sum_{j:Z_j=z}C_{n,A,z}(j)^2
E_s(r_{A,z,j}^2\mid Z_j,\mathbf D_z^\ast(\mathbf X_j))
\right]\\
&\le
\frac{w_{\max}^2}{M^2n^2}
E_s\!\left[
\sum_{j:Z_j=z}K_{M,z}(j)^2
E_s(r_{A,z,j}^2\mid Z_j,\mathbf D_z^\ast(\mathbf X_j))
\right]\\
&\le
\frac{C}{n}E_s\{\mathbf1(Z=z)r_{A,z}^2\}.
\end{aligned}
\]
For the last inequality, conditional on individual \(j\)'s treatment and matching score, its ordinary reuse count depends only on the other observations and is independent of its outcome and survey weight. The uniform conditional reuse-moment bound in \eqref{eq:stabilizer-inflated-reuse}, with \(a=0\), follows from the i.i.d.\ sampling, support and density conditions stated here and requires no additional residual-centering condition. Consequently, \(R_{n,A,z}^{\mathrm{or}}=O_p(n^{-1/2})\).

For a PATE arm, separate the reduced mean, observed residual, and matched residual:
\[
\begin{aligned}
&\frac1n\sum_iw_{A,Z_i}(\mathbf X_i)
\widetilde Y_i^{A,\mathrm{BC,or}}(z)\\
&\quad=
\frac1n\sum_iw_{A,Z_i}(\mathbf X_i)
m_z^A\{\mathbf D_z^\ast(\mathbf X_i)\}\\
&\qquad+
\frac1n\sum_{i:Z_i=z}w_{A,z}(\mathbf X_i)
\left[Y_i-m_z^A\{\mathbf D_z^\ast(\mathbf X_i)\}\right]
+R_{n,A,z}^{\mathrm{or}}.
\end{aligned}
\]
The weighted reduced-mean identity and the iid law of large numbers give
\[
\begin{aligned}
&\frac1n\sum_{i:Z_i=z}w_{A,z}(\mathbf X_i)
 \left[Y_i-m_z^A\{\mathbf D_z^\ast(\mathbf X_i)\}\right]\\
&\quad\to_p E_s\!\left[\mathbf1(Z=z)
 E_s\!\left[w_{A,z}\{Y-m_z^A(\mathbf D_z^\ast)\}
            \mid Z=z,\mathbf D_z^\ast\right]\right]\\
&\quad=0.
\end{aligned}
\]
The first term, divided by the outer denominator, has limit
\[
\begin{aligned}
&\frac{E_s[w_{A,Z}(\mathbf X)m_z^A\{\mathbf D_z^\ast(\mathbf X)\}]}
{E_s\{w_{A,Z}(\mathbf X)\}}\\
&\quad=E_\Omega[m_z^A\{\mathbf D_z^\ast(\mathbf X)\}]\\
&\quad=E_\Omega[E_\Omega\{Y(z)\mid\mathbf D_z^\ast(\mathbf X)\}]\\
&\quad=E_\Omega\{Y(z)\}.
\end{aligned}
\]
Subtracting the two arm limits proves PATE consistency.

For PATT, use the common-\(\mathbf D_0^\ast\) contrast \(\tau_{\mathrm{PATT}}(d)=m_{1\mid1}(d)-m_0^A(d)\). The centered numerator decomposes as
\[
\begin{aligned}
&\frac1n\sum_iZ_iw_{A,1}(\mathbf X_i)
\{Y_i-\widetilde Y_i^{A,\mathrm{BC,or}}(0)-\tau_{\mathrm{PATT}}\}\\
&\quad=
\frac1n\sum_iZ_iw_{A,1}(\mathbf X_i)
\{\tau_{\mathrm{PATT}}(\mathbf D_0^\ast(\mathbf X_i))
-\tau_{\mathrm{PATT}}\}\\
&\qquad+
\frac1n\sum_iZ_iw_{A,1}(\mathbf X_i)
\{Y_i-m_{1\mid1}(\mathbf D_0^\ast(\mathbf X_i))\}
-R_{n,A,0}^{\mathrm{or}}.
\end{aligned}
\]
Population recovery and the treated reduced-mean definition give
\[
\begin{aligned}
E_s\!\left[Zw_{A,1}
 \{\tau_{\mathrm{PATT}}(\mathbf D_0^\ast)-\tau_{\mathrm{PATT}}\}\right]
&=E_s(Zw_{A,1})
 E_\Omega\!\left[\tau_{\mathrm{PATT}}(\mathbf D_0^\ast)
                   -\tau_{\mathrm{PATT}}\mid Z=1\right]\\
&=0,\\
E_s\!\left[Zw_{A,1}\{Y-m_{1\mid1}(\mathbf D_0^\ast)\}\right]
&=E_s(Zw_{A,1})
 E_\Omega\!\left[Y(1)-m_{1\mid1}(\mathbf D_0^\ast)\mid Z=1\right]\\
&=0.
\end{aligned}
\]
By the iid law of large numbers, both averages converge to zero, while the treated denominator converges to \(E_s\{Zw_{A,1}(\mathbf X)\}>0\). Slutsky's theorem proves PATT consistency.
\end{proof}

\subsubsection{Auxiliary regressions for feasible estimation}
\label{app:stabilization-feasible}

For feasible estimation, we impose the following conditions.
\begin{enumerate}
\item[(i)] The iid sampling, support, density and bounded positive survey-weight conditions above hold, and \(E_s|Y|^p<\infty\) for \(p=2+\delta_0>2\).
\item[(ii)] The functions \(m_z^A\) and \(\rho_{A,z}\) are uniformly continuous on the compact score support. The data and covariate spaces are standard Borel spaces.
\item[(iii)] The population logistic and linear estimating equations have finite solutions and satisfy the positive-definite derivative conditions of Lemma~\ref{lem:primitive-logistic-linear-roots}. The regressor vector, including any polynomial features, is bounded.
\item[(iv)] Empirical standardization follows a fixed pooled or arm-specific rule with fixed bounded nonnegative standardization weights of positive expectation and positive limiting component standard deviations.
\end{enumerate}

Let \(\eta\) collect the score parameters and empirical centers and scales. Lemmas~\ref{lem:primitive-score-standardization} and~\ref{lem:primitive-logistic-linear-roots} give
\[
\widehat{\mathbf h}_{\eta,n}:=\sqrt n(\widehat\eta-\eta^\ast)=O_p(1).
\]
On a fixed neighborhood of \(\eta^\ast\), bounded regressors and positive limiting scales bound the first two derivatives of the standardized score map. With \(S_z(O)=\partial_\eta\mathbf D_z(\mathbf X;\eta^\ast)\), Taylor's formula gives, with probability tending to one,
\[
\begin{aligned}
&\sqrt n\max_i
\left\|\widehat{\mathbf D}_z(\mathbf X_i)
 -\mathbf D_z^\ast(\mathbf X_i)
 -S_z(O_i)(\widehat\eta-\eta^\ast)\right\|\\
&\le C\sqrt n\|\widehat\eta-\eta^\ast\|^2\\
&=O_p(n^{-1/2})=o_p(1).
\end{aligned}
\]
Thus
\begin{equation}\label{eq:stabilizer-derived-score-expansion}
\widehat{\mathbf D}_z(\mathbf X_i)
=
\mathbf D_z^\ast(\mathbf X_i)
+\frac1{\sqrt n}S_z(O_i)\widehat{\mathbf h}_{\eta,n}+\frac1{\sqrt n}\mathbf r_{n,z,i},
\qquad
\max_i\|\mathbf r_{n,z,i}\|\to_p0 .
\end{equation}
In particular, \(\max_i\|\widehat{\mathbf D}_z(\mathbf X_i)-\mathbf D_z^\ast(\mathbf X_i)\| =O_p(n^{-1/2})\).

\paragraph{Auxiliary kernel regressions.}

We use regression bandwidth \(\mathbf h_n=n^{-1/16}\) and define the ball-kernel auxiliary estimators by
\[
\widehat\rho_{A,z}(d)
:=
\frac{\sum_{j:Z_j=z}w_{A,z}(\mathbf X_j)
\mathbf1\{\|\widehat{\mathbf D}_z(\mathbf X_j)-d\|\le \mathbf h_n\}}
{\sum_{j:Z_j=z}
\mathbf1\{\|\widehat{\mathbf D}_z(\mathbf X_j)-d\|\le \mathbf h_n\}},
\]
\[
\widehat m_z^A(d)
:=
\frac{\sum_{j:Z_j=z}w_{A,z}(\mathbf X_j)Y_j
\mathbf1\{\|\widehat{\mathbf D}_z(\mathbf X_j)-d\|\le \mathbf h_n\}}
{\sum_{j:Z_j=z}w_{A,z}(\mathbf X_j)
\mathbf1\{\|\widehat{\mathbf D}_z(\mathbf X_j)-d\|\le \mathbf h_n\}}.
\]
For empty cells, we set \(\widehat\rho_{A,z}=w_{\min}\) and \(\widehat m_z^A=0\). Both regressions use the same sample. The following lemma shows that all neighborhoods evaluated at the observed scores are nonempty with probability tending to one.

\begin{lemma}[Uniform consistency at observed scores]
\label{lem:stabilizer-kernel-consistency}
Under the preceding conditions,
\[
\max_{1\le i\le n}
\left|\widehat m_z^A\{\widehat{\mathbf D}_z(\mathbf X_i)\}
-m_z^A\{\mathbf D_z^\ast(\mathbf X_i)\}\right|\to_p0,
\]
\[
\max_{1\le i\le n}
\left|\widehat\rho_{A,z}\{\widehat{\mathbf D}_z(\mathbf X_i)\}
-\rho_{A,z}\{\mathbf D_z^\ast(\mathbf X_i)\}\right|\to_p0.
\]
\end{lemma}
\begin{proof}
Write \(P_ng=n^{-1}\sum_i g(O_i)\) and \(I_{d,\mathbf h}=\mathbf1\{Z=z,\|\mathbf D_z^\ast-d\|\le \mathbf h\}\). The support and density bounds give \(\inf_{d\in\mathcal S_z}P_sI_{d,\mathbf h_n}\ge ch_n^2\), so the weighted denominators are bounded below by \(w_{\min} ch_n^2\).

To control the empirical numerators uniformly, take an \(n^{-1}\)-mesh of a fixed compact box and truncate \(Y\) at \(L_{Y,n}=n^{1/4}\). Hoeffding's inequality and a union bound apply. The mesh has \(O(n^2)\) points, and its boundary annuli have probability \(O(\mathbf h_n/n+n^{-2})\). The same finite-grid concentration bound controls their empirical masses. After truncation, their stochastic contribution is absorbed in \(L_{Y,n}\sqrt{\log n/n}\). Moreover, \(P_n(|Y|\mathbf1_{\{|Y|>L_{Y,n}\}})=O_p(L_{Y,n}^{1-p})\). Thus
\[
\sup_{d\in\mathcal S_z}
|(P_n-P_s)(w_{A,z}YI_{d,\mathbf h_n})|
=O_p\!\left(L_{Y,n}\sqrt{\frac{\log n}{n}}
 +\frac{L_{Y,n}\mathbf h_n}{n}+L_{Y,n}n^{-2}+L_{Y,n}^{1-p}\right)
=o_p(\mathbf h_n^2).
\]
The same bound holds for responses \(1\) and \(w_{A,z}\). Here \(\mathbf h_n=n^{-1/16}\) and \(p>2\), so \(L_{Y,n}\mathbf h_n^{-2}\sqrt{\log n/n}=n^{-1/8}\sqrt{\log n}\to0\) and \(L_{Y,n}^{1-p}\mathbf h_n^{-2}=n^{(3-2p)/8}\to0\).

By \eqref{eq:stabilizer-derived-score-expansion}, the largest generated-score error is at most \(a_n=n^{-1/2}\log n\) with probability tending to one. Changed memberships at observed centers lie in baseline annuli with radii \(\mathbf h_n\pm2a_n\). Their uniform empirical mass is \(O(\mathbf h_na_n+a_n^2)+O_p(\sqrt{\log n/n})\). Consequently the generated-score outcome-numerator error is bounded by
\[
O_p\!\left[L_{Y,n}\left(\mathbf h_na_n+a_n^2+
             \sqrt{\frac{\log n}{n}}\right)+L_{Y,n}^{1-p}\right]
=o_p(\mathbf h_n^2),
\]
since \(L_{Y,n}a_n/\mathbf h_n\to0\) and \(L_{Y,n}a_n^2/\mathbf h_n^2\to0\). The population smoothing errors satisfy
\[
\begin{aligned}
\left|\frac{P_s(w_{A,z}YI_{d,\mathbf h_n})}{P_s(w_{A,z}I_{d,\mathbf h_n})}
                 -m_z^A(d)\right|
&=\left|\frac{P_s[w_{A,z}\{m_z^A(\mathbf D_z^\ast)-m_z^A(d)\}I_{d,\mathbf h_n}]}
                   {P_s(w_{A,z}I_{d,\mathbf h_n})}\right|\\
&\le\sup_{u\in\mathcal S_z:\|u-d\|\le \mathbf h_n}|m_z^A(u)-m_z^A(d)|,\\
\left|\frac{P_s(w_{A,z}I_{d,\mathbf h_n})}{P_s I_{d,\mathbf h_n}}
                 -\rho_{A,z}(d)\right|
&\le\sup_{u\in\mathcal S_z:\|u-d\|\le \mathbf h_n}|\rho_{A,z}(u)-\rho_{A,z}(d)|.
\end{aligned}
\]
Uniform continuity makes both upper bounds converge to zero uniformly in \(d\). Together with the empirical bounds, the denominator lower bound gives all empirical denominators at least \(c'\mathbf h_n^2\) with probability tending to one. Dividing the numerator bounds by these denominators proves both ratio limits.
\end{proof}

\subsubsection{Feasible consistency}
\label{app:local-marked-consistency-tools}

For \(Z_i\ne z\), we define the feasible stabilized imputation by
\[
\begin{aligned}
\widetilde Y_i^{A,\mathrm{BC}}(z)
&=
\widehat m_z^A(\widehat{\mathbf D}_{z,i})\\
&\quad+
\frac1M\sum_{j\in\mathcal J_{\widehat{\mathbf D}_z}(i)}
\frac{w_{A,z}(\mathbf X_j)}
{\widehat\rho_{A,z}(\widehat{\mathbf D}_{z,j})}
\{Y_j-\widehat m_z^A(\widehat{\mathbf D}_{z,j})\},
\end{aligned}
\]
and retain \(\widetilde Y_i^{A,\mathrm{BC}}(Z_i)=Y_i\). We substitute these values into the original outer {H\'ajek} average for PATE or PATT, using the two auxiliary kernel regressions above.

\begin{theorem}[Feasible consistency]
\label{thm:stabilizer-feasible-consistency}
Suppose the conditions of Section~\ref{app:stabilization-feasible} and the geometry above hold, and the auxiliary fits are those of Lemma~\ref{lem:stabilizer-kernel-consistency}. Under the corresponding population recovery and mean-balancing route,
\[
\widetilde\tau_{\mathrm{PATE}}^{A,\mathrm{BC}}
\to_p\tau_{\mathrm{PATE}},
\qquad
\widetilde\tau_{\mathrm{PATT}}^{A,\mathrm{BC}}
\to_p\tau_{\mathrm{PATT}}.
\]
PATT uses only the control-side auxiliary regressions.
\end{theorem}
\begin{proof}
The fitted-score expansion \eqref{eq:stabilizer-derived-score-expansion} allows us to apply the local score and reuse bounds in Section~\ref{app:local-score-reuse} under the i.i.d.\ sampling, support, density and moment conditions stated here. The local marked limits in Section~\ref{app:local-score-marked-limits} first control matched residual averages at deterministic local parameters. The nearest-neighbor gap bounds then extend this result to fitted scores.

\emph{Matched residual averages.} For each individual requiring an imputed outcome, define
\[
\overline r_{n,A,z,i}(\mathbf h)
:=
\frac{\mathbf1\{Z_i=1-z\}w_{A,1-z}(\mathbf X_i)}{M}
\sum_{j\in\mathcal J_{z,n,\mathbf h}(i)}r_{A,z,j}.
\]
Conditional on the recipient's score \(d\), the limiting outcomes and survey weights of its \(M\) matches follow the arm-\(z\) conditional law at \(d\), independently of their local positions. By Proposition~\ref{prop:stabilizer-identification},
\[
E_s(r_{A,z}\mid Z=z,\mathbf D_z^\ast=d)=0.
\]
The limiting average therefore has conditional mean zero.

Let \(I_n,J_n\) be a uniformly sampled ordered pair of distinct indices, independent of the data. Conditional on those indices, the one-root and two-root calculations apply. Continuous tangent donor ranks remove dependence on the index-based tie convention. Jensen's inequality, bounded outer weights, reindexing, and the inflated reuse envelope give
\[
\begin{aligned}
E_s|\overline r_{n,A,z,I_n}(\mathbf h)|^p
&=
\frac1n E_s\sum_i|\overline r_{n,A,z,i}(\mathbf h)|^p\\
&\le
\frac{C}{nM}E_s\sum_{i:Z_i=1-z}
\sum_{j\in\mathcal J_{z,n,\mathbf h}(i)}|r_{A,z,j}|^p\\
&\le
\frac{C}{n}E_s\sum_{j:Z_j=z}
K_{M,z}^{+,a}(j)|r_{A,z,j}|^p\\
&\le C'E_s\{\mathbf1(Z=z)|r_{A,z}|^p\}<\infty .
\end{aligned}
\]
In the last step, conditional on an observation's treatment and baseline score, the inflated count depends only on other observations and is independent of its outcome and weight. The bound is uniform over \(\|\mathbf h\|\le C\). It also gives uniform integrability of products:
\[
E_s|\overline r_{n,A,z,I_n}(\mathbf h)\overline r_{n,A,z,J_n}(\mathbf h)|^{p/2}
\le
\{E_s|\overline r_{n,A,z,I_n}(\mathbf h)|^p
 E_s|\overline r_{n,A,z,J_n}(\mathbf h)|^p\}^{1/2}.
\]
The one- and two-query functional limits established above, together with uniform integrability, give
\[
E_s\overline r_{n,A,z,I_n}(\mathbf h)\to0,\qquad
E_s\{\overline r_{n,A,z,I_n}(\mathbf h)\overline r_{n,A,z,J_n}(\mathbf h)\}\to0.
\]
Averaging over the ordered root indices expresses the second moment of the sample average as
\[
\begin{aligned}
E_s\left[\left\{\frac1n\sum_i\overline r_{n,A,z,i}(\mathbf h)\right\}^2\right]
&=
\frac1nE_s\{\overline r_{n,A,z,I_n}(\mathbf h)^2\}\\
&\quad+
\frac{n-1}{n}
E_s\{\overline r_{n,A,z,I_n}(\mathbf h)\overline r_{n,A,z,J_n}(\mathbf h)\}\\
&\longrightarrow0 .
\end{aligned}
\]
Thus the sample average converges to zero at each deterministic local parameter.

\emph{Estimated-score matching.} On \(\{\|\widehat{\mathbf h}_{\eta,n}\|\le C\}\), choose \(\mathbf h'\in\mathcal L_{C,\delta}\) within distance \(\delta\) of \(\widehat{\mathbf h}_{\eta,n}\). By \eqref{eq:stabilizer-derived-score-expansion}, on an event of probability tending to one,
\[
\max_i
\|\widehat{\mathbf D}_{z,i}-\mathbf D_{z,n,\mathbf h'}(O_i)\|
\le (L_S+1)\delta/\sqrt n .
\]
Every donor-query distance changes by at most \(2(L_S+1)\delta/\sqrt n\). A changed set of \(M\) donors consequently requires a reference \(M\)-to-\((M+1)\) distance gap at most \(4(L_S+1)\delta/\sqrt n\). The finite-net bound in Section~\ref{app:local-score-marked-limits} shows that the fraction \(F_{n,z}\) of changed query sets obeys
\[
F_{n,z}\le C\delta+o_p(1)
\]
in the order \(n\to\infty\) with fixed \(\delta\), then \(\delta\downarrow0\).

To control residual magnitudes, note that the union of donors in the two graphs' changed sets has at most \(2MnF_{n,z}\) members. On the bounded-index event, both incoming counts are bounded by a fixed \(K_{M,z}^{+,a}\). Since \(p=2+\delta_0>2\), H\"older's inequality gives
\[
\begin{aligned}
&\left|
\frac1{nM}\sum_{i:Z_i=1-z}w_{A,1-z}(\mathbf X_i)
\left[
\sum_{j\in\mathcal J_{\widehat{\mathbf D}_z}(i)}r_{A,z,j}
-\sum_{j\in\mathcal J_{z,n,\mathbf h'}(i)}r_{A,z,j}
\right]\right|\\
&\quad\le
C F_{n,z}^{\,1/2-1/p}
\left\{\frac1n\sum_{j:Z_j=z}(K_{M,z}^{+,a}(j))^2\right\}^{1/2}
\left\{\frac1n\sum_{j:Z_j=z}|r_{A,z,j}|^p\right\}^{1/p}.
\end{aligned}
\]
The count average is \(O_p(1)\) by \eqref{eq:stabilizer-inflated-reuse}, and the residual average is \(O_p(1)\) by the law of large numbers. These bounds are uniform over the grid mesh because the envelope depends on \(C\), not \(\delta\). Thus this difference vanishes in the stated double-limit sense. For fixed \(\delta\), the finite maximum of the signed averages at the net points is \(o_p(1)\). Finally, we let \(\delta\downarrow0\), then \(C\uparrow\infty\), using tightness of \(\widehat{\mathbf h}_{\eta,n}\). The same-sample fitted graph therefore satisfies
\begin{equation}\label{eq:stabilizer-actual-graph-lln}
\frac1{nM}\sum_{i:Z_i=1-z}w_{A,1-z}(\mathbf X_i)
\sum_{j\in\mathcal J_{\widehat{\mathbf D}_z}(i)}
\frac{w_{A,z}(\mathbf X_j)}
{\rho_{A,z}(\mathbf D_{z,j}^\ast)}
\{Y_j-m_z^A(\mathbf D_{z,j}^\ast)\}
\to_p0 .
\end{equation}

\emph{Auxiliary-regression errors.} Let \(\Delta_{m,n}\) and \(\Delta_{\rho,n}\) be the two maximum errors in Lemma~\ref{lem:stabilizer-kernel-consistency}. Both are \(o_p(1)\). For each donor,
\[
\begin{aligned}
&\frac{w_{A,z}(\mathbf X_j)}{\widehat\rho_{A,z}(\widehat{\mathbf D}_{z,j})}
 \{Y_j-\widehat m_z^A(\widehat{\mathbf D}_{z,j})\}-r_{A,z,j}\\
&\quad=
\frac{w_{A,z}(\mathbf X_j)}{\widehat\rho_{A,z}(\widehat{\mathbf D}_{z,j})}
 \{m_z^A(\mathbf D_{z,j}^\ast)-\widehat m_z^A(\widehat{\mathbf D}_{z,j})\}\\
&\qquad+
w_{A,z}(\mathbf X_j)\{Y_j-m_z^A(\mathbf D_{z,j}^\ast)\}
 \left\{\frac1{\widehat\rho_{A,z}(\widehat{\mathbf D}_{z,j})}
              -\frac1{\rho_{A,z}(\mathbf D_{z,j}^\ast)}\right\}.
\end{aligned}
\]
Since \(\rho_{A,z},\widehat\rho_{A,z}\ge w_{\min}\) on nonempty cells, each donor residual changes by at most
\[
C\{\Delta_{m,n}
 +\Delta_{\rho,n}|Y_j-m_z^A(\mathbf D_{z,j}^\ast)|\}.
\]
The first term averages to \(O_p(\Delta_{m,n})\). For each bounded local-index event, the second averages to at most
\[
C\Delta_{\rho,n}\frac1n\sum_{j:Z_j=z}
 K_{M,z}^{+,a}(j)|Y_j-m_z^A(\mathbf D_{z,j}^\ast)|=o_p(1).
\]
Given an observation's treatment and baseline score, its outcome and weight are independent of its inflated reuse count. Thus \eqref{eq:stabilizer-inflated-reuse} gives
\[
\begin{aligned}
E_s\!\left[\frac1n\sum_{j:Z_j=z}
 K_{M,z}^{+,a}(j)|Y_j-m_z^A(\mathbf D_{z,j}^\ast)|\right]
&\le C E_s\!\left[\mathbf1(Z=z)|Y-m_z^A(\mathbf D_z^\ast)|\right]\\
&<\infty.
\end{aligned}
\]
By Markov's inequality, the average is \(O_p(1)\). Tightness of \(\widehat{\mathbf h}_{\eta,n}\) removes the bounded-index restriction. The target-side regression error is also \(o_p(1)\).

After these auxiliary-regression errors are removed, \eqref{eq:stabilizer-actual-graph-lln} shows that the remaining matched-residual average converges to zero. The direct outcomes and reduced means have the iid limits in Proposition~\ref{prop:stabilizer-oracle-consistency}. The outer denominators converge to \(E_sw_{A,Z}>0\) for PATE and \(E_sZw_{A,1}>0\) for PATT. Slutsky's theorem proves the claims.
\end{proof}

\subsection{Continuity of the local matching process}
\label{app:local-matching-process-proofs}

\begin{proof}[Proof of Proposition~\ref{prop:patt-poisson-local-process-equicontinuity}]
The mean-value theorem and the bounded derivatives give \eqref{eq:patt-poisson-bounded-score-displacement}. We write \(x_i=\sqrt n\,\mathbf D_0^\ast(\mathbf X_i)\). We first compare tie rules, then establish continuity on bounded spatial windows and remove the window restriction.

\emph{Comparison of tie rules.} We compare the specified rule with fixed original row-label priority. For \(m/n\in[1/2,2]\), consider a Poisson cloud of intensity \(mP_s\) with score scale \(\sqrt n\). For any \(B>0\), we choose \(R_n=L\sqrt{\log(n+1)}\) with \(L\) large enough. The ball-mass bound in Lemma~\ref{lem:oracle-score-radius} and the Poisson lower and upper tails, each summed over \(O(n)\) unit squares, give failure probability \(O(n^{-B})\) for the following event: every query has \(M\) baseline controls within \(R_n/(2\sqrt n)\), and each unit square contains at most \(C_B\log(n+1)\) rows.

Let \(d_K=p_A+2\). We use a parameter net of mesh \(n^{-2}\) with at most \(Cn^{2d_K}\) anchors. We call a square near-tied at an anchor if a query in it has two candidate controls within baseline scaled radius \(R_n\) whose squared current scaled distances differ by at most \(C_K(1+R_n)n^{-2}\). The three-point Mecke identity and the current donor-density bound give query, first-donor and second-donor factors bounded by \(C\), \(C(1+R_n)^2\) and \(C(1+R_n)n^{-2}\), respectively. Thus
\begin{equation}\label{eq:patt-binomial-tie-block-probability}
P(\text{specified square near-tied at specified anchor})
\le C(1+R_n)^3n^{-2}.
\end{equation}
The bound is uniform in the intensity ratio \(m/n\). We color the squares with period greater than \(2R_n+2\), so that their enlarged windows are independent within each of \(O(\log n)\) colors. The summed near-tie probabilities in a color are at most \(Cn(1+R_n)^3n^{-2}\le n^{-3/4}\). The factorial-moment bound for independent indicators therefore gives
\begin{equation}\label{eq:patt-binomial-tie-color-bound}
P(\text{at least }k\text{ near-tie squares in one color})
\le n^{-3k/4}/k! .
\end{equation}
We choose a fixed \(k\) with \(3k/4>2d_K+B+2\) and sum this bound over the net anchors and colors.

On the localization event, the score Lipschitz bound implies that every selection-relevant tie produces a near-tie at an anchor within \(n^{-2}\). The square-count bound gives
\begin{equation}\label{eq:patt-binomial-uniform-tied-query-count}
\sup_K\#\{\text{queries with a selection-relevant tie}\}
\le C_B\{\log(n+1)\}^2
\end{equation}
with failure probability \(O(n^{-B})\). Conditioning the Poisson count to equal \(m\) multiplies the probability bound by at most \(C\sqrt m\), since Stirling's formula gives \(P\{\operatorname{Poisson}(m)=m\}\ge c/\sqrt m\). Thus the same geometric event for \(m\) iid rows has failure \(O(n^{1/2-B})\), uniformly in \(m/n\in[1/2,2]\).

Two valid tie choices differ only at these queries, and each difference of normalized donor averages is bounded by twice the greatest absolute residual. On the good event,
\begin{equation}\label{eq:patt-binomial-tie-numerator-bound}
\sup_K\frac{|\mathcal I^{\rm rule}_{A,m;n}
                   -\mathcal I^{\rm priority}_{A,m;n}|}{\sqrt n}
\le\frac{C\log^2(n+1)}{\sqrt n}
                   \max_{j\le m}(1+|\zeta_{A,j}|).
\end{equation}
The maximum and exceptional-event bounds are
\[
\begin{aligned}
E_s\max_{j\le m}(1+|\zeta_{A,j}|)
&\le\left\{E_s\sum_{j\le m}(1+|\zeta_{A,j}|)^p\right\}^{1/p}\\
&\le Cn^{1/p},\\
\frac{C\log^2(n+1)}{\sqrt n}
 E_s\max_{j\le m}(1+|\zeta_{A,j}|)
&\le Cn^{1/p-1/2}\log^2(n+1)\longrightarrow0,\\
\frac{Cm}{\sqrt n}E_s\!\left[
 \mathbf1_{\mathrm{bad}}\max_{j\le m}(1+|\zeta_{A,j}|)\right]
&\le Cn^{1/2+1/p}P_s(\mathrm{bad})^{1-1/p}\\
&=o(1)
\end{aligned}
\]
for sufficiently large \(B\), since \(p>2\). These bounds are uniform over \(m/n\in[1/2,2]\). For the remaining Poisson counts, we use the global bound
\[
\sup_K|\mathcal I^{\rm rule}_{A,N_n;n}
       -\mathcal I^{\rm priority}_{A,N_n;n}|
\le C N_n\max_{j\le N_n}(1+|\zeta_{A,j}|),
\]
with the maximum defined as zero when \(N_n=0\). Independence of the count and rows gives
\[
\begin{aligned}
E\!\left[N_n^p\max_{j\le N_n}(1+|\zeta_{A,j}|)^p\right]
&\le E\!\left[N_n^p\sum_{j\le N_n}(1+|\zeta_{A,j}|)^p\right]\\
&=E(N_n^{p+1})E_s(1+|\zeta_A|)^p\\
&\le Cn^{p+1}.
\end{aligned}
\]
The Poisson tail bound and H\"older's inequality therefore yield
\[
\begin{aligned}
&E^*\!\left[
 \mathbf1\{N_n\notin[n/2,2n]\}
 \sup_K\frac{|\mathcal I^{\rm rule}_{A,N_n;n}
                  -\mathcal I^{\rm priority}_{A,N_n;n}|}{\sqrt n}
 \right]\\
&\le Cn^{-1/2}
 \left\{E\!\left[N_n^p\max_{j\le N_n}(1+|\zeta_{A,j}|)^p\right]\right\}^{1/p}
 P\{N_n\notin[n/2,2n]\}^{1-1/p}\\
&\le Cn^{1/2+1/p}e^{-cn}\longrightarrow0.
\end{aligned}
\]
Integrating the uniform fixed-count bound on \([n/2,2n]\), we obtain, for both \(m=n\) and \(m=N_n\),
\begin{equation}\label{eq:patt-binomial-tie-rule-transfer}
E^*\sup_K\frac{|\mathcal I^{\rm rule}_{A,m;n}
                   -\mathcal I^{\rm priority}_{A,m;n}|}{\sqrt n}
\longrightarrow0 .
\end{equation}

\emph{Continuity on a bounded window.} By \eqref{eq:patt-binomial-tie-rule-transfer}, it suffices to establish the modulus for fixed original row-label priority and then compare the centered processes. We therefore use this priority throughout the localization argument, retaining the same labels under every candidate restriction, and suppress its superscript.

For \(R>0\), we restrict a query's candidates to \(\|x_j-x_i\|\le R\) and retain its \(M\) nearest under \(\mathcal D_{n,\mathbf h,l}\), using the same normalized fractions. We set its contribution to zero if fewer than \(M\) candidates remain. Both the candidate restriction and this definition on insufficient samples are independent of \((\mathbf h,l)\). We denote the resulting numerator by \(\mathcal I^R_{A,m;n}\) and write
\begin{equation}\label{eq:patt-poisson-truncated-process}
\mathcal Z^{P,R}_{A,n}
=\{\mathcal I^R_{A,N_n;n}-E\mathcal I^R_{A,N_n;n}\}/\sqrt n.
\end{equation}

We partition the rescaled baseline support into \(O(n)\) unit squares \(Q_b\). Let \(B^R_{A,n,b}\) sum the truncated query contributions from \(Q_b\), and let \(N^R_{n,b}\) count all rows in \(Q_b^{+R}=Q_b+[-R,R]^2\). An envelope, including changes in every normalized donor fraction, is
\begin{equation}\label{eq:patt-poisson-block-envelope}
\sup_K|B^R_{A,n,b}|
\le F^R_{n,b}
:=C N^R_{n,b}\sum_{j:x_j\in Q_b^{+R}}(1+|\zeta_{A,j}|).
\end{equation}
By the power-of-a-sum inequality, the Mecke identity for intensity \(nP_s\), and independence of the restrictions to disjoint sets (Theorems~4.4 and~5.2 of Last and Penrose\citep{LastPenrose2017}),
\begin{equation}\label{eq:patt-poisson-summed-block-moment}
\begin{aligned}
\sum_b E(F^R_{n,b})^p
&\le C_p\sum_b\int_{x\in Q_b^{+R}}
 E(N^R_{n,b}+1)^{2p-1}(1+|\zeta_A|)^p\,nP_s(dO)\\
&\le C_p n(1+R)^{4p}E_s(1+|\zeta_A|)^p .
\end{aligned}
\end{equation}
The window intensity and overlap multiplicity are both \(O((1+R)^2)\). We color the squares modulo \(L_R>2R+1\), with \(L_R\le C(1+R)\). The windows within each of the \(O((1+R)^2)\) colors are disjoint. At deterministic grid anchors, ties have probability zero under the donor density, and the corresponding local blocks are independent.

Let \(u=(\mathbf h,l)\). In a radius-\(R\) window, scaled current distances are \(O_K(1+R)\). Their squares change by at most \(C_K(1+R)\eta\) when \(\|u-u'\|\le\eta\). Let \(\mathcal A^R_{n,b}(u;\eta)\) be the event that some query in \(Q_b\) and two distinct candidate controls have squared current scaled distances within this amount at \(u\). A change of the selected set implies this event. Since the weights are unchanged when the selected set is unchanged, we have
\begin{equation}\label{eq:patt-poisson-block-oscillation-envelope}
\sup_{\substack{u'\in K\\\|u-u'\|\le\eta}}
 |B^R_{A,n,b}(u')-B^R_{A,n,b}(u)|
\le D^R_{n,b}(u;\eta)
:=2F^R_{n,b}\mathbf1_{\mathcal A^R_{n,b}(u;\eta)} .
\end{equation}
In the three-point Mecke integral, the query, first donor and second donor have intensity bounds \(O(1)\), \(O((1+R)^2)\) and \(O_K((1+R)\eta)\), respectively. The last bound follows from \eqref{eq:patt-poisson-current-donor-density}: a squared-radius interval of half-width \(v=C_K(1+R)\eta/n\) has area at most \(2\pi v\). Hence
\begin{equation}\label{eq:patt-poisson-block-near-tie-probability}
P\{\mathcal A^R_{n,b}(u;\eta)\}
\le C_K(1+R)^3\eta.
\end{equation}
H\"older's inequality, summed over blocks, gives
\begin{equation}\label{eq:patt-poisson-summed-oscillation-moments}
\begin{aligned}
\sum_b E\{D^R_{n,b}(u;\eta)^r\}
&\le2^r\sum_b\{E(F^R_{n,b})^p\}^{r/p}
        P\{\mathcal A^R_{n,b}(u;\eta)\}^{1-r/p}\\
&\le2^r\left\{\sum_bE(F^R_{n,b})^p\right\}^{r/p}
        \left\{\sum_bP(\mathcal A^R_{n,b}(u;\eta))\right\}^{1-r/p}\\
&\le C_{R,K}n\eta^{1-r/p},\qquad r=1,2.
\end{aligned}
\end{equation}
For \(0<\eta\le1\), both exponents may be replaced by \(\beta=(p-2)/p>0\).

To bound the centered process, we truncate each block on \(F^R_{n,b}\le c_n=n^{a_p}\), where \(a_p=p/\{4(p-1)\}\). The cutoff does not depend on \(u\), and its error in expected supremum, for each color, is
\begin{equation}\label{eq:patt-poisson-cutoff-error}
\begin{aligned}
\frac2{\sqrt n}\sum_b
 E\{F^R_{n,b}\mathbf1(F^R_{n,b}>c_n)\}
&\le\frac{2c_n^{1-p}}{\sqrt n}\sum_b E(F^R_{n,b})^p\\
&\le C_{R,K}n^{1/2+a_p(1-p)}\\
&=C_{R,K}n^{-(p-2)/4}.
\end{aligned}
\end{equation}
We write \(\widetilde B=B\mathbf1(F\le c_n)\), \(\widetilde D=D\mathbf1(F\le c_n)\), and let \(\widetilde Z\) be the centered, \(n^{-1/2}\)-normalized sum of \(\widetilde B\) over one color. Its centered block increments are bounded by \(4c_n\). Bernstein's inequality (Theorem~3 of Boucheron et al.\citep{BoucheronLugosiBousquet2004}), a union bound and tail integration give, for \(m\) sums of independent centered variables satisfying \(|X_{b,j}|\le b_0\) and \(\sum_bEX_{b,j}^2\le n\sigma^2\),
\begin{equation}\label{eq:patt-poisson-finite-maximum-bound}
E\max_{j\le m}\left|n^{-1/2}\sum_bX_{b,j}\right|
\le C\{\sigma\sqrt{\log(2m)}+b_0\log(2m)/\sqrt n\}.
\end{equation}

We choose \(2^{-k}\)-nets \(G_k\) of \(K\) with \(|G_k|\le C_K2^{(p_A+2)k}\) and nearest-net maps \(\pi_k\). Applying \eqref{eq:patt-poisson-finite-maximum-bound} with the oscillation moments in \eqref{eq:patt-poisson-summed-oscillation-moments} gives
\begin{equation}\label{eq:patt-poisson-grid-increments}
E\max_{\substack{g\in G_k,\ g'\in G_{k-1}\\
                         \|g-g'\|\le C_K2^{-k}}}
 |\widetilde Z(g)-\widetilde Z(g')|
\le C_{R,K}\{2^{-k\beta/2}\sqrt{k+1}
             +n^{a_p-1/2}(k+1)\}.
\end{equation}
The same bound holds for nearby pairs in \(G_k\). For \(\eta_k=C_K2^{-k}\), the pointwise envelope gives
\[
\begin{aligned}
&\sup_{u\in K}|\widetilde Z(u)-\widetilde Z(\pi_k u)|\\
&\le\max_{v\in G_k}\frac1{\sqrt n}\sum_b
 \{\widetilde D_b(v;\eta_k)+E\widetilde D_b(v;\eta_k)\}\\
&\le\max_{v\in G_k}\left|\frac1{\sqrt n}\sum_b
 \{\widetilde D_b(v;\eta_k)-E\widetilde D_b(v;\eta_k)\}\right|
 +\frac2{\sqrt n}\max_{v\in G_k}\sum_bE\widetilde D_b(v;\eta_k).
\end{aligned}
\]
The finite-maximum bound and the first two oscillation moments yield
\begin{equation}\label{eq:patt-poisson-terminal-grid-envelope}
E^*\sup_{u\in K}|\widetilde Z(u)-\widetilde Z(\pi_k u)|
\le C_{R,K}\{2^{-k\beta/2}\sqrt{k+1}
+n^{a_p-1/2}(k+1)+\sqrt n\,2^{-k\beta}\}.
\end{equation}
With \(k_n=\lceil(2/\beta)\log_2(n+1)\rceil\), we telescope between levels \(k_0\) and \(k_n\) and apply the same-net bound at level \(k_0\). Adding the terminal errors and cutoff errors over the finitely many colors, and using \(\sum_{k\ge k_0}2^{-k\beta/2}\sqrt{k+1} \le C_\beta2^{-k_0\beta/2}\sqrt{k_0+1}\), gives
\begin{equation}\label{eq:patt-poisson-fixed-radius-modulus}
\begin{aligned}
&E^*\sup_{\substack{u,u'\in K\\\|u-u'\|\le2^{-k_0}}}
 |\mathcal Z^{P,R}_{A,n}(u)-\mathcal Z^{P,R}_{A,n}(u')|\\
&\le C_{R,K}\left\{
2^{-k_0\beta/2}\sqrt{k_0+1}
+n^{-(p-2)/\{4(p-1)\}}(1+\log n)^2
+n^{-3/2}+n^{-(p-2)/4}\right\}.
\end{aligned}
\end{equation}
Thus the fixed-radius modulus vanishes as \(n\to\infty\) and then \(k_0\to\infty\).

\emph{Removal of the window restriction.} To remove the radius cutoff \(R\), we use the ball-mass bound in Lemma~\ref{lem:oracle-score-radius}, which gives baseline control intensity between \(cR^2\) and \(CR^2\) in a rescaled \(R/2\)-ball, uniformly over the support for \(R\le c_0\sqrt n\). For \(R>8C_K\), if a query's baseline \(R/2\)-ball contains at least \(M\) controls, their current distances are at most \(R/2+2C_K<R-2C_K\). All selected donors therefore lie in the baseline \(R\)-window simultaneously on \(K\). Let \(\mathcal A^{\rm abs}_{n,b}(R)\) denote failure for some query in \(Q_b\). The one-point Mecke identity and the fixed-\(M\) Poisson lower tail give
\begin{equation}\label{eq:patt-poisson-radius-absence-bound}
P\{\mathcal A^{\rm abs}_{n,b}(R)\}
\le C\sum_{j=0}^{M-1}(CR^2)^j e^{-cR^2}
\le C'e^{-c'R^2}.
\end{equation}

Because both restrictions inherit the same row-label priority, the shell \(B^{2R}-B^R\) is zero off this event. With \(q=(p+2)/2\in(2,p)\) and \(H^R_{n,b}=2F^{2R}_{n,b}\mathbf1_{\mathcal A^{\rm abs}_{n,b}(R)}\), the block moment bound and H\"older's inequality give
\begin{equation}\label{eq:patt-poisson-shell-moment}
\begin{aligned}
\sum_bE(H^R_{n,b})^q
&\le2^q\left\{\sum_b E(F^{2R}_{n,b})^p\right\}^{q/p}
 \left\{\sum_bP(\mathcal A^{\rm abs}_{n,b}(R))\right\}^{1-q/p}\\
&\le Cn(1+R)^a e^{-cR^2}.
\end{aligned}
\end{equation}
Its oscillation is bounded by \(D^{R,\rm sh}_{n,b} =2H^R_{n,b}\mathbf1_{\mathcal A^R_{n,b}\cup\mathcal A^{2R}_{n,b}}\). A second application of H\"older and the near-tie bound gives
\begin{equation}\label{eq:patt-poisson-shell-oscillation-moment}
\sum_bE(D^{R,\rm sh}_{n,b})^r
\le Cn(1+R)^a e^{-cR^2}\eta^{1-r/q},
\qquad r=1,2.
\end{equation}

Let \(Y^R=\mathcal Z^{P,2R}_{A,n}-\mathcal Z^{P,R}_{A,n}\), and fix \(u_0\in K\). With \(c_R\le C(1+R)^2\) colors of period greater than \(4R+1\), independence within each color gives
\[
\begin{aligned}
E|Y^R(u_0)|
&\le\frac1{\sqrt n}\sum_{c=1}^{c_R}
 \left\{\sum_{b\text{ in color }c}E[(H^R_{n,b})^2]\right\}^{1/2}\\
&\le\left\{\frac{c_R}{n}\sum_bE[(H^R_{n,b})^2]\right\}^{1/2}\\
&\le C(1+R)^a e^{-cR^2}.
\end{aligned}
\]
The last inequality follows from \eqref{eq:patt-poisson-shell-moment}, \(q>2\), and the \(O(n)\) blocks. To control the supremum, we use
\[
E^*\sup_{u\in K}|Y^R(u)|
\le E|Y^R(u_0)|+E^*\sup_{u\in K}|Y^R(u)-Y^R(u_0)|.
\]
We apply \eqref{eq:patt-poisson-cutoff-error}--\eqref{eq:patt-poisson-fixed-radius-modulus} to the increments of \(Y^R\), with \(q\), \(\beta_q=(q-2)/q\), \(a_q=q/\{4(q-1)\}\), and the same coloring. The shell moment and oscillation bounds give, uniformly for \(R_0\le R\le c_0\sqrt n\),
\begin{equation}\label{eq:patt-poisson-finite-n-shell-bound}
E^*\sup_K|Y^R|
\le C(1+R)^a e^{-cR^2}
+C(1+R)^2n^{-\boldsymbol\gamma_q}(1+\log n)^2,
\qquad \boldsymbol\gamma_q=\frac{q-2}{4(q-1)}>0 .
\end{equation}
The constants in this bound are independent of \(n\) and \(R\) over the stated range.

A union over \(O(n)\) query squares gives
\[
P^*\{\sup_K|\mathcal I_{A,N_n;n}-\mathcal I^R_{A,N_n;n}|>0\}
\le Cn e^{-cR^2}.
\]
For the common envelope \(F_n=CN_n\sum_{j\le N_n}(1+|\zeta_{A,j}|)\), the Mecke identity gives
\[
\begin{aligned}
EF_n^p
&\le C E\left[N_n^{2p-1}\sum_{j\le N_n}(1+|\zeta_{A,j}|)^p\right]\\
&=Cn E(N_n+1)^{2p-1}E_s(1+|\zeta_A|)^p\\
&\le Cn^{2p}.
\end{aligned}
\]
H\"older's inequality yields
\begin{equation}\label{eq:patt-poisson-terminal-radius-tail}
\begin{aligned}
\frac1{\sqrt n}E^*\sup_K
 |\mathcal I_{A,N_n;n}-\mathcal I^R_{A,N_n;n}|
&\le\frac C{\sqrt n}(EF_n^p)^{1/p}
                       (Cn e^{-cR^2})^{1-1/p}\\
&\le Cn^{5/2-1/p}e^{-c(1-1/p)R^2}.
\end{aligned}
\end{equation}
Centering increases the bound by at most a factor of two. We choose \(R_n=L\sqrt{\log(n+1)}\), with \(L\) large enough that the bound is \(O(n^{-1})\) for all radii in \([R_n,2R_n]\). For fixed \(R\ge R_0\), let \(R_j=2^jR\) and stop at the first \(J_n\) with \(R_{J_n}\ge R_n\). The finite telescoping sum and \(\sum_{j<J_n}(1+R_j)^2=O(\log n)\) give
\begin{equation}\label{eq:patt-poisson-finite-shell-sum}
\begin{aligned}
E^*\sup_K|\mathcal Z^P_{A,n}-\mathcal Z^{P,R}_{A,n}|
&\le\sum_{j=0}^{J_n-1}E^*\sup_K
 |\mathcal Z^{P,2R_j}_{A,n}-\mathcal Z^{P,R_j}_{A,n}|
 +E^*\sup_K|\mathcal Z^P_{A,n}-\mathcal Z^{P,R_{J_n}}_{A,n}|\\
&\le\sum_{j\ge0}C(1+2^jR)^a e^{-c(2^jR)^2}\\
&\quad+Cn^{-\boldsymbol\gamma_q}(1+\log n)^3+O(n^{-1}).
\end{aligned}
\end{equation}
Hence
\begin{equation}\label{eq:patt-poisson-uniform-radius-removal}
\lim_{R\to\infty}\limsup_{n\to\infty}
 E^*\sup_K|\mathcal Z^P_{A,n}-\mathcal Z^{P,R}_{A,n}|=0.
\end{equation}
Finally,
\[
\begin{aligned}
&E^*\sup_{\substack{u,u'\in K\\\|u-u'\|\le\eta}}
       |\mathcal Z^P_{A,n}(u)-\mathcal Z^P_{A,n}(u')|\\
&\le E^*\sup_{\substack{u,u'\in K\\\|u-u'\|\le\eta}}
       |\mathcal Z^{P,R}_{A,n}(u)-\mathcal Z^{P,R}_{A,n}(u')|
       +2E^*\sup_K|\mathcal Z^P_{A,n}-\mathcal Z^{P,R}_{A,n}|.
\end{aligned}
\]
Equations~\eqref{eq:patt-poisson-fixed-radius-modulus} and \eqref{eq:patt-poisson-uniform-radius-removal}, with \(n\to\infty\), then \(\eta\downarrow0\), then \(R\to\infty\), prove \eqref{eq:patt-poisson-full-process-modulus} for the priority rule. It remains to return to the specified tie rule. We center each numerator at its own expectation. By \eqref{eq:patt-binomial-tie-rule-transfer},
\[
\begin{aligned}
&E^*\sup_K|\mathcal Z^{P,\rm rule}_{A,n}
                    -\mathcal Z^{P,\rm priority}_{A,n}|\\
&\le 2E^*\sup_K
 \frac{|\mathcal I^{\rm rule}_{A,N_n;n}
             -\mathcal I^{\rm priority}_{A,N_n;n}|}{\sqrt n}
\longrightarrow0.
\end{aligned}
\]
Consequently,
\[
\begin{aligned}
&E^*\sup_{\substack{u,u'\in K\\\|u-u'\|\le\eta}}
 |\mathcal Z^{P,\rm rule}_{A,n}(u)-\mathcal Z^{P,\rm rule}_{A,n}(u')|\\
&\le E^*\sup_{\substack{u,u'\in K\\\|u-u'\|\le\eta}}
 |\mathcal Z^{P,\rm priority}_{A,n}(u)-\mathcal Z^{P,\rm priority}_{A,n}(u')|\\
&\quad+2E^*\sup_K|\mathcal Z^{P,\rm rule}_{A,n}
                         -\mathcal Z^{P,\rm priority}_{A,n}|.
\end{aligned}
\]
Taking \(n\to\infty\) and then \(\eta\downarrow0\) proves the stated modulus for the specified tie rule.
\end{proof}

\begin{proof}[Proof of Proposition~\ref{prop:patt-binomial-local-process-transfer}]
We compare the fixed-size and Poissonized numerators by adding or deleting observations. We obtain uniform bounds on the moment and conditional mean of each insertion's contribution within a fixed neighborhood, then remove the neighborhood restriction.
\par\smallskip\noindent
\emph{Add-one decomposition.}\quad We first prove the result for fixed row-index priority. The comparison \eqref{eq:patt-binomial-tie-rule-transfer} extends the result to the specified tie rule. We retain the original row labels under every restriction and deletion. Let \(\mathcal C\) contain \(b\) iid rows, with \(b/n\in[1/2,2]\), and let \(q\) be an independent inserted row. Its priority is greater than every base index. We define the add-one cost by
\begin{equation}\label{eq:patt-binomial-addone-definition}
\begin{aligned}
A_{n,\mathbf h,l}(q;\mathcal C)
&:=\mathcal I_{A,b+1;n}(\mathbf h,l;\mathcal C\cup\{q\})\\
&\quad-\mathcal I_{A,b;n}(\mathbf h,l;\mathcal C).
\end{aligned}
\end{equation}
We write \(\mathcal J^{\mathcal C}_{\mathcal D_{n,\mathbf h,l}}(i)\) for the selected \(M\)-set when the donor cloud is \(\mathcal C\). On the event that the base contains at least \(M\) controls, a treated insertion has cost
\begin{equation}\label{eq:patt-binomial-treated-insertion}
A_{n,\mathbf h,l}(q;\mathcal C)
=w_{A,1}(\mathbf X_q)
 \sum_{j\in\mathcal J^{\mathcal C}_{\mathcal D_{n,\mathbf h,l}}(q)}
 \frac{w_{A,0}(\mathbf X_j)\zeta_{A,j}}
 {\displaystyle\sum_{k\in\mathcal J^{\mathcal C}_{\mathcal D_{n,\mathbf h,l}}(q)}
                 w_{A,0}(\mathbf X_k)}.
\end{equation}
A control insertion has cost
\begin{equation}\label{eq:patt-binomial-control-insertion}
\begin{aligned}
&A_{n,\mathbf h,l}(q;\mathcal C)\\
&=\sum_{i\in\mathcal C:Z_i=1}w_{A,1}(\mathbf X_i)
 \sum_{j\in\mathcal J^{\mathcal C\cup\{q\}}_{\mathcal D_{n,\mathbf h,l}}(i)}
 \frac{w_{A,0}(\mathbf X_j)\zeta_{A,j}}
 {\displaystyle\sum_{k\in\mathcal J^{\mathcal C\cup\{q\}}_{\mathcal D_{n,\mathbf h,l}}(i)}
                         w_{A,0}(\mathbf X_k)}\\
&\quad-\sum_{i\in\mathcal C:Z_i=1}w_{A,1}(\mathbf X_i)
 \sum_{j\in\mathcal J^{\mathcal C}_{\mathcal D_{n,\mathbf h,l}}(i)}
 \frac{w_{A,0}(\mathbf X_j)\zeta_{A,j}}
 {\displaystyle\sum_{k\in\mathcal J^{\mathcal C}_{\mathcal D_{n,\mathbf h,l}}(i)}
                         w_{A,0}(\mathbf X_k)}.
\end{aligned}
\end{equation}
When the base has fewer than \(M\) controls, we use \eqref{eq:patt-binomial-addone-definition} with the specified zero values.

The jointly Borel matching rules make the conditional integrals below well defined. We use outer probability and expectation for suprema over compact parameter sets.

\par\smallskip\noindent
\emph{Localization and moment bounds.}\quad Let \(d_i=\mathbf D_0^\ast(\mathbf X_i)\). The uniform displacement bound \eqref{eq:patt-poisson-bounded-score-displacement} gives a constant \(C_K\) such that each score moves by at most \(C_K/\sqrt n\). Let \(R_M(i;\mathcal C)\) be the baseline \(M\)-th control radius of a treated query, with value \(\infty\) if controls are insufficient. If a control insertion is selected by query \(i\) for some \((\mathbf h,l)\in K\), comparison with the old baseline \(M\)-donor set gives
\begin{equation}\label{eq:patt-binomial-incoming-radius}
\begin{aligned}
\|d_i-d_q\|
&\le\|\mathcal D_{n,\mathbf h,l}(\mathbf X_i)
             -\mathcal D_{n,\mathbf h,l}(\mathbf X_q)\|
                    +2C_K/\sqrt n\\
&\le R_M(i;\mathcal C)+4C_K/\sqrt n.
\end{aligned}
\end{equation}
To bound the number of affected queries uniformly over the local parameters, we define
\begin{equation}\label{eq:patt-binomial-incoming-counts}
\begin{aligned}
N(q;\mathcal C)
&:=\sum_{i\in\mathcal C:Z_i=1}
 \mathbf1\!\left\{R_M(i;\mathcal C)
       \ge(\|d_i-d_q\|-4C_K/\sqrt n)_+\right\},\\
N_L(q;\mathcal C)
&:=\sum_{i\in\mathcal C:Z_i=1}
 \mathbf1\!\left\{R_M(i;\mathcal C)
       \ge(\|d_i-d_q\|-4C_K/\sqrt n)_+\right\}\\
&\hspace{27mm}{}\times
              \mathbf1\{\sqrt n\|d_i-d_q\|>L\}.
\end{aligned}
\end{equation}
With fixed priority, every changed query selects the new control and is counted by \(N\).

For every fixed integer \(r\ge1\), uniformly in \(b/n\in[1/2,2]\) and almost every insertion location,
\begin{equation}\label{eq:patt-binomial-incoming-count-moments}
E_s\{(1+N)^r\mid d_q\}\le C_r,\qquad
E_s(N_L^r\mid d_q)\le C_r e^{-c_rL^2}.
\end{equation}
Here \(L\) ranges from a fixed constant to \(C\sqrt n\). To obtain these moment bounds, we condition on \(r\) distinct treated queries and let \(r_{\mathrm{match}}=\max_j\|d_j-d_q\|\). If all their incoming indicators are one, the farthest query's \((r_{\mathrm{match}}-4C_K/\sqrt n)_+\)-ball contains fewer than \(M\) controls. The remaining \(b-r\) rows are iid. The ball-mass lower bound and the fixed-\(M\) binomial tail bound this event by \(C_r e^{-c_rnr_{\mathrm{match}}^2}\), enlarging \(C_r\) for \(r_{\mathrm{match}}\le8C_K/\sqrt n\). Thus
\begin{equation}\label{eq:patt-binomial-farthest-query-integral}
E_s\{(N)_r\mid d_q\}
\le C_r(b)_r\int_{S^r}
 e^{-c_rn\max_j\|d_j-d_q\|^2}\,dd_1\cdots dd_r
\le C_r(b)_rn^{-r}\le C_r.
\end{equation}
The change of variables \(v_j=\sqrt n(d_j-d_q)\) gives the second inequality. Restricting the integral to \(\max_j\|v_j\|>L\) gives the tail bound. Expressing ordinary moments in terms of factorial moments then proves \eqref{eq:patt-binomial-incoming-count-moments}.

Let \(G_L\) be the event that \(N_L=0\), every old treated query within \(L/\sqrt n\) of \(q\) has baseline \(M\)-th control radius at most \(L/\sqrt n\), and \(q\) has \(M\) base controls within that radius. The latter requirement is also imposed on a control insertion. The incoming and deficient-ball bounds give
\begin{equation}\label{eq:patt-binomial-localization-event}
P_s(G_L^c)\le C(1+L)^{c_1}e^{-c_2L^2}.
\end{equation}
For nearby queries, we integrate the \(O(L^2/n)\) probability of lying in the neighborhood and bound the probability that the remaining observations supply fewer than \(M\) controls.

On \(G_L\), every affected query and every donor in its denominator lies within \((2L+4C_K)/\sqrt n\) of \(q\). We define
\begin{equation}\label{eq:patt-binomial-local-envelope}
\begin{aligned}
J_u(q;\mathcal C)&=\sum_{j\in\mathcal C}\mathbf1\{\sqrt n\|d_j-d_q\|\le u\},\\
F_u(q;\mathcal C)&=C\{1+J_u(q;\mathcal C)\}
 \sum_{\substack{j\in\mathcal C\cup\{q\}\\
                  \sqrt n\|d_j-d_q\|\le u}}(1+|\zeta_{A,j}|).
\end{aligned}
\end{equation}
The exact donor fractions are nonnegative and sum to one, so \(\sup_K|A_{n,\mathbf h,l}|\le F_{2L+4C_K}\) on \(G_L\). The integrated moment bound is
\begin{equation}\label{eq:patt-binomial-integrated-envelope-moment}
E_sF_u^p\le C_p(1+u)^{c_p}E_s(1+|\zeta_A|)^p.
\end{equation}
The power-of-a-sum inequality bounds \(F_u^p\) by \(C_p(1+J_u)^{2p-1}\) times the sum of the \(p\)-th powers of its marks. Conditional on the marked row and insertion, the remaining count has moments polynomial in \(1+u\). For the base marks,
\begin{equation}\label{eq:patt-binomial-mark-integration}
\begin{aligned}
\sum_{j=1}^bE_s[\mathbf1_{\{\sqrt n\|d_j-d_q\|\le u\}}
                   (1+|\zeta_{A,j}|)^p]
&=bE_s[(1+|\zeta_{A,1}|)^p
 P_s\{\sqrt n\|d_q-d_1\|\le u\mid O_1\}]\\
&\le C(1+u)^2E_s(1+|\zeta_A|)^p.
\end{aligned}
\end{equation}
The last bound uses the baseline density bound and \(b\le2n\).

Let \(q_0=(p+2)/2\in(2,p)\). We partition by the first integer \(L\ge L_0\) for which \(G_L\) holds. The envelope on insufficient-control samples is \(C(b+1)\sum_{j\in\mathcal C\cup\{q\}}(1+|\zeta_{A,j}|)\), with \(p\)-th moment \(O(n^{2p})\). H\"older's inequality gives
\[
\begin{aligned}
E_s\sup_K|A_{n,\mathbf h,l}|^{q_0}
&\le C+\sum_{L>L_0}(E_sF_{2L+4C_K}^p)^{q_0/p}
               P_s(G_{L-1}^c)^{1-q_0/p}
       +Cn^{2q_0}e^{-cn(1-q_0/p)}\\
&\le C+\sum_{L>L_0}C(1+L)^{c_3}e^{-c_4L^2}
       +Cn^{2q_0}e^{-cn(1-q_0/p)}\\
&\le C.
\end{aligned}
\]
Consequently,
\begin{equation}\label{eq:patt-binomial-full-addone-sup-moment}
\sup_{n,\ b/n\in[1/2,2]}
 E_s[\sup_K|A_{n,\mathbf h,l}(q;\mathcal C)|^{q_0}]<\infty .
\end{equation}

For \(R>8C_K+2\), let \(A^R_{n,\mathbf h,l}\) be the exact add-one difference when each query is restricted to baseline distance \(R/\sqrt n\), using the same normalized fractions and setting its contribution to zero if fewer than \(M\) controls lie in that ball. With fixed row-index priority this cost depends only on rows within \(2R/\sqrt n\) of the insertion. Consider the event that \(N_R=0\) and at least \(M\) base controls lie within \((R-4C_K-1)/\sqrt n\) of \(q\) and every old query within \(R/\sqrt n\) of \(q\). The complementary probability is at most \(C(1+R)^{c_1}e^{-c_2R^2}\). On this event, \eqref{eq:patt-binomial-incoming-radius} gives \(A=A^R\) on \(K\): all affected queries and both sets of denominator donors lie in their restriction windows. The truncated cost has envelope \(F_{2R}\). Applying H\"older's inequality with its polynomial \(p\)-moment and the unrestricted \(q_0\)-moment yields, for \(r=1,2\),
\begin{equation}\label{eq:patt-binomial-addone-radius-tail}
\limsup_{n\to\infty}\sup_{b/n\in[1/2,2]}
 E_s[\sup_K|A_{n,\mathbf h,l}-A^R_{n,\mathbf h,l}|^r]\le C_r e^{-c_rR^2}.
\end{equation}

\emph{Conditional mean of the insertion cost.} For an independent insertion \(Q\), we define
\begin{equation}\label{eq:patt-binomial-conditional-addone-mean}
a^R_{n,\mathbf h,l}(\mathcal C)
=\int A^R_{n,\mathbf h,l}(q;\mathcal C)\,P_s(dq).
\end{equation}
Suppose \(b/n\to1\) and fix \((\mathbf h,l)\). The one- and two-root marked Laplace-functional argument in the proof of Proposition~\ref{prop:primitive-residual-qv-limit}, starting from \eqref{eq:qv-marked-local-laplace}, applies with the enlarged row mark \((S_0,Z,w_{A,Z},\zeta_A)\). In a bounded baseline window about \(d\),
\begin{equation}\label{eq:patt-binomial-local-pair-displacement}
\begin{aligned}
\sqrt n\{\mathcal D_{n,\mathbf h,l}(\mathbf X)
                   -\mathcal D_{n,\mathbf h,l}(\mathbf X_{\rm root})\}
&=\sqrt n\{\mathbf D_0^\ast(\mathbf X)-d\}\\
&\quad+\{S_0(\mathbf X)-S_0(\mathbf X_{\rm root})\}\mathbf h+o(1).
\end{aligned}
\end{equation}
The extra diagonal contribution is \(O(n^{-1/2})\). Since \(b/n\to1\), the limiting intensity is unchanged. The truncated add-one functional is continuous outside null tie and window-boundary sets, with denominators at least \(Mw_{\min}\). The \(p\)-moment of \(F_{2R}\), \(p>2\), gives uniform integrability of the cost and its two-root product. For independent insertions \(Q,Q'\), conditioning on the entire base gives
\begin{equation}\label{eq:patt-binomial-fixed-index-mean-limit}
\begin{aligned}
E_sa^R_{n,\mathbf h,l}(\mathcal C)&\longrightarrow a_R(\mathbf h,l),\\
E_s\{a^R_{n,\mathbf h,l}(\mathcal C)^2\}
&=E_s\{A^R_{n,\mathbf h,l}(Q;\mathcal C)A^R_{n,\mathbf h,l}(Q';\mathcal C)\}\\
&\longrightarrow a_R(\mathbf h,l)^2.
\end{aligned}
\end{equation}
These two moment limits imply \(a^R_{n,\mathbf h,l}(\mathcal C)\to a_R(\mathbf h,l)\) in \(L^2\).

For an anchor \(u=(\mathbf h,l)\) and \(0<\eta<1\), let \(\mathcal N^R_{n,u,\eta}\) collect near-ties before and after insertion. This event includes the new treated query and old queries within \(R/\sqrt n\) of the insertion with candidates within \(2R/\sqrt n\). Squared rescaled distances change by at most \(C_R\eta\). With \(D^R_{n,u,\eta}=2F_{2R}\mathbf1_{\mathcal N^R_{n,u,\eta}}\),
\begin{equation}\label{eq:patt-binomial-addone-oscillation-envelope}
\sup_{\substack{u'\in K\\\|u'-u\|\le\eta}}
 |A^R_{n,u'}(q;\mathcal C)-A^R_{n,u}(q;\mathcal C)|
\le D^R_{n,u,\eta}(q;\mathcal C).
\end{equation}

By \eqref{eq:patt-poisson-block-near-tie-probability}, triples involving the insertion contribute at most \(C_R(b)_2n^{-2}\eta\). Triples in the base contribute \(C_R(b)_3n^{-3}\eta\) after integrating the insertion's nearness. These bounds use the current donor density for the annulus and the baseline density for nearness. Hence
\begin{equation}\label{eq:patt-binomial-addone-neartie-probability}
\begin{aligned}
P_s\{\mathcal N^R_{n,\mathbf h,l,\eta}(Q;\mathcal C)\}
&\le C_R\{(b)_2n^{-2}+(b)_3n^{-3}\}\eta\\
&\le C_R\eta.
\end{aligned}
\end{equation}
The integrated \(p\)-moment bound and H\"older's inequality give
\begin{equation}\label{eq:patt-binomial-addone-oscillation-moments}
\begin{aligned}
E_s\{D^R_{n,\mathbf h,l,\eta}(Q;\mathcal C)^r\}
&\le2^r(E_sF_{2R}^p)^{r/p}
       P_s\{\mathcal N^R_{n,\mathbf h,l,\eta}(Q;\mathcal C)\}^{1-r/p}\\
&\le C_R\eta^{1-r/p},\qquad r=1,2.
\end{aligned}
\end{equation}

For fixed \(\eta>0\), the same one-/two-window argument applies to \(D^R\): its threshold boundary is a donor circle of probability zero, and the local \(p\)-moment gives product uniform integrability. With \(d^R_{n,\mathbf h,l,\eta}(\mathcal C)=\int D^R_{n,\mathbf h,l,\eta}(q;\mathcal C)P_s(dq)\), this gives
\begin{equation}\label{eq:patt-binomial-envelope-mean-limit}
d^R_{n,\mathbf h,l,\eta}(\mathcal C)\longrightarrow_{L^2}d_R(\mathbf h,l,\eta),
\qquad 0\le d_R(\mathbf h,l,\eta)\le C_R\eta^{1-1/p}.
\end{equation}
The first-moment bound gives \(|a_R(u)-a_R(u')|\le C_R\|u-u'\|^{1-1/p}\). For the finite nets \(G_k\) from the preceding proof, with \(\eta=2^{-k}\), we obtain
\[
\begin{aligned}
\sup_{u\in K}|a^R_{n,u}(\mathcal C)-a_R(u)|
&\le\max_{v\in G_k}|a^R_{n,v}(\mathcal C)-a_R(v)|
 +\max_{v\in G_k}d^R_{n,v,\eta}(\mathcal C)
 +C_R\eta^{1-1/p}\\
&\le o_p(1)+2C_R\eta^{1-1/p}.
\end{aligned}
\]
Taking \(n\to\infty\), then \(\eta\downarrow0\), gives
\begin{equation}\label{eq:patt-binomial-uniform-conditional-mean}
\sup_{(\mathbf h,l)\in K}|a^R_{n,\mathbf h,l}(\mathcal C)-a_R(\mathbf h,l)|
\longrightarrow_p0\qquad(b/n\to1).
\end{equation}

We denote the enlarged conditional mark law by
\[
\mathsf K_{A,z\mid0}^\ast(d,\cdot)
:=\mathcal L_s\!\left\{
\bigl(S_0(\mathbf X),Z,w_{A,Z}(\mathbf X),\zeta_A\bigr)
\mid Z=z,\mathbf D_0^\ast(\mathbf X)=d
\right\}.
\]
We write \(\boldsymbol\eta=(s,z,w,\zeta)\) for these variables and \(s_{\mathrm{root}}\) for the root's score derivative. In the full homogeneous limiting field, the mark-dependent translations in \eqref{eq:patt-binomial-local-pair-displacement} preserve each treatment color's intensity:
\begin{equation}\label{eq:patt-binomial-mark-translation}
\int\!\int f\{y+(s-s_{\mathrm{root}})\mathbf h,z,\boldsymbol\eta\}\,dy\,
 \mathsf K_{A,z\mid0}^\ast(d,d\boldsymbol\eta)
=\int\!\int f(u,z,\boldsymbol\eta)\,du\,
 \mathsf K_{A,z\mid0}^\ast(d,d\boldsymbol\eta).
\end{equation}
Thus the full add-one limit \(A_A^\infty\) has the baseline law for every \((\mathbf h,l)\), with the root's conditional mark law at \(d\).

Conditional on the limiting-score locations, treatments and survey weights, the add-one cost is linear in centered residuals by \eqref{eq:patt-binomial-baseline-control-centering}. We first condition on bounded localization events determined by the geometry. The \(L^1\) bound in \eqref{eq:patt-binomial-addone-radius-tail} then permits removal of this restriction and gives \(E(A_A^\infty\mid d)=0\) for almost every \(d\). The one-window limit and the same tail yield
\begin{equation}\label{eq:patt-binomial-full-addone-mean-zero}
E_sA_{n,\mathbf h,l}(Q;\mathcal C)\longrightarrow0
\qquad\text{for fixed }(\mathbf h,l),\quad b/n\to1.
\end{equation}
Combining with the fixed-radius mean limit gives
\begin{equation}\label{eq:patt-binomial-cutoff-mean-bound}
\sup_{(\mathbf h,l)\in K}|a_R(\mathbf h,l)|\le Ce^{-cR^2},
\end{equation}
because the full-versus-cutoff \(L^1\) bound is uniform on \(K\).

\emph{Comparison of sample sizes.} We now insert iid \(Q_1,\ldots,Q_k\), independently of \(\mathcal C_b\), where \(b/n\to1\), \(k\to\infty\), and \(k\le C\sqrt n\). The common-base costs are conditionally iid, with identical priority relative to the base. Their integrated second-moment bound gives
\begin{equation}\label{eq:patt-binomial-commonbase-variance}
\begin{aligned}
E_s\left|\frac1k\sum_{j=1}^k
 \{A^R_{n,\mathbf h,l}(Q_j;\mathcal C_b)-a^R_{n,\mathbf h,l}(\mathcal C_b)\}
 \right|^2
&=\frac1kE_s\operatorname{Var}_s
                   \{A^R_{n,\mathbf h,l}(Q_1;\mathcal C_b)\mid\mathcal C_b\}\\
&\le\frac1kE_s|A^R_{n,\mathbf h,l}(Q_1;\mathcal C_b)|^2\\
&\le C_R/k.
\end{aligned}
\end{equation}
We apply the same variance bound to \(D^R\) on a finite \(\eta\)-net. The limit of its conditional integral bounds the off-grid error by \(C_R\eta^{1-1/p}\). The uniform conditional-mean law and \(\eta\downarrow0\) give
\begin{equation}\label{eq:patt-binomial-commonbase-uniform-law}
\sup_{(\mathbf h,l)\in K}
 \left|\frac1k\sum_{j=1}^k A^R_{n,\mathbf h,l}(Q_j;\mathcal C_b)
                  -a_R(\mathbf h,l)\right|\longrightarrow_p0.
\end{equation}

A truncated cost uses only rows within \(2R/\sqrt n\) of its insertion. If \(Q_j\) has no other insertion within \(4R/\sqrt n\), its successive and common-base costs agree on \(K\). The contrary event \(\mathcal H_j\) has probability at most \(C_Rk/n\). Both costs are bounded by a local envelope in the augmented iid cloud, which contains at most \(2n\) other rows. H\"older's inequality gives
\begin{equation}\label{eq:patt-binomial-insertion-interaction}
\begin{aligned}
&E_s^*\sup_K\frac1{\sqrt n}
 \left|\sum_{j=1}^k A^R_{n,\mathbf h,l}
   (Q_j;\mathcal C_b\cup\{Q_a:a<j\})
       -\sum_{j=1}^k A^R_{n,\mathbf h,l}(Q_j;\mathcal C_b)\right|\\
&\le C_R\frac{k}{\sqrt n}(k/n)^{1-1/p}\longrightarrow0 .
\end{aligned}
\end{equation}
Every intermediate base is iid, so the radius-tail bound gives
\begin{equation}\label{eq:patt-binomial-successive-radius-tail}
\begin{aligned}
&\limsup_n E_s^*\sup_K\frac1{\sqrt n}
 \left|\sum_{j=1}^k(A_{n,\mathbf h,l}-A^R_{n,\mathbf h,l})
       (Q_j;\mathcal C_b\cup\{Q_a:a<j\})\right|\\
&\le C\limsup_n(k/\sqrt n)e^{-cR^2}.
\end{aligned}
\end{equation}
The common-base law, the bound on \(\sup_K|a_R|\), and then \(R\to\infty\) prove
\begin{equation}\label{eq:patt-binomial-close-count-comparison}
\sup_K\frac{|\mathcal I_{A,b+k;n}-\mathcal I_{A,b;n}|}{\sqrt n}
\longrightarrow_p0
\qquad(b/n\to1,\ k\le C\sqrt n).
\end{equation}
For \(k<n^{1/4}\), the full add-one supremum moment instead bounds the expected error by \(Ck/\sqrt n\). For larger \(k\), the preceding argument applies. If the comparison were not uniform over these integer sequences, a violating subsequence would contradict the same bounds. Thus it is uniform over integers \(|b-n|\le C\sqrt n\) and \(0\le k\le C\sqrt n\).

We condition on \(N_n\) and take \(b=\min(N_n,n)\), \(k=|N_n-n|\). The uniform comparison and tightness of \((N_n-n)/\sqrt n\) imply de-Poissonization in probability.

For \(N_n\in[n/2,2n]\), the power-of-a-sum inequality and the full add-one moment give
\begin{equation}\label{eq:patt-binomial-count-difference-moment}
E\!\left[
 \left.\left\{\sup_K\frac{|\mathcal I_{A,N_n;n}
                    -\mathcal I_{A,n;n}|}{\sqrt n}\right\}^{q_0}
 \right|N_n\right]
\le C\{|N_n-n|/\sqrt n\}^{q_0}.
\end{equation}
The centered Poisson moment satisfies \(E|N_n-n|^{q_0}=O(n^{q_0/2})\). For an exceptional count \(N_n=m\notin[n/2,2n]\), a common \(\max(m,n)\)-row envelope bounds both numerators:
\begin{equation}\label{eq:patt-binomial-exceptional-count-envelope}
\begin{aligned}
\sup_K\frac{|\mathcal I_{A,m;n}-\mathcal I_{A,n;n}|}{\sqrt n}
&\le\frac{C(m+n+1)}{\sqrt n}
                  \sum_{i=1}^{\max(m,n)}(1+|\zeta_{A,i}|),\\
E\!\left[
 \left.\left\{\sup_K\frac{|\mathcal I_{A,N_n;n}
                   -\mathcal I_{A,n;n}|}{\sqrt n}\right\}^{q_0}
 \right|N_n=m\right]
&\le Cn^{-q_0/2}(m+n+1)^{2q_0}.
\end{aligned}
\end{equation}
The exponential count tail and these \(q_0\)-moment bounds give uniform integrability. The convergence in probability therefore also holds in \(L^1\). Consequently, for the fixed-priority rule,
\begin{equation}\label{eq:patt-binomial-center-triangle}
\sup_K\frac{|E\mathcal I_{A,N_n;n}-E\mathcal I_{A,n;n}|}{\sqrt n}
\le E^*\sup_K\frac{|\mathcal I_{A,N_n;n}
                         -\mathcal I_{A,n;n}|}{\sqrt n}
\longrightarrow0.
\end{equation}

The comparison \eqref{eq:patt-binomial-tie-rule-transfer} and the triangle inequality give
\[
\begin{aligned}
&E^*\sup_K\frac{|\mathcal I^{\rm rule}_{A,N_n;n}
                         -\mathcal I^{\rm rule}_{A,n;n}|}{\sqrt n}\\
&\le E^*\sup_K\frac{|\mathcal I^{\rm rule}_{A,N_n;n}
                         -\mathcal I^{\rm priority}_{A,N_n;n}|}{\sqrt n}\\
&\quad+E^*\sup_K\frac{|\mathcal I^{\rm priority}_{A,N_n;n}
                         -\mathcal I^{\rm priority}_{A,n;n}|}{\sqrt n}\\
&\quad+E^*\sup_K\frac{|\mathcal I^{\rm priority}_{A,n;n}
                         -\mathcal I^{\rm rule}_{A,n;n}|}{\sqrt n}\\
&\longrightarrow0.
\end{aligned}
\]
This proves de-Poissonization for the specified rule. Its expectation comparison follows from the same inequality as \eqref{eq:patt-binomial-center-triangle}.

Centering each numerator at the expectation for its sample size gives
\begin{equation}\label{eq:patt-binomial-centered-process-transfer}
E^*\sup_K|\mathcal Z^B_{A,n}-\mathcal Z^P_{A,n}|
\le2E^*\sup_K
 \frac{|\mathcal I_{A,n;n}-\mathcal I_{A,N_n;n}|}{\sqrt n}
\longrightarrow0 .
\end{equation}
Each fixed-size increment is bounded by the Poisson increment plus twice this supremum error. The Poisson modulus \eqref{eq:patt-poisson-full-process-modulus} therefore proves \eqref{eq:patt-binomial-centered-equicontinuity}.

The same comparison also gives
\begin{equation}
\label{eq:patt-poisson-binomial-increment-interface}
\begin{aligned}
&\lim_{\eta\downarrow0}\limsup_n
 P^\ast\Bigg\{
 \sup_{\substack{(\mathbf h,l),(\mathbf h',l')\in K\\
                  \|(\mathbf h,l)-(\mathbf h',l')\|\le\eta}}
 \frac1{\sqrt n}\Big|
 \{\mathcal I_{A,N_n;n}(\mathbf h,l)-\mathcal I_{A,n;n}(\mathbf h,l)\}\\
&\quad
 -\{\mathcal I_{A,N_n;n}(\mathbf h',l')
              -\mathcal I_{A,n;n}(\mathbf h',l')\}
 \Big|>\varepsilon\Bigg\}=0.
\end{aligned}
\end{equation}
Indeed, the supremum is bounded by twice \(\sup_K|\mathcal I_{A,N_n;n}-\mathcal I_{A,n;n}|/\sqrt n\), whose \(L^1\) norm tends to zero. Markov's inequality completes the proof.
\end{proof}

\clearpage
\section{Additional simulation results}\label{app:simulation-results}
\noindent Web Tables~{\let\enspace\relax\ref{tab:retrospective_good}--\ref{tab:prospective_poor}} report signed relative bias, relative efficiency, and coverage for all matching ratios considered in the simulation study.\par\medskip

\paragraph{{Simulation implementation.}}
{For WDSM, the matching coordinates were the fitted propensity probability and the corresponding arm-specific prognostic prediction. The propensity probability was neither clipped nor logit-transformed. Each coordinate was centered and scaled over the pooled sample using unweighted means and variances with divisor \(n\). The fixed six-term quadratic basis in Section~\ref{app:growing-series-bc} was fitted by survey-weighted least squares within each required treatment arm and evaluated for all sampled individuals. PATE required both arm-specific fits, whereas PATT required only the control-arm fit.}

{Each bootstrap replicate drew multinomial counts for individual observations, refitted the propensity, prognostic, and bias-correction models, and recomputed pooled scaling using these counts. The counts multiplied the corresponding original fitting weights. The original matching sets and incoming matching coefficients remained fixed. Nonpositive score scales, invalid or rank-deficient fits, and nonfinite predictions or replications were treated as numerical failures. Failed replications were not discarded when calculating variance. The saved simulation summaries contained successful results for all 1,000 simulated samples in each method and scenario, with no recorded numerical or sampling failures.}

{With 1,000 simulation replications, the Monte Carlo standard error of estimated coverage is approximately 0.69 percentage points at 95\% coverage and at most 1.58 percentage points. Under correct specification of both score models, the ratio of the average estimated standard error to the empirical standard deviation ranged from 0.944 to 1.015 across the WDSM configurations.}

\paragraph{{Tie handling and random seeds.}}
{For each recipient, donors strictly closer than the \(M\)th-distance boundary were retained. If the boundary group contained more donors than required, the remaining donors were selected uniformly without replacement from that group. The numerical tolerance for a tie was \(64\epsilon_{\mathrm{mach}}\max(1,d_{(M)}^2)\), where \(\epsilon_{\mathrm{mach}}\) is machine precision and \(d_{(M)}^2\) is the \(M\)th smallest squared Euclidean distance. This tie selection for each recipient used a separate random number stream, restored after matching. For simulation replication \(r\), the sampling, WDSM replication, and tie seeds were \(2025+r\), \(2000000+100r+10+M\), and \(3000000+100r+M\), respectively. The original tie allocations were retained during fixed-match replication.}

{The simulations and case studies also examine Gaussian covariates, clustered prospective sampling, and discrete score distributions beyond the sufficient conditions of the asymptotic theory.}

\begin{sidewaystable}[htbp]
\centering
\captionsetup{singlelinecheck=false,width=\linewidth}
\caption{Simulation results under retrospective sampling with good overlap ($\psi = 0.6$): relative bias (\%), relative efficiency, and coverage of 95\% Wald confidence intervals across 1,000 replications.}
\label{tab:retrospective_good}
\footnotesize
\renewcommand{\arraystretch}{1.15}
\setlength{\tabcolsep}{3.5pt}
\begin{tabular}{@{} >{\centering\arraybackslash}p{1.2cm} l c c *{4}{c} *{4}{c} *{4}{c} @{}}
\toprule
\multirow{2}{*}{Estimand} & \multirow{2}{*}{Method} & \multicolumn{2}{c}{{\shortstack{Survey-weighted\\score fit}}} & \multicolumn{4}{c}{Relative Bias (\%)} & \multicolumn{4}{c}{Relative Efficiency} & \multicolumn{4}{c}{Coverage (\%)} \\
\cmidrule(lr){3-4} \cmidrule(lr){5-8} \cmidrule(lr){9-12} \cmidrule(lr){13-16}
& & PS & PGS & Cor|Cor & Mis|Cor & Cor|Mis & Mis|Mis & Cor|Cor & Mis|Cor & Cor|Mis & Mis|Mis & Cor|Cor & Mis|Cor & Cor|Mis & Mis|Mis \\
\midrule
\multirow{7}{*}{PATE}
& PSM & \texttimes & \textendash & 6.4 & 14.7 & 6.4 & 14.7 & 1.00 & 0.85 & 1.00 & 0.85 & 97.0 & 84.0 & 97.0 & 84.0 \\[2pt]
& \cellcolor[rgb]{0.92,0.92,0.92}PGM & \cellcolor[rgb]{0.92,0.92,0.92}\textendash & \cellcolor[rgb]{0.92,0.92,0.92}\texttimes & \cellcolor[rgb]{0.92,0.92,0.92}6.3 & \cellcolor[rgb]{0.92,0.92,0.92}6.3 & \cellcolor[rgb]{0.92,0.92,0.92}11.4 & \cellcolor[rgb]{0.92,0.92,0.92}11.4 & \cellcolor[rgb]{0.92,0.92,0.92}2.60 & \cellcolor[rgb]{0.92,0.92,0.92}2.60 & \cellcolor[rgb]{0.92,0.92,0.92}1.82 & \cellcolor[rgb]{0.92,0.92,0.92}1.82 & \cellcolor[rgb]{0.92,0.92,0.92}81.2 & \cellcolor[rgb]{0.92,0.92,0.92}81.2 & \cellcolor[rgb]{0.92,0.92,0.92}68.5 & \cellcolor[rgb]{0.92,0.92,0.92}68.5 \\[2pt]
& DSM & \texttimes & \texttimes & 6.4 & 6.4 & 6.6 & 10.5 & 2.35 & 2.34 & 2.11 & 1.85 & 81.6 & 81.5 & 87.6 & 70.9 \\[2pt]
& \cellcolor[rgb]{0.92,0.92,0.92}SW-PSM & \cellcolor[rgb]{0.92,0.92,0.92}\checkmark & \cellcolor[rgb]{0.92,0.92,0.92}\textendash & \cellcolor[rgb]{0.92,0.92,0.92}$-$0.1 & \cellcolor[rgb]{0.92,0.92,0.92}$-$0.4 & \cellcolor[rgb]{0.92,0.92,0.92}$-$0.1 & \cellcolor[rgb]{0.92,0.92,0.92}$-$0.4 & \cellcolor[rgb]{0.92,0.92,0.92}1.75 & \cellcolor[rgb]{0.92,0.92,0.92}1.70 & \cellcolor[rgb]{0.92,0.92,0.92}1.75 & \cellcolor[rgb]{0.92,0.92,0.92}1.70 & \cellcolor[rgb]{0.92,0.92,0.92}93.2 & \cellcolor[rgb]{0.92,0.92,0.92}92.6 & \cellcolor[rgb]{0.92,0.92,0.92}93.2 & \cellcolor[rgb]{0.92,0.92,0.92}92.6 \\[2pt]
& WDSM ($M$=1) & \checkmark & \texttimes & $-$0.2 & 0.0 & $-$2.7 & 4.6 & 1.85 & 1.82 & 1.51 & 1.17 & 94.2 & 94.5 & 96.0 & 91.5 \\[2pt]
& WDSM ($M$=3) & \checkmark & \texttimes & $-$0.2 & $-$0.1 & $-$0.8 & 5.9 & 2.26 & 2.23 & 1.84 & 1.36 & 94.3 & 94.7 & 96.2 & 89.1 \\[2pt]
& WDSM ($M$=5) & \checkmark & \texttimes & $-$0.2 & $-$0.1 & $-$0.4 & 6.3 & 2.33 & 2.30 & 1.93 & 1.42 & 94.6 & 94.3 & 96.7 & 87.9 \\
\midrule
\multirow{7}{*}{PATT}
& PSM & \texttimes & \textendash & 6.0 & 11.4 & 6.0 & 11.4 & 1.00 & 0.86 & 1.00 & 0.86 & 96.1 & 83.5 & 96.1 & 83.5 \\[2pt]
& \cellcolor[rgb]{0.92,0.92,0.92}PGM & \cellcolor[rgb]{0.92,0.92,0.92}\textendash & \cellcolor[rgb]{0.92,0.92,0.92}\texttimes & \cellcolor[rgb]{0.92,0.92,0.92}5.8 & \cellcolor[rgb]{0.92,0.92,0.92}5.8 & \cellcolor[rgb]{0.92,0.92,0.92}12.9 & \cellcolor[rgb]{0.92,0.92,0.92}12.9 & \cellcolor[rgb]{0.92,0.92,0.92}1.98 & \cellcolor[rgb]{0.92,0.92,0.92}1.98 & \cellcolor[rgb]{0.92,0.92,0.92}1.32 & \cellcolor[rgb]{0.92,0.92,0.92}1.32 & \cellcolor[rgb]{0.92,0.92,0.92}80.3 & \cellcolor[rgb]{0.92,0.92,0.92}80.3 & \cellcolor[rgb]{0.92,0.92,0.92}53.1 & \cellcolor[rgb]{0.92,0.92,0.92}53.1 \\[2pt]
& DSM & \texttimes & \texttimes & 5.7 & 5.7 & 5.6 & 9.8 & 1.89 & 1.93 & 1.46 & 1.31 & 81.2 & 80.4 & 87.2 & 69.9 \\[2pt]
& \cellcolor[rgb]{0.92,0.92,0.92}SW-PSM & \cellcolor[rgb]{0.92,0.92,0.92}\checkmark & \cellcolor[rgb]{0.92,0.92,0.92}\textendash & \cellcolor[rgb]{0.92,0.92,0.92}$-$0.2 & \cellcolor[rgb]{0.92,0.92,0.92}$-$0.4 & \cellcolor[rgb]{0.92,0.92,0.92}$-$0.2 & \cellcolor[rgb]{0.92,0.92,0.92}$-$0.4 & \cellcolor[rgb]{0.92,0.92,0.92}2.19 & \cellcolor[rgb]{0.92,0.92,0.92}2.25 & \cellcolor[rgb]{0.92,0.92,0.92}2.19 & \cellcolor[rgb]{0.92,0.92,0.92}2.25 & \cellcolor[rgb]{0.92,0.92,0.92}90.9 & \cellcolor[rgb]{0.92,0.92,0.92}89.0 & \cellcolor[rgb]{0.92,0.92,0.92}90.9 & \cellcolor[rgb]{0.92,0.92,0.92}89.0 \\[2pt]
& WDSM ($M$=1) & \checkmark & \texttimes & $-$0.2 & $-$0.2 & 0.2 & 3.6 & 1.88 & 1.84 & 1.57 & 1.40 & 93.7 & 94.0 & 96.2 & 91.0 \\[2pt]
& WDSM ($M$=3) & \checkmark & \texttimes & $-$0.3 & $-$0.2 & 0.0 & 3.9 & 2.29 & 2.25 & 1.98 & 1.59 & 94.1 & 94.2 & 96.8 & 88.6 \\[2pt]
& WDSM ($M$=5) & \checkmark & \texttimes & $-$0.3 & $-$0.3 & $-$0.1 & 4.0 & 2.38 & 2.35 & 2.09 & 1.67 & 94.9 & 94.3 & 97.0 & 88.7 \\
\bottomrule
\end{tabular}
\vspace{0.5em}
\begin{minipage}{\textwidth}
\footnotesize
\textbf{Note:} PS = propensity score model, PGS = prognostic score model. In the PS and PGS columns, $\checkmark$ = survey-weighted estimation, $\times$ = unweighted estimation, \textendash\, = score not used by the method. Prognostic scores are estimated without survey weights for all methods, as described in the main manuscript. WDSM additionally incorporates survey weights in within-match imputation, bias correction, and H\'ajek normalization. Model specifications: Cor|Cor = both PS and PGS correctly specified, Mis|Cor = PS misspecified with PGS correct, Cor|Mis = PS correct with PGS misspecified, Mis|Mis = both misspecified. Correctly specified models include all six main effects plus the $X_{1i}X_{2i}$ interaction, while misspecified models omit this interaction. Relative efficiency is the ratio of the Monte Carlo variance of PSM under Cor|Cor to that of each method. PSM = propensity score matching, PGM = prognostic score matching, DSM = double score matching, SW-PSM = survey-weighted propensity score full matching, WDSM = weighted double score matching. PSM, PGM, and DSM use $M = 1$ nearest neighbor matching with replacement via the \texttt{dsmatch} package. SW-PSM uses optimal full matching via \texttt{MatchIt} with a correctly specified treatment-interacted outcome regression in all four scenarios. WDSM results are shown for $M \in \{1, 3, 5\}$ nearest neighbors with replacement. WDSM uses survey-weighted bias-correction regression with linear, squared, and interaction terms in the standardized double scores, and linearization-based bootstrap variance estimation with divisor $B$. PSM, PGM, and DSM use bootstrap sample variances with divisor $B-1$, and SW-PSM uses delta method standard errors based on a sandwich covariance matrix clustered by matched subclass. Bootstrap procedures use $B=200$, and all confidence intervals are centered normal Wald intervals.
\end{minipage}
\end{sidewaystable}

\begin{sidewaystable}[htbp]
\centering
\captionsetup{singlelinecheck=false,width=\linewidth}
\caption{Simulation results under retrospective sampling with poor overlap ($\psi = 2.0$): relative bias (\%), relative efficiency, and coverage of 95\% Wald confidence intervals across 1,000 replications.}
\label{tab:retrospective_poor}
\footnotesize
\renewcommand{\arraystretch}{1.15}
\setlength{\tabcolsep}{3.5pt}
\begin{tabular}{@{} >{\centering\arraybackslash}p{1.2cm} l c c *{4}{c} *{4}{c} *{4}{c} @{}}
\toprule
\multirow{2}{*}{Estimand} & \multirow{2}{*}{Method} & \multicolumn{2}{c}{{\shortstack{Survey-weighted\\score fit}}} & \multicolumn{4}{c}{Relative Bias (\%)} & \multicolumn{4}{c}{Relative Efficiency} & \multicolumn{4}{c}{Coverage (\%)} \\
\cmidrule(lr){3-4} \cmidrule(lr){5-8} \cmidrule(lr){9-12} \cmidrule(lr){13-16}
& & PS & PGS & Cor|Cor & Mis|Cor & Cor|Mis & Mis|Mis & Cor|Cor & Mis|Cor & Cor|Mis & Mis|Mis & Cor|Cor & Mis|Cor & Cor|Mis & Mis|Mis \\
\midrule
\multirow{7}{*}{PATE}
& PSM & \texttimes & \textendash & 10.5 & 38.1 & 10.5 & 38.1 & 1.00 & 1.06 & 1.00 & 1.06 & 94.8 & 78.4 & 94.8 & 78.4 \\[2pt]
& \cellcolor[rgb]{0.92,0.92,0.92}PGM & \cellcolor[rgb]{0.92,0.92,0.92}\textendash & \cellcolor[rgb]{0.92,0.92,0.92}\texttimes & \cellcolor[rgb]{0.92,0.92,0.92}6.1 & \cellcolor[rgb]{0.92,0.92,0.92}6.1 & \cellcolor[rgb]{0.92,0.92,0.92}22.7 & \cellcolor[rgb]{0.92,0.92,0.92}22.7 & \cellcolor[rgb]{0.92,0.92,0.92}22.01 & \cellcolor[rgb]{0.92,0.92,0.92}22.01 & \cellcolor[rgb]{0.92,0.92,0.92}13.67 & \cellcolor[rgb]{0.92,0.92,0.92}13.67 & \cellcolor[rgb]{0.92,0.92,0.92}87.9 & \cellcolor[rgb]{0.92,0.92,0.92}87.9 & \cellcolor[rgb]{0.92,0.92,0.92}39.1 & \cellcolor[rgb]{0.92,0.92,0.92}39.1 \\[2pt]
& DSM & \texttimes & \texttimes & 5.8 & 6.0 & 6.1 & 16.9 & 7.90 & 7.93 & 6.98 & 6.58 & 93.3 & 93.0 & 94.2 & 79.5 \\[2pt]
& \cellcolor[rgb]{0.92,0.92,0.92}SW-PSM & \cellcolor[rgb]{0.92,0.92,0.92}\checkmark & \cellcolor[rgb]{0.92,0.92,0.92}\textendash & \cellcolor[rgb]{0.92,0.92,0.92}$-$1.0 & \cellcolor[rgb]{0.92,0.92,0.92}$-$1.2 & \cellcolor[rgb]{0.92,0.92,0.92}$-$1.0 & \cellcolor[rgb]{0.92,0.92,0.92}$-$1.2 & \cellcolor[rgb]{0.92,0.92,0.92}3.22 & \cellcolor[rgb]{0.92,0.92,0.92}3.18 & \cellcolor[rgb]{0.92,0.92,0.92}3.22 & \cellcolor[rgb]{0.92,0.92,0.92}3.18 & \cellcolor[rgb]{0.92,0.92,0.92}81.3 & \cellcolor[rgb]{0.92,0.92,0.92}83.8 & \cellcolor[rgb]{0.92,0.92,0.92}81.3 & \cellcolor[rgb]{0.92,0.92,0.92}83.8 \\[2pt]
& WDSM ($M$=1) & \checkmark & \texttimes & 0.3 & 0.6 & $-$0.6 & 24.2 & 6.67 & 6.73 & 4.43 & 3.48 & 95.4 & 94.6 & 95.4 & 77.5 \\[2pt]
& WDSM ($M$=3) & \checkmark & \texttimes & 0.7 & 0.7 & 2.6 & 26.3 & 8.69 & 8.70 & 5.95 & 4.42 & 94.4 & 94.9 & 95.8 & 67.9 \\[2pt]
& WDSM ($M$=5) & \checkmark & \texttimes & 0.9 & 0.9 & 3.8 & 26.7 & 9.25 & 9.32 & 6.57 & 4.88 & 94.1 & 94.4 & 95.5 & 63.6 \\
\midrule
\multirow{7}{*}{PATT}
& PSM & \texttimes & \textendash & 4.5 & 21.0 & 4.5 & 21.0 & 1.00 & 0.91 & 1.00 & 0.91 & 96.1 & 80.8 & 96.1 & 80.8 \\[2pt]
& \cellcolor[rgb]{0.92,0.92,0.92}PGM & \cellcolor[rgb]{0.92,0.92,0.92}\textendash & \cellcolor[rgb]{0.92,0.92,0.92}\texttimes & \cellcolor[rgb]{0.92,0.92,0.92}3.4 & \cellcolor[rgb]{0.92,0.92,0.92}3.4 & \cellcolor[rgb]{0.92,0.92,0.92}22.6 & \cellcolor[rgb]{0.92,0.92,0.92}22.6 & \cellcolor[rgb]{0.92,0.92,0.92}14.11 & \cellcolor[rgb]{0.92,0.92,0.92}14.11 & \cellcolor[rgb]{0.92,0.92,0.92}8.96 & \cellcolor[rgb]{0.92,0.92,0.92}8.96 & \cellcolor[rgb]{0.92,0.92,0.92}90.1 & \cellcolor[rgb]{0.92,0.92,0.92}90.1 & \cellcolor[rgb]{0.92,0.92,0.92}7.1 & \cellcolor[rgb]{0.92,0.92,0.92}7.1 \\[2pt]
& DSM & \texttimes & \texttimes & 3.6 & 3.5 & 4.7 & 17.5 & 5.02 & 5.30 & 3.33 & 3.46 & 91.3 & 92.8 & 91.6 & 57.5 \\[2pt]
& \cellcolor[rgb]{0.92,0.92,0.92}SW-PSM & \cellcolor[rgb]{0.92,0.92,0.92}\checkmark & \cellcolor[rgb]{0.92,0.92,0.92}\textendash & \cellcolor[rgb]{0.92,0.92,0.92}$-$0.6 & \cellcolor[rgb]{0.92,0.92,0.92}$-$0.8 & \cellcolor[rgb]{0.92,0.92,0.92}$-$0.6 & \cellcolor[rgb]{0.92,0.92,0.92}$-$0.8 & \cellcolor[rgb]{0.92,0.92,0.92}4.47 & \cellcolor[rgb]{0.92,0.92,0.92}4.72 & \cellcolor[rgb]{0.92,0.92,0.92}4.47 & \cellcolor[rgb]{0.92,0.92,0.92}4.72 & \cellcolor[rgb]{0.92,0.92,0.92}85.4 & \cellcolor[rgb]{0.92,0.92,0.92}86.9 & \cellcolor[rgb]{0.92,0.92,0.92}85.4 & \cellcolor[rgb]{0.92,0.92,0.92}86.9 \\[2pt]
& WDSM ($M$=1) & \checkmark & \texttimes & 0.2 & 0.1 & 2.0 & 13.0 & 7.66 & 7.62 & 5.13 & 5.00 & 94.8 & 94.5 & 95.1 & 68.3 \\[2pt]
& WDSM ($M$=3) & \checkmark & \texttimes & 0.3 & 0.2 & 2.6 & 13.7 & 9.41 & 9.25 & 6.67 & 6.48 & 94.8 & 93.8 & 95.1 & 58.2 \\[2pt]
& WDSM ($M$=5) & \checkmark & \texttimes & 0.3 & 0.1 & 2.8 & 13.9 & 9.86 & 9.69 & 6.98 & 6.86 & 94.9 & 94.2 & 94.8 & 54.7 \\
\bottomrule
\end{tabular}
\vspace{0.5em}
\begin{minipage}{\textwidth}
\footnotesize
\textbf{Note:} PS = propensity score model, PGS = prognostic score model. In the PS and PGS columns, $\checkmark$ = survey-weighted estimation, $\times$ = unweighted estimation, \textendash\, = score not used by the method. Prognostic scores are estimated without survey weights for all methods, as described in the main manuscript. WDSM additionally incorporates survey weights in within-match imputation, bias correction, and H\'ajek normalization. Model specifications: Cor|Cor = both PS and PGS correctly specified, Mis|Cor = PS misspecified with PGS correct, Cor|Mis = PS correct with PGS misspecified, Mis|Mis = both misspecified. Correctly specified models include all six main effects plus the $X_{1i}X_{2i}$ interaction, while misspecified models omit this interaction. Relative efficiency is the ratio of the Monte Carlo variance of PSM under Cor|Cor to that of each method. PSM = propensity score matching, PGM = prognostic score matching, DSM = double score matching, SW-PSM = survey-weighted propensity score full matching, WDSM = weighted double score matching. PSM, PGM, and DSM use $M = 1$ nearest neighbor matching with replacement via the \texttt{dsmatch} package. SW-PSM uses optimal full matching via \texttt{MatchIt} with a correctly specified treatment-interacted outcome regression in all four scenarios. WDSM results are shown for $M \in \{1, 3, 5\}$ nearest neighbors with replacement. WDSM uses survey-weighted bias-correction regression with linear, squared, and interaction terms in the standardized double scores, and linearization-based bootstrap variance estimation with divisor $B$. PSM, PGM, and DSM use bootstrap sample variances with divisor $B-1$, and SW-PSM uses delta method standard errors based on a sandwich covariance matrix clustered by matched subclass. Bootstrap procedures use $B=200$, and all confidence intervals are centered normal Wald intervals.
\end{minipage}
\end{sidewaystable}

\begin{sidewaystable}[htbp]
\centering
\captionsetup{singlelinecheck=false,width=\linewidth}
\caption{Simulation results under prospective sampling with good overlap ($\psi = 0.6$): relative bias (\%), relative efficiency, and coverage of 95\% Wald confidence intervals across 1,000 replications.}
\label{tab:prospective_good}
\footnotesize
\renewcommand{\arraystretch}{1.15}
\setlength{\tabcolsep}{3.5pt}
\begin{tabular}{@{} >{\centering\arraybackslash}p{1.2cm} l c c *{4}{c} *{4}{c} *{4}{c} @{}}
\toprule
\multirow{2}{*}{Estimand} & \multirow{2}{*}{Method} & \multicolumn{2}{c}{{\shortstack{Survey-weighted\\score fit}}} & \multicolumn{4}{c}{Relative Bias (\%)} & \multicolumn{4}{c}{Relative Efficiency} & \multicolumn{4}{c}{Coverage (\%)} \\
\cmidrule(lr){3-4} \cmidrule(lr){5-8} \cmidrule(lr){9-12} \cmidrule(lr){13-16}
& & PS & PGS & Cor|Cor & Mis|Cor & Cor|Mis & Mis|Mis & Cor|Cor & Mis|Cor & Cor|Mis & Mis|Mis & Cor|Cor & Mis|Cor & Cor|Mis & Mis|Mis \\
\midrule
\multirow{7}{*}{PATE}
& PSM & \texttimes & \textendash & 6.4 & 14.6 & 6.4 & 14.6 & 1.00 & 0.87 & 1.00 & 0.87 & 95.8 & 82.4 & 95.8 & 82.4 \\[2pt]
& \cellcolor[rgb]{0.92,0.92,0.92}PGM & \cellcolor[rgb]{0.92,0.92,0.92}\textendash & \cellcolor[rgb]{0.92,0.92,0.92}\texttimes & \cellcolor[rgb]{0.92,0.92,0.92}5.9 & \cellcolor[rgb]{0.92,0.92,0.92}5.9 & \cellcolor[rgb]{0.92,0.92,0.92}11.4 & \cellcolor[rgb]{0.92,0.92,0.92}11.4 & \cellcolor[rgb]{0.92,0.92,0.92}2.35 & \cellcolor[rgb]{0.92,0.92,0.92}2.35 & \cellcolor[rgb]{0.92,0.92,0.92}1.78 & \cellcolor[rgb]{0.92,0.92,0.92}1.78 & \cellcolor[rgb]{0.92,0.92,0.92}81.3 & \cellcolor[rgb]{0.92,0.92,0.92}81.3 & \cellcolor[rgb]{0.92,0.92,0.92}67.6 & \cellcolor[rgb]{0.92,0.92,0.92}67.6 \\[2pt]
& DSM & \texttimes & \texttimes & 5.9 & 5.8 & 6.3 & 10.5 & 2.30 & 2.32 & 2.00 & 1.76 & 81.8 & 83.2 & 86.0 & 72.1 \\[2pt]
& \cellcolor[rgb]{0.92,0.92,0.92}SW-PSM & \cellcolor[rgb]{0.92,0.92,0.92}\checkmark & \cellcolor[rgb]{0.92,0.92,0.92}\textendash & \cellcolor[rgb]{0.92,0.92,0.92}$-$0.6 & \cellcolor[rgb]{0.92,0.92,0.92}$-$0.7 & \cellcolor[rgb]{0.92,0.92,0.92}$-$0.6 & \cellcolor[rgb]{0.92,0.92,0.92}$-$0.7 & \cellcolor[rgb]{0.92,0.92,0.92}1.89 & \cellcolor[rgb]{0.92,0.92,0.92}1.88 & \cellcolor[rgb]{0.92,0.92,0.92}1.89 & \cellcolor[rgb]{0.92,0.92,0.92}1.88 & \cellcolor[rgb]{0.92,0.92,0.92}92.5 & \cellcolor[rgb]{0.92,0.92,0.92}91.7 & \cellcolor[rgb]{0.92,0.92,0.92}92.5 & \cellcolor[rgb]{0.92,0.92,0.92}91.7 \\[2pt]
& WDSM ($M$=1) & \texttimes & \texttimes & $-$0.7 & $-$0.8 & $-$1.2 & 5.8 & 2.12 & 2.13 & 1.65 & 1.34 & 92.6 & 93.1 & 95.9 & 88.5 \\[2pt]
& WDSM ($M$=3) & \texttimes & \texttimes & $-$0.6 & $-$0.6 & $-$0.4 & 6.4 & 2.30 & 2.31 & 1.92 & 1.48 & 92.2 & 92.9 & 96.3 & 87.9 \\[2pt]
& WDSM ($M$=5) & \texttimes & \texttimes & $-$0.6 & $-$0.6 & $-$0.3 & 6.4 & 2.37 & 2.36 & 1.92 & 1.49 & 93.2 & 93.3 & 96.4 & 87.2 \\
\midrule
\multirow{7}{*}{PATT}
& PSM & \texttimes & \textendash & 5.9 & 11.4 & 5.9 & 11.4 & 1.00 & 0.81 & 1.00 & 0.81 & 95.7 & 82.8 & 95.7 & 82.8 \\[2pt]
& \cellcolor[rgb]{0.92,0.92,0.92}PGM & \cellcolor[rgb]{0.92,0.92,0.92}\textendash & \cellcolor[rgb]{0.92,0.92,0.92}\texttimes & \cellcolor[rgb]{0.92,0.92,0.92}5.4 & \cellcolor[rgb]{0.92,0.92,0.92}5.4 & \cellcolor[rgb]{0.92,0.92,0.92}12.8 & \cellcolor[rgb]{0.92,0.92,0.92}12.8 & \cellcolor[rgb]{0.92,0.92,0.92}1.73 & \cellcolor[rgb]{0.92,0.92,0.92}1.73 & \cellcolor[rgb]{0.92,0.92,0.92}1.15 & \cellcolor[rgb]{0.92,0.92,0.92}1.15 & \cellcolor[rgb]{0.92,0.92,0.92}80.7 & \cellcolor[rgb]{0.92,0.92,0.92}80.7 & \cellcolor[rgb]{0.92,0.92,0.92}52.7 & \cellcolor[rgb]{0.92,0.92,0.92}52.7 \\[2pt]
& DSM & \texttimes & \texttimes & 5.6 & 5.5 & 5.4 & 10.0 & 1.74 & 1.72 & 1.43 & 1.14 & 81.4 & 80.8 & 89.1 & 67.1 \\[2pt]
& \cellcolor[rgb]{0.92,0.92,0.92}SW-PSM & \cellcolor[rgb]{0.92,0.92,0.92}\checkmark & \cellcolor[rgb]{0.92,0.92,0.92}\textendash & \cellcolor[rgb]{0.92,0.92,0.92}$-$0.7 & \cellcolor[rgb]{0.92,0.92,0.92}$-$0.6 & \cellcolor[rgb]{0.92,0.92,0.92}$-$0.7 & \cellcolor[rgb]{0.92,0.92,0.92}$-$0.6 & \cellcolor[rgb]{0.92,0.92,0.92}1.57 & \cellcolor[rgb]{0.92,0.92,0.92}1.57 & \cellcolor[rgb]{0.92,0.92,0.92}1.57 & \cellcolor[rgb]{0.92,0.92,0.92}1.57 & \cellcolor[rgb]{0.92,0.92,0.92}90.5 & \cellcolor[rgb]{0.92,0.92,0.92}90.3 & \cellcolor[rgb]{0.92,0.92,0.92}90.5 & \cellcolor[rgb]{0.92,0.92,0.92}90.3 \\[2pt]
& WDSM ($M$=1) & \texttimes & \texttimes & $-$0.5 & $-$0.5 & $-$2.1 & 2.5 & 1.36 & 1.34 & 1.15 & 0.97 & 93.4 & 93.4 & 96.7 & 93.1 \\[2pt]
& WDSM ($M$=3) & \texttimes & \texttimes & $-$0.6 & $-$0.6 & $-$1.3 & 3.0 & 1.58 & 1.58 & 1.35 & 1.15 & 93.7 & 94.0 & 95.8 & 92.0 \\[2pt]
& WDSM ($M$=5) & \texttimes & \texttimes & $-$0.6 & $-$0.7 & $-$1.0 & 3.2 & 1.62 & 1.64 & 1.42 & 1.18 & 93.8 & 94.5 & 95.9 & 90.6 \\
\bottomrule
\end{tabular}
\vspace{0.5em}
\begin{minipage}{\textwidth}
\footnotesize
\textbf{Note:} PS = propensity score model, PGS = prognostic score model. In the PS and PGS columns, $\checkmark$ = survey-weighted estimation, $\times$ = unweighted estimation, \textendash\, = score not used by the method. Prognostic scores are estimated without survey weights for all methods, as described in the main manuscript. WDSM additionally incorporates survey weights in within-match imputation, bias correction, and H\'ajek normalization. Model specifications: Cor|Cor = both PS and PGS correctly specified, Mis|Cor = PS misspecified with PGS correct, Cor|Mis = PS correct with PGS misspecified, Mis|Mis = both misspecified. Correctly specified models include all six main effects plus the $X_{1i}X_{2i}$ interaction, while misspecified models omit this interaction. Relative efficiency is the ratio of the Monte Carlo variance of PSM under Cor|Cor to that of each method. PSM = propensity score matching, PGM = prognostic score matching, DSM = double score matching, SW-PSM = survey-weighted propensity score full matching, WDSM = weighted double score matching. PSM, PGM, and DSM use $M = 1$ nearest neighbor matching with replacement via the \texttt{dsmatch} package. SW-PSM uses optimal full matching via \texttt{MatchIt} with a correctly specified treatment-interacted outcome regression in all four scenarios. WDSM results are shown for $M \in \{1, 3, 5\}$ nearest neighbors with replacement. WDSM uses survey-weighted bias-correction regression with linear, squared, and interaction terms in the standardized double scores, and linearization-based bootstrap variance estimation with divisor $B$. PSM, PGM, and DSM use bootstrap sample variances with divisor $B-1$, and SW-PSM uses delta method standard errors based on a sandwich covariance matrix clustered by matched subclass. Bootstrap procedures use $B=200$, and all confidence intervals are centered normal Wald intervals.
\end{minipage}
\end{sidewaystable}

\begin{sidewaystable}[htbp]
\centering
\captionsetup{singlelinecheck=false,width=\linewidth}
\caption{Simulation results under prospective sampling with poor overlap ($\psi = 2.0$): relative bias (\%), relative efficiency, and coverage of 95\% Wald confidence intervals across 1,000 replications.}
\label{tab:prospective_poor}
\footnotesize
\renewcommand{\arraystretch}{1.15}
\setlength{\tabcolsep}{3.5pt}
\begin{tabular}{@{} >{\centering\arraybackslash}p{1.2cm} l c c *{4}{c} *{4}{c} *{4}{c} @{}}
\toprule
\multirow{2}{*}{Estimand} & \multirow{2}{*}{Method} & \multicolumn{2}{c}{{\shortstack{Survey-weighted\\score fit}}} & \multicolumn{4}{c}{Relative Bias (\%)} & \multicolumn{4}{c}{Relative Efficiency} & \multicolumn{4}{c}{Coverage (\%)} \\
\cmidrule(lr){3-4} \cmidrule(lr){5-8} \cmidrule(lr){9-12} \cmidrule(lr){13-16}
& & PS & PGS & Cor|Cor & Mis|Cor & Cor|Mis & Mis|Mis & Cor|Cor & Mis|Cor & Cor|Mis & Mis|Mis & Cor|Cor & Mis|Cor & Cor|Mis & Mis|Mis \\
\midrule
\multirow{7}{*}{PATE}
& PSM & \texttimes & \textendash & 10.6 & 39.7 & 10.6 & 39.7 & 1.00 & 0.91 & 1.00 & 0.91 & 96.3 & 82.5 & 96.3 & 82.5 \\[2pt]
& \cellcolor[rgb]{0.92,0.92,0.92}PGM & \cellcolor[rgb]{0.92,0.92,0.92}\textendash & \cellcolor[rgb]{0.92,0.92,0.92}\texttimes & \cellcolor[rgb]{0.92,0.92,0.92}5.6 & \cellcolor[rgb]{0.92,0.92,0.92}5.6 & \cellcolor[rgb]{0.92,0.92,0.92}23.4 & \cellcolor[rgb]{0.92,0.92,0.92}23.4 & \cellcolor[rgb]{0.92,0.92,0.92}21.74 & \cellcolor[rgb]{0.92,0.92,0.92}21.74 & \cellcolor[rgb]{0.92,0.92,0.92}12.82 & \cellcolor[rgb]{0.92,0.92,0.92}12.82 & \cellcolor[rgb]{0.92,0.92,0.92}89.1 & \cellcolor[rgb]{0.92,0.92,0.92}89.1 & \cellcolor[rgb]{0.92,0.92,0.92}48.9 & \cellcolor[rgb]{0.92,0.92,0.92}48.9 \\[2pt]
& DSM & \texttimes & \texttimes & 5.3 & 5.4 & 5.6 & 17.3 & 7.70 & 7.71 & 6.84 & 6.34 & 92.0 & 92.1 & 93.4 & 82.4 \\[2pt]
& \cellcolor[rgb]{0.92,0.92,0.92}SW-PSM & \cellcolor[rgb]{0.92,0.92,0.92}\checkmark & \cellcolor[rgb]{0.92,0.92,0.92}\textendash & \cellcolor[rgb]{0.92,0.92,0.92}$-$1.5 & \cellcolor[rgb]{0.92,0.92,0.92}$-$1.1 & \cellcolor[rgb]{0.92,0.92,0.92}$-$1.5 & \cellcolor[rgb]{0.92,0.92,0.92}$-$1.1 & \cellcolor[rgb]{0.92,0.92,0.92}4.57 & \cellcolor[rgb]{0.92,0.92,0.92}4.42 & \cellcolor[rgb]{0.92,0.92,0.92}4.57 & \cellcolor[rgb]{0.92,0.92,0.92}4.42 & \cellcolor[rgb]{0.92,0.92,0.92}82.8 & \cellcolor[rgb]{0.92,0.92,0.92}83.3 & \cellcolor[rgb]{0.92,0.92,0.92}82.8 & \cellcolor[rgb]{0.92,0.92,0.92}83.3 \\[2pt]
& WDSM ($M$=1) & \texttimes & \texttimes & $-$1.0 & $-$1.2 & $-$0.6 & 23.9 & 9.28 & 9.11 & 6.49 & 5.45 & 94.5 & 94.5 & 96.5 & 74.9 \\[2pt]
& WDSM ($M$=3) & \texttimes & \texttimes & $-$0.3 & $-$0.4 & 1.8 & 25.5 & 10.97 & 10.83 & 7.90 & 6.22 & 94.0 & 94.5 & 96.2 & 66.9 \\[2pt]
& WDSM ($M$=5) & \texttimes & \texttimes & 0.0 & $-$0.1 & 2.7 & 25.8 & 11.66 & 11.61 & 8.34 & 6.43 & 94.5 & 94.7 & 96.2 & 65.1 \\
\midrule
\multirow{7}{*}{PATT}
& PSM & \texttimes & \textendash & 6.7 & 24.3 & 6.7 & 24.3 & 1.00 & 0.90 & 1.00 & 0.90 & 95.7 & 71.5 & 95.7 & 71.5 \\[2pt]
& \cellcolor[rgb]{0.92,0.92,0.92}PGM & \cellcolor[rgb]{0.92,0.92,0.92}\textendash & \cellcolor[rgb]{0.92,0.92,0.92}\texttimes & \cellcolor[rgb]{0.92,0.92,0.92}5.0 & \cellcolor[rgb]{0.92,0.92,0.92}5.0 & \cellcolor[rgb]{0.92,0.92,0.92}24.0 & \cellcolor[rgb]{0.92,0.92,0.92}24.0 & \cellcolor[rgb]{0.92,0.92,0.92}11.62 & \cellcolor[rgb]{0.92,0.92,0.92}11.62 & \cellcolor[rgb]{0.92,0.92,0.92}8.02 & \cellcolor[rgb]{0.92,0.92,0.92}8.02 & \cellcolor[rgb]{0.92,0.92,0.92}81.9 & \cellcolor[rgb]{0.92,0.92,0.92}81.9 & \cellcolor[rgb]{0.92,0.92,0.92}5.2 & \cellcolor[rgb]{0.92,0.92,0.92}5.2 \\[2pt]
& DSM & \texttimes & \texttimes & 4.8 & 4.7 & 5.5 & 18.7 & 4.50 & 4.64 & 2.94 & 3.01 & 89.2 & 90.2 & 89.2 & 54.2 \\[2pt]
& \cellcolor[rgb]{0.92,0.92,0.92}SW-PSM & \cellcolor[rgb]{0.92,0.92,0.92}\checkmark & \cellcolor[rgb]{0.92,0.92,0.92}\textendash & \cellcolor[rgb]{0.92,0.92,0.92}$-$1.2 & \cellcolor[rgb]{0.92,0.92,0.92}$-$1.0 & \cellcolor[rgb]{0.92,0.92,0.92}$-$1.2 & \cellcolor[rgb]{0.92,0.92,0.92}$-$1.0 & \cellcolor[rgb]{0.92,0.92,0.92}2.96 & \cellcolor[rgb]{0.92,0.92,0.92}2.91 & \cellcolor[rgb]{0.92,0.92,0.92}2.96 & \cellcolor[rgb]{0.92,0.92,0.92}2.91 & \cellcolor[rgb]{0.92,0.92,0.92}84.0 & \cellcolor[rgb]{0.92,0.92,0.92}81.6 & \cellcolor[rgb]{0.92,0.92,0.92}84.0 & \cellcolor[rgb]{0.92,0.92,0.92}81.6 \\[2pt]
& WDSM ($M$=1) & \texttimes & \texttimes & $-$0.7 & $-$0.6 & $-$0.8 & 11.2 & 4.89 & 4.90 & 3.12 & 3.14 & 93.8 & 93.4 & 95.5 & 78.2 \\[2pt]
& WDSM ($M$=3) & \texttimes & \texttimes & $-$0.4 & $-$0.5 & 0.9 & 12.3 & 5.57 & 5.55 & 3.75 & 3.69 & 93.9 & 94.3 & 94.1 & 71.2 \\[2pt]
& WDSM ($M$=5) & \texttimes & \texttimes & $-$0.3 & $-$0.4 & 1.6 & 12.6 & 5.75 & 5.65 & 3.88 & 3.79 & 93.1 & 93.5 & 93.4 & 70.4 \\
\bottomrule
\end{tabular}
\vspace{0.5em}
\begin{minipage}{\textwidth}
\footnotesize
\textbf{Note:} PS = propensity score model, PGS = prognostic score model. In the PS and PGS columns, $\checkmark$ = survey-weighted estimation, $\times$ = unweighted estimation, \textendash\, = score not used by the method. Prognostic scores are estimated without survey weights for all methods, as described in the main manuscript. WDSM additionally incorporates survey weights in within-match imputation, bias correction, and H\'ajek normalization. Model specifications: Cor|Cor = both PS and PGS correctly specified, Mis|Cor = PS misspecified with PGS correct, Cor|Mis = PS correct with PGS misspecified, Mis|Mis = both misspecified. Correctly specified models include all six main effects plus the $X_{1i}X_{2i}$ interaction, while misspecified models omit this interaction. Relative efficiency is the ratio of the Monte Carlo variance of PSM under Cor|Cor to that of each method. PSM = propensity score matching, PGM = prognostic score matching, DSM = double score matching, SW-PSM = survey-weighted propensity score full matching, WDSM = weighted double score matching. PSM, PGM, and DSM use $M = 1$ nearest neighbor matching with replacement via the \texttt{dsmatch} package. SW-PSM uses optimal full matching via \texttt{MatchIt} with a correctly specified treatment-interacted outcome regression in all four scenarios. WDSM results are shown for $M \in \{1, 3, 5\}$ nearest neighbors with replacement. WDSM uses survey-weighted bias-correction regression with linear, squared, and interaction terms in the standardized double scores, and linearization-based bootstrap variance estimation with divisor $B$. PSM, PGM, and DSM use bootstrap sample variances with divisor $B-1$, and SW-PSM uses delta method standard errors based on a sandwich covariance matrix clustered by matched subclass. Bootstrap procedures use $B=200$, and all confidence intervals are centered normal Wald intervals.
\end{minipage}
\end{sidewaystable}

\clearpage
\section{{Additional case-study details}}\label{app:case-study-details}

\paragraph{{Sample construction.}}
{The NSDUH analysis retained all 17,471 respondents aged 18--29 in the released data. Analysis variables included SAMHSA's edited or imputed values. No additional analyst imputation was performed. Table~\ref{tab:nhanes-flow} gives the sequential NHANES exclusions. Missing analysis inputs were not imputed, and no additional item-nonresponse weight adjustment was applied. The sums of retained survey weights were 51,794,162.90 for NSDUH and 91,637,975.56 for NHANES.}

\begin{table}[!htbp]
\centering
\caption{{Sequential construction of the NHANES analytic sample.}}
\label{tab:nhanes-flow}
\fontsize{9}{11}\selectfont
{\begin{tabular}{@{}p{0.69\textwidth}rr@{}}
\toprule
Stage & Remaining & Removed \\
\midrule
2015--2018 participants aged 40--70 & 5,814 & --- \\
No reported diabetes diagnosis or borderline diabetes
(\texttt{DIQ010}=2 or 3) & 4,714 & 1,100 \\
Exclude recorded pregnancies (\texttt{RIDEXPRG}=1) & 4,706 & 8 \\
Observed sleep duration & 4,672 & 34 \\
Observed HbA1c & 4,304 & 368 \\
Observed education & 4,301 & 3 \\
Observed poverty-income ratio & 3,818 & 483 \\
Observed BMI & 3,784 & 34 \\
Observed smoking variable & 3,783 & 1 \\
\bottomrule
\end{tabular}}
\par\smallskip
\begin{minipage}{\textwidth}
\footnotesize
{Exclusion counts are sequential and do not overlap. Age, sex,
race, and positive examination weights caused no additional exclusions
after these stages. Among the 4,304 participants with observed sleep
duration and HbA1c, 521 (12.1\%) lacked at least one covariate.
Undetermined or unavailable pregnancy status was retained. Included
participants were not all confirmed nonpregnant.}
\end{minipage}
\end{table}

\paragraph{{Tie allocation and sensitivity.}}
{The case studies used the recipient-specific tie rule described above. The primary tie seed was 20260922 for NSDUH (\(M=5\)); the NHANES tie seeds were 20260918, 20260920, and 20260922 for \(M=1,3,5\), respectively. The replication seed was 2024. For each NSDUH outcome, 20 additional tie allocations used seeds 620001--620020, \(M=5\), and 200 replication draws. The replication random number stream was held constant across tie allocations. Table~\ref{tab:nsduh-tie-sensitivity} summarizes the resulting algorithmic sensitivity.}

{In the primary NSDUH analyses, boundary ties required random allocation for 1,517 of 1,877 treated recipients (80.8\%) and 7,097 of 15,594 control recipients (45.5\%) for AMIPY, and for 1,521 treated recipients (81.0\%) and 6,669 control recipients (42.8\%) for SMIPY. These are recipient counts, not numbers of tied donors.}

\begin{table}[!htbp]
\centering
\caption{{NSDUH WDSM results across 20 additional tie allocations.}}
\label{tab:nsduh-tie-sensitivity}
\fontsize{9}{11}\selectfont
{\begin{tabular}{@{}lcc@{}}
\toprule
Outcome & Estimate range (percentage points) &
\shortstack{Allocations with\\95\% CI including zero} \\
\midrule
AMIPY & 6.94--7.40 & 5/20 \\
SMIPY & 1.44--1.82 & 20/20 \\
\bottomrule
\end{tabular}}
\par\smallskip
\begin{minipage}{\textwidth}
\footnotesize
{AMIPY and SMIPY denote past-year any mental illness and serious
mental illness, respectively. Ranges summarize estimates across tie
allocations. They are not confidence intervals.}
\end{minipage}
\end{table}

\Needspace{12\baselineskip}
\paragraph{{NHANES matching ratios.}}
{Table~\ref{tab:nhanes-matching-ratios} reports WDSM estimates for the same 3,783 participants at \(M=1,3,5\). Each analysis used 2,000 replication draws. Intervals are centered 95\% normal Wald intervals, and all effects are HbA1c percentage-point differences. Only one treated recipient required random tie allocation at \(M=1\). No recipient required it at \(M=3\) or \(M=5\).}

\begin{table}[!htbp]
\centering
\caption{{NHANES WDSM PATE estimates by matching ratio.}}
\label{tab:nhanes-matching-ratios}
\fontsize{9}{11}\selectfont
{\begin{tabular}{@{}rcc@{}}
\toprule
\(M\) & Estimate & 95\% confidence interval \\
\midrule
1 & 0.06155 & (0.00932, 0.11377) \\
3 & 0.06482 & (0.01121, 0.11843) \\
5 & 0.08213 & (0.02384, 0.14043) \\
\bottomrule
\end{tabular}}
\end{table}

\clearpage

\bibliography{references}

%% file: preamble.tex
\usepackage[T1]{fontenc}
\usepackage[utf8]{inputenc}
\usepackage{newunicodechar}
\newunicodechar{—}{---}
\newunicodechar{–}{--}
\usepackage[letterpaper,margin=1in]{geometry}
\usepackage{amsmath,amssymb,amsthm,mathtools}
\usepackage[not1,notextcomp,lcgreekalpha]{stix}
\usepackage{bm}
\usepackage[table]{xcolor}
\usepackage{graphicx,graphics,float,rotating}
\usepackage{multirow,multicol,makecell,booktabs,array,tabularx}
\usepackage{caption,subcaption}
\usepackage{moreverb,url,needspace,enumitem}
\usepackage[super,sort&compress]{natbib}
\usepackage{etoolbox,ragged2e}
\AtBeginEnvironment{description}{\RaggedRight}
\usepackage[hidelinks]{hyperref}
\newcommand{\doi}[1]{\href{https://doi.org/#1}{\nolinkurl{#1}}}
\setlist[enumerate,1]{label=(\roman*),ref=\roman*,leftmargin=*}
\newtheoremstyle{wdsm}{6pt}{6pt}{\normalfont}{} {\bfseries}{.}{5pt}{}
\theoremstyle{wdsm}
\newtheorem{assumption}{Assumption}
\newtheorem{theorem}{Theorem}
\newtheorem{lemma}{Lemma}

\newcommand{\cmark}{\raisebox{0.5pt}{\checkmark}}